\documentclass[12pt]{article}
\usepackage{epsfig}
\usepackage{graphicx}
\usepackage{axodraw}
\usepackage{a4}
\usepackage{latexsym}
\usepackage{cite}

\usepackage{pslatex}
\usepackage[latin1]{inputenc}
\usepackage[T1]{fontenc}

\usepackage{color,colordvi}
\def\colour4colour#1{\Blue{#1}}

\renewcommand{\theequation}{\thesection.\arabic{equation}}
\newcommand{\gsim}{\raisebox{-0.07cm}{$\:\:\stackrel{>}{{\scriptstyle
 \sim}}\:\: $} }
\newcommand{\lsim}{\raisebox{-0.07cm}{$\:\:\stackrel{<}{{\scriptstyle
 \sim}}\:\: $} }
\newcommand{\beq}{\begin{equation}}
\newcommand{\eeq}{\end{equation}}
\newcommand{\bea}{\begin{eqnarray}}
\newcommand{\eea}{\end{eqnarray}}
\newcommand{\nn}{\nonumber}
\newcommand{\MSb}{$\overline{\mbox{MS}}$}
\newcommand{\as}{\alpha_{\rm s}}
\newcommand{\ar}{a_{\rm s}}
\newcommand{\ra}{\rightarrow}
\newcommand{\DD}{{\cal D}}
\newcommand{\ep}{\epsilon}

\begin{document}
\setlength{\parskip}{0.3cm}
\setlength{\baselineskip}{0.55cm}

\def\sign(#1){{(-1)^{#1}}}
\def\pow(#1,#2){{{#1}^{#2}}}
\def\plus{{\!+\!}}
\def\minus{{\!-\!}}
\def\z#1{{\zeta_{#1}}}
\def\zs{{\zeta_{2}^{\,2}}}
\def\ca{{C^{}_A}}
\def\cf{{C^{}_F}}
\def\nf{{n^{}_{\! f}}}
\def\n2f{{n^{\,2}_{\! f}}}
\def\dabcnc{{{d^{abc}d_{abc}}\over{n_c}}}
\def\dabcNA{{{d^{abc}d_{abc}}\over{N_A}}}
\def\fl11{fl_{11}}
\def\flg11{fl^g_{11}}

\def\S(#1){{{S}_{#1}}}
\def\Ss(#1,#2){{{S}_{#1,#2}}}
\def\Sss(#1,#2,#3){{{S}_{#1,#2,#3}}}
\def\Ssss(#1,#2,#3,#4){{{S}_{#1,#2,#3,#4}}}
\def\Sssss(#1,#2,#3,#4,#5){{{S}_{#1,#2,#3,#4,#5}}}
\def\Ssssss(#1,#2,#3,#4,#5,#6){{{S}_{#1,#2,#3,#4,#5,#6}}}
\def\gqq{g_{\rm{qq}}}
\def\gqg{g_{\rm{qg}}}
\def\Npm{{{\bf N_{\pm}}}}
\def\Npmi{{{\bf N_{\pm i}}}}
\def\Nminus{{{\bf N_{-}}}}
\def\Nplus{{{\bf N_{+}}}}
\def\Nminustwo{{{\bf N_{-2}}}}
\def\Nplustwo{{{\bf N_{+2}}}}
\def\Nminusthree{{{\bf N_{-3}}}}
\def\Nplusthree{{{\bf N_{+3}}}}
\def\Nminusfour{{{\bf N_{-4}}}}
\def\Nplusfour{{{\bf N_{+4}}}}

\def\pqq(#1){p_{\rm{qq}}(#1)}
\def\pqg(#1){p_{\rm{qg}}(#1)}
\def\pgq(#1){p_{\rm{gq}}(#1)}
\def\pgg(#1){p_{\rm{gg}}(#1)}
\def\gfunct#1{{g}_{#1}}
\def\H(#1){{\rm{H}}_{#1}}
\def\Hh(#1,#2){{\rm{H}}_{#1,#2}}
\def\Hhh(#1,#2,#3){{\rm{H}}_{#1,#2,#3}}
\def\Hhhh(#1,#2,#3,#4){{\rm{H}}_{#1,#2,#3,#4}}
\def\Hhhhh(#1,#2,#3,#4,#5){{\rm{H}}_{#1,#2,#3,#4,#5}}

\def\gqqz{\gamma_{\,\rm qq}^{\,(0)}}
\def\gnso{\gamma_{\,\rm ns}^{\,(1)}}
\def\gpso{\gamma_{\,\rm ps}^{\,(1)}}
\def\gnst{\gamma_{\,\rm ns}^{\,(2)}}
\def\gpst{\gamma_{\,\rm ps}^{\,(2)}}

\def\gqiz{\gamma_{\,\rm qi}^{\,(0)}}
\def\gqpz{\gamma_{\,\rm qp}^{\,(0)}}
\def\gipz{\gamma_{\,\rm ip}^{\,(0)}}
\def\gikz{\gamma_{\,\rm ik}^{\,(0)}}
\def\gkpz{\gamma_{\,\rm kp}^{\,(0)}}
\def\gqio{\gamma_{\,\rm qi}^{\,(1)}}
\def\gqpo{\gamma_{\,\rm qp}^{\,(1)}}
\def\giko{\gamma_{\,\rm ik}^{\,(1)}}
\def\gipo{\gamma_{\,\rm ip}^{\,(1)}}
\def\gqpt{\gamma_{\,\rm qp}^{\,(2)}}

\def\ctqo{c_{2,\rm q}^{(1)}}
\def\ctnt{c_{2,\rm ns}^{(2)}}
\def\ctpt{c_{2,\rm p}^{(2)}}
\def\ctnd{c_{2,\rm ns}^{(3)}}
\def\ctpd{c_{2,\rm p}^{(3)}}
\def\ctio{c_{2,\rm i}^{(1)}}
\def\ctit{c_{2,\rm i}^{(2)}}

\def\atqo{a_{2,\rm q}^{(1)}}
\def\atio{a_{2,\rm i}^{(1)}}
\def\atnt{a_{2,\rm ns}^{(2)}}
\def\atpt{a_{2,\rm p}^{(2)}}
\def\atit{a_{2,\rm i}^{(2)}}

\def\btqo{b_{2,\rm q}^{(1)}}
\def\btio{b_{2,\rm i}^{(1)}}

\def\clqo{c_{L,\rm q}^{(1)}}
\def\clnt{c_{L,\rm ns}^{(2)}}
\def\clpt{c_{L,\rm p}^{(2)}}
\def\clnd{c_{L,\rm n}^{(3)}}
\def\clpd{c_{L,\rm p}^{(3)}}
\def\clio{c_{L,\rm i}^{(1)}}
\def\clit{c_{L,\rm i}^{(2)}}

\def\alqo{a_{L,\rm q}^{(1)}}
\def\alio{a_{L,\rm i}^{(1)}}
\def\alnt{a_{L,\rm ns}^{(2)}}
\def\alpt{a_{L,\rm p}^{(2)}}
\def\alit{a_{L,\rm i}^{(2)}}

\def\blqo{b_{L,\rm q}^{(1)}}
\def\blio{b_{L,\rm i}^{(1)}}

\begin{titlepage}
\noindent
NIKHEF 05-006 \hfill {\tt hep-ph/0504242}\\
DCPT/05/28,\quad$\!$IPPP/05/14 \\
DESY 05-063, SFB/CPP-05-13 \\
April 2005 \\
\vspace{1.3cm}
\begin{center}
\Large
{\bf The third-order QCD corrections to } \\
\vspace{0.15cm}
{\bf deep-inelastic scattering by photon exchange} \\
\vspace{1.5cm}
\large
J.A.M. Vermaseren$^{\, a}$, A. Vogt$^{\, b}$ and S. Moch$^{\, c}$\\
\vspace{1.2cm}
\normalsize
{\it $^a$NIKHEF Theory Group \\
\vspace{0.1cm}
Kruislaan 409, 1098 SJ Amsterdam, The Netherlands} \\
\vspace{0.5cm}
{\it $^b$IPPP, Department of Physics, University of Durham\\
\vspace{0.1cm}
South Road, Durham DH1 3LE, United Kingdom}\\
\vspace{0.5cm}
{\it $^c$Deutsches Elektronensynchrotron DESY \\
\vspace{0.1cm}
Platanenallee 6, D--15738 Zeuthen, Germany}\\
\vfill
\large
{\bf Abstract}
\vspace{-0.2cm}
\end{center}
We compute the full three-loop coefficient functions for the structure
functions $F_{\,2}$ and $F_L$ in massless perturbative QCD. The results
for $F_L$ complete the next-to-next-to-leading order description of 
unpolarized electromagnetic deep-inelastic scattering.
The third-order coefficient functions for $F_{\,2}$ form, at not too 
small values of the Bjorken variable $x$, the dominant part of the 
next-to-next-to-next-to-leading order corrections, thus facilitating
improved determinations of the strong coupling $\as$ from scaling 
violations.
The three-loop corrections to $F_L$ are larger than those for 
$F_{\,2}$. Especially for the latter quantity the expansion in powers
of $\as$ is very stable, for photon virtualities $Q^{\,2}
\gg 1 \mbox{ GeV}^2$, over the full $x$-range accessible to 
fixed-target and collider measurements.   
\vfill
\end{titlepage}
%
%
\setcounter{equation}{0}
\section{Introduction}
\label{sec:introduction}
%
%
Structure functions in deep-inelastic scattering (DIS) and their scale 
evolution are closely related to the origins of Quantum Chromodynamics 
(QCD) and its very formulation as the gauge theory of the strong 
interaction~\cite
{Gross:1973id,Politzer:1973fx,Gross:1973ju,Georgi:1974sr,Gross:1974cs}.
In fact, ever since the pioneering measurements at SLAC~\cite
{Coward:1967au,Bloom:1969kc,Breidenbach:1969kd},
DIS structure functions have been the subject of detailed theoretical
and experimental investigations, see, e.g., the Review of Particle 
Properties~\cite{Eidelman:2004wy} and references therein. 
Today, with high-precision data from the electron--proton collider 
{\sc HERA} and in view of the outstanding importance of hard scattering 
processes at proton--(anti-)proton colliders like the {\sc Tevatron} 
and the forthcoming {\sc LHC}, a quantitative understanding of 
deep-inelastic processes is indispensable. 

For quantitatively reliable predictions of DIS and hard hadronic 
scattering processes, perturbative QCD corrections beyond the 
next-to-leading order (NLO) need to be taken into account. We have 
therefore calculated the three-loop splitting functions for the 
evolution of unpolarized parton distributions of hadrons~\cite
{Moch:2004pa,Vogt:2004mw}.
Together with the second-order coefficient functions~\cite 
{vanNeerven:1991nn,Zijlstra:1991qc,Zijlstra:1992kj,Zijlstra:1992qd,%
Moch:1999eb}, these recent results form the complete next-to-next-to-%
leading order (NNLO, N$^2$LO) approximation of massless perturbative 
QCD for the structure functions $F_1$, $F_{\,2}$ and $F_3$ in DIS.

In the present article, we extend the calculation of electromagnetic
(photon-exchange) DIS in perturbative QCD to the three-loop coefficient 
functions for both $F_{\,2}$ and $F_L = F_{\,2} - 2x F_1$. 
This represents the first calculation of third-order perturbative 
corrections to hard scattering observables depending on a dimensionless 
variable (Bjorken-$x$ in the case at hand) in the Standard Model. 
For the longitudinal structure function $F_L$ the third-order 
corrections are actually required to complete the NNLO predictions, 
since the leading contribution to the coefficient functions is of 
first order in the strong coupling constant $\as$. In a recent letter
\cite{Moch:2004xu} we have already presented the corresponding results
in a compact numerical form, and briefly discussed their 
phenomenological implications.

For the structure functions $F_1$ and $F_{\,2}$, on the other hand, the 
three-loop coefficient functions are part of the next-to-next-to-%
next-to-leading order (N$^3$LO) description of DIS in perturbative QCD.
In fact, due to the fast convergence of the splitting function series
\cite{Moch:2004pa,Vogt:2004mw}, these coefficient functions dominate 
the N$^3$LO corrections for not too small values of the Bjorken 
variable, $x \gsim 10^{-2}$. Thus the extraction of $\as$ from the 
scaling violations of structure functions can be effectively promoted
to N$^3$LO accuracy, reducing the (formerly dominant) uncertainty due 
to the truncation of the perturbation series to less than 1\%, see, 
e.g., Refs.~\cite{Santiago:2001mh,vanNeerven:2001pe,Kataev:2001kk,%
Alekhin:2002rk,Blumlein:2004ip}. 
The three-loop coefficient functions are also of considerable 
theoretical interest, for example facilitating the derivation of 
higher-order results for the resummation of threshold logarithms 
\cite{Sterman:1987aj,Catani:1989ne,Magnea:1990qg,Catani:1991rp,%
Vogt:2000ci,Catani:2003zt} and the quark form factor
\cite{Magnea:1990zb,Magnea:2000ss}. We will address these issues in a 
forthcoming publication~\cite{MVV7}.

As discussed in Refs.~\cite{Moch:2004pa,Vogt:2004mw}, see also 
Refs.~\cite{Moch:2002sn,Vermaseren:2002rn,Vogt:2004gi,Moch:2004sf}, 
the NNLO splitting functions have been determined via a Mellin-$N$ 
space calculation of physical matrix elements of electromagnetic DIS 
at three loops in dimensional regularization with $D = 4 - 2 \ep$.
While the splitting functions are extracted from the $1/\ep$ poles, 
the coefficient functions are obtained from the finite terms of the 
physical matrix elements corresponding to the structure functions 
$F_{\,2}$ and $F_L$. This is possible since we have taken care to control
all necessary three-loop integrals up to (and including) the finite 
contributions. In this respect our approach closely follows the 
third-order calculations of sum rules in DIS~\cite{Larin:1991zw,%
Larin:1991tj} and of low integer moments of structure functions
\cite{Larin:1991fx,Larin:1994vu,Larin:1997wd,Retey:2000nq,Moch:2001im},
however with the obvious distinction that we now derive the analytic 
dependence on $N$ and, consequently, on $x$.

The outline of this article is as follows.
In Section~\ref{sec:formalism} we briefly recall the formalism, based
on the operator product expansion, for calculating inclusive DIS in 
Mellin-$N$ space and discuss the extraction of the anomalous dimensions
(splitting functions) and coefficient functions.
Section~\ref{sec:method} explains selected details of the method
to calculate the analytic $N$-dependence of the diagrams and
addresses issues which did not occur in the calculation of the
splitting functions. 
In Section~\ref{sec:results} we present our results for $F_{\,2}$ in a
compact parametrized form and discuss the end-point behaviour of the
coefficient functions for $F_{\,2}$ and $F_L$. The lengthy full expressions
for both coefficient functions are deferred to Appendix A ($N$-space) 
and Appendix B ($x$-space).  
The numerical implications of these results are illustrated in Section
\ref{sec:sresults} before we summarize our findings in 
Section~\ref{sec:summary}. 
%
%
\setcounter{equation}{0}
\section{General formalism}
\label{sec:formalism}
%
%
The subject of our calculation is unpolarized inclusive deep-inelastic 
lepton-nucleon scattering,
\beq
\label{eq:dis}
  l(k) \:+\: {\rm nucl}(p) \:\:\ra\:\: l(k^{\,\prime}) \:+\:  X
\eeq
where $X$ stands for all hadronic states allowed by quantum number
conservation. Specifically, we here consider the lowest-order (one-%
photon exchange) QED contribution to this process. The hadronic part 
of the corresponding amplitude is given by the (spin-averaged) 
tensor
\bea
  W_{\mu\nu}(p,q) & = & \frac{1}{4\pi}
    \int \! d^{\,4}z\: {\rm{e}}^{{\rm{i}}q \cdot z} \, 
    \langle {\rm{nucl,}p}\vert J_{\mu}(z)J_{\nu}(0)\vert 
    {\rm{nucl,}p}\rangle  \nn \\[1mm]
  & = & 
     \bigg( \frac{q_{\mu} q_{\nu}}{q^2} - g_{\mu \nu} \bigg)
     \, F_{1}(x,Q^2) 
   - \big( q_{\mu} + 2x p_{\mu} \big) \big( q_{\nu} + 2x p_{\nu} \big)  
     \, \frac{1}{2xq^2}\, F_{2}(x,Q^2) \:\: . 
\label{eq:htensor}
\eea
Here $\vert {\rm nucl,} p \rangle$ denotes the nucleon state with 
momentum $p$, and $J_{\mu}$ represents the electromagnetic current.  
$q = k-k^{\,\prime}$ is the momentum transferred by the lepton, 
$Q^2=-q^2$, and $x = Q^2 / (2p \cdot q)$ is the Bjorken variable 
with $0 < x \leq 1$.
The longitudinal structure function $F_L$ is related to the structure
function $F_1$ in Eq.~(\ref{eq:htensor}) by $ F_L = F_{\,2} -2xF_1$.

The hadronic tensor $W_{\mu\nu}$ is connected by the optical theorem to 
the imaginary part of the forward amplitude $T_{\mu\nu}$ for the 
scattering of a virtual photon off the nucleon,
\beq
\label{eq:Tmunu}
  T_{\mu\nu}(p,q) \: = \: i \! \int \! d^{\, 4}z \: e^{iqz} \,
  \langle {\rm nucl,} p \vert \, T \big(J_{\mu}(z)J_{\nu}(0) 
  \big) \, \vert {\rm nucl,} p \rangle \:\: .
\eeq
This quantity represents a convenient starting point for practical
calculations, due to the presence of the time-ordered product of
currents to which standard perturbation theory applies. 

Approaching the Bjorken limit, $Q^2 \rightarrow \infty$ for fixed $x$, 
the integrations in Eq.~(\ref{eq:htensor}) and (\ref{eq:Tmunu}) are
dominated by the region near the light-cone, $z^2 \approx 0$, as only 
there the phase of the exponential factor becomes stationary.
In this situation, the operator-product expansion (OPE) can be applied
to the product of currents in Eq.~(\ref{eq:Tmunu}) together with a 
dispersion relation~\cite{Christ:1972ms}. This procedure is identical 
to that in previous lower-order and fixed-$N$ third-order calculations.
Thus we will recall it only briefly, referring the reader to 
Refs.~\cite{Larin:1997wd,Moch:1999eb} and the reviews 
\cite{Buras:1980yt,Reya:1981zk} for more details.

Disregarding contributions suppressed by powers of $1/Q^2$, the OPE
involves the standard set of the spin-$N$ twist-two irreducible 
flavour non-singlet quark, singlet quark and gluon operators,
\bea
\label{eq:loc-ops}
 O_{\rm ns}^{\{\mu^{\,}_1,...,\mu^{\,}_N\}} & = & 
   \overline{\psi}\,\lambda^{\alpha}\,\gamma^{\,\{\mu^{}_1}
   D^{\,\mu^{}_2} \ldots D^{\,\mu_N\}}\,\psi 
   \:\: , \quad\quad \alpha=3,8,...,(\n2f-1) \:\: , \nn \\
 O_{\rm q}^{\{\mu^{\,}_1,...,\mu^{\,}_N\}} & = & \overline{\psi}\,
   \gamma^{\,\{\mu^{}_1}D^{\,\mu^{}_2}\ldots D^{\,\mu^{}_N\}}\,\psi 
   \:\: , \nn \\
 O_{\rm g}^{\{\mu^{\,}_1,...,\mu^{\,}_N\}} & = & F^{\nu \{ \mu^{}_1} 
   D^{\,\mu^{}_2}\cdots D^{\,\mu_{N-1}}\, F^{\mu^{}_N \} \nu} \:\: ,
\eea
and their respective coefficient functions $C_{a,i}(N)$ for $a=2,\,L$. 
Here $\psi$ represents the quark field, $F^{\mu\nu}$ the gluon field
strength tensor, and $D^{\,\mu}$ the covariant derivative. The 
diagonal generators of the flavour group $SU(\nf)$ are denoted by 
$\lambda^{\alpha}$. The spin-averaged matrix elements of the 
(renormalized) operators in Eq.~(\ref{eq:loc-ops}) are given by
\bea
\label{eq:OME}
  \langle {\rm nucl,} p \vert \, O_{i}^{\{\mu^{\,}_1,...,\mu^{\,}_N\}}
  \,\vert {\rm nucl,} p \rangle 
  \: = \: p^{\{\mu^{\,}_1}...p^{\mu^{\,}_N\}}\, A_{i,\rm nucl}(N,\mu^2)
  \:\: , \quad\quad i \, = \, \mbox{ns, q, g} \:\: ,
\eea
where $\mu$ stands for the renormalization scale. It is understood in
Eqs.~(\ref{eq:loc-ops}) and (\ref{eq:OME}) that the symmetric and 
traceless part is taken with respect to the indices in curved brackets.

The application of the operator-product expansion to the forward
Compton amplitude (\ref{eq:Tmunu}), neglecting $1/Q^2$ power 
corrections, leads to the expansion
\bea
\label{eq:OPE}
   T_{\mu\nu}(p,q) & = &
   \sum_{N,i} \,\bigg(\,\frac{2p\cdot q}{Q^2}\,\bigg)^N A_{i,\rm nucl} 
        (N,\mu^2) \, 
   \left[ \left( g_{\mu \nu}+\frac{q_{\mu} q_{\nu}}{Q^2} \right) \,
   C_{L,i}\left(N,\frac{Q^2}{\mu^2},\alpha_s \right) \right.
   \nn \\ & & \left. \mbox{}
   - \left( g_{\mu \nu}-p_{\mu}p_{\nu}\frac{4x^2}{Q^2}
   - (p_{\mu}q_{\nu}+p_{\nu}q_{\mu})\frac{2x}{Q^2} \right)\,
   C_{2,i}\left(N,\frac{Q^2}{\mu^2},\alpha_s\right) \right]
   \:\: .
\eea
The continuation of this result to the physical region $0 < x \leq 1$
by a dispersion relation in the complex-$x$ plane finally yields the 
even-integer Mellin-$N$ moments of the structure functions 
$\frac{1}{x}F^{}_2$ and $\frac{1}{x}F^{}_L$ in Eq.~(\ref{eq:htensor}),
\beq
\label{eq:Mtrf}
  F_a(N,Q^2) \: = \:
  \int_0^1 \! dx\: x^{\,N-1} \:\frac{1}{x}\, F_a(x,Q^2) \:\: ,
\eeq
in terms of the matrix elements (\ref{eq:OME}) and the corresponding
coefficient functions, 
\beq
\label{eq:F2mellin}
  \frac{1 + (-1)^N}{2}\: F_a(N,Q^2) = \sum_{i = {\rm ns, q, g}}
  C_{a,i} \left(N, \frac{Q^2}{\mu^2}, \as \right) A_{i,\rm nucl}
  (N,\mu^2) \:\: , \quad\quad a = 2,L \:\: .
\eeq
Note that all (complex) moments $N$, and thus, by the inverse of the 
Mellin transformation (\ref{eq:Mtrf}), the complete $x$-dependence, are 
uniquely fixed by analytic continuation of these even-$N$ results.

The operators $O_{\rm q}$ and $O_{\rm g}$ in Eq.~(\ref{eq:loc-ops}) mix 
under renormalization. Expressing the renormalized operators in terms 
of their bare counterparts, this mixing can be written as
\beq
\label{eq:Oren}
  O_i \: = \: Z_{ik}\,O_k^{\,\rm bare} \:\: .
\eeq
The anomalous dimensions $\gamma_{ik}$ governing the scale dependence
of the operators $O_{i}$,
\beq
\label{eq:gamma}
  \frac{d}{d \ln \mu^2 }\, O_i \: = \: - \,\gamma_{ik}\, O_k
  \: \equiv \: P_{\,ik}\, O_k  \:\: ,
\eeq
are connected to the mixing matrix $Z_{ik}$ in Eq.~(\ref{eq:Oren}) by
\beq
\label{eq:gamZ}
 \gamma_{ik} \: = \: -\,\left( \frac{d }{d\ln\mu^2 }\, Z_{ij}
 \right) (Z^{-1})_{jk} \:\: .
\eeq
The summation convention is understood in Eqs.~(\ref{eq:Oren}) -- 
(\ref{eq:gamZ}), and the dependence on $N$ has been suppressed for
brevity. In Eq.~(\ref{eq:gamma}) we have taken the opportunity to 
recall the conventional relation between the anomalous dimensions and
the moments of the splitting functions $P_{ik}(x)$. 
Corresponding scalar relations, independent of the generator 
$\lambda^\alpha$ in Eq.~(\ref{eq:loc-ops}), hold for the non-singlet
operators collectively denoted by $O_{\rm ns}$.

In order to make practical use of Eq.~(\ref{eq:gamZ}) a regularization
procedure and a renormalization scheme need to be selected. We choose
dimensional regularization~\cite
{'tHooft:1972fi,Bollini:1972ui,Ashmore:1972uj,Cicuta:1972jf} 
and the modified~\cite{Bardeen:1978yd} minimal subtraction 
\cite{'tHooft:1973mm} scheme, \MSb, the standard choice for modern 
higher-order calculations in QCD. For this choice the running coupling 
in $ D = 4 - 2\ep $ dimensions evolves according to
\beq
\label{eq:arun}
  \frac{d}{d \ln \mu^2}\: \frac{\as}{4\pi} \:\: \equiv \:\: 
  \frac{d\,\ar}{d \ln \mu^2} \:\: = \:\: - \ep\, \ar 
  - \beta_0\, a_{\rm s}^2 - \beta_1\, a_{\rm s}^3 
  - \beta_2\, a_{\rm s}^4 - \ldots \:\: ,
\eeq
where $\beta_{\rm n}$ denote the usual four-dimensional expansion 
coefficients of the beta function in QCD 
\cite{Caswell:1974gg,Jones:1974mm,Tarasov:1980au,Larin:1993tp,
vanRitbergen:1997va,Czakon:2004bu}, $\,\beta_0 = 11 - 2/3\,\nf\,$ etc,
with $\nf$ representing the number of active quark flavours.

In this framework, the renormalization factors $Z_{ik}$ in   
Eq.~(\ref{eq:Oren}) and $Z_{\rm ns}$ are a series of poles in $1/\ep$,
expressed in terms of $\beta_{\rm n}$ and the expansion coefficients 
$\gamma^{\,(l)}$ of the anomalous dimensions in terms of $\ar$, 
\beq
\label{eq:gam-exp}
  \gamma(N) \: = \: \sum_{l=0}^{\infty}\,
  a_{\rm s}^{\,l+1}\, \gamma^{\,(l)}(N) \:\: . 
\eeq
For example, the expansion of $Z_{\rm ns}$ up to the third order in
the coupling constant reads
\bea
\label{eq:Zns3}
  Z_{\rm ns} & = & 
    1 \: + \: \:\ar\, \frac{1}{\ep}\,\gamma_{\,\rm ns}^{\,(0)} 
    \: + \: a_{\rm s}^2 \,\left[\, \frac{1}{2\ep^2}\, 
    \left\{ \left(\gamma_{\,\rm ns}^{\,(0)} - \beta_0 \right) 
    \gamma_{\,\rm ns}^{\,(0)} \right\}
    + \frac{1}{2\ep}\, \gamma_{\,\rm ns}^{\,(1)} \right] 
  \nn \\[1mm] & & \mbox{} + \:  
  a_{\rm s}^3 \,\left[\, \frac{1}{6\ep^3}\, 
    \left\{ \left( \gamma_{\,\rm ns}^{\,(0)} - 2 \beta_0 \right)
    \left( \gamma_{\,\rm ns}^{\,(0)} - \beta_0 \right) 
    \gamma_{\,\rm ns}^{\,(0)} \right\} \right.
  \nn \\[1mm] & & \left. \mbox{} \quad\quad \! + \:  
  \frac{1}{6\ep^2}\, \left\{ 3\, \gamma_{\,\rm ns}^{\,(0)}  
    \gamma_{\,\rm ns}^{\,(1)} - 2 \beta_0\, \gamma_{\,\rm ns}^{\,(1)}
    - 2 \beta_1\, \gamma_{\,\rm ns}^{\,(0)} \right\} \: + \: 
    \frac{1}{3\ep}\, \gamma_{\,\rm ns}^{\,(2)} \right] \:\: .
\eea
The anomalous dimensions $\gamma^{\,(l)}$ can thus be read off from the
$\ep^{-1}$ terms of the renormalization factors at order $a_{\rm s}
^{\,l+1}$, while the higher poles in $1/\ep$ can serve as checks for 
the calculation. The coefficient functions in Eq.~(\ref{eq:OPE}), on
the other hand, have an expansion in positive powers of $\ep$, viz
\beq
\label{eq:cf-exp}
  C_{a,i} \: = \: \delta_{a2}\, (1 - \delta_{\rm ig}) 
    + \sum_{l=1}^{\infty} \, a_{\rm s}^{\, l} \left( c_{a,i}^{\,(l)} 
    + \ep a_{a,i}^{\,(l)} + \ep^2 b_{a,i}^{\,(l)} + \ldots \right)
\eeq
where $a = 2,\, L$ and $i$ = ns, q, g, and we have again suppressed the
dependence on $N$ (and $Q^2/\mu^2$).

Due to the non-perturbative character of the nucleon state $\vert
{\rm nucl,}p \rangle$, Eqs.~(\ref{eq:Tmunu}) and (\ref{eq:F2mellin})
are not accessible to a perturbative computation. However, as the OPE 
represents an operator relation, the anomalous dimensions 
(\ref{eq:gam-exp}) and the coefficient functions (\ref{eq:cf-exp}) do 
not depend on this state. Hence the calculation can be performed using 
quark and gluon states $\vert k,p \rangle$. Instead of 
Eq.~(\ref{eq:Tmunu}) we thus consider
\beq
\label{eq:Tpart}
  T^{\,k}_{\mu\nu}(p,q) \: = \: i \! \int \! d^{\, 4}z \, e^{iqz} 
  \, \langle k, p \vert \, T \big(J_{\mu}(z)J_{\nu}(0) \big) \,
  \vert k, p \rangle \:\: , \quad k = {\rm ns,\,q,\,g} \:\: .
\eeq
At leading-twist accuracy the decomposition of $T^{\,k}_{\mu\nu}$ 
into $T_{2,k}$ and $T_{L,k}$ analogous to Eq.~(\ref{eq:htensor}) is
provided by
\bea
\label{eq:Tproj}
  T_{L,k}(p,q) & = & - \,\frac{q^2}{(p\cdot q)^2}\, p^\mu p^\nu 
    \: T^{\,k}_{\mu\nu}(p,q)
  \nn \\
  T_{2,k}(p,q) & = & - \left( \frac{3-2\ep}{2-2\ep}\: \frac{q^2}
    {(p\cdot q)^2}\, p^\mu p^\nu + \frac{1}{2-2\ep}\: g^{\mu\nu} 
    \right) T^{\,k}_{\mu\nu}(p,q)
\eea
with spin-averaging again being understood. The $N^{\rm th}$ moments are
obtained from Eqs.~(\ref{eq:Tproj}) by applying the projection operator 
\cite{Gorishnii:1983su,Gorishnii:1987xx}
\beq
\label{eq:PNop}
  T_{a,k} \left(N,\frac{Q^2}{\mu^2},\as,\ep\right) \: = \: 
  \left[ \frac{q^{ \{\,\mu^{}_1}\cdots q^{\,\mu^{}_N \}}}{2^N N !}\,
  \frac{\partial ^N}{\partial p^{\,\mu_1} \ldots  \partial p^{\,\mu_N}}
  \right] \, T_{a,k}(p,q,\as,\ep ) \Bigg|_{p=0} 
  \:\: ,
\eeq
where $q^{ \{\mu^{}_1}\cdots q^{\mu^{}_N \}}$ is the harmonic, i.e., 
the symmetric and traceless part of the tensor $q^{\mu^{}_1}\cdots 
q^{\mu^{}_N}$.

This operator does not act on the coefficient functions $C_{a,k}$ and
the renormalization constants $Z_{ik}$ in Eq.~(\ref{eq:Oren}), which 
are functions only of $N$, $a_{\rm s}$, and $\ep$. It does act, 
however, on the bare matrix elements $A_{i,k}$ (defined analogously 
to Eq.~(\ref{eq:OME})$\,$) and eliminates all diagrams containing 
loops, as the nullification of $p$ transform these diagrams to massless 
tadpole diagrams which are zero in dimensional regularization. Hence 
only the matrix elements $A_{k,k}^{\,\rm tree}(N,\ep)$ remain, leading 
to
\beq
\label{eq:T2LN}
  T_{a,k} \left (N,\frac{Q^2}{\mu^2}, \as, \ep \right)
  \:\: = \:\: C_{a,i} \left (N,\frac{Q^2}{\mu^2},\as ,\ep \right)
  \, Z_{ik} \left(N, \as ,\frac{1}{\ep} \right) 
  \, A_{k,k}^{\,\rm tree}(N, \ep)
\eeq
for $a=2,L$ and $k =$ ns, q, g. Here summation over $i = $ q, g is 
understood for the singlet cases $k =$ q, g, while $C_{a,i}$ and 
$Z_{ik}$ have to be replaced by $C_{a,\rm ns}$ and $Z_{\rm ns}$ of
Eq.~(\ref{eq:Zns3}), respectively, for the non-singlet case $k =$ ns.  
Expansion of (\ref{eq:T2LN}) in powers of $\as$ and $\ep$ provides
a system of equations which can be solved for the anomalous dimensions
(\ref{eq:gam-exp}) and coefficient functions (\ref{eq:cf-exp}).

For brevity suppressing the function arguments for the rest of this 
section, the expansion of the `master formula' (\ref{eq:T2LN}) to the 
third order in the strong coupling $\as $ can by written as
\beq
\label{eq:Texp}
  T_{a,k} \: = \: \sum_{l=0}^{3}\: a_{\rm s}^{\,l}
  \: S_\ep^{\, l} \bigg( \frac{\mu^2}{Q^2} \bigg)^{l\ep} \,
  \delta_k \: T^{(l)}_{a,k} \: A_{k,k}^{\,\rm tree} \:\: .
\eeq
The factor $S_\ep = \exp ( \ep \{\ln(4\pi)-\gamma_{\rm e}\} )$,
where $\gamma_{\rm e}$ denotes the Euler-Mascheroni constant, is an
artefact of dimensional regularization~\cite
{'tHooft:1972fi,Bollini:1972ui,Ashmore:1972uj,Cicuta:1972jf}
kept out of the coefficient functions and anomalous dimensions in the 
\MSb\ scheme \cite{Bardeen:1978yd}. $\delta_k$ collects the quark
charge factors,
\beq
\label{eq:charge}
  \delta_{\rm ns} \: = \: 1, \qquad 
  \delta_{\rm q}  \: = \: \delta_{\rm g} \: = \: \frac{1}{\nf} \:\sum
  _{i=1}^{\nf}\, e_{q^{}_i}^{\,2} \: \equiv \: \langle e^{\,2} \rangle
  \:\: .
\eeq
The $\as=0$ parts $T^{(0)}_{2,\rm ns}$ and $T^{(0)}_{2,\rm q}$ can be
rendered equal by a suitable normalization of non-singlet matrix 
elements $A_{\rm ns,ns}^{\,\rm tree}$. The amplitudes $T_{a,\rm ns}$
and $T_{a,\rm q}$ are then identical also at the first order in $\as$.
Consequently, the same holds for the anomalous dimensions and 
coefficient functions (recall the different counting of the 
superscripts in Eqs.~(\ref{eq:gam-exp}) and (\ref{eq:cf-exp})$\,$),
\beq
  \gamma^{\,(0)}_{\,\rm ns} \: = \: \gamma^{\,(0)}_{\,\rm qq}
  \:\: , \qquad
  \tilde{c}^{\,(1)}_{\,\rm ns} \: = \: \tilde{c}^{\,(1)}_{\,\rm q}
  \:\: , \quad \tilde{c} = c,\: a,\: b\, \ldots \:\: .
\eeq
In the expansions shown below, we will use these right-hand sides also 
in the results for $T^{(n>1)}_{a,\rm ns}$.

The zeroth-order contributions, with $T^{(0)}_{2,\rm q}$ being 
normalized by virtue of Eq.~(\ref{eq:charge}), read
\beq
\label{eq:T0}
  T_{2,\rm q}^{(0)} \: = \: c_{2,\rm q}^{(0)} \: = \: 1 
   \:\: , \qquad
  T_{2,\rm g}^{(0)} \: = \: T_{L,\rm q}^{(0)} 
  \: = \: T_{L,\rm g}^{(0)} \: = \: 0 \:\: .
\eeq
As will become clear below, the amplitudes at the first order in $\as$
need to be calculated up to order $\ep^2$ for our purposes, yielding
\beq
\label{eq:T2k1}
  T_{2, \rm p}^{(1)} \: = \: \frac{1}{\ep}\, \gamma_{\,\rm qp}^{\,(0)} 
  \: + \: c_{2,\rm p}^{(1)} \: + \: \ep\, a_{2,\rm p}^{(1)} 
  \: + \: \ep^2 b_{2,\rm p}^{(1)} 
\eeq 
and
\beq
\label{eq:TLk1}
  T_{L, \rm p}^{(1)} \: = \: c_{L,\rm p}^{(1)} 
  \: + \: \ep\, a_{L,\rm p}^{(1)} \: + \: \ep^2 b_{L,\rm p}^{(1)}
  \:\: , \\[0.5mm]
\eeq
with $\rm p = q,\, g$. Correspondingly the $\as^2$ contributions, where 
the non-singlet and singlet quark amplitudes differ for the first time, 
are required up to order $\ep$. These quantities are given by
\bea
\label{eq:T2n2}
  T_{2,\rm ns}^{(2)} & \!=\! & \frac{1}{2\ep^2}\, 
  \bigg\{ \left( \gqqz - \beta_0 \right) \gqqz \bigg\} 
  \: + \: \frac{1}{2\ep}\, \left\{ \gnso + 2\, \ctqo\, \gqqz \right\} 
\nn \\ & & \mbox{} 
  \: + \: \ctnt + \atqo\, \gqqz 
  \: + \: \ep\, \bigg\{ \atnt + \btqo\, \gqqz \bigg\} \:\: , 
  \\[2mm]
\label{eq:T2p2}
  T_{2,\rm p}^{(2)} & \!=\! & \frac{1}{2\ep^2}\,
  \bigg\{ \left( \gqiz - \beta_0 \delta_{\rm qi} \right) \gipz \bigg\}
  \: + \: \frac{1}{2\ep}\, \bigg\{ \gqpo + 2\, \ctio\, \gipz \bigg\}  
\nn \\ & & \mbox{} 
  \: + \: \ctpt\: + \atio\, \gipz 
  \: + \:  \ep\, \bigg\{ \atpt + \btio\, \gipz \bigg\} \:\: ,
\eea
where $\delta_{ik}$ is the Kronecker symbol, and
\bea
\label{eq:TLn2}
  T_{L,\rm ns}^{(2)} & \!=\! & \frac{1}{\ep}\,
  \bigg\{ \clqo\, \gqqz \bigg\} \, + \: \clnt + \alqo\, \gqqz
  \: + \: \ep\, \bigg\{ \alpt + \blqo\, \gqqz \bigg\} \:\; ,
  \\[2mm]
\label{eq:TLp2}
  T_{L,\rm p}^{(2)} & \!=\! & \frac{1}{\ep}\,
  \bigg\{ \clio\, \gipz \bigg\} \: + \: \clpt\: + \alio\, \gipz 
  \: + \: \ep\, \bigg\{ \alpt + \blio\, \gipz \bigg\}
  \:\: .
\eea

We are now finally ready to write down the third-order coefficients 
$T_{a, k}^{(3)}$ in Eq.~(\ref{eq:Texp}), reading
\bea
\label{eq:T2n3}
  T_{2,\rm ns}^{(3)} & \!=\! & \frac{1}{6\ep^3}\,
  \bigg\{ \left( \gqqz - 2\beta_0 \right) 
    \left( \gqqz - \beta_0 \right) \gqqz \bigg\}
\nn \\ & & \mbox{} 
  \: + \: \frac{1}{6\ep^2}\, \bigg\{ 3 \gnso\,\gqqz - 2 \beta_0\,\gnso
     - 2\beta_1\,\gqqz + 3\ctqo \left( \gqqz - \beta_0 \right) \gqqz 
       \bigg\}
\nn \\ & & \mbox{} 
  \: + \: \frac{1}{6\ep}\,\bigg\{ 2 \gnst + 3 \ctqo\,\gnso + 6 \ctnt\,
     \gqqz + 3 \atqo \left( \gqqz - \beta_0 \right) \gqqz \bigg\}
\nn \\ & & \mbox{} 
  \: + \: \ctnd + \frac{1}{2}\, \atqo\,\gnso + \atnt\,\gqqz + 
     \frac{1}{2}\, \btqo \left( \gqqz - \beta_0 \right) \gqqz \:\: ,
  \\[2mm]
\label{eq:T2p3}
  T_{2,\rm p}^{(3)} & \!=\! & \frac{1}{6\ep^3}\,
  \bigg\{ \gqiz\,\gikz\,\gkpz - 3 \beta_0 \gqiz\,\gipz + 2 \beta_0^2\, 
    \gqpz \bigg\} 
\nn \\ & & \mbox{}
  \: + \: \frac{1}{6\ep^2}\, \bigg\{ \gqiz\,\gipo + 2\gqio\,\gipz
    - 2 \beta_0 \gqpo - 2 \beta_1 \gqpz + 3 \ctio \left( \gikz - 
      \beta_0 \delta_{\rm ik}\right) \gkpz \bigg\}
\nn \\ & & \mbox{}
  \: + \: \frac{1}{6\ep}\, \bigg\{ 2 \gqpt + 3 \ctio\,\gipo + 6 \ctit\,
     \gipz + 3 \atio \left( \gikz - \beta_0 \delta_{\rm ik}\right) \gkpz
    \bigg\} \nn \\ & & \mbox{}
  \: + \: \ctpd + \frac{1}{2}\, \atio\,\gipo + \atit\,\gipz +
     \frac{1}{2}\, \btio \left( \gikz - \beta_0 \delta_{\rm ik}\right) 
     \gkpz
\eea
and
\bea
\label{eq:TLn3}
  T_{L,\rm ns}^{(3)} & \!=\! & \frac{1}{2\ep^2}\,
  \bigg\{ \clqo \left( \gqqz - \beta_0 \right) \gqqz \bigg\}
\nn \\ & & \mbox{} 
  \: + \: \frac{1}{2\ep}\, \bigg\{ \clqo \,\gnso + 2 \clnt\, \gqqz
     + \alqo \left( \gqqz - \beta_0 \right) \gqqz \bigg\}
\nn \\ & & \mbox{} 
  \: + \: \clnd + \frac{1}{2}\, \alqo\,\gnso + \alnt\,\gqqz
     +  \frac{1}{2}\, \blqo \left( \gqqz - \beta_0 \right) \gqqz
  \:\: ,
  \\[2mm]
\label{eq:TLp3}
  T_{L,\rm p}^{(3)} & \!=\! & \frac{1}{2\ep^2}\,
  \bigg\{ \clio \left( \gikz - \beta_0 \delta_{\rm ik} \right) \gkpz 
    \bigg\} \nn \\ & & \mbox{}
  \: + \: \frac{1}{2\ep}\, \bigg\{ \clio \,\gipo + 2 \clit\, \gipz
     + \alio \left( \gikz - \beta_0 \delta_{\rm ik} \right) \gkpz 
    \bigg\} \nn \\ & & \mbox{}
  \: + \: \clpd + \frac{1}{2}\, \alio\,\gipo + \alit\,\gipz
     +  \frac{1}{2}\, \blio \left( \gikz - \beta_0 \delta_{\rm ik} 
     \right) \gkpz \:\: .
\eea
Summation over $\rm i,k = q,g$ is understood in the singlet relations
(\ref{eq:T2p2}), (\ref{eq:TLp2}), (\ref{eq:T2p3}) and (\ref{eq:TLp3}).

The object of the present calculation, the coefficient functions 
$c_{a,\rm ns}^{(3)}$ and $c_{a,\rm p}^{(3)}$ with $a = 2,\, L$ and 
$\rm p = q,\,g$, can therefore be extracted from the projected 
three-loop contributions (\ref{eq:T2n3}) -- (\ref{eq:TLp3}) to the 
partonic forward-Compton amplitudes (\ref{eq:Tpart}), once the 
respective $\ep^2$ terms $b_{a,k}$ at one loop and $\ep^1$ pieces 
$a_{a,k}$ up to two loops have been determined using 
Eqs.~(\ref{eq:T2k1}) -- (\ref{eq:TLp2}).

The relations (\ref{eq:T2n3}) and (\ref{eq:T2p3}) are also part of the
system of equations from which we have determined the three-loop 
anomalous dimensions \cite{Moch:2004pa,Vogt:2004mw}.
The $\ep^{-1}$ term of Eq.~(\ref{eq:T2n3}) fixes one of the three 
non-singlet combinations, denoted by $\gamma_{\,\rm ns}^{\,(2)+}$ in
Ref.~\cite{Moch:2004pa}. To obtain the other two combinations,
quark--antiquark differences inaccessible in electromagnetic DIS,
we have also computed the $W$-exchange neutrino--nucleon structure 
function $F_3^{\,\nu N+\bar{\nu}N}$. The results for the corresponding
coefficient function will be presented elsewhere.
In the flavour-singlet sector, Eq.~(\ref{eq:T2p3}) includes 
$\gamma_{\,\rm qq}^{\,(2)}$ and $\gamma_{\,\rm qg}^{\,(2)}$, but, since
the gluon does not directly couple to the photon, not the lower row of
the anomalous dimension matrix, $\gamma_{\,\rm gq}^{\,(2)}$ and 
$\gamma_{\,\rm gg}^{\,(2)}$. These quantities have been computed in 
Ref.~\cite{Vogt:2004mw} via DIS by exchange of a (not entirely) 
fictitious scalar $\phi$ directly coupling only to gluons.

The forward Compton diagrams contributing to the present calculation 
of the electromagnetic three-loop coefficient functions, generated 
automatically with the diagram generator {\sc Qgraf} \cite
{Nogueira:1991ex}, are enumerated in Table \ref{tab:diagrams}. Among 
the partons $k$ in Eq.~(\ref{eq:Tpart}) we also include an external 
ghost $h$. 
This is a standard procedure, allowing us to take the sum over external
gluon spins by contracting with $-g_{\mu\nu}$ instead of the full 
physical expression which would, due to the presence of extra powers of
the gluon momentum $p$, lead to a considerable complication of our task.  
For the same reason our all-$N$ computations have been performed in the
Feynman gauge. We have however checked the gauge independence for a few
low values of $N$ using the {\sc Mincer} program
\cite{Gorishnii:1989gt,Larin:1991fz}. 
The latest version version of {\sc Form}~\cite{Vermaseren:2000nd,%
Vermaseren:2002rp} has been employed for all symbolic manipulations.

\begin{table}[htp]
\vspace{-4mm}
\begin{center}
\begin{tabular}{c r r r r}\\
\hline & & & & \\[-3mm]
{process} &{tree} &1-loop &2-loop &3-loop \\[1mm]
\hline & & & & \\[-3mm]
$q\,\gamma\:\ra\: q\,\gamma$ & 1 &  3 &  25 &  359 \\
$g\,\gamma\:\ra\: g\,\gamma$ &   &  2 &  17 &  345 \\
$h\,\gamma\:\ra\: h\,\gamma$ &   &    &   2 &   56 \\[1mm]
\hline & & & & \\[-3mm]
 sum & 2 & 10 & 88 & 1520 \\[1mm] \hline
\end{tabular}
\end{center}
\vspace{-3mm}
\caption{ \label{tab:diagrams}
 The number of diagrams for the amplitudes employed for the calculation 
 of the three-loop coefficient functions.  The sums includes a factor 
 of two from Lorentz projections to $F_{\,2}$ and $F_L$.}
\end{table}

\begin{figure}[ht]
\vspace{-4mm}
\begin{center}
  \includegraphics[width=3.0cm]{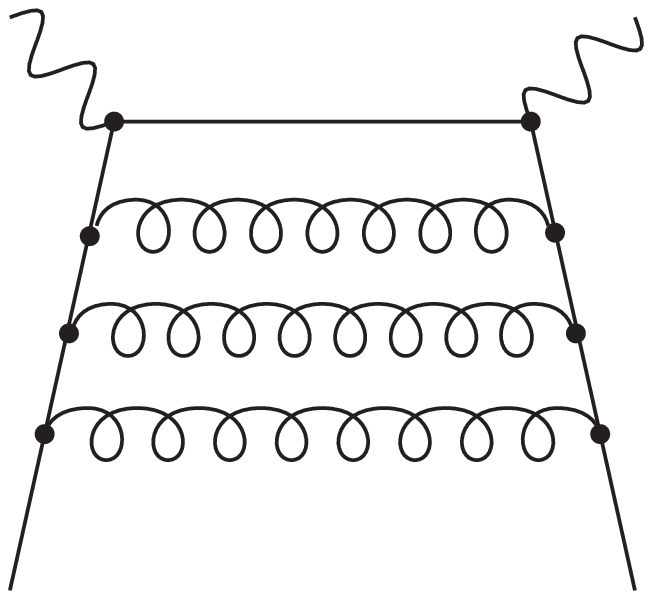}
  \includegraphics[width=3.0cm]{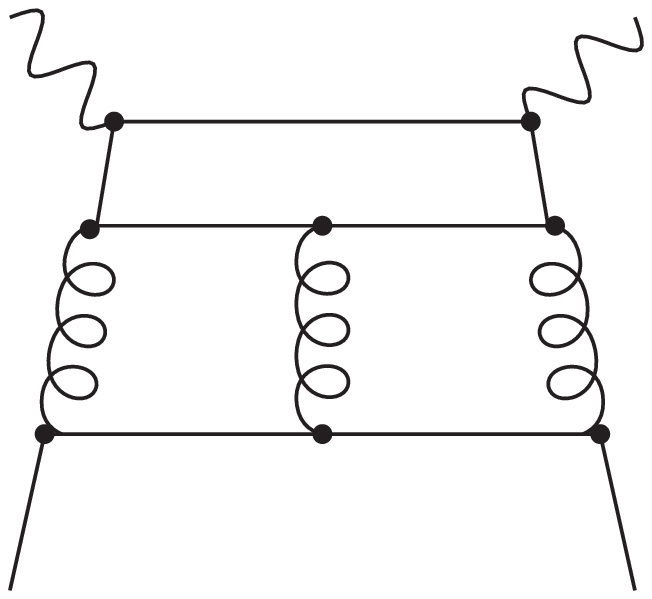}
  \includegraphics[width=3.0cm]{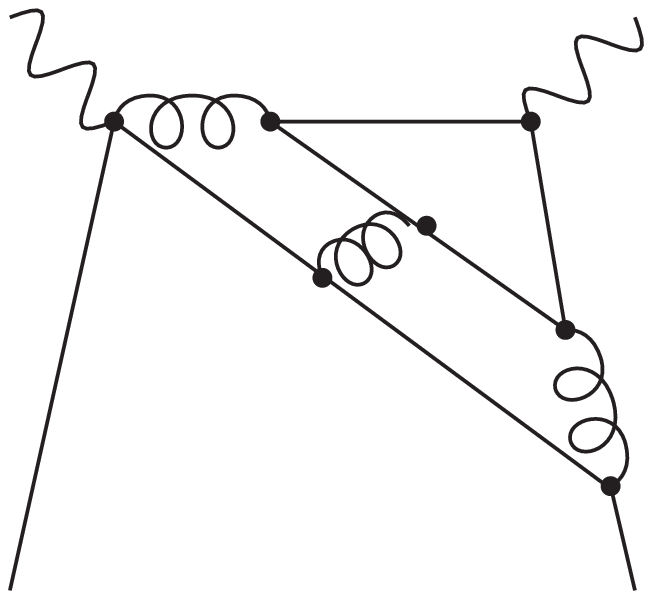}
  \includegraphics[width=3.0cm]{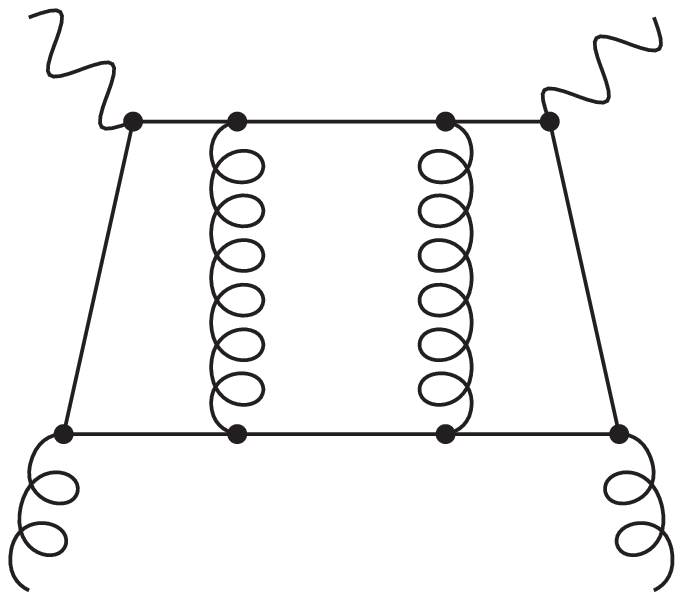}
  \includegraphics[width=3.0cm]{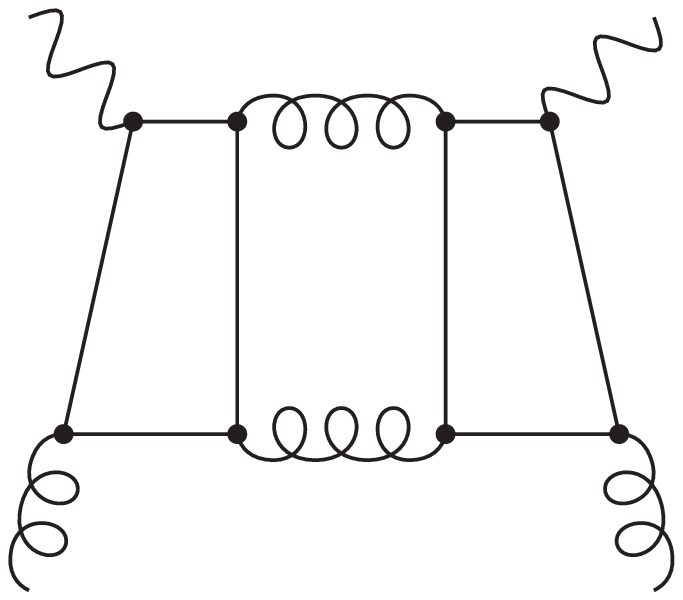}
\end{center}
\vspace{-5mm}
\caption{\label{fig:flavour}
 Representative three-loop diagrams of the flavour classes 
 (from left to right) $fl_{2}$, $fl_{02}$ and $fl_{11}$ for 
 photon--quark scattering and $fl^{\,g}_{2}$ and $fl^{\,g}_{11}$ 
 for photon--gluon scattering.}
\vspace*{3mm}
\end{figure}

We close this section by briefly noting that a new flavour structure
enters at the third order in $\as$. In this flavour structure, denoted
by $fl_{11}$ below, the in- and outgoing photons couple to different 
quark lines, see Fig.~\ref{fig:flavour}. The corresponding flavour 
factors are listed in Table \ref{tab:flavour}. Note that these diagrams
do not upset the $\lambda^{\alpha}$ independence of the non-singlet 
quantities, as discussed in Ref.~\cite{Larin:1994vu}.
 
\begin{table}[ht]
\label{tab:flavour}
\vspace{2mm}
\begin{center}
\begin{tabular}{c c c c c c}
\hline & & & & & \\[-3mm]
flavour factor &$fl_{2}$ &$fl_{02}$ &$fl_{11}$ &$fl^{\,g}_{2}$& 
$fl^{\,g}_{11}$
\\[1mm] \hline & & & & \\[-3mm]
non-singlet & 1 & 0 & $3 \langle e \rangle$ &  -- & -- \\[1mm]
singlet & 1 & 1 & $\displaystyle {\langle e \rangle^2 \over 
\langle e^{\,2} \rangle }$ & 1 & $\displaystyle {\langle e \rangle^2 
\over \langle e^{\,2} \rangle }$ \\[3mm] \hline
\end{tabular}
\end{center}
\vspace{-3mm}
\caption{The charge factors for the flavour topologies entering up to
 three loops, see also Ref.~\cite{Larin:1997wd}.}
\end{table}

%
\setcounter{equation}{0}
\section{Method of the calculation}
\label{sec:method}
%
%
In this section we discuss selected aspects relevant to our $N$-space 
calculation of the three-loop coefficient functions in DIS. Recalling
Eq.~(\ref{eq:OPE}) we need to extract, analytically, the coefficients 
of $(2p \cdot q)^N/Q^{2N}$ of the partonic forward-Compton amplitudes
(\ref{eq:Tpart}) up to the third order. While one of the two-loop 
topologies with a self-energy insertion is also not too simple, we will 
focus on genuine three-loop integrals, which are required to order 
${\cal O}(1)$ in the Laurent series in $\epsilon$. Various aspects of 
our approach have been discussed already in Refs.~\cite
{Moch:1999eb,Moch:2002sn,Vermaseren:2002rn,Moch:2004pa,Vogt:2004mw,%
Moch:2004sf,Vogt:2004gi}.
In particular, the key idea to systematically determine reduction 
identities based on sets of derivative equations for the $N$-th Mellin
moment of a given loop integral, the solution of which leads to 
harmonic sums, has been explained before in these articles.

\begin{figure}[ht]
\vspace{-1mm}
\begin{center}
  \includegraphics[width=5.4cm]{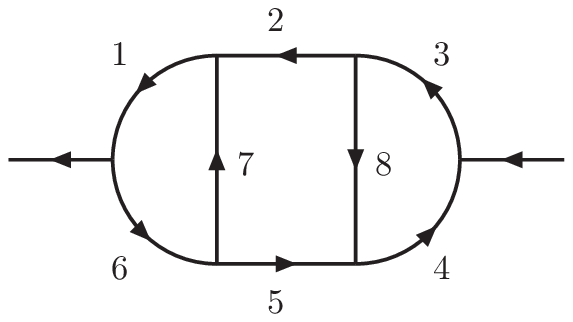}
  \includegraphics[width=5.4cm]{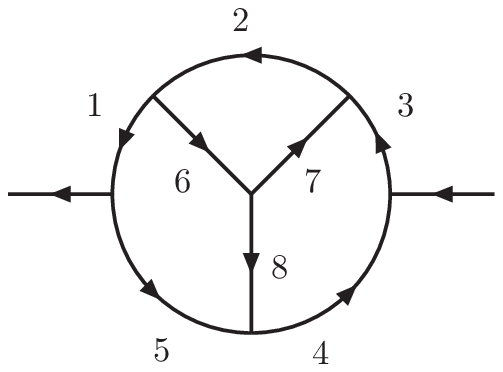}
  \includegraphics[width=5.4cm]{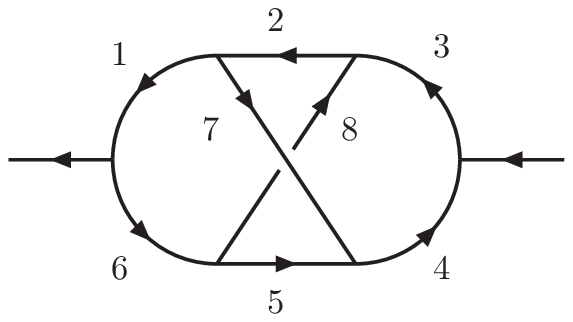}
\end{center}
\vspace{-4mm}
\caption{
\label{fig:bbbtop}
The top-level two-point topologies ${\rm LA}$ (left), ${\rm BE}$ 
(center) and ${\rm NO}$ (right) at three loops. The arrows indicate 
the assigned momentum flow. The external momentum is $q$ with 
$q^2 \neq 0$.
}
\end{figure}

\begin{figure}[ht]
\vspace{-1mm}
\begin{center}
  \includegraphics[width=5.4cm]{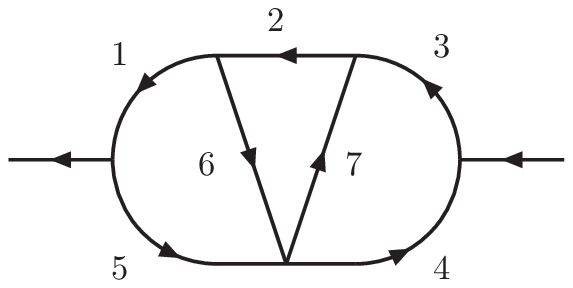}
  \includegraphics[width=5.4cm]{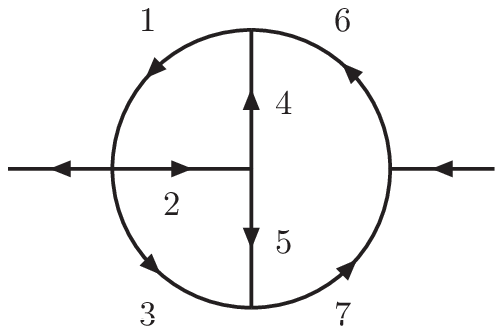}
\end{center}
\vspace{-4mm}
\caption{
\label{fig:bbbderived}
As Fig.~\ref{fig:bbbtop}, but for the second-level topologies 
${\rm FA}$ (left) and ${\rm BU}$ (right).  
}
\end{figure}

Let us start with a brief overview of the loop topologies.
We need to calculate massless four-point integrals with external momenta
$p$, $p^2 = 0$, and $q$, $q^2 \neq 0$. Classifying the topologies of 
these integrals is a two-stage process. 
It begins with two-point functions of the external momentum~$q$. Here 
we follow the notations of Refs.~\cite{Gorishnii:1989gt,Larin:1991fz}
recalled in Fig.~\ref{fig:bbbtop} for the top-level topologies, the 
ladder (${\rm LA}$), benz (${\rm BE}$) and non-planar (${\rm NO}$)
topologies. 
Other three-loop topologies are special cases of ${\rm LA}$, ${\rm BE}$
or ${\rm NO}$, with one or more of the lines $1,\dots,8$ missing. A 
complete list is provided in Table~\ref{tab:listoftopos} below. 
The most important examples denoted ${\rm FA}$ and ${\rm BU}$ are shown
in Fig.~\ref{fig:bbbderived}.


Subsequently all four-point functions can be constructed from these 
two-point functions by attaching two $p$-dependent external legs in 
all topologically independent ways to the various lines.
When the four-point functions have been constructed in this way, we
are referring to subtopologies. 
For instance, ${\rm NO}_{25}$ is a subtopology of type ${\rm NO}$ 
in which the momentum $p$ enters in line 2 and leaves in line 5, 
employing the numbering of the ${\rm NO}$ topology as in 
Fig.~\ref{fig:bbbtop}. 
Then we define basic building blocks (BBBs) as integrals in which both 
the incoming and the outgoing $p$-momentum are attached to the same 
line as, for instance, in ${\rm LA}_{11}$. Composite building blocks 
(CBBs), on the other hand, have incoming and outgoing $p$-momentum 
attached to different lines, as in the case of ${\rm NO}_{25}$ 
mentioned above. 
At the top level, there are 10 BBBs (3 ${\rm LA}$, 5 ${\rm BE}$ and 
2 ${\rm NO}$) and 32 CBBs (10 ${\rm LA}$, 16 ${\rm BE}$ and 
6 ${\rm NO}$), the smaller number of non-planar topologies being due 
to symmetries. 

Solving all four-point integrals in Mellin $N$-space in terms of 
harmonic sums \cite{Gonzalez-Arroyo:1979df,Gonzalez-Arroyo:1980he,%
Vermaseren:1998uu,Blumlein:1998if,Moch:2001zr} and the values 
$\zeta_{3}$, $\zeta_{4}$ and $\zeta_{5}$ of the Riemann 
\mbox{$\zeta$-function} requires an elaborate reduction scheme. This 
scheme is derived from algebraic relations based on integration by parts
\cite{'tHooft:1972fi,Tkachov:1981wb,Chetyrkin:1981qh,Tkachov:1984xk}, 
scaling equations, form-factor analysis~\cite{Passarino:1979jh} and 
some equations~\cite{Moch:2004pa} that fall in a special category 
because they involve higher twist and a careful study of the 
parton-momentum limit $p\cdot\! p \rightarrow 0$. 
 
Within this scheme, as a first step, we systematically reduce all CBB
integrals to such of BBB type. This is necessary because a direct 
application of the projection operator (\ref{eq:PNop}) on the CBBs is
not recommendable.  The resulting brute-force expansions would generate
sums which are not in the class of single-parameter nested sums, and 
could therefore not be solved with the algorithms used to express the 
result in harmonic sums \cite{Gonzalez-Arroyo:1979df,%
Gonzalez-Arroyo:1980he,Vermaseren:1998uu,Blumlein:1998if,Moch:2001zr}.
The second step of the reduction scheme consists of successively 
simplifying the topologies. For example, if one line is removed from 
the top-level diagrams of Fig.~\ref{fig:bbbtop}, its topology is 
reduced according to  
\beq
\label{eq:topomap}
  {\rm NO} \:\longrightarrow\: {\rm BU,\:  FA}
  \, \:\: , \qquad
  {\rm BE} \:\longrightarrow\: {\rm BU,\:  FA,}\:  \ldots
  \, \:\: , \qquad
  {\rm LA} \:\longrightarrow\: {\rm FA,}\: \ldots  \:\: ,
\eeq
where topologies below the level of Fig.~\ref{fig:bbbderived} have not
been written out.

The main problem we are faced with, as compared to the corresponding 
two-loop calculation~\cite{Moch:1999eb}, is that the reduction 
equations become much more complicated. This requires extensive 
automatization and the standard approach proceeds as follows.
One writes down all equations based on the relations between integrals
mentioned above and combines them to construct equations that can 
systematically bring the powers of the denominators in a given integral
down, either reducing them to zero or leaving them at a fixed unique 
value. When a line is eliminated a simpler topology or subtopology is 
reached. Then we can refer to the reduction equations for that topology
and so on. Eventually, this procedure will lead to an integral simple 
enough to be evaluated.

Different methods have emerged over the last years for practical 
implementations. 
One approach, commonly referred to as the Laporta algorithm~\cite
{vanRitbergen:1999fi,Laporta:2000dc,Laporta:2001dd,Schroder:2002re}, 
consists of systematically solving the system of linear equations of a 
given topology for a set of integrals with fixed numerical combinations
of powers of numerators and denominators.
The result for each integral of the set is one algebraic relation 
in terms of a unique (small) set of so-called master integrals.
This approach works well for a large variety of processes and has been 
successfully automated~\cite{Anastasiou:2004vj,Awramik:2004qv}.

Another approach follows the original {\sc Mincer} paper~\cite
{Gorishnii:1989gt,Larin:1991fz} and also attempts to solve the system 
of linear equations for a given topology, but using symbolic lowering 
(and raising) operators for all numerators and denominators occurring
in a given topology. This leads to a chain of algebraic relations for 
any given integral which maps it, for fixed numerical values of the
numerator and denominator powers, to the same unique set of master 
integrals.
 
We have adopted the {\sc Mincer} approach (although not in a fully
systematic manner for some very difficult subtopologies) for two 
reasons.  Firstly, we have encountered such a huge number of different
integrals that we have made no attempt at a fully complete tabulation.
Secondly, and more importantly, because we need to calculate the Mellin 
moments (coefficient of $(2p\cdot q)^N/Q^{2N\,}$) of the integrals for 
symbolic $N$, it is necessary to look for operator relations. 
For a given integral $I(N)$ these operator relations give rise to 
difference equations, which can generally be written as 
\beq
\label{eq:diffeq}
  a^{}_0(N)\, I(N) + a^{}_1(N)\,  I(N-1) + \ldots + a_m(N)\, I(N-m) 
  \: = \: G(N) \:\: ,
\eeq
where the inhomogeneous term $G(N)$ collects the simpler integrals
resulting, e.g., from removing one or more lines. 
Such recursion relations in $N$ (or, in the above terminology, $m$-th 
order difference equations) were introduced to loop calculations in 
Ref.~\cite{Kazakov:1988jk}. The solution of Eq.~(\ref{eq:diffeq}) 
requires $m$ boundary conditions $I(0), \dots, I(m-1)$, which can be
computed with the standard {\sc Mincer} techniques~\cite
{Gorishnii:1989gt,Larin:1991fz}.

Single-step difference equations can be summed analytically in a closed
form. The solution of Eq.~(\ref{eq:diffeq}) for $m=1$ reads
\beq
\label{eq:firstsol}
  I(N) \: = \: 
  \frac{\prod_{j=1}^N a^{}_1(j)}{\prod_{j=1}^N a^{}_0(j)}\: I(0)
  \: + \: \sum_{i=1}^N\frac{\prod_{j=i+1}^N a^{}_1(j)}{\prod_{j=i}^N 
  a^{}_0(j)}\: G(i) \:\: .
\eeq
In case that the functions $a_i(N)$ can be factorized in linear 
polynomials in $N$ of the type $N + m + n\, \epsilon$ with integer 
$m,n$, the products can be written as combinations of 
\mbox{$\Gamma$-functions}. 
In the presence of parametric dependence on $\epsilon$ the 
\mbox{$\Gamma$-functions} should be expanded around $\epsilon = 0$. 
This will lead to factorials and harmonic sums. 
If the function $G(N)$ is expressed as a Laurent series in $\epsilon$ 
with the coefficients being combinations of harmonic sums in $N+m$ and 
powers of $N+m$, with $m$ a fixed integer, the sum in 
Eq.~(\ref{eq:firstsol}) can be performed, and $I(N)$ is expressed as a 
combination of harmonic sums in $N+k$ and powers of $N+k$, where $k$ 
is a fixed integer. A condition for a solution in terms of harmonic
sums, to which we will refer later below Eq.~(\ref{eq:la27box}), is 
that the highest powers in $N$ in the polynomials $a^{}_0$ and 
$a^{}_1$ have prefactors with the same modulus. 

Higher-order difference equations require a completely different 
approach. Under conditions which are fulfilled by all cases we have 
encountered in the present calculation, their solution can be expressed 
in terms of harmonic sums, and hence a corresponding ansatz can by 
used. Suppose we have for some integral $I(N)$ a relation like 
Eq.~(\ref{eq:diffeq}) with $m \ge 2$. The inhomogeneous function $G(N)$ 
is again assumed to be a Laurent series in $\epsilon$ with the 
coefficients being combinations of harmonic sums in $N+k$ with a fixed
integer $k$. The functions $a_i(N)$ in Eq.~(\ref{eq:diffeq}) are 
polynomials in $N$ and $\epsilon$ subject to certain conditions.
In order to solve for $I(N)$ under these assumptions, we write down an 
ansatz in powers of $\epsilon$, harmonic sums $S_{\vec{m}}$ of given 
weight, possibly in combination with values of the 
\mbox{$\zeta$-function} and factors $(-1)^N$.
The harmonic sums have arguments $N+l$, where the integer $l$ samples 
the various offsets. One may have to introduce positive powers of $N^k$ 
multiplying the harmonic sums as well (see also the discussion below).
Thus $I(N)$ is written as 
\begin{eqnarray}
\label{eq:ansatz}
I(N) & = &
  \sum\, c_+(j,k,l,{\vec{m}})\, \ep^j\, (N-l)^k\, S_{\vec{m}}(N-l) + 
\nn \\[0.5mm]
& &
  \sum\, c_-(j,k,l,{\vec{m}})\, \ep^j\, (-1)^N\, (N-l)^k\, 
  S_{\vec{m}}(N-l) + 
\nn \\[0.5mm]
& &
  \sum\, c_+(j,k,l,{\vec{m}},n)\, \ep^j\, (N-l)^k\, \zeta_n\, 
  S_{\vec{m}}(N-l) + 
\nn \\[0.5mm]
& &
  \sum\, c_-(j,k,l,{\vec{m}},n)\, \ep^j\, (-1)^N\, (N-l)^k\, 
  \zeta_n\, S_{\vec{m}}(N-l) \:\: .
\end{eqnarray}
The sum runs over a suitable set of the parameter space spanned by the 
powers $\epsilon^j$, the weight ${\vec m}$ of the harmonic sums, 
positive powers $N^k$, and the offset $l$ in the argument of the 
harmonic sums.
For efficiency, it is important to take into account the correlation 
between the loop order and the weight of harmonic sums in the ansatz, 
i.e. in the choice of the parameter set $j,k,l,{\vec{m}},n$. 
In the case of DIS structure functions, one- (two-, three-) loop 
integrals can be expressed, at order $\epsilon^0$, in terms of harmonic
sums up to weight two (four, six). Accordingly, the single (double, 
triple) pole terms in $\epsilon$ are expressed through harmonic sums 
with maximal weights decreased by one (two, three).

The solution for the integral $I(N)$ is obtained by determining the 
coefficients $c_-$ and $c_+$. To that end, we insert the ansatz 
(\ref{eq:ansatz}) into Eq.~(\ref{eq:diffeq}) and normalize the left 
hand side by pulling all expressions back to the unique basis in 
harmonic sums. 
This synchronization can be performed with the algorithms of the 
{\sc Summer} package~\cite{Vermaseren:1998uu} in {\sc Form}. The 
coefficients of all individual terms such as $\ep^j\, (-1)^N\, 
(N-l)^k\, S_{\vec{m}}(N-l)\, \zeta_n$ then determine a set of linear 
equations for the unknown $c_-$ and $c_+$ in our ansatz 
(\ref{eq:ansatz}), which can be solved by standard means if the 
chosen parameter set $j,k,l,{\vec{m}},n$ was large enough.
In practice, an iterative procedure for the determination of the 
coefficients $c_-$ and $c_+$ is advantageous, since an improved ansatz 
reduces the size of the system of equations. We will give an explicit 
example for a two-step difference equation below.

\begin{figure}[ht]
\begin{center}
\begin{tabular}{cccc}
\begin{minipage}{3.0cm}
\begin{center}
  \includegraphics[width=3.0cm]{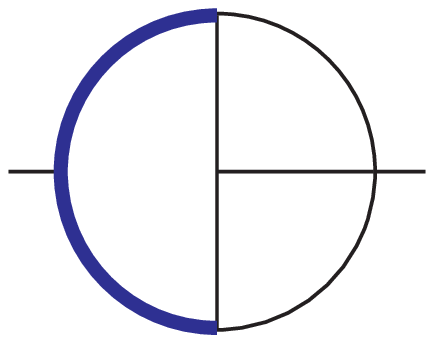}
\end{center}
\end{minipage}
&
$\displaystyle
= {\:\:(2 p \cdot q)^N \over (Q^2)^{N+\alpha}}\: {\rm BE}_{15}(N)\:\: ,
$
&
\begin{minipage}{3.0cm}
\begin{center}
  \includegraphics[width=3.0cm]{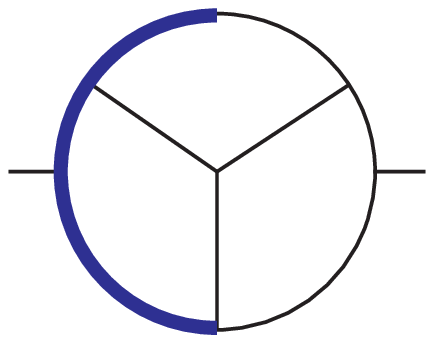}
\end{center}
\end{minipage}
&
$\displaystyle
= {\:\:(2 p \cdot q)^N \over (Q^2)^{N+\alpha}}\: {\rm BE}_{25}(N)
$
\end{tabular}
\end{center}
\vspace{-3mm}
\caption{
\label{fig:basicbe}
Generic ${\rm BE}_{15}$ (${\rm BU}$-type) and ${\rm BE}_{25}$ 
subtopologies. The momenta $q$ and $p$ (fat lines) flow from right to 
left and from top to bottom through the diagram, respectively.
}
\end{figure}
\begin{figure}[ht]
\begin{center}
\begin{tabular}{cccc}
\begin{minipage}{3.5cm}
\begin{center}
  \includegraphics[width=3.5cm]{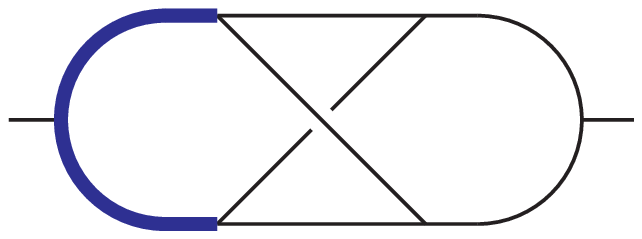}
\end{center}
\end{minipage}
&
$\displaystyle
= {\:\:(2 p \cdot q)^N \over (Q^2)^{N+\alpha}}\: {\rm NO}_{16}(N)\:\: ,
$
&
\begin{minipage}{3.5cm}
\begin{center}
  \includegraphics[width=3.5cm]{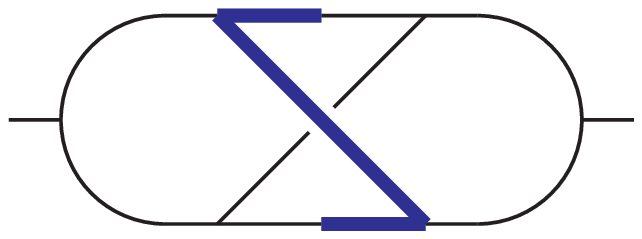}
\end{center}
\end{minipage}
&
$\displaystyle
= {\:\:(2 p \cdot q)^N \over (Q^2)^{N+\alpha}}\: {\rm NO}_{25}(N)
$
\end{tabular}
\end{center}
\vspace{-3mm}
\caption{
\label{fig:basicno}
As Fig.~\ref{fig:basicbe}, but for the generic ${\rm NO}_{16}$ and 
${\rm NO}_{25}$ topologies.
}
\end{figure}

A crucial issue in the derivation of reduction relations is the 
implementation of symmetries of the Feynman integrals under 
consideration, because the discrete symmetries are reflected in the 
recursion relations.  In the presence of an even-$N$ symmetry, the odd 
coefficients $a^{}_1, a^{}_3, \dots $ vanish in Eq.~(\ref{eq:diffeq}).
Particular examples which we like to mention here are the even-$N$ 
symmetry of the ${\rm BE}_{15}$ (${\rm BU}$-type), ${\rm BE}_{25}$, 
${\rm NO}_{16}$ and ${\rm NO}_{25}$ topologies shown in 
Figs.~\ref{fig:basicbe} and~\ref{fig:basicno}.
Here and below, the fat lines indicate the flow of the parton momentum
$p$ through the diagram. 
The equations in Figs.~\ref{fig:basicbe} and~\ref{fig:basicno} indicate
that we calculate the $N$-th Mellin moment of the respective diagram,
given precisely by the dimensionless functions of $N$ written on the
right-hand sides. 

The discrete symmetry in $N$ of the examples displayed in 
Figs.~\ref{fig:basicbe} and~\ref{fig:basicno} is realized as follows.
${\rm BE}_{15}$ which is actually a certain ${\rm BU}$-type since the 
line 3 is missing, can be turned upside down under interchange of the 
pairs of lines (1,5), (2,4) and (6,8) (see the labeling in 
Fig.~\ref{fig:bbbtop}) and $p \to -p$, which introduces a factor of 
$(-1)^N$. 
${\rm BE}_{25}$ can be flipped around the vertical axis, interchanging
the pairs of lines (1,3), (4,5) and (6,7) and $q \to -q$, which again 
leads to a factor of $(-1)^N$. In addition, the $p$-momentum flow has 
to be rerouted internally. The situation is similar for ${\rm NO}_{16}$,
which can be turned upside down, and for ${\rm NO}_{25}$. In the latter
case two symmetry operations are possible, either turning it upside 
down or flipping around the vertical axis, both choices requiring also 
the $p$-momentum flow to be rerouted internally.

To illustrate the discussion of recursion relations and symmetries, let 
us present as an example the basic integral of the ${\rm NO}_{25}$ 
topology. The general scalar ${\rm NO}_{25}$ integral is defined by
\bea
\label{eq:basicno25}
{\lefteqn{
{\rm NO}_{25}(N;n_1,\dots,n_6,n_8,m_2,m_5,m_7) \:\: = \:\: 
(Q^2)^{n_1+\dots+n_6+n_8+m_2+m_5+m_7-6+3\epsilon} \:\cdot }}
\nn \\
&& \cdot \: P_N\: \int\, \prod_{n=1}^3\, d^Dl_n \:
\Big[ (l_1)^{2\, n_1}\, (l_2)^{2\, n_2}\, (l_2+p)^{2\, m_2}\, 
(l_3)^{2\, n_3}\, (l_3-q)^{2\, n_4}\, (l_1-l_2+l_3-q)^{2\, n_5} \:\cdot
\nn \\[-1mm]
&& \qquad\qquad \cdot \:
(l_1-l_2+l_3-q-p)^{2\, m_5}\, (l_1-q)^{2\, n_6}\,
(l_2-l_1+p)^{2\, m_7}\, (l_2-l_3)^{2\, n_8} \Big]^{-1} \:\: ,
\eea
where $P_N$ represents the Mellin-$N$ projection (\ref{eq:PNop}), 
and $l_{1,2,3}$ denote the loop momenta in the notation of 
Fig.~\ref{fig:bbbtop}.  The integral we now consider was among the most
complicated ones of the whole calculation,
\beq
\label{eq:basicdef}
   {\rm NO}_{25}(N;1,1,1,1,1,1,1,1,1,1) 
   \:\ \equiv \:\: {\rm NO}_{25}(N;{\bf 1}_{10})\:\: ,
\eeq
where we have introduced a short-hand notation for $m$ identical 
arguments $i$,
\beq
\label{eq:shorthand}
{\bf i}_m \:\: = \:\: \underbrace{i,\dots,i}_{m} \:\: .
\eeq
We obtain a two-step recursion relation, $m=2$ in Eq.~(\ref{eq:diffeq}),
for ${\rm NO}_{25}(N;{\bf 1}_{10})$ with the coefficients 
\begin{eqnarray}
\label{eq:no25coeffs}
a_0(N) &=& - (N+4+4 \* \epsilon) \* (N+2) \* (N+1-2 \* \epsilon)\, , 
\nonumber \\
a_1(N) &=& 0\, ,\nonumber \\
a_2(N) &=& (N+4+6 \* \epsilon) \* (N+3+3 \* \epsilon) \* (N+2+3 \* 
\epsilon) \, .
\end{eqnarray}
Using the short-hand notation (\ref{eq:shorthand}) the corresponding 
function $G_{\,\rm NO}(N)$, expressed in terms of simpler (sub-) 
topologies, reads 
\begin{eqnarray}
\label{eq:basicNO25rec}
{\lefteqn{G_{\,\rm NO}(N) \:\: = \:\:}} 
\nonumber\\
& &
(N+4+4 \* \epsilon) \* (N+2-2 \* \epsilon) \* (N+1-2 \* \epsilon) \* 
\bigl\{
     {\rm NO}_{25}(N+1;1,0,{\bf 1}_{8}) 
    -{\rm NO}_{25}(N+1;{\bf 1}_{4},0,{\bf 1}_{5})
\nonumber\\
&&
    -{\rm NO}_{25}(N+1;{\bf 1}_{7},0,{\bf 1}_{2})
    +{\rm NO}_{25}(N+1;{\bf 1}_{8},0,1)
\bigr\}
\nonumber\\
&&
+(N+4+4 \* \epsilon) \* (N+1-2 \* \epsilon) \* 
\bigl\{
     {\rm NO}_{25}(N;2,{\bf 1}_{8},0)
    -{\rm NO}_{25}(N;1,0,2,{\bf 1}_{7})
    -{\rm NO}_{25}(N;{\bf 1}_{3},2,{\bf 1}_{5},0)
\nonumber\\
&&
    -2 \* {\rm NO}_{25}(N;1,0,{\bf 1}_{5},2,{\bf 1}_{2})
    -{\rm NO}_{25}(N;{\bf 1}_{5},2,0,{\bf 1}_{3})
    +{\rm NO}_{25}(N;{\bf 1}_{3},2,{\bf 1}_{4},0,1)
\nonumber\\
&&
    +{\rm NO}_{25}(N;{\bf 1}_{2},2,{\bf 1}_{3},0,{\bf 1}_{3})
    +{\rm NO}_{25}(N;{\bf 1}_{4},0,2,{\bf 1}_{4})
\bigr\}
\nonumber\\
&&
+(N+4+4 \* \epsilon) \* (N+3+3 \* \epsilon) \* (N+1-2 \* \epsilon) \* 
\bigl\{
     {\rm NO}_{25}(N;{\bf 1}_{2},0,{\bf 1}_{7})
    -{\rm NO}_{25}(N;{\bf 1}_{5},0,{\bf 1}_{4})
\nonumber\\
&&
    -{\rm NO}_{25}(N;{\bf 1}_{3},0,{\bf 1}_{6})
    +{\rm NO}_{25}(N;0,{\bf 1}_{9})
\bigr\}
+2 \* (N+2+\epsilon) \* (N+1-2 \* \epsilon) \* 
\bigl\{
     {\rm NO}_{25}(N;2,{\bf 1}_{3},0,{\bf 1}_{5})
\nonumber\\
&&
    -{\rm NO}_{25}(N;2,{\bf 1}_{7},0,1)
\bigr\}
+(N+3+3 \* \epsilon) \* (N+1-2 \* \epsilon) \* (N-2 \* \epsilon) \* 
\bigl\{
     {\rm NO}_{25}(N;{\bf 1}_{4},0,{\bf 1}_{5})
\nonumber\\
&&
    -{\rm NO}_{25}(N;{\bf 1}_{8},0,1)
    +{\rm NO}_{25}(N;{\bf 1}_{7},0,{\bf 1}_{2})
    -{\rm NO}_{25}(N;1,0,{\bf 1}_{8})
\bigr\}
\nonumber\\
&&
-(N+1-2 \* \epsilon) \* (N-2 \* \epsilon) \* 
{\rm NO}_{25}(N;2,{\bf 1}_{6},0,{\bf 1}_{2})
-2 \* (2+3 \* \epsilon) \* (N+1-2 \* \epsilon) \* 
{\rm NO}_{25}(N;2,0,{\bf 1}_{8})
\nonumber\\
&&
+2 \* (N+2+\epsilon) \* 
\bigl\{
      {\rm NO}_{25}(N-1;2,{\bf 1}_{4},2,0,{\bf 1}_{3})
     -{\rm NO}_{25}(N-1;2,0,2,{\bf 1}_{7})
\nonumber\\
&&
     -{\rm NO}_{25}(N-1;2,{\bf 1}_{3},0,2,{\bf 1}_{4})
     +{\rm NO}_{25}(N-1;2,1,2,{\bf 1}_{3},0,{\bf 1}_{3})
\bigr\}
\nonumber\\
&&
+2 \* (2+3 \* \epsilon) \* 
\bigl\{
      {\rm NO}_{25}(N-1;2,{\bf 1}_{2},2,{\bf 1}_{4},0,1)
     -{\rm NO}_{25}(N-1;2,{\bf 1}_{2},2,{\bf 1}_{5},0)
\nonumber\\
&&
     +2 \* {\rm NO}_{25}(N-1;3,{\bf 1}_{6},0,{\bf 1}_{2})
     -2 \* {\rm NO}_{25}(N-1;3,{\bf 1}_{8},0)
\bigr\}
\nonumber\\
&&
+(N+3+3 \* \epsilon) \* (N-2 \* \epsilon) \* 
\bigl\{
      2 \* {\rm NO}_{25}(N-1;{\bf 1}_{8},0,2)
     -2 \* {\rm NO}_{25}(N-1;2,{\bf 1}_{5},0,{\bf 1}_{3})
\nonumber\\
&&
     +{\rm NO}_{25}(N-1;1,0,2,{\bf 1}_{7})
     -{\rm NO}_{25}(N-1;{\bf 1}_{5},2,0,{\bf 1}_{3})
     +{\rm NO}_{25}(N-1;{\bf 1}_{3},2,{\bf 1}_{5},0)
\nonumber\\
&&
     +{\rm NO}_{25}(N-1;{\bf 1}_{4},0,2,{\bf 1}_{4})
     -2 \* {\rm NO}_{25}(N-1;{\bf 1}_{3},0,{\bf 1}_{5},2)
     -{\rm NO}_{25}(N-1;{\bf 1}_{3},2,{\bf 1}_{4},0,1)
\nonumber\\
&&
     -{\rm NO}_{25}(N-1;{\bf 1}_{2},2,{\bf 1}_{3},0,{\bf 1}_{3})
     +{\rm NO}_{25}(N-1;2,{\bf 1}_{6},0,{\bf 1}_{2})
     +2 \* {\rm NO}_{25}(N-1;2,{\bf 1}_{7},0,1)
\nonumber\\
&&
     -{\rm NO}_{25}(N-1;2,{\bf 1}_{8},0)
\bigr\}
+2 \* (2+3 \* \epsilon) \* (N+3+3 \* \epsilon) \* 
\bigl\{
      {\rm NO}_{25}(N-1;2,{\bf 1}_{2},0,{\bf 1}_{6})
\nonumber\\
&&
     -{\rm NO}_{25}(N-1;2,1,0,{\bf 1}_{7})
     -{\rm NO}_{25}(N-1;2,{\bf 1}_{4},0,{\bf 1}_{4})
\bigr\}
\nonumber\\
&&
+2 \* (N+3+3 \* \epsilon) \* 
\bigl\{
      {\rm NO}_{25}(N-2;2,{\bf 1}_{3},0,2,{\bf 1}_{4})
     -{\rm NO}_{25}(N-2;2,{\bf 1}_{4},2,0,{\bf 1}_{3})
\nonumber\\
&&
     -{\rm NO}_{25}(N-2;2,1,2,{\bf 1}_{3},0,{\bf 1}_{3})
     +{\rm NO}_{25}(N-2;2,0,2,{\bf 1}_{7})
\bigr\}
\nonumber\\
&&
+(N+3+3 \* \epsilon) \* (N+2+3 \* \epsilon) \* (N-2 \* \epsilon) \* 
\bigl\{
      {\rm NO}_{25}(N-1;{\bf 1}_{5},0,{\bf 1}_{4})
     -{\rm NO}_{25}(N-1;0,{\bf 1}_{9})
\nonumber\\
&&
     -{\rm NO}_{25}(N-1;{\bf 1}_{2},0,{\bf 1}_{7})
     +{\rm NO}_{25}(N-1;{\bf 1}_{3},0,{\bf 1}_{6})
\bigr\}
\nonumber\\
&&
+2 \* (N+3+3 \* \epsilon) \* (N+2+3 \* \epsilon) \* 
\bigl\{
      {\rm NO}_{25}(N-2;1,0,{\bf 1}_{5},2,{\bf 1}_{2})
     -{\rm NO}_{25}(N-2;{\bf 1}_{4},0,2,{\bf 1}_{4})
\nonumber\\
&&
     +{\rm NO}_{25}(N-2;{\bf 1}_{3},0,{\bf 1}_{5},2)
     -{\rm NO}_{25}(N-2;{\bf 1}_{8},0,2)
     +{\rm NO}_{25}(N-2;{\bf 1}_{5},2,0,{\bf 1}_{3})
\nonumber\\
&&
     -{\rm NO}_{25}(N-2;2,{\bf 1}_{3},0,{\bf 1}_{5})
     +{\rm NO}_{25}(N-2;2,{\bf 1}_{5},0,{\bf 1}_{3})
\bigr\} 
\, .
\end{eqnarray}
Each term of $G_{\,\rm NO}(N)$ has to be determined in terms of harmonic
sums by finding and solving the appropriate reduction equations (which 
in turn involve simpler integrals which have to be solved by more 
reduction equations, etc). Once all that has been done, $G_{\,\rm NO}
(N)$ can finally be used to solve the difference equation
(\ref{eq:diffeq}) with the coefficients~(\ref{eq:no25coeffs}) for 
${\rm NO}_{25}(N;{\bf 1}_{10})$, using an ansatz~(\ref{eq:ansatz}).
 
Since the complete results for both $G_{\,\rm NO}(N)$ and 
${\rm NO}_{25}(N;{\bf 1}_{10})$ contain of the order of $1000$
terms, they are too long to be presented here. However we write down 
the two leading poles in $\ep$ for illustration. We choose a compact 
representation for the harmonic sums, employing the abbreviation 
$S_{\vec{m}}\,\equiv\, S_{\vec{m}}(N)$, together with the notation
\beq
\label{eq:shiftN}
  \Npm \, S_{\vec{m}} \: = \: S_{\vec{m}}(N \pm 1) \:\: , \quad\quad
  \Npmi\, S_{\vec{m}} \: = \: S_{\vec{m}}(N \pm i) \:\: ,
\eeq
for arguments shifted by $\pm 1$ or a larger integer $i$.
In the G-scheme~\cite{Gorishnii:1989gt,Chetyrkin:1980pr}, in which 
$l$-loop integrals are divided by the $l$-th power of the basic 
massless one-loop integral, $G_{\,\rm NO}(N)$ then reads 
\bea
&& G_{\,\rm NO}(N) \:\: = \:\:
        {16 \over 3 \* \epsilon^3} \* \bigl( 1 + \sign(N) \bigr) \* 
	\biggl(
            6
	  - ( 
	    8
	  + 6 \* N
	  + N^2
	  ) \* 
	  \bigg[
	      3 \* \S(-3)
	    - 2 \* \Ss(-2,1)
            + \S(3)
	  \bigg]
          - ( 
            13
	  + 4 \* N
	  ) \* 
	  \bigg[
	      \S(-2)
\nonumber\\&& \mbox{} \quad
            + \S(2)
	  \bigg]
          - 7 \* \S(1)
          - 3 \* \Nplustwo \* \S(1)
          + (
	    3
	  - 3 \* \Nplus
          - \Nplustwo
          + \Nplusthree ) \* 
	  \bigg[
	      \Ss(1,-2)
	    + \Ss(1,1)
	    + \Ss(1,2)
	    + 2 \* \Ss(2,1)
            - \S(2)
	    - 2 \* \S(3)
	  \bigg]
\nonumber\\&& \mbox{} \quad
          + 11 \* \Ss(1,1)
          - 11 \* \S(2)
	  - \Nplus \*
 	  \bigg[
	      13 \* \Ss(1,1)
            - 17 \* \S(2)
	  \bigg]
	  +  2 \* \Nplustwo \*
 	  \bigg[
	      \Ss(1,1)
            - 3 \* \S(2)
	  \bigg]
          \biggr)
       + {16 \over 3 \* \epsilon^2} \* \bigl( 1 + \sign(N) \bigr) \* 
         ( - 24
\nonumber\\&& \mbox{} \quad
	  + ( 
	    8
	  + 6 \* N
	  + N^2
	  ) \* 
	  \bigg[
              {79 \over 4} \* \S(-4)
            + {37 \over 4} \* \S(-3)
            - {85 \over 4} \* \S(-3,1)
	    - \Ss(-2,-2)
            - {11 \over 2} \* \Ss(-2,1)
	    + 10 \* \Sss(-2,1,1)
            - {13 \over 2} \* \Ss(-2,2)
\nonumber\\&& \mbox{} \quad
            + {11 \over 2} \* \Sss(1,-2,1)
            - {37 \over 4} \* \Ss(1,-3)
            - {7 \over 2} \* \Ss(1,3)
            + 2 \* \Ss(2,-2)
            + \Ss(2,2)
            + {7 \over 2} \* \S(3)
            - {1 \over 2} \* \Ss(3,1)
            + {17 \over 4} \* \S(4)
	  \bigg]
	  - N \*
 	  \bigg[
            {93 \over 8} \* \S(-3)
\nonumber\\&& \mbox{} \quad
          - {17 \over 2} \* \S(-2)
          - {7 \over 4} \* \Ss(-2,1)
          + 9 \* \S(1,-2)
          + {21 \over 2} \* \S(1,2)
          - {41 \over 4} \* \S(2)
          + {11 \over 2} \* \S(2,1)
          + {11 \over 4} \* \S(3)
	  \bigg]
          - {87 \over 2} \* \S(-3)
          + {39 \over 4} \* \S(-2)
\nonumber\\&& \mbox{} \quad
          + 16 \* \Ss(-2,1)
          + 9 \* \S(1)
          - {111 \over 4} \* \Ss(1,-2)
          - {101 \over 2} \* \Ss(1,1)
          + 55 \* \Sss(1,1,1)
          - {437 \over 8} \* \Ss(1,2)
          + {329 \over 8} \* \S(2)
          - {393 \over 4} \* \Ss(2,1)
          + {345 \over 8} \* \S(3)
\nonumber\\&& \mbox{} \quad
          + (
	    3
	  - 3 \* \Nplus
          - \Nplustwo
          + \Nplusthree ) \* 
	  \bigg[
              {7 \over 4} \* \Ss(1,-3)
            + {7 \over 4} \* \Ss(1,-2)
	    - 2 \* \Sss(1,-2,1)
            - \Ss(1,1)
            + 5 \* \Sss(1,1,1)
            + {3 \over 2} \* \Sss(1,1,-2)
\nonumber\\&& \mbox{} \quad
            + {9 \over 4} \* \Sss(1,1,2)
            - {1 \over 4} \* \Ss(1,2)
            + {7 \over 4} \* \Sss(1,2,1)
            - {11 \over 4} \* \Ss(1,3)
            + \S(2)
            + 10 \* \Sss(2,1,1)
	    - {31 \over 4} \* \Ss(2,2)
            - 3 \* \Ss(2,-2)
            - {7 \over 2} \* \Ss(2,1)
            + {1 \over 2} \* \S(3)
\nonumber\\&& \mbox{} \quad
            - {83 \over 4} \* \Ss(3,1)
            + {31 \over 2} \* \S(4)
	  \bigg]  
	  + \Nplus \*
 	  \bigg[
            14 \* \S(1)
          + {1 \over 4} \* \Ss(1,-2)
          + {163 \over 8} \* \Ss(1,1)
          - 65 \* \Sss(1,1,1)
          + {227 \over 8} \* \Ss(1,2)
          - {113 \over 8} \* \S(2)
\nonumber\\&& \mbox{} \quad
          + {381 \over 4} \* \Ss(2,1)
          - {443 \over 8} \* \S(3)
	  \bigg]
	  +  \Nplustwo \* \
 	  \bigg[
            16 \* \S(1)
          + {1 \over 2} \* \Ss(1,-2)
          - {159 \over 8} \* \Ss(1,1)
          + 10 \* \Sss(1,1,1)
          - {27 \over 4} \* \Ss(1,2)
          + {57 \over 8} \* \S(2)
\nonumber\\&& \mbox{} \quad
          - 16 \* \Ss(2,1)
          + {41 \over 4} \* \S(3)
	  \bigg]
	\biggr)
\:\: + \:\: {\cal O}\! \left( \frac{1}{\ep} \right) \:\: .
\label{eq:no25GN}
\eea
Note the positive powers of $N$ multiplying some harmonic sums in 
Eq.~(\ref{eq:no25GN}). We will return to this issue below.
The boundary conditions for the ${\rm NO}_{25}(N;{\bf 1}_{10})$ 
integral of Eq.~(\ref{eq:basicdef}) are shorter,
\begin{eqnarray}
\label{eq:basicNO25boundary}
   {\rm NO}_{25}(0;{\bf 1}_{10}) &=&
      - \: {10 \over 3} \* {1 \over \epsilon^3}
   \: + \: {1 \over 3} \* {1 \over \epsilon^2}
   \: + \: {\cal O}\! \left( \frac{1}{\ep} \right)
   \:\:  , \nn\\
   {\rm NO}_{25}(1;{\bf 1}_{10}) &=& 0 \:\: .
\end{eqnarray}
Finally the following expression for ${\rm NO}_{25}(N;{\bf 1}_{10})$ 
in the G-scheme is obtained:
\bea
&& {\rm NO}_{25}(N;{\bf 1}_{10}) \:\: = \:\:
  {16 \over 3} \* \bigl( 1 + \sign(N) \bigr) \* {1 \over \epsilon^3} \* 
	\biggl(
          - {3 \over 2}  \* \S(-3)
          + \Ss(-2,1)
          + (
	    \Nplustwo
          - 2 \* \Nplusthree
          + \Nplusfour ) \* 
	  \bigg[
	    \S(1) 
	  + \Ss(1,1) 
          - \S(2)
	  \bigg]
\nonumber\\&& \mbox{} \quad
          + {1 \over 2}  \* (
            \Nplustwo
          - \Nplusfour ) \* 
	  \bigg[
	    \Ss(1,-2) 
	  + \Ss(1,2)
	  \bigg]
          + (
	    1
          - 2 \* \Nplus
          + 2 \* \Nplustwo ) \* 
	  \bigg[
	    \Ss(2,1) 
	    - {3 \over 2} \* \S(3)
	  \bigg]
          - \Nplusthree \* \Ss(2,1) 
          + \Nplusfour \* \S(3)
          \biggr)
\nonumber\\&& \mbox{} \quad
        + {16 \over 3} \* \bigl( 1 + \sign(N) \bigr) \* 
          {1 \over \epsilon^2}  \* 
	\biggl(
            {79 \over 8} \* \S(-4)
          + {15 \over 2} \* \S(-3)
          - {85 \over 8} \* \Ss(-3,1)
          - {1 \over 2} \* \Ss(-2,-2)
          - 5 \* \Ss(-2,1)
          + 5 \* \Sss(-2,1,1)
          - {13 \over 4} \* \Ss(-2,2)
\nonumber\\&& \mbox{} \quad
          + (
	    \Nplustwo
          - 2 \* \Nplusthree
          + \Nplusfour ) \* 
	  \bigg[
            {27 \over 16} \* \Ss(1,-3)
          - {3 \over 2} \* \Ss(1,-2)
          - {9 \over 4} \* \Sss(1,-2,1)
          - 11 \* \S(1)
          - 4 \* \S(1,1)
          - {3 \over 2} \* \Sss(1,1,-2)
          + 5 \* \Sss(1,1,1)
\nonumber\\&& \mbox{} \quad
          - {3 \over 2} \* \Sss(1,1,2)
          - {3 \over 2} \* \Ss(1,2)
          - {1 \over 2} \* \Sss(1,2,1)
          - {1 \over 2} \* \Ss(1,3)
          - 9 \* \Ss(2,1)
          + {7 \over 2} \* \Ss(2,2)
          + 5 \* \S(3)
	  \bigg]
          - (
	    1
          - 2 \* \Nplus
          + 2 \* \Nplustwo ) \* 
	  \bigg[
            {1 \over 2} \* \Ss(2,-2)
\nonumber\\&& \mbox{} \quad
          + 5 \* \S(2,1)
          - 5 \* \Sss(2,1,1)
          + {13 \over 4} \* \S(2,2)
          - {15 \over 2} \* \S(3)
          + {85 \over 8} \* \Ss(3,1)
          - {79 \over 8} \* \S(4)
	  \bigg]
          + (
            \Nplustwo
          - \Nplusthree ) \* 
	  \bigg[
            {9 \over 4} \* \Sss(1,1,-2)
          + {21 \over 8} \* \Sss(1,1,2)
\nonumber\\&& \mbox{} \quad
          - 4 \* \Ss(1,2)
          + {11 \over 8} \* \Sss(1,2,1)
          + 7 \* \S(2)
          + 6 \* \Ss(2,1)
	  \bigg]
          - \Nplustwo \* 
	  \bigg[
            {87 \over 16} \* \Ss(1,-3)
          + 2 \* \Ss(1,-2)
          - 4 \* \Sss(1,-2,1)
          + {21 \over 8} \* \S(1,3)
          + {29 \over 8} \* \Ss(2,2)
\nonumber\\&& \mbox{} \quad
          + 5 \* \S(3)
	  \bigg]
          + \Nplusthree \*
	  \bigg[
           {7 \over 8} \* \Ss(1,3)
          - 2 \* \Ss(2,-2)
          + 5 \* \Ss(2,1)
          - 5 \* \Sss(2,1,1)
          + {59 \over 8} \* \Ss(2,2)
          + {51 \over 8} \* \Ss(3,1)
          - {7 \over 4} \* \S(4)
	  \bigg]
          + \Nplusfour \* 
	  \bigg[
            {13 \over 16} \* \Ss(1,-3)
\nonumber\\&& \mbox{} \quad
          + 2 \* \Ss(1,-2)
          - {5 \over 4} \* \Sss(1,-2,1)
          + {7 \over 2} \* \Ss(2,-2)
          + 4 \* \Ss(3,1)
          - 6 \* \S(4)
	  \bigg]
          \biggr)
\:\: + \:\: {\cal O}\! \left( \frac{1}{\ep} \right) \:\: .
\label{eq:resno25}
\eea

In the remainder of this section we briefly address three issues which 
are, to a varying extent, special to the calculation of the coefficient 
functions. The first is the control of the expansion in powers of the 
dimensional offset $\ep$. This is more critical here than for the 
computation of the anomalous dimensions~\cite{Moch:2004pa,Vogt:2004mw},
as we now rely on the last coefficients in $\ep$ kept in 
Eqs.~(\ref{eq:T2k1}) -- (\ref{eq:TLp3}).
In other words, if some three-loop integrals entering the diagram 
calculation were actually not accurate to order $\ep^0$ (something, 
in fact, our extensive fixed-$N$ checks using the {\sc Mincer} 
program~\cite {Gorishnii:1989gt,Larin:1991fz} would have indicated), 
then that would not have affected the results for the anomalous 
dimensions, but spoiled the present calculation of the coefficient 
functions.

The one- (two-, three-) loop integrals are required to order $\ep^2$ 
($\ep^1$, $\ep^0$) for the calculation of the respective diagrams
entering Eqs.~(\ref{eq:T2k1}) -- (\ref{eq:TLp3}). 
However, all but the top-level topologies are also required in the
reduction schemes for higher-level topologies. Guided by the rule of
the triangle~\cite{Tkachov:1981wb,Chetyrkin:1981qh,Tkachov:1984xk}, a 
factor $1/\ep$ is expected for each line (completely) removed, which
then has to be compensated by controlling the lower-level topologies
to one more power in $\ep$. If this pattern holds, the various 
topologies are maximally required to the accuracies shown in Table 
\ref{tab:listoftopos}. For instance, the LA topology in Fig.\
\ref{fig:bbbtop} is reduced to FA case in Fig.~\ref{fig:bbbderived} 
by removing line 2 or line 5.

\begin{table}[ht]
\begin{center}
\begin{tabular}{c l c}
number of loops$\:\:$ & topologies\quad & expansion depth \\[1mm]
\hline & & \\[-3mm]
1  &  ${\rm L1}$                         & $\ep^5$ \\
2  &  ${\rm T2}$, ${\rm T3}$             & $\ep^4$ \\
2  &  ${\rm T1}$                         & $\ep^3$ \\
3  &  ${\rm Y1}$, \dots,   ${\rm Y5}$    & $\ep^3$ \\
3  &  ${\rm O1}$, \dots,   ${\rm O4}$    & $\ep^2$ \\
3  &  ${\rm FA}$, ${\rm BU}$             & $\ep^1$ \\
3  &  ${\rm LA}$, ${\rm BE}$, ${\rm NO}$ & $\ep^0$ \\[1mm]
\hline
\end{tabular}
\end{center}
\vspace{-2mm}
\caption{\label{tab:listoftopos}
The two-point topologies up to three loops, using the notation of 
Refs.~\cite{Gorishnii:1989gt,Larin:1991fz}, with the maximal power of 
$\epsilon$ kept for the determination of the third-order coefficient 
functions.}
\end{table}

If a reduction equation introduces a factor $\ep^{-2}$ while removing
a line, or a factor $\ep^{-1}$ in a simplification not removing a line,
the equation is said to contain a spurious pole. The lower-level
topologies are worked out to a finite accuracy as well, either for
efficiency or since the underlying integrals, for instance providing 
the boundary conditions for Eq.~(\ref{eq:diffeq}), are known only to a
certain accuracy in $\ep$. Therefore spurious poles endanger the 
integrity of the reduction procedure. Indeed, one of the greatest 
difficulties in constructing reduction schemes is to avoid such poles. 

Actually,
for some cases we have not found expressions free of spurious poles. 
Consequently these integrals have not been calculated to the design
order in $\ep$. For instance, most ${\rm O1}_{15}$ integrals, generated 
by removing line 2 from the FA$_{17}$ subtopology, have only been 
computed to order $\ep$ instead of including the $\ep^2$ terms as 
indicated in Table \ref{tab:listoftopos}. This was only permissible 
since in fact many more reductions are actually more benign than the
triangle reduction, removing lines without introducing any $\ep^{-1}$
pole, see, e.g., Eq.~(\ref{eq:basicNO25rec}). Carefully utilizing this 
fact, we were able to reach the required $\ep^0$ accuracy for all 
three-loop integrals entering the diagram calculations. 

\begin{figure}[htb]
\vspace{3mm}
\begin{center}
\begin{tabular}{cc}
\begin{minipage}{4.0cm}
\begin{center}
  \includegraphics[width=4.0cm]{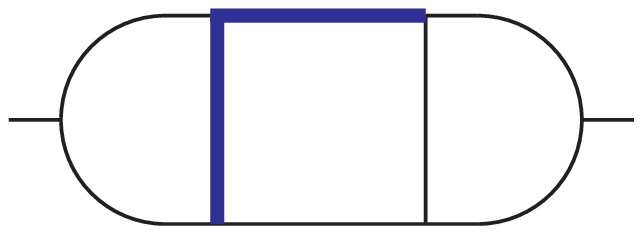}
\end{center}
\end{minipage}
&
$\displaystyle
\:\: =\:\: {(2 p \cdot q)^N \over (Q^2)^{N-2-3\epsilon}}\: 
 {\rm LA}_{27}{\rm box}(N)
$
\end{tabular}
\end{center}
\vspace{-2mm}
\caption{ \label{fig:labox27}
The integral ${\rm LA}_{27}{\rm box}$. All propagators have unit power. 
The momenta $q$ and $p$ (fat lines) flow from right to left and top to 
bottom through the diagram, respectively.
}
\end{figure}

The second issue specific for the calculation of the coefficient 
functions is the appearance of one integral (finite for $\ep \ra 0$) 
which can not be expressed in terms of harmonic sums. This integral, 
denoted by  ${\rm LA}_{27} {\rm box}$ and graphically illustrated in 
Fig.~\ref{fig:labox27}, is given by
\bea
\label{eq:la27box}
{\lefteqn{
{\rm LA}_{27}{\rm box}(N) \:\: = \:\:  (Q^2)^{2+3\epsilon}\: P_N\:
\int\, \prod_{n=1}^3\, d^Dl_n \: \cdot}}\\
&&\cdot \:\:
\Big[ (l_1)^2\, (l_2+p)^2\, (l_3)^2\, (l_3-q)^2\, (l_2-q)^2\, (l_1-q)^2
\, (l_1-l_2-p)^2\, (l_3-l_2)^2 \Big]^{-1}\:\: ,
\nn
\eea
where $P_N$ again stands for the Mellin-$N$ projection. 
${\rm LA}_{27}{\rm box}$ is subject to a first-order difference 
equation of the form
\beq
\label{eq:la27rec}
  {\rm LA}_{27}{\rm box}(N) \: - \:\frac{1}{2}\: 
  {\rm LA}_{27}{\rm box}(N-1) \:\: = \:\: G_{\,\rm LA}(N) \:\: .
\eeq
This equation does not fulfill, due to the factor $1/2$, the condition 
for a solution in terms of harmonic sums specified at the end of the 
paragraph below Eq.~(\ref{eq:firstsol}).

The solution of Eq.~(\ref{eq:la27rec}) can be written in terms of 
generalized harmonic sums \cite{Moch:2001zr} given by
\begin{eqnarray}
\label{eq:S-definition}
S(n) & = & \left\{ \begin{array}{cc}
1, & n  >  0 \:\: , \\
0, & n \le 0 \:\: , 
\end{array} \right.
\nn \\[1mm]
S(n;m_1,...,m_k;x_1,...,x_k) & = & \sum\limits_{i=1}^n \:
\frac{x_1^{\,i}}{i^{m_1}} \: S(i;m_2,...,m_k;x_2,...,x_k) \:\: .
\end{eqnarray}
Using the short-hand notation of Eq.~(\ref{eq:shorthand}) the result 
reads
\bea
&& {\rm LA}_{27}{\rm box}(N) \:\: = \:\:
         \sign(N) \* \pow(2, - N-1) \* \biggl(
            20 \* S(N+1;1;2) \* \z5
          + 4 \* S(N+1;{\bf 1}_{3};2,{\bf 1}_{2}) \* \z3
  \nonumber\\&& \mbox{} \quad
          + 6 \* S(N+1;{\bf 1}_{4},2;2,{\bf 1}_{3},-1)
          + 3 \* S(N+1;{\bf 1}_{4},2;2,{\bf 1}_{4})
          - S(N+1;{\bf 1}_{3},2,1;2,{\bf 1}_{4})
  \nonumber\\&& \mbox{} \quad
          - 9 \* S(N+1;{\bf 1}_{3},3;2,{\bf 1}_{2},-1)
          - 3 \* S(N+1;{\bf 1}_{3},3;2,{\bf 1}_{3})
          - 2 \* S(N+1;{\bf 1}_{2},2,{\bf 1}_{2};2,{\bf 1}_{4})
  \nonumber\\&& \mbox{} \quad
          - 6 \* S(N+1;{\bf 1}_{2},{\bf 2}_{2};2,{\bf 1}_{2},-1)
          + 6 \* S(N+1;{\bf 1}_{2},3,1;2,{\bf 1}_{3})
          + 6 \* S(N+1;{\bf 1}_{2},4;2,1,-1)
  \nonumber\\&& \mbox{} \quad
          - 6 \* S(N+1;{\bf 1}_{2},4;2,{\bf 1}_{2})
          + 12 \* S(N+1;1,2;2,-1) \* \z3
          - 6 \* S(N+1;1,2,1,2;2,{\bf 1}_{2},-1)
  \nonumber\\&& \mbox{} \quad
          - 3 \* S(N+1;1,2,1,2;2,{\bf 1}_{3})
          - 12 \* S(N+1;1,{\bf 2}_{2},1;2,-1,-1,1)
          - 3 \* S(N+1;1,{\bf 2}_{2},1;2,{\bf 1}_{3})
  \nonumber\\&& \mbox{} \quad
          + 18 \* S(N+1;1,2,3;2,-1,-1)
          + 9 \* S(N+1;1,2,3;2,1,-1)
          + 9 \* S(N+1;1,2,3;2,{\bf 1}_{2})
  \nonumber\\&& \mbox{} \quad
          + 6 \* S(N+1;1,3,{\bf 1}_{2};2,{\bf 1}_{3})
          + 6 \* S(N+1;1,3,2;2,1,-1)
          - 6 \* S(N+1;1,3,2;2,{\bf 1}_{2})
  \nonumber\\&& \mbox{} \quad
          - 4 \* S(N+1;2,1;2,1) \* \z3
          - 6 \* S(N+1;2,{\bf 1}_{2},2;2,{\bf 1}_{2},-1)
          - 3 \* S(N+1;2,{\bf 1}_{2},2;2,{\bf 1}_{3})
  \nonumber\\&& \mbox{} \quad
          + S(N+1;2,1,2,1;2,{\bf 1}_{3})
          + 9 \* S(N+1;2,1,3;2,1,-1)
          + 3 \* S(N+1;2,1,3;2,{\bf 1}_{2})
  \nonumber\\&& \mbox{} \quad
          + 2 \* S(N+1;{\bf 2}_{2},{\bf 1}_{2};2,{\bf 1}_{3})
          + 6 \* S(N+1;{\bf 2}_{3};2,1,-1)
          - 6 \* S(N+1;2,3,1;2,{\bf 1}_{2})
  \nonumber\\&& \mbox{} \quad
          - 6 \* S(N+1;2,4;2,-1)
          + 6 \* S(N+1;2,4;2,1)
          - 12 \* S(N+1;3;-2) \* \z3
          - 9 \* S(N+1;{\bf 3}_{2};2,1)
  \nonumber\\&& \mbox{} \quad
          + 6 \* S(N+1;3,1,2;2,1,-1)
          + 3 \* S(N+1;3,1,2;2,{\bf 1}_{2})
          + 12 \* S(N+1;3,2,1;-2,-1,1)
  \nonumber\\&& \mbox{} \quad
          + 3 \* S(N+1;3,2,1;2,{\bf 1}_{2})
          - 18 \* S(N+1;{\bf 3}_{2};-2,-1)
          - 9 \* S(N+1;{\bf 3}_{2};2,-1)
  \nonumber\\&& \mbox{} \quad
          - 6 \* S(N+1;4,{\bf 1}_{2};2,{\bf 1}_{2})
          - 6 \* S(N+1;4,2;2,-1)
          + 6 \* S(N+1;4,2;2,1)
          \biggr)
\:\: ,\label{eq:resla27box}
\eea
including terms with $x_j = \pm 2$ in the sums 
(\ref{eq:S-definition}). 
Note also the overall factor of $\pow(2,-N)$, which keeps also sums 
over ${\rm LA}_{27}{\rm box}$ within the class of generalized $S$-sums.

We did not need sums over ${\rm LA}_{27}{\rm box}$, for which all 
algorithms are however known from Ref.~\cite{Moch:2001zr}, in our
reduction procedures for higher-level subtopologies, e.g., 
${\rm LA}_{78}$. Thus we actually kept this integral as an `unknown' 
function in all intermediate expressions, only using 
Eq.~(\ref{eq:la27rec}) to shift the argument $N$. Finally, while 
terms $N^{k}\: {\rm LA}_{27}{\rm box}(N)$ occurred in the results 
of individual diagrams, all dependence on ${\rm LA}_{27}{\rm box}$ 
cancelled in the final results for the coefficient functions. 
 
The final issue we need to mention is that certain combinations of 
harmonic sums multiplied with positive powers of $N$ occur in the 
three-loop coefficient functions. Such structures are encountered in 
very many integrals also at the level of the $1/\ep$ poles, see 
Eq.~(\ref{eq:no25GN}) above, but they cancel in the final results for
the anomalous dimensions. On the other hand, the following
combinations are present in the final result for the coefficient 
functions:
\bea
&& g^{}_1(N) \:\: = \:\:
        N \* f(N) 
\:\ ,\label{eq:n1fac1}\\
&& g^{}_2(N) \:\: = \:\:
       N^2 \* f(N)
\:\: ,\label{eq:n2fac1}\\
&& g^{}_3(N) \:\: = \:\:
       N^3  \* f(N)
       - 2 \* N  \*  \bigl(  
         \z3 
       - \S(-3)
       - \S(-2)
       + 2 \* \Ss(-2,1)
       \bigr)
\:\: ,\label{eq:n3fac1}
\eea
with the function $f(N)$ given by 
\bea
&& f(N) \:\: = \:\:
         5 \* \z5 
       - 2 \* \S(-5) 
       + 4 \* \S(-2) \* \z3 
       - 4 \* \Ss(-2,-3) 
       + 8 \* \Sss(-2,-2,1) 
       + 4 \* \Ss(3,-2) 
       - 4 \* \Ss(4,1) 
       + 2 \* \S(5)
\:\: .\label{eq:fac1}
\eea
This function vanishes sufficiently fast for $N \to \infty$ for
Eqs.~(\ref{eq:n1fac1}) -- (\ref{eq:n3fac1}) to behave at most as 
constants in this limit. Thus the standard asymptotic behaviour
$\ln^{\,k}(N)$, $k = 1,\,\ldots ,\, 6$ of the three-loop quark 
coefficient functions $c^{(3)}_{2,\rm ns}$ and $c^{(3)}_{2,q}$ is 
unaffected by these new structures.

Positive powers as in Eqs.~(\ref{eq:n1fac1}) -- (\ref{eq:n3fac1}),
unlike negative powers of $N$, cannot be written entirely in terms of 
harmonic sums. Consequently a larger class of functions is required 
also in $x$-space, as a one-to-one relation exists between the set of 
harmonic sums of weight $w$ and the harmonic polylogarithms
\cite{Goncharov,Borwein,Remiddi:1999ew} $H_{\vec{m}}(x)/(1\pm x)$ where
$\vec{m}$ has weight $w-1$. The Mellin inverse of $g^{}_1(N)$ in 
Eq.~(\ref{eq:n1fac1}) can be derived by partial integration from that 
of $f(N)$ in Eq.~(\ref{eq:fac1}),
\bea
\label{eq:gNposdef}
  g^{}_1(N) & = & N f(N) \:\: = \:\: N \int_0^1 dx\: x^{\,N-1} f(x)
  \:\: =  \:\: f(1) \: - \: \int_0^1 \! dx \: x^{\,N-1} 
  \: xf^{\,\prime}(x) \nn \\
  & = & \int_0^1 \! dx\: x^{\, N-1} \biggl\{ \,\delta(1-x)
        f(1) - \: x f^{\,\prime} (x) \biggr\} \:\: ,
\eea
where $g^{}_1(x)$ is given by the expression in curved brackets in the 
second line. This procedure is then repeated for the functions $g^{}_2$
and $g^{}_3$, leading to the $x$-space expressions
\bea
&& g^{}_1(x) \:\: = \:\:
         {1 \over (1-x)^2}  \*  \biggl(
          - {4 \over 5} \* \z2^2
          + 8 \* \H(-2) \* \z2
          - 4 \* \Hhh(-2,0,0)
          - 8 \* \Hh(-2,2)
          - 6 \* \H(0) \* \z3
          - 6 \* \Hh(0,0) \* \z2
          + 2 \* \Hhhh(0,0,0,0)
  \nonumber\\&& \mbox{} \quad
          + 4 \* \H(4)
          \biggr)
       + {1 \over {1-x}}  \*  \biggl(
            {4 \over 5} \* \z2^2
          - 6 \* \z3
          - 8 \* \H(-2) \* \z2
          + 4 \* \Hhh(-2,0,0)
          + 8 \* \Hh(-2,2)
          + 8 \* \H(-1) \* \z2
          - 4 \* \Hhh(-1,0,0)
  \nonumber\\&& \mbox{} \quad
          - 8 \* \Hh(-1,2)
          - 6 \* \H(0) \* \z2
          + 6 \* \H(0) \* \z3
          + 6 \* \Hh(0,0) \* \z2
          + 2 \* \Hhh(0,0,0)
          - 2 \* \Hhhh(0,0,0,0)
          + 4 \* \H(3)
          - 4 \* \H(4)
          \biggr)
  \nonumber\\&& \mbox{} \quad
       + {1 \over (1+x)^2}  \*  \biggl(
            {21 \over 5} \* \z2^2
          + 4 \* \Hh(-3,0)
          + 4 \* \H(0) \* \z3
          + 2 \* \Hh(0,0) \* \z2
          - 2 \* \Hhhh(0,0,0,0)
          \biggr)
       + {1 \over {1+x}}  \*  \biggl(
          - {21 \over 5} \* \z2^2
          + 4 \* \z3
  \nonumber\\&& \mbox{} \quad
          - 4 \* \Hh(-3,0)
          + 4 \* \Hh(-2,0)
          + 2 \* \H(0) \* \z2
          - 4 \* \H(0) \* \z3
          - 2 \* \Hh(0,0) \* \z2
          - 2 \* \Hhh(0,0,0)
          + 2 \* \Hhhh(0,0,0,0)
          \biggr)
       + 2 \* \z3
          - 4 \* \Hh(-2,0)
  \nonumber\\&& \mbox{} \quad
          - 8 \* \H(-1) \* \z2
          + 4 \* \Hhh(-1,0,0)
          + 8 \* \Hh(-1,2)
          + 4 \* \H(0) \* \z2
          - 4 \* \H(3)
\:\ ,\label{eq:xn1fac1} \\[2mm]
&& g^{}_2(x) \:\: = \:\:
         {1 \over (1-x)^3}  \*  \biggl(
            {8 \over 5} \* \z2^2
          - 16 \* \H(-2) \* \z2
          + 8 \* \Hhh(-2,0,0)
          + 16 \* \Hh(-2,2)
          + 12 \* \H(0) \* \z3
          + 12 \* \Hh(0,0) \* \z2
          - 4 \* \Hhhh(0,0,0,0)
  \nonumber\\&& \mbox{} \quad
          - 8 \* \H(4)
          \biggr)
       + {1 \over (1-x)^2}  \*  \biggl(
          - {12 \over 5} \* \z2^2
          + 12 \* \z3
          + 24 \* \H(-2) \* \z2
          - 12 \* \Hhh(-2,0,0)
          - 24 \* \Hh(-2,2)
          - 16 \* \H(-1) \* \z2
  \nonumber\\&& \mbox{} \quad
          + 8 \* \Hhh(-1,0,0)
          + 16 \* \Hh(-1,2)
          + 12 \* \H(0) \* \z2
          - 18 \* \H(0) \* \z3
          - 18 \* \Hh(0,0) \* \z2
          - 4 \* \Hhh(0,0,0)
          + 6 \* \Hhhh(0,0,0,0)
          - 8 \* \H(3)
  \nonumber\\&& \mbox{} \quad
          + 12 \* \H(4)
          \biggr)
       + {1 \over {1-x}}  \*  \biggl(
            2 \* \z2
          + {4 \over 5} \* \z2^2
          - 12 \* \z3
          - 8 \* \H(-2) \* \z2
          + 4 \* \Hhh(-2,0,0)
          + 8 \* \Hh(-2,2)
          + 16 \* \H(-1) \* \z2
  \nonumber\\&& \mbox{} \quad
          - 8 \* \Hhh(-1,0,0)
          - 16 \* \Hh(-1,2)
          - 12 \* \H(0) \* \z2
          + 6 \* \H(0) \* \z3
          + 6 \* \Hh(0,0) \* \z2
          + 4 \* \Hhh(0,0,0)
          - 2 \* \Hhhh(0,0,0,0)
          + 8 \* \H(3)
          - 4 \* \H(4)
          \biggr)
  \nonumber\\&& \mbox{} \quad
       + {1 \over (1+x)^3}  \*  \biggl(
          - {42 \over 5} \* \z2^2
          - 8 \* \Hh(-3,0)
          - 8 \* \H(0) \* \z3
          - 4 \* \Hh(0,0) \* \z2
          + 4 \* \Hhhh(0,0,0,0)
          \biggr)
       + {1 \over (1+x)^2}  \*  \biggl(
            {63 \over 5} \* \z2^2
          - 8 \* \z3
  \nonumber\\&& \mbox{} \quad
          + 12 \* \Hh(-3,0)
          - 8 \* \Hh(-2,0)
          - 4 \* \H(0) \* \z2
          + 12 \* \H(0) \* \z3
          + 6 \* \Hh(0,0) \* \z2
          + 4 \* \Hhh(0,0,0)
          - 6 \* \Hhhh(0,0,0,0)
          \biggr)
       + {1 \over {1+x}}  \*  \biggl(
          - 6 \* \z2
  \nonumber\\&& \mbox{} \quad
          - {21 \over 5} \* \z2^2
          + 8 \* \z3
          - 4 \* \Hh(-3,0)
          + 8 \* \Hh(-2,0)
          - 4 \* \Hh(-1,0)
          + 4 \* \H(0) \* \z2
          - 4 \* \H(0) \* \z3
          + 4 \* \Hh(0,0)
          - 2 \* \Hh(0,0) \* \z2
  \nonumber\\&& \mbox{} \quad
          - 4 \* \Hhh(0,0,0)
          + 2 \* \Hhhh(0,0,0,0)
          + 4 \* \H(2)
          \biggr)
       + \delta(1-x)  \*  \bigl(
            \z2
          + \z3
          \bigr)
       + 4 \* \z2
          + 4 \* \Hh(-1,0)
          - 4 \* \Hh(0,0)
          - 4 \* \H(2)
\:\ ,\label{eq:xn2fac1} \\[3mm]
&& g^{}_3(x) \:\: = \:\:
         {1 \over (1-x)^4}  \*  \biggl(
          - {24 \over 5} \* \z2^2
          + 48 \* \H(-2) \* \z2
          - 24 \* \Hhh(-2,0,0)
          - 48 \* \Hh(-2,2)
          - 36 \* \H(0) \* \z3
          - 36 \* \Hh(0,0) \* \z2
  \nonumber\\&& \mbox{} \quad
          + 12 \* \Hhhh(0,0,0,0)
          + 24 \* \H(4)
          \biggr)
       + {1 \over (1-x)^3}  \*  \biggl(
            {48 \over 5} \* \z2^2
          - 36 \* \z3
          - 96 \* \H(-2) \* \z2
          + 48 \* \Hhh(-2,0,0)
          + 96 \* \Hh(-2,2)
  \nonumber\\&& \mbox{} \quad
          + 48 \* \H(-1) \* \z2
          - 24 \* \Hhh(-1,0,0)
          - 48 \* \Hh(-1,2)
          - 36 \* \H(0) \* \z2
          + 72 \* \H(0) \* \z3
          + 72 \* \Hh(0,0) \* \z2
          + 12 \* \Hhh(0,0,0)
          - 24 \* \Hhhh(0,0,0,0)
  \nonumber\\&& \mbox{} \quad
          + 24 \* \H(3)
          - 48 \* \H(4)
          \biggr)
       + {1 \over (1-x)^2}  \*  \biggl(
          - 6 \* \z2
          - {28 \over 5} \* \z2^2
          + 54 \* \z3
          + 56 \* \H(-2) \* \z2
          - 28 \* \Hhh(-2,0,0)
          - 56 \* \Hh(-2,2)
  \nonumber\\&& \mbox{} \quad
          - 72 \* \H(-1) \* \z2
          + 36 \* \Hhh(-1,0,0)
          + 72 \* \Hh(-1,2)
          + 54 \* \H(0) \* \z2
          - 42 \* \H(0) \* \z3
          - 42 \* \Hh(0,0) \* \z2
          - 18 \* \Hhh(0,0,0)
          + 14 \* \Hhhh(0,0,0,0)
  \nonumber\\&& \mbox{} \quad
          - 36 \* \H(3)
          + 28 \* \H(4)
          \biggr)
       + {1 \over {1-x}}  \*  \biggl(
            2 \* \z2
          + {4 \over 5} \* \z2^2
          - 18 \* \z3
          - 8 \* \H(-2) \* \z2
          + 4 \* \Hhh(-2,0,0)
          + 8 \* \Hh(-2,2)
          + 24 \* \H(-1) \* \z2
  \nonumber\\&& \mbox{} \quad
          - 12 \* \Hhh(-1,0,0)
          - 24 \* \Hh(-1,2)
          - 18 \* \H(0) \* \z2
          + 6 \* \H(0) \* \z3
          + 2 \* \Hh(0,0)
          + 6 \* \Hh(0,0) \* \z2
          + 6 \* \Hhh(0,0,0)
          - 2 \* \Hhhh(0,0,0,0)
          + 4 \* \H(2)
  \nonumber\\&& \mbox{} \quad
          + 12 \* \H(3)
          - 4 \* \H(4)
          \biggr)
       + {1 \over (1+x)^4}  \*  \biggl(
            {126 \over 5} \* \z2^2
          + 24 \* \Hh(-3,0)
          + 24 \* \H(0) \* \z3
          + 12 \* \Hh(0,0) \* \z2
          - 12 \* \Hhhh(0,0,0,0)
          \biggr)
  \nonumber\\&& \mbox{} \quad
       + {1 \over (1+x)^3}  \*  \biggl(
          - {252 \over 5} \* \z2^2
          + 24 \* \z3
          - 48 \* \Hh(-3,0)
          + 24 \* \Hh(-2,0)
          + 12 \* \H(0) \* \z2
          - 48 \* \H(0) \* \z3
          - 24 \* \Hh(0,0) \* \z2
  \nonumber\\&& \mbox{} \quad
          - 12 \* \Hhh(0,0,0)
          + 24 \* \Hhhh(0,0,0,0)
          \biggr)
       + {1 \over (1+x)^2}  \*  \biggl(
            6 \* \z2
          + {147 \over 5} \* \z2^2
          - 36 \* \z3
          + 28 \* \Hh(-3,0)
          - 36 \* \Hh(-2,0)
  \nonumber\\&& \mbox{} \quad 
          + 12 \* \Hh(-1,0)
          - 6 \* \H(0)
          - 18 \* \H(0) \* \z2
          + 28 \* \H(0) \* \z3
          - 6 \* \Hh(0,0)
          + 14 \* \Hh(0,0) \* \z2
          + 18 \* \Hhh(0,0,0)
          - 14 \* \Hhhh(0,0,0,0)
          \biggr)
  \nonumber\\&& \mbox{} \quad 
       + {1 \over {1+x}}  \*  \biggl(
          - 2
          - 2 \* \z2
          - {21 \over 5} \* \z2^2
          + 12 \* \z3
          - 4 \* \Hh(-3,0)
          + 12 \* \Hh(-2,0)
          - 12 \* \Hh(-1,0)
          + 8 \* \H(0)
          + 6 \* \H(0) \* \z2
  \nonumber\\&& \mbox{} \quad 
          - 4 \* \H(0) \* \z3
          + 4 \* \Hh(0,0)
          - 2 \* \Hh(0,0) \* \z2
          - 6 \* \Hhh(0,0,0)
          + 2 \* \Hhhh(0,0,0,0)
          - 4 \* \H(2)
          \biggr)
       - \delta(1-x)  \*  \bigl(
            \z2
          + \z3
          \bigr)
  \nonumber\\&& \mbox{} \quad
       + 2
          - 2 \* \H(0)
\:\: . \label{eq:xn3fac1}
\eea
 
The above equations are not suitable for a numerical implementation
at $x$-values very close to $x=1$, a region which contributes to all 
numerical calculations of moments and, more importantly, Mellin 
convolutions. For application in this region we instead provide the 
expansions 
\bea
\label{eq:g1xto1}
g^{}_1(x) &\simeq &
         \z2
          + \z3
       - (1-x)  \*  \bigl(
            \z2
          + \z3
          \bigr)
       + (1-x)^2  \*  \biggl(
            {5 \over 8}
          - {1 \over 4} \* \z2
          - {1 \over 2} \* \z3
          - {1 \over 2} \* \ln(1-x)
          \biggr)
\nonumber \\
&& \quad \mbox{}
       + {\cal O}\bigl( (1-x)^3 \bigr)
         \, ,
\\[1mm]
\label{eq:g2xto1}
g_2(x) &\simeq &
          \delta(1-x)  \* \bigl(
             \z2
           + \z3
           \bigr)
        - \z2
           - \z3
        + (1-x)  \* \biggl(
             {3 \over 4}
           + {1 \over 2} \* \z2
           - \ln(1-x)
           \biggr)
\nonumber \\
&& \quad \mbox{}
        - (1-x)^2  \*  \biggl(
             {9 \over 8}
           - {1 \over 4} \* \z2
           - {1 \over 2} \* \z3
           - {1 \over 2} \* \ln(1-x)
           \biggr)
        + {\cal O}\bigl( (1-x)^3 \bigr)
          \:\: ,
\\[1mm]
\label{eq:g3xto1}
g_3(x) &\simeq &
        - \delta(1-x)  \*  \bigl(
             \z2
           + \z3
           \bigr)
        + {3 \over 4}
           + {1 \over 2} \* \z2
           + \ln(1-x)
        - (1-x)  \*  \biggl(
             {1 \over 2}
           - \z3
           \biggr)
\nonumber \\
&& \quad \mbox{}
        - (1-x)^2  \*  \biggl(
             {7 \over 24}
           + {1 \over 12} \* \z2
           - {1 \over 2} \* \z3
           + {1 \over 2} \* \ln(1-x)
           \biggr)
        + {\cal O}\bigl( (1-x)^3 \bigr)
          \:\: .
\eea
These expansions have been derived by expanding the harmonic 
polylogarithms sufficiently deep in $(1-x)$ via the transformation 
$t = (1-x)/(1+x)$ and an expansion around $t=0$ as described in 
Ref.~\cite{Remiddi:1999ew} and implemented in {\sc Form}~\cite
{Vermaseren:2000nd,Vermaseren:2002rp}.
%
%
\setcounter{equation}{0}
\section{Results and discussion}
\label{sec:results}
%
%
We are now ready to present the third-order contributions 
$c_{a,i}^{(3)}$ to the coefficient functions $C_{a,i}$ for the 
structure functions $F_{a=2,L}$ in electromagnetic DIS,
\beq
\label{eq:Fa-dec}
 x^{-1} F_a \: = \: C_{a,\rm ns} \otimes q_{\rm ns}
 + \langle e^2 \rangle \left( C_{a,\rm q} \otimes q_{\rm s}
                            + C_{a,\rm g} \otimes g \right) \:\: .
\eeq
Recall that $q_i^{}$ and $g$ represent the number distributions of 
quarks and gluons, respectively, in the fractional hadron momentum,
with $q_{\rm s}$ standing for the flavour-singlet quark distribution,
$q_{\rm s} = \sum_{i=1}^{\nf} ( q_i^{} + \bar{q}_i^{} )$
where $\nf$ denotes the number of effectively massless flavours.
The normalization of the corresponding non-singlet combination 
$q_{\rm ns}$ is defined via Eq.~(\ref{eq:c20}) below. Again $\langle e^2
\rangle$ represents the average squared charge, and $\otimes$ denotes 
the Mellin convolution which turns into a simple multiplication in 
$N$-space. Below the singlet-quark coefficient function is decomposed 
into the non-singlet and a `pure singlet' contribution, $c^{\,(n)}_
{a,\rm q} = c^{\,(n)}_{a,\rm ns} + c^{\,(n)}_{a,\rm ps}$, and the
results are given in the \MSb\ scheme for the standard choice
$\mu_r^{\,2} = \mu_{\!f}^{\,2} = Q^{\,2}$ of the renormalization and 
factorization scales.  
The complete expressions for the dependence on $\mu_r$ and $\mu_f$ up 
to the third order in our expansion parameter $\ar \equiv \as /(4\pi)$
can be found, for example, in Eqs.~(2.16) -- (2.18) of 
Ref.~\cite{vanNeerven:2000uj}.

As discussed above, our calculation via the optical theorem and a 
dispersion relation directly determines the coefficient functions for
all even-integer moments $N$ in terms of harmonic sums
\cite{Gonzalez-Arroyo:1979df,Gonzalez-Arroyo:1980he,%
Vermaseren:1998uu,Blumlein:1998if,Moch:2001zr}.
From these results the $x$-space expressions can be reconstructed
algebraically \cite{Moch:1999eb,Remiddi:1999ew} in terms of harmonic 
polylogarithms~\cite{Goncharov,Borwein,Remiddi:1999ew}. Unfortunately,
but not entirely unexpectedly, the exact results are very lengthy.
The complete expressions in both $N$-space and $x$-space are therefore
deferred to the appendices of this article. Here we confine ourselves
to (sufficiently accurate) approximations for $c_{2,i}^{\,(3)}(x)$, 
quite analogous to those already presented for $c_{L,i}^{\,(3)}(x)$ in 
Ref.~\cite{Moch:2004xu}.

For the convenience of the reader we first recall the known results up
to the second order. The coefficient functions at zeroth and first 
order \cite{Bardeen:1978yd} are given by 
\beq
\label{eq:c20}
  c_{2, \rm ns}^{\,(0)}(x) \: = \: \delta(x_1) \:\: , \quad
  c_{2, \rm ps}^{\,(0)}(x) \: = \: c_{2, \rm g}^{\,(0)}(x)
  \: = \: c_{2, \rm ps}^{\,(1)}(x) \: = \: 0 
\eeq
and
\bea
\label{eq:c2ns1}
  c_{2, \rm ns}^{\,(1)}(x) &\! =\! &
  \cf \{ 4\, \DD_1 - 3\, \DD_0 - (9 + 4\,\z2)\,\delta(x_1)
  - 2\,(1+x) (L_1 - L_0) 
  \nn \\ & & \mbox{} \quad
  - 4\, x_1^{-1} L_0 + 6 + 4\,x \} \:\: , \\[1mm]
\label{eq:c2gl1}
  c_{2, \rm g}^{\,(1)}(x)  & = \!&
  \nf\: \{ (2-4\, xx_1) (L_1 - L_0) - 2 + 16\, xx_1 \}
\eea
with $\cf = (N_c^{\,2}-1)/(2N_c) = 4/3$ in QCD. Here and below we use the 
abbreviations
\beq
\label{eq:abbr}
  x_1 \: = \: 1-x            \:\: ,\quad
  L_0 \: = \: \ln\, x        \:\: ,\quad
  L_1 \: = \: \ln\, x_1      \:\: ,\quad
  \DD_k \: = \: [ x_1^{-1} L_1^k ]_+ 
\eeq
where, as usual, the +-distributions are defined via 
\beq
\label{eq:plus}
  \int_0^1 \! dx \, a(x)_+ f(x) \: = \: \int_0^1 \! dx \, a(x)
  \,\{ f(x) - f(1) \} 
\eeq
for regular functions $f(x)$. Convolutions with the distributions 
$\DD_k$ in Eq.~(\ref{eq:abbr}) can be written as
\beq
\label{eq:Dkconv}
  x[\DD_k \otimes f](x) = 
  \int_x^1 \! dy \: \frac{\ln^k (1-x)}{1-x} \left\{ \frac{x}{y}\,
  f\!\left( \frac{x}{y} \right) - xf(x) \right\}
  + \, xf(x) \frac{1}{k+1} \ln^{k+1}(1-x) \:\: .
\eeq

With an error of 0.1\% or less, the two-loop coefficient functions
\cite{vanNeerven:1991nn,Zijlstra:1991qc,Zijlstra:1992kj,%
Zijlstra:1992qd,Moch:1999eb} can be represented by
\bea
\label{eq:c2ns2}
  c_{\,2, \rm ns}^{\,(2)}(x)\!\! & \cong &  \quad
       128/9\: \DD_3 - 184/3\: \DD_2 - 31.1052\: \DD_1 + 188.641\: \DD_0
     - 338.513\: \delta (x_1) 
  \nn \\ & & \mbox{} \quad
     - 17.74\: L_1^3 + 72.24\: L_1^2 - 628.8\: L_1 
     - 181.0 - 806.7\: x + 0.719\: x L_0^4 
  \nn \\ & & \mbox{} \quad
     + L_0 L_1 ( 37.75\: L_0 - 147.1\: L_1) 
     - 28.384\: L_0 - 20.70\: L_0^2 - 80/27\: L_0^3
  \nn \\[1mm] &+& \mbox{} \nf \:\big\{ \,
       16/9\: \DD_2 - 232/27\: \DD_1 + 6.34888\: \DD_0 
     + 46.8531\: \delta (x_1) - 1.500\: L_1^2 
  \nn \\ & & \mbox{} \quad
     + 24.87\: L_1 - 7.8109 - 17.82\: x - 12.97\: x^2 - 0.185\: xL_0^3
     + 8.113\: L_0 L_1 
  \nn \\ & & \mbox{} \quad
     + 16/3\: L_0 + 20/9\: L_0^2 \,\big\}
\:\: ,
  \\[2mm]
\label{eq:c2ps2}
  c_{\,2, \rm ps}^{\,(3)}(x)\!\! & \cong & \nf \:\big\{ \,
      (8/3\: L_1^2 - 32/3\: L_1 + 9.8937)\, x_1 
     + (9.57 - 13.41\: x + 0.08\: L_1^3)\, x_1^2 
  \nn \\ & & \mbox{} \quad
     + 5.667\: xL_0^3 - L_0^2 L_1 (20.26 - 33.93\: x) + 43.36\: x_1 L_0 
     - 1.053\: L_0^2 
  \nn \\ & & \mbox{} \quad
     + 40/9\: L_0^3 + 5.2903\: x^{-1}x_1^2 \,\big\}
\: , \:\:
  \\[2mm]
\label{eq:c2gl2}
  c_{\,2, \rm g}^{\,(3)}(x)\! & \cong & \nf \:\big\{ \,
       58/9\: L_1^3 - 24\: L_1^2 - 34.88\: L_1 + 30.586 
     - ( 25.08 + 760.3\: x 
  \nn \\ & & \mbox{} \quad
     + 29.65\: L_1^3 ) \, x_1 + 1204\: xL_0^2
     + L_0 L_1 ( 293.8 + 711.2\: x + 1043\: L_0 ) 
  \nn \\ & & \mbox{} \quad
     + 115.6\: L_0 - 7.109\: L_0^2 + 70/9\: L_0^3 
     + 11.9033\: x^{-1} x_1 \,\big\} \:\: .
\eea
Eqs.~(\ref{eq:c2ns2}) -- (\ref{eq:c2gl2}) are less compact, 
but more accurate than the previous parametrizations \cite
{vanNeerven:1999ca,vanNeerven:2000uj}.

Now we present our three-loop results. As in Eqs.~(\ref{eq:c2ns2})%
--(\ref{eq:c2gl2}) inserting the numerical values of the 
$\nf$-independent colour factors, the non-singlet coefficient function 
can be parametrized as
\bea
\label{eq:c2ns3}
  c_{\,2, \rm ns}^{\,(3)}(x)\!\! & \cong &  \quad
       512/27\: \DD_5 - 5440/27\: \DD_4 + 501.099\: \DD_3 
     + 1171.54\: \DD_2 - 7328.45\: \DD_1 
  \nn \\ & & \mbox{} \quad
     + 4442.76\: \DD_0 - 9170.38\: \delta (x_1) 
     - 512/27\: L_1^5 + 704/3\: L_1^4 - 3368\: L_1^3 
  \nn \\ & & \mbox{} \quad
     - 2978\: L_1^2 + 18832\: L_1
     - 4926 + 7725\: x + 57256\: x^2 + 12898\: x^3
  \nn \\ & & \mbox{} \quad
     - 56000\: x_1 L_1^2 - L_0 L_1 (6158 + 1836\: L_0) 
     + 4.719\: xL_0^5
     - 775.8\: L_0 
  \nn \\ & & \mbox{} \quad
     - 899.6\: L_0^2 - 309.1\: L_0^3 - 2932/81\: L_0^4 - 32/27\: L_0^5 
  \nn \\[1mm] &+& \mbox{} \nf \:\big\{ \,
     640/81\: \DD_4 - 6592/81\: \DD_3 + 220.573\: \DD_2
     + 294.906\: \DD_1 - 729.359\: \DD_0 
  \nn \\ & & \mbox{} \quad
     + 2574.687\: \delta (x_1) 
     - 640/81\: L_1^4 + 153.5\: L_1^3 - 828.7\: L_1^2 - 501.1\: L_1
     + 831.6
  \nn \\ & & \mbox{} \quad
     - 6752\: x - 2778\: x^2 + 171.0\: x_1 L_1^4
     + L_0 L_1\, (4365 + 716.2\: L_0 - 5983\: L_1) 
  \nn \\ & & \mbox{} \quad
     + 4.102\: xL_0^4
     + 275.6\: L_0 + 187.3\: L_0^2 + 12224/243\: L_0^3 
     + 728/243\: L_0^4 \, \big\}
  \nn \\[1mm] &+& \mbox{} \nf^{\!\!\! 2} \:\big\{ \,
     64/81\: \DD_3 - 464/81\: \DD_2 + 7.67505\: \DD_1 + 1.00830\: \DD_0
     - 103.2366\: \delta (x_1)
  \nn \\ & & \mbox{} \quad
     - 64/81\: L_1^3 + 18.21\: L_1^2 - 19.09\: L_1
     + 129.2\: x + 102.5\: x^2 
     + L_0 L_1\, (- 96.07 
  \nn \\ & & \mbox{} \quad
     - 12.46\: L_0 + 85.88\: L_1)
     - 8.042\: L_0 - 1984/243\: L_0^2 - 368/243\: L_0^3 \, \big\}
  \nn \\ &+& \mbox{} \!\! fl_{11}^{\:\rm ns} \: \nf \: \big\{
     ( 126.42 - 50.29\, x - 50.15\, x^2) x_1  
     - 11.888\, \delta (x_1) - 26.717 
     - 9.075\, xx_1 L_1
  \nn \\ & & \mbox{} \quad
     - xL_0^2 (101.8 + 34.79\: L_0 + 3.070\: L_0^2)
     + 59.59\: L_0 - 320/81\: L_0^2 (5 + L_0) 
     \,\big\} \: x
\:\: .
  \nn \\[-2mm] & &
\eea
Slightly less accurate parametrizations of the (non-$fl_{11}$) 
$\nf$-contributions were already presented in Ref.~\cite{Moch:2002sn}.
The corresponding pure-singlet coefficient function can be approximated 
by
\bea
\label{eq:c2ps3}
  c_{\,2, \rm ps}^{\,(3)}(x)\!\! & \cong & \nf \:\big\{ \,
     ( 856/81 \, L_1^4 - 6032/81\, L_1^3 + 130.57\, L_1^2
     - 542\, L_1 + 8501\, - 4714\, x + 61.5\, x^2 )
  \nn \\ & & \mbox{} \quad
     \cdot x_1 + L_0 L_1 (8831\: L_0 + 4162\: x_1) 
     - 15.44\: xL_0^5 + 3333\: xL_0^2 + 1615\: L_0 + 1208\: L_0^2
  \nn \\ & & \mbox{} \quad
     - 333.73\: L_0^3 + 4244/81\: L_0^4 - 40/9\: L_0^5
     - x^{-1} (2731.82\: x_1 + 414.262\: L_0) \big\} 
  \nn \\[1mm] &+& \mbox{} \n2f \:\big\{ \,
     ( - 64/81\: L_1^3 + 208/81\: L_1^2 + 23.09\: L_1
     - 220.27 + 59.80\: x - 177.6 x^2)\, x_1 
  \nn \\ & & \mbox{} \quad
     -  L_0 L_1 (160.3\: L_0 + 135.4\: x_1) - 24.14\: xL_0^3
     - 215.4\: xL_0^2 - 209.8\: L_0 - 90.38 L_0^2
  \nn \\ & & \mbox{} \quad
     - 3568/243\: L_0^3 - 184/81\: L_0^4 + 40.2426\: x_1 x^{-1}
  \, \big\}
  \nn \\ &+& \mbox{} \!\! fl_{11}^{\:\rm ps} \: \nf \: \big\{
     ( 126.42 - 50.29\, x - 50.15\, x^2) x_1
     - 11.888\, \delta (x_1) - 26.717
     - 9.075\, xx_1 L_1
  \nn \\ & & \mbox{} \quad
     - xL_0^2 (101.8 + 34.79\: L_0 + 3.070\: L_0^2)
     + 59.59\: L_0 - 320/81\: L_0^2 (5 + L_0)
     \, \big\} \: x 
\:\: .
  \nn \\[-2mm] & &
\eea
Finally the third-order gluon coefficient function can be written as
\bea
\label{eq:c2gl3}
  c_{\,2, \rm g}^{\,(3)}(x)\!\! & \cong & \nf \:\big\{ \,
     966/81\: L_1^5 - 1871/18\: L_1^4 + 89.31\: L_1^3
     + 979.2\: L_1^2 - 2405\: L_1 + 1372\: x_1 L_1^4
  \nn \\ & & \mbox{} \quad
     - 15729 - 310510\: x + 331570\: x^2 
     - 244150\: xL_0^2 - 253.3\: xL_0^5
  \nn \\ & & \mbox{} \quad
     + L_0 L_1 (138230 - 237010\: L_0) - 11860\: L_0
     - 700.8\: L_0^2 - 1440\: L_0^3 
  \nn \\ & & \mbox{} \quad
     + 4961/162\: L_0^4 - 134/9\: L_0^5 
     - x^{-1} (6362.54 - 932.089\: L_0) 
     + 0.625\: \delta(x_1) \, \big\}
  \nn \\[1mm] &+& \mbox{} \n2f \:\big\{ \,
     131/81\: L_1^4 - 14.72\: L_1^3 + 3.607\: L_1^2
     - 226.1\: L_1 + 4.762 - 190\: x - 818.4\: x^2
  \nn \\ & & \mbox{} \quad
     - 4019\: xL_0^2 - L_0 L_1 (791.5 + 4646\: L_0)
     + 739.0\: L_0 + 418.0\: L_0^2 + 104.3\: L_0^3
  \nn \\ & & \mbox{} \quad
     + 809/81\: L_0^4 + 12/9\: L_0^5 + 84.423\: x^{-1}
     \, \big\} 
  \nn \\ &+& \mbox{} \!\! fl_{11}^{\:\rm g} \: \n2f \: \big\{
     3.211\: L_1^2 + 19.04\: xL_1 + 0.623\: x_1 L_1^3
     - 64.47\: x + 121.6\: x^2 - 45.82\: x^3 
  \nn \\ & & \mbox{} \quad
     - x L_0 L_1 ( 31.68 + 37.24\: L_0) 
     + 11.27\: x^2 L_0^3 - 82.40\: xL_0 - 16.08\: xL_0^2
  \nn \\ & & \mbox{} \quad
     + 520/81\: x L_0^3 + 20/27\: x L_0^4 \, \big\} 
\:\: .
\eea
The new charge factors $fl_{11}$ have been specified in Table
\ref{tab:flavour} above, and $fl_{11}^{\:\rm ps}$ is given by
$fl_{11}^{\:\rm s} - fl_{11}^{\:\rm ns}$.

The coefficients of $\DD_k$ and of $x^{-1}$ in Eqs.~(\ref{eq:c2ns2}) 
-- (\ref{eq:c2gl3}) are exact up to a truncation of irrational 
numbers. Also exact are those coefficients of $L_0^k \equiv\ln^{\,k} x$ 
and $L_1^k \equiv \ln^{\,k} (1-x)$ given as fractions. Most of the 
remaining coefficients have been obtained by fits to the exact 
coefficient functions at $10^{-6} \leq x \leq 1\! -\! 10^{-6}$ which we 
evaluated using a weight-five extension of the program \cite
{Gehrmann:2001pz} for the evaluation of the harmonic polylogarithms 
\cite{Remiddi:1999ew}.
Finally the coefficients of $\delta (1-x)$ have been slightly adjusted 
from their exact values using the lowest integer moments, as discussed
in Ref.~\cite{Vogt:2004mw}. Like their second-order counterparts  
(\ref{eq:c2ns2}) -- (\ref{eq:c2gl2}), the three-loop parametrizations 
(\ref{eq:c2ns3}) -- (\ref{eq:c2gl3}) deviate from the exact results by
less than one part in a thousand.

For use with $N$-space evolution programs (see, e.g., Refs.~\cite
{Weinzierl:2002mv,Vogt:2004ns}) for parton distributions and structure 
functions, the above approximations can be readily transformed to 
Mellin space for any complex value of $N$. For the time being, this is 
especially important for our new results for which (unlike the two-loop 
coefficient functions and three-loop splitting functions \cite
{Blumlein:2000hw,Blumlein:2005jg}) the analytic continuations of the 
exact expressions to $N \neq 2k$, $k = 1,\,2,\,3\,\ldots$ are not yet 
known. 

We now address the end-point behaviour of the third-order coefficient 
functions for $F_{\,2}$. The leading terms at large $x$ are the soft-gluon 
+-distributions $\DD_k$, $ k = 0,\:\ldots ,\: 2n\!-\!1$ of $c_{\,2,
\rm ns}^{\,(n)}(x)$ in Eqs.~(\ref{eq:c2ns1}), (\ref{eq:c2ns2}) and 
(\ref{eq:c2ns3}). For the highest four coefficients at three loops, our 
exact results 
\bea
\label{eq:c2qd5}
  c_{\,2,\rm ns}^{\,(3)} \Big|_{\,\DD^{}_5} \! & = \! & 
    8\: C_F^{\,3} \:\: , \\[2mm]
\label{eq:c2qd4}
  c_{\,2,\rm ns}^{\,(3)} \Big|_{\,\DD^{}_4} \! & = \! &
  - \frac{220}{9}\: C_A C_F^{\,2}  - 30\: C_F^{\,3} 
  + \frac{40}{9}\: C_F^{\,2}\nf  \:\: , \\[2mm]
\label{eq:c2qd3}
  c_{\,2,\rm ns}^{\,(3)} \Big|_{\,\DD^{}_3} \! & = \! &
  \frac{484}{27}\: C_A^{\,2} C_F \: + \:
  C_A C_F^{\,2} \left[ \frac{1732}{9} - 32\:\z2 \right] \: + \:
  C_F^{\,3} \Big[ - 36 - 96\:\z2 \Big] \nn \\[0.5mm] & & \mbox{} - \:
  \frac{176}{27}\: C_F C_A \nf \: - \: \frac{280}{9}\: C_F^{\,2} \nf 
  \: + \: \frac{16}{27}\: C_F \n2f \:\: , \\[2mm]
\label{eq:c2qd2}
  c_{\,2,\rm ns}^{\,(3)} \Big|_{\,\DD^{}_2} \! & = \! &
  C_A^{\,2} C_F \,\left[ -\frac{4649}{27} + \frac{88}{3}\:\z2 \right]
     \: + \:
  C_A  C_F^{\,2} \,\left[ -\frac{8425}{18} + \frac{724}{3}\z2 
     + 240\:\z3 \right] 
     \nn \\[0.5mm] & & \mbox{} + \:
  C_F^{\,3} \,\left[ \frac{279}{2} + 288\:\z2 + 16\:\z3 \right]
     \: + \:
  C_A C_F \nf \,\left[ \frac{1552}{27} - \frac{16}{3}\:\z2 \right]
     \nn \\[0.5mm] & & \mbox{} + \:
  C_F^{\,2} \nf \,\left[ \frac{683}{9} - \frac{112}{3}\:\z2 \right]
     \: - \:  \frac{116}{27}\: C_F \n2f
\eea
completely agree with the prediction \cite{Vogt:1999xa} of the 
next-to-leading logarithmic threshold resummation \cite{Sterman:1987aj,%
Catani:1989ne,Magnea:1990qg,Catani:1991rp}. The remaining two terms read
\bea
\label{eq:c2qd1}
  c_{\,2,\rm ns}^{\,(3)} \Big|_{\,\DD^{}_1} \! & = \! &
  C_A^{\,2} C_F \,\left[ \frac{50689}{81} - \frac{680}{3}\:\z2 
      - 264\:\z3 + \frac{176}{5}\:\zs \right] \: + \: 
  C_A C_F^{\,2} \,\left[ - \frac{5563}{18} 
  \right. \nn \\[0.5mm] && \left. \: \mbox{} 
      - 972\:\z2 - \frac{160}{3}\:\z3 + \frac{764}{5}\:\zs \right] 
      \: + \:
  C_F^{\,3} \,\left[ \frac{187}{2} + 240\:\z2 - 360\:\z3 
      + \frac{376}{5}\:\zs \right] \nn  \\[0.5mm] & & + \:
  C_A C_F \nf\, \left[ - \frac{15062}{81} + \frac{512}{9}\:\z2
      + 16\:\z3 \right] \: + \:
  C_F^{\,2} \nf\, \left[ \frac{83}{9} + 168\:\z2 + \frac{112}{3}\:\z3 
      \right]  \nn \\[0.5mm] & & \mbox{} + \:
  C_F \n2f \left[ \frac{940}{81} - \frac{32}{9}\:\z2 \right]
      \:\: , \\[2mm]
  c_{\,2,\rm ns}^{\,(3)} \Big|_{\,\DD^{}_0} \! & = \! &
  C_A^{\,2} C_F \,\left[ - \frac{599375}{729} + \frac{32126}{81}\:\z2
      + \frac{21032}{27}\:\z3 - \frac{652}{15}\:\zs 
      - \frac{176}{3}\:\z2\z3 + 232\:\z5 \right]  \nn \\[0.5mm] 
      & & \mbox{} + \:
  C_A C_F^{\,2} \,\left[ \frac{16981}{24} + \frac{26885}{27}\:\z2 
      - \frac{3304}{9}\:\z3 - 209\:\zs - 400\:\z2\z3 - 120\:\z5
      \right]  \nn \\[0.5mm] & & \mbox{} + \:
  C_F^{\,3} \,\left[ - \frac{1001}{8} - 429\:\z2 + 274\:\z3 - 210\:\zs
      + 32\:\z2\z3 + 432\:\z5 \right]  \nn \\[0.5mm] & & \mbox{} + \:
  C_A C_F \nf\, \left[ \frac{160906}{729} - \frac{9920}{81}\:\z2
      - \frac{776}{9}\:\z3 + \frac{208}{15}\:\z2^{\!\! 2}\right] \:+\:
  C_F^{\,2} \nf\, \left[ - \frac{2003}{108} \right. \nn \\[0.5mm] 
      & & \mbox{} \left. - \frac{4226}{27}\:\z2
      - 60\:\z3 + 16\:\zeta_2^{\, 2}\right] \: + \:
  C_F \n2f \left[ - \frac{8714}{729} + \frac{232}{27}\:\z2
      - \frac{32}{27}\:\z3 \right] \:\: .
\label{eq:c2qd0}
\eea
The fermionic ($\nf$) contributions in Eqs.~(\ref{eq:c2qd1}) and 
(\ref{eq:c2qd0}) were presented already in Ref.~\cite{Moch:2002sn}.
From this part of Eq.~(\ref{eq:c2qd1}), the non-fermionic part is
actually predicted \cite{Moch:2002sn} by the next-to-next-to-leading 
logarithmic threshold resummation \cite{Vogt:2000ci} in terms of the 
leading large-$x$ coefficient $A_3$ of the three-loop quark-quark 
splitting function \cite{Moch:2004pa}, cf.~the three-loop prediction
for the Drell-Yan coefficient function in Ref.~\cite{Vogt:2000ci}.
Our result (\ref{eq:c2qd1}) agrees with this prediction, thus 
constituting the first verification of the next-to-next-to-leading
logarithmic soft-gluon resummation by a full calculation at third order.
The final coefficient (\ref{eq:c2qd0}) of $\DD^{}_0$ (of which the
leading-$\nf$ part could have been inferred already from Ref.~\cite
{Gardi:2002xm}) can in turn be employed for the next order of the 
soft-gluon resummation which we will present in Ref.~\cite{MVV7}.

The analytic expression for the $\delta(1-x)$ term of $c_{\,2,\rm ns}
^{\,(3)}(x)$ can be read off, with a bit of patience, from Eq.~(B.8) 
in the appendix together with Eqs.~(\ref{eq:g1xto1}) and 
(\ref{eq:g2xto1}). Also this coefficient is relevant for the prediction 
of higher-order +-distributions by means of the threshold resummation 
\cite{MVV7}.

The subleading class of large-$x$ terms in $c_{\,2,\rm ns}^{\,(n)}(x)$
(and the leading one in $c_{\,2,\rm g}^{\,(n)}(x)\,$) is formed by the 
logarithms $L_1^k$ with $ k = 0,\:\ldots ,\: 2n\!-\!1$. For brevity we 
refrain from writing down the corresponding coefficients. There is, 
however, a relation between coefficients of the +-distributions and the 
logarithms in $c_{\,2,\rm ns}^{\,(n)}(x)$ which we would like to 
mention: as predicted in Ref.~\cite{Kramer:1996iq}, 
the coefficient of the highest power $L_1^{k}$ for a given colour 
factor equals, up to a sign, that of the leading term $\DD_{k}$. 
This means that the coefficients of $L_1^5$ for the $C_F^{\,3}$ term, 
those of $L_1^4$ for the $C_A C_F^{\,2}$ and $C_F^{\,2}\nf $ terms, and 
those of $L_1^3$ for the remaining contributions can be directly read
off from Eqs.~(\ref{eq:c2qd5}) -- (\ref{eq:c2qd3}). 

The small-$x$ limit of the non-singlet coefficient functions
$c_{\,2, \rm ns}^{\,(n)}(x)$ is dominated by the contributions $L_0^k$ 
with again $ k = 0,\:\ldots ,\: 2n\!-\!1$. The corresponding three-loop
coefficients are
\bea
\label{eq:c2nl5}
  c_{\,2,\rm ns}^{\,(3)} \Big|_{\,L_0^5} \! & = \! &
  - \frac{1}{2} \, C_F^{\,3} \:\: , \\[1mm]
\label{eq:c2nl4}
  c_{\,2,\rm ns}^{\,(3)} \Big|_{\,L_0^4} \! & = \! &
  - \frac{1001}{108}\, C_A C_F^{\,2}  + \frac{67}{12}\, C_F^{\,3}
  + \frac{91}{54}\, C_F^{\,2}\nf  \:\: , \\[2mm]
\label{eq:c2nl3}
  c_{\,2,\rm ns}^{\,(3)} \Big|_{\,L_0^3} \! & = \! &
  C_A^{\,2} C_F \left[ - \frac{2783}{81} + 20\:\z2 \right] \: + \:
  C_A C_F^{\,2} \left[ - \frac{835}{54} - 64\:\z2 \right] \: + \:
  C_F^{\,3} \left[ 5 + \frac{262}{3}\:\z2 \right] \nn \\[0.5mm] 
  & & \mbox{} + \:
  \frac{1012}{81}\: C_F C_A \nf \: + \: \frac{5}{27}\: C_F^{\,2} \nf 
  \: - \: \frac{92}{81}\: C_F \n2f \:\: , \\[2mm]
\label{eq:c2nl2}
  c_{\,2,\rm ns}^{\,(3)} \Big|_{\,L_0^2} \! & = \! &
  C_A^{\,2} C_F \,\left[ -\frac{23062}{81} + 84\:\z2 \right] \: + \:
  C_A  C_F^{\,2} \,\left[ \frac{17315}{162} - \frac{265}{9}\:\z2 -  
     64\:\z3 \right]
     \nn \\[0.5mm] & & \mbox{} + \:
  C_F^{\,3} \,\left[ - \frac{113}{6} + \frac{299}{3}\:\z2 
     + \frac{646}{3}\:\z3 \right] \: + \:
  C_A C_F \nf \,\left[ \frac{7196}{81} - 8\:\z2 \right]
     \nn \\[0.5mm] & & \mbox{} + \:
  C_F^{\,2} \nf \,\left[ - \frac{1315}{81} - \frac{266}{9}\:\z2 \right]
     \: - \: \frac{496}{81}\: C_F \n2f \:\: , \\[2mm]
\label{eq:c2nl1}
  c_{\,2,\rm ns}^{\,(3)} \Big|_{\,L_0^1} \! & = \! &
  C_A^{\,2} C_F \,\left[ - \frac{78338}{81} + \frac{3058}{9}\:\z2
      +32\:\z3 - 24\:\zs \right] \: + \:
  C_A C_F^{\,2} \,\left[ \frac{106801}{324}
  \right. \nn \\[0.5mm] && \left. \: \mbox{}
      + \frac{599}{9}\:\z2 + \frac{418}{3}\:\z3 + \frac{656}{15}\:\zs 
      \right] \: + \:
  C_F^{\,3} \,\left[ \frac{1619}{12} + \frac{764}{3}\:\z2 + 154\:\z3
      - \frac{556}{3}\:\zs \right] \nn  \\[0.5mm] & & + \:
  C_A C_F \nf\, \left[ \frac{7076}{27} - \frac{688}{9}\:\z2
      + \frac{128}{3}\:\z3 \right] \: + \:
  C_F^{\,2} \nf\, \left[ - \frac{2999}{162} - \frac{482}{9}\:\z2
      - 132\:\z3
      \right]  \nn \\[0.5mm] & & \mbox{} + \:
  C_F \n2f \left[ - \frac{1204}{81} + \frac{16}{3}\:\z2 \right]
      \:\: , \\[2mm]
\label{eq:c2nl0}
  c_{\,2,\rm ns}^{\,(3)} \Big|_{\,L_0^0} \! & = \! &
  C_A^{\,2} C_F \,\left[ - \frac{1779023}{1458} + \frac{14917}{27}\:\z2
      - \frac{1960}{27}\:\z3 - \frac{148}{3}\:\zs 
      - \frac{436}{3}\:\z2\z3 - \frac{152}{3}\:\z5 \right] \nn\\[0.5mm] 
      & & \mbox{} + \:
  C_A C_F^{\,2} \,\left[ \frac{193961}{648} + \frac{6014}{81}\:\z2 
      + \frac{13189}{27}\:\z3 - \frac{3434}{45}\:\zs + 520\:\z2\z3 
      + \frac{1970}{3}\:\z5 \right]  \nn \\[0.5mm] & & \mbox{} + \:
  C_F^{\,3} \,\left[ \frac{5603}{24} + \frac{533}{3}\:\z2 
      + \frac{1730}{3}\:\z3 - \frac{68}{15}\:\zs
      - 904\:\z2\z3 - 872\:\z5 \right]  \nn \\[0.5mm] & & \mbox{} + \:
  C_A C_F \nf\, \left[ \frac{224219}{729} - \frac{4520}{27}\:\z2
      + \frac{1412}{27}\:\z3 + \frac{352}{15}\:\z2^{\!\! 2}\right] \:+\:
  C_F^{\,2} \nf\, \left[ - \frac{2881}{324} \right. \nn \\[0.5mm] 
      & & \mbox{} \left. + \frac{427}{81}\:\z2 - \frac{5902}{27}\:\z3 
      - \frac{424}{45}\:\zeta_2^{\, 2}\right] \: + \:
  C_F \n2f \left[ - \frac{11170}{729} + \frac{232}{27}\:\z2
      + \frac{32}{27}\:\z3 \right] \:\: ,
\eea
or, after inserting $C_A=N_c=3$ and $C_F=4/3$ and the numerical values 
of the $\zeta$-function
\bea
\label{eq:c2nln}
  c_{\,2,\rm ns}^{\,(3)} \big|_{\,L_0^5} \! & \cong \! & 
    - 1.18519 \nn\\
  c_{\,2,\rm ns}^{\,(3)} \big|_{\,L_0^4} \! & \cong \! & 
    - 36.1975 + 2.99588\:\nf \nn\\
  c_{\,2,\rm ns}^{\,(3)} \big|_{\,L_0^3} \! & \cong \! & 
    - 309.079 + 50.3045\:\nf - 1.51440\:\n2f \nn\\
  c_{\,2,\rm ns}^{\,(3)} \big|_{\,L_0^2} \! & \cong \! & 
    - 899.553 + 187.429\:\nf - 8.16461\:\n2f \nn\\
  c_{\,2,\rm ns}^{\,(3)} \big|_{\,L_0^1} \! & \cong \! & 
    - 787.175 + 278.856\:\nf - 8.12162\:\n2f \nn\\
  c_{\,2,\rm ns}^{\,(3)} \big|_{\,L_0^0} \! & \cong \! & 
    - 591.159 + 123.002\:\nf + 0.31540\:\n2f \:\: .
\eea

\begin{figure}[p]
\vspace{-4mm}
\centerline{\epsfig{file=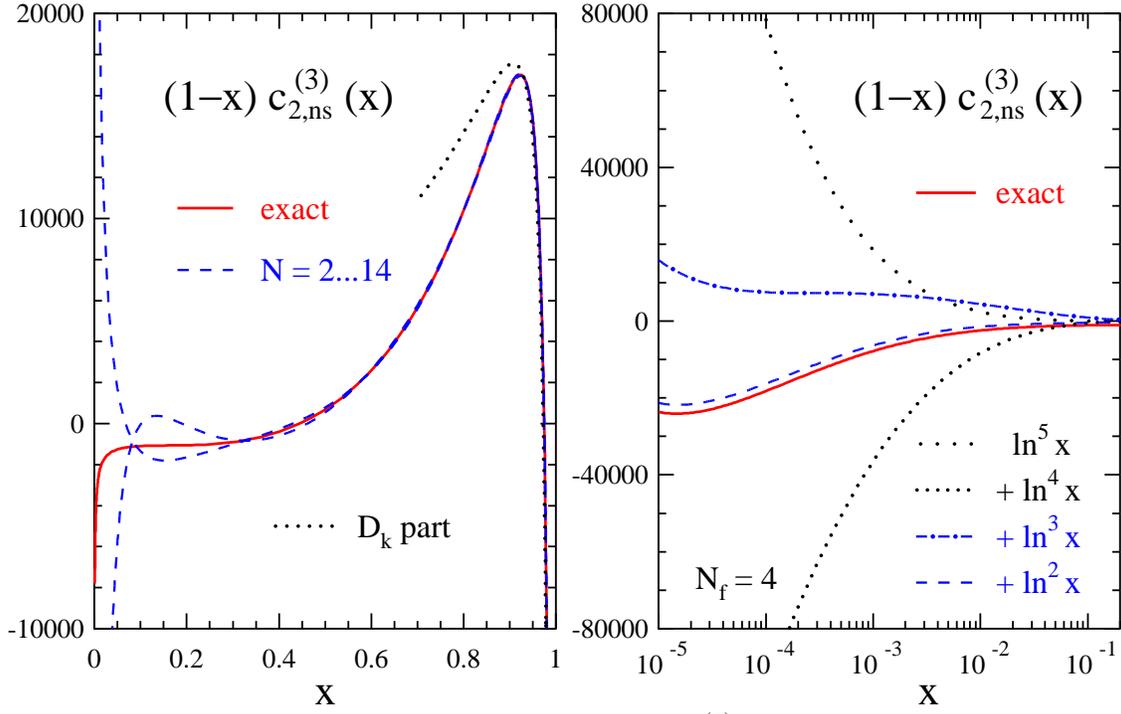,width=15.0cm,angle=0}}
\vspace{-3mm}
\caption{ \label{pic:c2ns3}
  The three-loop non-singlet coefficient function 
 $c_{\,2,\rm ns}^{\,(3)}(x)$ for four flavours, multiplied by 
 $(1\!-\!x)$ for display purposes. 
 Also shown (left) are the large-$x$ approximation by all soft-gluon
 $\DD_k$ terms (\ref{eq:c2qd5}) -- (\ref{eq:c2qd1}), the (dashed) 
 uncertainty band of Ref.~\cite{vanNeerven:2001pe}, and (right) the 
 small-$x$ approximations obtained by successively including 
 Eqs.~(\ref{eq:c2nl5}) -- (\ref{eq:c2nl2}).}
\end{figure}
\begin{figure}[p]
\vspace{-2mm}
\centerline{\epsfig{file=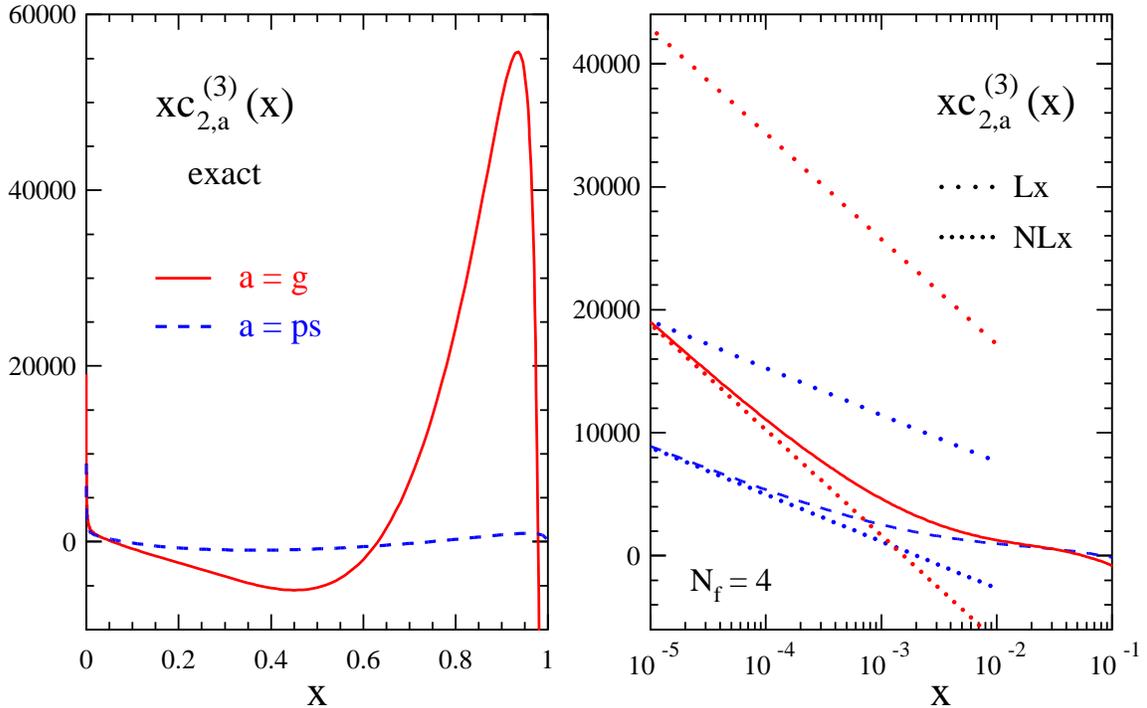,width=15.0cm,angle=0}}
\vspace{-3mm}
\caption{ \label{pic:c2sg3}
 The three-loop pure-singlet and gluon coefficient functions
 $xc_{\,2,\rm ps}^{\,(3)}(x)$ and $xc_{\,2,\rm g}^{\,(3)}(x)$.
 Also shown (right) are the leading \cite{Catani:1994sq} and 
 next-to-leading (besides Eqs.~(\ref{eq:c2px1}) and (\ref{eq:c2gx1}) 
 also including Eqs.~(\ref{eq:c2px0}) and (\ref{eq:c2gx0})$\,$) 
 small-$x$ approximations, respectively denoted by L$x$ and NL$x$.}
\end{figure}

In Fig.~\ref{pic:c2ns3} the non-singlet coefficient function 
$c_{\,2,\rm ns}^{\,(3)}(x)$ of Eqs.~(B.8) and (\ref{eq:c2ns3}) is 
compared for $\nf = 4$ with the large-$x$ and small-$x$ approximations 
specified above and with the previous uncertainty band \cite
{vanNeerven:2001pe} based on the lowest seven even-integer moments 
$N = 2\ldots 14$ of Refs.~\cite{Larin:1994vu,Larin:1997wd,Retey:2000nq}
and the four large-$x$ coefficients (\ref{eq:c2qd5})--(\ref{eq:c2qd2}) 
predicted by the threshold resummation~\cite{Vogt:1999xa}.
 
The complete soft-gluon contribution including $\DD_0 \ldots \DD_5$
deviates from the full coefficient function by less than 20\% only at
$x \geq 0.85$. The corresponding $x$-range reads $x \leq 0.09$ for the
small-$x$ approximation by the terms $\ln x \ldots \ln^5 x$. Note that 
this range only arises when all \mbox{small-$x$} logarithms are taken 
into account. As obvious from Eq.~(\ref{eq:c2nln}), 
small-$x$ approximations by only the first ($\ln^5 x)$, the first two, 
and even the first three logarithms qualitatively fail in the 
$x$-region shown in the figure. In fact, a 20\% accuracy is reached 
with one, two and three small-$x$ logarithms only at $x < 10^{-50}$ 
(sic), $x < 10^{-14}$ and  $x < 10^{-8}$, respectively.
 
What physically matters, of course, is not Fig.~\ref{pic:c2ns3} but 
the contribution to the structure function, given by the convolution 
with the parton distributions. Using the schematic, but sufficiently 
typical form $xq^{}_{\rm ns} = x^{\,0.5} (1-x)^3$ one finds that the 
effect of the soft-gluon $\DD_k$ part approximates the full result to 
better than 20\% only for $x>0.87$. The approximation by all small-$x$
logarithms actually never reaches this accuracy. The large-$x$ range can
be improved to $x > 0.78$ by instead using the large-$N$ version of the 
threshold expansion, keeping only the $\ln^{\,k} N$, $k = 1,\, \ldots\, 
2n$ terms in $c_{\,2,\rm ns}^{\,(n=3)}(N)$. Both versions do however
cover a significantly smaller $x$-range, by about a factor of two, 
than the corresponding soft-gluon approximations at two loops, which 
provide good approximations to $c_{\,2,\rm ns}^{\,(2)} \otimes 
q^{}_{\rm ns}$ for $x \geq 0.7$ using the $\DD_k$ terms and for 
$x \geq 0.55$ keeping only the $\ln\, N$ contributions.

At large $x$ the contributions of the flavour-singlet quantities 
$C_{2,\rm ps}$ and $C_{2,\rm g}$ are small compared to the non-singlet
coefficient functions discussed so far. We therefore do not write out
the coefficient of the (leading) large-$x$ terms $(1-x)L_1^k$, $k = 1,
\,\ldots ,4$ of $c_{\,2,\rm ps}^{\,(3)}(x)$ and $L_1^{k\,'}$, $k\,' = 
1, \,\ldots ,5$ of $c_{\,2,\rm g}^{\,(3)}(x)$ for brevity. 
The leading small-$x$ contributions to these functions are, at $n \geq 
2$ loops, of the form $x^{-1} \ln^{\, k} x$, $k = 1,\,\ldots ,n-2$. The 
numerical QCD values of the three-loop coefficients can be read off 
from Eqs.~(\ref{eq:c2ps3}) and Eqs.~(\ref{eq:c2gl3}) above. The 
corresponding analytical results read  
\bea
\label{eq:c2px1}
  c_{\,2,\rm ps}^{\,(3)} \Big|_{\,L_0/x} \! & = \! &
  C_A C_F \nf \,\left[ - \frac{39488}{243} + \frac{416}{9}\:\z2 
    - \frac{128}{9}\:\z3 \right] \:\: , \\[1mm] 
\label{eq:c2px0}
  c_{\,2,\rm ps}^{\,(3)} \Big|_{\,1/x} \:\: & = \! &
  C_A C_F \nf \,\left[ - \frac{971284}{729} + \frac{15040}{81}\:\z2
    + \frac{752}{9}\:\z3 +  \frac{3968}{45}\:\zs \right] 
     \nn \\[0.5mm] &&  \mbox{} \: + \: 
  C_F^{\,2} \nf \,\left[ \frac{1090}{27} - 16\:\z2
    - \frac{800}{9}\:\z3 +  \frac{192}{5}\:\zs \right]  
     \nn \\[0.5mm] &&  \mbox{} \: + \: 
  C_F \n2f \,\left[ \frac{22112}{729} - \frac{32}{9}\:\z2
    + \frac{128}{27}\:\z3 \right]
\eea
and 
\bea
\label{eq:c2gx1}
  c_{\,2,\rm g}^{\,(3)} \Big|_{\,L_0/x} \! & = \! &
  C_A^{\,2} \nf \,\left[ - \frac{39488}{243} + \frac{416}{9}\:\z2
    - \frac{128}{9}\:\z3 \right] \:\:\: = \:\:\: 
   \frac{C_A}{C_F}\: c_{\,2,\rm ps}^{\,(3)} \Big|_{\,L_0/x}
   \:\: , \\[1mm]
\label{eq:c2gx0}
  c_{\,2,\rm g}^{\,(3)} \Big|_{\,1/x} \:\: & = \! &
  C_A^{\,2} \nf \,\left[ - \frac{1002332}{729} + \frac{16096}{81}\:\z2
    + \frac{2192}{9}\:\z3 +  \frac{3968}{45}\:\zs \right]
     \nn \\[0.5mm] &&  \mbox{} \: + \:
  C_A C_F \nf \,\left[ \frac{1090}{27} - 16\:\z2
    - \frac{800}{9}\:\z3 +  \frac{192}{5}\:\zs \right] \: + \:
  C_A \n2f \,\left[ - \frac{572}{729} \right. 
     \nn \\[0.5mm] &&  \left. \mbox{} 
    + \frac{160}{27}\:\z2 + \frac{64}{27}\:\z3 \right] \: + \:
  C_F \n2f \,\left[ \frac{45368}{729} - \frac{512}{27}\:\z2
    + \frac{128}{27}\:\z3 \right] \:\: .
\eea
The leading contributions (\ref{eq:c2px1}) and (\ref{eq:c2gx1}) were
derived already ten years ago in Ref.~\cite{Catani:1994sq} in the
framework of the small-$x$ resummation. As illustrated in the right 
part of Fig.~\ref{pic:c2sg3}, these leading terms alone do not provide
a useful approximation at $x$-values relevant to collider measurements.
At $x=10^{-4}$, for example, they overshoot the respective full results
for $c_{\,2,\rm ps}^{\,(3)}(x)$ and $c_{\,2,\rm g}^{\,(3)}(x)$ in 
Eqs.~(B.10), (\ref{eq:c2ps3}) and (B.9), (\ref{eq:c2gl3}) by a factor 
of about three. This situation is completely analogous to, if somewhat 
worse than that for the three-loop splitting functions discussed in 
Ref.~\cite{Vogt:2004mw}.

It should be noted that also the singlet coefficient functions receive
contributions from non-$1/x$ logarithms up to $\ln^{2k-1} x$ at  
order $\as^{\,k}$. In fact, the $1/x$ terms (\ref{eq:c2px1}) -- (\ref
{eq:c2gx0}) contribute more than 80\% of $c_{\,2,\rm ps}^{\,(3)}(x)$ 
and $c_{\,2,\rm g}^{\,(3)}(x)$ only at $ x \leq 3\cdot 10^{-4}$. One
may expect this range to shrink further at higher orders due to the
double-logarithmic enhancement of the non-$1/x$ terms. However, as the
above third-order range is rather similar to that for the second-order
coefficient functions, our results do not provide evidence for this
effect.

\begin{figure}[p]
\vspace{-4mm}
\centerline{\epsfig{file=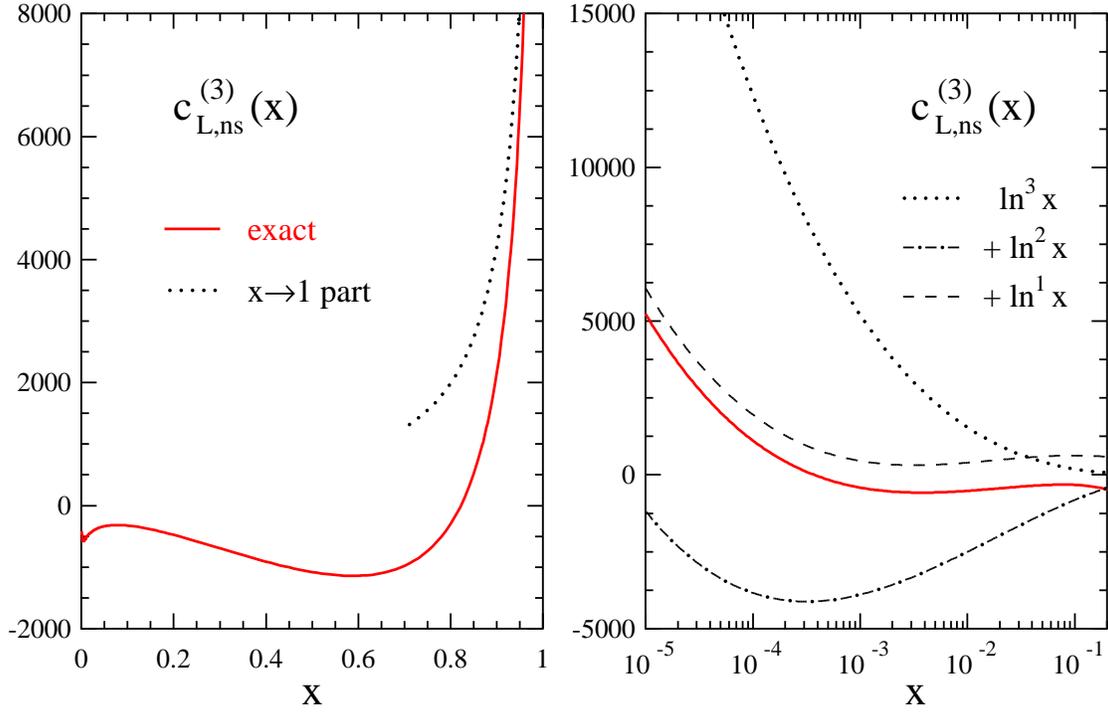,width=15.0cm,angle=0}}
\vspace{-3mm}
\caption{ \label{pic:clns3}
 The three-loop non-singlet coefficient function
 $c_{L,\rm ns}^{\,(3)}(x)$ for four flavours.
 Also shown (left) are the large-$x$ approximation by all terms 
(\ref{eq:clql14}) -- (\ref{eq:clql10}) not vanishing for $x \ra 1$, 
 and (right) the small-$x$ approximations obtained by successively 
 including Eqs.~(\ref{eq:clnl3}) -- (\ref{eq:clnl1}).}
\end{figure}

For the rest of this section we turn to the third-order coefficient
functions for the longitudinal structure function $F_L$ which we only 
briefly discussed in Ref.~\cite{Moch:2004xu}. The behaviour of the 
coefficient functions $c_{L, \rm i}^{\,(3)}(x)$ for $ x\ra 1 $ is 
given by $\,\ln^{\,k}(1-x) \equiv L_1^{k}$, $k = 0, \,\ldots ,4\,$ for 
$\,\rm i = ns\,$, by $\,(1-x) L_1^{k}$, $k = 0, \,\ldots ,4\,$ for 
$\,\rm i = g\,$ and $\,(1-x)^2 L_1^{k}$, $k = 0, \,\ldots ,3\,$ for 
$\,\rm i = ps$.
The coefficients for the dominant non-singlet contribution read
\bea
\label{eq:clql14}
  c_{L,\rm ns}^{\,(3)} \Big|_{\,L_1^{4}} \! & = \! &
    8\: C_F^{\,3} \:\: , \\[2mm]
\label{eq:clql13}
  c_{L,\rm ns}^{\,(3)} \Big|_{\,L_1^{3}} \! & = \! &
  C_A C_F^{\,2} \left[ - \frac{640}{9} + 32\:\z2 \right] \: + \:
  C_F^{\,3} \Big[ 72 - 64\:\z2 \Big] \: + \: \frac{64}{9}\: C_F \n2f 
  \:\: , \\[2mm]
\label{eq:clql12}
  c_{L,\rm ns}^{\,(3)} \Big|_{\,L_1^{2}} \! & = \! &
  C_A^{\,2} C_F \,\left[ \frac{1276}{9} - 56\:\z2 - 32\:\z3 \right]
     \: + \:
  C_A  C_F^{\,2} \,\left[ -\frac{530}{9} + 80\:\z2 + 80\:\z3 \right]
     \nn \\[0.5mm] & & \mbox{} + \:
  C_F^{\,3} \,\Big[ -34 - 32\:\z2 -32\:\z3 \Big]
     \: + \:
  C_A C_F \nf \,\left[ -\frac{320}{9} + 16\:\z2 \right]
     \nn \\[0.5mm] & & \mbox{} + \:
  C_F^{\,2} \nf \,\left[ \frac{92}{9} - 32\:\z2 \right]
     \: + \:  \frac{16}{9}\: C_F \n2f  \:\: , \\[2mm]
\label{eq:clql11}
  c_{L,\rm ns}^{\,(3)} \Big|_{\,L_1^{1}} \! & = \! &
  C_A^{\,2} C_F \,\left[ - \frac{25756}{27} + \frac{3008}{9}\:\z2 
     + \frac{880}{3}\:\z3 - \frac{128}{5}\:\zs\right] 
  \nn \\[0.5mm] & & \mbox{} + \:
  C_A C_F^{\,2} \,\left[ \frac{32732}{27} - \frac{4720}{9}\:\z2 
     + \frac{472}{3}\:\z3 - \frac{1152}{5}\:\zs \right] \: + \: 
  C_F^{\,3} \,\Bigg[ - 264 
     \nn \\[0.5mm] & & \left. \mbox{} 
     + 16\:\z2 - 752\:\z3 - \frac{2816}{5}\:\zs \right] \: + \:
  C_A C_F \nf \,\left[ \frac{6640}{27} - \frac{320}{9}\:\z2 
     - \frac{256}{3}\:\z3 \right]
     \nn \\[0.5mm] & & \mbox{} + \:
  C_F^{\,2} \nf \,\left[ -\frac{4736}{27} + \frac{352}{9}\:\z2 
     + \frac{320}{3}\:\z3 \right]
     \: - \:  \frac{304}{27}\: C_F \n2f \:\: , \\[2mm]
\label{eq:clql10}
  c_{L,\rm ns}^{\,(3)} \Big|_{\,L_1^{0}} \! & = \! &
  C_A^{\,2} C_F \,\left[ \frac{67312}{81} + \frac{824}{3}\:\z2
     - \frac{1264}{3}\:\z3 + 56\:\zs - 80\:\z2\z3 - 160\:\z5 \right] 
     \nn \\[0.5mm] & & \mbox{} + \:
  C_A C_F^{\,2} \,\left[ -\frac{5255}{6} - \frac{10988}{9}\:\z2
     + \frac{3280}{3}\:\z3 - 516\:\zs + 416\:\z2\z3 + 1200\:\z5 \right] 
     \nn \\[0.5mm] & & \mbox{} + \:
  C_F^{\,3} \,\left[ \frac{1937}{6} - 
     + 508\:\z2 - 88\:\z3 + \frac{3384}{5}\:\zs - 512\:\z2\z3 
     - 1760\:\z5 \right] 
     \nn \\[0.5mm] & & \mbox{} + \:
  C_A C_F \nf \,\left[ -\frac{21488}{81} + \frac{32}{9}\:\z2
     + \frac{64}{3}\:\z3 - \frac{32}{5}\:\zs \right]
     \nn \\[0.5mm] & & \mbox{} + \:
  C_F^{\,2} \nf \,\left[ 79 + \frac{1064}{9}\:\z2 
     - \frac{400}{3}\:\z3 + \frac{256}{5}\zs \right]
     \: + \:  
  C_F \n2f \,\left[ \frac{1624}{81} - \frac{32}{9}\:\z2 \right]
     \nn \\[1mm] & & \mbox{} + \:
  9\: fl_{11}^{\:\rm ns} \, \nf \, \left[ - 320 - 1120\:\z2 - 1760\:\z3 
     + 32\:\zs +320\:\z2\z3 +3200\:\z5 \right] \:\: . \quad
\eea
Except for Eq.~(\ref{eq:clql14}) addressed already in Ref.~\cite{Moch:2004xu}, 
these large-$x$ coefficients do not exhibit any obvious
relation to those of $c_{\,2,\rm ns}^{\,(3)}$ in Eqs.~(\ref{eq:c2qd5})
-- (\ref{eq:c2qd0}). Thus the above coefficients should provide 
important checks and inputs for an explicit higher-order threshold 
resummation for $F_L$ along the lines of Refs.~\cite{Akhoury:1998gs,%
Akhoury:2003fw}. Such a resummation might not be of a large 
phenomenological relevance in view of the rather narrow region of
validity of the large-$x$ approximation, see the left part of Fig.~%
\ref{pic:clns3}, and the experimental status of $F_L$ at large $x$. 
It would definitely be useful, however, in conjunction with a possible
future four-loop generalization of the fixed-$N$ calculations of 
Ref.~\cite{Larin:1994vu}.

The leading small-$x$ contributions to $c_{L,\rm ns}^{\,(m)}(x)$ are
given by the terms $L_0^k$ with $k = 0, \ldots, 2m\!-\!3$. The 
corresponding three-loop coefficients read
\bea
\label{eq:clnl3}
  c_{L,\rm ns}^{\,(3)} \Big|_{\,L_0^3} \! & = \! &
  - \frac{20}{3} \, C_F^{\,3} \:\: , \\[1mm]
\label{eq:clnl2}
  c_{L,\rm ns}^{\,(3)} \Big|_{\,L_0^2} \! & = \! &
  - 66\: C_A C_F^{\,2}  + 32\: C_F^{\,3}
  + 12\: C_F^{\,2}\nf  \:\: , \\[2mm]
\label{eq:clnl1}
  c_{L,\rm ns}^{\,(3)} \Big|_{\,L_0^1} \! & = \! &
  C_A^{\,2} C_F \left[ - \frac{968}{9} + 120\:\z2 \right] \: + \:
  C_A C_F^{\,2} \left[ - \frac{1832}{9} - 384\:\z2 \right] 
  \nn \\[0.5mm] & & \mbox{} + \:
  C_F^{\,3} \Big[ 168 + 416\:\z2 \Big] \: + \: 
  \frac{352}{9}\: C_F C_A \nf \: + \: \frac{224}{9}\: C_F^{\,2} \nf
  \: - \: \frac{32}{9}\: C_F \n2f \:\: , \\[2mm]
\label{eq:clnl0}
  c_{L,\rm ns}^{\,(3)} \Big|_{\,L_0^0} \! & = \! &
  C_A^{\,2} C_F \,\left[ -\frac{13060}{27} + 244\:\z2 \right] \: + \:
  C_A  C_F^{\,2} \,\left[ \frac{5800}{27} - \frac{1696}{3}\:\z2 
     - 96\:\z3 \right]
     \nn \\[0.5mm] & & \mbox{} + \:
  C_F^{\,3} \,\Big[ 288 + 608\:\z2 \Big] \: + \:
  C_A C_F \nf \,\left[ \frac{4184}{27} - 16\:\z2 \right]
     \nn \\[0.5mm] & & \mbox{} + \:
  C_F^{\,2} \nf \,\left[ - \frac{1144}{27} - \frac{32}{3}\:\z2 \right]
     \: - \: \frac{304}{27}\: C_F \n2f \:\: .
\eea
In the flavour-singlet sector, the dominant $n$-loop small-$x$ terms 
for $F_L$ are of the same form as those for $F_{\,2}$ discussed above. The 
respective $1/x$ contributions to $c_{L,\rm i}^{\,(3)}$, 
$\,\rm i = ps, g\,$ are given by
\begin{figure}[p]
\vspace{-2mm}
\centerline{\epsfig{file=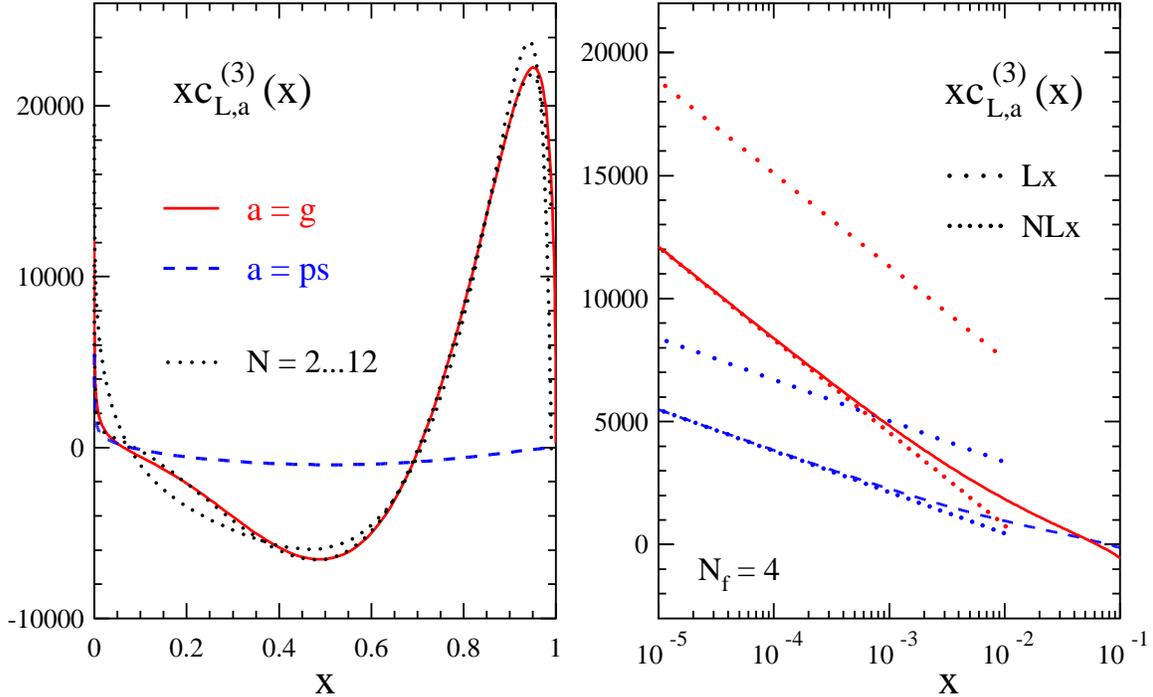,width=15.0cm,angle=0}}
\vspace{-3mm}
\caption{ \label{pic:clsg3}
 The three-loop pure-singlet and gluon coefficient functions
 $xc_{L,\rm ps}^{\,(3)}(x)$ and $xc_{L,\rm g}^{\,(3)}(x)$.
 Also shown (left) are the previous uncertainty band for the latter 
 quantity \cite{Martin:2000gq,Thorne:2004ci} inferred from the results 
 of Refs.~\cite{Retey:2000nq,Catani:1994sq}, and (right) the leading 
 (Lx) \cite{Catani:1994sq} and next-to-leading (NLx) small-$x$ 
 approximations as given by Eqs.~(\ref{eq:clpx1}) -- (\ref{eq:clgx0})
 analogous to the right part of Fig.~\ref{pic:c2sg3}.}
\end{figure}
\bea
\label{eq:clpx1}
  c_{L,\rm ps}^{\,(3)} \Big|_{\,L_0/x} \! & = \! &
  C_A C_F \nf \,\left[ - \frac{2176}{27} + \frac{64}{3}\:\z2 \right]
    \:\: , \\[2mm]
\label{eq:clpx0}
  c_{L,\rm ps}^{\,(3)} \Big|_{\,1/x} \:\: & = \! &
  C_A C_F \nf \,\left[ - \frac{14384}{27} + 112\:\z2
    + \frac{320}{3}\:\z3 \right]
     \nn \\[0.5mm] &&  \mbox{} + \:
  C_F^{\,2} \nf \,\left[ \frac{1792}{27} - \frac{32}{3}\:\z2
    - \frac{128}{3}\:\z3 \right] \: + \:
  C_F \n2f \,\left[ \frac{3392}{81} - \frac{64}{9}\:\z2 \right]
\eea
and
\bea
\label{eq:clgx1}
  c_{L,\rm g}^{\,(3)} \Big|_{\,L_0/x} \! & = \! &
  C_A^{\,2} \nf \,\left[ - \frac{2176}{27} + \frac{64}{3}\:\z2 \right] 
   \:\:\: = \:\:\:
   \frac{C_A}{C_F}\: c_{\,\rm L,ps}^{\,(3)} \Big|_{\,L_0/x}
   \:\: , \\[2mm]
\label{eq:clgx0}
  c_{L,\rm g}^{\,(3)} \Big|_{\,1/x} \:\: & = \! &
  C_A^{\,2} \nf \,\left[ - \frac{44096}{81} + \frac{1040}{9}\:\z2
    + \frac{320}{3}\:\z3 \right] \: + \:
  C_A \n2f \,\left[ \frac{808}{27} - \frac{32}{9}\:\z2 \right] 
     \nn \\[0.5mm] &&  \mbox{} \: + \:
  C_A C_F \nf \,\left[ \frac{1792}{27} - \frac{32}{3}\:\z2 
    - \frac{128}{3}\:\z3 \right] \: + \:
  C_F \n2f \,\left[ \frac{1936}{81} - \frac{64}{9}\:\z2 \right]
    \:\: .
\eea

The exact $C_F/C_A$ relation between the leading small-$x$ terms \cite
{Catani:1994sq} of $c_{a,\rm ps}^{\,(n)}$ and $c_{a,\rm g}^{\,(n)}$
does not hold for the subleading $1/x$-contributions at three loops.
However, most coefficients (actually those for all colour-factor$\,/\,
\zeta$-function combinations not occurring in the leading terms) are 
still closely related. Besides the relations obvious from Eqs.\
(\ref{eq:c2px0}) and (\ref{eq:c2gx0}) for $c^{}_{\,2,\rm g}$ and
Eqs.~(\ref{eq:clpx0}) and (\ref{eq:clgx0}) for $c^{}_{L,\rm g}$, we 
note that in both cases the sum of half the coefficient of $C_F\n2f$
and the coefficient of $C_A\n2f$ in the gluonic coefficient function
equals that of $C_F\n2f$ in the pure-singlet coefficient function.
Numerically the $C_F/C_A$ relation is violated by less than 5\% for
realistic values of $\nf$.

Eqs.~(\ref{eq:clnl3}) -- (\ref{eq:clgx0}) lead to the following 
numerical values for QCD,
\bea
\label{eq:clnln}
  c_{L,\rm ns}^{\,(3)} \big|_{\,L_0^3} \! & \cong \! &
    - 15.8025 \nn\\
  c_{L,\rm ns}^{\,(3)} \big|_{\,L_0^2} \! & \cong \! &
    - 276.148 + 21.3333\:\nf \nn\\
  c_{L,\rm ns}^{\,(3)} \big|_{\,L_0^1} \! & \cong \! &
    - 1356.17 + 200.691\:\nf - 4.74074\:\n2f \nn\\
  c_{L,\rm ns}^{\,(3)} \big|_{\,L_0^0} \! & \cong \! &
    - 2226.25 + 408,058\:\nf + 15.0123\:\n2f 
\eea
and
\bea
\label{eq:clsln}
  c_{L,\rm ps}^{\,(3)} \Big|_{\,L_0/x} \! & \cong \! &
    - 182.003\:\nf \nn\\
  c_{L,\rm ps}^{\,(3)} \Big|_{\,1/x} \:\: & \cong \! &
    - 885.534\:\nf + 40.2390\:\n2f\:\: , \\[2mm]
  c_{L,\rm g}^{\,(3)} \Big|_{\,L_0/x}  \! & \cong \! &
    - 409.506\:\nf \nn\\
  c_{L,\rm g}^{\,(3)} \Big|_{\,1/x}  \:\: & \cong \! &
    - 2044.70\:\nf + 88.5037\:\n2f \:\: .
\eea
The coefficients in both the non-singlet and singlet cases exhibit the
pattern by now familiar from the three-loop splitting functions
\cite{Moch:2004pa,Vogt:2004mw} and the coefficient functions for 
$F_{\,2}$ discussed above. The successive approximations of 
$c_{L,\rm i}^{\,(3)}$ by the leading small-$x$ terms are compared 
in the right parts of Figs.~\ref{pic:clns3} and \ref{pic:clsg3} to the
complete results (B.16) -- (B.18). Also here the dominant $\,x \ra 0$ 
contributions, $\,\ln^3 x\,$ for $\,\rm i = ns\,$ and $\,x^{-1}\ln x\,$
for $\,\rm i = ps,\,g\,$, alone do not provide useful approximations 
for practically relevant values of $x$.
Such endpoint constraints are however phenomenologically important when 
combined with other partial results as, e.g., in Refs.~\cite
{Martin:2000gq,Thorne:2004ci} for the previously used approximations
illustrated in the left part of Fig.~\ref{pic:clsg3} for 
$c_{L,\rm g}^{\,(3)}(x)$. 
%
%
\setcounter{equation}{0}
\section{Numerical implications}
\label{sec:sresults}
%
%
In this section we finally discuss the size and convergence of the
perturbative corrections to the structure functions $F_{\,2}$ and $F_L$ in
electromagnetic DIS. For brevity we confine ourselves to one physical 
scale $Q^{\,2} = Q_0^{\,2}$ for almost all illustrations, and fix 
the renormalization and factorization scales by $\mu_r^{\,2} = 
\mu_{\!f}^{\,2} = Q^{\,2}$.  The scale $Q_0^{\,2}$ is specified by an 
order-independent value of the strong coupling in the \MSb\ scheme,
\beq
\label{eq:asref}
  \as (Q_{0}^{\,2}) \: = \: 0.2 \quad \mbox{for} \quad
  \nf \: = \: 4 \:\: .
\eeq
Depending on the precise value of $\as$ at the $Z$-boson mass, this
choice corresponds (beyond the leading order) to a scale $Q_0^{\,2} 
\approx 30 \ldots 50 \mbox{ GeV}^2$, where especially $F_{\,2}$ has been
measured over a wide range in $x$ by fixed-target experiments and at
the $ep$ collider HERA~\cite{Eidelman:2004wy}.
  
We also assume, for a straightforward comparison of the effects of the
various orders in Eqs.~(\ref{eq:cf-exp}) (at $\ep = 0$) and (\ref
{eq:Fa-dec}), that the operator matrix elements (\ref{eq:OME}) and 
their all-$N/$all-$x$ generalizations, the parton distributions (PDFs),
do not depend on the perturbative order in $\as$. We are aware that 
this is not the case in practical analyses in perturbative QCD, where 
these non-perturbative quantities are fitted to data.
Our choice can be viewed as an idealization, representing a situation
in which both $\as(Q_{0}^{\,2})$ and the PDFs at this scale have been
determined, independent of the order in $\as$, by a non-perturbative
solution of QCD. Specifically we choose
\beq
\label{eq:q-ns}
  xq_{\rm ns}(x,Q_{0}^{\,2}) \: = \: x^{\, 0.5} (1-x)^3 
\eeq
for the flavour non-singlet combination and
\bea
\label{eq:p-sg}
  xq_{\rm s}(x,Q_0^{\,2}) &\! = \! &
  0.6\: x^{\, -0.3} (1-x)^{3.5}\, (1 + 5.0\: x^{\, 0.8\,}) \:\: , 
\nn \\
  xg (x,Q_0^{\,2})\:\: &\! = \! &
  1.6\: x^{\, -0.3} (1-x)^{4.5}\, (1 - 0.6\: x^{\, 0.3\,})
\eea
for the singlet quark and gluon distributions. The same schematic, but
sufficiently realistic input distributions have also been employed in
Refs.~\cite{Moch:2004pa,Vogt:2004mw}.

\begin{figure}[p]
\vspace{-4mm}
\centerline{\epsfig{file=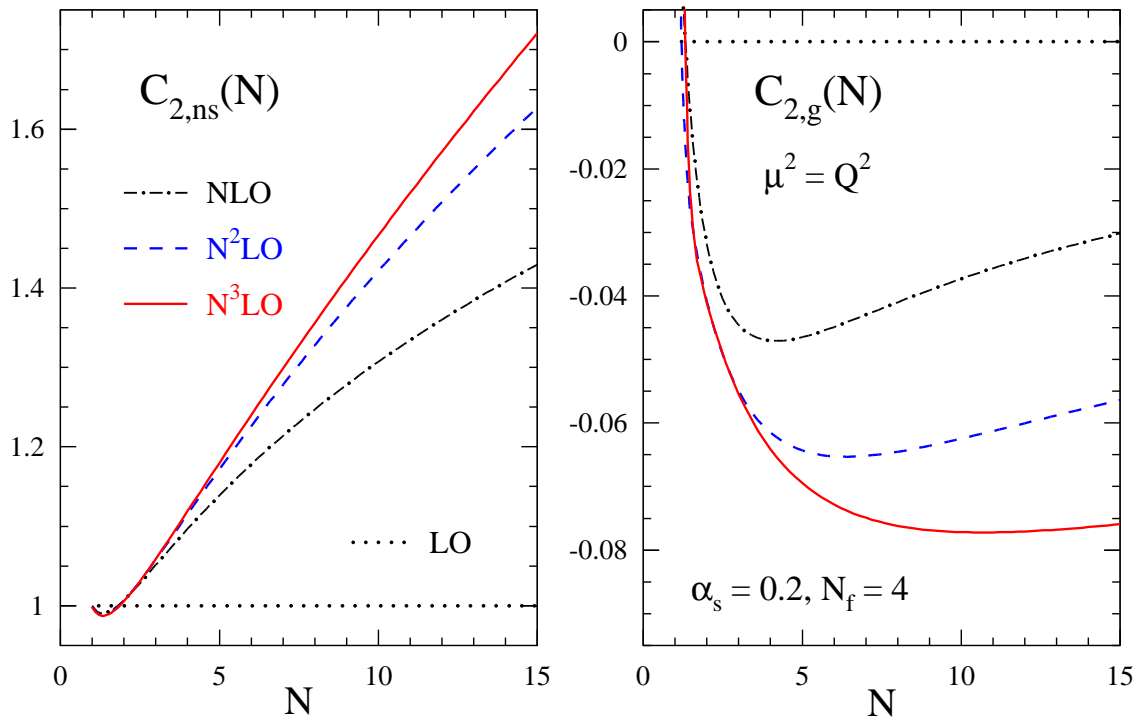,width=15.5cm,angle=0}}
\vspace{-2mm}
\caption{ \label{pic:c2ns-n}
 The perturbative expansion of the non-singlet (left) and gluon (right)
 $N$-space coefficient functions for $x^{-1}F_{\,2}$ at our reference point
 (\ref{eq:asref}). The N$^n$LO curves include the $\ep\!=\!0$ terms up 
 to order $\ar^{\,n}$ in Eq.~(\ref{eq:cf-exp}), obtained by Mellin
 inverting the results in Section 4 and Appendix B.}
\end{figure}

\begin{figure}[p]
\vspace{-2mm}
\centerline{\epsfig{file=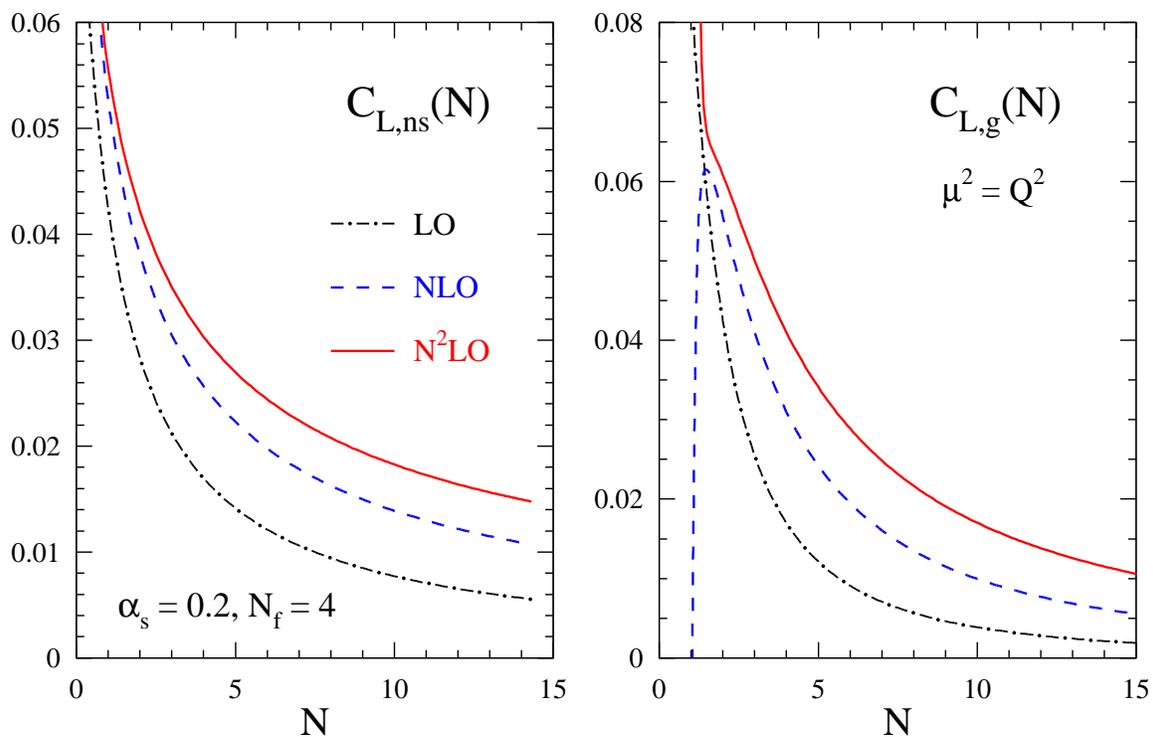,width=15.5cm,angle=0}}
\vspace{-1mm}
\caption{ \label{pic:clns-n}
 As Fig.~\ref{pic:c2ns-n}, but for $x^{-1}F_L$
 where the N$^n$LO results include the terms up to order $\ar^{\,n+1}$.
 }
\end{figure}

We start our illustrations of the perturbative expansion in Mellin-$N$
space. As discussed above, the results for $N \neq 2,\,4,\,6,\,\ldots$ 
cannot be computed directly from the expressions in Appendix A, but are
obtained by Mellin inverting (either analytically or numerically) the 
$x$-space expressions (\ref{eq:c2ns1}), (\ref{eq:c2gl1}) and 
(\ref{eq:c2ns2}) -- (\ref{eq:c2gl3}) or (B.3) -- (B.18). 
The expansions of $C_{a,\rm ns}(\as,N)$ and $C_{a,\rm g}(\as,N)$, 
$a = 2,L$, up to order $\as^{\,3}$ are shown in Figs.~\ref{pic:c2ns-n} 
and \ref{pic:clns-n} at our reference point (\ref{eq:asref}). Here and 
below we use a linear scale up to $N = 15$, hence the main parts of 
these figures correspond to rather large values of $x$, recall 
Eq.~(\ref{eq:F2mellin}) or Eq.~(\ref{eq:gNposdef}). 
The small-$x$ region will be addressed below. 


The largest absolute corrections are found for $C_{2,\rm ns}$ at large
$N$. Here the coefficient functions at order $\as^{\,n}$ behave as
$\ar^{\,n} \ln^{\,k} N$, $k = 1,\,\ldots,\,2n$, corresponding to the
+-distributions in Eqs.~(\ref{eq:c2qd5}) -- (\ref{eq:c2qd0}). Hence the
expansion in powers of $\as$ breaks down for $N \ra \infty$ with
$\,c_{2,\rm ns}^{(n+1)} / c_{2,\rm ns}^{(n)} \sim \ar \ln^2 N$.
In the region of $N$ shown in the figures, however, the three-loop
effect is always smaller than half the two-loop contribution.
$C_{2,\rm g}$ and $C_{L,\rm ns}$, on the other hand, vanish as $N^{-1}
\ln^{\,l}N$ for $N\ra\infty$, and $C_{L,\rm g}$ as $N^{-2} \ln^{\,l}N$,
a behaviour that is not yet relevant, in particular for $C_{2,\rm g}$, 
at practically important values of $N$ either. All these coefficient 
functions receive considerably larger relative $\as^{\,3}$ corrections
than $C_{2,\rm ns}$. In general the perturbative stability of $F_L$ is 
worse than that of $F_{\,2}$.

Having calculated, for the first time, the complete third-order 
corrections to a one-scale process, inclusive DIS, we are in a unique 
situation to study the behaviour of the perturbation series. 
In order to illustrate the $N$-dependent convergence (or the lack 
thereof) of the corresponding coefficient functions, we introduce the 
quantity
\beq
\label{eq:ashat}
  \widehat{\alpha}_{a,i}^{\,(n)}(N) \:\: = \:\: 4\,\pi\: 
  \frac{c_{a,i}^{(n-1)}(N)}{2\,c_{a,i}^{(n)}(N)} \:\: .
\eeq
Recalling the normalization (\ref{eq:arun}), $\,\ar\equiv\as/(4\pi)\,$,
of our expansion parameter, $\widehat{\alpha}^{\,(n)}(N)$ represents 
the value of $\as$ for which the $n$-th order correction is half as 
large as that of the previous order. $\as \lsim \widehat 
{\alpha}_{a,i}^{\,(n)}(N)$ therefore defines, somewhat arbitrarily due
to the choice of a factor of two, a region of good convergence of
$C_{a,i}(\as,N)$. 
Obviously, the (absolute) size of the $n$-th and $(n\!-\!1)$-th order 
effects are equal for $\as= 2\,\widehat{\alpha}^{\,(n)}(N)$. Thus the 
quantity (\ref{eq:ashat}) also indicates where the expansion appears 
not to be reliable any more for a given value of the Mellin variable, 
$\as \gsim 2\,\widehat{\alpha}^{\,(n)}(N)$.

\begin{figure}[p]
\vspace{-4mm}
\centerline{\epsfig{file=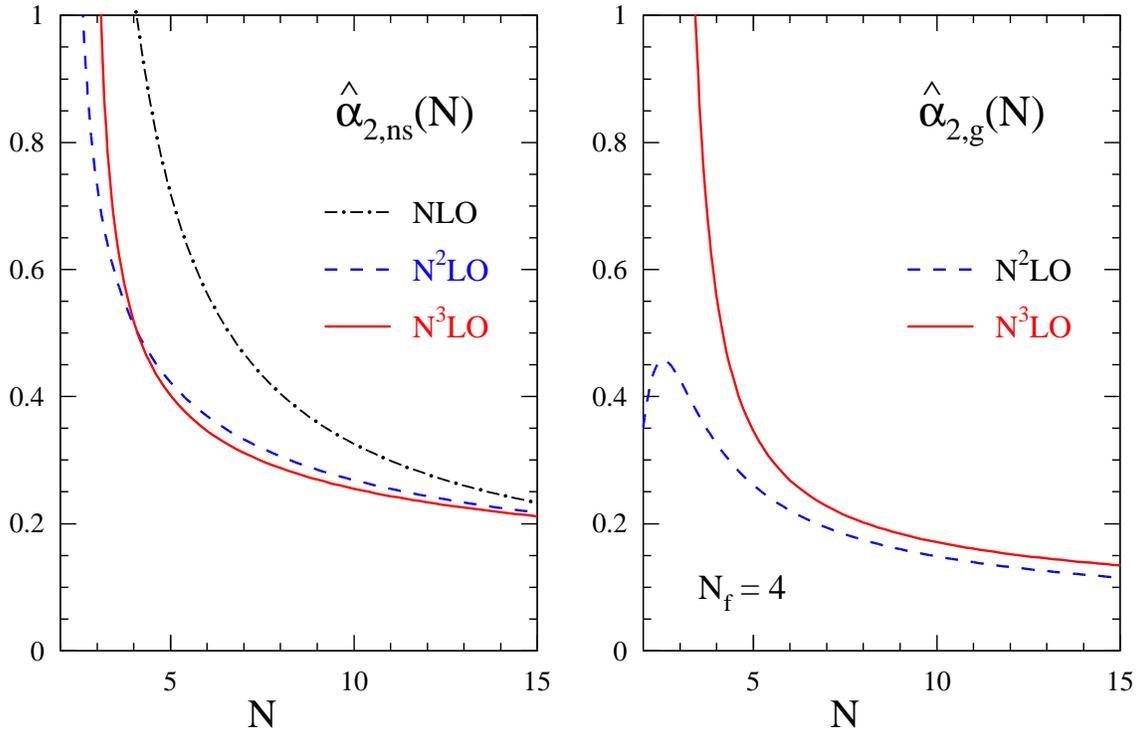,width=15.5cm,angle=0}}
\vspace{-2mm}
\caption{ \label{pic:ahat-2}
 The $N$-dependent values (\ref{eq:ashat}) of $\as$ at which the effect
 of the $n$-th order (N$^n$LO) non-singlet and gluon coefficient 
 functions for $F_{\,2}$ is half as large as that of the previous order. 
 A NLO curve can only be shown for the non-singlet since only here the 
 LO contribution does not vanish.}
\end{figure}

\begin{figure}[p]
\vspace{-2mm}
\centerline{\epsfig{file=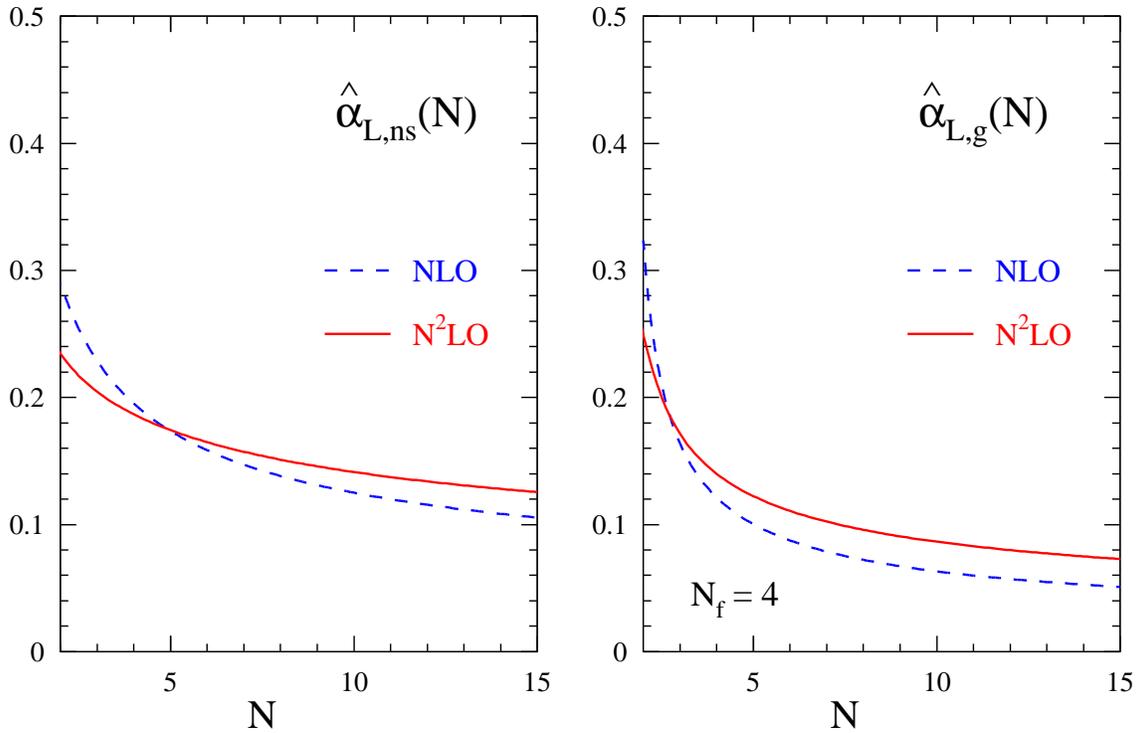,width=15.5cm,angle=0}}
\vspace{-1mm}
\caption{ \label{pic:ahat-l}
 As Fig.~\ref{pic:ahat-2}, but for $F_L$ where the terms up to order 
 $\as^{\,n+1}$ form the N$^n$LO approximation.}
\end{figure}

The function $\widehat{\alpha}^{\,(n)}(N)$ is shown in Fig.~\ref
{pic:ahat-2} for $C_{2,\rm ns}$ and $C_{2,\rm g}$ and in Fig.~\ref
{pic:ahat-l} for $C_{L,\rm ns}$ and $C_{L,\rm g}$ at $N \geq 2$.
For the coefficient function $C_{2,\rm ns}$ dominating the corrections
to $F_{\,2}$ in the large-$N\,/\,$large-$x$ region, $\widehat{\alpha}
^{\,(n)}(N)$ is always smaller than 0.2 at $N \leq 17$, and even 
smaller than 0.35 at $N \leq 6$. It is also important to note that 
the resulting safe region $\as \leq \widehat{\alpha}_{2,\rm ns}
^{\,(n)}(N)$ only marginally shrinks at three loops (N$^3$LO, $n=3$ in
Eq.~(\ref{eq:ashat})$\,$) with respect to the previous order. Thus we 
do not observe any sign of a breakdown of the perturbative expansion at 
phenomenologically relevant values of $N$. This also holds for the
other cases shown in Figs.~\ref{pic:ahat-2} and \ref{pic:ahat-l}. In 
fact, while being considerably smaller than for $C_{2,\rm ns}$, the
regions of fast convergence actually increase for the third-order
(N$^2$LO for $F_L$) results, except for $C_{L,i}$ at small $N$ where 
the stability is relatively best anyhow with, for instance, 
$\widehat{\alpha}^{(3)}_ {L,i} = 0.20$ and 0.17 for $i=\rm ns$ and 
$i=\rm g$ at $N=3$.

We end our $N$-space illustrations by estimating the size of the 
presently (and, presumably, in the near future) uncalculated fourth-%
order corrections to the non-singlet coefficient function at large $N$.
For this purpose we make use of the Pad\'e summation of the 
perturbation series, discussed in detail for QCD, e.g., in Refs.~\cite
{Pade1,Pade2,Pade3}. 
In this approach $C_{2,\rm ns}(N)$ in Eq.~(\ref{eq:cf-exp}) (for 
$\ep = 0$) is replaced by a rational function in $\ar$, 
\beq
\label{eq:pade}
 \widetilde{C}_{2,\rm ns}^{\:[\cal{N}/\cal{D}]}(N) \:\: = \:\: 
 \frac{1 + \ar p_{1}(N) + \ldots + \ar^{\cal N} p_{\cal N}(N)}
 {1 + \ar q_{1}(N) + \ldots + \ar^{\cal D} q_{\cal D}(N)} \:\: .
\eeq
Here ${\cal D} \,\geq\, 1$ and ${\cal N} + {\cal D} \, = \, n$, where 
$n$ stands for the maximal order in $\as$ at which the expansion 
coefficients $c_{2,\rm ns}^{(k)}(N)$ have been determined from an exact
calculation. 
The functions $p_i(N)$ and $q_{\!j}(N)$ are determined from these known
coefficients by expanding Eq.~(\ref{eq:pade}) in powers of $\as$. This
expansion then also provides the $[\cal{N}/\cal{D}]$ Pad\'e approximant
for the ($n\!+\!1$)-th order quantities $c_{2,\rm ns}^{(n\!+\!1)}$.
 

\begin{figure}[htb]
\vspace*{2mm}
\centerline{\epsfig{file=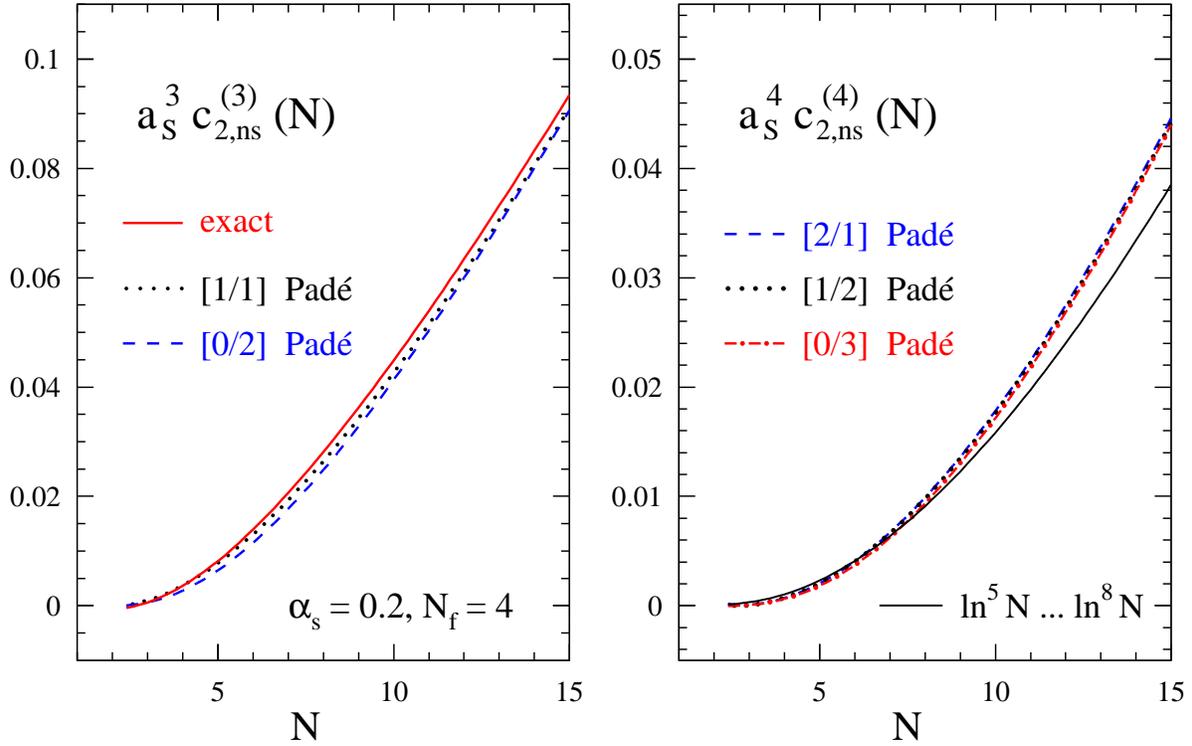,width=16.0cm,angle=0}}
\vspace{-1mm}
\caption{ \label{pic:cns-pade}
 Pad\'{e} estimates for the large-$N$ behaviour of the three-loop
 (left) and four-loop (right) contributions to the non-singlet
 coefficient function $C_{2,\rm ns}(\as,N)$ at the reference point
 (\ref{eq:asref}). The three-loop approximants are compared with our
 exact results. Also shown at four loops is the estimate by the sum of
 the four leading $\ln^{\,k}N$ terms fixed by the soft-gluon 
 resummation \cite{Vogt:1999xa}. 
 }
\vspace{1mm}
\end{figure}

In the left part of Fig.~\ref{pic:cns-pade} the corresponding [1/1]
and [0/2] Pad\'e predictions for the three-loop coefficient function
are compared to our new exact results. Obviously the Pad\'e 
approximants provide a fair estimate of the true corrections.
Hence it seems reasonable to expect that, at not too small values of 
$N$, the very similar [2/1], [1/2] and [0/3] fourth-order approximants
shown in the right part of Fig.~\ref{pic:cns-pade} correctly indicate
at least the rough size of the four-loop corrections. 
This expectation is corroborated by a comparison (also shown in the 
figure) with the estimate by the four highest $\ln^{\,k} N$ 
contributions, $k = 5,\,\ldots,\,8$ known from the next-to-leading 
logarithmic threshold resummation~\cite{Vogt:1999xa}.

\begin{figure}[p]
\vspace{-4mm}
\centerline{\epsfig{file=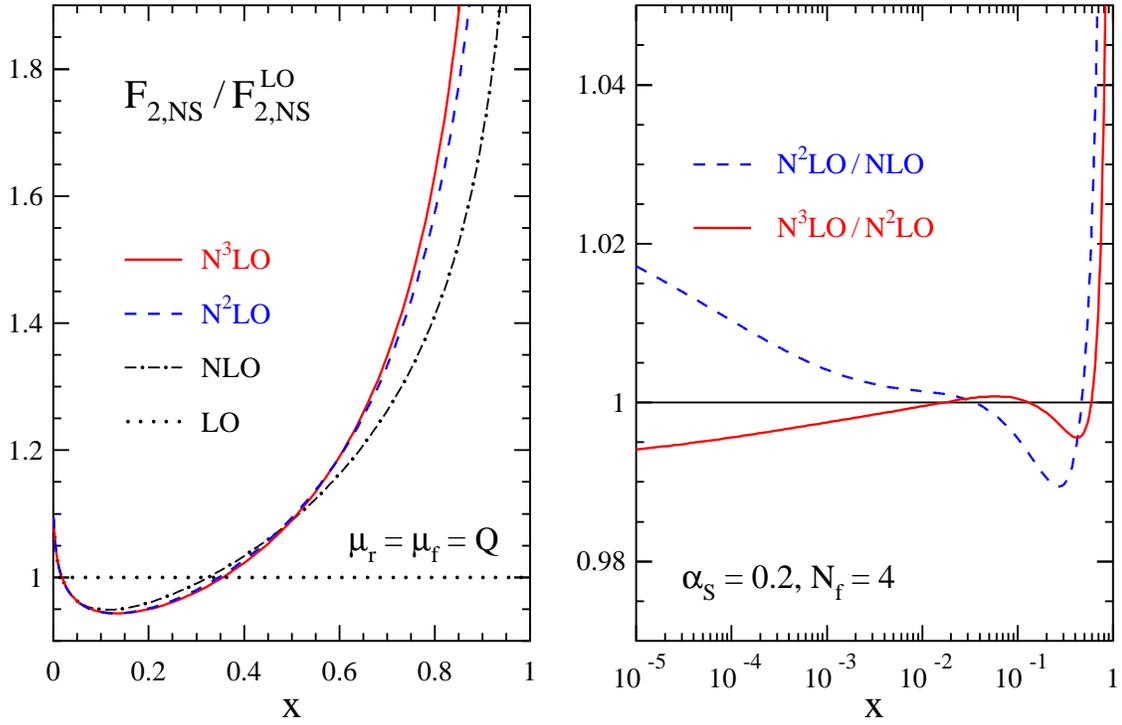,width=15.5cm,angle=0}}
\vspace{-2mm}
\caption{ \label{pic:F2ns-x}
 The perturbative expansion of the non-singlet structure function 
 $F_{2,\rm ns}$ up to three loops (N$^3$LO). On the left all curves are 
 normalized to the leading-order result $F_{2,\rm ns}^{\,\rm LO} = 
 q_{\rm ns}$ given by Eq.~(\ref{eq:q-ns}), on the right we show the 
 relative effects of the two-loop and three-loop corrections.}
\end{figure}

\begin{figure}[p]
\vspace{-2mm}
\centerline{\epsfig{file=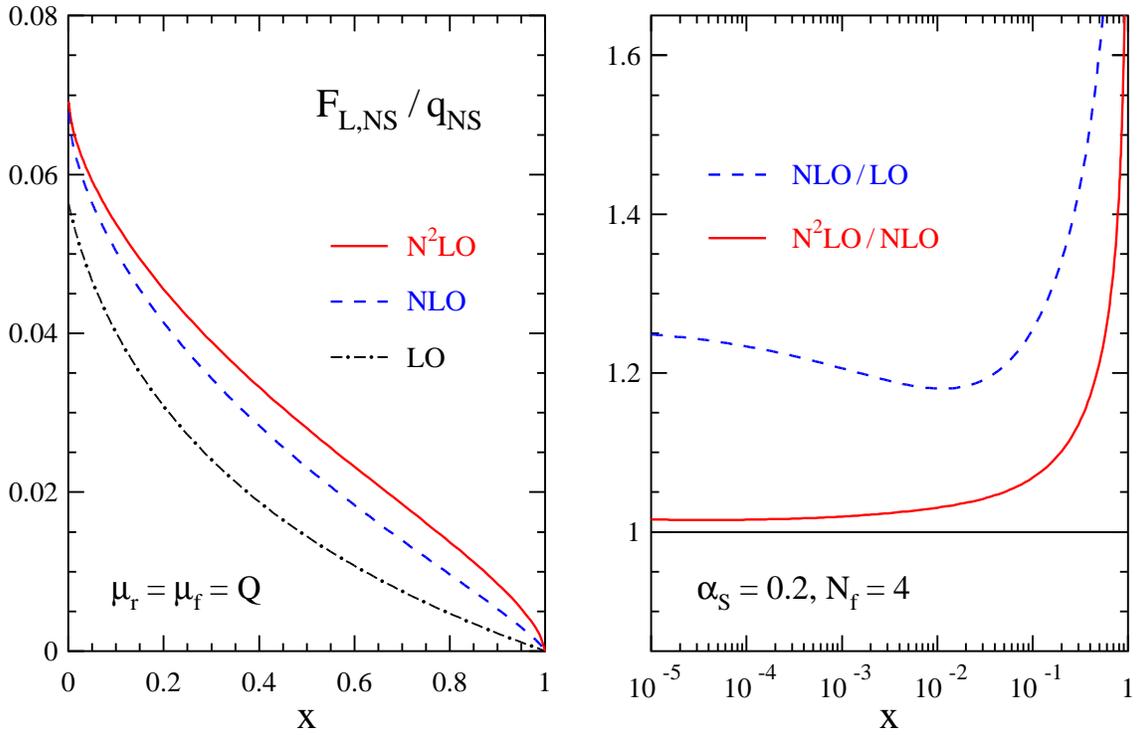,width=15.5cm,angle=0}}
\vspace{-1mm}
\caption{ \label{pic:FLns-x}
 As Fig.~\ref{pic:F2ns-x}, but for $F_L$ where the terms up to order
 $\as^{\,n+1}$ form the N$^n$LO approximation. Also here the left plot
 is normalized to $q_{\rm ns}$, facilitating a direct comparison
 with $F_{2,\rm ns}$.}
\end{figure}

The $x$-space results for the non-singlet quantities $F_{2,\rm ns}$ and
$F_{L,\rm ns}$ are shown in Figs.~\ref{pic:F2ns-x} and \ref
{pic:FLns-x}, respectively, for our reference input (\ref{eq:asref}) 
and (\ref{eq:q-ns}). 
In accordance with the left parts of Figs.~\ref{pic:c2ns-n} and
\ref{pic:clns-n}, the relative large-$x$ corrections are much larger 
for $F_L$ than for $F_{\,2}$ (note the rather different scales of the right
parts of Figs.~\ref{pic:F2ns-x} and \ref{pic:FLns-x}). Nevertheless
the third-order (N$^2$LO) corrections to $F_L$ amount to less than 10\%
for $x< 0.2$ and 3\% at $x< 10^{-2}$, constituting a clear improvement
over the NLO results. The three-loop corrections for $F_{2,\rm ns}$, on
the other hand, even contribute less than 0.5\% at $4\cdot 10^{-5}\leq x
\leq 0.65$, and exceed 3\% only at very large $x$-values, $x\geq 0.78$,
outside the measured region at scales $Q^2 > 30 \mbox{ GeV}^2$, see Ref.~\cite 
{Eidelman:2004wy}.


\begin{figure}[bth]
\centerline{\epsfig{file=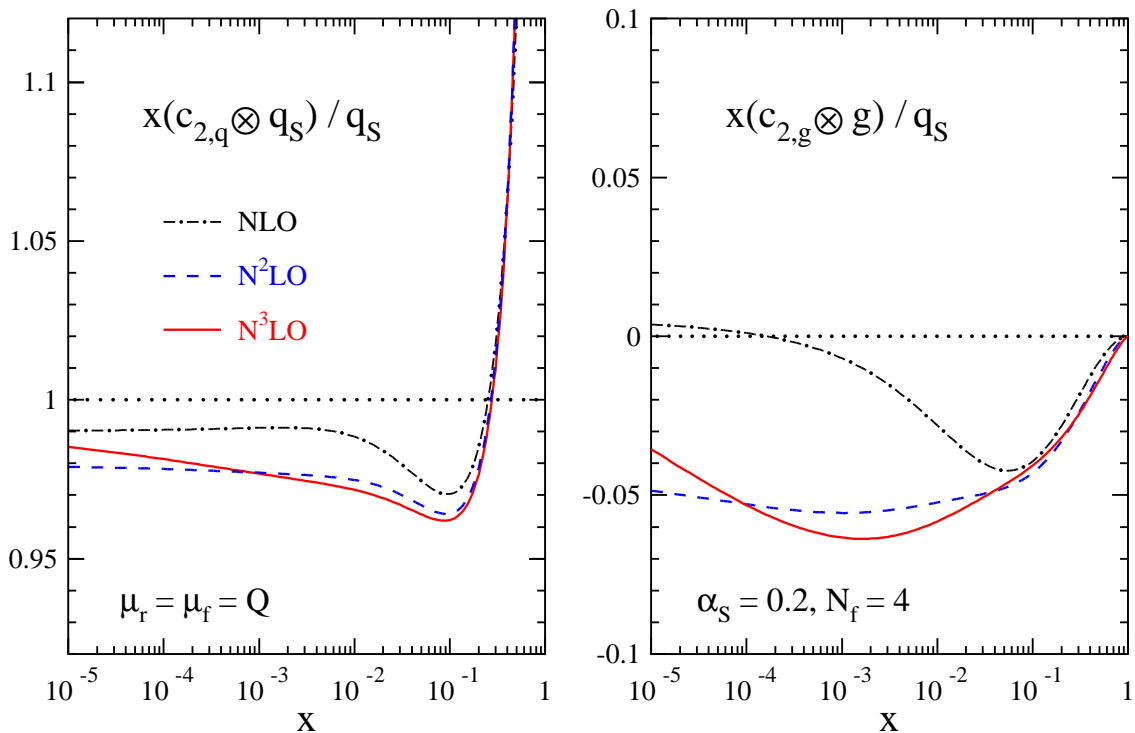,width=15.5cm,angle=0}}
\vspace{-1mm}
\caption{ \label{pic:c2sg-c}
 The perturbative expansion up to three loops (N$^3$LO) of the quark 
 (left) and gluon (right) contributions to singlet structure function
 $F_{2,\rm s}$ at our reference point (\ref{eq:asref}). All curves 
 have been normalized to the leading-order result $\,F_{2,\rm s}^
 {\,\rm LO} = \langle e^2 \rangle \: q_{\rm s}\,$ given by 
 Eq.~(\ref{eq:p-sg}).}
\end{figure}

The corresponding contributions of the (typical) quark and gluon 
distributions (\ref{eq:p-sg}) to the flavour-singlet part of the
structure function $F_{\,2}$ are displayed in Fig.~\ref{pic:c2sg-c}. 
The three-loop pure-singlet quark coefficient function $c_{2,\rm ps}
^{(3)} \equiv c_{2,\rm q}^{(3)} - c_{2,\rm ns}^{(3)}$ contributes less
than 0.1\% at $x \geq 0.13$. Hence the large-$x$ behaviour in the quark
part (left graph) is completely analogous to that of $F_{2,\rm ns}$ 
just discussed.
Likewise the effect of the third-order gluon coefficient function
$c_{2,\rm g}^{(3)}$ (right graph) amounts to less than 0.2\% of the
lowest order result $q_{\rm s}$ at $x \geq 0.03$.

The situation is more interesting at small $x$, as at $x < 10^{-2}$ 
the two-loop corrections are larger, in both cases, than the one-loop 
contributions. Without our new three-loop results, this behaviour 
might be interpreted as indicative of an early breakdown of the 
expansion in $\as$. 
Our results show, however, that this is not the case. In fact,
at $x$-values relevant to collider experiments, the third-order 
contributions are always (considerably) smaller than their second-order
counterparts, exceeding 1\% only at $\,x < 4\cdot 10^{-7}\,$ for 
$\,c_{\,2,\rm q}\otimes q_{\rm s}\,$ and at $\,x <2\cdot 10^{-5}\,$ for
$\,c_{\,2,\rm g}\otimes g\,$. As illustrated already in Ref.~\cite
{Moch:2004xu}, the perturbative stability of $F_L$ is worse also in 
this region of $x$.

\begin{figure}[tbh]
\centerline{\epsfig{file=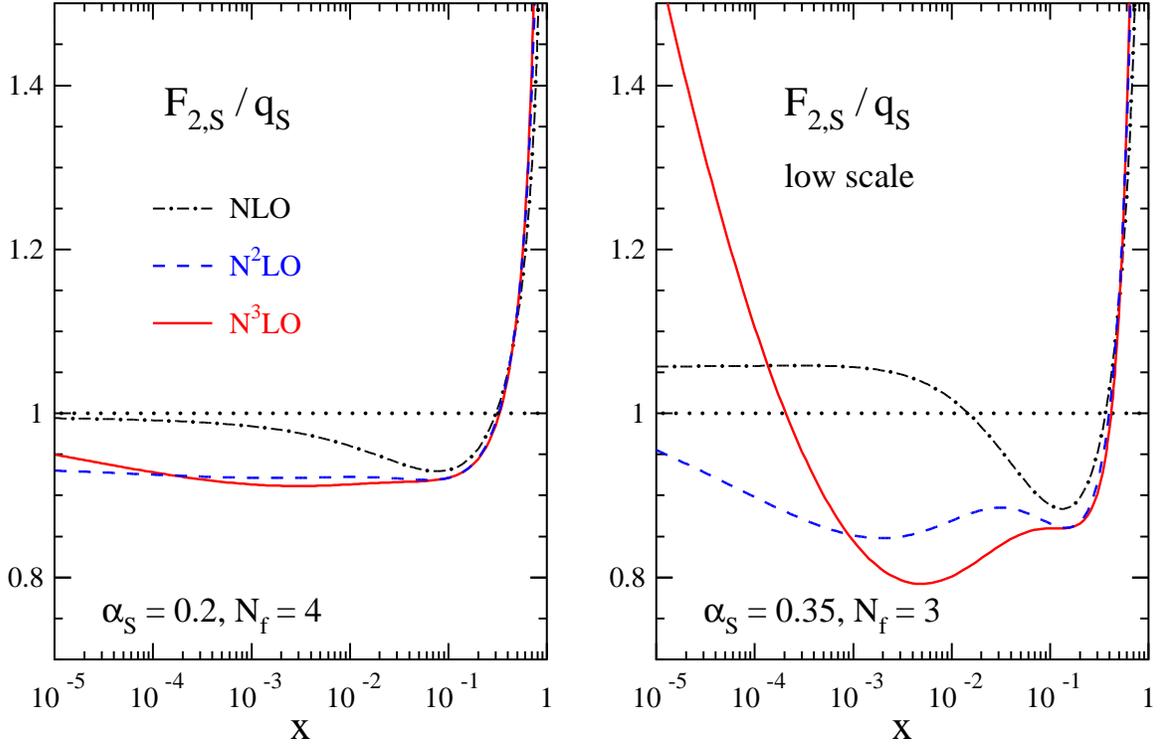,width=16.0cm,angle=0}}
\vspace{-2mm}
\caption{ \label{pic:F2sg-x}
 The flavour-singlet structure function $F_{2,\rm s}(x,Q^2)$ at our
 standard reference point $Q_0^{\,2} \,\approx\, 40 \mbox{ GeV}^2$ 
 (left) and at the low scale $Q_1^{\,2} \,\approx\, 2 \mbox{ GeV}^2$ 
 (right) up to the third order. All curves have been normalized to 
 the respective leading-order results $\,F_{2,\rm s}^{\,\rm LO} = 
 \langle e^2 \rangle \: q_{\rm s}\,$ given by Eqs.~(\ref{eq:p-sg}) 
 and (\ref{eq:p-sg2}).}
\end{figure}

The above quark and gluon contributions are combined in the left part
of Fig.~\ref{pic:F2sg-x}. The total third-order correction is larger 
than 1\% only outside the range $4\cdot 10^{-5} \leq x \leq 0.65$. 
It rises towards $x \ra 0$ and exceeds the size of the (opposite-sign)
second-order contribution below $x \simeq 10^{-8}$.
As all illustrations and numerical values presented in this section so
far, these results refer to a point in the safely deep-inelastic 
region, $Q^2 \,=\, Q_0^{\,2} \,\approx\, 30 \ldots 50 \mbox{ GeV}^2$, 
specified by Eqs.~(\ref{eq:asref}) -- (\ref{eq:p-sg}).   
 
The corresponding results for the singlet structure function 
$F_{2,\rm s}$ at a low scale, $Q_1^{\,2} \,\approx\, 2 \mbox{ GeV}^2$
with $\as = 0.35$ and $\nf= 3$ active flavours, are shown in the right
part of Fig.~\ref{pic:F2sg-x} for the (again order-independent, see
above) quark and gluon distributions~\cite{Vogt:2004mw,Moch:2004xu}
\bea
\label{eq:p-sg2}
  xq_{\rm s}(x,Q_1^{\,2}) &\! = \! &
  0.6\: x^{\, -0.1} (1-x)^{3}\, (1 + 10\: x^{\, 0.8\,}) \:\: , \nn \\
  xg (x,Q_1^{\,2})\:\: &\! = \! &
  1.2\: x^{\, -0.1} (1-x)^{4}\, (1 + 1.5\: x) \:\: .
\eea
If all other parameters were kept equal, the N$^3$LO corrections (with
respect to $\,F_{2,\rm s}^{\,\rm LO}=\langle e^2\rangle\: q_{\rm s}\,$
as shown in the figure) would be larger by a factor of about five here
simply due to the increase in the coupling constant. The modified quark 
and gluon distributions, though, especially their much flatter 
small-$x$ behaviour --- $x^{\, -0.1}$ in Eq.~(\ref{eq:p-sg2}) instead of 
$x^{\, -0.3}$ in Eq.~(\ref{eq:p-sg}), lead to a qualitatively different
pattern at small $x$. While the three-loop corrections remain below 
2\% in the range $0.07 < x < 0.57$ and below 10\% at $3\cdot 10^{-4}
< x < 0.73$, they rise sharply towards lower $x$ at $x \lsim 10^{-3}$.
Consequently, the perturbative expansion of $\,F_{2,\rm s}$ at low
scales appears to be out of control at $x < 10^{-4}$. This rise for 
$x \ra 0$ is very similar to that of $\,F_{L,\rm s}$ in Ref.~\cite
{Moch:2004xu} where the relative third-order (N$^2$LO) corrections are 
however much larger over the full $x$-range.

Finally we need to address the relative importance of our new 
three-loop coefficient functions $c_{2,i}^{(3)}$ and the yet unknown
four-loop splitting functions $P^{(3)}$. Together these two sets of 
quantities form the N$^3$LO approximation for $F_{\,2}$ once, as usual 
in order to resum large $Q^2/\mu_{\!f}^{\,2}$ logarithms, the 
factorization scale $\mu_{\! f}^{}$ is not kept fixed, but varied with 
the physical hard scale $Q^2$. 
The corresponding issue at N$^2$LO has been considered in 
Refs.~\cite{vanNeerven:2000uj,vanNeerven:1999ca}. It was found that
the effect of the three-loop splitting functions on the scaling 
violations of $F_{\,2}$ is small at $x >0.01$ for both the non-singlet 
and singlet structure functions.

Here we confine ourselves to the non-singlet case, which is most
important for the determination of $\as$ from the $Q^2$-dependence of
the structure functions. Following Ref.~\cite{vanNeerven:2001pe} we
express the scaling violation of $F_{2,\rm ns}$ in terms of the 
`physical' N$^n$LO evolution kernel,
\beq
\label{eq:k2ns}
  \frac{d}{d \ln Q^2} \:\, x^{-1} F^{}_{2,\rm ns} \:\: =  \:\: 
  \Bigg\{ \ar P_{\,\rm ns}^{(0)} + \sum_{l=1}^{n}  \ar^{\, l+1}
  \Bigg( P_{\,\rm ns}^{(l)} - \sum_{k=0}^{l-1} \beta_k \,
  \tilde{c}_{2,\rm ns}^{\,(l-k)} \Bigg) \Bigg\} 
  \otimes \Big( x^{-1} F^{}_{2,\rm ns} \Big) 
\eeq
with
\bea
\label{eq:ctilde}
  \tilde{c}_{2,\rm ns}^{\,(1)} & = & c_{2,\rm ns}^{\,(1)} \:\: ,
  \nn \\
  \tilde{c}_{2,\rm ns}^{\,(2)} & = & 2\, c_{2,\rm ns}^{\,(2)} 
     - c_{2,\rm ns}^{\,(1)} \otimes c_{2,\rm ns}^{\,(1)} \:\: ,
  \nn \\
  \tilde{c}_{2,\rm ns}^{\,(3)} & = & 3\, c_{2,\rm ns}^{\,(3)} 
     - 3\, c_{2,\rm ns}^{\,(2)} \otimes c_{2,\rm ns}^{\,(1)}
     + c_{2,\rm ns}^{\,(1)} \otimes c_{2,\rm ns}^{\,(1)} \otimes
      c_{2,\rm ns}^{\,(1)} \:\: , \quad \ldots \quad .
\eea
Here $P^{(l)}$ are the $l$-loop splitting functions, recall Eq.~%
(\ref{eq:gamma}), and $\beta_k$ the coefficient of the $\beta$-function
of QCD in Eq.~(\ref{eq:arun}). The coefficients (\ref{eq:ctilde}) are
given for $\mu_r = Q$, the explicit generalization to $\mu_r \neq Q$
up to N$^4$LO can be found in Ref.~\cite{vanNeerven:2001pe}. 
 
Given the small effect of the three-loop splitting functions $P^{(2)}$
at large $x$, we expect that a rough estimate of $P_{\rm ns}^{(3)}(x)$ 
is sufficient in Eq.~(\ref{eq:k2ns}). We choose the Mellin inverse of
\beq
\label{eq:P3pade}
  P_{\rm ns,\eta}^{(3)}(N) \:\: = \:\: \eta\:
  \Big[ P_{\rm ns}^{(3)}(N) \Big]^{}_{\, [1/1]\:\rm Pad\acute{e}}
  \:\: , \quad \eta \: =\: 0,\: 2 \:\: ,
\eeq
i.e., we assign a 100\% error to the four-loop prediction of the [1/1]
Pad\'e summation. 

\begin{figure}[tbh]
\centerline{\epsfig{file=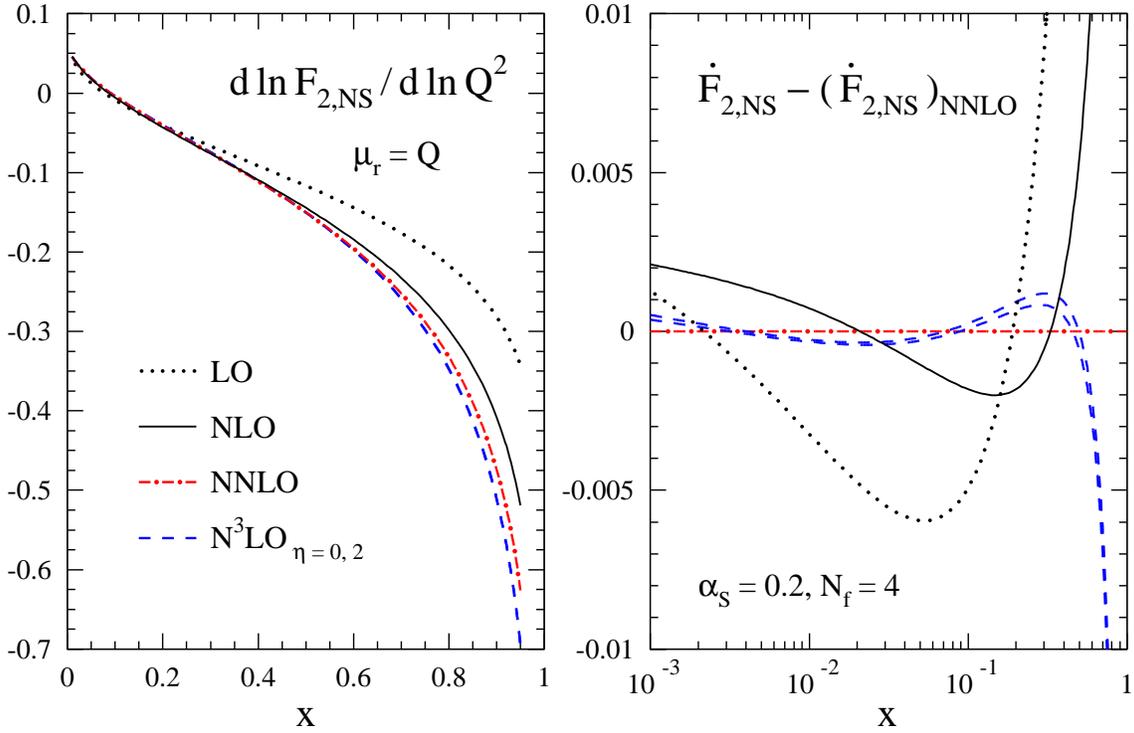,width=15.5cm,angle=0}}
\vspace{-1mm}
\caption{ \label{pic:dF2ns-x}
 The perturbative expansion (\ref{eq:k2ns}) of the scale derivative 
 $\dot{F}_{2,\rm ns} \equiv d\ln F_{2,\rm ns\,} / d\ln Q^2$  for the 
 initial condition $F_{2,\rm ns} = x^{\, 0.5} (1-x)^3$ at the reference
 point (\ref{eq:asref}). 
 The two N$^3$LO curves indicate the uncertainty due to the four-loop 
 splitting functions as estimated in Eq.~(\ref{eq:P3pade}). }
\vspace{2mm}
\end{figure}

The $Q^2$-derivative of $F_{2,\rm ns}$ is illustrated in Fig.~\ref
{pic:dF2ns-x}, again assuming (now for the structure function) the 
non-singlet shape (\ref{eq:q-ns}) at our reference point 
(\ref{eq:asref}). Also for this quantity the N$^3$LO corrections are 
sizeable only at very large values of $x$. They rapidly decrease with 
decreasing $x$, for example from 6\% at $x\! =\! 0.85$ to 2\% at 
$x\! =\! 0.65$, and are smaller by a factor of two and three, 
respectively, than the N$^2$LO contributions at these points. 
As shown in the right part of the figure, the uncertainty due to the 
unknown four-loop splitting function is indeed very small over the 
whole $x$-range considered here. Therefore QCD analyses of the scaling 
violations can be extended to the N$^3$LO outside the small-$x$ region.
An idealized fit to $d\ln F_{2,\rm ns\,} / d\ln Q^2$ at $Q_0^2 \approx
40 \mbox{ GeV}^2$, as described in more detail in Ref.~\cite
{vanNeerven:2001pe}, yields the following order-dependence of the 
central values: 
\bea
\label{eq:asres}
 \alpha_s(Q_0^2)^{\,}_{\rm NLO} \:\:\: & = & \: 0.208  \:\: , \quad  
 \alpha_s(Q_0^2)^{\,}_{\rm N^2LO}   \:\: = \:\: 0.201  \:\: ,
   \nn\\[0.5mm]
 \alpha_s(Q_0^2)^{\,}_{\rm N^3LO}      & = & \: 0.200  \:\: , \quad 
 \alpha_s(Q_0^2)^{\,}_{\rm N^4LO}   \:\: = \:\: 0.200  \:\: ,
\eea
where the N$^4$LO kernel has been estimated by the Pad\'e summation.
Thus, even if the idealized fit underestimates the shifts by a factor
of two, an $\as$-uncertainty of 1\% or less from the truncation of the 
perturbation series has been reached by the calculation
of the  N$^3$LO coefficient functions. 

%
%
\section{Summary}
\label{sec:summary}
%
%
We have calculated the complete third-order coefficient functions for 
the electromagnetic structure functions $F_{\,2}$ and $F_L$ in massless
perturbative QCD. Our calculation has been performed in 
\mbox{Mellin-$N$} space, as previous fixed-$N$ computations of deep-%
inelastic scattering \cite{Larin:1994vu,Larin:1997wd,Retey:2000nq,%
Moch:2001im} using the optical theorem and a dispersion relation in the
Bjorken variable $x$. 
However, generalizing the two-loop calculation of Ref.~\cite
{Moch:1999eb} and our previous computation of the fermionic non-singlet
corrections at three loops~\cite{Moch:2002sn}, we have now obtained the 
complete third-order coefficient functions for all even values of $N$ 
by deriving the analytic $N$-dependence of all required integrals 
through an elaborate iterative system of reduction equations.
The full dependence of the coefficient functions on $x$ and $N$ has 
then been reconstructed from the even-$N$ results by analytic 
continuation, making use of the relation between the harmonic sums
$S_{\vec{n}}(N)$ \cite{Vermaseren:1998uu} and the harmonic 
polylogarithms $H_{\vec{m}}(x)$ \cite{Remiddi:1999ew} in which the
respective results can be expressed.

Our coefficient functions agree with all partial results available in
the literature. The coefficients of the leading small-$x$ terms 
$\,x^{\,-1} \ln x\,$ of the singlet quark and gluon coefficient 
functions were derived already in Ref.~\cite{Catani:1994sq} in the 
framework of the small-$x$ resummation. No corresponding result is
known for the non-singlet case, in contrast to the splitting functions
\cite{Blumlein:1995jp}. The first four large-$x$ terms, $\ln^{\,k} N$
with $k = 3,\ldots 6$ in Mellin space, were predicted by the 
soft-gluon threshold resummation \cite{Vogt:1999xa}. Most importantly,
the even moments $\,N = 2,\ldots,12\,$ for the singlet case and $\,N = 
2,\ldots,14\,$ for the non-singlet coefficient function were computed 
before \cite{Larin:1994vu,Larin:1997wd,Retey:2000nq} using
the {\sc Form} \cite{Vermaseren:2000nd,Vermaseren:2002rp} version of
the {\sc Mincer} program \cite{Gorishnii:1989gt,Larin:1991fz}. 
In fact, we have made extensive use of this program for checks at
intermediate stages of our calculations.
Finally our results also agree with the recent computation of the 
$N\!=\!16$ non-singlet moments of $F_{\,2}$ and $F_L$ \cite
{Blumlein:2004xt}, which was specifically performed as an independent 
simultaneous check of our calculations.

We have investigated the convergence of the perturbative expansion of
the coefficient functions $C_{a,i}(N,\as)$ by determining, at $2 \leq 
N \leq 20$, the $N$-dependent range of $\as$ for which the $n$-th order 
corrections are at most half as large as the $(n\!-\!1)$-th order 
contributions.
For $C_{\,2,\rm ns}$ this $\as$-region shrinks significantly from the
first to the second order, but only marginally from the second to the
third. The coefficient functions $C_{\,2,\rm g}$ and $C_{L,i}$ exhibit 
larger corrections than $C_{\,2,\rm ns}$ --- at $N=6$, for example, 
the above $\as$-region is $\as < 0.35$ for $C_{\,2,\rm ns}$, but 
$\as < 0.17$ for $C_{L,\rm ns}$ --- however this smaller region 
actually increases from the second to the third order for most values
of $N$. Thus, up to the third order, we find no sign of the supposed 
asymptotic character of the perturbative expansion.

Besides the phenomenologically less relevant limit $x \ra 1$, the 
above region of $N$ also excludes the small-$x$ region opened up 
experimentally by HERA. The expansion of the coefficient functions is 
unstable for $x \ra 0$ as $\,c^{(n)}/c^{(n\!-\!1)}\sim \as \ln^{\,\xi}
x\,$ with $\xi=2$ ($\xi=1$) for the non-singlet (singlet) cases.
This behaviour does not spoil the convergence of the $\as$-expansion
for $x$-values relevant to collider measurements in the safely 
deep-inelastic regime $Q^{\,2} \gg 1 \mbox{ GeV}^2$, however, due to an
apparently systematic suppression of the coefficients of leading terms
(see also Refs.~\cite{Moch:2004pa,Vogt:2004mw}) and the Mellin 
convolution with the parton distributions. At $Q^{\,2} \approx 30 
\mbox{ GeV}^2$, for instance, the total third-order correction to 
$F_{\,2}$ is larger than 1\% only outside the wide range $4\cdot 
10^{-5} \leq x \leq 0.65$, and exceeds 5\% only at $x \lsim 10^{-7}$ 
and $x > 0.8$. At low scales $Q^{\,2} \approx 2 \mbox{ GeV}^2$, on the
other hand, the perturbative expansion appears to be out of control at
$x < 10^{-4}$.

Our three-loop results for the splitting functions~\cite
{Moch:2004pa,Vogt:2004mw} and $F_L$ (briefly discussed already in 
Ref.~\cite{Moch:2004xu}$\,$) facilitate NNLO analyses of deep-inelastic
scattering over the full $x$-range covered by data. The present 
additional results for $F_{\,2}$ can be employed to effectively extend
the main part of DIS analyses to the N$^3$LO at $x > 10^{-2}$ where the
effect of the unknown fourth-order splitting functions is expected to 
be very small, for example leading to determinations of $\as(M_Z)$ with 
an error of less than 1\% from the truncation of the perturbation 
series.  For use in such analyses we have provided compact and accurate
parametrizations of our very lengthy exact results.
{\sc Form} files of these results, and {\sc Fortran} subroutines of the 
exact and approximate coefficient functions can be obtained from the
preprint server \ {\tt HTTP://arXiv.org} by downloading the source of 
this article. Furthermore they are available from the authors upon 
request.

Finally our three-loop results for a one-scale process also are of
theoretical interest, as they open up a new order in the study of
perturbative QCD. As first further steps they facilitate the extension
of the Sudakov resummations of threshold logarithms in DIS and of the
quark form factor beyond the orders obtained so far \cite{Vogt:2000ci,%
Catani:2003zt,Magnea:1990zb,Magnea:2000ss,Moch:2002sn}.
We have performed the required additional calculations for both cases
and will report the results in a forthcoming publication~\cite{MVV7}.
%
%
\subsection*{Acknowledgments}
We would like to thank S.A.~Larin, 
F.~J.~Yndurain, E.~Remiddi, E.~Laenen, W.L.~van Neerven, P.~Uwer, 
S.~Weinzierl and J.~Bl\"umlein for stimulating discussions.  M.~Zhou 
has contributed some {\sc Form} routines in an early stage of this 
project. We are grateful to T.~Gehrmann for providing a weight-five 
extension of the {\sc Fortran} package~\cite{Gehrmann:2001pz} for the 
harmonic polylogarithms. The Feynman diagrams in this article have been 
drawn using the packages {\sc Axodraw} \cite{Vermaseren:1994je} and 
{\sc Jaxo\-draw} \cite{Binosi:2003yf}.
The work of S.M. has been supported in part by the Deutsche 
Forschungsgemeinschaft in Sonderforschungsbereich/Transregio 9.
The work of J.V. has been part of the research program of the
Dutch Foundation for Fundamental Research of Matter (FOM).
%

\renewcommand{\theequation}{A.\arabic{equation}}
\setcounter{equation}{0}
%
\section*{Appendix A: The exact Mellin-space results}
Here we provide the exact even-$N$ expressions for the coefficient 
functions $C_{2,i}$ and $C_{L,i}$ up to the third order in $\ar =
\as / (4\pi)$. These results are expressed in terms of harmonic sums
\cite{Gonzalez-Arroyo:1979df,Gonzalez-Arroyo:1980he,Vermaseren:1998uu,%
Blumlein:1998if}, following Ref.~\cite{Vermaseren:1998uu} recursively 
defined by
\beq
\label{eq:Hsum2}
  S_{\pm m_1,m_2,\ldots,m_k}(N) \:\: = \:\: \sum_{i=1}^{N}\:
  \frac{(\pm 1)^{i}}{i^{\, m_1}}\: S_{m_2,\ldots,m_k}(i) 
  \:\: , \quad S(N) \:\:  = \:\: 1 \:\: .
\eeq
The sum of the absolute values of the indices $m_k$ defines the weight
of the harmonic sum. Sums with weight up to $2n$ occur in the $n$-loop 
coefficient functions for both $F_{\,2}$ and $F_L$.  
As in Section~3 we employ the notation (\ref{eq:shiftN}),
$$
  \Npm \, S_{\vec{m}} \: = \: S_{\vec{m}}(N \pm 1) \:\: , \quad\quad
  \Npmi\, S_{\vec{m}} \: = \: S_{\vec{m}}(N \pm i) \:\: ,
$$
together with the abbreviations
\bea
  \gqq &\! =\! & \Nplus + \Nminus\, ,\nn \\
  \gqg &\! =\! & 2 \* \Nplustwo - 4 \* \Nplus - \Nminus + 3 \:\: .
\eea

The $N$-independent zeroth-order quark coefficient function for 
$F_{\,2}$ is set to unity, recall Eq.~(\ref{eq:c20}). In the above
 notation the well-known first-order results for $F_{\,2}$ read
%
%
\bea
c^{(1)}_{2,\rm{q}}(N) & \! = \! &  
         \colour4colour{\cf}  \*  (
            9 \* (\S(1) - 1)
          + 2 \* \gqq \* (\Ss(1,1) + 2 \* \S(1) - \S(2))
          - 7 \* (\Nminus + 1) \* \S(1)
          )
\:\: , \\[1mm]
c^{(1)}_{2,\rm{g}}(N) & \! = \! &  
         \colour4colour{\nf}  \*  (
          - 2 \* \gqg \* (\Ss(1,1) + 4 \* \S(1) - \S(2))
          - 6 \* (\Nminus - 1) \* \S(1) 
          )
\:\: .
\eea
The corresponding expressions for the two-loop coefficient functions 
\cite{vanNeerven:1991nn,Zijlstra:1991qc,Moch:1999eb} are given by
\small
\bea
&& c^{(2)}_{2,\rm ns}(N)  \:\: = \:\: 
         \delta(N-2) \* \biggl\{
         \colour4colour{\cf \* \nf}  \*  (
          - 4
          )
       + \colour4colour{\cf^2}  \*  \biggl(
          - {4189 \over 810}
          + {96 \over 5} \* \z3
          \biggr)
       + \colour4colour{\ca \* \cf}  \*  \biggl(
            {3677 \over 135}
          - {128 \over 5} \* \z3
          \biggr)\biggr\}
  \nonumber\\&& \mbox{} 
       + \theta(N-4) \* \biggl\{
         \colour4colour{\cf \* \biggl(\cf-{\ca \over 2}\biggr)}  \*  \biggl(
            16 \* \Ss(-3,1)
          + 48 \* \Ss(-2,-2)
          + 144 \* \Ss(1,-2)
          - {16 \over 5} \* (\Nminusthree - \Nminustwo) \* \Ss(1,-2) 
  \nonumber\\&& \mbox{} 
          - 80 \* \Ss(2,-2)
          - 24 \* \S(-4)
          - 16 \* \S(-2)
          + {144 \over 5} \* (\Nplusthree - \Nplustwo) \* (\Ss(1,-2) + \S(3))
          + {144 \over 5} \* (\Nplustwo - 3) \* (\S(1) - \S(2))
  \nonumber\\&& \mbox{} 
          - {16 \over 5} \* (\Nminustwo - \Nminus) \* (\S(1) + \S(2))
          - 16 \* (2 \* \Nminus + 3) \* (\Ss(1,-3) - \Ss(1,3) - 2 \* \Sss(1,1,-2) + 3 \* \S(1) \* \z3)
          \biggr)
       + \colour4colour{\cf \* \nf}  \*  \biggl(
            {457 \over 36}
  \nonumber\\&& \mbox{} 
          - {2 \over 3} \* \Ss(1,1)
          - {2 \over 27} \* \gqq \* (102 \* \Ss(1,1) - 18 \* \Ss(1,2) - 36 \* \Ss(2,1) 
	    + 18 \* \Sss(1,1,1) + 244 \* \S(1) - 171 \* \S(2) + 45 \* \S(3))
          - {23 \over 9} \* \S(1)
  \nonumber\\&& \mbox{}
          + {14 \over 3} \* \S(2)
          + {1 \over 9} \* (\Nminus + 1) \* (42 \* \Ss(1,1) + 133 \* \S(1) - 78 \* \S(2))
          \biggr)
       + \colour4colour{\cf^2}  \*  \biggl(
            59 \* \Ss(1,1)
          - 28 \* \Ss(1,2)
          - 40 \* \Ss(2,1)
          + 8 \* \Ss(2,2)
  \nonumber\\&& \mbox{}
          + 20 \* \Sss(1,1,1)
          - 8 \* \Sss(2,1,1)
          + {1 \over 5} \* \gqq \* (400 \* \Ss(1,-3) - 200 \* \Ss(1,-2) 
            - 340 \* \Ss(1,1) - 160 \* \Ss(1,2) - 140 \* \Ss(1,3) - 240 \* \Ss(2,1) 
  \nonumber\\&& \mbox{}
	    + 140 \* \Ss(2,2) + 120 \* \Ss(3,1) - 80 \* \Sss(1,-2,1) - 560 \* \Sss(1,1,-2) 
	    + 180 \* \Sss(1,1,1) - 160 \* \Sss(1,1,2) - 120 \* \Sss(1,2,1) - 160 \* \Sss(2,1,1) 
  \nonumber\\&& \mbox{}
	    + 120 \* \Ssss(1,1,1,1) - 605 \* \S(1) + 720 \* \S(1) \* \z3 + 646 \* \S(2) 
	    - 20 \* \S(3) - 30 \* \S(4)
         )
          + {67 \over 2} \* \S(1)
          - {169 \over 5} \* \S(2)
          + 42 \* \S(3)
          + 4 \* \S(4)
  \nonumber\\&& \mbox{}
          - {1 \over 10} \* (\Nminus + 1) \* (320 \* \Ss(1,-2) - 250 \* \Ss(1,1) - 280 \* \Ss(1,2) 
            - 640 \* \Ss(2,-2) - 560 \* \Ss(2,1) + 80 \* \Ss(2,2) + 280 \* \Sss(1,1,1) 
  \nonumber\\&& \mbox{}
	    - 80 \* \Sss(2,1,1) - 1203 \* \S(1) + 1106 \* \S(2) + 140 \* \S(3)
            + 40 \* \S(4))
          + {331 \over 8} - 72 \* \z3
          \biggr)
       + \colour4colour{\ca \* \cf}  \*  \biggl(
          - {85 \over 3} \* \Ss(1,1)
          + 8 \* \Ss(3,1)
  \nonumber\\&& \mbox{}
          - {1 \over 135} \* \gqq \* (5400 \* \Ss(1,-3) - 2700 \* \Ss(1,-2) 
	    - 10020 \* \Ss(1,1) + 990 \* \Ss(1,2) - 3780 \* \Ss(1,3) 
	    + 1980 \* \Ss(2,1) - 1080 \* \Ss(3,1) 
  \nonumber\\&& \mbox{}
            - 1080 \* \Sss(1,-2,1) - 7560 \* \Sss(1,1,-2) - 990 \* \Sss(1,1,1) 
	    - 540 \* \Sss(1,1,2) + 540 \* \Sss(1,2,1) - 19570 \* \S(1) 
	    + 11340 \* \S(1) \* \z3 
  \nonumber\\&& \mbox{}
	    + 17181 \* \S(2) - 5175 \* \S(3) + 1620 \* \S(4))
          - {581 \over 18} \* \S(1)
          + {221 \over 15} \* \S(2)
          - 12 \* \S(4)
          + {1 \over 90} \* (\Nminus + 1) \* (1440 \* \Ss(1,-2) 
  \nonumber\\&& \mbox{}
	    - 3570 \* \Ss(1,1) - 2880 \* \Ss(2,-2) - 720 \* \Ss(3,1) - 10261 \* \S(1) 
	    + 8502 \* \S(2) - 1800 \* \S(3) + 1080 \* \S(4))
          - {5465 \over 72} 
  \nonumber\\&& \mbox{}
	  + 54 \* \z3
          \biggr) \biggr\}
\:\: , 
\eea
\bea
&& c^{(2)}_{2,\rm{g}}(N)  \:\: = \:\: 
         \delta(N-2) \* \biggl\{
         \colour4colour{\cf \* \nf}  \*  \biggl(
          - {4799 \over 810}
          + {16 \over 5} \* \z3
          \biggr)
       + \colour4colour{\ca \* \nf}  \*  \biggl(
            {115 \over 324}
          - 2 \* \z3
          \biggr)\biggr\}
  \nonumber\\&& \mbox{}
       + \theta(N-4) \* \biggl\{ \colour4colour{\cf \* \nf}  \*  \biggl(
          - {8 \over 15} \* (\Nminusthree - \Nminustwo) \* \Ss(1,-2)
          - {2 \over 5} \* \gqg \* (20 \* \Ss(1,-3) - 24 \* \Ss(1,-2) + 30 \* \Ss(1,1) - 90 \* \Ss(1,2) 
  \nonumber\\&& \mbox{}
	    + 30 \* \Ss(1,3) + 40 \* \Ss(2,-2) - 90 \* \Ss(2,1) + 40 \* \Ss(2,2)
            + 60 \* \Ss(3,1) - 40 \* \Sss(1,1,-2) + 90 \* \Sss(1,1,1) - 40 \* \Sss(1,1,2)
            - 60 \* \Sss(1,2,1) 
  \nonumber\\&& \mbox{}
            - 50 \* \Sss(2,1,1) + 50 \* \Ssss(1,1,1,1) + 9 \* \S(1)
            - 54 \* \S(2) + 66 \* \S(3) - 50 \* \S(4))
          - {96 \over 5} \* (\Nplusthree - 1) \* (\Ss(1,-2) + \S(3))
          + {1 \over 30} \* (2 \* \Nplus 
  \nonumber\\&& \mbox{}
                                         + \Nminus - 3) \* (240 \* \Ss(1,-3) + 256 \* \Ss(1,-2) 
	    + 240 \* \Ss(1,1) - 120 \* \Ss(1,2) - 240 \* \Ss(1,3) - 480 \* \Ss(2,-2) 
	    + 240 \* \Ss(2,1) 
  \nonumber\\&& \mbox{}
	    - 120 \* \Ss(2,2) - 240 \* \Ss(3,1) - 480 \* \Sss(1,1,-2) 
	    + 120 \* \Sss(1,1,1) + 120 \* \Sss(2,1,1) + 609 \* \S(1) + 720 \* \S(1) \* \z3
            + 309 \* \S(2) - 284 \* \S(3) 
  \nonumber\\&& \mbox{}
	    + 300 \* \S(4))
          - {1 \over 30} \* (\Nminus - 1) \* (1408 \* \Ss(1,-2) + 180 \* \Ss(1,1) 
	    - 420 \* \Ss(1,2) - 960 \* \Ss(2,-2) - 360 \* \Ss(2,1) + 420 \* \Sss(1,1,1) 
  \nonumber\\&& \mbox{}
	    - 593 \* \S(1) + 117 \* \S(2) + 418 \* \S(3))
          - {8 \over 15} \* (\Nminustwo - 1) \* (\S(1) + \S(2))
          \biggr)
       + \colour4colour{\ca \* \nf}  \*  \biggl(
          - {1 \over 54} \* \gqg \* (648 \* \Ss(1,-3) - 720 \* \Ss(1,-2) 
  \nonumber\\&& \mbox{}
	    + 4710 \* \Ss(1,1) - 2412 \* \Ss(1,2) + 432 \* \Ss(1,3) + 432 \* \Ss(2,-2) 
	    - 3996 \* \Ss(2,1) + 432 \* \Ss(2,2) + 432 \* \Ss(3,1) - 432 \* \Sss(1,-2,1) 
  \nonumber\\&& \mbox{}	    
	    + 2196 \* \Sss(1,1,1) - 648 \* \Sss(1,1,2) - 216 \* \Sss(1,2,1) - 432 \* \Sss(2,1,1) 
	    + 216 \* \Ssss(1,1,1,1) + 4493 \* \S(1) - 6270 \* \S(2) + 3492 \* \S(3))
  \nonumber\\&& \mbox{}
          - {1 \over 54} \* (2 \* \Nplus + \Nminus - 3) \* (216 \* \Ss(1,-3) 
	    - 720 \* \Ss(1,-2) + 624 \* \Ss(1,1) - 252 \* \Ss(1,2) - 216 \* \Ss(1,3) 
	    + 432 \* \Ss(2,-2) 
  \nonumber\\&& \mbox{}         
	    - 108 \* \Ss(2,1) - 864 \* \Ss(2,2) - 1296 \* \Ss(3,1) 
	    - 432 \* \Sss(1,1,-2) + 252 \* \Sss(1,1,1) + 864 \* \Sss(2,1,1) + 1277 \* \S(1) 
	    + 648 \* \S(1) \* \z3 
  \nonumber\\&& \mbox{}    
	    - 1014 \* \S(2) - 1260 \* \S(3) + 1512 \* \S(4))
          + {1 \over 27} \* (\Nminus - 1) \* (792 \* \Ss(1,-2) - 2913 \* \Ss(1,1) 
	  + 828 \* \Ss(1,2) + 1728 \* \Ss(2,1) 
  \nonumber\\&& \mbox{}
	  - 648 \* \Ss(2,2) - 1080 \* \Ss(3,1) - 720 \* \Sss(1,1,1) + 648 \* \Sss(2,1,1) 
	  - 1981 \* \S(1) + 4194 \* \S(2) - 2322 \* \S(3) + 1296 \* \S(4))
  \nonumber\\&& \mbox{}
          - {8 \over 27} \* (\Nminustwo - 1) \* (18 \* \Ss(1,-2) - 39 \* \Ss(1,1) - 18
          \* \Ss(1,2) + 18 \* \Sss(1,1,1) + 43 \* \S(1))
          \biggr)\biggr\}
\eea
\normalsize
and 
\small
\bea
&& c^{(2)}_{2,\rm{ps}}(N) \:\: = \:\: 
         \colour4colour{\cf \* \nf}  \*  \biggl(
          - {128 \over 3} \* \Ss(1,-2)
          + {688 \over 9} \* \Ss(1,1)
          + 32 \* \Ss(2,1)
          + 16 \* \Ss(2,2)
          + 32 \* \Ss(3,1)
          - 16 \* \Sss(2,1,1)
  \nonumber\\&& \mbox{}
          + {2 \over 27} \* \gqq \* (288 \* \Ss(1,-2) - 408 \* \Ss(1,1) - 18 \* \Ss(1,2) 
	    - 216 \* \Ss(2,1) + 108 \* \Ss(2,2) + 216 \* \Ss(3,1) + 18 \* \Sss(1,1,1) 
  \nonumber\\&& \mbox{}
	    - 108 \* \Sss(2,1,1)          
	    - 857 \* \S(1) + 204 \* \S(2) + 693 \* \S(3) - 270 \* \S(4))
          + {1844 \over 27} \* \S(1)
          - {752 \over 9} \* \S(2)
          - {140 \over 3} \* \S(3)
          - 40 \* \S(4)
  \nonumber\\&& \mbox{}         
          - {16 \over 27} \* (\Nplustwo - 3) \* (9 \* \Ss(1,-2) 
	    - 6 \* \Ss(1,1) - 9 \* \Ss(1,2) - 27 \* \Ss(2,1) + 9 \* \Sss(1,1,1) - 28 \* \S(1) 
	 - 24 \* \S(2) + 36 \* \S(3))
  \nonumber\\&& \mbox{}
          - {8 \over 27} \* (\Nminustwo - \Nminus) \* (18 \* \Ss(1,-2) - 39 \* \Ss(1,1) 
	  - 18 \* \Ss(1,2) + 18 \* \Sss(1,1,1) + 43 \* \S(1))
          - {4 \over 27} \* (\Nminus + 1) \* (36 \* \Ss(1,-2) 
  \nonumber\\&& \mbox{}
	    + 30 \* \Ss(1,1) - 108 \* \Ss(2,1) + 108 \* \Ss(2,2) + 216 \* \Ss(3,1) 
	    - 108 \* \Sss(2,1,1) - 310 \* \S(1) - 276 \* \S(2) + 333 \* \S(3) - 270 \* \S(4))
  \nonumber\\&& \mbox{}
          + {8 \over 3} \* (2 \* \Nminus + 3) \* (\Ss(1,2) - \Sss(1,1,1))
          \biggr)
\eea
\normalsize
 
\noindent
Our new three-loop non-singlet quark coefficient function for the 
structure function $F_{\,2}$ reads
\small
\bea
&& c^{(3)}_{2,\rm ns}(N) \:\: = \:\: 
         \delta(N-2) \* \biggl\{
         \colour4colour{\dabcnc} \* \fl11  \*  \biggl(
            {496 \over 5}
          - {256 \over 3} \* \z5
          - {1216 \over 15} \* \z3
          \biggr)
       + \colour4colour{\cf \* \nf^2}  \*  \biggl(
            {7814 \over 2187}
          + {64 \over 81} \* \z3
          \biggr)
  \nonumber\\&& \mbox{}
       + \colour4colour{\cf^2 \* \nf}  \*  \biggl(
          - {341701 \over 21870}
          + {32 \over 3} \* \z4
          - {1352 \over 45} \* \z3
          \biggr)
       + \colour4colour{\cf^3}  \*  \biggl(
          - {201577 \over 7290}
          + {416 \over 3} \* \z5
          + {64 \over 3} \* \z4
          - {50864 \over 405} \* \z3
          \biggr)
  \nonumber\\&& \mbox{}
       + \colour4colour{\ca \* \cf \* \nf}  \*  \biggl(
          - {998153 \over 10935}
          + {80 \over 3} \* \z5
          - {32 \over 3} \* \z4
          + {17432 \over 405} \* \z3
          \biggr)
       + \colour4colour{\ca \* \cf^2}  \*  \biggl(
          - {1013578 \over 10935}
          - {1552 \over 3} \* \z5
          - 32 \* \z4
  \nonumber\\&& \mbox{}
          + {30776 \over 45} \* \z3
          \biggr)
       + \colour4colour{\ca^2 \* \cf}  \*  \biggl(
            {3667498 \over 10935}
          + 296 \* \z5
          + {32 \over 3} \* \z4
          - {46160 \over 81} \* \z3
          \biggr)\biggr\}
  \nonumber\\&& \mbox{}
       + \theta(N-4) \* \biggl\{ \colour4colour{\dabcnc} \* \fl11  \*  \biggl(
            128 \* \Ss(-2,-2)
          - 512 \* \Ss(-2,1)
          + {21248 \over 15} \* \Ss(1,-3)
          - {7168 \over 3} \* \Ss(1,-2)
          - {2048 \over 15} \* \Ss(1,3)
  \nonumber\\&& \mbox{}
          - 2048 \* \Ss(1,-2) \* \z3
          + {2048 \over 15} \* \Ss(1,1)
          - 1024 \* \Ss(1,1) \* \z3
          + 2816 \* \Ss(2,-2)
          + 2048 \* \Ss(2,1) \* \z3
          + 1536 \* \Ss(2,3)
          - {22784 \over 15} \* \Ss(2,1)
  \nonumber\\&& \mbox{}
          - {15104 \over 15} \* \Ss(3,1)
          - 1536 \* \Ss(4,1)
          - 1024 \* \Sss(1,-4,1)
          - {4096 \over 15} \* \Sss(1,-2,1)
          + 1024 \* \Sss(1,-2,3)
          - 2560 \* \Sss(1,1,-2)
  \nonumber\\&& \mbox{}
          + 512 \* \Sss(1,1,3)
          - 512 \* \Sss(1,3,1)
          - 1024 \* \Sss(2,1,3)
          + 1024 \* \Sss(2,3,1)
          - {64 \over 15} \* \gqq \* (30 \* \Ss(1,-4)
          + 286 \* \Ss(1,-3)
          - 430 \* \Ss(1,-2)
  \nonumber\\&& \mbox{}
          - 480 \* \Ss(1,-2) \* \z3
          + 216 \* \Ss(1,1)
          + 160 \* \Ss(1,1) \* \z3
          - 66 \* \Ss(1,3)
          - 30 \* \Ss(1,4)
          + 20 \* \Ss(2,-3)
          + 280 \* \Ss(2,-2)
          - 612 \* \Ss(2,1)
  \nonumber\\&& \mbox{}
          - 480 \* \Ss(2,1) \* \z3
          + 40 \* \Ss(2,3)
          - 346 \* \Ss(3,1)
          - 240 \* \Sss(1,-4,1)
          - 60 \* \Ss(4,1)
          - 80 \* \Sss(1,-3,1)
          + 60 \* \Sss(1,-2,-2)
          - 202 \* \Sss(1,-2,1)
  \nonumber\\&& \mbox{}
          - 80 \* \Sss(1,1,-3)
          - 370 \* \Sss(1,1,-2)
          - 40 \* \Sss(1,1,3)
          + 240 \* \Sss(1,-2,3)
          + 40 \* \Sss(1,3,1)
          - 100 \* \Sss(2,-2,1)
          + 60 \* \Sss(2,1,-2)
  \nonumber\\&& \mbox{}
          + 240 \* \Sss(2,1,3)
          - 240 \* \Sss(2,3,1)
          + 160 \* \Ssss(1,1,-2,1)
          + 135 \* \S(1)
          - 1200 \* \S(1) \* \z5
          + 801 \* \S(1) \* \z3
          - 252 \* \S(2)
          - 160 \* \S(2) \* \z3
  \nonumber\\&& \mbox{}
          + 272 \* \S(3)
          + 10 \* \S(4))
          + 64 \* \S(-4)
          + 256 \* \S(-3)
          - 320 \* \S(-2)
          + 384 \* \S(1)
          - 5120 \* \S(1) \* \z5
          + {55808 \over 15} \* \S(1) \* \z3
          - {29696 \over 15} \* \S(2)
  \nonumber\\&& \mbox{}
          - 3072 \* \S(2) \* \z3
          + {4480 \over 3} \* \S(3)
          - 64 \* \S(4)
          - {384 \over 5} \* (\Nplusthree - \Nplustwo) \* (10 \* \Ss(1,-3)
          - 13 \* \Ss(1,-2)
          + 4 \* \Ss(1,1) \* \z3
          - 4 \* \Ss(2,-3)
  \nonumber\\&& \mbox{}
          + 10 \* \Ss(2,-2)
          + 4 \* \Ss(2,3)
          - 10 \* \Ss(3,1)
          + 4 \* \Sss(1,-3,1)
          - 10 \* \Sss(1,-2,1)
          + 4 \* \Sss(1,1,-3)
          - 10 \* \Sss(1,1,-2)
          - 4 \* \Sss(1,1,3)
          + 4 \* \Sss(1,3,1)
  \nonumber\\&& \mbox{}
          + 8 \* \Sss(2,-2,1)
          - 8 \* \Ssss(1,1,-2,1)
          - 4 \* \S(2) \* \z3
          - 13 \* \S(3)
          + 10 \* \S(4))
          + {128 \over 15} \* (\Nminusthree - \Nminustwo) \* (10 \* \Ss(1,-3)
          - 13 \* \Ss(1,-2)
  \nonumber\\&& \mbox{}
          + 4 \* \Ss(1,1) \* \z3
          + 4 \* \Sss(1,-3,1)
          - 10 \* \Sss(1,-2,1)
          + 4 \* \Sss(1,1,-3)
          - 10 \* \Sss(1,1,-2)
          - 4 \* \Sss(1,1,3)
          + 4 \* \Sss(1,3,1)
          - 8 \* \Ssss(1,1,-2,1))
  \nonumber\\&& \mbox{}
          + {384 \over 5} \* (\Nplustwo - 3) \* (4 \* \Ss(1,-3)
          - 10 \* \Ss(1,-2)
          + 10 \* \Ss(1,1)
          + 20 \* \Ss(1,1) \* \z3
          - 4 \* \Ss(1,3)
          + 15 \* \Ss(2,-2)
          - 10 \* \Ss(2,1)
          + 10 \* \Ss(2,3)
  \nonumber\\&& \mbox{}
          - 7 \* \Ss(3,1)
          - 10 \* \Ss(4,1)
          + 7 \* \Sss(1,-2,1)
          - 15 \* \Sss(1,1,-2)
          - 10 \* \Sss(1,1,3)
          + 10 \* \Sss(1,3,1)
          + 3 \* \S(1)
          + 19 \* \S(1) \* \z3
          - 13 \* \S(2)
          - 20 \* \S(2) \* \z3
  \nonumber\\&& \mbox{}
          + 10 \* \S(3))
          - {128 \over 15} \* (\Nminustwo - \Nminus) \* (4 \* \Ss(1,-3)
          - 10 \* \Ss(1,-2)
          + 10 \* \Ss(1,1)
          - 4 \* \Ss(1,3)
          + 10 \* \Ss(2,1)
          - 8 \* \Sss(1,-2,1)
          + 3 \* \S(1)
  \nonumber\\&& \mbox{}
          + 4 \* \S(1) \* \z3
          + 3 \* \S(2)
          - 10 \* \S(3))
          + {64 \over 15} \* (\Nminus + 1) \* (30 \* \Ss(1,-4)
          + 192 \* \Ss(1,-3)
          - 330 \* \Ss(1,-2)
          - 240 \* \Ss(1,-2) \* \z3
  \nonumber\\&& \mbox{}
          + 380 \* \Ss(1,1)
          + 640 \* \Ss(1,1) \* \z3
          - 122 \* \Ss(1,3)
          - 30 \* \Ss(1,4)
          + 20 \* \Ss(2,-3)
          + 220 \* \Ss(2,-2)
          - 614 \* \Ss(2,1)
          - 720 \* \Ss(2,1) \* \z3
  \nonumber\\&& \mbox{}
          + 40 \* \Ss(2,3)
          - 354 \* \Ss(3,1)
          - 60 \* \Ss(4,1)
          - 120 \* \Sss(1,-4,1)
          - 80 \* \Sss(1,-3,1)
          + 60 \* \Sss(1,-2,-2)
          - 44 \* \Sss(1,-2,1)
          + 120 \* \Sss(1,-2,3)
  \nonumber\\&& \mbox{}
          - 80 \* \Sss(1,1,-3)
          - 340 \* \Sss(1,1,-2)
          - 280 \* \Sss(1,1,3)
          + 280 \* \Sss(1,3,1)
          - 100 \* \Sss(2,-2,1)
          + 60 \* \Sss(2,1,-2)
          + 360 \* \Sss(2,1,3)
  \nonumber\\&& \mbox{}
          - 360 \* \Sss(2,3,1)
          + 160 \* \Ssss(1,1,-2,1)
          + 144 \* \S(1)
          - 600 \* \S(1) \* \z5
          + 707 \* \S(1) \* \z3
          - 254 \* \S(2)
          - 160 \* \S(2) \* \z3
          + 262 \* \S(3)
  \nonumber\\&& \mbox{}
          + 10 \* \S(4))
          + {32 \over 3} \* (6 - 40 \* \z5 + 7 \* \z3)
          \biggr)
  \nonumber\\&& \mbox{}
       + \colour4colour{\cf \* \biggl(\cf-{\ca \over 2}\biggr)^2}  \*  \biggl(
            {92 \over 3} \* \gfunct1(N)
          - {4 \over 3} \* \gfunct2(N)
          \biggr)
  \nonumber\\&& \mbox{}
       + \colour4colour{\cf \* \nf^2}  \*  \biggl(
          - {116 \over 27} \* \Ss(1,1)
          - {8 \over 9} \* \Ss(1,2)
          - {8 \over 3} \* \Ss(2,1)
          + {8 \over 9} \* \Sss(1,1,1)
          + {4 \over 729} \* \gqq \* (5040 \* \Ss(1,1)
          - 1836 \* \Ss(1,2)
          + 324 \* \Ss(1,3)
  \nonumber\\&& \mbox{}
          - 3672 \* \Ss(2,1)
          + 648 \* \Ss(2,2)
          + 972 \* \Ss(3,1)
          + 1836 \* \Sss(1,1,1)
          - 324 \* \Sss(1,1,2)
          - 324 \* \Sss(1,2,1)
          - 648 \* \Sss(2,1,1)
          + 324 \* \Ssss(1,1,1,1)
  \nonumber\\&& \mbox{}
          + 9187 \* \S(1)
          + 108 \* \S(1) \* \z3 - 9864 \* \S(2)
          + 5310 \* \S(3)
          - 1242 \* \S(4))
          - {1514 \over 81} \* \S(1)
          - {172 \over 81} \* \S(2)
          + {152 \over 27} \* \S(3)
  \nonumber\\&& \mbox{}
          - {2 \over 81} \* (\Nminus + 1) \* (798 \* \Ss(1,1)
          - 252 \* \Ss(1,2)
          - 468 \* \Ss(2,1)
          + 252 \* \Sss(1,1,1)
          + 1421 \* \S(1)
          - 1590 \* \S(2)
          + 684 \* \S(3))
  \nonumber\\&& \mbox{}
          - {1 \over 486} \* (9517 + 432 \* \z3) \biggr)
  \nonumber\\&& \mbox{}
       + \colour4colour{\cf^2 \* \biggl(\cf-{\ca \over 2}\biggr)}  \*  \biggl(
          - 384 \* \gqq \* (5 \* \S(1) \* \z5 + 3 \* \S(3) \* \z3)
          + 384 \* (\Nminus + 1) \* (5 \* \S(1) \* \z5 + 3 \* \S(3) \* \z3)
          - 168 \* \Ss(-3,2)
  \nonumber\\&& \mbox{}
          + 64 \* \Ss(-2,2)
          - 64 \* \Sss(-2,1,1)
          - 16 \* \Sss(-2,1,2)
          + 16 \* \Sss(-2,2,1)
          + 192 \* \Sss(1,2,3)
          + 192 \* \Ssss(-3,1,1,1)
          - 448 \* \Ssss(1,1,1,3)
  \nonumber\\&& \mbox{}
          + 576 \* \Ssss(1,1,3,1)
          \biggr)
  \nonumber\\&& \mbox{}
       + \colour4colour{\cf^2 \* \nf}  \*  \biggl(
            {128 \over 9} \* \Ss(-4,1)
          + {512 \over 3} \* \Ss(-3,-2)
          - {640 \over 27} \* \Ss(-3,1)
          + {512 \over 3} \* \Ss(-2,-3)
          - {416 \over 3} \* \Ss(-2,-2)
          + {6656 \over 9} \* \Ss(1,-3)
  \nonumber\\&& \mbox{}
          - {121856 \over 135} \* \Ss(1,-2)
          - {752 \over 3} \* \Ss(1,1)
          + {8662 \over 27} \* \Ss(1,2)
          - {940 \over 9} \* \Ss(1,3)
          - 320 \* \Ss(2,-3)
          + {9280 \over 9} \* \Ss(2,-2)
          + {3034 \over 135} \* \Ss(2,1)
  \nonumber\\&& \mbox{}
          + {112 \over 3} \* \Ss(2,3)
          - {896 \over 3} \* \Ss(3,-2)
          - {13904 \over 27} \* \Ss(3,1)
          + {112 \over 3} \* \Ss(3,2)
          + {424 \over 9} \* \Ss(4,1)
          + {128 \over 9} \* \Sss(-3,1,1)
          - {256 \over 3} \* \Sss(-2,-2,1)
  \nonumber\\&& \mbox{}
          + {512 \over 9} \* \Sss(1,-2,1)
          - {5888 \over 9} \* \Sss(1,1,-2)
          - {1114 \over 9} \* \Sss(1,1,1)
          - {824 \over 9} \* \Sss(1,1,2)
          + {304 \over 3} \* \Sss(1,2,1)
          + 128 \* \Sss(2,-2,1)
          + {800 \over 9} \* \Sss(2,1,1)
  \nonumber\\&& \mbox{}
          - 16 \* \Sss(2,1,2)
          - 16 \* \Sss(2,2,1)
          - {464 \over 9} \* \Sss(3,1,1)
          + 16 \* \Ssss(2,1,1,1)
          + {1 \over 4050} \* \gqq \* (1137600 \* \Ss(1,-4)
          - 1824000 \* \Ss(1,-3)
  \nonumber\\&& \mbox{}
          + 1658240 \* \Ss(1,-2)
          + 973875 \* \Ss(1,1)
          - 1144800 \* \Ss(1,1) \* \z3
          - 800 \* \Ss(1,2)
          + 42000 \* \Ss(1,3)
          - 457200 \* \Ss(1,4)
  \nonumber\\&& \mbox{}
          + 360000 \* \Ss(2,-3)
          - 628800 \* \Ss(2,-2)
          + 416020 \* \Ss(2,1)
          - 1107600 \* \Ss(2,2)
          + 72000 \* \Ss(2,3)
          + 14400 \* \Ss(3,-2)
  \nonumber\\&& \mbox{}
          - 592500 \* \Ss(3,1)
          + 424800 \* \Ss(3,2)
          + 430200 \* \Ss(4,1)
          - 28800 \* \Sss(1,-3,1)
          - 604800 \* \Sss(1,-2,-2)
          - 364800 \* \Sss(1,-2,1)
  \nonumber\\&& \mbox{}
          - 1785600 \* \Sss(1,1,-3)
          + 2534400 \* \Sss(1,1,-2)
          - 217800 \* \Sss(1,1,1)
          + 982800 \* \Sss(1,1,2)
          - 327600 \* \Sss(1,1,3)
  \nonumber\\&& \mbox{}
          - 1180800 \* \Sss(1,2,-2)
          + 622800 \* \Sss(1,2,1)
          - 439200 \* \Sss(1,2,2)
          + 414000 \* \Sss(1,3,1)
          + 576000 \* \Sss(2,-2,1)
  \nonumber\\&& \mbox{}
          - 1180800 \* \Sss(2,1,-2)
          + 1083600 \* \Sss(2,1,1)
          - 399600 \* \Sss(2,1,2)
          - 306000 \* \Sss(2,2,1)
          - 468000 \* \Sss(3,1,1)
  \nonumber\\&& \mbox{}
          - 57600 \* \Ssss(1,-2,1,1)
          + 1152000 \* \Ssss(1,1,1,-2)
          - 680400 \* \Ssss(1,1,1,1)
          + 403200 \* \Ssss(1,1,1,2)
          + 108000 \* \Ssss(1,1,2,1)
  \nonumber\\&& \mbox{}
          + 244800 \* \Ssss(1,2,1,1)
          + 327600 \* \Ssss(2,1,1,1)
          - 216000 \* \Sssss(1,1,1,1,1)
          + 1751525 \* \S(1)
          + 32400 \* \S(1) \* \z4
  \nonumber\\&& \mbox{}
          - 2415600 \* \S(1) \* \z3
          - 2741907 \* \S(2)
          + 907200 \* \S(2) \* \z3
          + 2449340 \* \S(3)
          + 226500 \* \S(4)
          - 160200 \* \S(5))
  \nonumber\\&& \mbox{}
          - {736 \over 9} \* \S(-5)
          + {1328 \over 27} \* \S(-4)
          - {6112 \over 81} \* \S(-3)
          + {464 \over 9} \* \S(-2)
          + {448 \over 3} \* \S(-2) \* \z3
          - {88169 \over 324} \* \S(1)
          + {2024 \over 3} \* \S(1) \* \z3
  \nonumber\\&& \mbox{}
          + {118694 \over 2025} \* \S(2)
          - 352 \* \S(2) \* \z3
          - {108566 \over 405} \* \S(3)
          + {3208 \over 27} \* \S(4)
          + {8 \over 9} \* \S(5)
          + {16 \over 25} \* (\Nplusthree - \Nplustwo) \* (180 \* \Ss(1,-3)
  \nonumber\\&& \mbox{}
          - 159 \* \Ss(1,-2)
          + 60 \* \Ss(2,-2)
          + 125 \* \Ss(2,2)
          - 185 \* \Ss(3,1)
          - 60 \* \Sss(1,-2,1)
          - 60 \* \Sss(1,1,-2)
          - 125 \* \Sss(1,1,2)
  \nonumber\\&& \mbox{}
          + 125 \* \Sss(1,2,1)
          - 570 \* \S(1) \* \z3
          - 159 \* \S(3)
          + 180 \* \S(4))
          - {16 \over 225} \* (\Nminusthree - \Nminustwo) \* (180 \* \Ss(1,-3)
          - 119 \* \Ss(1,-2)
  \nonumber\\&& \mbox{}
          + 120 \* \Ss(2,-2)
          - 60 \* \Sss(1,-2,1)
          - 60 \* \Sss(1,1,-2)
          + 180 \* \S(1) \* \z3)
          + {16 \over 25} \* (\Nplustwo - 3) \* (60 \* \Ss(1,-2)
          + 65 \* \Ss(1,1)
  \nonumber\\&& \mbox{}
          + 125 \* \Ss(1,2)
          + 200 \* \Ss(2,-2)
          - 65 \* \Ss(2,1)
          + 50 \* \Ss(2,2)
          - 250 \* \Ss(3,1)
          + 200 \* \Sss(1,-2,1)
          - 200 \* \Sss(1,1,-2)
          - 50 \* \Sss(1,1,2)
  \nonumber\\&& \mbox{}
          + 50 \* \Sss(1,2,1)
          - 219 \* \S(1)
          - 100 \* \S(1) \* \z3
          + 154 \* \S(2)
          - 180 \* \S(3))
          - {16 \over 225} \* (\Nminustwo - \Nminus) \* (60 \* \Ss(1,-2)
          - 60 \* \Ss(1,1)
  \nonumber\\&& \mbox{}
          - 60 \* \Ss(2,1)
          - 179 \* \S(1)
          - 59 \* \S(2)
          + 180 \* \S(3))
          - {1 \over 8100} \* (\Nminus + 1) \* (115200 \* \Ss(1,-3)
          - 915840 \* \Ss(1,-2)
  \nonumber\\&& \mbox{}
          + 558240 \* \Ss(1,1)
          - 117000 \* \Ss(1,2)
          + 414000 \* \Ss(1,3)
          - 1555200 \* \Ss(2,-3)
          + 2361600 \* \Ss(2,-2)
          + 896520 \* \Ss(2,1)
  \nonumber\\&& \mbox{}
          - 1641600 \* \Ss(2,2)
          - 216000 \* \Ss(2,3)
          - 1900800 \* \Ss(3,-2)
          - 751200 \* \Ss(3,1)
          + 302400 \* \Ss(3,2)
          + 381600 \* \Ss(4,1)
  \nonumber\\&& \mbox{}
          - 1728000 \* \Sss(1,-2,1)
          + 2880000 \* \Sss(1,1,-2)
          - 322200 \* \Sss(1,1,1)
          + 885600 \* \Sss(1,1,2)
          + 475200 \* \Sss(1,2,1)
  \nonumber\\&& \mbox{}
          + 1728000 \* \Sss(2,-2,1)
          + 1598400 \* \Sss(2,1,1)
          - 129600 \* \Sss(2,1,2)
          - 129600 \* \Sss(2,2,1)
          - 417600 \* \Sss(3,1,1)
  \nonumber\\&& \mbox{}
          - 604800 \* \Ssss(1,1,1,1)
          + 129600 \* \Ssss(2,1,1,1)
          + 3461121 \* \S(1)
          - 1317600 \* \S(1) \* \z3
          - 5325512 \* \S(2)
  \nonumber\\&& \mbox{}
          + 86400 \* \S(2) \* \z3
          + 4583840 \* \S(3)
          + 444000 \* \S(4)
          + 7200 \* \S(5))
  \nonumber\\&& \mbox{}
          - {32 \over 3} \* (2 \* \Nminus + 3) \* (6 \* \Ss(1,-4)
          - 10 \* \Ss(1,1) \* \z3
          - 6 \* \Ss(1,4)
          - 4 \* \Sss(1,-2,-2)
          - 12 \* \Sss(1,1,-3)
          - 8 \* \Sss(1,2,-2)
          - \Sss(1,2,2)
  \nonumber\\&& \mbox{}
          + 5 \* \Sss(1,3,1)
          - 8 \* \Sss(2,1,-2)
          + 8 \* \Ssss(1,1,1,-2)
          + \Ssss(1,1,1,2)
          - \Ssss(1,1,2,1))
          - {1 \over 36} \* (341 + 432 \* \z4 - 13344 \* \z3)
          \biggr)
  \nonumber\\&& \mbox{}
       + \colour4colour{\cf^3}  \*  \biggl(
            1120 \* \Ss(-5,1)
          + {2560 \over 3} \* \Ss(-4,-2)
          + 104 \* \Ss(-4,1)
          + {1952 \over 3} \* \Ss(-4,2)
          + 1920 \* \Ss(-3,-3)
          - 288 \* \Ss(-3,-2)
  \nonumber\\&& \mbox{}
          + {976 \over 3} \* \Ss(-3,1)
          + {1024 \over 3} \* \Ss(-3,3)
          + 1088 \* \Ss(-2,-4)
          - {1616 \over 3} \* \Ss(-2,-3)
          - 704 \* \Ss(-2,1) \* \z3
          + {16888 \over 3} \* \Ss(1,-4)
  \nonumber\\&& \mbox{}
          - {296 \over 3} \* \Ss(-2,-2)
          + {208 \over 3} \* \Ss(-2,1)
          - 128 \* \Ss(-2,4)
          - 6000 \* \Ss(1,-3)
          - {60856 \over 75} \* \Ss(1,-2)
          - {25336 \over 15} \* \Ss(1,1)
          - 1600 \* \Ss(1,1) \* \z3
  \nonumber\\&& \mbox{}
          + 799 \* \Ss(1,2)
          - 320 \* \Ss(1,2) \* \z3
          + {9538 \over 15} \* \Ss(1,3)
          - {6064 \over 3} \* \Ss(1,4)
          - 1120 \* \Ss(2,-4)
          + {14560 \over 3} \* \Ss(2,-3)
          + {37136 \over 15} \* \Ss(2,-2)
  \nonumber\\&& \mbox{}
          + {35643 \over 25} \* \Ss(2,1)
          - 320 \* \Ss(2,1) \* \z3
          - {5072 \over 15} \* \Ss(2,2)
          - {5264 \over 3} \* \Ss(2,3)
          + 560 \* \Ss(2,4)
          - {4096 \over 3} \* \Ss(3,-3)
          + {2144 \over 3} \* \Ss(3,-2)
  \nonumber\\&& \mbox{}
          - {18206 \over 5} \* \Ss(3,1)
          - 24 \* \Ss(3,2)
          + {32 \over 3} \* \Ss(3,3)
          - {256 \over 3} \* \Ss(4,-2)
          + 1236 \* \Ss(4,1)
          - {640 \over 3} \* \Ss(4,2)
          - 664 \* \Ss(5,1)
          - 864 \* \Sss(-4,1,1)
  \nonumber\\&& \mbox{}
          - {2048 \over 3} \* \Sss(-3,-2,1)
          - 2304 \* \Sss(-3,1,-2)
          + 256 \* \Sss(-3,1,1)
          - 192 \* \Sss(-3,1,2)
          - {4672 \over 3} \* \Sss(-2,-3,1)
          + {128 \over 3} \* \Sss(-2,-2,-2)
  \nonumber\\&& \mbox{}
          - {832 \over 3} \* \Sss(-3,2,1)
          + {1744 \over 3} \* \Sss(-2,-2,1)
          - 192 \* \Sss(-2,-2,2)
          - 2240 \* \Sss(-2,1,-3)
          + 1008 \* \Sss(-2,1,-2)
          + 128 \* \Sss(-2,1,3)
  \nonumber\\&& \mbox{}
          - 704 \* \Sss(-2,2,-2)
          - {128 \over 3} \* \Sss(-2,2,2)
          + 192 \* \Sss(-2,3,1)
          - 9344 \* \Sss(1,-3,1)
          + 2256 \* \Sss(1,-2,-2)
          + {22288 \over 3} \* \Sss(1,-2,1)
  \nonumber\\&& \mbox{}
          - {5056 \over 3} \* \Sss(1,-2,2)
          - {28432 \over 3} \* \Sss(1,1,-3)
          + {12464 \over 5} \* \Sss(1,1,-2)
          - {3731 \over 3} \* \Sss(1,1,1)
          + 832 \* \Sss(1,1,1) \* \z3
          - {1220 \over 3} \* \Sss(1,1,2)
  \nonumber\\&& \mbox{}
          + {3368 \over 3} \* \Sss(1,1,3)
          - {5296 \over 3} \* \Sss(1,2,-2)
          - {1408 \over 3} \* \Sss(1,2,1)
          + {64 \over 3} \* \Sss(1,2,2)
          - {592 \over 3} \* \Sss(1,3,1)
          - 768 \* \Sss(1,4,1)
          + {7328 \over 3} \* \Sss(2,-3,1)
  \nonumber\\&& \mbox{}
          - {896 \over 3} \* \Sss(2,-2,-2)
          - 4000 \* \Sss(2,-2,1)
          + {640 \over 3} \* \Sss(2,-2,2)
          + {9632 \over 3} \* \Sss(2,1,-3)
          - {13216 \over 3} \* \Sss(2,1,-2)
          + {2044 \over 5} \* \Sss(2,1,1)
  \nonumber\\&& \mbox{}
          + {1448 \over 3} \* \Sss(2,1,2)
          - {304 \over 3} \* \Sss(2,1,3)
          + 832 \* \Sss(2,2,-2)
          + {1112 \over 3} \* \Sss(2,2,1)
          - 224 \* \Sss(2,2,2)
          - 816 \* \Sss(2,3,1)
          + 576 \* \Sss(3,-2,1)
  \nonumber\\&& \mbox{}
          + {6848 \over 3} \* \Sss(3,1,-2)
          - {496 \over 3} \* \Sss(3,1,1)
          - {832 \over 3} \* \Sss(3,1,2)
          - {224 \over 3} \* \Sss(3,2,1)
          + {1232 \over 3} \* \Sss(4,1,1)
          + 192 \* \Ssss(-2,-2,1,1)
  \nonumber\\&& \mbox{}
          + {6464 \over 3} \* \Ssss(-2,1,-2,1)
          + {2624 \over 3} \* \Ssss(-2,1,1,-2)
          + {64 \over 3} \* \Ssss(-2,1,1,2)
          - {64 \over 3} \* \Ssss(-2,1,2,1)
          + 2112 \* \Ssss(1,-2,1,1)
  \nonumber\\&& \mbox{}
          + {34880 \over 3} \* \Ssss(1,1,-2,1)
          + {6496 \over 3} \* \Ssss(1,1,1,-2)
          + 468 \* \Ssss(1,1,1,1)
          - {272 \over 3} \* \Ssss(1,1,1,2)
          - 32 \* \Ssss(1,1,2,1)
          - {424 \over 3} \* \Ssss(1,2,1,1)
  \nonumber\\&& \mbox{}
          - 192 \* \Ssss(2,-2,1,1)
          - 3840 \* \Ssss(2,1,-2,1)
          - {3392 \over 3} \* \Ssss(2,1,1,-2)
          - 360 \* \Ssss(2,1,1,1)
          + {704 \over 3} \* \Ssss(2,1,1,2)
          + {736 \over 3} \* \Ssss(2,1,2,1)
  \nonumber\\&& \mbox{}
          + 192 \* \Ssss(2,2,1,1)
          + 192 \* \Ssss(3,1,1,1)
          + 48 \* \Sssss(1,1,1,1,1)
          - 192 \* \Sssss(2,1,1,1,1)
          + {1 \over 900} \* \gqq \* (
            1857600 \* \Ss(1,-5)
  \nonumber\\&& \mbox{}
          - 2388000 \* \Ss(1,-4)
          + 2661600 \* \Ss(1,-3)
          - 140144 \* \Ss(1,-2)
          - 1843200 \* \Ss(1,-2) \* \z3 
          + 1463355 \* \Ss(1,1)
  \nonumber\\&& \mbox{}
          + 2040000 \* \Ss(1,1) \* \z3 
          + 177000 \* \Ss(1,2)
          - 288000 \* \Ss(1,2) \* \z3 
          - 146040 \* \Ss(1,3)
          - 964800 \* \Ss(1,5)
  \nonumber\\&& \mbox{}
          + 825600 \* \Ss(1,4)
          + 1915200 \* \Ss(2,-4)
          - 2289600 \* \Ss(2,-3)
          + 906240 \* \Ss(2,-2)
          - 1267200 \* \Ss(2,1) \* \z3 
  \nonumber\\&& \mbox{}
          - 867912 \* \Ss(2,1)
          - 1210920 \* \Ss(2,2)
          + 973200 \* \Ss(2,3)
          - 525600 \* \Ss(2,4)
          + 662400 \* \Ss(3,-3)
          - 432000 \* \Ss(3,-2)
  \nonumber\\&& \mbox{}
          - 4748760 \* \Ss(3,1)
          - 82800 \* \Ss(3,2)
          + 105600 \* \Ss(3,3)
          + 115200 \* \Ss(4,-2)
          + 237000 \* \Ss(4,1)
          + 163200 \* \Ss(4,2)
  \nonumber\\&& \mbox{}
          - 274800 \* \Ss(5,1)
          - 3470400 \* \Sss(1,-4,1)
          - 124800 \* \Sss(1,-3,-2)
          + 4502400 \* \Sss(1,-3,1)
          - 984000 \* \Sss(1,-3,2)
  \nonumber\\&& \mbox{}
          + 249600 \* \Sss(1,-2,-3)
          - 1166400 \* \Sss(1,-2,-2)
          - 4711200 \* \Sss(1,-2,1)
          + 816000 \* \Sss(1,-2,2)
          - 76800 \* \Sss(1,-2,3)
  \nonumber\\&& \mbox{}
          - 4344000 \* \Sss(1,1,-4)
          + 4644000 \* \Sss(1,1,-3)
          + 553440 \* \Sss(1,1,-2)
          + 215400 \* \Sss(1,1,1)
          - 230400 \* \Sss(1,1,1) \* \z3
  \nonumber\\&& \mbox{}
          + 717000 \* \Sss(1,1,2)
          - 1288800 \* \Sss(1,1,3)
          + 1219200 \* \Sss(1,1,4)
          - 4416000 \* \Sss(1,2,-3)
          + 1089600 \* \Sss(1,2,-2)
  \nonumber\\&& \mbox{}
          + 678600 \* \Sss(1,2,1)
          + 585600 \* \Sss(1,2,2)
          - 340800 \* \Sss(1,2,3)
          - 1507200 \* \Sss(1,3,-2)
          + 1492800 \* \Sss(1,3,1)
  \nonumber\\&& \mbox{}
          - 350400 \* \Sss(1,3,2)
          + 1339200 \* \Sss(1,4,1)
          - 2092800 \* \Sss(2,-3,1)
          + 259200 \* \Sss(2,-2,-2)
          + 3912000 \* \Sss(2,-2,1)
  \nonumber\\&& \mbox{}
          - 652800 \* \Sss(2,-2,2)
          - 2457600 \* \Sss(2,1,-3)
          - 648000 \* \Sss(2,1,-2)
          + 1526520 \* \Sss(2,1,1)
          + 861600 \* \Sss(2,1,2)
  \nonumber\\&& \mbox{}
          - 184800 \* \Sss(2,1,3)
          - 950400 \* \Sss(2,2,-2)
          + 808800 \* \Sss(2,2,1)
          - 624000 \* \Sss(2,2,2)
          - 472800 \* \Sss(2,3,1)
  \nonumber\\&& \mbox{}
          - 835200 \* \Sss(3,-2,1)
          - 86400 \* \Sss(3,1,-2)
          + 175200 \* \Sss(3,1,1)
          - 638400 \* \Sss(3,1,2)
          - 518400 \* \Sss(3,2,1)
  \nonumber\\&& \mbox{}
          - 194400 \* \Sss(4,1,1)
          + 1228800 \* \Ssss(1,-3,1,1)
          - 1814400 \* \Ssss(1,-2,-2,1)
          + 1200000 \* \Ssss(1,-2,1,-2)
  \nonumber\\&& \mbox{}
          - 1065600 \* \Ssss(1,-2,1,1)
          + 163200 \* \Ssss(1,-2,1,2)
          + 259200 \* \Ssss(1,-2,2,1)
          + 6470400 \* \Ssss(1,1,-3,1)
  \nonumber\\&& \mbox{}
          - 1027200 \* \Ssss(1,1,-2,-2)
          - 6028800 \* \Ssss(1,1,-2,1)
          + 1478400 \* \Ssss(1,1,-2,2)
          + 6681600 \* \Ssss(1,1,1,-3)
  \nonumber\\&& \mbox{}
          - 1876800 \* \Ssss(1,1,1,-2)
          - 577800 \* \Ssss(1,1,1,1)
          - 756000 \* \Ssss(1,1,1,2)
          + 648000 \* \Ssss(1,1,1,3)
          + 1526400 \* \Ssss(1,1,2,-2)
  \nonumber\\&& \mbox{}
          - 616800 \* \Ssss(1,1,2,1)
          + 604800 \* \Ssss(1,1,2,2)
          - 57600 \* \Ssss(1,1,3,1)
          + 4665600 \* \Ssss(1,2,-2,1)
          + 1555200 \* \Ssss(1,2,1,-2)
  \nonumber\\&& \mbox{}
          - 571200 \* \Ssss(1,2,1,1)
          + 595200 \* \Ssss(1,2,1,2)
          + 518400 \* \Ssss(1,2,2,1)
          + 364800 \* \Ssss(1,3,1,1)
          + 844800 \* \Ssss(2,-2,1,1)
  \nonumber\\&& \mbox{}
          + 1900800 \* \Ssss(2,1,-2,1)
          + 1305600 \* \Ssss(2,1,1,-2)
          - 842400 \* \Ssss(2,1,1,1)
          + 753600 \* \Ssss(2,1,1,2)
  \nonumber\\&& \mbox{}
          - 7315200 \* \Sssss(1,1,1,-2,1)
          + 662400 \* \Ssss(2,1,2,1)
          + 614400 \* \Ssss(2,2,1,1)
          + 619200 \* \Ssss(3,1,1,1)
  \nonumber\\&& \mbox{}
          - 172800 \* \Sssss(1,-2,1,1,1)
          - 1785600 \* \Sssss(1,1,-2,1,1)
          - 1555200 \* \Sssss(1,1,1,1,-2)
          + 604800 \* \Sssss(1,1,1,1,1)
  \nonumber\\&& \mbox{}
          - 604800 \* \Sssss(1,1,1,1,2)
          - 518400 \* \Sssss(1,1,1,2,1)
          - 475200 \* \Sssss(1,1,2,1,1)
          - 475200 \* \Sssss(1,2,1,1,1)
  \nonumber\\&& \mbox{}
          - 604800 \* \Sssss(2,1,1,1,1)
          + 979578 \* \S(1)
          - 3888000 \* \S(1) \* \z5
          + 43200 \* \S(1) \* \z4
          - 603360 \* \S(1) \* \z3)
  \nonumber\\&& \mbox{}
          - {1 \over 900} \* \gqq \* (2271219 \* \S(2)
          - 43200 \* \S(2) \* \z4
          + 1327200 \* \S(2) \* \z3
          + 173168 \* \S(3)
          - 1089600 \* \S(3) \* \z3
  \nonumber\\&& \mbox{}
          - 4587720 \* \S(4)
          + 969000 \* \S(5)
          - 356400 \* \S(6)
          - 432000 \* \Ssssss(1,1,1,1,1,1))
          - 544 \* \S(-6)
          - {1192 \over 3} \* \S(-5)
          + {956 \over 3} \* \S(-4)
  \nonumber\\&& \mbox{}
          - {440 \over 3} \* \S(-3)
          + 1216 \* \S(-3) \* \z3
          + {964 \over 3} \* \S(-2)
          - 96 \* \S(-2) \* \z4
          + {224 \over 3} \* \S(-2) \* \z3
          - {395957 \over 1800} \* \S(1)
          + 3040 \* \S(1) \* \z5
  \nonumber\\&& \mbox{}
          - 96 \* \S(1) \* \z4
          - {19688 \over 15} \* \S(1) \* \z3
          + {260401 \over 225} \* \S(2)
          + 96 \* \S(2) \* \z4
          + 4880 \* \S(2) \* \z3
          - {29433 \over 25} \* \S(3)
          - 992 \* \S(3) \* \z3
          - {9814 \over 15} \* \S(4)
  \nonumber\\&& \mbox{}
          + {1564 \over 3} \* \S(5)
          + 184 \* \S(6)
          + {24 \over 25} \* (\Nplusthree - \Nplustwo) \* (400 \* \Ss(1,-4)
          - 948 \* \Ss(1,-3)
          + 337 \* \Ss(1,-2)
          + 1820 \* \Ss(1,1) \* \z3 
  \nonumber\\&& \mbox{}
          - 40 \* \Ss(1,4)
          + 1000 \* \Ss(2,-3)
          - 788 \* \Ss(2,-2)
          + 770 \* \Ss(2,3)
          + 200 \* \Ss(3,-2)
          + 788 \* \Ss(3,1)
          - 120 \* \Ss(3,2)
          - 1770 \* \Ss(4,1)
  \nonumber\\&& \mbox{}
          - 1000 \* \Sss(1,-3,1)
          + 80 \* \Sss(1,-2,-2)
          + 788 \* \Sss(1,-2,1)
          - 120 \* \Sss(1,-2,2)
          - 1000 \* \Sss(1,1,-3)
          + 788 \* \Sss(1,1,-2)
  \nonumber\\&& \mbox{}
          - 770 \* \Sss(1,1,3)
          - 120 \* \Sss(1,2,-2)
          + 770 \* \Sss(1,3,1)
          - 1400 \* \Sss(2,-2,1)
          - 120 \* \Sss(2,1,-2)
          + 120 \* \Sss(3,1,1)
          + 120 \* \Ssss(1,-2,1,1)
  \nonumber\\&& \mbox{}
          + 1400 \* \Ssss(1,1,-2,1)
          + 120 \* \Ssss(1,1,1,-2)
          - 240 \* \S(1) \* \z3 - 1820 \* \S(2) \* \z3 + 337 \* \S(3)
          - 948 \* \S(4)
          + 440 \* \S(5))
  \nonumber\\&& \mbox{}
          - {8 \over 225} \* (\Nminusthree - \Nminustwo) \* (1200 \* \Ss(1,-4)
          - 2604 \* \Ss(1,-3)
          + 1501 \* \Ss(1,-2)
          + 5460 \* \Ss(1,1) \* \z3 - 120 \* \Ss(1,4)
          + 720 \* \Ss(2,-3)
  \nonumber\\&& \mbox{}
          - 480 \* \Ss(2,-2)
          - 3000 \* \Sss(1,-3,1)
          + 240 \* \Sss(1,-2,-2)
          + 2124 \* \Sss(1,-2,1)
          - 360 \* \Sss(1,-2,2)
          - 3000 \* \Sss(1,1,-3)
  \nonumber\\&& \mbox{}
          + 2124 \* \Sss(1,1,-2)
          - 2310 \* \Sss(1,1,3)
          - 360 \* \Sss(1,2,-2)
          + 2310 \* \Sss(1,3,1)
          - 720 \* \Sss(2,-2,1)
          - 720 \* \Sss(2,1,-2)
  \nonumber\\&& \mbox{}
          + 360 \* \Ssss(1,-2,1,1)
          + 4200 \* \Ssss(1,1,-2,1)
          + 360 \* \Ssss(1,1,1,-2)
          - 720 \* \S(1) \* \z3)
          + {8 \over 25} \* (\Nplustwo - 3) \* (3000 \* \Ss(1,-3)
  \nonumber\\&& \mbox{}
          - 2124 \* \Ss(1,-2)
          - 1506 \* \Ss(1,1)
          - 5000 \* \Ss(1,1) \* \z3
          - 360 \* \Ss(1,2)
          + 2310 \* \Ss(1,3)
          + 4620 \* \Ss(2,-2)
          + 1146 \* \Ss(2,1)
  \nonumber\\&& \mbox{}
          + 360 \* \Ss(2,2)
          - 2500 \* \Ss(2,3)
          - 3810 \* \Ss(3,1)
          + 2500 \* \Ss(4,1)
          + 300 \* \Sss(1,-2,1)
          - 4860 \* \Sss(1,1,-2)
          + 360 \* \Sss(1,1,1)
  \nonumber\\&& \mbox{}
          + 2500 \* \Sss(1,1,3)
          - 2500 \* \Sss(1,3,1)
          - 360 \* \Sss(2,1,1)
          - 1353 \* \S(1)
          - 960 \* \S(1) \* \z3
          + 2859 \* \S(2)
          + 5000 \* \S(2) \* \z3
  \nonumber\\&& \mbox{}
          - 1386 \* \S(3)
          - 1080 \* \S(4))
          - {8 \over 225} \* (\Nminustwo - \Nminus) \* (3000 \* \Ss(1,-3)
          - 2244 \* \Ss(1,-2)
          - 1746 \* \Ss(1,1)
          - 360 \* \Ss(1,2)
  \nonumber\\&& \mbox{}
          + 2310 \* \Ss(1,3)
          + 600 \* \Ss(2,-2)
          - 2466 \* \Ss(2,1)
          - 360 \* \Ss(2,2)
          - 5310 \* \Ss(3,1)
          - 4200 \* \Sss(1,-2,1)
          - 360 \* \Sss(1,1,-2)
  \nonumber\\&& \mbox{}
          + 360 \* \Sss(1,1,1)
          + 360 \* \Sss(2,1,1)
          - 983 \* \S(1)
          - 5460 \* \S(1) \* \z3 
          - 743 \* \S(2)
          - 2004 \* \S(3)
          + 1320 \* \S(4))
  \nonumber\\&& \mbox{}
          - {1 \over 1800} \* (\Nminus + 1) \* (225600 \* \Ss(1,-4)
          - 1425600 \* \Ss(1,-3)
          + 270464 \* \Ss(1,-2)
          + 2189856 \* \Ss(1,1)
          + 923160 \* \Ss(1,2)
  \nonumber\\&& \mbox{}
          + 5520000 \* \Ss(1,1) \* \z3
          - 230400 \* \Ss(1,2) \* \z3 
          - 717360 \* \Ss(1,3)
          - 38400 \* \Ss(1,4)
          - 518400 \* \Ss(2,-4)
          - 182400 \* \Ss(2,-3)
  \nonumber\\&& \mbox{}
          + 1787520 \* \Ss(2,-2)
          - 1350072 \* \Ss(2,1)
          - 2361600 \* \Ss(2,1) \* \z3
          - 2267520 \* \Ss(2,2)
          + 1756800 \* \Ss(2,3)
          - 144000 \* \Ss(2,4)
  \nonumber\\&& \mbox{}
          - 2592000 \* \Ss(3,-3)
          - 288000 \* \Ss(3,-2)
          - 9933840 \* \Ss(3,1)
          + 28800 \* \Ss(3,2)
          - 115200 \* \Ss(3,3)
          - 768000 \* \Ss(4,-2)
  \nonumber\\&& \mbox{}
          + 832800 \* \Ss(4,1)
          - 499200 \* \Ss(4,2)
          - 1492800 \* \Ss(5,1)
          - 124800 \* \Sss(1,-3,1)
          - 432000 \* \Sss(1,-2,-2)
  \nonumber\\&& \mbox{}
          - 2611200 \* \Sss(1,-2,1)
          - 230400 \* \Sss(1,-2,2)
          + 326400 \* \Sss(1,1,-3)
          + 5688960 \* \Sss(1,1,-2)
          - 644760 \* \Sss(1,1,1)
  \nonumber\\&& \mbox{}
          - 230400 \* \Sss(1,1,1) \* \z3
          + 621600 \* \Sss(1,1,2)
          - 3028800 \* \Sss(1,1,3)
          + 374400 \* \Sss(1,2,-2)
          + 590400 \* \Sss(1,2,1)
  \nonumber\\&& \mbox{}
          + 614400 \* \Sss(1,2,2)
          + 3571200 \* \Sss(1,3,1)
          + 2073600 \* \Sss(1,4,1)
          + 2841600 \* \Sss(2,-3,1)
          - 307200 \* \Sss(2,-2,-2)
  \nonumber\\&& \mbox{}
          + 3792000 \* \Sss(2,-2,1)
          + 268800 \* \Sss(2,-2,2)
          + 2496000 \* \Sss(2,1,-3)
          - 5462400 \* \Sss(2,1,-2)
          + 3177120 \* \Sss(2,1,1)
  \nonumber\\&& \mbox{}
          + 1617600 \* \Sss(2,1,2)
          + 105600 \* \Sss(2,1,3)
          + 1512000 \* \Sss(2,2,1)
          - 403200 \* \Sss(2,2,2)
          - 720000 \* \Sss(2,3,1)
  \nonumber\\&& \mbox{}
          + 1497600 \* \Sss(3,-2,1)
          + 3302400 \* \Sss(3,1,-2)
          - 259200 \* \Sss(3,1,1)
          - 460800 \* \Sss(3,1,2)
          - 172800 \* \Sss(3,2,1)
  \nonumber\\&& \mbox{}
          + 854400 \* \Sss(4,1,1)
          + 230400 \* \Ssss(1,-2,1,1)
          - 902400 \* \Ssss(1,1,-2,1)
          - 1286400 \* \Ssss(1,1,1,-2)
          - 540000 \* \Ssss(1,1,1,1)
  \nonumber\\&& \mbox{}
          - 816000 \* \Ssss(1,1,1,2)
          + 230400 \* \Ssss(1,1,1,3)
          - 614400 \* \Ssss(1,1,2,1)
          - 691200 \* \Ssss(1,1,3,1)
          - 686400 \* \Ssss(1,2,1,1)
  \nonumber\\&& \mbox{}
          - 230400 \* \Ssss(2,-2,1,1)
          - 4377600 \* \Ssss(2,1,-2,1)
          + 153600 \* \Ssss(2,1,1,-2)
          - 1425600 \* \Ssss(2,1,1,1)
          + 422400 \* \Ssss(2,1,1,2)
  \nonumber\\&& \mbox{}
          + 441600 \* \Ssss(2,1,2,1)
          + 345600 \* \Ssss(2,2,1,1)
          + 345600 \* \Ssss(3,1,1,1)
          + 604800 \* \Sssss(1,1,1,1,1)
          - 345600 \* \Sssss(2,1,1,1,1)
  \nonumber\\&& \mbox{}
          + 2427893 \* \S(1)
          - 4032000 \* \S(1) \* \z5
          - 1266240 \* \S(1) \* \z3
          - 4786568 \* \S(2)
          + 172800 \* \S(2) \* \z4
  \nonumber\\&& \mbox{}
          - 1488000 \* \S(2) \* \z3
          - 506488 \* \S(3)
          + 921600 \* \S(3) \* \z3
          + 9086880 \* \S(4)
          - 1696800 \* \S(5)
          + 820800 \* \S(6))
  \nonumber\\&& \mbox{}
          - 64 \* (2 \* \Nminus + 3) \* (5 \* \Ss(1,-5)
          - 9 \* \Ss(1,-2) \* \z3
          - 5 \* \Ss(1,5)
          - 11 \* \Sss(1,-4,1)
          + \Sss(1,-3,-2)
          - \Sss(1,-3,2)
          + 3 \* \Sss(1,-2,-3)
  \nonumber\\&& \mbox{}
          + 2 \* \Sss(1,-2,3)
          - 12 \* \Sss(1,1,-4)
          + 6 \* \Sss(1,1,4)
          - 15 \* \Sss(1,2,-3)
          - 5 \* \Sss(1,3,-2)
          + \Sss(1,3,2)
          + \Ssss(1,-3,1,1)
          - 8 \* \Ssss(1,-2,-2,1)
  \nonumber\\&& \mbox{}
          + 2 \* \Ssss(1,-2,1,-2)
          + 19 \* \Ssss(1,1,-3,1)
          - 4 \* \Ssss(1,1,-2,-2)
          + 2 \* \Ssss(1,1,-2,2)
          + 25 \* \Ssss(1,1,1,-3)
          + 6 \* \Ssss(1,1,2,-2)
  \nonumber\\&& \mbox{}
          + 16 \* \Ssss(1,2,-2,1)
          + 6 \* \Ssss(1,2,1,-2)
          - \Ssss(1,3,1,1)
          - 2 \* \Sssss(1,1,-2,1,1)
          - 28 \* \Sssss(1,1,1,-2,1)
          - 6 \* \Sssss(1,1,1,1,-2))
  \nonumber\\&& \mbox{}
          - {1 \over 24} \* (7255 - 48320 \* \z5 + 1440 \* \z4 + 9824 \* \z3)
          \biggr)
  \nonumber\\&& \mbox{}
       + \colour4colour{\ca \* \cf \* \nf}  \*  \biggl(
          - {64 \over 9} \* \Ss(-4,1)
          - {256 \over 3} \* \Ss(-3,-2)
          + {320 \over 27} \* \Ss(-3,1)
          - {256 \over 3} \* \Ss(-2,-3)
          + {208 \over 3} \* \Ss(-2,-2)
  \nonumber\\&& \mbox{}
          - {3328 \over 9} \* \Ss(1,-3)
          + {60928 \over 135} \* \Ss(1,-2)
          + {4438 \over 27} \* \Ss(1,1)
          - {544 \over 3} \* \Ss(1,1) \* \z3
          - {1672 \over 9} \* \Ss(1,2)
          + {1424 \over 9} \* \Ss(1,3)
  \nonumber\\&& \mbox{}
          + 160 \* \Ss(2,-3)
          - {4640 \over 9} \* \Ss(2,-2)
          + {1252 \over 15} \* \Ss(2,1)
          - {4 \over 3} \* \Ss(2,2)
          - {224 \over 3} \* \Ss(2,3)
          + {448 \over 3} \* \Ss(3,-2)
          + {1588 \over 27} \* \Ss(3,1)
  \nonumber\\&& \mbox{}
          + {640 \over 9} \* \Ss(4,1)
          - {64 \over 9} \* \Sss(-3,1,1)
          + {128 \over 3} \* \Sss(-2,-2,1)
          - {256 \over 9} \* \Sss(1,-2,1)
          + {2944 \over 9} \* \Sss(1,1,-2)
          + {928 \over 9} \* \Sss(1,1,1)
  \nonumber\\&& \mbox{}
          + {184 \over 9} \* \Sss(1,1,2)
          + 64 \* \Sss(1,1,3)
          - {184 \over 9} \* \Sss(1,2,1)
          - {112 \over 3} \* \Sss(1,3,1)
          - 64 \* \Sss(2,-2,1)
          - {80 \over 3} \* \Sss(2,1,1)
          + {64 \over 9} \* \Sss(3,1,1)
  \nonumber\\&& \mbox{}
          - {1 \over 18225} \* \gqq \* (2559600 \* \Ss(1,-4)
          - 4104000 \* \Ss(1,-3)
          + 3731040 \* \Ss(1,-2)
          + 7996950 \* \Ss(1,1)
          - 4441500 \* \Ss(1,2)
  \nonumber\\&& \mbox{}
          - 4276800 \* \Ss(1,1) \* \z3
          + 3272400 \* \Ss(1,3)
          - 1717200 \* \Ss(1,4)
          + 810000 \* \Ss(2,-3)
          - 1414800 \* \Ss(2,-2)
  \nonumber\\&& \mbox{}
          - 4264380 \* \Ss(2,1)
          + 729000 \* \Ss(2,2)
          - 1377000 \* \Ss(2,3)
          + 32400 \* \Ss(3,-2)
          + 1382400 \* \Ss(3,1)
  \nonumber\\&& \mbox{}
          - 194400 \* \Ss(3,2)
          + 405000 \* \Ss(4,1)
          - 64800 \* \Sss(1,-3,1)
          - 1360800 \* \Sss(1,-2,-2)
          - 820800 \* \Sss(1,-2,1)
  \nonumber\\&& \mbox{}
          - 4017600 \* \Sss(1,1,-3)
          + 5702400 \* \Sss(1,1,-2)
          + 3744900 \* \Sss(1,1,1)
          - 64800 \* \Sss(1,1,2)
          + 567000 \* \Sss(1,1,3)
  \nonumber\\&& \mbox{}
          - 2656800 \* \Sss(1,2,-2)
          - 648000 \* \Sss(1,2,1)
          - 388800 \* \Sss(1,2,2)
          + 972000 \* \Sss(1,3,1)
          + 1296000 \* \Sss(2,-2,1)
  \nonumber\\&& \mbox{}
          - 2656800 \* \Sss(2,1,-2)
          - 469800 \* \Sss(2,1,1)
          - 113400 \* \Sss(2,1,2)
          + 113400 \* \Sss(2,2,1)
          + 113400 \* \Sss(3,1,1)
  \nonumber\\&& \mbox{}
          - 129600 \* \Ssss(1,-2,1,1)
          + 2592000 \* \Ssss(1,1,1,-2)
          + 356400 \* \Ssss(1,1,1,1)
          + 469800 \* \Ssss(1,1,1,2)
          - 324000 \* \Ssss(1,1,2,1)
  \nonumber\\&& \mbox{}
          - 145800 \* \Ssss(1,2,1,1)
          + 16836575 \* \S(1)
          + 145800 \* \S(1) \* \z4
          - 10597500 \* \S(1) \* \z3
          + 4276800 \* \S(2) \* \z3
  \nonumber\\&& \mbox{}
          - 17163522 \* \S(2)
          + 12371940 \* \S(3)
          - 3159000 \* \S(4)
          + 745200 \* \S(5))
          + {368 \over 9} \* \S(-5)
          - {664 \over 27} \* \S(-4)
          + {3056 \over 81} \* \S(-3)
  \nonumber\\&& \mbox{}
          - {232 \over 9} \* \S(-2)
          - {224 \over 3} \* \S(-2) \* \z3
          + {43324 \over 81} \* \S(1)
          - 720 \* \S(1) \* \z3
          - {46222 \over 2025} \* \S(2)
          + {896 \over 3} \* \S(2) \* \z3
          - {10372 \over 405} \* \S(3)
  \nonumber\\&& \mbox{}
          + {664 \over 27} \* \S(4)
          - {368 \over 9} \* \S(5)
          - {24 \over 25} \* (\Nplusthree - \Nplustwo) \* (60 \* \Ss(1,-3)
          - 53 \* \Ss(1,-2)
          - 100 \* \Ss(1,1) \* \z3
          + 20 \* \Ss(2,-2)
          + 50 \* \Ss(2,2)
  \nonumber\\&& \mbox{}
          - 50 \* \Ss(2,3)
          - 70 \* \Ss(3,1)
          + 50 \* \Ss(4,1)
          - 20 \* \Sss(1,-2,1)
          - 20 \* \Sss(1,1,-2)
          - 50 \* \Sss(1,1,2)
          + 50 \* \Sss(1,1,3)
          + 50 \* \Sss(1,2,1)
  \nonumber\\&& \mbox{}
          - 50 \* \Sss(1,3,1)
          - 240 \* \S(1) \* \z3
          + 100 \* \S(2) \* \z3
          - 53 \* \S(3)
          + 60 \* \S(4))
          + {8 \over 225} \* (\Nminusthree - \Nminustwo) \* (180 \* \Ss(1,-3)
          - 119 \* \Ss(1,-2)
  \nonumber\\&& \mbox{}
          - 300 \* \Ss(1,1) \* \z3 + 120 \* \Ss(2,-2)
          - 60 \* \Sss(1,-2,1)
          - 60 \* \Sss(1,1,-2)
          + 150 \* \Sss(1,1,3)
          - 150 \* \Sss(1,3,1)
          + 180 \* \S(1) \* \z3)
  \nonumber\\&& \mbox{}
          - {8 \over 25} \* (\Nplustwo - 3) \* (60 \* \Ss(1,-2)
          + 240 \* \Ss(1,1)
          + 200 \* \Ss(1,1) \* \z3 
          + 150 \* \Ss(1,2)
          - 150 \* \Ss(1,3)
          + 200 \* \Ss(2,-2)
  \nonumber\\&& \mbox{}
          - 240 \* \Ss(2,1)
          + 100 \* \Ss(2,3)
          - 50 \* \Ss(3,1)
          - 100 \* \Ss(4,1)
          + 200 \* \Sss(1,-2,1)
          - 200 \* \Sss(1,1,-2)
          - 100 \* \Sss(1,1,3)
  \nonumber\\&& \mbox{}
          + 100 \* \Sss(1,3,1)
          - 219 \* \S(1)
          + 500 \* \S(1) \* \z3
          - 21 \* \S(2)
          - 200 \* \S(2) \* \z3
          - 30 \* \S(3))
  \nonumber\\&& \mbox{}
          + {8 \over 225} \* (\Nminustwo - \Nminus) \* (60 \* \Ss(1,-2)
          + 90 \* \Ss(1,1)
          - 150 \* \Ss(1,3)
          + 90 \* \Ss(2,1)
          + 150 \* \Ss(3,1)
          - 179 \* \S(1)
  \nonumber\\&& \mbox{}
          + 300 \* \S(1) \* \z3
          - 59 \* \S(2)
          + 180 \* \S(3))
          + {2 \over 2025} \* (\Nminus + 1) \* (7200 \* \Ss(1,-3)
          - 57240 \* \Ss(1,-2)
          + 189165 \* \Ss(1,1)
  \nonumber\\&& \mbox{}
          - 194400 \* \Ss(1,1) \* \z3 - 147600 \* \Ss(1,2)
          + 117000 \* \Ss(1,3)
          - 97200 \* \Ss(2,-3)
          + 147600 \* \Ss(2,-2)
          - 64980 \* \Ss(2,1)
  \nonumber\\&& \mbox{}
          - 7200 \* \Ss(2,2)
          - 64800 \* \Ss(2,3)
          - 118800 \* \Ss(3,-2)
          + 2100 \* \Ss(3,1)
          - 2700 \* \Ss(3,2)
          + 38700 \* \Ss(4,1)
  \nonumber\\&& \mbox{}
          - 108000 \* \Sss(1,-2,1)
          + 180000 \* \Sss(1,1,-2)
          + 97650 \* \Sss(1,1,1)
          - 3150 \* \Sss(1,1,2)
          + 43200 \* \Sss(1,1,3)
          + 3150 \* \Sss(1,2,1)
  \nonumber\\&& \mbox{}
          + 10800 \* \Sss(1,3,1)
          + 108000 \* \Sss(2,-2,1)
          + 27000 \* \Sss(2,1,1)
          + 2700 \* \Sss(2,1,2)
          - 2700 \* \Sss(2,2,1)
          - 7200 \* \Sss(3,1,1)
  \nonumber\\&& \mbox{}
          + 623806 \* \S(1)
          - 332100 \* \S(1) \* \z3
          - 686307 \* \S(2)
          + 113400 \* \S(2) \* \z3
          + 506890 \* \S(3)
          - 99600 \* \S(4)
  \nonumber\\&& \mbox{}
          + 41400 \* \S(5))
          + {16 \over 3} \* (2 \* \Nminus + 3) \* (6 \* \Ss(1,-4)
          - 6 \* \Ss(1,4)
          - 4 \* \Sss(1,-2,-2)
          - 12 \* \Sss(1,1,-3)
          - 8 \* \Sss(1,2,-2)
          - \Sss(1,2,2)
  \nonumber\\&& \mbox{}
          - 8 \* \Sss(2,1,-2)
          + 8 \* \Ssss(1,1,1,-2)
          + \Ssss(1,1,1,2)
          - \Ssss(1,1,2,1))
          + {1 \over 486} \* (142883 + 5832 \* \z4 - 113508 \* \z3)
          \biggr)
  \nonumber\\&& \mbox{}
       + \colour4colour{\ca \* \cf^2}  \*  \biggl(
          - 880 \* \Ss(-5,1)
          - {1472 \over 3} \* \Ss(-4,-2)
          - {3764 \over 9} \* \Ss(-4,1)
          - {1360 \over 3} \* \Ss(-4,2)
          - 1088 \* \Ss(-3,-3)
          - {1280 \over 3} \* \Ss(-3,-2)
  \nonumber\\&& \mbox{}
          - {3848 \over 27} \* \Ss(-3,1)
          - {832 \over 3} \* \Ss(-3,3)
          - {2272 \over 3} \* \Ss(-2,-4)
          - 8 \* \Ss(-2,-3)
          + 572 \* \Ss(-2,-2)
          - {784 \over 3} \* \Ss(-2,1)
          + 608 \* \Ss(-2,1) \* \z3
  \nonumber\\&& \mbox{}
          - 32 \* \Ss(-2,3)
          + {448 \over 3} \* \Ss(-2,4)
          - {11884 \over 3} \* \Ss(1,-4)
          + {6472 \over 9} \* \Ss(1,-3)
          + {4004332 \over 675} \* \Ss(1,-2)
          + {72551 \over 45} \* \Ss(1,1)
  \nonumber\\&& \mbox{}
          + {7000 \over 3} \* \Ss(1,1) \* \z3
          - {61603 \over 27} \* \Ss(1,2)
          + 416 \* \Ss(1,2) \* \z3
          - {16126 \over 45} \* \Ss(1,3)
          + {5164 \over 3} \* \Ss(1,4)
          + {2480 \over 3} \* \Ss(2,-4)
          - {6832 \over 3} \* \Ss(2,-3)
  \nonumber\\&& \mbox{}
          - {325448 \over 45} \* \Ss(2,-2)
          - {980213 \over 675} \* \Ss(2,1)
          + 912 \* \Ss(2,1) \* \z3
          + {1072 \over 5} \* \Ss(2,2)
          + 2520 \* \Ss(2,3)
          - {1280 \over 3} \* \Ss(2,4)
          + {1984 \over 3} \* \Ss(3,-3)
  \nonumber\\&& \mbox{}
          + {2128 \over 3} \* \Ss(3,-2)
          + {693448 \over 135} \* \Ss(3,1)
          - {1756 \over 3} \* \Ss(3,2)
          + {880 \over 3} \* \Ss(3,3)
          + {160 \over 3} \* \Ss(4,-2)
          - {26896 \over 9} \* \Ss(4,1)
          + {1648 \over 3} \* \Ss(4,2)
  \nonumber\\&& \mbox{}
          + 1072 \* \Ss(5,1)
          + {1936 \over 3} \* \Sss(-4,1,1)
          - 256 \* \Sss(-3,-2,1)
          + 2176 \* \Sss(-3,1,-2)
          - {1856 \over 9} \* \Sss(-3,1,1)
          + {160 \over 3} \* \Sss(-3,1,2)
  \nonumber\\&& \mbox{}
          + {544 \over 3} \* \Sss(-3,2,1)
          + {2912 \over 3} \* \Sss(-2,-3,1)
          - 64 \* \Sss(-2,-2,-2)
          - 376 \* \Sss(-2,-2,1)
          - {224 \over 3} \* \Sss(-2,-2,2)
          + {5984 \over 3} \* \Sss(-2,1,-3)
  \nonumber\\&& \mbox{}
          - 1080 \* \Sss(-2,1,-2)
          - {448 \over 3} \* \Sss(-2,1,3)
          + {2080 \over 3} \* \Sss(-2,2,-2)
          + 64 \* \Sss(-2,2,2)
          + 8576 \* \Sss(1,-3,1)
          - {8408 \over 3} \* \Sss(1,-2,-2)
  \nonumber\\&& \mbox{}
          - {736 \over 3} \* \Sss(-2,3,1)
          - {51032 \over 9} \* \Sss(1,-2,1)
          + {3808 \over 3} \* \Sss(1,-2,2)
          + {23464 \over 3} \* \Sss(1,1,-3)
          + {28552 \over 45} \* \Sss(1,1,-2)
          + {15619 \over 9} \* \Sss(1,1,1)
  \nonumber\\&& \mbox{}
          - 928 \* \Sss(1,1,1) \* \z3
          + {7124 \over 9} \* \Sss(1,1,2)
          - {8344 \over 3} \* \Sss(1,1,3)
          + {2392 \over 3} \* \Sss(1,2,-2)
          - {184 \over 3} \* \Sss(1,2,1)
          + {8 \over 3} \* \Sss(1,2,2)
          + {6848 \over 3} \* \Sss(1,3,1)
  \nonumber\\&& \mbox{}
          + 640 \* \Sss(1,4,1)
          - 1888 \* \Sss(2,-3,1)
          + {1408 \over 3} \* \Sss(2,-2,-2)
          + {7792 \over 3} \* \Sss(2,-2,1)
          + {160 \over 3} \* \Sss(2,-2,2)
          - {8512 \over 3} \* \Sss(2,1,-3)
  \nonumber\\&& \mbox{}
          + {9616 \over 3} \* \Sss(2,1,-2)
          - {41128 \over 45} \* \Sss(2,1,1)
          + {164 \over 3} \* \Sss(2,1,2)
          - 96 \* \Sss(2,1,3)
          - {2048 \over 3} \* \Sss(2,2,-2)
          + {316 \over 3} \* \Sss(2,2,1)
          - {128 \over 3} \* \Sss(2,2,2)
  \nonumber\\&& \mbox{}
          + {1568 \over 3} \* \Sss(2,3,1)
          + {1280 \over 3} \* \Sss(3,-2,1)
          - {6080 \over 3} \* \Sss(3,1,-2)
          + {8168 \over 9} \* \Sss(3,1,1)
          + {16 \over 3} \* \Sss(3,1,2)
          - 240 \* \Sss(3,2,1)
          - {2416 \over 3} \* \Sss(4,1,1)
  \nonumber\\&& \mbox{}
          + {352 \over 3} \* \Ssss(-2,-2,1,1)
          - {5792 \over 3} \* \Ssss(-2,1,-2,1)
          - {2720 \over 3} \* \Ssss(-2,1,1,-2)
          - 32 \* \Ssss(-2,1,1,2)
          + 32 \* \Ssss(-2,1,2,1)
          - 384 \* \Ssss(1,1,1,1)
  \nonumber\\&& \mbox{}
          - 1696 \* \Ssss(1,-2,1,1)
          - {36128 \over 3} \* \Ssss(1,1,-2,1)
          - {3952 \over 3} \* \Ssss(1,1,1,-2)
          + {104 \over 3} \* \Ssss(1,1,1,2)
          - {272 \over 3} \* \Ssss(1,1,2,1)
          + 56 \* \Ssss(1,2,1,1)
  \nonumber\\&& \mbox{}
          - {352 \over 3} \* \Ssss(2,-2,1,1)
          + 3744 \* \Ssss(2,1,-2,1)
          + 928 \* \Ssss(2,1,1,-2)
          - 88 \* \Ssss(2,1,1,1)
          - 32 \* \Ssss(2,1,2,1)
          + 32 \* \Ssss(2,2,1,1)
          + 96 \* \Ssss(3,1,1,1)
  \nonumber\\&& \mbox{}
          - {1 \over 8100} \* \gqq \* (13456800 \* \Ss(1,-5)
          - 5878800 \* \Ss(1,-4)
          - 123600 \* \Ss(1,-3)
          + 16658632 \* \Ss(1,-2)
  \nonumber\\&& \mbox{}
          - 18662400 \* \Ss(1,-2) \* \z3
          + 21666165 \* \Ss(1,1)
          + 13284000 \* \Ss(1,1) \* \z3
          - 679600 \* \Ss(1,2)
          - 907200 \* \Ss(1,2) \* \z3
  \nonumber\\&& \mbox{}
          - 3225840 \* \Ss(1,3)
          + 6526800 \* \Ss(1,4)
          - 11253600 \* \Ss(1,5)
          + 11901600 \* \Ss(2,-4)
          - 10231200 \* \Ss(2,-3)
  \nonumber\\&& \mbox{}
          + 2061600 \* \Ss(2,-2)
          - 17987404 \* \Ss(2,1)
          - 12765600 \* \Ss(2,1) \* \z3
          - 18621840 \* \Ss(2,2)
          + 19011600 \* \Ss(2,3)
  \nonumber\\&& \mbox{}
          - 6782400 \* \Ss(2,4)
          + 5788800 \* \Ss(3,-3)
          - 3211200 \* \Ss(3,-2)
          - 39614460 \* \Ss(3,1)
          + 8647200 \* \Ss(3,2)
  \nonumber\\&& \mbox{}
          - 3823200 \* \Ss(3,3)
          + 734400 \* \Ss(4,-2)
          + 2539800 \* \Ss(4,1)
          - 3153600 \* \Ss(4,2)
          - 6328800 \* \Ss(5,1)
  \nonumber\\&& \mbox{}
          - 28317600 \* \Sss(1,-4,1)
          + 4104000 \* \Sss(1,-3,-2)
          + 35064000 \* \Sss(1,-3,1)
          - 6544800 \* \Sss(1,-3,2)
  \nonumber\\&& \mbox{}
          + 9504000 \* \Sss(1,-2,-3)
          - 20196000 \* \Sss(1,-2,-2)
          - 38749200 \* \Sss(1,-2,1)
          + 5400000 \* \Sss(1,-2,2)
  \nonumber\\&& \mbox{}
          - 259200 \* \Sss(1,-2,3)
          - 28706400 \* \Sss(1,1,-4)
          + 16700400 \* \Sss(1,1,-3)
          + 27661680 \* \Sss(1,1,-2)
  \nonumber\\&& \mbox{}
          + 1119600 \* \Sss(1,1,1)
          - 4017600 \* \Sss(1,1,1) \* \z3
          + 14871600 \* \Sss(1,1,2)
          - 23864400 \* \Sss(1,1,3)
  \nonumber\\&& \mbox{}
          + 14320800 \* \Sss(1,1,4)
          - 30758400 \* \Sss(1,2,-3)
          - 5623200 \* \Sss(1,2,-2)
          + 9928800 \* \Sss(1,2,1)
          - 4647600 \* \Sss(1,2,2)
  \nonumber\\&& \mbox{}
          + 4492800 \* \Sss(1,2,3)
          - 9720000 \* \Sss(1,3,-2)
          + 17427600 \* \Sss(1,3,1)
          + 12484800 \* \Sss(1,4,1)
          + 345600 \* \Sss(1,3,2)
  \nonumber\\&& \mbox{}
          - 18165600 \* \Sss(2,-3,1)
          + 4190400 \* \Sss(2,-2,-2)
          + 36943200 \* \Sss(2,-2,1)
          - 6091200 \* \Sss(2,-2,2)
  \nonumber\\&& \mbox{}
          - 13888800 \* \Sss(2,1,-3)
          - 21650400 \* \Sss(2,1,-2)
          + 21270240 \* \Sss(2,1,1)
          - 4352400 \* \Sss(2,1,2)
          + 6739200 \* \Sss(2,1,3)
  \nonumber\\&& \mbox{}
          - 3024000 \* \Sss(2,2,-2)
          - 3236400 \* \Sss(2,2,1)
          + 324000 \* \Sss(2,2,2)
          - 4168800 \* \Sss(2,3,1)
          - 11577600 \* \Sss(3,-2,1)
  \nonumber\\&& \mbox{}
          + 3542400 \* \Sss(3,1,-2)
          - 9792000 \* \Sss(3,1,1)
          + 324000 \* \Sss(3,1,2)
          + 1576800 \* \Sss(3,2,1)
          + 4190400 \* \Sss(4,1,1)
  \nonumber\\&& \mbox{}
          + 8985600 \* \Ssss(1,-3,1,1)
          - 25444800 \* \Ssss(1,-2,-2,1)
          + 9028800 \* \Ssss(1,-2,1,-2)
          - 8020800 \* \Ssss(1,-2,1,1)
  \nonumber\\&& \mbox{}
          + 302400 \* \Ssss(1,-2,1,2)
          + 1598400 \* \Ssss(1,-2,2,1)
          + 55555200 \* \Ssss(1,1,-3,1)
          - 14126400 \* \Ssss(1,1,-2,-2)
  \nonumber\\&& \mbox{}
          - 50976000 \* \Ssss(1,1,-2,1)
          + 11491200 \* \Ssss(1,1,-2,2)
          + 49248000 \* \Ssss(1,1,1,-3)
          - 655200 \* \Ssss(1,1,1,-2)
  \nonumber\\&& \mbox{}
          - 10918800 \* \Ssss(1,1,1,1)
          + 3614400 \* \Ssss(1,1,1,2)
          - 5356800 \* \Ssss(1,1,1,3)
          + 6350400 \* \Ssss(1,1,2,-2)
  \nonumber\\&& \mbox{}
          + 1544400 \* \Ssss(1,1,2,1)
          + 129600 \* \Ssss(1,1,2,2)
          + 2505600 \* \Ssss(1,1,3,1)
          + 41990400 \* \Ssss(1,2,-2,1)
          + 129600 \* \Ssss(1,2,1,2)
  \nonumber\\&& \mbox{}
          + 6739200 \* \Ssss(1,2,1,-2)
          + 3157200 \* \Ssss(1,2,1,1)
          - 475200 \* \Ssss(1,2,2,1)
          - 129600 \* \Ssss(1,3,1,1)
          + 8121600 \* \Ssss(2,-2,1,1)
  \nonumber\\&& \mbox{}
          + 15940800 \* \Ssss(2,1,-2,1)
          + 4406400 \* \Ssss(2,1,1,-2)
          + 3603600 \* \Ssss(2,1,1,1)
          + 669600 \* \Ssss(2,1,1,2)
          - 129600 \* \Ssss(2,1,2,1)
  \nonumber\\&& \mbox{}
          - 540000 \* \Ssss(2,2,1,1)
          - 777600 \* \Ssss(3,1,1,1)
          - 777600 \* \Sssss(1,-2,1,1,1)
          - 14947200 \* \Sssss(1,1,-2,1,1)
  \nonumber\\&& \mbox{}
          - 69206400 \* \Sssss(1,1,1,-2,1)
          - 6998400 \* \Sssss(1,1,1,1,-2)
          - 2376000 \* \Sssss(1,1,1,1,1)
          - 777600 \* \Sssss(1,1,1,1,2)
  \nonumber\\&& \mbox{}
          + 86400 \* \Sssss(1,1,1,2,1)
          + 302400 \* \Sssss(1,1,2,1,1)
          + 388800 \* \Sssss(1,2,1,1,1)
          + 34803466 \* \S(1)
          - 32400000 \* \S(1) \* \z5
  \nonumber\\&& \mbox{}
          + 583200 \* \S(1) \* \z4
          - 22037400 \* \S(1) \* \z3)
          + {1 \over 8100} \* \gqq \* (48164505 \* \S(2)
          - 583200 \* \S(2) \* \z4
          + 9979200 \* \S(2) \* \z3
  \nonumber\\&& \mbox{}
          - 38377324 \* \S(3)
          - 4082400 \* \S(3) \* \z3
          - 30815580 \* \S(4)
          + 9311400 \* \S(5)
          - 4600800 \* \S(6))
          + 400 \* \S(-6)
  \nonumber\\&& \mbox{}
          + {7732 \over 9} \* \S(-5)
          - {19118 \over 27} \* \S(-4)
          + {55048 \over 81} \* \S(-3)
          - 1312 \* \S(-3) \* \z3
          - {6170 \over 9} \* \S(-2)
  \nonumber\\&& \mbox{}
          + 144 \* \S(-2) \* \z4
          - 608 \* \S(-2) \* \z3
          + {37427779 \over 16200} \* \S(1)
          - 2640 \* \S(1) \* \z5
          + 144 \* \S(1) \* \z4
  \nonumber\\&& \mbox{}
          - 1058 \* \S(1) \* \z3
          - {5525594 \over 2025} \* \S(2)
          - 144 \* \S(2) \* \z4
          - 5048 \* \S(2) \* \z3
          + {6207649 \over 2025} \* \S(3)
          + 1072 \* \S(3) \* \z3
          + {38818 \over 135} \* \S(4)
  \nonumber\\&& \mbox{}
          - {3728 \over 9} \* \S(5)
          - 400 \* \S(6)
          - {4 \over 25} \* (\Nplusthree - \Nplustwo) \* (1440 \* \Ss(1,-4)
          - 914 \* \Ss(1,-3)
          - 2727 \* \Ss(1,-2)
          + 13800 \* \Ss(1,1) \* \z3 
  \nonumber\\&& \mbox{}
          - 360 \* \Ss(1,4)
          + 5400 \* \Ss(2,-3)
          - 3074 \* \Ss(2,-2)
          + 2750 \* \Ss(2,2)
          + 5280 \* \Ss(2,3)
          + 1080 \* \Ss(3,-2)
          + 324 \* \Ss(3,1)
  \nonumber\\&& \mbox{}
          - 360 \* \Ss(3,2)
          - 10680 \* \Ss(4,1)
          - 5400 \* \Sss(1,-3,1)
          + 720 \* \Sss(1,-2,-2)
          + 3074 \* \Sss(1,-2,1)
          - 360 \* \Sss(1,-2,2)
          - 5400 \* \Sss(1,1,-3)
  \nonumber\\&& \mbox{}
          + 3074 \* \Sss(1,1,-2)
          - 2750 \* \Sss(1,1,2)
          - 5280 \* \Sss(1,1,3)
          - 360 \* \Sss(1,2,-2)
          + 2750 \* \Sss(1,2,1)
          + 5280 \* \Sss(1,3,1)
          - 9000 \* \Sss(2,-2,1)
  \nonumber\\&& \mbox{}
          - 360 \* \Sss(2,1,-2)
          + 360 \* \Sss(3,1,1)
          + 360 \* \Ssss(1,-2,1,1)
          + 9000 \* \Ssss(1,1,-2,1)
          + 360 \* \Ssss(1,1,1,-2)
          - 13260 \* \S(1) \* \z3
  \nonumber\\&& \mbox{}
          - 13800 \* \S(2) \* \z3
          - 2727 \* \S(3)
          - 914 \* \S(4)
          + 1800 \* \S(5))
          + {4 \over 225} \* (\Nminusthree - \Nminustwo) \* (1440 \* \Ss(1,-4)
          - 674 \* \Ss(1,-3)
  \nonumber\\&& \mbox{}
          - 1357 \* \Ss(1,-2)
          + 13800 \* \Ss(1,1) \* \z3 
          - 360 \* \Ss(1,4)
          + 720 \* \Ss(2,-3)
          + 2160 \* \Ss(2,-2)
          - 5400 \* \Sss(1,-3,1)
          + 720 \* \Sss(1,-2,-2)
  \nonumber\\&& \mbox{}
          + 2834 \* \Sss(1,-2,1)
          - 360 \* \Sss(1,-2,2)
          - 5400 \* \Sss(1,1,-3)
          - 5280 \* \Sss(1,1,3)
          - 360 \* \Sss(1,2,-2)
          + 5280 \* \Sss(1,3,1)
          + 3240 \* \S(1) \* \z3
  \nonumber\\&& \mbox{}
          + 360 \* \Ssss(1,1,1,-2)
          + 2834 \* \Sss(1,1,-2)
          - 720 \* \Sss(2,-2,1)
          - 720 \* \Sss(2,1,-2)
          + 360 \* \Ssss(1,-2,1,1)
          + 9000 \* \Ssss(1,1,-2,1)
              )
  \nonumber\\&& \mbox{}
          - {4 \over 25} \* (\Nplustwo - 3) \* (5400 \* \Ss(1,-3)
          - 2354 \* \Ss(1,-2)
          - 3416 \* \Ss(1,1)
          - 16800 \* \Ss(1,1) \* \z3 
          + 2390 \* \Ss(1,2)
          + 5280 \* \Ss(1,3)
  \nonumber\\&& \mbox{}
          + 10240 \* \Ss(2,-2)
          + 3056 \* \Ss(2,1)
          + 1460 \* \Ss(2,2)
          - 8400 \* \Ss(2,3)
          - 11580 \* \Ss(3,1)
          + 8400 \* \Ss(4,1)
          + 1600 \* \Sss(1,-2,1)
  \nonumber\\&& \mbox{}
          - 10960 \* \Sss(1,1,-2)
          + 360 \* \Sss(1,1,1)
          - 1100 \* \Sss(1,1,2)
          + 8400 \* \Sss(1,1,3)
          + 1100 \* \Sss(1,2,1)
          - 8400 \* \Sss(1,3,1)
          - 360 \* \Sss(2,1,1)
  \nonumber\\&& \mbox{}
          - 7961 \* \S(1)
          - 9800 \* \S(1) \* \z3 
          + 11377 \* \S(2)
          + 16800 \* \S(2) \* \z3 
          - 8686 \* \S(3)
          - 1080 \* \S(4))
  \nonumber\\&& \mbox{}
          + {4 \over 225} \* (\Nminustwo - \Nminus) \* (5400 \* \Ss(1,-3)
          - 2474 \* \Ss(1,-2)
          - 6406 \* \Ss(1,1)
          - 360 \* \Ss(1,2)
          + 5280 \* \Ss(1,3)
          + 1080 \* \Ss(2,-2)
  \nonumber\\&& \mbox{}
          - 7126 \* \Ss(2,1)
          - 360 \* \Ss(2,2)
          - 10680 \* \Ss(3,1)
          - 9000 \* \Sss(1,-2,1)
          - 360 \* \Sss(1,1,-2)
          + 360 \* \Sss(1,1,1)
  \nonumber\\&& \mbox{}
          + 360 \* \Sss(2,1,1)
          - 6711 \* \S(1)
          - 13800 \* \S(1) \* \z3
          - 3831 \* \S(2)
          + 406 \* \S(3)
          + 1800 \* \S(4))
  \nonumber\\&& \mbox{}
          + {1 \over 16200} \* (\Nminus + 1) \* (9396000 \* \Ss(1,-4)
          - 9021600 \* \Ss(1,-3)
          - 11780352 \* \Ss(1,-2)
          + 36624072 \* \Ss(1,1)
  \nonumber\\&& \mbox{}
          + 53481600 \* \Ss(1,1) \* \z3
          + 1287720 \* \Ss(1,2)
          + 1036800 \* \Ss(1,2) \* \z3
          - 8051760 \* \Ss(1,3)
          - 6285600 \* \Ss(1,4)
  \nonumber\\&& \mbox{}
          - 5616000 \* \Ss(2,-4)
          - 12916800 \* \Ss(2,-3)
          + 44516160 \* \Ss(2,-2)
          - 36332904 \* \Ss(2,1)
          - 29030400 \* \Ss(2,1) \* \z3
  \nonumber\\&& \mbox{}
          - 31695840 \* \Ss(2,2)
          + 32140800 \* \Ss(2,3)
          + 691200 \* \Ss(2,4)
          - 11923200 \* \Ss(3,-3)
          - 19526400 \* \Ss(3,-2)
  \nonumber\\&& \mbox{}
          - 78133440 \* \Ss(3,1)
          + 14774400 \* \Ss(3,2)
          - 5356800 \* \Ss(3,3)
          - 3974400 \* \Ss(4,-2)
          + 6033600 \* \Ss(4,1)
  \nonumber\\&& \mbox{}
          - 9417600 \* \Ss(4,2)
          - 18878400 \* \Ss(5,1)
          - 1944000 \* \Sss(1,-3,1)
          - 14212800 \* \Sss(1,-2,-2)
          - 34704000 \* \Sss(1,-2,1)
  \nonumber\\&& \mbox{}
          - 1036800 \* \Sss(1,-2,2)
          - 21124800 \* \Sss(1,1,-3)
          + 68800320 \* \Sss(1,1,-2)
          - 5168520 \* \Sss(1,1,1)
          - 5184000 \* \Sss(1,1,1) \* \z3
  \nonumber\\&& \mbox{}
          + 13413600 \* \Sss(1,1,2)
          - 37368000 \* \Sss(1,1,3)
          - 11707200 \* \Sss(1,2,-2)
          + 6804000 \* \Sss(1,2,1)
          - 2030400 \* \Sss(1,2,2)
  \nonumber\\&& \mbox{}
          + 3110400 \* \Sss(1,2,3)
          + 40651200 \* \Sss(1,3,1)
          + 14515200 \* \Sss(1,4,1)
          + 19440000 \* \Sss(2,-3,1)
          - 4492800 \* \Sss(2,-2,-2)
  \nonumber\\&& \mbox{}
          + 51537600 \* \Sss(2,-2,1)
          - 1382400 \* \Sss(2,-2,2)
          + 27043200 \* \Sss(2,1,-3)
          - 63201600 \* \Sss(2,1,-2)
          - 1360800 \* \Sss(2,1,2)
  \nonumber\\&& \mbox{}
          + 39255840 \* \Sss(2,1,1)
          + 8035200 \* \Sss(2,1,3)
          + 4320000 \* \Sss(2,2,-2)
          - 1144800 \* \Sss(2,2,1)
          + 691200 \* \Sss(2,2,2)
  \nonumber\\&& \mbox{}
          - 10281600 \* \Sss(2,3,1)
          - 3283200 \* \Sss(3,-2,1)
          + 29203200 \* \Sss(3,1,-2)
          - 19713600 \* \Sss(3,1,1)
          + 259200 \* \Sss(3,1,2)
  \nonumber\\&& \mbox{}
          + 3542400 \* \Sss(3,2,1)
          + 13564800 \* \Sss(4,1,1)
          + 1036800 \* \Ssss(1,-2,1,1)
          - 1296000 \* \Ssss(1,1,-2,1)
          + 5356800 \* \Ssss(1,1,1,-2)
  \nonumber\\&& \mbox{}
          - 9374400 \* \Ssss(1,1,1,1)
          + 1036800 \* \Ssss(1,1,1,2)
          - 5184000 \* \Ssss(1,1,1,3)
          - 1252800 \* \Ssss(1,1,2,1)
          + 3110400 \* \Ssss(1,1,3,1)
  \nonumber\\&& \mbox{}
          + 216000 \* \Ssss(1,2,1,1)
          + 2419200 \* \Ssss(2,-2,1,1)
          - 40953600 \* \Ssss(2,1,-2,1)
          + 1425600 \* \Ssss(2,1,1,1)
          + 518400 \* \Ssss(2,1,2,1)
  \nonumber\\&& \mbox{}
          - 5184000 \* \Ssss(2,1,1,-2)
          - 518400 \* \Ssss(2,2,1,1)
          - 1555200 \* \Ssss(3,1,1,1)
          + 65796867 \* \S(1)
          - 38880000 \* \S(1) \* \z5
  \nonumber\\&& \mbox{}
          + 10800 \* \S(1) \* \z3
          - 90885568 \* \S(2)
          + 2332800 \* \S(2) \* \z4
          - 31233600 \* \S(2) \* \z3
          + 71824864 \* \S(3)
  \nonumber\\&& \mbox{}
          - 3110400 \* \S(3) \* \z3
          + 59972640 \* \S(4)
          - 15019200 \* \S(5)
          + 9201600 \* \S(6) )
  \nonumber\\&& \mbox{}
          + 32 \* (2 \* \Nminus + 3) \* (9 \* \Ss(1,-5)
          - 21 \* \Ss(1,-2) \* \z3
          - 9 \* \Ss(1,5)
          - 21 \* \Sss(1,-4,1)
          + 5 \* \Sss(1,-3,-2)
          - \Sss(1,-3,2)
  \nonumber\\&& \mbox{}
          + 11 \* \Sss(1,-2,-3)
          + 4 \* \Sss(1,-2,3)
          - 16 \* \Sss(1,1,-4)
          + 10 \* \Sss(1,1,4)
          - 23 \* \Sss(1,2,-3)
          - 9 \* \Sss(1,3,-2)
          + \Sss(1,3,2)
          + \Ssss(1,-3,1,1)
  \nonumber\\&& \mbox{}
          - 24 \* \Ssss(1,-2,-2,1)
          + 2 \* \Ssss(1,-2,1,-2)
          + 35 \* \Ssss(1,1,-3,1)
          - 12 \* \Ssss(1,1,-2,-2)
          + 2 \* \Ssss(1,1,-2,2)
          + 41 \* \Ssss(1,1,1,-3)
  \nonumber\\&& \mbox{}
          + 6 \* \Ssss(1,1,2,-2)
          + 32 \* \Ssss(1,2,-2,1)
          + 6 \* \Ssss(1,2,1,-2)
          - \Ssss(1,3,1,1)
          - 2 \* \Sssss(1,1,-2,1,1)
          - 60 \* \Sssss(1,1,1,-2,1)
          - 6 \* \Sssss(1,1,1,1,-2))
  \nonumber\\&& \mbox{}
          + {1 \over 12} \* (9161 - 15520 \* \z5 + 1080 \* \z4 - 20084 \* \z3)
          \biggr)
  \nonumber\\&& \mbox{}
       + \colour4colour{\ca^2 \* \cf}  \*  \biggl(
            160 \* \Ss(-5,1)
          + 32 \* \Ss(-4,-2)
          + {1648 \over 9} \* \Ss(-4,1)
          + 64 \* \Ss(-4,2)
          + 64 \* \Ss(-3,-3)
          + {856 \over 3} \* \Ss(-3,-2)
  \nonumber\\&& \mbox{}
          - {272 \over 27} \* \Ss(-3,1)
          + {160 \over 3} \* \Ss(-3,3)
          + {320 \over 3} \* \Ss(-2,-4)
          + {416 \over 3} \* \Ss(-2,-3)
          - {784 \over 3} \* \Ss(-2,-2)
          + {340 \over 3} \* \Ss(-2,1)
          - 128 \* \Ss(-2,1) \* \z3
  \nonumber\\&& \mbox{}
          + 16 \* \Ss(-2,3)
          - {128 \over 3} \* \Ss(-2,4)
          + {1720 \over 3} \* \Ss(1,-4)
          + {10264 \over 9} \* \Ss(1,-3)
          - {373048 \over 135} \* \Ss(1,-2)
          - {94801 \over 135} \* \Ss(1,1)
  \nonumber\\&& \mbox{}
          - {1280 \over 3} \* \Ss(1,1) \* \z3
          + {3674 \over 3} \* \Ss(1,2)
          - 128 \* \Ss(1,2) \* \z3
          - {19696 \over 45} \* \Ss(1,3)
          - {1144 \over 3} \* \Ss(1,4)
          - {400 \over 3} \* \Ss(2,-4)
          - {224 \over 3} \* \Ss(2,-3)
  \nonumber\\&& \mbox{}
          + {134872 \over 45} \* \Ss(2,-2)
          + {9454 \over 45} \* \Ss(2,1)
          - 320 \* \Ss(2,1) \* \z3
          - {436 \over 3} \* \Ss(2,2)
          - 760 \* \Ss(2,3)
          + {352 \over 3} \* \Ss(2,4)
          + {32 \over 3} \* \Ss(3,-3)
  \nonumber\\&& \mbox{}
          - {1600 \over 3} \* \Ss(3,-2)
          - {133892 \over 135} \* \Ss(3,1)
          + 100 \* \Ss(3,2)
          - 56 \* \Ss(3,3)
          - {16 \over 3} \* \Ss(4,-2)
          + {6500 \over 9} \* \Ss(4,1)
          - 88 \* \Ss(4,2)
  \nonumber\\&& \mbox{}
          - 208 \* \Ss(5,1)
          - {320 \over 3} \* \Sss(-4,1,1)
          + {896 \over 3} \* \Sss(-3,-2,1)
          - 512 \* \Sss(-3,1,-2)
          + {352 \over 9} \* \Sss(-3,1,1)
          + {64 \over 3} \* \Sss(-3,1,2)
          - {64 \over 3} \* \Sss(-3,2,1)
  \nonumber\\&& \mbox{}
          - 96 \* \Sss(-2,-3,1)
          + {64 \over 3} \* \Sss(-2,-2,-2)
          + {128 \over 3} \* \Sss(-2,-2,1)
          + {256 \over 3} \* \Sss(-2,-2,2)
          - {1312 \over 3} \* \Sss(-2,1,-3)
          + 288 \* \Sss(-2,1,-2)
  \nonumber\\&& \mbox{}
          + {128 \over 3} \* \Sss(-2,1,3)
          - {512 \over 3} \* \Sss(-2,2,-2)
          - {64 \over 3} \* \Sss(-2,2,2)
          + {224 \over 3} \* \Sss(-2,3,1)
          - 1952 \* \Sss(1,-3,1)
          + {2512 \over 3} \* \Sss(1,-2,-2)
  \nonumber\\&& \mbox{}
          + {8800 \over 9} \* \Sss(1,-2,1)
          - {640 \over 3} \* \Sss(1,-2,2)
          - {4624 \over 3} \* \Sss(1,1,-3)
          - {8464 \over 9} \* \Sss(1,1,-2)
          - {2650 \over 3} \* \Sss(1,1,1)
          + 256 \* \Sss(1,1,1) \* \z3
  \nonumber\\&& \mbox{}
          - {856 \over 9} \* \Sss(1,1,2)
          + 952 \* \Sss(1,1,3)
          + {128 \over 3} \* \Sss(1,2,-2)
          + {856 \over 9} \* \Sss(1,2,1)
          + {32 \over 3} \* \Sss(1,2,2)
          - {2584 \over 3} \* \Sss(1,3,1)
          - 128 \* \Sss(1,4,1)
  \nonumber\\&& \mbox{}
          + {1000 \over 3} \* \Sss(2,-3,1)
          - 160 \* \Sss(2,-2,-2)
          - {896 \over 3} \* \Sss(2,-2,1)
          - 80 \* \Sss(2,-2,2)
          + 616 \* \Sss(2,1,-3)
          - {1504 \over 3} \* \Sss(2,1,-2)
  \nonumber\\&& \mbox{}
          + {1120 \over 3} \* \Sss(2,1,1)
          + {4 \over 3} \* \Sss(2,1,2)
          + {40 \over 3} \* \Sss(2,1,3)
          + {400 \over 3} \* \Sss(2,2,-2)
          + {20 \over 3} \* \Sss(2,2,1)
          + {56 \over 3} \* \Sss(2,2,2)
          - {344 \over 3} \* \Sss(2,3,1)
  \nonumber\\&& \mbox{}
          - {1072 \over 3} \* \Sss(3,-2,1)
          + {1328 \over 3} \* \Sss(3,1,-2)
          - {1864 \over 9} \* \Sss(3,1,1)
          - {80 \over 3} \* \Sss(3,1,2)
          + {80 \over 3} \* \Sss(3,2,1)
          - {320 \over 3} \* \Ssss(-2,-2,1,1)
  \nonumber\\&& \mbox{}
          + {440 \over 3} \* \Sss(4,1,1)
          + {1280 \over 3} \* \Ssss(-2,1,-2,1)
          + {704 \over 3} \* \Ssss(-2,1,1,-2)
          + {32 \over 3} \* \Ssss(-2,1,1,2)
          - {32 \over 3} \* \Ssss(-2,1,2,1)
          + 320 \* \Ssss(1,-2,1,1)
  \nonumber\\&& \mbox{}
          + {9344 \over 3} \* \Ssss(1,1,-2,1)
          + {352 \over 3} \* \Ssss(1,1,1,-2)
          - {56 \over 3} \* \Ssss(1,1,1,2)
          + {40 \over 3} \* \Ssss(1,1,2,1)
          + {16 \over 3} \* \Ssss(1,2,1,1)
          + {320 \over 3} \* \Ssss(2,-2,1,1)
  \nonumber\\&& \mbox{}
          - 912 \* \Ssss(2,1,-2,1)
          - {544 \over 3} \* \Ssss(2,1,1,-2)
          - {32 \over 3} \* \Ssss(2,1,1,2)
          + {32 \over 3} \* \Ssss(2,1,2,1)
          + {1 \over 36450} \* \gqq \* (11469600 \* \Ss(1,-5)
  \nonumber\\&& \mbox{}
          + 10951200 \* \Ss(1,-4)
          - 27226800 \* \Ss(1,-3)
          + 38900880 \* \Ss(1,-2)
          - 23328000 \* \Ss(1,-2) \* \z3
          + 68981310 \* \Ss(1,1)
  \nonumber\\&& \mbox{}
          - 6220800 \* \Ss(1,1) \* \z3
          - 40178700 \* \Ss(1,2)
          + 3499200 \* \Ss(1,2) \* \z3
          + 22654080 \* \Ss(1,3)
          - 5184000 \* \Ss(1,4)
  \nonumber\\&& \mbox{}
          - 11469600 \* \Ss(1,5)
          + 7387200 \* \Ss(2,-4)
          + 162000 \* \Ss(2,-3)
          - 4537080 \* \Ss(2,-2)
          - 69701040 \* \Ss(2,1)
  \nonumber\\&& \mbox{}
          - 12247200 \* \Ss(2,1) \* \z3
          + 1717200 \* \Ss(2,2)
          + 13235400 \* \Ss(2,3)
          - 5346000 \* \Ss(2,4)
          + 6318000 \* \Ss(3,-3)
  \nonumber\\&& \mbox{}
          - 2851200 \* \Ss(3,-2)
          - 13247280 \* \Ss(3,1)
          + 631800 \* \Ss(3,2)
          - 3159000 \* \Ss(3,3)
          + 486000 \* \Ss(4,-2)
  \nonumber\\&& \mbox{}
          - 13672800 \* \Ss(4,1)
          - 1020600 \* \Ss(4,2)
          - 3402000 \* \Ss(5,1)
          - 28576800 \* \Sss(1,-4,1)
          + 10497600 \* \Sss(1,-3,-2)
  \nonumber\\&& \mbox{}
          + 33307200 \* \Sss(1,-3,1)
          - 4762800 \* \Sss(1,-3,2)
          + 18856800 \* \Sss(1,-2,-3)
          - 33631200 \* \Sss(1,-2,-2)
  \nonumber\\&& \mbox{}
          - 39484800 \* \Sss(1,-2,1)
          + 3888000 \* \Sss(1,-2,2)
          + 194400 \* \Sss(1,-2,3)
          - 20606400 \* \Sss(1,1,-4)
          - 9444600 \* \Sss(1,1,-3)
  \nonumber\\&& \mbox{}
          + 56635200 \* \Sss(1,1,-2)
          + 35213400 \* \Sss(1,1,1)
          - 9331200 \* \Sss(1,1,1) \* \z3
          + 2170800 \* \Sss(1,1,2)
          - 24526800 \* \Sss(1,1,3)
  \nonumber\\&& \mbox{}
          + 15940800 \* \Sss(1,1,4)
          - 24494400 \* \Sss(1,2,-3)
          - 23684400 \* \Sss(1,2,-2)
          - 6139800 \* \Sss(1,2,1)
          - 3985200 \* \Sss(1,2,2)
  \nonumber\\&& \mbox{}
          - 6609600 \* \Sss(1,3,-2)
          + 34894800 \* \Sss(1,3,1)
          - 2527200 \* \Sss(1,3,2)
          + 6706800 \* \Sss(1,4,1)
          - 19683000 \* \Sss(2,-3,1)
  \nonumber\\&& \mbox{}
          + 8748000 \* \Sss(1,2,3)
          + 6804000 \* \Sss(2,-2,-2)
          + 43513200 \* \Sss(2,-2,1)
          - 7095600 \* \Sss(2,-2,2)
          - 6366600 \* \Sss(2,1,-3)
  \nonumber\\&& \mbox{}
          - 42152400 \* \Sss(2,1,-2)
          + 2932200 \* \Sss(2,1,1)
          - 1927800 \* \Sss(2,1,2)
          + 11518200 \* \Sss(2,1,3)
          + 2818800 \* \Sss(2,2,-2)
  \nonumber\\&& \mbox{}
          + 2073600 \* \Sss(2,2,1)
          + 631800 \* \Sss(2,2,2)
          - 12490200 \* \Sss(2,3,1)
          - 17593200 \* \Sss(3,-2,1)
          + 8845200 \* \Sss(3,1,-2)
  \nonumber\\&& \mbox{}
          - 1814400 \* \Sss(3,1,1)
          - 291600 \* \Sss(3,1,2)
          + 291600 \* \Sss(3,2,1)
          + 1701000 \* \Sss(4,1,1)
          + 7776000 \* \Ssss(1,-3,1,1)
  \nonumber\\&& \mbox{}
          - 38880000 \* \Ssss(1,-2,-2,1)
          + 8164800 \* \Ssss(1,-2,1,-2)
          - 7257600 \* \Ssss(1,-2,1,1)
          - 972000 \* \Ssss(1,-2,1,2)
  \nonumber\\&& \mbox{}
          + 59486400 \* \Ssss(1,1,-3,1)
          + 972000 \* \Ssss(1,-2,2,1)
          - 21384000 \* \Ssss(1,1,-2,-2)
          - 53654400 \* \Ssss(1,1,-2,1)
  \nonumber\\&& \mbox{}
          + 10886400 \* \Ssss(1,1,-2,2)
          + 43156800 \* \Ssss(1,1,1,-3)
          + 17528400 \* \Ssss(1,1,1,-2)
          + 1960200 \* \Ssss(1,1,1,1)
  \nonumber\\&& \mbox{}
          + 4973400 \* \Ssss(1,1,1,2)
          - 14774400 \* \Ssss(1,1,1,3)
          - 1166400 \* \Ssss(1,1,2,-2)
          - 3515400 \* \Ssss(1,1,2,1)
          - 583200 \* \Ssss(1,1,2,2)
  \nonumber\\&& \mbox{}
          + 13024800 \* \Ssss(1,1,3,1)
          + 47239200 \* \Ssss(1,2,-2,1)
          - 583200 \* \Ssss(1,2,1,-2)
          - 1458000 \* \Ssss(1,2,1,1)
  \nonumber\\&& \mbox{}
          - 194400 \* \Ssss(1,2,1,2)
          + 194400 \* \Ssss(1,2,2,1)
          + 3888000 \* \Ssss(1,3,1,1)
          + 9720000 \* \Ssss(2,-2,1,1)
          + 16621200 \* \Ssss(2,1,-2,1)
  \nonumber\\&& \mbox{}
          - 3304800 \* \Ssss(2,1,1,-2)
          - 97200 \* \Ssss(2,1,1,2)
          + 97200 \* \Ssss(2,2,1,1)
          - 15552000 \* \Sssss(1,1,-2,1,1)
  \nonumber\\&& \mbox{}
          - 81648000 \* \Sssss(1,1,1,-2,1)
          + 388800 \* \Sssss(1,1,1,2,1)
          - 388800 \* \Sssss(1,1,2,1,1)
          + 136806035 \* \S(1)
          - 29160000 \* \S(1) \* \z5
  \nonumber\\&& \mbox{}
          + 874800 \* \S(1) \* \z4
          - 94664160 \* \S(1) \* \z3
          - 143170722 \* \S(2)
          + 874800 \* \S(2) \* \z4
          + 12733200 \* \S(2) \* \z3
  \nonumber\\&& \mbox{}
          + 132482160 \* \S(3)
          - 5248800 \* \S(3) \* \z3
          - 8115660 \* \S(4)
          + 4212000 \* \S(5)
          + 2916000 \* \S(6))
          - 64 \* \S(-6)
  \nonumber\\&& \mbox{}
          - {2972 \over 9} \* \S(-5)
          + {7408 \over 27} \* \S(-4)
          - {24554 \over 81} \* \S(-3)
          + 352 \* \S(-3) \* \z3
          + {2362 \over 9} \* \S(-2)
          - 48 \* \S(-2) \* \z4
          + {856 \over 3} \* \S(-2) \* \z3
  \nonumber\\&& \mbox{}
          - {1068388 \over 405} \* \S(1)
          + 560 \* \S(1) \* \z5
          - 48 \* \S(1) \* \z4
          + {38104 \over 15} \* \S(1) \* \z3
          + {2573501 \over 2025} \* \S(2)
          + 48 \* \S(2) \* \z4
          + {3040 \over 3} \* \S(2) \* \z3
  \nonumber\\&& \mbox{}
          - {314078 \over 405} \* \S(3)
          - 352 \* \S(3) \* \z3
          - {7408 \over 27} \* \S(4)
          + {2972 \over 9} \* \S(5)
          + 64 \* \S(6)
          + {4 \over 25} \* (\Nplusthree - \Nplustwo) \* (120 \* \Ss(1,-4)
  \nonumber\\&& \mbox{}
          + 965 \* \Ss(1,-3)
          - 1869 \* \Ss(1,-2)
          + 2190 \* \Ss(1,1) \* \z3 - 120 \* \Ss(1,4)
          + 1200 \* \Ss(2,-3)
          - 355 \* \Ss(2,-2)
  \nonumber\\&& \mbox{}
          + 1650 \* \Ss(2,2)
          + 495 \* \Ss(2,3)
          + 240 \* \Ss(3,-2)
          - 1295 \* \Ss(3,1)
          - 1695 \* \Ss(4,1)
          - 1200 \* \Sss(1,-3,1)
          + 240 \* \Sss(1,-2,-2)
  \nonumber\\&& \mbox{}
          + 355 \* \Sss(1,-2,1)
          - 1200 \* \Sss(1,1,-3)
          + 355 \* \Sss(1,1,-2)
          - 1650 \* \Sss(1,1,2)
          - 495 \* \Sss(1,1,3)
          + 1650 \* \Sss(1,2,1)
          + 495 \* \Sss(1,3,1)
  \nonumber\\&& \mbox{}
          - 2400 \* \Sss(2,-2,1)
          + 2400 \* \Ssss(1,1,-2,1)
          - 7920 \* \S(1) \* \z3
          - 2190 \* \S(2) \* \z3
          - 1869 \* \S(3)
          + 965 \* \S(4)
          + 240 \* \S(5))
  \nonumber\\&& \mbox{}
          - {4 \over 225} \* (\Nminusthree - \Nminustwo) \* (120 \* \Ss(1,-4)
          + 965 \* \Ss(1,-3)
          - 1429 \* \Ss(1,-2)
          + 2190 \* \Ss(1,1) \* \z3 
          - 120 \* \Ss(1,4)
          + 1320 \* \Ss(2,-2)
  \nonumber\\&& \mbox{}
          - 1200 \* \Sss(1,-3,1)
          + 240 \* \Sss(1,-2,-2)
          + 355 \* \Sss(1,-2,1)
          - 1200 \* \Sss(1,1,-3)
          + 355 \* \Sss(1,1,-2)
          - 495 \* \Sss(1,1,3)
          + 495 \* \Sss(1,3,1)
  \nonumber\\&& \mbox{}
          + 2400 \* \Ssss(1,1,-2,1)
          + 1980 \* \S(1) \* \z3)
          + {4 \over 25} \* (\Nplustwo - 3) \* (1200 \* \Ss(1,-3)
          - 115 \* \Ss(1,-2)
          + 310 \* \Ss(1,1)
          - 4700 \* \Ss(1,1) \* \z3
  \nonumber\\&& \mbox{}
          + 1650 \* \Ss(1,2)
          + 495 \* \Ss(1,3)
          + 2810 \* \Ss(2,-2)
          - 310 \* \Ss(2,1)
          - 2350 \* \Ss(2,3)
          - 2345 \* \Ss(3,1)
          + 2350 \* \Ss(4,1)
          - 3304 \* \S(1)
  \nonumber\\&& \mbox{}
          + 650 \* \Sss(1,-2,1)
          - 3050 \* \Sss(1,1,-2)
          + 2350 \* \Sss(1,1,3)
          - 2350 \* \Sss(1,3,1)
          + 860 \* \S(1) \* \z3
          + 2994 \* \S(2)
          + 4700 \* \S(2) \* \z3
  \nonumber\\&& \mbox{}
          - 2660 \* \S(3))
          - {4 \over 225} \* (\Nminustwo - \Nminus) \* (1200 \* \Ss(1,-3)
          - 115 \* \Ss(1,-2)
          - 1340 \* \Ss(1,1)
          + 495 \* \Ss(1,3)
          + 240 \* \Ss(2,-2)
  \nonumber\\&& \mbox{}
          - 1340 \* \Ss(2,1)
          - 1695 \* \Ss(3,1)
          - 2400 \* \Sss(1,-2,1)
          - 2864 \* \S(1)
          - 2190 \* \S(1) \* \z3
          - 1544 \* \S(2)
          + 1205 \* \S(3)
          + 240 \* \S(4))
  \nonumber\\&& \mbox{}
          - {1 \over 2025} \* (\Nminus + 1) \* (523800 \* \Ss(1,-4)
          - 162900 \* \Ss(1,-3)
          - 812340 \* \Ss(1,-2)
          + 2387235 \* \Ss(1,1)
  \nonumber\\&& \mbox{}
          + 1101600 \* \Ss(1,1) \* \z3
          - 1231650 \* \Ss(1,2)
          + 129600 \* \Ss(1,2) \* \z3
          + 408420 \* \Ss(1,3)
          - 523800 \* \Ss(1,4)
  \nonumber\\&& \mbox{}
          - 205200 \* \Ss(2,-4)
          - 756000 \* \Ss(2,-3)
          + 2279520 \* \Ss(2,-2)
          - 2710350 \* \Ss(2,1)
          - 1036800 \* \Ss(2,1) \* \z3
  \nonumber\\&& \mbox{}
          - 349200 \* \Ss(2,2)
          + 788400 \* \Ss(2,3)
          + 172800 \* \Ss(2,4)
          - 16200 \* \Ss(3,-3)
          - 1139400 \* \Ss(3,-2)
  \nonumber\\&& \mbox{}
          - 1411320 \* \Ss(3,1)
          + 221400 \* \Ss(3,2)
          - 113400 \* \Ss(3,3)
          - 32400 \* \Ss(4,-2)
          - 419400 \* \Ss(4,1)
          - 178200 \* \Ss(4,2)
  \nonumber\\&& \mbox{}
          - 432000 \* \Ss(5,1)
          - 86400 \* \Sss(1,-3,1)
          - 766800 \* \Sss(1,-2,-2)
          - 1434600 \* \Sss(1,-2,1)
          - 1412100 \* \Sss(1,1,-3)
  \nonumber\\&& \mbox{}
          + 2700000 \* \Sss(1,1,-2)
          + 713250 \* \Sss(1,1,1)
          - 259200 \* \Sss(1,1,1) \* \z3
          + 12600 \* \Sss(1,1,2)
          - 1031400 \* \Sss(1,1,3)
  \nonumber\\&& \mbox{}
          - 837000 \* \Sss(1,2,-2)
          - 15300 \* \Sss(1,2,1)
          - 129600 \* \Sss(1,2,2)
          + 194400 \* \Sss(1,2,3)
          + 1509300 \* \Sss(1,3,1)
  \nonumber\\&& \mbox{}
          + 324000 \* \Sss(1,4,1)
          + 415800 \* \Sss(2,-3,1)
          - 194400 \* \Sss(2,-2,-2)
          + 2154600 \* \Sss(2,-2,1)
          - 162000 \* \Sss(2,-2,2)
  \nonumber\\&& \mbox{}
          + 988200 \* \Sss(2,1,-3)
          - 2413800 \* \Sss(2,1,-2)
          + 758700 \* \Sss(2,1,1)
          - 18900 \* \Sss(2,1,2)
          + 351000 \* \Sss(2,1,3)
  \nonumber\\&& \mbox{}
          + 270000 \* \Sss(2,2,-2)
          + 27000 \* \Sss(2,2,1)
          + 37800 \* \Sss(2,2,2)
          - 556200 \* \Sss(2,3,1)
          - 626400 \* \Sss(3,-2,1)
  \nonumber\\&& \mbox{}
          + 896400 \* \Sss(3,1,-2)
          - 411300 \* \Sss(3,1,1)
          - 43200 \* \Sss(3,1,2)
          + 43200 \* \Sss(3,2,1)
          + 297000 \* \Sss(4,1,1)
  \nonumber\\&& \mbox{}
          + 172800 \* \Ssss(1,1,-2,1)
          + 696600 \* \Ssss(1,1,1,-2)
          + 118800 \* \Ssss(1,1,1,2)
          - 388800 \* \Ssss(1,1,1,3)
          - 132300 \* \Ssss(1,1,2,1)
  \nonumber\\&& \mbox{}
          + 388800 \* \Ssss(1,1,3,1)
          + 13500 \* \Ssss(1,2,1,1)
          + 216000 \* \Ssss(2,-2,1,1)
          - 1328400 \* \Ssss(2,1,-2,1)
          - 367200 \* \Ssss(2,1,1,-2)
  \nonumber\\&& \mbox{}
          - 21600 \* \Ssss(2,1,1,2)
          + 21600 \* \Ssss(2,1,2,1)
          + 5167396 \* \S(1)
          - 1296000 \* \S(1) \* \z5
          - 2036340 \* \S(1) \* \z3
  \nonumber\\&& \mbox{}
          - 5979747 \* \S(2)
          + 97200 \* \S(2) \* \z4
          - 550800 \* \S(2) \* \z3
          + 6159140 \* \S(3)
          - 583200 \* \S(3) \* \z3
          - 33420 \* \S(4)
  \nonumber\\&& \mbox{}
          + 234000 \* \S(5)
          + 162000 \* \S(6) )
          - 32 \* (2 \* \Nminus + 3) \* (2 \* \Ss(1,-5)
          - 6 \* \Ss(1,-2) \* \z3
          - 2 \* \Ss(1,5)
          - 5 \* \Sss(1,-4,1)
  \nonumber\\&& \mbox{}
          + 2 \* \Sss(1,-3,-2)
          + 4 \* \Sss(1,-2,-3)
          + \Sss(1,-2,3)
          - 2 \* \Sss(1,1,-4)
          + 2 \* \Sss(1,1,4)
          - 4 \* \Sss(1,2,-3)
          - 2 \* \Sss(1,3,-2)
  \nonumber\\&& \mbox{}
          - 8 \* \Ssss(1,-2,-2,1)
          + 8 \* \Ssss(1,1,-3,1)
          - 4 \* \Ssss(1,1,-2,-2)
          + 8 \* \Ssss(1,1,1,-3)
          + 8 \* \Ssss(1,2,-2,1)
          - 16 \* \Sssss(1,1,1,-2,1))
  \nonumber\\&& \mbox{}
          - {1 \over 1944} \* (1909753 + 71280 \* \z5 + 58320 \* \z4
          - 2524176 \* \z3)
          \biggr)\biggr\}
\:\: ,
\eea
\normalsize
where the functions $g_i(N)$ collecting the terms with positive powers
of $N$ have been defined in Eqs.~(\ref{eq:n1fac1})--(\ref{eq:fac1}).
The $\nf$ and $\n2f$ contributions to Eq.~(A.8) were presented already,
in a slightly different notation, in Ref.~\cite{Moch:2002sn}. 
The corresponding third-order gluon coefficient function is 
\small
\bea
&& c^{(3)}_{2,\rm{g}}(N) \:\: = \:\: 
         \delta(N-2) \* \biggl\{
         \colour4colour{\dabcNA} \* \flg11  \*  \biggl(
          - 8
          - {128 \over 3} \* \z5
          + {544 \over 15} \* \z3
          \biggr)
       + \colour4colour{\cf \* \nf^2}  \*  \biggl(
            {23291 \over 4860}
          - {104 \over 405} \* \z3
          \biggr)
  \nonumber\\&& \mbox{}
       + \colour4colour{\cf^2 \* \nf}  \*  \biggr(
            {28403 \over 4860}
          - {160 \over 3} \* \z5
          + {16 \over 3} \* \z4
          + {4148 \over 405} \* \z3
          \biggr)
       + \colour4colour{\ca \* \cf \* \nf}  \*  \biggl(
          - {161284 \over 3645}
          + 40 \* \z5
          - {52 \over 3} \* \z4
  \nonumber\\&& \mbox{}
          + {274 \over 15} \* \z3
          \biggr)
       + \colour4colour{\ca \* \nf^2}  \*  \biggl(
            {130219 \over 21870}
          + {622 \over 405} \* \z3
          \biggr)
       + \colour4colour{\ca^2 \* \nf}  \*  \biggl(
          - {3444493 \over 87480}
          + 12 \* \z5
          + 12 \* \z4
          + {1828 \over 405} \* \z3
          \biggr)\biggr\}
  \nonumber\\&& \mbox{}
       + \theta(N-4) \* \biggl\{ \colour4colour{\dabcNA} \* \flg11  \*  \biggl(
            {1472 \over 45} \* \gfunct1(N)
          - {64 \over 15} \* \gfunct2(N)
          + {64 \over 45} \* \gfunct3(N)
          + {1792 \over 15} \* \Ss(-2,-3)
          + {1024 \over 45} \* \Ss(-2,1)
  \nonumber\\&& \mbox{}
          - {1792 \over 15} \* \Ss(3,-2)
          + {1792 \over 15} \* \Ss(4,1)
          - {3584 \over 15} \* \Sss(-2,-2,1)
          - {32 \over 225} \* \gqg \* (3300 \* \Ss(1,-4)
          - 4014 \* \Ss(1,-3)
          + 4749 \* \Ss(1,-2)
  \nonumber\\&& \mbox{}
          - 7734 \* \Ss(1,1)
          + 35520 \* \Ss(1,1) \* \z3
          + 6825 \* \Ss(1,2)
          - 16200 \* \Ss(1,2) \* \z3
          - 5310 \* \Ss(1,3)
          - 3300 \* \Ss(1,4)
          + 3900 \* \Ss(2,-3)
  \nonumber\\&& \mbox{}
          - 12829 \* \Ss(2,-2)
          + 7854 \* \Ss(2,1)
          - 16200 \* \Ss(2,1) \* \z3
          + 19710 \* \Ss(2,3)
          - 4560 \* \Ss(3,-2)
          + 15889 \* \Ss(3,1)
          - 17850 \* \Ss(4,1)
  \nonumber\\&& \mbox{}
          + 2700 \* \Sss(1,-2,-2)
          - 4801 \* \Sss(1,-2,1)
          - 3900 \* \Sss(1,1,-3)
          + 12829 \* \Sss(1,1,-2)
          - 13650 \* \Sss(1,1,1)
          + 32400 \* \Sss(1,1,1) \* \z3
  \nonumber\\&& \mbox{}
          - 19710 \* \Sss(1,1,3)
          - 3900 \* \Sss(1,2,-2)
          + 8100 \* \Sss(1,2,3)
          + 23610 \* \Sss(1,3,1)
          - 8100 \* \Sss(1,4,1)
          - 5760 \* \Sss(2,-2,1)
  \nonumber\\&& \mbox{}
          - 2040 \* \Sss(2,1,-2)
          + 8100 \* \Sss(2,1,3)
          - 8100 \* \Sss(2,3,1)
          + 7800 \* \Ssss(1,1,1,-2)
          - 16200 \* \Ssss(1,1,1,3)
          + 16200 \* \Ssss(1,1,3,1)
  \nonumber\\&& \mbox{}
          - 2436 \* \S(1)
          + 40500 \* \S(1) \* \z5
          + 2480 \* \S(1) \* \z3
          + 6780 \* \S(2)
          - 41280 \* \S(2) \* \z3
          - 2394 \* \S(3)
          - 1764 \* \S(4))
  \nonumber\\&& \mbox{}
          + {896 \over 15} \* \S(-5)
          - {512 \over 45} \* \S(-3)
          - {512 \over 45} \* \S(-2)
          - {1792 \over 15} \* \S(-2) \* \z3
          - {896 \over 15} \* \S(5)
          - {128 \over 25} \* (\Nplusthree - 1) \* (98 \* \Ss(1,-3)
  \nonumber\\&& \mbox{}
          - 45 \* \Ss(1,-2)
          - 840 \* \Ss(1,1) \* \z3 
          + 98 \* \Ss(2,-2)
          - 420 \* \Ss(2,3)
          - 98 \* \Ss(3,1)
          + 420 \* \Ss(4,1)
          - 98 \* \Sss(1,-2,1)
          - 98 \* \Sss(1,1,-2)
  \nonumber\\&& \mbox{}
          + 420 \* \Sss(1,1,3)
          - 420 \* \Sss(1,3,1)
          + 840 \* \S(2) \* \z3
          - 45 \* \S(3)
          + 98 \* \S(4))
          - {32 \over 225} \* (\Nminusthree - \Nminustwo) \* (
            98 \* \Ss(1,-3)
  \nonumber\\&& \mbox{}
          - 45 \* \Ss(1,-2)
          - 840 \* \Ss(1,1) \* \z3
          - 98 \* \Sss(1,-2,1)
          - 98 \* \Sss(1,1,-2)
          + 420 \* \Sss(1,1,3)
          - 420 \* \Sss(1,3,1))
  \nonumber\\&& \mbox{}
          - {32 \over 225} \* (2 \* \Nplus + \Nminus - 3) \* (300 \* \Ss(1,-4)
          - 768 \* \Ss(1,-3)
          - 3334 \* \Ss(1,-2)
          + 1642 \* \Ss(1,1)
          + 30840 \* \Ss(1,1) \* \z3
  \nonumber\\&& \mbox{}
          - 975 \* \Ss(1,2)
          + 1800 \* \Ss(1,2) \* \z3
          - 3690 \* \Ss(1,3)
          - 300 \* \Ss(1,4)
          + 6060 \* \Ss(2,-3)
          - 12733 \* \Ss(2,-2)
          + 1307 \* \Ss(2,1)
  \nonumber\\&& \mbox{}
          + 1800 \* \Ss(2,1) \* \z3
          + 23370 \* \Ss(2,3)
          - 4200 \* \Ss(3,-2)
          + 11023 \* \Ss(3,1)
          - 23670 \* \Ss(4,1)
          + 900 \* \Sss(1,-2,-2)
  \nonumber\\&& \mbox{}
          + 2603 \* \Sss(1,-2,1)
          + 300 \* \Sss(1,1,-3)
          - 1067 \* \Sss(1,1,-2)
          + 1950 \* \Sss(1,1,1)
          - 3600 \* \Sss(1,1,1) \* \z3
          - 15270 \* \Sss(1,1,3)
  \nonumber\\&& \mbox{}
          + 300 \* \Sss(1,2,-2)
          - 900 \* \Sss(1,2,3)
          + 14970 \* \Sss(1,3,1)
          + 900 \* \Sss(1,4,1)
          - 12120 \* \Sss(2,-2,1)
          - 900 \* \Sss(2,1,3)
  \nonumber\\&& \mbox{}
          + 900 \* \Sss(2,3,1)
          - 600 \* \Ssss(1,1,1,-2)
          + 1800 \* \Ssss(1,1,1,3)
          - 1800 \* \Ssss(1,1,3,1)
          - 318 \* \S(1)
          - 4500 \* \S(1) \* \z5
  \nonumber\\&& \mbox{}
          + 8540 \* \S(1) \* \z3
          + 5432 \* \S(2)
          - 52800 \* \S(2) \* \z3
          - 8832 \* \S(3)
          - 1668 \* \S(4)
          + 360 \* \S(5) )
  \nonumber\\&& \mbox{}
          - {32 \over 225} \* (\Nminus - 1) \* (1800 \* \Ss(1,-4)
          + 5324 \* \Ss(1,-3)
          - 6666 \* \Ss(1,-2)
          - 7176 \* \Ss(1,1)
          - 12720 \* \Ss(1,1) \* \z3
  \nonumber\\&& \mbox{}
          + 4950 \* \Ss(1,2)
          - 10800 \* \Ss(1,2) \* \z3
          - 3660 \* \Ss(1,3)
          - 1800 \* \Ss(1,4)
          - 5400 \* \Ss(2,-3)
          + 4404 \* \Ss(2,-2)
          + 6379 \* \Ss(2,1)
  \nonumber\\&& \mbox{}
          - 10800 \* \Ss(2,1) \* \z3
          - 3660 \* \Ss(2,3)
          - 360 \* \Ss(3,-2)
          + 4656 \* \Ss(3,1)
          + 5820 \* \Ss(4,1)
          + 1800 \* \Sss(1,-2,-2)
  \nonumber\\&& \mbox{}
          - 13724 \* \Sss(1,-2,1)
          - 1800 \* \Sss(1,1,-3)
          + 3076 \* \Sss(1,1,-2)
          - 9900 \* \Sss(1,1,1)
          + 21600 \* \Sss(1,1,1) \* \z3
          + 5460 \* \Sss(1,1,3)
  \nonumber\\&& \mbox{}
          - 1800 \* \Sss(1,2,-2)
          + 5400 \* \Sss(1,2,3)
          - 3660 \* \Sss(1,3,1)
          - 5400 \* \Sss(1,4,1)
          + 9600 \* \Sss(2,-2,1)
          + 1200 \* \Sss(2,1,-2)
  \nonumber\\&& \mbox{}
          + 5400 \* \Sss(2,1,3)
          - 5400 \* \Sss(2,3,1)
          + 3600 \* \Ssss(1,1,1,-2)
          - 10800 \* \Ssss(1,1,1,3)
          + 10800 \* \Ssss(1,1,3,1)
          - 2305 \* \S(1)
  \nonumber\\&& \mbox{}
          + 27000 \* \S(1) \* \z5
          + 270 \* \S(1) \* \z3
          + 1349 \* \S(2)
          + 11520 \* \S(2) \* \z3
          + 6501 \* \S(3)
          - 96 \* \S(4)
          - 360 \* \S(5) )
  \nonumber\\&& \mbox{}
          - {32 \over 225} \* (\Nminustwo - 1) \* (98 \* \Ss(1,-2)
          + 322 \* \Ss(1,1)
          - 420 \* \Ss(1,3)
          + 322 \* \Ss(2,1)
          + 420 \* \Ss(3,1)
          + 53 \* \S(1)
  \nonumber\\&& \mbox{}
          + 840 \* \S(1) \* \z3
          + 53 \* \S(2)
          + 98 \* \S(3))
          - {64 \over 45} \* (105 \* \z5 - 8 \* \z3)
          \biggr)
  \nonumber\\&& \mbox{}
       + \colour4colour{\cf \* \nf^2}  \*  \biggl(
          - {1 \over 36450} \* \gqg \* (583200 \* \Ss(1,-4)
          + 1697760 \* \Ss(1,-3)
          - 1991088 \* \Ss(1,-2)
          - 7615380 \* \Ss(1,1)
  \nonumber\\&& \mbox{}
          - 907200 \* \Ss(1,1) \* \z3
          + 6013800 \* \Ss(1,2)
          - 3607200 \* \Ss(1,3)
          + 907200 \* \Ss(1,4)
          + 2527200 \* \Ss(2,-3)
  \nonumber\\&& \mbox{}
          - 589680 \* \Ss(2,-2)
          + 3734280 \* \Ss(2,1)
          - 3758400 \* \Ss(2,2)
          + 2332800 \* \Ss(3,-2)
          - 1030320 \* \Ss(3,1)
  \nonumber\\&& \mbox{}
          - 194400 \* \Ss(3,2)
          - 972000 \* \Ss(4,1)
          - 194400 \* \Sss(1,-3,1)
          - 388800 \* \Sss(1,-2,-2)
          - 771120 \* \Sss(1,-2,1)
  \nonumber\\&& \mbox{}
          - 1360800 \* \Sss(1,1,-3)
          + 460080 \* \Sss(1,1,-2)
          - 5261400 \* \Sss(1,1,1)
          + 3888000 \* \Sss(1,1,2)
          - 388800 \* \Sss(1,1,3)
  \nonumber\\&& \mbox{}
          - 777600 \* \Sss(1,2,-2)
          + 2656800 \* \Sss(1,2,1)
          - 972000 \* \Sss(1,2,2)
          - 1166400 \* \Sss(1,3,1)
          - 777600 \* \Sss(2,-2,1)
  \nonumber\\&& \mbox{}
          - 1166400 \* \Sss(2,1,-2)
          + 2667600 \* \Sss(2,1,1)
          - 194400 \* \Sss(2,1,2)
          - 194400 \* \Sss(2,2,1)
          + 194400 \* \Sss(3,1,1)
  \nonumber\\&& \mbox{}
          + 388800 \* \Ssss(1,1,-2,1)
          + 777600 \* \Ssss(1,1,1,-2)
          - 2797200 \* \Ssss(1,1,1,1)
          + 583200 \* \Ssss(1,1,1,2)
          + 583200 \* \Ssss(1,1,2,1)
  \nonumber\\&& \mbox{}
          + 972000 \* \Ssss(1,2,1,1)
          + 162000 \* \Ssss(2,1,1,1)
          - 550800 \* \Sssss(1,1,1,1,1)
          - 16071037 \* \S(1)
          + 10856160 \* \S(1)\* \z3
  \nonumber\\&& \mbox{}
          + 9442692 \* \S(2)
          + 2332800 \* \S(2) \* \z3
          + 2095992 \* \S(3)
          - 570240 \* \S(4)
          + 1879200 \* \S(5))
  \nonumber\\&& \mbox{}
          - {32 \over 75} \* (\Nplusthree - 1) \* (90 \* \Ss(1,-3)
          - 107 \* \Ss(1,-2)
          + 30 \* \Ss(2,-2)
          + 75 \* \Ss(2,2)
          - 105 \* \Ss(3,1)
          - 30 \* \Sss(1,-2,1)
  \nonumber\\&& \mbox{}
          - 30 \* \Sss(1,1,-2)
          - 75 \* \Sss(1,1,2)
          + 75 \* \Sss(1,2,1)
          - 360 \* \S(1) \* \z3
          - 107 \* \S(3)
          + 90 \* \S(4))
  \nonumber\\&& \mbox{}
          - {8 \over 675} \* (\Nminusthree - \Nminustwo) \* (90 \* \Ss(1,-3)
          - 127 \* \Ss(1,-2)
          + 60 \* \Ss(2,-2)
          - 30 \* \Sss(1,-2,1)
          - 30 \* \Sss(1,1,-2)
          + 90 \* \S(1) \* \z3 )
  \nonumber\\&& \mbox{}
          + {1 \over 72900} \* (2 \* \Nplus + \Nminus - 3) \* (
            1166400 \* \Ss(1,-4)
          - 6013440 \* \Ss(1,-3)
          + 12625632 \* \Ss(1,-2)
          - 39675780 \* \Ss(1,1)
  \nonumber\\&& \mbox{}
          - 1166400 \* \Ss(1,1) \* \z3
          + 14299200 \* \Ss(1,2)
          - 1522800 \* \Ss(1,3)
          - 1166400 \* \Ss(1,4)
          - 11275200 \* \Ss(2,-3)
  \nonumber\\&& \mbox{}
          + 10458720 \* \Ss(2,-2)
          + 10993320 \* \Ss(2,1)
          + 3823200 \* \Ss(2,2)
          - 6998400 \* \Ss(3,-2)
          - 13131720 \* \Ss(3,1)
  \nonumber\\&& \mbox{}
          + 6220800 \* \Ss(3,2)
          - 1166400 \* \Ss(3,3)
          + 18079200 \* \Ss(4,1)
          - 3499200 \* \Ss(4,2)
          - 6415200 \* \Ss(5,1)
  \nonumber\\&& \mbox{}
          - 388800 \* \Sss(1,-3,1)
          - 777600 \* \Sss(1,-2,-2)
          + 1594080 \* \Sss(1,-2,1)
          - 2721600 \* \Sss(1,1,-3)
          + 4056480 \* \Sss(1,1,-2)
  \nonumber\\&& \mbox{}
          - 14477400 \* \Sss(1,1,1)
          - 939600 \* \Sss(1,1,2)
          + 388800 \* \Sss(1,1,3)
          - 1555200 \* \Sss(1,2,-2)
          + 6447600 \* \Sss(1,2,1)
  \nonumber\\&& \mbox{}
          - 388800 \* \Sss(1,2,2)
          + 777600 \* \Sss(1,3,1)
          + 3888000 \* \Sss(2,-2,1)
          + 3110400 \* \Sss(2,1,-2)
          - 129600 \* \Sss(2,1,1)
  \nonumber\\&& \mbox{}
          - 388800 \* \Sss(2,1,2)
          - 388800 \* \Sss(2,2,1)
          - 6220800 \* \Sss(3,1,1)
          + 1166400 \* \Sss(3,1,2)
          + 1166400 \* \Sss(3,2,1)
  \nonumber\\&& \mbox{}
          + 3499200 \* \Sss(4,1,1)
          + 777600 \* \Ssss(1,1,-2,1)
          + 1555200 \* \Ssss(1,1,1,-2)
          - 2754000 \* \Ssss(1,1,1,1)
  \nonumber\\&& \mbox{}
          + 388800 \* \Ssss(1,1,1,2)
          - 388800 \* \Ssss(1,1,2,1)
          + 388800 \* \Ssss(2,1,1,1)
          - 1166400 \* \Ssss(3,1,1,1)
  \nonumber\\&& \mbox{}
          - 85078177 \* \S(1)
          - 28801440 \* \S(1) \* \z3
          + 32969748 \* \S(2)
          - 10756800 \* \S(2) \* \z3
          - 30867588 \* \S(3)
  \nonumber\\&& \mbox{}
          + 777600 \* \S(3) \* \z3
          + 39260160 \* \S(4)
          - 33048000 \* \S(5)
          + 8748000 \* \S(6))
  \nonumber\\&& \mbox{}
          - {1 \over 18225} \* (\Nminus - 1) \* (3946320 \* \Ss(1,-3)
          - 7632576 \* \Ss(1,-2)
          + 7131915 \* \Ss(1,1)
          - 668250 \* \Ss(1,2)
  \nonumber\\&& \mbox{}
          + 24300 \* \Ss(1,3)
          - 4665600 \* \Ss(2,-3)
          + 7341840 \* \Ss(2,-2)
          - 8095410 \* \Ss(2,1)
          + 4009500 \* \Ss(2,2)
  \nonumber\\&& \mbox{}
          - 437400 \* \Ss(2,3)
          - 3499200 \* \Ss(3,-2)
          + 5128110 \* \Ss(3,1)
          - 145800 \* \Ss(3,2)
          + 1385100 \* \Ss(4,1)
  \nonumber\\&& \mbox{}
          - 1315440 \* \Sss(1,-2,1)
          - 1315440 \* \Sss(1,1,-2)
          + 1001700 \* \Sss(1,1,1)
          - 542700 \* \Sss(1,1,2)
          + 234900 \* \Sss(1,2,1)
  \nonumber\\&& \mbox{}
          + 1555200 \* \Sss(2,-2,1)
          + 1555200 \* \Sss(2,1,-2)
          - 3685500 \* \Sss(2,1,1)
          + 437400 \* \Sss(2,1,2)
          + 437400 \* \Sss(2,2,1)
  \nonumber\\&& \mbox{}
          + 145800 \* \Sss(3,1,1)
          + 218700 \* \Ssss(1,1,1,1)
          - 437400 \* \Ssss(2,1,1,1)
          + 10832806 \* \S(1)
          + 2288520 \* \S(1) \* \z3
  \nonumber\\&& \mbox{}
          - 18838809 \* \S(2)
          - 4374000 \* \S(2) \* \z3
          + 10648989 \* \S(3)
          - 2472930 \* \S(4)
          - 4009500 \* \S(5))
  \nonumber\\&& \mbox{}
          - {8 \over 18225} \* (\Nminustwo - 1) \* (48600 \* \Ss(1,-3)
          + 44010 \* \Ss(1,-2)
          - 78210 \* \Ss(1,1)
          - 43200 \* \Ss(1,2)
          - 16200 \* \Ss(1,3)
  \nonumber\\&& \mbox{}
          - 810 \* \Ss(2,1)
          - 16200 \* \Sss(1,-2,1)
          - 16200 \* \Sss(1,1,-2)
          + 43200 \* \Sss(1,1,1)
          + 16200 \* \Sss(1,1,2)
          + 16200 \* \Sss(1,2,1)
  \nonumber\\&& \mbox{}
          - 16200 \* \Ssss(1,1,1,1)
          + 137536 \* \S(1)
          + 59400 \* \S(1) \* \z3
          - 2619 \* \S(2)
          + 2430 \* \S(3))
          \biggr)
  \nonumber\\&& \mbox{}
       + \colour4colour{\cf^2 \* \nf}  \*  \biggl(
          - {1 \over 75} \* \gqg \* (2000 \* \Ss(1,-5)
          - 4720 \* \Ss(1,-4)
          + 24700 \* \Ss(1,-3)
          - 29788 \* \Ss(1,-2)
          - 120000 \* \Ss(1,-2) \* \z3
  \nonumber\\&& \mbox{}
          + 111885 \* \Ss(1,1)
          + 22320 \* \Ss(1,1) \* \z3
          - 44500 \* \Ss(1,2)
          - 5600 \* \Ss(1,2) \* \z3
          + 19520 \* \Ss(1,3)
          - 41280 \* \Ss(1,4)
  \nonumber\\&& \mbox{}
          + 16000 \* \Ss(1,5)
          + 28800 \* \Ss(2,-4)
          + 9260 \* \Ss(2,-3)
          - 26820 \* \Ss(2,-2)
          - 36110 \* \Ss(2,1)
          + 89600 \* \Ss(2,1) \* \z3
  \nonumber\\&& \mbox{}
          + 40850 \* \Ss(2,2)
          - 9220 \* \Ss(2,3)
          + 11600 \* \Ss(2,4)
          + 58400 \* \Ss(3,-3)
          - 1240 \* \Ss(3,-2)
          + 106650 \* \Ss(3,1)
  \nonumber\\&& \mbox{}
          - 48000 \* \Ss(3,2)
          + 19600 \* \Ss(3,3)
          + 51200 \* \Ss(4,-2)
          - 91240 \* \Ss(4,1)
          + 31600 \* \Ss(4,2)
          + 32800 \* \Ss(5,1)
  \nonumber\\&& \mbox{}
          - 48000 \* \Sss(1,-4,1)
          - 30400 \* \Sss(1,-3,-2)
          - 3360 \* \Sss(1,-3,1)
          - 2400 \* \Sss(1,-3,2)
          - 32000 \* \Sss(1,-2,-3)
  \nonumber\\&& \mbox{}
          + 20960 \* \Sss(1,-2,-2)
          - 71100 \* \Sss(1,-2,1)
          + 3600 \* \Sss(1,-2,2)
          + 48800 \* \Sss(1,-2,3)
          - 28800 \* \Sss(1,1,-4)
  \nonumber\\&& \mbox{}
          - 9260 \* \Sss(1,1,-3)
          + 32580 \* \Sss(1,1,-2)
          + 54350 \* \Sss(1,1,1)
          - 3200 \* \Sss(1,1,1) \* \z3
          - 40850 \* \Sss(1,1,2)
          + 9220 \* \Sss(1,1,3)
  \nonumber\\&& \mbox{}
          - 11600 \* \Sss(1,1,4)
          - 26400 \* \Sss(1,2,-3)
          + 1400 \* \Sss(1,2,-2)
          - 37950 \* \Sss(1,2,1)
          + 44400 \* \Sss(1,2,2)
          - 25600 \* \Sss(1,2,3)
  \nonumber\\&& \mbox{}
          - 20800 \* \Sss(1,3,-2)
          + 73480 \* \Sss(1,3,1)
          - 29200 \* \Sss(1,3,2)
          - 27600 \* \Sss(1,4,1)
          - 7200 \* \Sss(2,-3,1)
          - 16000 \* \Sss(2,-2,-2)
  \nonumber\\&& \mbox{}
          - 15120 \* \Sss(2,-2,1)
          - 4800 \* \Sss(2,-2,2)
          - 24000 \* \Sss(2,1,-3)
          - 3000 \* \Sss(2,1,-2)
          - 45450 \* \Sss(2,1,1)
          + 47500 \* \Sss(2,1,2)
  \nonumber\\&& \mbox{}
          - 77600 \* \Sss(2,1,3)
          - 16000 \* \Sss(2,2,-2)
          + 47300 \* \Sss(2,2,1)
          - 25600 \* \Sss(2,2,2)
          + 30000 \* \Sss(2,3,1)
          - 19200 \* \Sss(3,-2,1)
  \nonumber\\&& \mbox{}
          - 27200 \* \Sss(3,1,-2)
          + 56400 \* \Sss(3,1,1)
          - 30000 \* \Sss(3,1,2)
          - 32400 \* \Sss(3,2,1)
          - 39200 \* \Sss(4,1,1)
          + 2400 \* \Ssss(1,-3,1,1)
  \nonumber\\&& \mbox{}
          + 19200 \* \Ssss(1,-2,-2,1)
          + 9600 \* \Ssss(1,-2,1,-2)
          - 3600 \* \Ssss(1,-2,1,1)
          + 7200 \* \Ssss(1,1,-3,1)
          + 16000 \* \Ssss(1,1,-2,-2)
  \nonumber\\&& \mbox{}
          + 24720 \* \Ssss(1,1,-2,1)
          + 4800 \* \Ssss(1,1,-2,2)
          + 24000 \* \Ssss(1,1,1,-3)
          - 6600 \* \Ssss(1,1,1,-2)
          + 45450 \* \Ssss(1,1,1,1)
  \nonumber\\&& \mbox{}
          - 47500 \* \Ssss(1,1,1,2)
          + 34800 \* \Ssss(1,1,1,3)
          + 16000 \* \Ssss(1,1,2,-2)
          - 47300 \* \Ssss(1,1,2,1)
          + 25600 \* \Ssss(1,1,2,2)
  \nonumber\\&& \mbox{}
          + 12800 \* \Ssss(1,1,3,1)
          + 17600 \* \Ssss(1,2,1,-2)
          - 52800 \* \Ssss(1,2,1,1)
          + 29600 \* \Ssss(1,2,1,2)
          + 32400 \* \Ssss(1,2,2,1)
  \nonumber\\&& \mbox{}
          + 37200 \* \Ssss(1,3,1,1)
          + 4800 \* \Ssss(2,-2,1,1)
          - 9600 \* \Ssss(2,1,-2,1)
          + 16000 \* \Ssss(2,1,1,-2)
          - 51200 \* \Ssss(2,1,1,1)
  \nonumber\\&& \mbox{}
          + 27200 \* \Ssss(2,1,1,2)
          + 28400 \* \Ssss(2,1,2,1)
          + 32800 \* \Ssss(2,2,1,1)
          + 35200 \* \Ssss(3,1,1,1)
          - 4800 \* \Sssss(1,1,-2,1,1)
  \nonumber\\&& \mbox{}
          + 9600 \* \Sssss(1,1,1,-2,1)
          - 16000 \* \Sssss(1,1,1,1,-2)
          + 51200 \* \Sssss(1,1,1,1,1)
          - 26800 \* \Sssss(1,1,1,1,2)
          - 28400 \* \Sssss(1,1,1,2,1)
  \nonumber\\&& \mbox{}
          - 33200 \* \Sssss(1,1,2,1,1)
          - 35200 \* \Sssss(1,2,1,1,1)
          - 30000 \* \Sssss(2,1,1,1,1)
          + 57793 \* \S(1)
          - 240000 \* \S(1) \* \z5
  \nonumber\\&& \mbox{}
          + 14280 \* \S(1) \* \z3
          - 92853 \* \S(2)
          - 12720 \* \S(2) \* \z3
          + 75392 \* \S(3)
          + 39200 \* \S(3) \* \z3
          - 53340 \* \S(4)
  \nonumber\\&& \mbox{}
          + 51760 \* \S(5)
          - 18000 \* \S(6)
          + 30000 \* \Ssssss(1,1,1,1,1,1))
  \nonumber\\&& \mbox{}
          + {16 \over 25} \* (\Nplusthree - 1) \* (120 \* \Ss(1,-4)
          - 460 \* \Ss(1,-3)
          + 533 \* \Ss(1,-2)
          - 1320 \* \Ss(1,1) \* \z3
          - 120 \* \Ss(1,4)
          + 240 \* \Ss(2,-3)
  \nonumber\\&& \mbox{}
          - 220 \* \Ss(2,-2)
          - 780 \* \Ss(2,3)
          + 240 \* \Ss(3,-2)
          + 220 \* \Ss(3,1)
          + 540 \* \Ss(4,1)
          - 240 \* \Sss(1,-3,1)
          + 240 \* \Sss(1,-2,-2)
  \nonumber\\&& \mbox{}
          + 220 \* \Sss(1,-2,1)
          - 240 \* \Sss(1,1,-3)
          + 220 \* \Sss(1,1,-2)
          + 780 \* \Sss(1,1,3)
          - 780 \* \Sss(1,3,1)
          - 480 \* \Sss(2,-2,1)
  \nonumber\\&& \mbox{}
          + 480 \* \Ssss(1,1,-2,1)
          - 360 \* \S(1) \* \z3
          + 1320 \* \S(2) \* \z3
          + 533 \* \S(3)
          - 460 \* \S(4)
          + 240 \* \S(5) )
  \nonumber\\&& \mbox{}
          + {4 \over 225} \* (\Nminusthree - \Nminustwo) \* (120 \* \Ss(1,-4)
          - 460 \* \Ss(1,-3)
          + 463 \* \Ss(1,-2)
          - 1320 \* \Ss(1,1) \* \z3
          - 120 \* \Ss(1,4)
          - 240 \* \Ss(2,-2)
  \nonumber\\&& \mbox{}
          - 240 \* \Sss(1,-3,1)
          + 240 \* \Sss(1,-2,-2)
          + 220 \* \Sss(1,-2,1)
          - 240 \* \Sss(1,1,-3)
          + 220 \* \Sss(1,1,-2)
          + 780 \* \Sss(1,1,3)
  \nonumber\\&& \mbox{}
          - 780 \* \Sss(1,3,1)
          + 480 \* \Ssss(1,1,-2,1)
          - 360 \* \S(1) \* \z3)
          - {1 \over 1800} \* (2 \* \Nplus + \Nminus - 3) \* (
            115200 \* \Ss(1,-5)
  \nonumber\\&& \mbox{}
          + 882240 \* \Ss(1,-4)
          - 4009280 \* \Ss(1,-3)
          + 5220416 \* \Ss(1,-2)
          + 576000 \* \Ss(1,-2) \* \z3
          + 722160 \* \Ss(1,1)
  \nonumber\\&& \mbox{}
          - 1559040 \* \Ss(1,1) \* \z3
          + 27000 \* \Ss(1,2)
          - 230400 \* \Ss(1,2) \* \z3
          - 90480 \* \Ss(1,3)
          - 474240 \* \Ss(1,4)
          - 115200 \* \Ss(1,5)
  \nonumber\\&& \mbox{}
          + 873600 \* \Ss(2,-4)
          - 551520 \* \Ss(2,-3)
          - 1365920 \* \Ss(2,-2)
          + 1421820 \* \Ss(2,1)
          + 3868800 \* \Ss(2,1) \* \z3
  \nonumber\\&& \mbox{}
          - 452400 \* \Ss(2,2)
          + 1295040 \* \Ss(2,3)
          - 465600 \* \Ss(2,4)
          + 1516800 \* \Ss(3,-3)
          - 237120 \* \Ss(3,-2)
  \nonumber\\&& \mbox{}
          + 3551000 \* \Ss(3,1)
          - 588000 \* \Ss(3,2)
          + 28800 \* \Ss(3,3)
          + 1228800 \* \Ss(4,-2)
          - 1738320 \* \Ss(4,1)
  \nonumber\\&& \mbox{}
          + 451200 \* \Ss(4,2)
          + 588000 \* \Ss(5,1)
          + 230400 \* \Sss(1,-4,1)
          + 115200 \* \Sss(1,-3,-2)
          - 353280 \* \Sss(1,-3,1)
  \nonumber\\&& \mbox{}
          + 115200 \* \Sss(1,-2,-3)
          - 126720 \* \Sss(1,-2,-2)
          + 2036960 \* \Sss(1,-2,1)
          - 57600 \* \Sss(1,-2,2)
          - 345600 \* \Sss(1,-2,3)
  \nonumber\\&& \mbox{}
          - 115200 \* \Sss(1,1,-4)
          - 1325280 \* \Sss(1,1,-3)
          + 2500640 \* \Sss(1,1,-2)
          + 45000 \* \Sss(1,1,1)
          + 460800 \* \Sss(1,1,1) \* \z3
  \nonumber\\&& \mbox{}
          + 388800 \* \Sss(1,1,2)
          + 132960 \* \Sss(1,1,3)
          + 115200 \* \Sss(1,1,4)
          - 115200 \* \Sss(1,2,-3)
          - 763200 \* \Sss(1,2,-2)
  \nonumber\\&& \mbox{}
          + 417600 \* \Sss(1,2,1)
          - 211200 \* \Sss(1,2,2)
          + 172800 \* \Sss(1,2,3)
          - 115200 \* \Sss(1,3,-2)
          - 960 \* \Sss(1,3,1)
  \nonumber\\&& \mbox{}
          - 57600 \* \Sss(1,4,1)
          - 115200 \* \Sss(2,-3,1)
          - 268800 \* \Sss(2,-2,-2)
          + 455040 \* \Sss(2,-2,1)
          - 57600 \* \Sss(2,-2,2)
  \nonumber\\&& \mbox{}
          - 1171200 \* \Sss(2,1,-3)
          + 849600 \* \Sss(2,1,-2)
          + 490800 \* \Sss(2,1,1)
          + 211200 \* \Sss(2,1,2)
          - 2808000 \* \Sss(2,1,3)
  \nonumber\\&& \mbox{}
          - 768000 \* \Sss(2,2,-2)
          + 177600 \* \Sss(2,2,1)
          - 211200 \* \Sss(2,2,2)
          + 2827200 \* \Sss(2,3,1)
          - 172800 \* \Sss(3,-2,1)
  \nonumber\\&& \mbox{}
          - 1171200 \* \Sss(3,1,-2)
          + 592800 \* \Sss(3,1,1)
          - 297600 \* \Sss(3,1,2)
          - 326400 \* \Sss(3,2,1)
          - 230400 \* \Ssss(1,-2,-2,1)
  \nonumber\\&& \mbox{}
          - 513600 \* \Sss(4,1,1)
          + 57600 \* \Ssss(1,-2,1,1)
          + 230400 \* \Ssss(1,1,-3,1)
          - 230400 \* \Ssss(1,1,-2,-2)
          + 418560 \* \Ssss(1,1,-2,1)
  \nonumber\\&& \mbox{}
          + 230400 \* \Ssss(1,1,1,-3)
          + 936000 \* \Ssss(1,1,1,-2)
          - 393600 \* \Ssss(1,1,1,1)
          + 199200 \* \Ssss(1,1,1,2)
          - 345600 \* \Ssss(1,1,1,3)
  \nonumber\\&& \mbox{}
          + 160800 \* \Ssss(1,1,2,1)
          + 345600 \* \Ssss(1,1,3,1)
          + 230400 \* \Ssss(1,2,-2,1)
          + 211200 \* \Ssss(1,2,1,1)
          + 57600 \* \Ssss(2,-2,1,1)
  \nonumber\\&& \mbox{}
          - 57600 \* \Ssss(2,1,-2,1)
          + 787200 \* \Ssss(2,1,1,-2)
          - 182400 \* \Ssss(2,1,1,1)
          + 187200 \* \Ssss(2,1,1,2)
          + 177600 \* \Ssss(2,1,2,1)
  \nonumber\\&& \mbox{}
          + 206400 \* \Ssss(2,2,1,1)
          + 321600 \* \Ssss(3,1,1,1)
          - 460800 \* \Sssss(1,1,1,-2,1)
          - 177600 \* \Sssss(1,1,1,1,1)
          - 177600 \* \Sssss(2,1,1,1,1)
  \nonumber\\&& \mbox{}
          + 188829 \* \S(1)
          + 2448000 \* \S(1) \* \z5
          - 129600 \* \S(1) \* \z4
          - 3763200 \* \S(1) \* \z3
          - 1776576 \* \S(2)
          - 43200 \* \S(2) \* \z4
  \nonumber\\&& \mbox{}
          - 2708160 \* \S(2) \* \z3
          + 5801216 \* \S(3)
          + 2016000 \* \S(3) \* \z3
          - 3034640 \* \S(4)
          + 1307280 \* \S(5)
          - 487200 \* \S(6) )
  \nonumber\\&& \mbox{}
          - {1 \over 450} \* (\Nminus - 1) \* (39120 \* \Ss(1,-4)
          - 572920 \* \Ss(1,-3)
          + 761688 \* \Ss(1,-2)
          - 345600 \* \Ss(1,-2) \* \z3
          + 825910 \* \Ss(1,1)
  \nonumber\\&& \mbox{}
          + 1050480 \* \Ss(1,1) \* \z3
          - 237900 \* \Ss(1,2)
          + 101520 \* \Ss(1,3)
          - 142920 \* \Ss(1,4)
          - 134400 \* \Ss(2,-4)
          + 882240 \* \Ss(2,-3)
  \nonumber\\&& \mbox{}
          - 990160 \* \Ss(2,-2)
          - 612015 \* \Ss(2,1)
          - 748800 \* \Ss(2,1) \* \z3
          + 288600 \* \Ss(2,2)
          - 400080 \* \Ss(2,3)
          + 134400 \* \Ss(2,4)
  \nonumber\\&& \mbox{}
          - 312000 \* \Ss(3,-3)
          + 575040 \* \Ss(3,-2)
          - 288860 \* \Ss(3,1)
          - 61200 \* \Ss(3,2)
          + 76800 \* \Ss(3,3)
          - 240000 \* \Ss(4,-2)
  \nonumber\\&& \mbox{}
          - 70260 \* \Ss(4,1)
          - 14400 \* \Ss(4,2)
          - 28800 \* \Ss(5,1)
          - 172800 \* \Sss(1,-4,1)
          + 111360 \* \Sss(1,-3,1)
          - 212160 \* \Sss(1,-2,-2)
  \nonumber\\&& \mbox{}
          + 1840 \* \Sss(1,-2,1)
          + 7200 \* \Sss(1,-2,2)
          + 172800 \* \Sss(1,-2,3)
          - 167040 \* \Sss(1,1,-3)
          + 663520 \* \Sss(1,1,-2)
  \nonumber\\&& \mbox{}
          + 306600 \* \Sss(1,1,1)
          - 125400 \* \Sss(1,1,2)
          - 699720 \* \Sss(1,1,3)
          - 182400 \* \Sss(1,2,-2)
          - 120000 \* \Sss(1,2,1)
  \nonumber\\&& \mbox{}
          + 98400 \* \Sss(1,2,2)
          + 1009320 \* \Sss(1,3,1)
          - 57600 \* \Sss(2,-3,1)
          + 144000 \* \Sss(2,-2,-2)
          - 485280 \* \Sss(2,-2,1)
  \nonumber\\&& \mbox{}
          + 249600 \* \Sss(2,1,-3)
          - 712800 \* \Sss(2,1,-2)
          - 301200 \* \Sss(2,1,1)
          + 87600 \* \Sss(2,1,2)
          + 595200 \* \Sss(2,1,3)
  \nonumber\\&& \mbox{}
          + 211200 \* \Sss(2,2,-2)
          + 103200 \* \Sss(2,2,1)
          + 4800 \* \Sss(2,2,2)
          - 724800 \* \Sss(2,3,1)
          + 28800 \* \Sss(3,-2,1)
          + 78600 \* \Sss(3,1,1)
  \nonumber\\&& \mbox{}
          + 307200 \* \Sss(3,1,-2)
          + 14400 \* \Sss(4,1,1)
          - 7200 \* \Ssss(1,-2,1,1)
          - 186720 \* \Ssss(1,1,-2,1)
          + 194400 \* \Ssss(1,1,1,-2)
  \nonumber\\&& \mbox{}
          + 136200 \* \Ssss(1,1,1,1)
          - 107400 \* \Ssss(1,1,1,2)
          - 108600 \* \Ssss(1,1,2,1)
          - 121800 \* \Ssss(1,2,1,1)
          + 115200 \* \Ssss(2,1,-2,1)
  \nonumber\\&& \mbox{}
          - 230400 \* \Ssss(2,1,1,-2)
          - 97800 \* \Ssss(2,1,1,1)
          - 4800 \* \Ssss(2,1,1,2)
          + 4800 \* \Ssss(2,1,2,1)
          + 113400 \* \Sssss(1,1,1,1,1)
  \nonumber\\&& \mbox{}
          + 417102 \* \S(1)
          - 864000 \* \S(1) \* \z5
          - 16200 \* \S(1) \* \z4
          - 389040 \* \S(1) \* \z3
          - 136298 \* \S(2)
          + 21600 \* \S(2) \* \z4
  \nonumber\\&& \mbox{}
          + 1245120 \* \S(2) \* \z3
          - 1043312 \* \S(3)
          - 518400 \* \S(3) \* \z3
          + 467240 \* \S(4)
          - 9660 \* \S(5)
          + 40800 \* \S(6) )
  \nonumber\\&& \mbox{}
          + {4 \over 225} \* (\Nminustwo - 1) \* (240 \* \Ss(1,-3)
          + 20 \* \Ss(1,-2)
          + 760 \* \Ss(1,1)
          - 780 \* \Ss(1,3)
          + 240 \* \Ss(2,-2)
          + 760 \* \Ss(2,1)
  \nonumber\\&& \mbox{}
          + 540 \* \Ss(3,1)
          - 480 \* \Sss(1,-2,1)
          + 723 \* \S(1)
          + 1320 \* \S(1) \* \z3
          + 483 \* \S(2)
          - 220 \* \S(3)
          + 240 \* \S(4) )
          \biggr)
  \nonumber\\&& \mbox{}
       + \colour4colour{\ca \* \nf^2}  \*  \biggl(
            {4 \over 15} \* \Ss(1,-2) \* (\Nminusthree - \Nminustwo)
          - {1 \over 7290} \* \gqg \* (149040 \* \Ss(1,-4)
          - 238680 \* \Ss(1,-3)
          + 261612 \* \Ss(1,-2)
  \nonumber\\&& \mbox{}
          - 2855640 \* \Ss(1,1)
          + 142560 \* \Ss(1,1) \* \z3
          + 1031400 \* \Ss(1,2)
          - 328320 \* \Ss(1,3)
          + 77760 \* \Ss(1,4)
          + 97200 \* \Ss(2,-3)
  \nonumber\\&& \mbox{}
          - 110160 \* \Ss(2,-2)
          + 1312200 \* \Ss(2,1)
          - 340200 \* \Ss(2,2)
          + 58320 \* \Ss(2,3)
          + 38880 \* \Ss(3,-2)
          - 385560 \* \Ss(3,1)
  \nonumber\\&& \mbox{}
          + 38880 \* \Ss(3,2)
          + 38880 \* \Ss(4,1)
          - 97200 \* \Sss(1,-3,1)
          + 38880 \* \Sss(1,-2,-2)
          + 110160 \* \Sss(1,-2,1)
  \nonumber\\&& \mbox{}
          - 38880 \* \Sss(1,-2,2)
          + 19440 \* \Sss(1,1,-3)
          - 12960 \* \Sss(1,1,-2)
          - 1052640 \* \Sss(1,1,1)
          + 448200 \* \Sss(1,1,2)
  \nonumber\\&& \mbox{}
          - 129600 \* \Sss(1,1,3)
          + 38880 \* \Sss(1,2,-2)
          + 115560 \* \Sss(1,2,1)
          - 51840 \* \Sss(1,2,2)
          + 32400 \* \Sss(1,3,1)
          - 38880 \* \Sss(2,-2,1)
  \nonumber\\&& \mbox{}
          + 340200 \* \Sss(2,1,1)
          - 77760 \* \Sss(2,1,2)
          - 38880 \* \Sss(3,1,1)
          + 38880 \* \Ssss(1,-2,1,1)
          - 38880 \* \Ssss(1,1,1,-2)
  \nonumber\\&& \mbox{}
          - 266760 \* \Ssss(1,1,1,1)
          + 64800 \* \Ssss(1,1,1,2)
          + 51840 \* \Ssss(1,1,2,1)
          - 6480 \* \Ssss(1,2,1,1)
          + 38880 \* \Ssss(2,1,1,1)
  \nonumber\\&& \mbox{}
          - 45360 \* \Sssss(1,1,1,1,1)
          - 4708987 \* \S(1)
          + 193320 \* \S(1) \* \z3
          + 2569392 \* \S(2)
          - 880668 \* \S(3)
          + 253800 \* \S(4) )
  \nonumber\\&& \mbox{}
          + {48 \over 5} \* (\Nplusthree - 1) \* (\Ss(1,-2)
          + \S(3))
          - {1 \over 7290} \* (2 \* \Nplus + \Nminus - 3) \* (58320 \* \Ss(1,-4)
          - 255960 \* \Ss(1,-3)
  \nonumber\\&& \mbox{}
          + 416124 \* \Ss(1,-2)
          - 138240 \* \Ss(1,1)
          - 58320 \* \Ss(1,1) \* \z3
          + 63720 \* \Ss(1,2)
          + 38880 \* \Ss(1,3)
          - 58320 \* \Ss(1,4)
  \nonumber\\&& \mbox{}
          + 136080 \* \Ss(2,-3)
          - 165240 \* \Ss(2,-2)
          - 748980 \* \Ss(2,1)
          + 289980 \* \Ss(2,2)
          - 97200 \* \Ss(2,3)
          + 38880 \* \Ss(3,-2)
  \nonumber\\&& \mbox{}
          + 691740 \* \Ss(3,1)
          - 155520 \* \Ss(3,2)
          - 252720 \* \Ss(4,1)
          - 19440 \* \Sss(1,-3,1)
          - 38880 \* \Sss(1,-2,-2)
  \nonumber\\&& \mbox{}
          + 64800 \* \Sss(1,-2,1)
          - 136080 \* \Sss(1,1,-3)
          + 187920 \* \Sss(1,1,-2)
          - 70200 \* \Sss(1,1,1)
          + 22680 \* \Sss(1,1,2)
  \nonumber\\&& \mbox{}
          + 19440 \* \Sss(1,1,3)
          - 77760 \* \Sss(1,2,-2)
          + 22680 \* \Sss(1,2,1)
          - 19440 \* \Sss(1,2,2)
          + 38880 \* \Sss(1,3,1)
          - 38880 \* \Sss(2,-2,1)
  \nonumber\\&& \mbox{}
          - 77760 \* \Sss(2,1,-2)
          - 298080 \* \Sss(2,1,1)
          + 77760 \* \Sss(2,1,2)
          + 77760 \* \Sss(2,2,1)
          + 155520 \* \Sss(3,1,1)
  \nonumber\\&& \mbox{}
          + 38880 \* \Ssss(1,1,-2,1)
          + 77760 \* \Ssss(1,1,1,-2)
          - 22680 \* \Ssss(1,1,1,1)
          + 19440 \* \Ssss(1,1,1,2)
          - 19440 \* \Ssss(1,1,2,1)
  \nonumber\\&& \mbox{}
          - 77760 \* \Ssss(2,1,1,1)
          - 385579 \* \S(1)
          - 422280 \* \S(1) \* \z3
          - 1521924 \* \S(2)
          + 226800 \* \S(2) \* \z3 
          + 2088144 \* \S(3)
  \nonumber\\&& \mbox{}
          - 1134000 \* \S(4)
          + 330480 \* \S(5) )
          + {1 \over 3645} \* (\Nminus - 1) \* (191160 \* \Ss(1,-3)
          - 209304 \* \Ss(1,-2)
          + 1356570 \* \Ss(1,1)
  \nonumber\\&& \mbox{}
          - 419040 \* \Ss(1,2)
          + 77760 \* \Ss(1,3)
          + 108540 \* \Ss(2,-2)
          - 990090 \* \Ss(2,1)
          + 305370 \* \Ss(2,2)
          - 58320 \* \Ss(2,3)
  \nonumber\\&& \mbox{}
          - 38880 \* \Ss(3,-2)
          + 549990 \* \Ss(3,1)
          - 97200 \* \Ss(3,2)
          - 136080 \* \Ss(4,1)
          - 61560 \* \Sss(1,-2,1)
          - 61560 \* \Sss(1,1,-2)
  \nonumber\\&& \mbox{}
          + 419580 \* \Sss(1,1,1)
          - 116640 \* \Sss(1,1,2)
          - 32400 \* \Sss(1,2,1)
          - 309420 \* \Sss(2,1,1)
          + 58320 \* \Sss(2,1,2)
          + 58320 \* \Sss(2,2,1)
  \nonumber\\&& \mbox{}
          + 97200 \* \Sss(3,1,1)
          + 77760 \* \Ssss(1,1,1,1)
          - 58320 \* \Ssss(2,1,1,1)
          + 2392019 \* \S(1)
          + 73440 \* \S(1) \* \z3
          - 2221866 \* \S(2)
  \nonumber\\&& \mbox{}
          + 38880 \* \S(2) \* \z3
          + 1533366 \* \S(3)
          - 765720 \* \S(4)
          + 168480 \* \S(5) )
          - {4 \over 3645} \* (\Nminustwo - 1) \* (
            9720 \* \Ss(1,-3)
  \nonumber\\&& \mbox{}
          - 5400 \* \Ss(1,-2)
          + 2700 \* \Ss(1,1)
          + 5400 \* \Ss(1,2)
          - 3240 \* \Ss(1,3)
          - 3240 \* \Sss(1,-2,1)
          - 3240 \* \Sss(1,1,-2)
          - 5400 \* \Sss(1,1,1)
  \nonumber\\&& \mbox{}
          + 3240 \* \Sss(1,1,2)
          + 3240 \* \Sss(1,2,1)
          - 3240 \* \Ssss(1,1,1,1)
          - 958 \* \S(1)
          + 11880 \* \S(1) \* \z3
          - 243 \* \S(2) )
          \biggr)
  \nonumber\\&& \mbox{}
       + \colour4colour{\ca \* \cf \* \nf}  \*  \biggl(
          - {1 \over 2025} \* \gqg \* (729000 \* \Ss(1,-5)
          - 1751220 \* \Ss(1,-4)
          + 498564 \* \Ss(1,-3)
          + 188460 \* \Ss(1,-2)
  \nonumber\\&& \mbox{}
          + 3110400 \* \Ss(1,-2) \* \z3
          - 4195196 \* \Ss(1,1)
          + 48600 \* \Ss(1,1) \* \z4
          + 853020 \* \Ss(1,1) \* \z3
          + 453215 \* \Ss(1,2)
  \nonumber\\&& \mbox{}
          - 151200 \* \Ss(1,2) \* \z3
          + 770520 \* \Ss(1,3)
          + 283140 \* \Ss(1,4)
          - 243000 \* \Ss(1,5)
          + 891000 \* \Ss(2,-4)
  \nonumber\\&& \mbox{}
          - 3669300 \* \Ss(2,-3)
          + 1667214 \* \Ss(2,-2)
          + 522961 \* \Ss(2,1)
          - 2905200 \* \Ss(2,1) \* \z3
          + 1796685 \* \Ss(2,2)
  \nonumber\\&& \mbox{}
          - 1797120 \* \Ss(2,3)
          + 91800 \* \Ss(2,4)
          + 313200 \* \Ss(3,-3)
          - 2411640 \* \Ss(3,-2)
          - 98379 \* \Ss(3,1)
  \nonumber\\&& \mbox{}
          - 2084040 \* \Ss(3,2)
          + 399600 \* \Ss(3,3)
          - 604800 \* \Ss(4,-2)
          - 1309680 \* \Ss(4,1)
          + 162000 \* \Ss(4,2)
  \nonumber\\&& \mbox{}
          + 54000 \* \Ss(5,1)
          + 280800 \* \Sss(1,-4,1)
          + 464400 \* \Sss(1,-3,-2)
          + 1665900 \* \Sss(1,-3,1)
          - 356400 \* \Sss(1,-3,2)
  \nonumber\\&& \mbox{}
          + 421200 \* \Sss(1,-2,-3)
          - 211680 \* \Sss(1,-2,-2)
          + 1016766 \* \Sss(1,-2,1)
          + 326160 \* \Sss(1,-2,2)
          - 1544400 \* \Sss(1,-2,3)
  \nonumber\\&& \mbox{}
          - 1117800 \* \Sss(1,1,-4)
          + 3528900 \* \Sss(1,1,-3)
          - 1949094 \* \Sss(1,1,-2)
          - 1423215 \* \Sss(1,1,1)
          + 691200 \* \Sss(1,1,1) \* \z3
  \nonumber\\&& \mbox{}
          - 977925 \* \Sss(1,1,2)
          + 1377720 \* \Sss(1,1,3)
          + 124200 \* \Sss(1,1,4)
          - 1236600 \* \Sss(1,2,-3)
          + 1616760 \* \Sss(1,2,-2)
  \nonumber\\&& \mbox{}
          - 557175 \* \Sss(1,2,1)
          + 1307700 \* \Sss(1,2,2)
          - 172800 \* \Sss(1,2,3)
          - 216000 \* \Sss(1,3,-2)
          + 181080 \* \Sss(1,3,1)
  \nonumber\\&& \mbox{}
          - 118800 \* \Sss(1,3,2)
          + 351000 \* \Sss(1,4,1)
          - 1776600 \* \Sss(2,-3,1)
          + 540000 \* \Sss(2,-2,-2)
          + 2230200 \* \Sss(2,-2,1)
  \nonumber\\&& \mbox{}
          - 378000 \* \Sss(2,-2,2)
          - 1744200 \* \Sss(2,1,-3)
          + 2334960 \* \Sss(2,1,-2)
          - 1205310 \* \Sss(2,1,1)
          + 1671300 \* \Sss(2,1,2)
  \nonumber\\&& \mbox{}
          + 1339200 \* \Sss(2,1,3)
          - 410400 \* \Sss(2,2,-2)
          + 1906200 \* \Sss(2,2,1)
          - 329400 \* \Sss(2,2,2)
          - 1701000 \* \Sss(2,3,1)
  \nonumber\\&& \mbox{}
          - 820800 \* \Sss(3,-2,1)
          - 540000 \* \Sss(3,1,-2)
          + 2285640 \* \Sss(3,1,1)
          + 421200 \* \Ssss(1,-3,1,1)
          - 475200 \* \Ssss(1,-2,-2,1)
  \nonumber\\&& \mbox{}
          - 378000 \* \Sss(3,1,2)
          - 270000 \* \Sss(3,2,1)
          + 237600 \* \Ssss(1,-2,1,-2)
          - 358560 \* \Ssss(1,-2,1,1)
          + 1873800 \* \Ssss(1,1,-3,1)
  \nonumber\\&& \mbox{}
          - 183600 \* \Sss(4,1,1)
          + 162000 \* \Ssss(1,-2,1,2)
          + 183600 \* \Ssss(1,-2,2,1)
          - 561600 \* \Ssss(1,1,-2,-2)
          - 2489400 \* \Ssss(1,1,-2,1)
  \nonumber\\&& \mbox{}
          + 378000 \* \Ssss(1,1,-2,2)
          + 2116800 \* \Ssss(1,1,1,-3)
          - 1794960 \* \Ssss(1,1,1,-2)
          + 400050 \* \Ssss(1,1,1,1)
          - 108000 \* \Ssss(1,1,1,3)
  \nonumber\\&& \mbox{}
          - 1083600 \* \Ssss(1,1,1,2)
          + 529200 \* \Ssss(1,1,2,-2)
          - 1075500 \* \Ssss(1,1,2,1)
          + 372600 \* \Ssss(1,1,2,2)
          + 453600 \* \Ssss(1,1,3,1)
  \nonumber\\&& \mbox{}
          + 1490400 \* \Ssss(1,2,-2,1)
          + 486000 \* \Ssss(1,2,1,-2)
          - 1047600 \* \Ssss(1,2,1,1)
          + 259200 \* \Ssss(1,2,1,2)
          + 232200 \* \Ssss(1,2,2,1)
  \nonumber\\&& \mbox{}
          - 135000 \* \Ssss(1,3,1,1)
          + 453600 \* \Ssss(2,-2,1,1)
          + 2300400 \* \Ssss(2,1,-2,1)
          + 648000 \* \Ssss(2,1,1,-2)
          - 1665000 \* \Ssss(2,1,1,1)
  \nonumber\\&& \mbox{}
          + 329400 \* \Ssss(2,1,1,2)
          + 313200 \* \Ssss(2,1,2,1)
          + 178200 \* \Ssss(2,2,1,1)
          + 248400 \* \Ssss(3,1,1,1)
          - 172800 \* \Sssss(1,-2,1,1,1)
  \nonumber\\&& \mbox{}
          - 453600 \* \Sssss(1,1,-2,1,1)
          - 2494800 \* \Sssss(1,1,1,-2,1)
          - 572400 \* \Sssss(1,1,1,1,-2)
          + 766800 \* \Sssss(1,1,1,1,1)
  \nonumber\\&& \mbox{}
          - 275400 \* \Sssss(1,1,1,1,2)
          - 210600 \* \Sssss(1,1,1,2,1)
          - 64800 \* \Sssss(1,1,2,1,1)
          - 162000 \* \Sssss(2,1,1,1,1)
          - 4855533 \* \S(1)
  \nonumber\\&& \mbox{}
          + 6318000 \* \S(1) \* \z5
          + 251100 \* \S(1) \* \z4 )
          + {1 \over 2025} \* \gqg \* (1837260 \* \S(1) \* \z3
          - 2771054 \* \S(2)
          - 48600 \* \S(2) \* \z4
  \nonumber\\&& \mbox{}
          + 1182420 \* \S(2) \* \z3
          + 1256337 \* \S(3)
          + 1069200 \* \S(3) \* \z3
          + 811116 \* \S(4)
          - 1266840 \* \S(5))
  \nonumber\\&& \mbox{}
          - {16 \over 75} \* (\Nplusthree - 1) \* (1620 \* \Ss(1,-4)
          - 4104 \* \Ss(1,-3)
          + 3671 \* \Ss(1,-2)
          - 5220 \* \Ss(1,1) \* \z3 
          - 540 \* \Ss(1,4)
  \nonumber\\&& \mbox{}
          + 2700 \* \Ss(2,-3)
          - 2424 \* \Ss(2,-2)
          - 825 \* \Ss(2,2)
          - 2880 \* \Ss(2,3)
          + 1440 \* \Ss(3,-2)
          + 3249 \* \Ss(3,1)
          - 360 \* \Ss(3,2)
  \nonumber\\&& \mbox{}
          + 180 \* \Ss(4,1)
          - 2700 \* \Sss(1,-3,1)
          + 1080 \* \Sss(1,-2,-2)
          + 2424 \* \Sss(1,-2,1)
          - 360 \* \Sss(1,-2,2)
          - 2700 \* \Sss(1,1,-3)
  \nonumber\\&& \mbox{}
          + 2424 \* \Sss(1,1,-2)
          + 825 \* \Sss(1,1,2)
          + 2880 \* \Sss(1,1,3)
          - 360 \* \Sss(1,2,-2)
          - 825 \* \Sss(1,2,1)
          - 2880 \* \Sss(1,3,1)
  \nonumber\\&& \mbox{}
          - 3600 \* \Sss(2,-2,1)
          - 360 \* \Sss(2,1,-2)
          + 360 \* \Sss(3,1,1)
          + 360 \* \Ssss(1,-2,1,1)
          + 3600 \* \Ssss(1,1,-2,1)
          + 360 \* \Ssss(1,1,1,-2)
  \nonumber\\&& \mbox{}
          + 2430 \* \S(1) \* \z3
          + 5220 \* \S(2) \* \z3
          + 3671 \* \S(3)
          - 4104 \* \S(4)
          + 2160 \* \S(5))
  \nonumber\\&& \mbox{}
          - {4 \over 675} \* (\Nminusthree - \Nminustwo) \* (1620 \* \Ss(1,-4)
          - 3594 \* \Ss(1,-3)
          + 3481 \* \Ss(1,-2)
          - 5220 \* \Ss(1,1) \* \z3
          - 540 \* \Ss(1,4)
  \nonumber\\&& \mbox{}
          + 720 \* \Ss(2,-3)
          - 1680 \* \Ss(2,-2)
          - 2700 \* \Sss(1,-3,1)
          + 1080 \* \Sss(1,-2,-2)
          + 1914 \* \Sss(1,-2,1)
          - 360 \* \Sss(1,-2,2)
  \nonumber\\&& \mbox{}
          - 2700 \* \Sss(1,1,-3)
          + 1914 \* \Sss(1,1,-2)
          + 2880 \* \Sss(1,1,3)
          - 360 \* \Sss(1,2,-2)
          - 2880 \* \Sss(1,3,1)
          - 720 \* \Sss(2,-2,1)
  \nonumber\\&& \mbox{}
          - 720 \* \Sss(2,1,-2)
          + 360 \* \Ssss(1,-2,1,1)
          + 3600 \* \Ssss(1,1,-2,1)
          + 360 \* \Ssss(1,1,1,-2)
          - 2520 \* \S(1) \* \z3)
  \nonumber\\&& \mbox{}
          + {1 \over 48600} \* (2 \* \Nplus + \Nminus - 3) \* (
            9331200 \* \Ss(1,-5)
          + 14398560 \* \Ss(1,-4)
          - 121401072 \* \Ss(1,-3)
  \nonumber\\&& \mbox{}
          + 163294992 \* \Ss(1,-2)
          + 12441600 \* \Ss(1,-2) \* \z3
          + 24916708 \* \Ss(1,1)
          - 42953760 \* \Ss(1,1) \* \z3
          - 2294520 \* \Ss(1,2)
  \nonumber\\&& \mbox{}
          - 15552000 \* \Ss(1,2) \* \z3
          + 3566880 \* \Ss(1,3)
          - 13849920 \* \Ss(1,4)
          - 9331200 \* \Ss(1,5)
          + 16977600 \* \Ss(2,-4)
  \nonumber\\&& \mbox{}
          + 32983200 \* \Ss(2,-3)
          - 58795632 \* \Ss(2,-2)
          + 57317208 \* \Ss(2,1)
          + 92145600 \* \Ss(2,1) \* \z3
          - 53764920 \* \Ss(2,2)
  \nonumber\\&& \mbox{}
          + 66108960 \* \Ss(2,3)
          - 27475200 \* \Ss(2,4)
          + 5184000 \* \Ss(3,-3)
          + 49403520 \* \Ss(3,-2)
          + 48005352 \* \Ss(3,1)
  \nonumber\\&& \mbox{}
          + 28321920 \* \Ss(3,2)
          - 2332800 \* \Ss(3,3)
          + 15552000 \* \Ss(4,-2)
          + 22645440 \* \Ss(4,1)
          + 14644800 \* \Ss(4,2)
  \nonumber\\&& \mbox{}
          + 15552000 \* \Ss(5,1)
          - 3110400 \* \Sss(1,-4,1)
          + 3110400 \* \Sss(1,-3,-2)
          + 6609600 \* \Sss(1,-3,1)
          + 60653232 \* \Sss(1,-2,1)
  \nonumber\\&& \mbox{}
          - 1555200 \* \Sss(1,-3,2)
          + 3110400 \* \Sss(1,-2,-3)
          - 4095360 \* \Sss(1,-2,-2)
          + 1710720 \* \Sss(1,-2,2)
          - 9331200 \* \Sss(1,-2,3)
  \nonumber\\&& \mbox{}
          - 20217600 \* \Sss(1,1,-4)
          - 16588800 \* \Sss(1,1,-3)
          + 70062192 \* \Sss(1,1,-2)
          + 393720 \* \Sss(1,1,1)
          + 26438400 \* \Sss(1,1,1) \* \z3
  \nonumber\\&& \mbox{}
          + 17433000 \* \Sss(1,1,2)
          + 1499040 \* \Sss(1,1,3)
          + 10886400 \* \Sss(1,1,4)
          - 21772800 \* \Sss(1,2,-3)
          - 15266880 \* \Sss(1,2,-2)
  \nonumber\\&& \mbox{}
          - 5117400 \* \Sss(1,2,1)
          - 1965600 \* \Sss(1,2,2)
          + 6220800 \* \Sss(1,2,3)
          - 9331200 \* \Sss(1,3,-2)
          + 4570560 \* \Sss(1,3,1)
  \nonumber\\&& \mbox{}
          + 1555200 \* \Sss(1,3,2)
          + 6220800 \* \Sss(1,4,1)
          + 24753600 \* \Sss(2,-3,1)
          - 11923200 \* \Sss(2,-2,-2)
          + 1684800 \* \Sss(2,-2,1)
  \nonumber\\&& \mbox{}
          + 6998400 \* \Sss(2,-2,2)
          - 32270400 \* \Sss(2,1,-3)
          - 3991680 \* \Sss(2,1,-2)
          + 40875120 \* \Sss(2,1,1)
          - 15163200 \* \Sss(2,1,2)
  \nonumber\\&& \mbox{}
          - 84110400 \* \Sss(2,1,3)
          - 32659200 \* \Sss(2,2,-2)
          - 17949600 \* \Sss(2,2,1)
          - 11275200 \* \Sss(2,2,2)
          + 86443200 \* \Sss(2,3,1)
  \nonumber\\&& \mbox{}
          + 32400000 \* \Sss(3,-2,1)
          - 21513600 \* \Sss(3,1,-2)
          - 34974720 \* \Sss(3,1,1)
          - 10238400 \* \Sss(3,1,2)
          - 14644800 \* \Sss(3,2,1)
  \nonumber\\&& \mbox{}
          - 16718400 \* \Sss(4,1,1)
          + 1555200 \* \Ssss(1,-3,1,1)
          - 9331200 \* \Ssss(1,-2,-2,1)
          + 3110400 \* \Ssss(1,-2,1,-2)
  \nonumber\\&& \mbox{}
          - 2877120 \* \Ssss(1,-2,1,1)
          + 26438400 \* \Ssss(1,1,-3,1)
          - 9331200 \* \Ssss(1,1,-2,-2)
          - 12571200 \* \Ssss(1,1,-2,1)
  \nonumber\\&& \mbox{}
          + 3110400 \* \Ssss(1,1,-2,2)
          + 35769600 \* \Ssss(1,1,1,-3)
          + 18636480 \* \Ssss(1,1,1,-2)
          - 7318800 \* \Ssss(1,1,1,1)
  \nonumber\\&& \mbox{}
          + 2548800 \* \Ssss(1,1,1,2)
          - 10886400 \* \Ssss(1,1,1,3)
          + 9331200 \* \Ssss(1,1,2,-2)
          + 3067200 \* \Ssss(1,1,2,1)
  \nonumber\\&& \mbox{}
          + 7776000 \* \Ssss(1,1,3,1)
          + 21772800 \* \Ssss(1,2,-2,1)
          + 9331200 \* \Ssss(1,2,1,-2)
          + 3758400 \* \Ssss(1,2,1,1)
  \nonumber\\&& \mbox{}
          - 1555200 \* \Ssss(1,3,1,1)
          - 8294400 \* \Ssss(2,-2,1,1)
          - 23587200 \* \Ssss(2,1,-2,1)
          + 36288000 \* \Ssss(2,1,1,-2)
  \nonumber\\&& \mbox{}
          + 16351200 \* \Ssss(2,1,1,1)
          + 11275200 \* \Ssss(2,1,1,2)
          + 10497600 \* \Ssss(2,1,2,1)
          + 12571200 \* \Ssss(2,2,1,1)
  \nonumber\\&& \mbox{}
          + 13867200 \* \Ssss(3,1,1,1)
          - 3110400 \* \Sssss(1,1,-2,1,1)
          - 37324800 \* \Sssss(1,1,1,-2,1)
          - 9331200 \* \Sssss(1,1,1,1,-2)
  \nonumber\\&& \mbox{}
          - 3607200 \* \Sssss(1,1,1,1,1)
          - 12052800 \* \Sssss(2,1,1,1,1)
          + 29287105 \* \S(1)
          + 56376000 \* \S(1) \* \z5
  \nonumber\\&& \mbox{}
          - 7192800 \* \S(1) \* \z4
          - 51576480 \* \S(1) \* \z3
          - 3072980 \* \S(2)
          - 6998400 \* \S(2) \* \z4
          - 76321440 \* \S(2) \* \z3
  \nonumber\\&& \mbox{}
          + 130853940 \* \S(3)
          + 67780800 \* \S(3) \* \z3
          - 35726472 \* \S(4)
          - 24239520 \* \S(5)
          - 9460800 \* \S(6) )
  \nonumber\\&& \mbox{}
          - {1 \over 12150} \* (\Nminus - 1) \* (3800520 \* \Ss(1,-4)
          + 5176116 \* \Ss(1,-3)
          - 12721788 \* \Ss(1,-2)
          + 9331200 \* \Ss(1,-2) \* \z3
  \nonumber\\&& \mbox{}
          - 30019849 \* \Ss(1,1)
          - 29547720 \* \Ss(1,1) \* \z3
          + 6683910 \* \Ss(1,2)
          - 1406880 \* \Ss(1,3)
          + 1978560 \* \Ss(1,4)
  \nonumber\\&& \mbox{}
          + 3628800 \* \Ss(2,-4)
          - 17690400 \* \Ss(2,-3)
          + 23125176 \* \Ss(2,-2)
          + 22469436 \* \Ss(2,1)
          + 17107200 \* \Ss(2,1) \* \z3
  \nonumber\\&& \mbox{}
          - 2932740 \* \Ss(2,2)
          + 10021320 \* \Ss(2,3)
          - 7711200 \* \Ss(2,4)
          + 6998400 \* \Ss(3,-3)
          - 12357360 \* \Ss(3,-2)
  \nonumber\\&& \mbox{}
          + 11758284 \* \Ss(3,1)
          - 4435560 \* \Ss(3,2)
          + 583200 \* \Ss(3,3)
          + 9072000 \* \Ss(4,-2)
          - 1046520 \* \Ss(4,1)
  \nonumber\\&& \mbox{}
          + 6544800 \* \Ss(4,2)
          + 6868800 \* \Ss(5,1)
          + 4665600 \* \Sss(1,-4,1)
          - 16216200 \* \Sss(1,-3,1)
          + 10018080 \* \Sss(1,-2,-2)
  \nonumber\\&& \mbox{}
          + 7736904 \* \Sss(1,-2,1)
          - 2507760 \* \Sss(1,-2,2)
          - 4665600 \* \Sss(1,-2,3)
          - 2835000 \* \Sss(1,1,-3)
          - 14151456 \* \Sss(1,1,-2)
  \nonumber\\&& \mbox{}
          - 10583610 \* \Sss(1,1,1)
          - 1254600 \* \Sss(1,1,2)
          + 25726680 \* \Sss(1,1,3)
          + 6920640 \* \Sss(1,2,-2)
          - 4632300 \* \Sss(1,2,1)
  \nonumber\\&& \mbox{}
          + 2727000 \* \Sss(1,2,2)
          - 26903880 \* \Sss(1,3,1)
          + 11793600 \* \Sss(2,-3,1)
          - 5832000 \* \Sss(2,-2,-2)
          + 10238400 \* \Sss(2,-2,1)
  \nonumber\\&& \mbox{}
          + 2592000 \* \Sss(2,-2,2)
          - 7646400 \* \Sss(2,1,-3)
          + 13854240 \* \Sss(2,1,-2)
          + 2646540 \* \Sss(2,1,1)
          + 4179600 \* \Sss(2,1,2)
  \nonumber\\&& \mbox{}
          - 22809600 \* \Sss(2,1,3)
          - 10627200 \* \Sss(2,2,-2)
          + 5119200 \* \Sss(2,2,1)
          + 22096800 \* \Sss(2,3,1)
          + 6220800 \* \Sss(3,-2,1)
  \nonumber\\&& \mbox{}
          - 4471200 \* \Sss(2,2,2)
          - 9331200 \* \Sss(3,1,-2)
          + 3528360 \* \Sss(3,1,1)
          - 5119200 \* \Sss(3,1,2)
          + 20055600 \* \Ssss(1,1,-2,1)
  \nonumber\\&& \mbox{}
          - 6415200 \* \Sss(3,2,1)
          - 7322400 \* \Sss(4,1,1)
          + 2604960 \* \Ssss(1,-2,1,1)
          - 8346240 \* \Ssss(1,1,1,-2)
          - 2851200 \* \Ssss(2,-2,1,1)
  \nonumber\\&& \mbox{}
          + 1837800 \* \Ssss(1,1,1,1)
          - 2241000 \* \Ssss(1,1,1,2)
          - 2046600 \* \Ssss(1,1,2,1)
          - 1765800 \* \Ssss(1,2,1,1)
          - 12960000 \* \Ssss(2,1,-2,1)
  \nonumber\\&& \mbox{}
          - 4131000 \* \Ssss(2,1,1,1)
          + 4471200 \* \Ssss(2,1,1,2)
          + 4082400 \* \Ssss(2,1,2,1)
          + 4665600 \* \Ssss(2,2,1,1)
          + 5961600 \* \Ssss(3,1,1,1)
  \nonumber\\&& \mbox{}
          + 11145600 \* \Ssss(2,1,1,-2)
          + 1290600 \* \Sssss(1,1,1,1,1)
          - 4471200 \* \Sssss(2,1,1,1,1)
          - 36057811 \* \S(1)
          + 23328000 \* \S(1) \* \z5
  \nonumber\\&& \mbox{}
          + 1652400 \* \S(1) \* \z4
          + 21648600 \* \S(1) \* \z3
          + 20322285 \* \S(2)
          - 2624400 \* \S(2) \* \z4
          - 42693480 \* \S(2) \* \z3
  \nonumber\\&& \mbox{}
          + 23324085 \* \S(3)
          + 23328000 \* \S(3) \* \z3
          - 15582564 \* \S(4)
          - 667440 \* \S(5)
          - 3823200 \* \S(6) )
  \nonumber\\&& \mbox{}
          + {2 \over 6075} \* (\Nminustwo - 1) \* (259200 \* \Ss(1,-4)
          + 189000 \* \Ss(1,-3)
          - 27108 \* \Ss(1,-2)
          - 33487 \* \Ss(1,1)
  \nonumber\\&& \mbox{}
          - 216000 \* \Ss(1,1) \* \z3
          + 45780 \* \Ss(1,2)
          - 57960 \* \Ss(1,3)
          - 140400 \* \Ss(1,4)
          - 25920 \* \Ss(2,-2)
          - 50652 \* \Ss(2,1)
  \nonumber\\&& \mbox{}
          + 6480 \* \Ss(2,2)
          - 3240 \* \Ss(3,1)
          - 129600 \* \Sss(1,-3,1)
          - 32400 \* \Sss(1,-2,1)
          - 518400 \* \Sss(1,1,-3)
          - 371520 \* \Sss(1,1,-2)
  \nonumber\\&& \mbox{}
          - 17430 \* \Sss(1,1,1)
          - 39600 \* \Sss(1,1,2)
          - 216000 \* \Sss(1,1,3)
          - 259200 \* \Sss(1,2,-2)
          - 136800 \* \Sss(1,2,1)
  \nonumber\\&& \mbox{}
          - 237600 \* \Sss(1,2,2)
          - 86400 \* \Sss(1,3,1)
          - 6480 \* \Sss(2,1,1)
          + 259200 \* \Ssss(1,1,-2,1)
          + 259200 \* \Ssss(1,1,1,-2)
  \nonumber\\&& \mbox{}
          + 88200 \* \Ssss(1,1,1,1)
          + 237600 \* \Ssss(1,1,1,2)
          + 237600 \* \Ssss(1,1,2,1)
          - 41166 \* \S(2)
          + 259200 \* \Ssss(1,2,1,1)
  \nonumber\\&& \mbox{}
          - 248400 \* \Sssss(1,1,1,1,1)
          - 181071 \* \S(1)
          - 97200 \* \S(1) \* \z4
          + 316440 \* \S(1) \* \z3
          + 38772 \* \S(3)
          - 38880 \* \S(4) )
          \biggr)
  \nonumber\\&& \mbox{}
       + \colour4colour{\ca^2 \* \nf}  \*  \biggl(
          - {1 \over 14580} \* \gqg \* (1321920 \* \Ss(1,-5)
          - 10885104 \* \Ss(1,-4)
          + 7457508 \* \Ss(1,-3)
          - 9110196 \* \Ss(1,-2)
  \nonumber\\&& \mbox{}
          - 5365440 \* \Ss(1,-2) \* \z3
          + 77840592 \* \Ss(1,1)
          - 349920 \* \Ss(1,1) \* \z4
          - 11577168 \* \Ss(1,1) \* \z3
          - 39153960 \* \Ss(1,2)
  \nonumber\\&& \mbox{}
          + 2177280 \* \Ss(1,2) \* \z3
          + 27245700 \* \Ss(1,3)
          - 11017296 \* \Ss(1,4)
          + 2177280 \* \Ss(1,5)
          + 1244160 \* \Ss(2,-4)
  \nonumber\\&& \mbox{}
          - 10854648 \* \Ss(2,-3)
          + 1207332 \* \Ss(2,-2)
          - 44635932 \* \Ss(2,1)
          + 6065280 \* \Ss(2,1) \* \z3
          + 35764740 \* \Ss(2,2)
  \nonumber\\&& \mbox{}
          - 8861400 \* \Ss(2,3)
          + 855360 \* \Ss(2,4)
          + 1088640 \* \Ss(3,-3)
          - 6289488 \* \Ss(3,-2)
          + 47422368 \* \Ss(3,1)
  \nonumber\\&& \mbox{}
          - 9447840 \* \Ss(3,2)
          + 233280 \* \Ss(3,3)
          + 622080 \* \Ss(4,-2)
          - 11070432 \* \Ss(4,1)
          + 1166400 \* \Ss(4,2)
  \nonumber\\&& \mbox{}
          + 1399680 \* \Ss(5,1)
          - 4471200 \* \Sss(1,-4,1)
          - 1477440 \* \Sss(1,-3,-2)
          + 10073808 \* \Sss(1,-3,1)
          - 1866240 \* \Sss(1,-3,2)
  \nonumber\\&& \mbox{}
          - 1166400 \* \Sss(1,-2,-3)
          + 903312 \* \Sss(1,-2,-2)
          - 6627636 \* \Sss(1,-2,1)
          + 4237920 \* \Sss(1,-2,2)
          + 1866240 \* \Sss(1,-2,3)
  \nonumber\\&& \mbox{}
          - 1632960 \* \Sss(1,1,-4)
          + 2871288 \* \Sss(1,1,-3)
          + 1341900 \* \Sss(1,1,-2)
          + 43750620 \* \Sss(1,1,1)
          - 4354560 \* \Sss(1,1,1) \* \z3
  \nonumber\\&& \mbox{}
          - 29059020 \* \Sss(1,1,2)
          + 11433960 \* \Sss(1,1,3)
          - 2954880 \* \Sss(1,1,4)
          + 894240 \* \Sss(1,2,-3)
          - 1218240 \* \Sss(1,2,-2)
  \nonumber\\&& \mbox{}
          - 26818020 \* \Sss(1,2,1)
          + 9259920 \* \Sss(1,2,2)
          - 1788480 \* \Sss(1,2,3)
          - 466560 \* \Sss(1,3,-2)
          + 10925280 \* \Sss(1,3,1)
  \nonumber\\&& \mbox{}
          - 1944000 \* \Sss(1,3,2)
          - 3188160 \* \Sss(1,4,1)
          - 1905120 \* \Sss(2,-3,1)
          - 1477440 \* \Sss(2,-2,-2)
          + 6345216 \* \Sss(2,-2,1)
  \nonumber\\&& \mbox{}
          - 1010880 \* \Sss(2,-2,2)
          - 894240 \* \Sss(2,1,-3)
          + 2935440 \* \Sss(2,1,-2)
          - 37847520 \* \Sss(2,1,1)
          + 12214800 \* \Sss(2,1,2)
  \nonumber\\&& \mbox{}
          - 4510080 \* \Sss(2,1,3)
          - 77760 \* \Sss(2,2,-2)
          + 9438120 \* \Sss(2,2,1)
          - 1283040 \* \Sss(2,2,2)
          + 1127520 \* \Sss(2,3,1)
  \nonumber\\&& \mbox{}
          - 622080 \* \Sss(3,-2,1)
          - 1555200 \* \Sss(3,1,-2)
          + 10967400 \* \Sss(3,1,1)
          - 1244160 \* \Sss(3,1,2)
          + 2177280 \* \Ssss(1,-3,1,1)
  \nonumber\\&& \mbox{}
          - 1555200 \* \Sss(3,2,1)
          - 2099520 \* \Sss(4,1,1)
          + 544320 \* \Ssss(1,-2,-2,1)
          + 544320 \* \Ssss(1,-2,1,-2)
          - 4652640 \* \Ssss(1,-2,1,1)
  \nonumber\\&& \mbox{}
          + 544320 \* \Ssss(1,-2,1,2)
          + 388800 \* \Ssss(1,-2,2,1)
          + 894240 \* \Ssss(1,1,-3,1)
          + 1166400 \* \Ssss(1,1,-2,-2)
  \nonumber\\&& \mbox{}
          - 1783296 \* \Ssss(1,1,-2,1)
          + 1632960 \* \Ssss(1,1,-2,2)
          - 2877120 \* \Ssss(1,1,1,-3)
          + 2766960 \* \Ssss(1,1,1,-2)
  \nonumber\\&& \mbox{}
          + 26653320 \* \Ssss(1,1,1,1)
          - 9434880 \* \Ssss(1,1,1,2)
          + 2877120\* \Ssss(1,1,1,3)
          - 1088640 \* \Ssss(1,1,2,-2)
          - 9185400 \* \Ssss(1,1,2,1)
  \nonumber\\&& \mbox{}
          + 1671840 \* \Ssss(1,1,2,2)
          + 1710720 \* \Ssss(1,1,3,1)
          - 1166400 \* \Ssss(1,2,-2,1)
          - 1088640 \* \Ssss(1,2,1,-2)
          - 8472600 \* \Ssss(1,2,1,1)
  \nonumber\\&& \mbox{}
          + 1710720 \* \Ssss(1,2,1,2)
          + 1360800 \* \Ssss(1,2,2,1)
          + 2604960 \* \Ssss(1,3,1,1)
          + 1244160 \* \Ssss(2,-2,1,1)
          - 388800 \* \Ssss(2,1,-2,1)
  \nonumber\\&& \mbox{}
          - 466560 \* \Ssss(2,1,1,-2)
          - 8754480 \* \Ssss(2,1,1,1)
          + 1671840 \* \Ssss(2,1,1,2)
          + 1555200 \* \Ssss(2,1,2,1)
          + 1671840 \* \Ssss(2,2,1,1)
  \nonumber\\&& \mbox{}
          + 1399680 \* \Ssss(3,1,1,1)
          - 155520 \* \Sssss(1,-2,1,1,1)
          - 1710720 \* \Sssss(1,1,-2,1,1)
          + 3499200 \* \Sssss(1,1,1,-2,1)
  \nonumber\\&& \mbox{}
          + 1632960 \* \Sssss(1,1,1,1,-2)
          + 7464960 \* \Sssss(1,1,1,1,1)
          - 1205280 \* \Sssss(1,1,1,1,2)
          - 1360800 \* \Sssss(1,1,1,2,1)
  \nonumber\\&& \mbox{}
          - 1477440 \* \Sssss(1,1,2,1,1)
          - 1555200 \* \Sssss(1,2,1,1,1)
          - 1399680 \* \Sssss(2,1,1,1,1)
          + 79819747\* \S(1)
          - 11664000 \* \S(1) \* \z5 )
  \nonumber\\&& \mbox{}
          + {1 \over 1215} \* \gqg \* (150660 \* \S(1) \* \z4 
          + 2064699 \* \S(1) \* \z3 
          + 6629957 \* \S(2)
          + 29160 \* \S(2) \* \z4
          - 314604 \* \S(2) \* \z3
  \nonumber\\&& \mbox{}
          - 3076389 \* \S(3)
          - 174960\* \S(3) \* \z3
          + 3423240 \* \S(4)
          - 265356 \* \S(5)
          - 97200 \* \Ssssss(1,1,1,1,1,1) )
  \nonumber\\&& \mbox{}
          + {12 \over 5} \* (\Nplusthree - 1) \* (16 \* \Ss(1,-4)
          + 10 \* \Ss(1,-3)
          - 19 \* \Ss(1,-2)
          - 128 \* \Ss(1,1) \* \z3 
          - 16 \* \Ss(1,4)
          + 32 \* \Ss(2,-3)
          + 10 \* \Ss(2,-2)
  \nonumber\\&& \mbox{}
          - 80 \* \Ss(2,3)
          + 32 \* \Ss(3,-2)
          - 10 \* \Ss(3,1)
          + 48 \* \Ss(4,1)
          - 32 \* \Sss(1,-3,1)
          + 32 \* \Sss(1,-2,-2)
          - 10 \* \Sss(1,-2,1)
          - 32 \* \Sss(1,1,-3)
  \nonumber\\&& \mbox{}
          - 10 \* \Sss(1,1,-2)
          + 80 \* \Sss(1,1,3)
          - 80 \* \Sss(1,3,1)
          - 64 \* \Sss(2,-2,1)
          + 64 \* \Ssss(1,1,-2,1)
          + 128 \* \S(2) \* \z3
          - 19 \* \S(3)
          + 10 \* \S(4)
          + 32 \* \S(5) )
  \nonumber\\&& \mbox{}
          + {1 \over 15} \* (\Nminusthree - \Nminustwo) \* (16 \* \Ss(1,-4)
          + 10 \* \Ss(1,-3)
          - 19 \* \Ss(1,-2)
          - 128 \* \Ss(1,1) \* \z3
          - 16 \* \Ss(1,4)
  \nonumber\\&& \mbox{}
          - 32 \* \Sss(1,-3,1)
          + 32 \* \Sss(1,-2,-2)
          - 10 \* \Sss(1,-2,1)
          - 32 \* \Sss(1,1,-3)
          - 10 \* \Sss(1,1,-2)
          + 80 \* \Sss(1,1,3)
  \nonumber\\&& \mbox{}
          - 80 \* \Sss(1,3,1)
          + 64 \* \Ssss(1,1,-2,1) )
          - {1 \over 14580} \* (2 \* \Nplus + \Nminus - 3) \* (
            1166400 \* \Ss(1,-5)
          - 3114288 \* \Ss(1,-4)
  \nonumber\\&& \mbox{}
          - 411588 \* \Ss(1,-3)
          + 3880008 \* \Ss(1,-2)
          + 699840 \* \Ss(1,-2) \* \z3
          - 1755828 \* \Ss(1,1)
          - 2742336 \* \Ss(1,1) \* \z3
  \nonumber\\&& \mbox{}
          + 362700 \* \Ss(1,2)
          - 1866240 \* \Ss(1,2) \* \z3
          + 2451060 \* \Ss(1,3)
          - 2588112 \* \Ss(1,4)
          - 1166400 \* \Ss(1,5)
  \nonumber\\&& \mbox{}
          - 1905120 \* \Ss(2,-4)
          - 11859696 \* \Ss(2,-3)
          - 2980044 \* \Ss(2,-2)
          + 23221008 \* \Ss(2,1)
          + 1321920 \* \Ss(2,1) \* \z3
  \nonumber\\&& \mbox{}
          + 698220 \* \Ss(2,2)
          + 13134960 \* \Ss(2,3)
          - 7931520 \* \Ss(2,4)
          + 583200 \* \Ss(3,-3)
          - 5729616 \* \Ss(3,-2)
  \nonumber\\&& \mbox{}
          + 3579984 \* \Ss(3,1)
          + 14725800 \* \Ss(3,2)
          - 8378640 \* \Ss(3,3)
          - 349920 \* \Ss(4,-2)
          + 21194136 \* \Ss(4,1)
  \nonumber\\&& \mbox{}
          - 7989840 \* \Ss(4,2)
          - 9136800 \* \Ss(5,1)
          - 933120 \* \Sss(1,-4,1)
          + 233280 \* \Sss(1,-3,-2)
          + 4935816 \* \Sss(1,-3,1)
  \nonumber\\&& \mbox{}
          - 233280 \* \Sss(1,-3,2)
          + 233280 \* \Sss(1,-2,-3)
          - 364176 \* \Sss(1,-2,-2)
          + 425196 \* \Sss(1,-2,1)
          + 874800 \* \Sss(1,-2,2)
  \nonumber\\&& \mbox{}
          - 699840 \* \Sss(1,-2,3)
          - 2799360 \* \Sss(1,1,-4)
          + 3720816 \* \Sss(1,1,-3)
          + 1789020 \* \Sss(1,1,-2)
          - 1959120 \* \Sss(1,1,1)
  \nonumber\\&& \mbox{}
          + 3032640 \* \Sss(1,1,1) \* \z3
          - 2002860 \* \Sss(1,1,2)
          + 1367280 \* \Sss(1,1,3)
          + 1399680 \* \Sss(1,1,4)
          - 3032640 \* \Sss(1,2,-3)
  \nonumber\\&& \mbox{}
          - 71280 \* \Sss(1,2,-2)
          - 896940 \* \Sss(1,2,1)
          + 1292760 \* \Sss(1,2,2)
          + 583200 \* \Sss(1,2,3)
          - 1166400 \* \Sss(1,3,-2)
  \nonumber\\&& \mbox{}
          + 1960200 \* \Sss(1,3,1)
          + 233280 \* \Sss(1,3,2)
          + 1049760 \* \Sss(1,4,1)
          + 2080080 \* \Sss(2,-3,1)
          - 1632960 \* \Sss(2,-2,-2)
  \nonumber\\&& \mbox{}
          + 8782992 \* \Sss(2,-2,1)
          + 1205280 \* \Sss(2,-2,2)
          - 5423760 \* \Sss(2,1,-3)
          + 8223120 \* \Sss(2,1,-2)
          - 273240 \* \Sss(2,1,1)
  \nonumber\\&& \mbox{}
          - 7264080 \* \Sss(2,1,2)
          - 1302480 \* \Sss(2,1,3)
          - 4237920 \* \Sss(2,2,-2)
          - 8641080 \* \Sss(2,2,1)
          + 3751920 \* \Sss(2,2,2)
  \nonumber\\&& \mbox{}
          + 12150000 \* \Sss(2,3,1)
          + 1905120 \* \Sss(3,-2,1)
          - 5248800 \* \Sss(3,1,-2)
          - 16102800 \* \Sss(3,1,1)
          + 7426080 \* \Sss(3,1,2)
  \nonumber\\&& \mbox{}
          + 6570720 \* \Sss(3,2,1)
          + 8339760 \* \Sss(4,1,1)
          + 233280 \* \Ssss(1,-3,1,1)
          - 933120 \* \Ssss(1,-2,-2,1)
          + 466560 \* \Ssss(1,-2,1,-2)
  \nonumber\\&& \mbox{}
          - 803520 \* \Ssss(1,-2,1,1)
          + 3499200 \* \Ssss(1,1,-3,1)
          - 933120 \* \Ssss(1,1,-2,-2)
          + 466560 \* \Ssss(1,1,-2,2)
          - 965520 \* \Ssss(1,1,1,2)
  \nonumber\\&& \mbox{}
          - 4882032 \* \Ssss(1,1,-2,1)
          + 4898880 \* \Ssss(1,1,1,-3)
          + 550800 \* \Ssss(1,1,1,-2)
          + 1074600 \* \Ssss(1,1,1,1)
          - 933120 \* \Ssss(1,1,1,3)
  \nonumber\\&& \mbox{}
          + 1399680 \* \Ssss(1,1,2,-2)
          - 839160 \* \Ssss(1,1,2,1)
          + 466560 \* \Ssss(1,1,3,1)
          + 2799360 \* \Ssss(1,2,-2,1)
          + 1399680 \* \Ssss(1,2,1,-2)
  \nonumber\\&& \mbox{}
          - 826200 \* \Ssss(1,2,1,1)
          - 233280 \* \Ssss(1,3,1,1)
          - 1399680 \* \Ssss(2,-2,1,1)
          + 1360800 \* \Ssss(2,1,-2,1)
          + 5365440 \* \Ssss(2,1,1,-2)
  \nonumber\\&& \mbox{}
          + 8897040 \* \Ssss(2,1,1,1)
          - 3032640 \* \Ssss(2,1,1,2)
          - 3188160 \* \Ssss(2,1,2,1)
          - 2799360 \* \Ssss(2,2,1,1)
          - 6376320 \* \Ssss(3,1,1,1)
  \nonumber\\&& \mbox{}
          - 466560 \* \Sssss(1,1,-2,1,1)
          - 4665600 \* \Sssss(1,1,1,-2,1)
          - 1399680 \* \Sssss(1,1,1,1,-2)
          + 589680 \* \Sssss(1,1,1,1,1)
  \nonumber\\&& \mbox{}
          + 2021760 \* \Sssss(2,1,1,1,1)
          - 4564193 \* \S(1)
          + 3499200 \* \S(1) \* \z5
          - 1108080 \* \S(1) \* \z4
          - 7680852 \* \S(1) \* \z3
  \nonumber\\&& \mbox{}
          - 1749600 \* \S(2) \* \z4
          + 365064 \* \S(2)
          - 23462784 \* \S(2) \* \z3
          - 21096612 \* \S(3)
          + 13141440 \* \S(3) \* \z3
          - 778896 \* \S(4)
  \nonumber\\&& \mbox{}
          - 23530176 \* \S(5)
          + 7698240 \* \S(6))
          + {1 \over 7290} \* (\Nminus - 1) \* (4931928 \* \Ss(1,-4)
          - 7855164 \* \Ss(1,-3)
  \nonumber\\&& \mbox{}
          - 4812048 \* \Ss(1,-2)
          + 1399680 \* \Ss(1,-2) \* \z3
          - 24931458 \* \Ss(1,1)
          - 2376864 \* \Ss(1,1) \* \z3
          + 22178430 \* \Ss(1,2)
  \nonumber\\&& \mbox{}
          - 13081500 \* \Ss(1,3)
          + 3488832 \* \Ss(1,4)
          - 2643840 \* \Ss(2,-4)
          - 1882764 \* \Ss(2,-3)
          - 2956176 \* \Ss(2,-2)
  \nonumber\\&& \mbox{}
          + 36800442 \* \Ss(2,1)
          - 933120 \* \Ss(2,1) \* \z3
          - 15864660 \* \Ss(2,2)
          + 11212020 \* \Ss(2,3)
          - 5520960 \* \Ss(2,4)
  \nonumber\\&& \mbox{}
          - 913680 \* \Ss(3,-3)
          - 2519424 \* \Ss(3,-2)
          - 16461144 \* \Ss(3,1)
          + 12057660 \* \Ss(3,2)
          - 5423760 \* \Ss(3,3)
  \nonumber\\&& \mbox{}
          - 1283040 \* \Ss(4,-2)
          + 16365564 \* \Ss(4,1)
          - 4840560 \* \Ss(4,2)
          - 4860000 \* \Ss(5,1)
          + 699840 \* \Sss(1,-4,1)
  \nonumber\\&& \mbox{}
          - 6622236 \* \Sss(1,-3,1)
          + 1712016 \* \Sss(1,-2,-2)
          + 5334768 \* \Sss(1,-2,1)
          - 1487160 \* \Sss(1,-2,2)
          - 699840 \* \Sss(1,-2,3)
  \nonumber\\&& \mbox{}
          - 4046436 \* \Sss(1,1,-3)
          - 1614600 \* \Sss(1,1,-2)
          - 23254290 \* \Sss(1,1,1)
          + 12641940 \* \Sss(1,1,2)
          + 458460 \* \Sss(1,1,3)
  \nonumber\\&& \mbox{}
          + 482760 \* \Sss(1,2,-2)
          + 11611080 \* \Sss(1,2,1)
          - 3027780 \* \Sss(1,2,2)
          - 8464500 \* \Sss(1,3,1)
          + 3557520 \* \Sss(2,-3,1)
  \nonumber\\&& \mbox{}
          - 1399680 \* \Sss(2,-2,-2)
          + 868968 \* \Sss(2,-2,1)
          + 1283040 \* \Sss(2,-2,2)
          - 1846800 \* \Sss(2,1,-3)
          + 3188160 \* \Sss(2,1,-2)
  \nonumber\\&& \mbox{}
          + 16079580 \* \Sss(2,1,1)
          - 9175680 \* \Sss(2,1,2)
          + 19440 \* \Sss(2,1,3)
          - 2527200 \* \Sss(2,2,-2)
          - 8553600 \* \Sss(2,2,1)
  \nonumber\\&& \mbox{}
          + 2818800 \* \Sss(2,2,2)
          + 7678800 \* \Sss(2,3,1)
          + 1866240 \* \Sss(3,-2,1)
          - 2604960 \* \Sss(3,1,-2)
          - 13292100 \* \Sss(3,1,1)
  \nonumber\\&& \mbox{}
          + 5054400 \* \Sss(3,1,2)
          + 4276800 \* \Sss(3,2,1)
          + 1451520 \* \Ssss(1,-2,1,1)
          + 6084072 \* \Ssss(1,1,-2,1)
          - 1276560 \* \Ssss(1,1,1,-2)
  \nonumber\\&& \mbox{}
          + 5307120 \* \Sss(4,1,1)
          - 11043000 \* \Ssss(1,1,1,1)
          + 2883600 \* \Ssss(1,1,1,2)
          + 2980800 \* \Ssss(1,1,2,1)
          + 2935440 \* \Ssss(1,2,1,1)
  \nonumber\\&& \mbox{}
          - 1399680 \* \Ssss(2,-2,1,1)
          - 1283040 \* \Ssss(2,1,-2,1)
          + 8411040 \* \Ssss(2,1,1,1)
          - 2293920 \* \Ssss(2,1,1,2)
  \nonumber\\&& \mbox{}
          - 2371680 \* \Ssss(2,1,2,1)
          + 3032640 \* \Ssss(2,1,1,-2)
          - 2099520 \* \Ssss(2,2,1,1)
          - 4121280 \* \Ssss(3,1,1,1)
          + 729000 \* \S(1) \* \z4
  \nonumber\\&& \mbox{}
          - 2404080 \* \Sssss(1,1,1,1,1)
          + 1516320 \* \Sssss(2,1,1,1,1)
          - 56294033 \* \S(1)
          + 3499200 \* \S(1) \* \z5
          + 11314944 \* \S(1) \* \z3
  \nonumber\\&& \mbox{}
          + 21814674 \* \S(2)
          - 1224720 \* \S(2) \* \z4
          - 17536176 \* \S(2) \* \z3
          - 36642258 \* \S(3)
          + 7931520 \* \S(3) \* \z3
  \nonumber\\&& \mbox{}
          + 11327580 \* \S(4)
          - 14520384 \* \S(5)
          + 2099520 \* \S(6))
          + {1 \over 3645} \* (\Nminustwo - 1) \* (272160 \* \Ss(1,-4)
  \nonumber\\&& \mbox{}
          + 1811376 \* \Ss(1,-3)
          + 1267326 \* \Ss(1,-2)
          - 2756286 \* \Ss(1,1)
          + 531360 \* \Ss(1,1) \* \z3
          - 1128780 \* \Ss(1,2)
  \nonumber\\&& \mbox{}
          - 165240 \* \Ss(1,3)
          + 440640 \* \Ss(1,4)
          - 311040 \* \Ss(2,-3)
          - 160704 \* \Ss(2,-2)
          + 542034 \* \Ss(2,1)
          + 336960 \* \Ss(2,2)
  \nonumber\\&& \mbox{}
          + 155520 \* \Ss(2,3)
          + 11664 \* \Ss(3,1)
          - 155520 \* \Sss(1,-3,1)
          + 51840 \* \Sss(1,-2,-2)
          - 918432 \* \Sss(1,-2,1)
  \nonumber\\&& \mbox{}
          - 155520 \* \Sss(1,-2,2)
          + 311040 \* \Sss(1,1,-3)
          - 483840 \* \Sss(1,1,-2)
          + 1210140 \* \Sss(1,1,1)
          + 130680 \* \Sss(1,1,2)
  \nonumber\\&& \mbox{}
          - 362880 \* \Sss(1,1,3)
          + 181440 \* \Sss(1,2,-2)
          + 258120 \* \Sss(1,2,1)
          - 311040 \* \Sss(1,2,2)
          - 362880 \* \Sss(1,3,1)
  \nonumber\\&& \mbox{}
          + 155520 \* \Sss(2,-2,1)
          + 155520 \* \Sss(2,1,-2)
          - 442800 \* \Sss(2,1,1)
          - 194400 \* \Sss(2,1,2)
          - 194400 \* \Sss(2,2,1)
  \nonumber\\&& \mbox{}
          + 207360 \* \Ssss(1,-2,1,1)
          - 259200 \* \Ssss(1,1,-2,1)
          - 207360 \* \Ssss(1,1,1,-2)
          - 292680 \* \Ssss(1,1,1,1)
          + 259200 \* \Ssss(1,1,1,2)
  \nonumber\\&& \mbox{}
          + 259200 \* \Ssss(1,1,2,1)
          + 233280 \* \Ssss(1,2,1,1)
          + 207360 \* \Ssss(2,1,1,1)
          - 168480 \* \Sssss(1,1,1,1,1)
          + 5017249 \* \S(1)
  \nonumber\\&& \mbox{}
          + 116640 \* \S(1) \* \z4
          + 1212624 \* \S(1) \* \z3
          - 586731 \* \S(2)
          - 285120 \* \S(2) \* \z3
          + 10206 \* \S(3)
          + 7776 \* \S(4))
          \biggr)\biggr\}
\:\: .
\eea
\normalsize
Finally the $\ar^{\, 3}$ contribution to the pure-singlet coefficient 
function for $F_{\,2}$, defined in the paragraph below Eq.~(\ref
{eq:Fa-dec}), is given by
\small
\bea
&& c^{(3)}_{2,\rm{ps}}(N) \:\: = \:\: 
         \delta(N-2) \* \biggl\{
         \colour4colour{\cf \* \nf^2}  \*  \biggl(
          - {77623 \over 10935}
          + {3232 \over 405} \* \z3
          \biggr)
       + \colour4colour{\cf^2 \* \nf}  \*  \biggl(
          - {28249 \over 2430}
          + {128 \over 3} \* \z5
  \nonumber\\&& \mbox{}
          + {32 \over 3} \* \z4
          - {17296 \over 405} \* \z3
          \biggr)
       + \colour4colour{\ca \* \cf \* \nf}  \*  \biggl(
            {1177679 \over 21870}
          - {64 \over 3} \* \z5
          - {32 \over 3} \* \z4
          - {1376 \over 45} \* \z3
          \biggr)\biggr\}
  \nonumber\\&& \mbox{}
       + \theta(N-4) \* \biggl\{ \colour4colour{\cf \* \nf^2}  \*  \biggl(
          - {1024 \over 9} \* \Ss(1,-3)
          + {448 \over 3} \* \Ss(1,-2)
          + {32 \over 45} \* \Ss(1,-2) \* (\Nminusthree - \Nminustwo)
          - {496 \over 9} \* \Ss(1,1)
  \nonumber\\&& \mbox{}
          - {176 \over 27} \* \Ss(1,2)
          - {1024 \over 9} \* \Ss(2,-2)
          + {496 \over 81} \* \Ss(2,1)
          - {320 \over 9} \* \Ss(2,2)
          - {1888 \over 27} \* \Ss(3,1)
          + {64 \over 9} \* \Ss(4,1)
          + {5488 \over 81} \* \Sss(1,1,1)
  \nonumber\\&& \mbox{}
          + {1072 \over 27} \* \Sss(2,1,1)
          + {32 \over 3} \* \Sss(2,1,2)
          + {32 \over 3} \* \Sss(2,2,1)
          + {224 \over 9} \* \Sss(3,1,1)
          - {128 \over 9} \* \Ssss(2,1,1,1)
  \nonumber\\&& \mbox{}
          + {8 \over 3645} \* \gqq \* (25920 \* \Ss(1,-3)
          - 40500 \* \Ss(1,-2)
          + 405 \* \Ss(1,1)
          + 3915 \* \Ss(1,2)
          + 6480 \* \Ss(2,-2)
          + 28125 \* \Ss(2,1)
  \nonumber\\&& \mbox{}
          - 7290 \* \Ss(2,2)
          - 15120 \* \Ss(3,1)
          + 1620 \* \Ss(4,1)
          - 15435 \* \Sss(1,1,1)
          - 405 \* \Sss(1,1,2)
          - 405 \* \Sss(1,2,1)
  \nonumber\\&& \mbox{}
          - 1485 \* \Sss(2,1,1)
          + 2430 \* \Sss(2,1,2)
          + 2430 \* \Sss(2,2,1)
          + 5670 \* \Sss(3,1,1)
          + 540 \* \Ssss(1,1,1,1)
          - 3240 \* \Ssss(2,1,1,1)
  \nonumber\\&& \mbox{}
          + 65150 \* \S(1)
          + 39420 \* \S(1) \* \z3 
          + 71781 \* \S(2)
          - 3240 \* \S(2) \* \z3 
          - 131130 \* \S(3)
          + 86805 \* \S(4)
          - 18630 \* \S(5) )
  \nonumber\\&& \mbox{}
          - {313024 \over 729} \* \S(1)
          - {4672 \over 27} \* \S(1) \* \z3
          + {159536 \over 1215} \* \S(2)
          - {128 \over 9} \* \S(2) \* \z3
          - {14752 \over 81} \* \S(3)
          + {2032 \over 27} \* \S(4)
          - {736 \over 9} \* \S(5)
  \nonumber\\&& \mbox{}
          - {32 \over 5} \* (\Nplusthree - \Nplustwo) \* (\Ss(1,-2)
          + \S(3))
          - {16 \over 3645} \* (\Nplustwo - 3) \* (3240 \* \Ss(1,-3)
          - 4050 \* \Ss(1,-2)
          - 5760 \* \Ss(1,1)
  \nonumber\\&& \mbox{}
          + 2565 \* \Ss(1,2)
          + 3240 \* \Ss(2,-2)
          + 4455 \* \Ss(2,1)
          + 810 \* \Ss(2,2)
          + 810 \* \Ss(3,1)
          - 2655 \* \Sss(1,1,1)
          - 810 \* \Sss(1,1,2)
  \nonumber\\&& \mbox{}
          - 810 \* \Sss(1,2,1)
          - 3240 \* \Sss(2,1,1)
          + 1080 \* \Ssss(1,1,1,1)
          - 9587 \* \S(1)
          + 5940 \* \S(1) \* \z3 
          + 26262 \* \S(2)
          - 14715 \* \S(3)
  \nonumber\\&& \mbox{}
          + 6480 \* \S(4))
          - {16 \over 3645} \* (\Nminustwo - \Nminus) \* (3240 \* \Ss(1,-3)
          + 810 \* \Ss(1,-2)
          + 315 \* \Ss(1,1)
          + 1350 \* \Ss(1,2)
  \nonumber\\&& \mbox{}
          - 2655 \* \Sss(1,1,1)
          - 810 \* \Sss(1,1,2)
          - 810 \* \Sss(1,2,1)
          + 1080 \* \Ssss(1,1,1,1)
          + 6748 \* \S(1)
          + 5940 \* \S(1) \* \z3
          - 162 \* \S(2) )
  \nonumber\\&& \mbox{}
          - {16 \over 3645} \* (\Nminus + 1) \* (3240 \* \Ss(1,-3)
          - 7290 \* \Ss(1,-2)
          - 11835 \* \Ss(1,1)
          + 3780 \* \Ss(1,2)
          - 6480 \* \Ss(2,-2)
  \nonumber\\&& \mbox{}
          + 19215 \* \Ss(2,1)
          - 6885 \* \Ss(2,2)
          - 14715 \* \Ss(3,1)
          + 1620 \* \Ss(4,1)
          - 2655 \* \Sss(1,1,1)
          + 540 \* \Sss(2,1,1)
  \nonumber\\&& \mbox{}
          + 2430 \* \Sss(2,1,2)
          + 2430 \* \Sss(2,2,1)
          + 5670 \* \Sss(3,1,1)
          - 3240 \* \Ssss(2,1,1,1)
          - 25922 \* \S(1)
          + 5940 \* \S(1) \* \z3
  \nonumber\\&& \mbox{}
          + 77109 \* \S(2)
          - 3240 \* \S(2) \* \z3
          - 101025 \* \S(3)
          + 58455 \* \S(4)
          - 18630 \* \S(5) )
  \nonumber\\&& \mbox{}
          + {16 \over 27} \* (2 \* \Nminus + 3) \* (3 \* \Sss(1,1,2)
          + 3 \* \Sss(1,2,1)
          - 4 \* \Ssss(1,1,1,1))
          \biggr)
  \nonumber\\&& \mbox{}
       + \colour4colour{\cf^2 \* \nf}  \*  \biggl(
          - {2560 \over 3} \* \Ss(1,-4)
          + {11360 \over 3} \* \Ss(1,-3)
          - {90208 \over 45} \* \Ss(1,-2)
          + {1163168 \over 405} \* \Ss(1,1)
          - {256 \over 9} \* \Ss(1,1) \* \z3
  \nonumber\\&& \mbox{}
          - {135208 \over 81} \* \Ss(1,2)
          + {128528 \over 135} \* \Ss(1,3)
          + {3344 \over 9} \* \Ss(1,4)
          + 512 \* \Ss(2,-4)
          - {3776 \over 3} \* \Ss(2,-3)
          + {18400 \over 9} \* \Ss(2,-2)
  \nonumber\\&& \mbox{}
          - {624232 \over 405} \* \Ss(2,1)
          - {896 \over 3} \* \Ss(2,1) \* \z3
          + {30112 \over 27} \* \Ss(2,2)
          - {512 \over 3} \* \Ss(2,3)
          - {992 \over 3} \* \Ss(2,4)
          + 1152 \* \Ss(3,-3)
  \nonumber\\&& \mbox{}
          - {1472 \over 3} \* \Ss(3,-2)
          + {7736 \over 5} \* \Ss(3,1)
          + {3152 \over 9} \* \Ss(3,2)
          + {1088 \over 3} \* \Ss(3,3)
          + 768 \* \Ss(4,-2)
          + {8576 \over 9} \* \Ss(4,1)
  \nonumber\\&& \mbox{}
          + {2144 \over 3} \* \Ss(4,2)
          + {2656 \over 3} \* \Ss(5,1)
          + {1280 \over 3} \* \Sss(1,-3,1)
          - {19040 \over 9} \* \Sss(1,-2,1)
          + {512 \over 3} \* \Sss(1,-2,2)
          + {2816 \over 3} \* \Sss(1,1,-3)
  \nonumber\\&& \mbox{}
          - {21856 \over 9} \* \Sss(1,1,-2)
          + {164648 \over 81} \* \Sss(1,1,1)
          - {23096 \over 27} \* \Sss(1,1,2)
          + {5632 \over 9} \* \Sss(1,1,3)
          + 512 \* \Sss(1,2,-2)
  \nonumber\\&& \mbox{}
          - {34040 \over 27} \* \Sss(1,2,1)
          - {8000 \over 9} \* \Sss(1,3,1)
          - 128 \* \Sss(2,-3,1)
          + {2368 \over 3} \* \Sss(2,-2,1)
          - 896 \* \Sss(2,1,-3)
          + {1856 \over 3} \* \Sss(2,1,-2)
  \nonumber\\&& \mbox{}
          - {37144 \over 27} \* \Sss(2,1,1)
          - {368 \over 3} \* \Sss(2,1,2)
          - {1408 \over 3} \* \Sss(2,1,3)
          - 512 \* \Sss(2,2,-2)
          - 144 \* \Sss(2,2,1)
          - {896 \over 3} \* \Sss(2,2,2)
  \nonumber\\&& \mbox{}
          + {512 \over 3} \* \Sss(2,3,1)
          - 512 \* \Sss(3,-2,1)
          - 768 \* \Sss(3,1,-2)
          - {1696 \over 9} \* \Sss(3,1,1)
          - {1472 \over 3} \* \Sss(3,1,2)
  \nonumber\\&& \mbox{}
          - 576 \* \Sss(3,2,1)
          - 800 \* \Sss(4,1,1)
          - {512 \over 3} \* \Ssss(1,-2,1,1)
          - 512 \* \Ssss(1,1,1,-2)
          + {2560 \over 3} \* \Ssss(1,1,1,1)
  \nonumber\\&& \mbox{}
          + 256 \* \Ssss(2,1,-2,1)
          + 512 \* \Ssss(2,1,1,-2)
          + {256 \over 9} \* \Ssss(2,1,1,1)
          + {800 \over 3} \* \Ssss(2,1,1,2)
          + {800 \over 3} \* \Ssss(2,1,2,1)
  \nonumber\\&& \mbox{}
          + 288 \* \Ssss(2,2,1,1)
          + {1408 \over 3} \* \Ssss(3,1,1,1)
          - {704 \over 3} \* \Sssss(2,1,1,1,1)
          + {2 \over 2025} \* \gqq \* (432000 \* \Ss(1,-4)
          - 1971000 \* \Ss(1,-3)
  \nonumber\\&& \mbox{}
          + 960240 \* \Ss(1,-2)
          - 1293985 \* \Ss(1,1)
          + 100800 \* \Ss(1,1) \* \z3
          + 1053850 \* \Ss(1,2)
          - 616980 \* \Ss(1,3)
  \nonumber\\&& \mbox{}
          - 188100 \* \Ss(1,4)
          + 259200 \* \Ss(2,-4)
          - 75600 \* \Ss(2,-3)
          - 869400 \* \Ss(2,-2)
          + 552950 \* \Ss(2,1)
  \nonumber\\&& \mbox{}
          - 151200 \* \Ss(2,1) \* \z3
          - 213600 \* \Ss(2,2)
          + 243000 \* \Ss(2,3)
          - 167400 \* \Ss(2,4)
          + 64800 \* \Ss(3,-3)
  \nonumber\\&& \mbox{}
          - 54000 \* \Ss(3,-2)
          + 1221030 \* \Ss(3,1)
          - 593100 \* \Ss(3,2)
          + 183600 \* \Ss(3,3)
          - 1384200 \* \Ss(4,1)
  \nonumber\\&& \mbox{}
          + 361800 \* \Ss(4,2)
          + 448200 \* \Ss(5,1)
          - 216000 \* \Sss(1,-3,1)
          + 1128600 \* \Sss(1,-2,1)
          - 86400 \* \Sss(1,-2,2)
  \nonumber\\&& \mbox{}
          - 475200 \* \Sss(1,1,-3)
          + 1222200 \* \Sss(1,1,-2)
          - 1150100 \* \Sss(1,1,1)
          + 416850 \* \Sss(1,1,2)
          - 360000 \* \Sss(1,1,3)
  \nonumber\\&& \mbox{}
          - 259200 \* \Sss(1,2,-2)
          + 654450 \* \Sss(1,2,1)
          + 25200 \* \Sss(1,2,2)
          + 493200 \* \Sss(1,3,1)
          - 64800 \* \Sss(2,-3,1)
  \nonumber\\&& \mbox{}
          + 226800 \* \Sss(2,-2,1)
          - 453600 \* \Sss(2,1,-3)
          + 54000 \* \Sss(2,1,-2)
          + 87600 \* \Sss(2,1,1)
          + 240300 \* \Sss(2,1,2)
  \nonumber\\&& \mbox{}
          - 237600 \* \Sss(2,1,3)
          - 259200 \* \Sss(2,2,-2)
          + 315900 \* \Sss(2,2,1)
          - 151200 \* \Sss(2,2,2)
          + 86400 \* \Sss(2,3,1)
  \nonumber\\&& \mbox{}
          + 129600 \* \Sss(3,-2,1)
          - 259200 \* \Sss(3,1,-2)
          + 716400 \* \Sss(3,1,1)
          - 248400 \* \Sss(3,1,2)
          - 291600 \* \Sss(3,2,1)
  \nonumber\\&& \mbox{}
          - 405000 \* \Sss(4,1,1)
          + 86400 \* \Ssss(1,-2,1,1)
          + 259200 \* \Ssss(1,1,1,-2)
          - 432000 \* \Ssss(1,1,1,1)
          - 22500 \* \Ssss(1,1,1,2)
  \nonumber\\&& \mbox{}
          - 22500 \* \Ssss(1,1,2,1)
          - 24300 \* \Ssss(1,2,1,1)
          + 129600 \* \Ssss(2,1,-2,1)
          + 259200 \* \Ssss(2,1,1,-2)
          - 262800 \* \Ssss(2,1,1,1)
  \nonumber\\&& \mbox{}
          + 135000 \* \Ssss(2,1,1,2)
          + 135000 \* \Ssss(2,1,2,1)
          + 145800 \* \Ssss(2,2,1,1)
          + 237600 \* \Ssss(3,1,1,1)
          + 19800 \* \Sssss(1,1,1,1,1)
  \nonumber\\&& \mbox{}
          - 118800 \* \Sssss(2,1,1,1,1)
          - 959530 \* \S(1)
          + 8100 \* \S(1) \* \z4
          - 978840 \* \S(1) \* \z3
          + 1782744 \* \S(2)
          - 48600 \* \S(2) \* \z4
  \nonumber\\&& \mbox{}
          - 1029600 \* \S(2) \* \z3
          + 53225 \* \S(3)
          + 237600 \* \S(3) \* \z3
          - 1303650 \* \S(4)
          + 725400 \* \S(5)
          - 243000 \* \S(6) )
  \nonumber\\&& \mbox{}
          + {406724 \over 405} \* \S(1)
          + {115808 \over 45} \* \S(1) \* \z3
          - {545324 \over 675} \* \S(2)
          - 96 \* \S(2) \* \z4
          - {4352 \over 9} \* \S(2) \* \z3
          - {4652 \over 81} \* \S(3)
  \nonumber\\&& \mbox{}
          + {2944 \over 3} \* \S(3) \* \z3
          - {2152 \over 3} \* \S(4)
          - {10256 \over 9} \* \S(5)
          - 480 \* \S(6)
          + {32 \over 75} \* (\Nplusthree - \Nplustwo) \* (45 \* \Ss(1,-3)
          - 17 \* \Ss(1,-2)
  \nonumber\\&& \mbox{}
          + 360 \* \Ss(1,1) \* \z3 
          - 15 \* \Ss(2,-2)
          + 180 \* \Ss(2,3)
          + 15 \* \Ss(3,1)
          - 180 \* \Ss(4,1)
          + 15 \* \Sss(1,-2,1)
          + 15 \* \Sss(1,1,-2)
          - 180 \* \Sss(1,1,3)
  \nonumber\\&& \mbox{}
          + 180 \* \Sss(1,3,1)
          + 90 \* \S(1) \* \z3 
          - 360 \* \S(2) \* \z3 
          - 17 \* \S(3)
          + 45 \* \S(4) )
          - {32 \over 675} \* (\Nminusthree - \Nminustwo) \* (45 \* \Ss(1,-3)
          - 22 \* \Ss(1,-2)
  \nonumber\\&& \mbox{}
          + 360 \* \Ss(1,1) \* \z3 
          + 60 \* \Ss(2,-2)
          + 15 \* \Sss(1,-2,1)
          + 15 \* \Sss(1,1,-2)
          - 180 \* \Sss(1,1,3)
          + 180 \* \Sss(1,3,1)
          + 90 \* \S(1) \* \z3 )
  \nonumber\\&& \mbox{}
          + {4 \over 2025} \* (\Nplustwo - 3) \* (10800 \* \Ss(1,-4)
          + 48600 \* \Ss(1,-3)
          + 23760 \* \Ss(1,-2)
          - 15440 \* \Ss(1,1)
  \nonumber\\&& \mbox{}
          - 126000 \* \Ss(1,1) \* \z3
          - 68750 \* \Ss(1,2)
          + 52080 \* \Ss(1,3)
          - 12600 \* \Ss(1,4)
          + 54000 \* \Ss(2,-3)
          + 55800 \* \Ss(2,-2)
  \nonumber\\&& \mbox{}
          + 72890 \* \Ss(2,1)
          - 69600 \* \Ss(2,2)
          + 32400 \* \Ss(2,3)
          + 54000 \* \Ss(3,-2)
          - 148680 \* \Ss(3,1)
          + 187200 \* \Ss(3,2)
  \nonumber\\&& \mbox{}
          + 327600 \* \Ss(4,1)
          + 10800 \* \Sss(1,-3,1)
          - 41400 \* \Sss(1,-2,1)
          + 10800 \* \Sss(1,-2,2)
          - 54000 \* \Sss(1,1,-3)
  \nonumber\\&& \mbox{}
          - 55800 \* \Sss(1,1,-2)
          + 26050 \* \Sss(1,1,1)
          + 2400 \* \Sss(1,1,2)
          + 18000 \* \Sss(1,1,3)
          - 32400 \* \Sss(1,2,-2)
  \nonumber\\&& \mbox{}
          - 30000 \* \Sss(1,2,1)
          - 50400 \* \Sss(1,2,2)
          - 90000 \* \Sss(1,3,1)
          - 10800 \* \Sss(2,-2,1)
          - 54000 \* \Sss(2,1,-2)
  \nonumber\\&& \mbox{}
          + 80400 \* \Sss(2,1,1)
          - 113400 \* \Sss(2,1,2)
          - 135000 \* \Sss(2,2,1)
          - 196200 \* \Sss(3,1,1)
          - 10800 \* \Ssss(1,-2,1,1)
  \nonumber\\&& \mbox{}
          + 32400 \* \Ssss(1,1,-2,1)
          + 32400 \* \Ssss(1,1,1,-2)
          - 3600 \* \Ssss(1,1,1,1)
          + 45000 \* \Ssss(1,1,1,2)
          + 45000 \* \Ssss(1,1,2,1)
  \nonumber\\&& \mbox{}
          + 48600 \* \Ssss(1,2,1,1)
          + 104400 \* \Ssss(2,1,1,1)
          - 39600 \* \Sssss(1,1,1,1,1)
          + 207453 \* \S(1)
          - 16200 \* \S(1) \* \z4
  \nonumber\\&& \mbox{}
          - 106560 \* \S(1) \* \z3
          - 75838 \* \S(2)
          + 169200 \* \S(2) \* \z3
          + 5450 \* \S(3)
          + 109800 \* \S(4)
          - 214200 \* \S(5) )
  \nonumber\\&& \mbox{}
          + {2 \over 2025} \* (\Nminustwo - \Nminus) \* (21600 \* \Ss(1,-4)
          + 64800 \* \Ss(1,-3)
          - 15480 \* \Ss(1,-2)
          + 129095 \* \Ss(1,1)
  \nonumber\\&& \mbox{}
          + 7200 \* \Ss(1,1) \* \z3
          + 71300 \* \Ss(1,2)
          - 30840 \* \Ss(1,3)
          - 25200 \* \Ss(1,4)
          + 7920 \* \Ss(2,1)
          + 8640 \* \Ss(3,1)
  \nonumber\\&& \mbox{}
          + 21600 \* \Sss(1,-3,1)
          - 18000 \* \Sss(1,-2,1)
          + 21600 \* \Sss(1,-2,2)
          - 108000 \* \Sss(1,1,-3)
          - 111600 \* \Sss(1,1,-2)
  \nonumber\\&& \mbox{}
          - 68950 \* \Sss(1,1,1)
          - 11400 \* \Sss(1,1,2)
          - 93600 \* \Sss(1,1,3)
          - 64800 \* \Sss(1,2,-2)
          - 43800 \* \Sss(1,2,1)
  \nonumber\\&& \mbox{}
          - 100800 \* \Sss(1,2,2)
          - 50400 \* \Sss(1,3,1)
          - 21600 \* \Ssss(1,-2,1,1)
          + 64800 \* \Ssss(1,1,-2,1)
          + 64800 \* \Ssss(1,1,1,-2)
  \nonumber\\&& \mbox{}
          - 7200 \* \Ssss(1,1,1,1)
          + 90000 \* \Ssss(1,1,1,2)
          + 90000 \* \Ssss(1,1,2,1)
          + 97200 \* \Ssss(1,2,1,1)
          - 79200 \* \Sssss(1,1,1,1,1)
  \nonumber\\&& \mbox{}
          - 36219 \* \S(1)
          - 32400 \* \S(1) \* \z4
          + 154080 \* \S(1) \* \z3
          + 1776 \* \S(2)
          - 2160 \* \S(3) )
  \nonumber\\&& \mbox{}
          + {2 \over 2025} \* (\Nminus + 1) \* (21600 \* \Ss(1,-4)
          + 151200 \* \Ss(1,-3)
          + 102120 \* \Ss(1,-2)
          - 190855 \* \Ss(1,1)
          - 346300 \* \Ss(1,2)
  \nonumber\\&& \mbox{}
          - 338400 \* \Ss(1,1) \* \z3
          + 239160 \* \Ss(1,3)
          - 25200 \* \Ss(1,4)
          - 518400 \* \Ss(2,-4)
          + 820800 \* \Ss(2,-3)
          - 54000 \* \Ss(2,-2)
  \nonumber\\&& \mbox{}
          + 373120 \* \Ss(2,1)
          + 302400 \* \Ss(2,1) \* \z3
          - 490200 \* \Ss(2,2)
          - 91800 \* \Ss(2,3)
          + 334800 \* \Ss(2,4)
          - 648000 \* \Ss(3,-3)
  \nonumber\\&& \mbox{}
          + 410400 \* \Ss(3,-2)
          - 2301660 \* \Ss(3,1)
          + 790200 \* \Ss(3,2)
          - 367200 \* \Ss(3,3)
          - 388800 \* \Ss(4,-2)
  \nonumber\\&& \mbox{}
          + 1557000 \* \Ss(4,1)
          - 723600 \* \Ss(4,2)
          - 896400 \* \Ss(5,1)
          + 21600 \* \Sss(1,-3,1)
          - 140400 \* \Sss(1,-2,1)
  \nonumber\\&& \mbox{}
          + 21600 \* \Sss(1,-2,2)
          - 108000 \* \Sss(1,1,-3)
          - 104400 \* \Sss(1,1,-2)
          + 173150 \* \Sss(1,1,1)
          + 21000 \* \Sss(1,1,2)
  \nonumber\\&& \mbox{}
          + 79200 \* \Sss(1,1,3)
          - 64800 \* \Sss(1,2,-2)
          - 76200 \* \Sss(1,2,1)
          - 223200 \* \Sss(1,3,1)
          + 129600 \* \Sss(2,-3,1)
  \nonumber\\&& \mbox{}
          - 648000 \* \Sss(2,-2,1)
          + 907200 \* \Sss(2,1,-3)
          - 475200 \* \Sss(2,1,-2)
          + 769650 \* \Sss(2,1,1)
          - 405000 \* \Sss(2,1,2)
  \nonumber\\&& \mbox{}
          + 475200 \* \Sss(2,1,3)
          + 518400 \* \Sss(2,2,-2)
          - 513000 \* \Sss(2,2,1)
          + 302400 \* \Sss(2,2,2)
          - 172800 \* \Sss(2,3,1)
  \nonumber\\&& \mbox{}
          + 129600 \* \Sss(3,-2,1)
          + 648000 \* \Sss(3,1,-2)
          - 1013400 \* \Sss(3,1,1)
          + 496800 \* \Sss(3,1,2)
          + 583200 \* \Sss(3,2,1)
  \nonumber\\&& \mbox{}
          + 810000 \* \Sss(4,1,1)
          - 21600 \* \Ssss(1,-2,1,1)
          + 64800 \* \Ssss(1,1,-2,1)
          + 64800 \* \Ssss(1,1,1,-2)
          - 7200 \* \Ssss(1,1,1,1)
  \nonumber\\&& \mbox{}
          - 259200 \* \Ssss(2,1,-2,1)
          - 518400 \* \Ssss(2,1,1,-2)
          + 457200 \* \Ssss(2,1,1,1)
          - 270000 \* \Ssss(2,1,1,2)
          - 270000 \* \Ssss(2,1,2,1)
  \nonumber\\&& \mbox{}
          - 291600 \* \Ssss(2,2,1,1)
          - 475200 \* \Ssss(3,1,1,1)
          + 237600 \* \Sssss(2,1,1,1,1)
          + 866031 \* \S(1)
          - 537120 \* \S(1) \* \z3
  \nonumber\\&& \mbox{}
          - 1525427 \* \S(2)
          + 97200 \* \S(2) \* \z4
          + 1612800 \* \S(2) \* \z3
          - 13250 \* \S(3)
          - 734400 \* \S(3) \* \z3
          + 1886400 \* \S(4)
  \nonumber\\&& \mbox{}
          - 576900 \* \S(5)
          + 486000 \* \S(6))
  \nonumber\\&& \mbox{}
          - {16 \over 9} \* (2 \* \Nminus + 3) \* (28 \* \Sss(1,2,2)
          - 25 \* \Ssss(1,1,1,2)
          - 25 \* \Ssss(1,1,2,1)
          - 27 \* \Ssss(1,2,1,1)
          + 22 \* \Sssss(1,1,1,1,1)
          + 9 \* \S(1) \* \z4)
          \biggr)
  \nonumber\\&& \mbox{}
       + \colour4colour{\ca \* \cf \* \nf}  \*  \biggl(
            {1120 \over 3} \* \Ss(1,-4)
          + {30416 \over 27} \* \Ss(1,-3)
          + {8696 \over 45} \* \Ss(1,-2)
          - {3471092 \over 1215} \* \Ss(1,1)
          + {2032 \over 9} \* \Ss(1,1) \* \z3
  \nonumber\\&& \mbox{}
          - {67808 \over 81} \* \Ss(1,2)
          - {15448 \over 45} \* \Ss(1,3)
          - {1856 \over 9} \* \Ss(1,4)
          + {640 \over 3} \* \Ss(2,-4)
          + {10400 \over 9} \* \Ss(2,-3)
          + {41392 \over 27} \* \Ss(2,-2)
  \nonumber\\&& \mbox{}
          - {301912 \over 135} \* \Ss(2,1)
          + {800 \over 3} \* \Ss(2,1) \* \z3
          - {13336 \over 9} \* \Ss(2,2)
          + {2728 \over 9} \* \Ss(2,3)
          + {1408 \over 3} \* \Ss(2,4)
          + {640 \over 3} \* \Ss(3,-3)
  \nonumber\\&& \mbox{}
          + {8752 \over 9} \* \Ss(3,-2)
          - {387176 \over 135} \* \Ss(3,1)
          + {344 \over 3} \* \Ss(3,2)
          + 208 \* \Ss(3,3)
          + {992 \over 3} \* \Ss(4,-2)
          - {2096 \over 9} \* \Ss(4,1)
          + 16 \* \Ss(4,2)
  \nonumber\\&& \mbox{}
          - {640 \over 3} \* \Ss(5,1)
          - {184 \over 3} \* \Sss(1,-3,1)
          + {352 \over 9} \* \Sss(1,-2,-2)
          - {8624 \over 27} \* \Sss(1,-2,1)
          - 144 \* \Sss(1,-2,2)
          - {328 \over 3} \* \Sss(1,1,-3)
  \nonumber\\&& \mbox{}
          + {1232 \over 27} \* \Sss(1,1,-2)
          + {28648 \over 81} \* \Sss(1,1,1)
          + {2464 \over 27} \* \Sss(1,1,2)
          - {3880 \over 9} \* \Sss(1,1,3)
          - {1456 \over 9} \* \Sss(1,2,-2)
          - 24 \* \Sss(1,2,2)
  \nonumber\\&& \mbox{}
          + {12848 \over 27} \* \Sss(1,2,1)
          + {4184 \over 9} \* \Sss(1,3,1)
          - {688 \over 3} \* \Sss(2,-3,1)
          + {64 \over 3} \* \Sss(2,-2,-2)
          - {1648 \over 3} \* \Sss(2,-2,1)
          - {352 \over 3} \* \Sss(2,-2,2)
  \nonumber\\&& \mbox{}
          + {688 \over 3} \* \Sss(2,1,-3)
          - {1232 \over 3} \* \Sss(2,1,-2)
          + {44704 \over 27} \* \Sss(2,1,1)
          - {1744 \over 9} \* \Sss(2,1,2)
          - 112 \* \Sss(2,1,3)
          + {608 \over 3} \* \Sss(2,2,-2)
  \nonumber\\&& \mbox{}
          - 176 \* \Sss(2,2,1)
          - {592 \over 3} \* \Sss(2,2,2)
          - 432 \* \Sss(2,3,1)
          - {352 \over 3} \* \Sss(3,-2,1)
          + {352 \over 3} \* \Sss(3,1,-2)
          - {2344 \over 9} \* \Sss(3,1,1)
  \nonumber\\&& \mbox{}
          - {736 \over 3} \* \Sss(3,1,2)
          - {416 \over 3} \* \Sss(3,2,1)
          - {176 \over 3} \* \Sss(4,1,1)
          + {1792 \over 9} \* \Ssss(1,-2,1,1)
          - {3536 \over 9} \* \Ssss(1,1,-2,1)
          + {1568 \over 9} \* \Ssss(1,1,1,-2)
  \nonumber\\&& \mbox{}
          - {7456 \over 27} \* \Ssss(1,1,1,1)
          + 128 \* \Ssss(2,-2,1,1)
          - {224 \over 3} \* \Ssss(2,1,-2,1)
          - {704 \over 3} \* \Ssss(2,1,1,-2)
          + 176 \* \Ssss(2,1,1,1)
  \nonumber\\&& \mbox{}
          + {544 \over 3} \* \Ssss(2,1,1,2)
          + {544 \over 3} \* \Ssss(2,1,2,1)
          + 160 \* \Ssss(2,2,1,1)
          + {608 \over 3} \* \Ssss(3,1,1,1)
          - {448 \over 3} \* \Sssss(2,1,1,1,1)
  \nonumber\\&& \mbox{}
          - {2 \over 3645} \* \gqq \* (340200 \* \Ss(1,-4)
          + 585900 \* \Ss(1,-3)
          - 554526 \* \Ss(1,-2)
          - 1689369 \* \Ss(1,1)
          + 283500 \* \Ss(1,1) \* \z3
  \nonumber\\&& \mbox{}
          - 446130 \* \Ss(1,2)
          - 653022 \* \Ss(1,3)
          - 187920 \* \Ss(1,4)
          - 194400 \* \Ss(2,-4)
          + 61560 \* \Ss(2,-3)
          + 107460 \* \Ss(2,-2)
  \nonumber\\&& \mbox{}
          + 94644 \* \Ss(2,1)
          - 243000 \* \Ss(2,1) \* \z3
          - 738720 \* \Ss(2,2)
          + 757350 \* \Ss(2,3)
          - 427680 \* \Ss(2,4)
          - 68040 \* \Ss(3,-3)
  \nonumber\\&& \mbox{}
          + 341820 \* \Ss(3,-2)
          - 1056078 \* \Ss(3,1)
          + 444690 \* \Ss(3,2)
          - 403380 \* \Ss(3,3)
          - 68040 \* \Ss(4,-2)
          + 495720 \* \Ss(4,1)
  \nonumber\\&& \mbox{}
          - 422820 \* \Ss(4,2)
          - 486000 \* \Ss(5,1)
          - 55890 \* \Sss(1,-3,1)
          + 35640 \* \Sss(1,-2,-2)
          + 3780 \* \Sss(1,-2,1)
          - 60840 \* \Sss(1,1,1)
  \nonumber\\&& \mbox{}
          - 131220 \* \Sss(1,-2,2)
          - 99630 \* \Sss(1,1,-3)
          + 44820 \* \Sss(1,1,-2)
          + 228960 \* \Sss(1,1,2)
          - 431730 \* \Sss(1,1,3)
  \nonumber\\&& \mbox{}
          - 147420 \* \Sss(1,2,-2)
          + 637740 \* \Sss(1,2,1)
          - 21870 \* \Sss(1,2,2)
          + 462510 \* \Sss(1,3,1)
          + 208980 \* \Sss(2,-3,1)
  \nonumber\\&& \mbox{}
          - 19440 \* \Sss(2,-2,-2)
          + 247860 \* \Sss(2,-2,1)
          + 106920 \* \Sss(2,-2,2)
          - 208980 \* \Sss(2,1,-3)
          + 121500 \* \Sss(2,1,-2)
  \nonumber\\&& \mbox{}
          + 559710 \* \Sss(2,1,1)
          - 234900 \* \Sss(2,1,2)
          + 102060 \* \Sss(2,1,3)
          - 184680 \* \Sss(2,2,-2)
          - 160380 \* \Sss(2,2,1)
  \nonumber\\&& \mbox{}
          + 179820 \* \Sss(2,2,2)
          + 393660 \* \Sss(2,3,1)
          + 165240 \* \Sss(3,-2,1)
          - 223560 \* \Sss(3,1,-2)
          - 386370 \* \Sss(3,1,1)
  \nonumber\\&& \mbox{}
          + 379080 \* \Sss(3,1,2)
          + 320760 \* \Sss(3,2,1)
          + 461700 \* \Sss(4,1,1)
          + 181440 \* \Ssss(1,-2,1,1)
          - 358020 \* \Ssss(1,1,-2,1)
  \nonumber\\&& \mbox{}
          + 158760 \* \Ssss(1,1,1,-2)
          - 426600 \* \Ssss(1,1,1,1)
          + 27540 \* \Ssss(1,1,1,2)
          + 27540 \* \Ssss(1,1,2,1)
          + 24300 \* \Ssss(1,2,1,1)
  \nonumber\\&& \mbox{}
          - 116640 \* \Ssss(2,-2,1,1)
          + 68040 \* \Ssss(2,1,-2,1)
          + 213840 \* \Ssss(2,1,1,-2)
          + 197640 \* \Ssss(2,1,1,1)
          - 165240 \* \Ssss(2,1,1,2)
  \nonumber\\&& \mbox{}
          - 165240 \* \Ssss(2,1,2,1)
          - 145800 \* \Ssss(2,2,1,1)
          - 359640 \* \Ssss(3,1,1,1)
          - 22680 \* \Sssss(1,1,1,1,1)
          + 136080 \* \Sssss(2,1,1,1,1)
  \nonumber\\&& \mbox{}
          + 726001 \* \S(1)
          + 14580 \* \S(1) \* \z4
          + 850824 \* \S(1) \* \z3
          + 3069669 \* \S(2)
          - 87480 \* \S(2) \* \z4
  \nonumber\\&& \mbox{}
          - 1475820 \* \S(2) \* \z3
          - 2009556 \* \S(3)
          + 690120 \* \S(3) \* \z3
          + 2752110 \* \S(4)
          - 1085400 \* \S(5)
          + 437400 \* \S(6) )
  \nonumber\\&& \mbox{}
          + {13630024 \over 3645} \* \S(1)
          + {22816 \over 45} \* \S(1) \* \z3
          + {1886512 \over 405} \* \S(2)
          + 96 \* \S(2) \* \z4
          + {4976 \over 9} \* \S(2) \* \z3
          + {1648048 \over 405} \* \S(3)
  \nonumber\\&& \mbox{}
          - {1120 \over 3} \* \S(3) \* \z3
          + {97088 \over 27} \* \S(4)
          + {6832 \over 9} \* \S(5)
          + 480 \* \S(6)
          + {16 \over 5} \* (\Nplusthree - \Nplustwo) \* (4 \* \Ss(1,-3)
          - \Ss(1,-2)
  \nonumber\\&& \mbox{}
          - 24 \* \Ss(1,1) \* \z3
          + 4 \* \Ss(2,-2)
          - 12 \* \Ss(2,3)
          - 4 \* \Ss(3,1)
          + 12 \* \Ss(4,1)
          - 4 \* \Sss(1,-2,1)
          - 4 \* \Sss(1,1,-2)
          + 12 \* \Sss(1,1,3)
  \nonumber\\&& \mbox{}
          - 12 \* \Sss(1,3,1)
          + 24 \* \S(2) \* \z3 - \S(3)
          + 4 \* \S(4))
          - {16 \over 45} \* (\Nminusthree - \Nminustwo) \* (4 \* \Ss(1,-3)
          - \Ss(1,-2)
          - 24 \* \Ss(1,1) \* \z3
  \nonumber\\&& \mbox{}
          - 4 \* \Sss(1,-2,1)
          - 4 \* \Sss(1,1,-2)
          + 12 \* \Sss(1,1,3)
          - 12 \* \Sss(1,3,1))
          + {2 \over 3645} \* (\Nplustwo - 3) \* (252720 \* \Ss(1,-4)
  \nonumber\\&& \mbox{}
          + 428760 \* \Ss(1,-3)
          - 121932 \* \Ss(1,-2)
          - 592494 \* \Ss(1,1)
          + 356400 \* \Ss(1,1) \* \z3
          - 304920 \* \Ss(1,2)
  \nonumber\\&& \mbox{}
          - 411804 \* \Ss(1,3)
          + 181440 \* \Ss(1,4)
          + 362880 \* \Ss(2,-3)
          + 361800 \* \Ss(2,-2)
          - 556776 \* \Ss(2,1)
  \nonumber\\&& \mbox{}
          - 785700 \* \Ss(2,2)
          + 317520 \* \Ss(2,3)
          + 336960 \* \Ss(3,-2)
          - 1238436 \* \Ss(3,1)
          + 155520 \* \Ss(3,2)
  \nonumber\\&& \mbox{}
          - 25920 \* \Ss(4,1)
          - 194400 \* \Sss(1,-3,1)
          + 25920 \* \Sss(1,-2,-2)
          - 137160 \* \Sss(1,-2,1)
          - 116640 \* \Sss(1,-2,2)
  \nonumber\\&& \mbox{}
          + 38880 \* \Sss(1,1,-3)
          - 219240 \* \Sss(1,1,-2)
          + 277200 \* \Sss(1,1,1)
          + 181980 \* \Sss(1,1,2)
          - 259200 \* \Sss(1,1,3)
  \nonumber\\&& \mbox{}
          + 51840 \* \Sss(1,2,-2)
          + 304020 \* \Sss(1,2,1)
          - 116640 \* \Sss(1,2,2)
          - 25920 \* \Sss(1,3,1)
          - 194400 \* \Sss(2,-2,1)
  \nonumber\\&& \mbox{}
          - 38880 \* \Sss(2,1,-2)
          + 712260 \* \Sss(2,1,1)
          - 200880 \* \Sss(2,1,2)
          - 136080 \* \Sss(2,2,1)
          - 168480 \* \Sss(3,1,1)
  \nonumber\\&& \mbox{}
          + 142560 \* \Ssss(1,-2,1,1)
          - 90720 \* \Ssss(1,1,-2,1)
          - 64800 \* \Ssss(1,1,1,-2)
          - 227340 \* \Ssss(1,1,1,1)
          + 110160 \* \Ssss(1,1,1,2)
  \nonumber\\&& \mbox{}
          + 110160 \* \Ssss(1,1,2,1)
          + 97200 \* \Ssss(1,2,1,1)
          + 162000 \* \Ssss(2,1,1,1)
          - 90720 \* \Sssss(1,1,1,1,1)
          - 255239 \* \S(1)
  \nonumber\\&& \mbox{}
          + 58320 \* \S(1) \* \z4
          + 940248 \* \S(1) \* \z3
          + 1798638 \* \S(2)
          - 369360 \* \S(2) \* \z3
          + 100836 \* \S(3)
          + 1696140 \* \S(4) )
  \nonumber\\&& \mbox{}
          + {4 \over 3645} \* (\Nminustwo - \Nminus) \* (126360 \* \Ss(1,-4)
          + 447660 \* \Ss(1,-3)
          + 301104 \* \Ss(1,-2)
          - 753222 \* \Ss(1,1)
  \nonumber\\&& \mbox{}
          + 61560 \* \Ss(1,1) \* \z3
          - 310815 \* \Ss(1,2)
          - 35802 \* \Ss(1,3)
          + 90720 \* \Ss(1,4)
          - 77760 \* \Ss(2,-3)
          - 42120 \* \Ss(2,-2)
  \nonumber\\&& \mbox{}
          + 130608 \* \Ss(2,1)
          + 84240 \* \Ss(2,2)
          + 38880 \* \Ss(2,3)
          - 3888 \* \Ss(3,1)
          - 97200 \* \Sss(1,-3,1)
          + 12960 \* \Sss(1,-2,-2)
  \nonumber\\&& \mbox{}
          - 228960 \* \Sss(1,-2,1)
          - 58320 \* \Sss(1,-2,2)
          + 19440 \* \Sss(1,1,-3)
          - 124200 \* \Sss(1,1,-2)
          + 330165 \* \Sss(1,1,1)
  \nonumber\\&& \mbox{}
          + 18090 \* \Sss(1,1,2)
          - 71280 \* \Sss(1,1,3)
          + 25920 \* \Sss(1,2,-2)
          + 49950 \* \Sss(1,2,1)
          - 58320 \* \Sss(1,2,2)
  \nonumber\\&& \mbox{}
          - 71280 \* \Sss(1,3,1)
          + 38880 \* \Sss(2,-2,1)
          + 38880 \* \Sss(2,1,-2)
          - 110700 \* \Sss(2,1,1)
          - 48600 \* \Sss(2,1,2)
  \nonumber\\&& \mbox{}
          - 48600 \* \Sss(2,2,1)
          + 71280 \* \Ssss(1,-2,1,1)
          - 45360 \* \Ssss(1,1,-2,1)
          - 32400 \* \Ssss(1,1,1,-2)
          - 26190 \* \Ssss(1,1,1,1)
  \nonumber\\&& \mbox{}
          + 55080 \* \Ssss(1,1,1,2)
          + 55080 \* \Ssss(1,1,2,1)
          + 48600 \* \Ssss(1,2,1,1)
          + 51840 \* \Ssss(2,1,1,1)
          - 45360 \* \Sssss(1,1,1,1,1)
  \nonumber\\&& \mbox{}
          + 1213133 \* \S(1)
          + 29160 \* \S(1) \* \z4
          + 275724 \* \S(1) \* \z3
          - 149052 \* \S(2)
          - 71280 \* \S(2) \* \z3
          - 1296 \* \S(3))
  \nonumber\\&& \mbox{}
          + {2 \over 3645} \* (\Nminus + 1) \* (252720 \* \Ss(1,-4)
          - 11880 \* \Ss(1,-3)
          - 852552 \* \Ss(1,-2)
          + 321456 \* \Ss(1,1)
  \nonumber\\&& \mbox{}
          + 434160 \* \Ss(1,1) \* \z3
          + 11790 \* \Ss(1,2)
          - 752004 \* \Ss(1,3)
          + 181440 \* \Ss(1,4)
          - 388800 \* \Ss(2,-4)
          - 628560 \* \Ss(2,-3)
  \nonumber\\&& \mbox{}
          - 927720 \* \Ss(2,-2)
          + 1575774 \* \Ss(2,1)
          - 486000 \* \Ss(2,1) \* \z3
          - 174150 \* \Ss(2,2)
          + 798660 \* \Ss(2,3)
  \nonumber\\&& \mbox{}
          - 855360 \* \Ss(2,4)
          - 262440 \* \Ss(3,-3)
          - 207360 \* \Ss(3,-2)
          + 318924 \* \Ss(3,1)
          + 495720 \* \Ss(3,2)
  \nonumber\\&& \mbox{}
          - 592920 \* \Ss(3,3)
          - 369360 \* \Ss(4,-2)
          + 682020 \* \Ss(4,1)
          - 437400 \* \Ss(4,2)
          - 291600 \* \Ss(5,1)
  \nonumber\\&& \mbox{}
          - 194400 \* \Sss(1,-3,1)
          + 25920 \* \Sss(1,-2,-2)
          + 157680 \* \Sss(1,-2,1)
          - 116640 \* \Sss(1,-2,2)
          + 38880 \* \Sss(1,1,-3)
  \nonumber\\&& \mbox{}
          - 216000 \* \Sss(1,1,-2)
          - 105930 \* \Sss(1,1,1)
          + 327780 \* \Sss(1,1,2)
          - 298080 \* \Sss(1,1,3)
          + 51840 \* \Sss(1,2,-2)
  \nonumber\\&& \mbox{}
          + 508140 \* \Sss(1,2,1)
          - 116640 \* \Sss(1,2,2)
          + 12960 \* \Sss(1,3,1)
          + 417960 \* \Sss(2,-3,1)
          - 38880 \* \Sss(2,-2,-2)
  \nonumber\\&& \mbox{}
          + 554040 \* \Sss(2,-2,1)
          + 213840 \* \Sss(2,-2,2)
          - 417960 \* \Sss(2,1,-3)
          + 456840 \* \Sss(2,1,-2)
          - 236790 \* \Sss(2,1,1)
  \nonumber\\&& \mbox{}
          - 259200 \* \Sss(2,1,2)
          + 204120 \* \Sss(2,1,3)
          - 369360 \* \Sss(2,2,-2)
          - 136080 \* \Sss(2,2,1)
          + 359640 \* \Sss(2,2,2)
  \nonumber\\&& \mbox{}
          + 787320 \* \Sss(2,3,1)
          + 272160 \* \Sss(3,-2,1)
          - 330480 \* \Sss(3,1,-2)
          - 317520 \* \Sss(3,1,1)
          + 602640 \* \Sss(3,1,2)
  \nonumber\\&& \mbox{}
          + 447120 \* \Sss(3,2,1)
          + 515160 \* \Sss(4,1,1)
          + 142560 \* \Ssss(1,-2,1,1)
          - 90720 \* \Ssss(1,1,-2,1)
          - 64800 \* \Ssss(1,1,1,-2)
  \nonumber\\&& \mbox{}
          - 402300 \* \Ssss(1,1,1,1)
          - 233280 \* \Ssss(2,-2,1,1)
          + 136080 \* \Ssss(2,1,-2,1)
          + 427680 \* \Ssss(2,1,1,-2)
          + 199260 \* \Ssss(2,1,1,1)
  \nonumber\\&& \mbox{}
          - 330480 \* \Ssss(2,1,1,2)
          - 330480 \* \Ssss(2,1,2,1)
          - 291600 \* \Ssss(2,2,1,1)
          - 544320 \* \Ssss(3,1,1,1)
          + 272160 \* \Sssss(2,1,1,1,1)
  \nonumber\\&& \mbox{}
          - 2936744 \* \S(1)
          + 1329048 \* \S(1) \* \z3
          + 623655 \* \S(2)
          - 174960 \* \S(2) \* \z4
          - 2349000 \* \S(2) \* \z3
  \nonumber\\&& \mbox{}
          - 5616828 \* \S(3)
          + 1030320 \* \S(3) \* \z3
          + 1171530 \* \S(4)
          - 1777140 \* \S(5) )
  \nonumber\\&& \mbox{}
          + {16 \over 9} \* (2 \* \Nminus + 3) \* (17 \* \Ssss(1,1,1,2)
          + 17 \* \Ssss(1,1,2,1)
          + 15 \* \Ssss(1,2,1,1)
          - 14 \* \Sssss(1,1,1,1,1)
          + 9 \* \S(1) \* \z4)
          \biggr)\biggr\}
\:\: .
\eea
\normalsize

\noindent
The leading-order coefficient functions for the logitudinal structure
function are given by
\bea
c^{(1)}_{L,\rm{q}}(N) & \! = \! &  
        \colour4colour{\cf}  \*  (
            4 \* (\Nplus - 1) \* \S(1)
          )
\:\: , \\[1mm]
c^{(1)}_{L,\rm{g}}(N) & \! = \! &  
        \colour4colour{\nf}  \*  (
          - 8 \* (\Nplustwo - 2 \* \Nplus + 1) \* \S(1) 
          )
\:\: .
\eea
The corresponding second-order (NLO) corrections~\cite
{SanchezGuillen:1991iq,Zijlstra:1991qc,Moch:1999eb} read
\small
\bea
&& c^{(2)}_{L,\rm ns}(N) \:\: = \:\: 
         \delta(N-2) \* \biggl\{
         \colour4colour{\cf \* \nf}  \*  \biggl(
          - {92 \over 27}
          \biggr)
       + \colour4colour{\cf^2}  \*  \biggl(
          - {1906 \over 135}
          + {64 \over 5} \* \z3
          \biggr)
       + \colour4colour{\ca \* \cf}  \*  \biggl(
            {2878 \over 135}
          - {32 \over 5} \* \z3
          \biggr)\biggr\}
  \nonumber\\&& \mbox{}
       + \theta(N-4) \* \biggl\{ \colour4colour{\cf \* \biggl(
         \cf-{\ca \over 2}\biggr)}  \*  \biggl(
            128 \* \Ss(1,-2)
          - {64 \over 5} \* (\Nminusthree - \Nminustwo) \* \Ss(1,-2) 
          + {96 \over 5} \* (\Nplusthree - \Nplustwo) \* (\Ss(1,-2) 
  \nonumber\\&& \mbox{}
            + \S(3))
          + {96 \over 5} \* (\Nplustwo - 3) \* (\S(1)
          - \S(2))
          - {64 \over 5} \* (\Nminustwo - \Nminus) \* (\S(1)
          + \S(2))
          \biggr)
       + \colour4colour{\cf \* \nf}  \*  \biggl(
          - {4 \over 9} \* \gqq \* (6 \* \Ss(1,1)
          + 25 \* \S(1) 
  \nonumber\\&& \mbox{}
	    - 12 \* \S(2))
          + {16 \over 3} \* \S(1)
          + {4 \over 9} \* (\Nminus + 1) \* (6 \* \Ss(1,1)
          + 19 \* \S(1)
          - 12 \* \S(2))
          \biggr)
       + \colour4colour{\cf^2}  \*  \biggl(
          - 16 \* \Ss(1,1)
          + {2 \over 5} \* \gqq \* (80 \* \Ss(1,-3) 
  \nonumber\\&& \mbox{}
	    - 80 \* \Ss(1,-2)
          - 70 \* \Ss(1,1)
          - 40 \* \Ss(1,2)
          - 80 \* \Ss(1,3)
          - 60 \* \Ss(2,1) 
	    - 160 \* \Sss(1,1,-2)
          + 40 \* \Sss(1,1,1)
          - 235 \* \S(1)
          + 240 \* \S(1) \* \z3 
  \nonumber\\&& \mbox{}
	    + 152 \* \S(2)
          - 40 \* \S(3))
          + 56 \* \S(1)
          - {176 \over 5} \* \S(2)
          - {2 \over 5} \* (\Nminus + 1) \* (80 \* \Ss(1,-3)
          + 80 \* \Ss(1,-2) 
	    - 90 \* \Ss(1,1)
          - 40 \* \Ss(1,2)
  \nonumber\\&& \mbox{} 
          - 80 \* \Ss(1,3) 
	     - 60 \* \Ss(2,1)
          - 160 \* \Sss(1,1,-2)
          + 40 \* \Sss(1,1,1)
          - 213 \* \S(1) 
	    + 240 \* S(1) \* \z3 + 156 \* \S(2)
          - 40 \* \S(3))
          \biggr)
  \nonumber\\&& \mbox{}
       + \colour4colour{\ca \* \cf}  \*  \biggl(
          - {2 \over 45} \* \gqq \* (360 \* \Ss(1,-3) 
	    - 360 \* \Ss(1,-2)
          - 690 \* \Ss(1,1)
          - 360 \* \Ss(1,3)
          - 720 \* \Sss(1,1,-2) 
	    - 1945 \* \S(1) 
  \nonumber\\&& \mbox{}
	    + 1080 \* \S(1) \* \z3 + 1344 \* \S(2)
          - 360 \* \S(3))
          - {136 \over 3} \* \S(1)
          + {128 \over 5} \* \S(2)
          + {2 \over 45} \* (\Nminus + 1) \* (360 \* \Ss(1,-3)
          + 360 \*
         \Ss(1,-2) 
  \nonumber\\&& \mbox{}
	    - 690 \* \Ss(1,1)
          - 360 \* \Ss(1,3) 
	    - 720 \* \Sss(1,1,-2)
          - 1651 \* \S(1)
          + 1080
          \* \S(1) \* \z3 + 1272 \* \S(2)
          - 360 \* \S(3))
          \biggr)\biggr\}
\:\: , 
\eea
\bea
&& c^{(2)}_{L,\rm{g}}(N) \:\: = \:\:  
         \delta(N-2) \* \biggl\{
         \colour4colour{\cf \* \nf}  \*  \biggl(
          - {116 \over 135}
          - {16 \over 5} \* \z3
          \biggr)
       + \colour4colour{\ca \* \nf}  \*  \biggl(
            {173 \over 27}
          \biggr)\bigg\}
  \nonumber\\&& \mbox{}
       + \theta(N-4) \* \biggl\{ \colour4colour{\cf \* \nf}  \*  \biggl(
            - {32 \over 15} \* (\Nminusthree - \Nminustwo) \* \Ss(1,-2) 
            + {8 \over 5} \* \gqg \* (4 \* \Ss(1,-2)
          + 10 \* \Ss(1,1)
          + 21 \* \S(1)
          - 6 \* \S(2) 
	    + 4 \* \S(3))
  \nonumber\\&& \mbox{}
          - {64 \over 5} \* (\Nplusthree - 1) \* (\Ss(1,-2)
          + \S(3))
          + {4 \over 15} \* (2 \* \Nplus + \Nminus - 3) \* (68 \* \Ss(1,-2)
          + 15 \* \Ss(1,1) 
	    + 30 \* \Ss(2,1)
          + 12 \* \S(1)
          + 42 \* \S(2) 
  \nonumber\\&& \mbox{}
	    + 8 \* \S(3))
          - {4 \over 15} \* (\Nminus - 1) \* (44 \* \Ss(1,-2)
          - 75 \* \Ss(1,1)
          + 30 \* \Ss(2,1) 
            - 154 \* \S(1)
          + 96 \* \S(2) 
	    - 16 \* \S(3))
          - {32 \over 15} \* (\Nminustwo - 1) \* (\S(1) 
  \nonumber\\&& \mbox{}
	    + \S(2))
          \biggr)
       + \colour4colour{\ca \* \nf}  \*  \biggl(
            {8 \over 9} \* \gqg \* (18 \* \Ss(1,-2)
          - 87 \* \Ss(1,1)
          + 18 \* \Ss(1,2) 
	    + 18 \* \Ss(2,1)
          - 18 \* \Sss(1,1,1)
          - 53 \* \S(1)
          + 117 \* \S(2)) 
  \nonumber\\&& \mbox{}
            - {8 \over 9} \* (2 \* \Nplus + \Nminus - 3) \* (6 \* \Ss(1,1)
          + 36 \* \Ss(2,1) 
	    + 2 \* \S(1)
          - 45 \* \S(2) 
	    - 54 \* \S(3))
          + {8 \over 9} \* (\Nminus - 1) \* (18 \* \Ss(1,-2)
          - 105 \* \Ss(1,1) 
  \nonumber\\&& \mbox{} 
	    + 18 \* \Ss(1,2)
          + 54 \* \Ss(2,1)
          - 18 \* \Sss(1,1,1)
          - 59 \* \S(1)
          + 90 \* \S(2) 
	   - 54 \* \S(3))
          + {16 \over 9} \* (\Nminustwo - 1) \* (3 \* \Ss(1,1)
          + \S(1))
          \biggr)\biggr\}
\eea
\normalsize
and
\small
\bea
&& c^{(2)}_{L,\rm{ps}}(N) \:\: = \:\: 
         \colour4colour{\cf \* \nf}  \*  \biggl(
            {32 \over 3} \* \Ss(1,1)
          + {16 \over 9} \* \gqq \* (6 \* \Ss(1,1)
          - 9 \* \Ss(2,1)
          - 22 \* \S(1)
          - 9 \* \S(2) 
	    + 18 \* \S(3))
          + {416 \over 9} \* \S(1)
  \nonumber\\&& \mbox{}
          + 32 \* \S(2)
          - {32 \over 9} \* (\Nplustwo - 3) \* (3 \* \Ss(1,1)
          - 5 \* \S(1)
          - 9 \* \S(2))
          + {16 \over 9} \* (\Nminustwo - \Nminus) \* (3 \* \Ss(1,1)
          + \S(1))
  \nonumber\\&& \mbox{}
          - {16 \over 9} \* (\Nminus + 1) \* (15 \* \Ss(1,1) 
	    - 9 \* \Ss(2,1)
          - 19 \* \S(1)
          - 18 \* \S(2)
          + 18 \* \S(3))
          \biggr)
\:\: .
\eea
\normalsize
For the third-order (N$^2$LO) non-singlet coefficient function of 
$F_L$ we obtain
\small
\bea
&& c^{(3)}_{L,\rm ns}(N) \:\: = \:\: 
         \delta(N-2) \* \biggl\{
         \colour4colour{\dabcnc} \* \fl11  \*  \biggl(
            {1312 \over 15}
          + 512 \* \z5
          - {2688 \over 5} \* \z3
          \biggr)
       + \colour4colour{\cf \* \nf^2}  \*  \biggl(
            {2168 \over 243}
          \biggr)
  \nonumber\\&& \mbox{}
       + \colour4colour{\cf^2 \* \nf}  \*  \biggl(
            {25534 \over 405}
          - {2848 \over 45} \* \z3
          \biggr)
       + \colour4colour{\cf^3}  \*  \biggl(
          - {232798 \over 1215}
          + {3584 \over 3} \* \z5
          - {39424 \over 45} \* \z3
          \biggr)
  \nonumber\\&& \mbox{}
       + \colour4colour{\ca \* \cf \* \nf}  \*  \biggl(
          - {204548 \over 1215}
          + {320 \over 3} \* \z5
          - {1568 \over 45} \* \z3
          \biggr)
       + \colour4colour{\ca \* \cf^2}  \*  \biggl(
          - {41536 \over 405}
          - {5248 \over 3} \* \z5
          + {73504 \over 45} \* \z3
          \biggr)
  \nonumber\\&& \mbox{}
       + \colour4colour{\ca^2 \* \cf}  \*  \biggl(
            {548668 \over 1215}
          - {3680 \over 9} \* \z3
          + 224 \* \z5
          \biggr)\biggr\}
  \nonumber\\&& \mbox{}
       + \theta(N-4) \* \biggl\{ \colour4colour{\dabcnc} \* \fl11  \*  \biggl(
            {512 \over 15} \* \Ss(1,-3)
          - {2560 \over 3} \* \Ss(1,-2)
          - {3328 \over 15} \* \Ss(1,1)
          - {512 \over 15} \* \Ss(1,3)
          + 1536 \* \Ss(2,-2)
  \nonumber\\&& \mbox{}
          - {33536 \over 15} \* \Ss(2,1)
          + 1024 \* \Ss(2,3)
          - {14336 \over 15} \* \Ss(3,1)
          - 1024 \* \Ss(4,1)
          + {14336 \over 15} \* \Sss(1,-2,1)
          - 1024 \* \Sss(1,1,-2)
  \nonumber\\&& \mbox{}
          - {64 \over 15} \* \gqq \* (30 \* \Ss(1,-4)
          + 124 \* \Ss(1,-3)
          - 250 \* \Ss(1,-2)
          - 240 \* \Ss(1,-2) \* \z3
          + 174 \* \Ss(1,1)
          + 280 \* \Ss(1,1) \* \z3
          - 54 \* \Ss(1,3)
  \nonumber\\&& \mbox{}
          - 30 \* \Ss(1,4)
          + 20 \* \Ss(2,-3)
          + 160 \* \Ss(2,-2)
          - 408 \* \Ss(2,1)
          - 240 \* \Ss(2,1) \* \z3
          + 100 \* \Ss(2,3)
          - 154 \* \Ss(3,1)
  \nonumber\\&& \mbox{}
          - 120 \* \Ss(4,1)
          - 120 \* \Sss(1,-4,1)
          - 80 \* \Sss(1,-3,1)
          + 60 \* \Sss(1,-2,-2)
          - 58 \* \Sss(1,-2,1)
          + 120 \* \Sss(1,-2,3)
          - 80 \* \Sss(1,1,-3)
  \nonumber\\&& \mbox{}
          - 190 \* \Sss(1,1,-2)
          - 100 \* \Sss(1,1,3)
          + 100 \* \Sss(1,3,1)
          - 100 \* \Sss(2,-2,1)
          + 60 \* \Sss(2,1,-2)
          + 120 \* \Sss(2,1,3)
          - 120 \* \Sss(2,3,1)
  \nonumber\\&& \mbox{}
          + 160 \* \Ssss(1,1,-2,1)
          + 90 \* \S(1)
          - 600 \* \S(1) \* \z5
          + 489 \* \S(1) \* \z3
          - 198 \* \S(2)
          - 280 \* \S(2) \* \z3
          + 218 \* \S(3)
          - 20 \* \S(4))
  \nonumber\\&& \mbox{}
          + {15872 \over 15} \* \S(1) \* \z3
          - {18944 \over 15} \* \S(2)
          - 2048 \* \S(2) \* \z3
          + {4096 \over 3} \* \S(3)
  \nonumber\\&& \mbox{}
          - {256 \over 5} \* (\Nplusthree - \Nplustwo) \* (10 \* \Ss(1,-3)
          - 13 \* \Ss(1,-2)
          + 4 \* \Ss(1,1) \* \z3 
          - 4 \* \Ss(2,-3)
          + 10 \* \Ss(2,-2)
          + 4 \* \Ss(2,3)
          - 10 \* \Ss(3,1)
  \nonumber\\&& \mbox{}
          + 4 \* \Sss(1,-3,1)
          - 10 \* \Sss(1,-2,1)
          + 4 \* \Sss(1,1,-3)
          - 10 \* \Sss(1,1,-2)
          - 4 \* \Sss(1,1,3)
          + 4 \* \Sss(1,3,1)
          + 8 \* \Sss(2,-2,1)
          - 8 \* \Ssss(1,1,-2,1)
  \nonumber\\&& \mbox{}
          - 4 \* \S(2) \* \z3
          - 13 \* \S(3)
          + 10 \* \S(4) )
          + {512 \over 15} \* (\Nminusthree - \Nminustwo) \* (10 \* \Ss(1,-3)
          - 13 \* \Ss(1,-2)
          + 4 \* \Ss(1,1) \* \z3
          + 4 \* \Sss(1,-3,1)
  \nonumber\\&& \mbox{}
          - 10 \* \Sss(1,-2,1)
          + 4 \* \Sss(1,1,-3)
          - 10 \* \Sss(1,1,-2)
          - 4 \* \Sss(1,1,3)
          + 4 \* \Sss(1,3,1)
          - 8 \* \Ssss(1,1,-2,1))
          + {256 \over 5} \* (\Nplustwo - 3) \* (4 \* \Ss(1,-3)
  \nonumber\\&& \mbox{}
          - 10 \* \Ss(1,-2)
          + 10 \* \Ss(1,1)
          + 20 \* \Ss(1,1) \* \z3 
          - 4 \* \Ss(1,3)
          + 15 \* \Ss(2,-2)
          - 10 \* \Ss(2,1)
          + 10 \* \Ss(2,3)
          - 7 \* \Ss(3,1)
          - 10 \* \Ss(4,1)
  \nonumber\\&& \mbox{}
          + 7 \* \Sss(1,-2,1)
          - 15 \* \Sss(1,1,-2)
          - 10 \* \Sss(1,1,3)
          + 10 \* \Sss(1,3,1)
          + 3 \* \S(1)
          + 19 \* \S(1) \* \z3
          - 13 \* \S(2)
          - 20 \* \S(2) \* \z3
          + 10 \* \S(3) )
  \nonumber\\&& \mbox{}
          - {128 \over 15} \* (\Nminustwo - \Nminus) \* (16 \* \Ss(1,-3)
          - 40 \* \Ss(1,-2)
          + 40 \* \Ss(1,1)
          + 60 \* \Ss(1,1) \* \z3 
          - 16 \* \Ss(1,3)
          + 40 \* \Ss(2,1)
  \nonumber\\&& \mbox{}
          + 13 \* \Sss(1,-2,1)
          - 45 \* \Sss(1,1,-2)
          - 30 \* \Sss(1,1,3)
          + 30 \* \Sss(1,3,1)
          + 12 \* \S(1)
          + 61 \* \S(1) \* \z3 
          + 12 \* \S(2)
          - 40 \* \S(3) )
  \nonumber\\&& \mbox{}
          + {64 \over 15} \* (\Nminus + 1) \* (30 \* \Ss(1,-4)
          + 168 \* \Ss(1,-3)
          - 270 \* \Ss(1,-2)
          - 240 \* \Ss(1,-2) \* \z3
          + 320 \* \Ss(1,1)
          - 98 \* \Ss(1,3)
          - 30 \* \Ss(1,4)
  \nonumber\\&& \mbox{}
          + 520 \* \Ss(1,1) \* \z3
          + 20 \* \Ss(2,-3)
          + 160 \* \Ss(2,-2)
          - 266 \* \Ss(2,1)
          - 240 \* \Ss(2,1) \* \z3
          + 100 \* \Ss(2,3)
          - 126 \* \Ss(3,1)
          - 120 \* \Ss(4,1)
  \nonumber\\&& \mbox{}
          - 120 \* \Sss(1,-4,1)
          - 80 \* \Sss(1,-3,1)
          + 60 \* \Sss(1,-2,-2)
          - 86 \* \Sss(1,-2,1)
          + 120 \* \Sss(1,-2,3)
          - 80 \* \Sss(1,1,-3)
          - 250 \* \Sss(1,1,-2)
  \nonumber\\&& \mbox{}
          - 220 \* \Sss(1,1,3)
          + 220 \* \Sss(1,3,1)
          - 100 \* \Sss(2,-2,1)
          + 60 \* \Sss(2,1,-2)
          + 120 \* \Sss(2,1,3)
          - 120 \* \Sss(2,3,1)
          + 160 \* \Ssss(1,1,-2,1)
  \nonumber\\&& \mbox{}
          + 126 \* \S(1)
          - 600 \* \S(1) \* \z5
          + 593 \* \S(1) \* \z3
          - 206 \* \S(2)
          - 280 \* \S(2) \* \z3
          + 178 \* \S(3)
          - 20 \* \S(4) )
          \biggr)
  \nonumber\\&& \mbox{}
       + \colour4colour{\cf \* \biggl(\cf-{\ca \over 2}\biggr)^2}  \*  \biggl(
            32 \* \gfunct1(N)
          - 1024 \* \Ss(1,4)
          \biggr)
  \nonumber\\&& \mbox{}
       + \colour4colour{\cf \* \nf^2}  \*  \biggl(
          - {64 \over 9} \* \Ss(1,1)
          + {8 \over 81} \* \gqq \* (150 \* \Ss(1,1)
          - 36 \* \Ss(1,2)
          - 72 \* \Ss(2,1)
          + 36 \* \Sss(1,1,1)
          + 317 \* \S(1)
          - 300 \* \S(2)
  \nonumber\\&& \mbox{}
          + 108 \* \S(3))
          - {608 \over 27} \* \S(1)
          + {64 \over 9} \* \S(2)
          - {8 \over 81} \* (\Nminus + 1) \* (
            114 \* \Ss(1,1)
          - 36 \* \Ss(1,2)
          - 72 \* \Ss(2,1)
          + 36 \* \Sss(1,1,1)
  \nonumber\\&& \mbox{}
          + 203 \* \S(1)
          - 264 \* \S(2)
          + 108 \* \S(3))
          \biggr)
  \nonumber\\&& \mbox{}
       + \colour4colour{\cf^2 \* \biggl(\cf-{\ca \over 2}\biggr)}  \*  \biggl(
          - 256 \* \gqq \* (5 \* \S(1) \* \z5 
          + 3 \* \S(3) \* \z3)
          + 256 \* (\Nminus + 1) \* (5 \* \S(1) \* \z5 
          + 3 \* \S(3) \* \z3)
          + 2048 \* \Ss(2,-3)
  \nonumber\\&& \mbox{}
          + 512 \* \Ss(3,-2)
          - 512 \* \Sss(1,-2,2)
          - 1024 \* \Sss(1,2,-2)
          - 1536 \* \Sss(2,-2,1)
          - 1536 \* \Sss(2,1,-2)
          + 512 \* \Ssss(1,-2,1,1)
  \nonumber\\&& \mbox{}
          + 1024 \* \Ssss(1,1,1,-2)
          \biggr)
  \nonumber\\&& \mbox{}
       + \colour4colour{\cf^2 \* \nf}  \*  \biggl(
            512 \* \Ss(1,-3)
          - {30848 \over 45} \* \Ss(1,-2)
          + {32 \over 9} \* \Ss(1,1)
          + {160 \over 3} \* \Ss(1,2)
          + 512 \* \Ss(2,-2)
          - {2176 \over 15} \* \Ss(2,1)
  \nonumber\\&& \mbox{}
          + {128 \over 3} \* \Ss(2,2)
          - {640 \over 3} \* \Ss(3,1)
          - {1024 \over 3} \* \Sss(1,1,-2)
          + {160 \over 3} \* \Sss(1,1,1)
          - {128 \over 3} \* \Sss(1,1,2) \* (2 \* \Nminus + 3)
          + {128 \over 3} \* \Sss(1,2,1)
  \nonumber\\&& \mbox{}
          + {1 \over 675} \* \gqq \* (86400 \* \Ss(1,-4)
          - 163200 \* \Ss(1,-3)
          + 180960 \* \Ss(1,-2)
          + 110000 \* \Ss(1,1)
          - 144000 \* \Ss(1,1) \* \z3 
  \nonumber\\&& \mbox{}
          - 22200 \* \Ss(1,2)
          + 58800 \* \Ss(1,3)
          - 86400 \* \Ss(1,4)
          + 28800 \* \Ss(2,-3)
          - 72000 \* \Ss(2,-2)
          + 37080 \* \Ss(2,1)
  \nonumber\\&& \mbox{}
          - 61200 \* \Ss(2,2)
          - 28800 \* \Ss(2,3)
          + 25200 \* \Ss(3,1)
          - 57600 \* \Sss(1,-2,-2)
          - 57600 \* \Sss(1,-2,1)
          - 172800 \* \Sss(1,1,-3)
  \nonumber\\&& \mbox{}
          + 268800 \* \Sss(1,1,-2)
          - 4200 \* \Sss(1,1,1)
          + 57600 \* \Sss(1,1,2)
          - 115200 \* \Sss(1,2,-2)
          + 7200 \* \Sss(1,2,1)
          - 14400 \* \Sss(1,2,2)
  \nonumber\\&& \mbox{}
          + 72000 \* \Sss(1,3,1)
          + 57600 \* \Sss(2,-2,1)
          - 115200 \* \Sss(2,1,-2)
          + 46800 \* \Sss(2,1,1)
          + 115200 \* \Ssss(1,1,1,-2)
  \nonumber\\&& \mbox{}
          - 28800 \* \Ssss(1,1,1,1)
          + 14400 \* \Ssss(1,1,1,2)
          - 14400 \* \Ssss(1,1,2,1)
          + 236725 \* \S(1)
          - 237600 \* \S(1) \* \z3 
          - 290248 \* \S(2)
  \nonumber\\&& \mbox{}
          + 144000 \* \S(2) \* \z3 
          + 270960 \* \S(3)
          - 30600 \* \S(4) )
          - {6416 \over 27} \* \S(1)
          + {1280 \over 3} \* \S(1) \* \z3
          + {22144 \over 225} \* \S(2)
          - {784 \over 5} \* \S(3)
  \nonumber\\&& \mbox{}
          + {32 \over 75} \* (\Nplusthree - \Nplustwo) \* (180 \* \Ss(1,-3)
          - 159 \* \Ss(1,-2)
          + 60 \* \Ss(2,-2)
          + 125 \* \Ss(2,2)
          - 185 \* \Ss(3,1)
          - 60 \* \Sss(1,-2,1)
  \nonumber\\&& \mbox{}
          - 60 \* \Sss(1,1,-2)
          - 125 \* \Sss(1,1,2)
          + 125 \* \Sss(1,2,1)
          - 570 \* \S(1) \* \z3
          - 159 \* \S(3)
          + 180 \* \S(4) )
  \nonumber\\&& \mbox{}
          - {64 \over 225} \* (\Nminusthree - \Nminustwo) \* (180 \* \Ss(1,-3)
          - 149 \* \Ss(1,-2)
          + 120 \* \Ss(2,-2)
          - 60 \* \Sss(1,-2,1)
          - 60 \* \Sss(1,1,-2)
          + 180 \* \S(1) \* \z3 )
  \nonumber\\&& \mbox{}
          + {32 \over 75} \* (\Nplustwo - 3) \* (60 \* \Ss(1,-2)
          + 65 \* \Ss(1,1)
          + 125 \* \Ss(1,2)
          + 200 \* \Ss(2,-2)
          - 65 \* \Ss(2,1)
          + 50 \* \Ss(2,2)
          - 250 \* \Ss(3,1)
  \nonumber\\&& \mbox{}
          + 200 \* \Sss(1,-2,1)
          - 200 \* \Sss(1,1,-2)
          - 50 \* \Sss(1,1,2)
          + 50 \* \Sss(1,2,1)
          - 219 \* \S(1)
          - 100 \* \S(1) \* \z3 
          + 154 \* \S(2)
          - 180 \* \S(3) )
  \nonumber\\&& \mbox{}
          - {64 \over 225} \* (\Nminustwo - \Nminus) \* (60 \* \Ss(1,-2)
          - 60 \* \Ss(1,1)
          - 60 \* \Ss(2,1)
          + 150 \* \Sss(1,-2,1)
          - 150 \* \Sss(1,1,-2)
          - 209 \* \S(1)
  \nonumber\\&& \mbox{}
          + 150 \* \S(1) \* \z3
          - 89 \* \S(2)
          + 180 \* \S(3) )
          - {1 \over 675} \* (\Nminus + 1) \* (86400 \* \Ss(1,-4)
          + 9600 \* \Ss(1,-3)
          - 67680 \* \Ss(1,-2)
  \nonumber\\&& \mbox{}
          + 92480 \* \Ss(1,1)
          - 144000 \* \Ss(1,1) \* \z3
          - 40200 \* \Ss(1,2)
          + 58800 \* \Ss(1,3)
          - 86400 \* \Ss(1,4)
          + 28800 \* \Ss(2,-3)
  \nonumber\\&& \mbox{}
          + 43200 \* \Ss(2,-2)
          + 6840 \* \Ss(2,1)
          - 61200 \* \Ss(2,2)
          - 28800 \* \Ss(2,3)
          + 25200 \* \Ss(3,1)
          - 57600 \* \Sss(1,-2,-2)
  \nonumber\\&& \mbox{}
          - 115200 \* \Sss(1,-2,1)
          - 172800 \* \Sss(1,1,-3)
          + 211200 \* \Sss(1,1,-2)
          + 13800 \* \Sss(1,1,1)
          - 115200 \* \Sss(1,2,-2)
  \nonumber\\&& \mbox{}
          + 7200 \* \Sss(1,2,1)
          - 14400 \* \Sss(1,2,2)
          + 72000 \* \Sss(1,3,1)
          + 57600 \* \Sss(2,-2,1)
          - 115200 \* \Sss(2,1,-2)
          + 46800 \* \Sss(2,1,1)
  \nonumber\\&& \mbox{}
          + 115200 \* \Ssss(1,1,1,-2)
          - 28800 \* \Ssss(1,1,1,1)
          + 14400 \* \Ssss(1,1,1,2)
          - 14400 \* \Ssss(1,1,2,1)
          + 219597 \* \S(1)
  \nonumber\\&& \mbox{}
          - 64800 \* \S(1) \* \z3
          - 301384 \* \S(2)
          + 144000 \* \S(2) \* \z3
          + 269880 \* \S(3)
          - 30600 \* \S(4) )
          \biggr)
  \nonumber\\&& \mbox{}
       + \colour4colour{\cf^3}  \*  \biggl(
            2560 \* \Ss(1,-4)
          - {9472 \over 3} \* \Ss(1,-3)
          - {111104 \over 75} \* \Ss(1,-2)
          - {43984 \over 15} \* \Ss(1,1)
          - {640 \over 3} \* \Ss(1,1) \* \z3
          - 512 \* \Ss(1,2)
  \nonumber\\&& \mbox{}
          + {4384 \over 5} \* \Ss(1,3)
          + {22144 \over 15} \* \Ss(2,-2)
          + {115936 \over 75} \* \Ss(2,1)
          + {224 \over 5} \* \Ss(2,2)
          - {3200 \over 3} \* \Ss(2,3)
          - {27392 \over 15} \* \Ss(3,1)
          + {3200 \over 3} \* \Ss(4,1)
  \nonumber\\&& \mbox{}
          - 4736 \* \Sss(1,-3,1)
          + 1024 \* \Sss(1,-2,-2)
          + {12416 \over 3} \* \Sss(1,-2,1)
          - 5504 \* \Sss(1,1,-3)
          + {8576 \over 5} \* \Sss(1,1,-2)
          + 448 \* \Sss(1,1,1)
  \nonumber\\&& \mbox{}
          + 256 \* \Sss(1,1,2)
          + {1280 \over 3} \* \Sss(1,1,3)
          + 192 \* \Sss(1,2,1)
          - {512 \over 3} \* \Sss(1,3,1)
          - {64 \over 5} \* \Sss(2,1,1)
          + 6912 \* \Ssss(1,1,-2,1)
          - 192 \* \Ssss(1,1,1,1)
  \nonumber\\&& \mbox{}
          + {1 \over 450} \* \gqq \* (288000 \* \Ss(1,-5)
          - 494400 \* \Ss(1,-4)
          + 782400 \* \Ss(1,-3)
          + 102752 \* \Ss(1,-2)
          - 518400 \* \Ss(1,-2) \* \z3
  \nonumber\\&& \mbox{}
          + 1023360 \* \Ss(1,1)
          + 628800 \* \Ss(1,1) \* \z3
          + 207000 \* \Ss(1,2)
          - 57600 \* \Ss(1,2) \* \z3
          - 52080 \* \Ss(1,3)
          + 249600 \* \Ss(1,4)
  \nonumber\\&& \mbox{}
          - 288000 \* \Ss(1,5)
          + 172800 \* \Ss(2,-4)
          - 465600 \* \Ss(2,-3)
          + 161280 \* \Ss(2,-2)
          + 335496 \* \Ss(2,1)
  \nonumber\\&& \mbox{}
          - 547200 \* \Ss(2,1) \* \z3
          - 195840 \* \Ss(2,2)
          + 381600 \* \Ss(2,3)
          - 172800 \* \Ss(2,4)
          + 28800 \* \Ss(3,-3)
  \nonumber\\&& \mbox{}
          - 96000 \* \Ss(3,-2)
          - 1002720 \* \Ss(3,1)
          - 50400 \* \Ss(3,2)
          - 28800 \* \Ss(3,3)
          - 88800 \* \Ss(4,1)
          - 633600 \* \Sss(1,-4,1)
  \nonumber\\&& \mbox{}
          + 57600 \* \Sss(1,-3,-2)
          + 1012800 \* \Sss(1,-3,1)
          - 57600 \* \Sss(1,-3,2)
          + 172800 \* \Sss(1,-2,-3)
          - 412800 \* \Sss(1,-2,-2)
  \nonumber\\&& \mbox{}
          - 1540800 \* \Sss(1,-2,1)
          + 57600 \* \Sss(1,-2,2)
          + 115200 \* \Sss(1,-2,3)
          - 691200 \* \Sss(1,1,-4)
          + 1214400 \* \Sss(1,1,-3)
  \nonumber\\&& \mbox{}
          + 348480 \* \Sss(1,1,-2)
          - 189000 \* \Sss(1,1,1)
          - 57600 \* \Sss(1,1,1) \* \z3 
          + 136800 \* \Sss(1,1,2)
          - 518400 \* \Sss(1,1,3)
  \nonumber\\&& \mbox{}
          + 345600 \* \Sss(1,1,4)
          - 864000 \* \Sss(1,2,-3)
          + 259200 \* \Sss(1,2,-2)
          + 172800 \* \Sss(1,2,1)
          - 288000 \* \Sss(1,3,-2)
  \nonumber\\&& \mbox{}
          + 86400 \* \Sss(1,2,2)
          + 504000 \* \Sss(1,3,1)
          + 57600 \* \Sss(1,3,2)
          + 518400 \* \Sss(1,4,1)
          - 86400 \* \Sss(2,-3,1)
          + 988800 \* \Sss(2,-2,1)
  \nonumber\\&& \mbox{}
          - 374400 \* \Sss(2,1,-3)
          - 230400 \* \Sss(2,1,-2)
          + 260640 \* \Sss(2,1,1)
          + 144000 \* \Sss(2,1,2)
          + 144000 \* \Sss(2,1,3)
  \nonumber\\&& \mbox{}
          - 230400 \* \Sss(2,2,-2)
          + 129600 \* \Sss(2,2,1)
          + 28800 \* \Sss(2,3,1)
          + 57600 \* \Sss(3,-2,1)
          - 115200 \* \Sss(3,1,-2)
  \nonumber\\&& \mbox{}
          + 86400 \* \Sss(3,1,1)
          + 57600 \* \Ssss(1,-3,1,1)
          - 460800 \* \Ssss(1,-2,-2,1)
          + 115200 \* \Ssss(1,-2,1,-2)
          - 57600 \* \Ssss(1,-2,1,1)
  \nonumber\\&& \mbox{}
          + 1094400 \* \Ssss(1,1,-3,1)
          - 230400 \* \Ssss(1,1,-2,-2)
          - 1737600 \* \Ssss(1,1,-2,1)
          - 151200 \* \Ssss(1,1,1,1)
          + 57600 \* \Ssss(1,1,1,3)
  \nonumber\\&& \mbox{}
          + 115200 \* \Ssss(1,1,-2,2)
          + 1440000 \* \Ssss(1,1,1,-3)
          - 460800 \* \Ssss(1,1,1,-2)
          - 115200 \* \Ssss(1,1,1,2)
          + 345600 \* \Ssss(1,1,2,-2)
  \nonumber\\&& \mbox{}
          - 86400 \* \Ssss(1,1,2,1)
          - 172800 \* \Ssss(1,1,3,1)
          + 921600 \* \Ssss(1,2,-2,1)
          + 345600 \* \Ssss(1,2,1,-2)
          - 100800 \* \Ssss(1,2,1,1)
  \nonumber\\&& \mbox{}
          - 57600 \* \Ssss(1,3,1,1)
          + 172800 \* \Ssss(2,1,-2,1)
          + 345600 \* \Ssss(2,1,1,-2)
          - 129600 \* \Ssss(2,1,1,1)
          - 115200 \* \Sssss(1,1,-2,1,1)
  \nonumber\\&& \mbox{}
          - 1612800 \* \Sssss(1,1,1,-2,1)
          - 345600 \* \Sssss(1,1,1,1,-2)
          + 86400 \* \Sssss(1,1,1,1,1)
          + 426651 \* \S(1)
          - 1296000 \* \S(1) \* \z5
  \nonumber\\&& \mbox{}
          - 258720 \* \S(1)\* \z3
          - 1243248 \* \S(2)
          - 340800 \* \S(2) \* \z3
          - 401656 \* \S(3)
          + 489600 \* \S(3) \* \z3
          + 1082640 \* \S(4)
  \nonumber\\&& \mbox{}
          - 157200 \* \S(5) )
          - {84736 \over 225} \* \S(1)
          - {15232 \over 15} \* \S(1) \* \z3
          + {202384 \over 225} \* \S(2)
          + {8704 \over 3} \* \S(2) \* \z3
          - {63616 \over 75} \* \S(3)
          - {2672 \over 5} \* \S(4)
  \nonumber\\&& \mbox{}
          + {16 \over 25} \* (\Nplusthree - \Nplustwo) \* (400 \* \Ss(1,-4)
          - 948 \* \Ss(1,-3)
          + 337 \* \Ss(1,-2)
          + 1820 \* \Ss(1,1) \* \z3
          - 40 \* \Ss(1,4)
          + 1000 \* \Ss(2,-3)
  \nonumber\\&& \mbox{}
          - 788 \* \Ss(2,-2)
          + 770 \* \Ss(2,3)
          + 200 \* \Ss(3,-2)
          + 788 \* \Ss(3,1)
          - 120 \* \Ss(3,2)
          - 1770 \* \Ss(4,1)
          - 1000 \* \Sss(1,-3,1)
  \nonumber\\&& \mbox{}
          + 80 \* \Sss(1,-2,-2)
          + 788 \* \Sss(1,-2,1)
          - 120 \* \Sss(1,-2,2)
          - 1000 \* \Sss(1,1,-3)
          + 788 \* \Sss(1,1,-2)
          - 770 \* \Sss(1,1,3)
  \nonumber\\&& \mbox{}
          - 120 \* \Sss(1,2,-2)
          + 770 \* \Sss(1,3,1)
          - 1400 \* \Sss(2,-2,1)
          - 120 \* \Sss(2,1,-2)
          + 120 \* \Sss(3,1,1)
          + 120 \* \Ssss(1,-2,1,1)
  \nonumber\\&& \mbox{}
          + 1400 \* \Ssss(1,1,-2,1)
          + 120 \* \Ssss(1,1,1,-2)
          - 240 \* \S(1) \* \z3
          - 1820 \* \S(2) \* \z3
          + 337 \* \S(3)
          - 948 \* \S(4)
          + 440 \* \S(5) )
  \nonumber\\&& \mbox{}
          - {32 \over 225} \* (\Nminusthree - \Nminustwo) \* (1200 \* \Ss(1,-4)
          - 2784 \* \Ss(1,-3)
          + 1621 \* \Ss(1,-2)
          + 5460 \* \Ss(1,1) \* \z3 
          - 120 \* \Ss(1,4)
  \nonumber\\&& \mbox{}
          + 720 \* \Ss(2,-3)
          - 480 \* \Ss(2,-2)
          - 3000 \* \Sss(1,-3,1)
          + 240 \* \Sss(1,-2,-2)
          + 2304 \* \Sss(1,-2,1)
          - 360 \* \Sss(1,-2,2)
  \nonumber\\&& \mbox{}
          - 3000 \* \Sss(1,1,-3)
          + 2304 \* \Sss(1,1,-2)
          - 2310 \* \Sss(1,1,3)
          - 360 \* \Sss(1,2,-2)
          + 2310 \* \Sss(1,3,1)
          - 720 \* \Sss(2,-2,1)
  \nonumber\\&& \mbox{}
          - 720 \* \Sss(2,1,-2)
          + 360 \* \Ssss(1,-2,1,1)
          + 4200 \* \Ssss(1,1,-2,1)
          + 360 \* \Ssss(1,1,1,-2)
          - 720 \* \S(1) \* \z3 )
  \nonumber\\&& \mbox{}
          + {16 \over 75} \* (\Nplustwo - 3) \* (3000 \* \Ss(1,-3)
          - 2124 \* \Ss(1,-2)
          - 1506 \* \Ss(1,1)
          - 5000 \* \Ss(1,1) \* \z3
          - 360 \* \Ss(1,2)
  \nonumber\\&& \mbox{}
          + 2310 \* \Ss(1,3)
          + 4620 \* \Ss(2,-2)
          + 1146 \* \Ss(2,1)
          + 360 \* \Ss(2,2)
          - 2500 \* \Ss(2,3)
          - 3810 \* \Ss(3,1)
  \nonumber\\&& \mbox{}
          + 2500 \* \Ss(4,1)
          + 300 \* \Sss(1,-2,1)
          - 4860 \* \Sss(1,1,-2)
          + 360 \* \Sss(1,1,1)
          + 2500 \* \Sss(1,1,3)
          - 2500 \* \Sss(1,3,1)
  \nonumber\\&& \mbox{}
          - 360 \* \Sss(2,1,1)
          - 1353 \* \S(1)
          - 960 \* \S(1) \* \z3
          + 2859 \* \S(2)
          + 5000 \* \S(2) \* \z3
          - 1386 \* \S(3)
          - 1080 \* \S(4) )
  \nonumber\\&& \mbox{}
          - {32 \over 225} \* (\Nminustwo - \Nminus) \* (3000 \* \Ss(1,-3)
          - 2424 \* \Ss(1,-2)
          - 1566 \* \Ss(1,1)
          - 3750 \* \Ss(1,1) \* \z3 
          - 360 \* \Ss(1,2)
          + 2310 \* \Ss(1,3)
  \nonumber\\&& \mbox{}
          + 600 \* \Ss(2,-2)
          - 2286 \* \Ss(2,1)
          - 360 \* \Ss(2,2)
          - 5310 \* \Ss(3,1)
          - 825 \* \Sss(1,-2,1)
          - 3735 \* \Sss(1,1,-2)
          + 360 \* \Sss(1,1,1)
  \nonumber\\&& \mbox{}
          + 1875 \* \Sss(1,1,3)
          - 1875 \* \Sss(1,3,1)
          + 360 \* \Sss(2,1,1)
          - 1043 \* \S(1)
          - 2085 \* \S(1) \* \z3
          - 803 \* \S(2)
          - 2184 \* \S(3)
          + 1320 \* \S(4) )
  \nonumber\\&& \mbox{}
          - {1 \over 450} \* (\Nminus + 1) \* (288000 \* \Ss(1,-5)
          + 81600 \* \Ss(1,-4)
          - 216000 \* \Ss(1,-3)
          - 26656 \* \Ss(1,-2)
          - 518400 \* \Ss(1,-2) \* \z3
  \nonumber\\&& \mbox{}
          + 508176 \* \Ss(1,1)
          + 1060800 \* \Ss(1,1) \* \z3
          + 126360 \* \Ss(1,2)
          - 57600 \* \Ss(1,2) \* \z3
          - 76560 \* \Ss(1,3)
          + 19200 \* \Ss(1,4)
  \nonumber\\&& \mbox{}
          - 288000 \* \Ss(1,5)
          + 172800 \* \Ss(2,-4)
          - 4800 \* \Ss(2,-3)
          + 49920 \* \Ss(2,-2)
          + 573288 \* \Ss(2,1)
          - 547200 \* \Ss(2,1) \* \z3
  \nonumber\\&& \mbox{}
          - 220320 \* \Ss(2,2)
          + 381600 \* \Ss(2,3)
          - 172800 \* \Ss(2,4)
          + 28800 \* \Ss(3,-3)
          + 19200 \* \Ss(3,-2)
          - 1047840 \* \Ss(3,1)
  \nonumber\\&& \mbox{}
          - 50400 \* \Ss(3,2)
          - 28800 \* \Ss(3,3)
          - 88800 \* \Ss(4,1)
          - 633600 \* \Sss(1,-4,1)
          + 57600 \* \Sss(1,-3,-2)
          - 52800 \* \Sss(1,-3,1)
  \nonumber\\&& \mbox{}
          - 57600 \* \Sss(1,-3,2)
          + 172800 \* \Sss(1,-2,-3)
          - 182400 \* \Sss(1,-2,-2)
          - 638400 \* \Sss(1,-2,1)
          - 57600 \* \Sss(1,-2,2)
  \nonumber\\&& \mbox{}
          + 115200 \* \Sss(1,-2,3)
          - 691200 \* \Sss(1,1,-4)
          - 24000 \* \Sss(1,1,-3)
          + 1200960 \* \Sss(1,1,-2)
          - 122760 \* \Sss(1,1,1)
  \nonumber\\&& \mbox{}
          - 57600 \* \Sss(1,1,1) \* \z3
          + 194400 \* \Sss(1,1,2)
          - 662400 \* \Sss(1,1,3)
          + 345600 \* \Sss(1,1,4)
          - 864000 \* \Sss(1,2,-3)
  \nonumber\\&& \mbox{}
          + 28800 \* \Sss(1,2,-2)
          + 216000 \* \Sss(1,2,1)
          + 86400 \* \Sss(1,2,2)
          - 288000 \* \Sss(1,3,-2)
          + 705600 \* \Sss(1,3,1)
  \nonumber\\&& \mbox{}
          + 57600 \* \Sss(1,3,2)
          + 518400 \* \Sss(1,4,1)
          - 86400 \* \Sss(2,-3,1)
          + 643200 \* \Sss(2,-2,1)
          - 374400 \* \Sss(2,1,-3)
  \nonumber\\&& \mbox{}
          - 576000 \* \Sss(2,1,-2)
          + 292320 \* \Sss(2,1,1)
          + 144000 \* \Sss(2,1,2)
          + 144000 \* \Sss(2,1,3)
          - 230400 \* \Sss(2,2,-2)
  \nonumber\\&& \mbox{}
          + 129600 \* \Sss(2,2,1)
          + 28800 \* \Sss(2,3,1)
          + 57600 \* \Sss(3,-2,1)
          - 115200 \* \Sss(3,1,-2)
          + 86400 \* \Sss(3,1,1)
  \nonumber\\&& \mbox{}
          + 57600 \* \Ssss(1,-3,1,1)
          - 460800 \* \Ssss(1,-2,-2,1)
          + 115200 \* \Ssss(1,-2,1,-2)
          + 57600 \* \Ssss(1,-2,1,1)
          + 1094400 \* \Ssss(1,1,-3,1)
  \nonumber\\&& \mbox{}
          - 230400 \* \Ssss(1,1,-2,-2)
          - 182400 \* \Ssss(1,1,-2,1)
          + 115200 \* \Ssss(1,1,-2,2)
          - 230400 \* \Ssss(1,1,1,-2)
          - 194400 \* \Ssss(1,1,1,1)
  \nonumber\\&& \mbox{}
          + 1440000 \* \Ssss(1,1,1,-3)
          - 115200 \* \Ssss(1,1,1,2)
          + 57600 \* \Ssss(1,1,1,3)
          + 345600 \* \Ssss(1,1,2,-2)
          - 86400 \* \Ssss(1,1,2,1)
  \nonumber\\&& \mbox{}
          - 172800 \* \Ssss(1,1,3,1)
          + 921600 \* \Ssss(1,2,-2,1)
          + 345600 \* \Ssss(1,2,1,-2)
          - 100800 \* \Ssss(1,2,1,1)
          - 57600 \* \Ssss(1,3,1,1)
  \nonumber\\&& \mbox{}
          + 172800 \* \Ssss(2,1,-2,1)
          + 345600 \* \Ssss(2,1,1,-2)
          - 129600 \* \Ssss(2,1,1,1)
          - 115200 \* \Sssss(1,1,-2,1,1)
          - 345600 \* \Sssss(1,1,1,1,-2)
  \nonumber\\&& \mbox{}
          - 1612800 \* \Sssss(1,1,1,-2,1)
          + 86400 \* \Sssss(1,1,1,1,1)
          + 471803 \* \S(1)
          - 1296000 \* \S(1) \* \z5
          - 395040 \* \S(1) \* \z3
  \nonumber\\&& \mbox{}
          - 1315328 \* \S(2)
          - 168000 \* \S(2) \* \z3
          - 459448 \* \S(3)
          + 489600 \* \S(3) \* \z3
          + 1066080 \* \S(4)
          - 157200 \* \S(5) )
          \biggr)
  \nonumber\\&& \mbox{}
       + \colour4colour{\ca \* \cf \* \nf}  \*  \biggl(
          - 256 \* \Ss(1,-3)
          + {15424 \over 45} \* \Ss(1,-2)
          + {704 \over 9} \* \Ss(1,1)
          - 64 \* \Ss(1,2)
          + 64 \* \Ss(1,3)
          - 256 \* \Ss(2,-2)
  \nonumber\\&& \mbox{}
          + {576 \over 5} \* \Ss(2,1)
          - {128 \over 3} \* \Ss(2,3)
          + {64 \over 3} \* \Ss(3,1)
          + {128 \over 3} \* \Ss(4,1)
          + {512 \over 3} \* \Sss(1,1,-2)
          + {128 \over 3} \* \Sss(1,1,3)
          - {128 \over 3} \* \Sss(1,3,1)
  \nonumber\\&& \mbox{}
          - {8 \over 2025} \* \gqq \* (16200 \* \Ss(1,-4)
          - 30600 \* \Ss(1,-3)
          + 33930 \* \Ss(1,-2)
          + 54075 \* \Ss(1,1)
          - 43200 \* \Ss(1,1) \* \z3
  \nonumber\\&& \mbox{}
          - 26100 \* \Ss(1,2)
          + 25200 \* \Ss(1,3)
          - 16200 \* \Ss(1,4)
          + 5400 \* \Ss(2,-3)
          - 13500 \* \Ss(2,-2)
          - 8460 \* \Ss(2,1)
  \nonumber\\&& \mbox{}
          - 13500 \* \Ss(2,3)
          + 5400 \* \Ss(3,1)
          + 8100 \* \Ss(4,1)
          - 10800 \* \Sss(1,-2,-2)
          - 10800 \* \Sss(1,-2,1)
          - 32400 \* \Sss(1,1,-3)
  \nonumber\\&& \mbox{}
          + 50400 \* \Sss(1,1,-2)
          + 18000 \* \Sss(1,1,1)
          + 2700 \* \Sss(1,1,2)
          + 8100 \* \Sss(1,1,3)
          - 21600 \* \Sss(1,2,-2)
  \nonumber\\&& \mbox{}
          - 2700 \* \Sss(1,2,1)
          - 2700 \* \Sss(1,2,2)
          + 5400 \* \Sss(1,3,1)
          + 10800 \* \Sss(2,-2,1)
          - 21600 \* \Sss(2,1,-2)
          + 2700 \* \Ssss(1,1,1,2)
  \nonumber\\&& \mbox{}
          + 21600 \* \Ssss(1,1,1,-2)
          - 2700 \* \Ssss(1,1,2,1)
          + 137425 \* \S(1)
          - 81000 \* \S(1) \* \z3
          - 124584 \* \S(2)
  \nonumber\\&& \mbox{}
          + 43200 \* \S(2) \* \z3
          + 75780 \* \S(3)
          - 13500 \* \S(4) )
          + {10864 \over 27} \* \S(1)
          - {1408 \over 3} \* \S(1) \* \z3
          - {2208 \over 25} \* \S(2)
          + {256 \over 15} \* \S(3)
  \nonumber\\&& \mbox{}
          - {16 \over 25} \* (\Nplusthree - \Nplustwo) \* (60 \* \Ss(1,-3)
          - 53 \* \Ss(1,-2)
          - 100 \* \Ss(1,1) \* \z3 
          + 20 \* \Ss(2,-2)
          + 50 \* \Ss(2,2)
          - 50 \* \Ss(2,3)
          - 70 \* \Ss(3,1)
  \nonumber\\&& \mbox{}
          + 50 \* \Ss(4,1)
          - 20 \* \Sss(1,-2,1)
          - 20 \* \Sss(1,1,-2)
          - 50 \* \Sss(1,1,2)
          + 50 \* \Sss(1,1,3)
          + 50 \* \Sss(1,2,1)
          - 50 \* \Sss(1,3,1)
          - 240 \* \S(1) \* \z3
  \nonumber\\&& \mbox{}
          + 100 \* \S(2) \* \z3
          - 53 \* \S(3)
          + 60 \* \S(4) )
          + {32 \over 225} \* (\Nminusthree - \Nminustwo) \* (180 \* \Ss(1,-3)
          - 149 \* \Ss(1,-2)
          - 300 \* \Ss(1,1) \* \z3
  \nonumber\\&& \mbox{}
          + 120 \* \Ss(2,-2)
          - 60 \* \Sss(1,-2,1)
          - 60 \* \Sss(1,1,-2)
          + 150 \* \Sss(1,1,3)
          - 150 \* \Sss(1,3,1)
          + 180 \* \S(1) \* \z3 )
  \nonumber\\&& \mbox{}
          - {16 \over 75} \* (\Nplustwo - 3) \* (60 \* \Ss(1,-2)
          + 240 \* \Ss(1,1)
          + 200 \* \Ss(1,1) \* \z3
          + 150 \* \Ss(1,2)
          - 150 \* \Ss(1,3)
          + 200 \* \Ss(2,-2)
          - 240 \* \Ss(2,1)
  \nonumber\\&& \mbox{}
          + 100 \* \Ss(2,3)
          - 50 \* \Ss(3,1)
          - 100 \* \Ss(4,1)
          + 200 \* \Sss(1,-2,1)
          - 200 \* \Sss(1,1,-2)
          - 100 \* \Sss(1,1,3)
          + 100 \* \Sss(1,3,1)
          - 219 \* \S(1)
  \nonumber\\&& \mbox{}
          + 500 \* \S(1) \* \z3
          - 21 \* \S(2)
          - 200 \* \S(2) \* \z3
          - 30 \* \S(3) )
          + {32 \over 225} \* (\Nminustwo - \Nminus) \* (60 \* \Ss(1,-2)
          + 90 \* \Ss(1,1)
          + 150 \* \Ss(1,1) \* \z3
  \nonumber\\&& \mbox{}
          - 150 \* \Ss(1,3)
          + 90 \* \Ss(2,1)
          + 150 \* \Ss(3,1)
          + 150 \* \Sss(1,-2,1)
          - 150 \* \Sss(1,1,-2)
          - 75 \* \Sss(1,1,3)
          + 75 \* \Sss(1,3,1)
          - 209 \* \S(1)
  \nonumber\\&& \mbox{}
          + 450 \* \S(1) \* \z3
          - 89 \* \S(2)
          + 180 \* \S(3) )
          + {8 \over 2025} \* (\Nminus + 1) \* (16200 \* \Ss(1,-4)
          + 1800 \* \Ss(1,-3)
          - 12690 \* \Ss(1,-2)
  \nonumber\\&& \mbox{}
          + 31215 \* \Ss(1,1)
          - 26100 \* \Ss(1,2)
          + 25200 \* \Ss(1,3)
          - 16200 \* \Ss(1,4)
          + 5400 \* \Ss(2,-3)
          + 8100 \* \Ss(2,-2)
          - 10080 \* \Ss(2,1)
  \nonumber\\&& \mbox{}
          - 13500 \* \Ss(2,3)
          + 5400 \* \Ss(3,1)
          + 8100 \* \Ss(4,1)
          - 10800 \* \Sss(1,-2,-2)
          - 21600 \* \Sss(1,-2,1)
          - 32400 \* \Sss(1,1,-3)
  \nonumber\\&& \mbox{}
          + 39600 \* \Sss(1,1,-2)
          + 18000 \* \Sss(1,1,1)
          + 2700 \* \Sss(1,1,2)
          + 8100 \* \Sss(1,1,3)
          - 21600 \* \Sss(1,2,-2)
          - 2700 \* \Sss(1,2,1)
  \nonumber\\&& \mbox{}
          - 2700 \* \Sss(1,2,2)
          + 5400 \* \Sss(1,3,1)
          + 10800 \* \Sss(2,-2,1)
          - 21600 \* \Sss(2,1,-2)
          + 21600 \* \Ssss(1,1,1,-2)
          + 2700 \* \Ssss(1,1,1,2)
  \nonumber\\&& \mbox{}
          - 2700 \* \Ssss(1,1,2,1)
          + 98326 \* \S(1)
          - 48600 \* \S(1) \* \z3
          - 112272 \* \S(2)
          + 75240 \* \S(3)
          - 13500 \* \S(4) )
  \nonumber\\&& \mbox{}
          - {256 \over 3} \* (2 \* \Nminus + 3) \* (\Ss(1,1) \* \z3 
          - \S(2) \* \z3)
          \biggr)
  \nonumber\\&& \mbox{}
       + \colour4colour{\ca \* \cf^2}  \*  \biggl(
          - 1792 \* \Ss(1,-4)
          - {704 \over 3} \* \Ss(1,-3)
          + {404032 \over 75} \* \Ss(1,-2)
          + {119344 \over 45} \* \Ss(1,1)
          + 832 \* \Ss(1,1) \* \z3
          - {304 \over 3} \* \Ss(1,2)
  \nonumber\\&& \mbox{}
          - {15328 \over 15} \* \Ss(1,3)
          - {57664 \over 15} \* \Ss(2,-2)
          - {439264 \over 225} \* \Ss(2,1)
          - {5056 \over 15} \* \Ss(2,2)
          + 1792 \* \Ss(2,3)
          + {35008 \over 15} \* \Ss(3,1)
  \nonumber\\&& \mbox{}
          - 1792 \* \Ss(4,1)
          + 4416 \* \Sss(1,-3,1)
          - 1536 \* \Sss(1,-2,-2)
          - 2944 \* \Sss(1,-2,1)
          + 4800 \* \Sss(1,1,-3)
          - {1664 \over 15} \* \Sss(1,1,-2)
  \nonumber\\&& \mbox{}
          - {1456 \over 3} \* \Sss(1,1,1)
          + {608 \over 3} \* \Sss(1,1,2)
          - 1600 \* \Sss(1,1,3)
          - {608 \over 3} \* \Sss(1,2,1)
          + 1472 \* \Sss(1,3,1)
          + {512 \over 5} \* \Sss(2,1,1)
          - 7552 \* \Ssss(1,1,-2,1)
  \nonumber\\&& \mbox{}
          - {1 \over 1350} \* \gqq \* (777600 \* \Ss(1,-5)
          - 410400 \* \Ss(1,-4)
          + 240000 \* \Ss(1,-3)
          + 2157648 \* \Ss(1,-2)
          - 1814400 \* \Ss(1,-2) \* \z3
  \nonumber\\&& \mbox{}
          + 4516360 \* \Ss(1,1)
          + 1130400 \* \Ss(1,1) \* \z3
          - 53400 \* \Ss(1,2)
          + 86400 \* \Ss(1,2) \* \z3
          + 57840 \* \Ss(1,3)
          + 151200 \* \Ss(1,4)
  \nonumber\\&& \mbox{}
          - 777600 \* \Ss(1,5)
          + 259200 \* \Ss(2,-4)
          - 684000 \* \Ss(2,-3)
          + 26400 \* \Ss(2,-2)
          - 199656 \* \Ss(2,1)
          - 1684800 \* \Ss(2,1) \* \z3
  \nonumber\\&& \mbox{}
          - 988560 \* \Ss(2,2)
          + 1612800 \* \Ss(2,3)
          - 259200 \* \Ss(2,4)
          + 43200 \* \Ss(3,-3)
          - 331200 \* \Ss(3,-2)
          - 2018640 \* \Ss(3,1)
  \nonumber\\&& \mbox{}
          + 86400 \* \Ss(3,2)
          - 43200 \* \Ss(3,3)
          - 864000 \* \Ss(4,1)
          - 1814400 \* \Sss(1,-4,1)
          + 432000 \* \Sss(1,-3,-2)
          + 2786400 \* \Sss(1,-3,1)
  \nonumber\\&& \mbox{}
          - 86400 \* \Sss(1,-3,2)
          + 950400 \* \Sss(1,-2,-3)
          - 2318400 \* \Sss(1,-2,-2)
          - 4533600 \* \Sss(1,-2,1)
          + 86400 \* \Sss(1,-2,2)
  \nonumber\\&& \mbox{}
          + 345600 \* \Sss(1,-2,3)
          - 1382400 \* \Sss(1,1,-4)
          + 1360800 \* \Sss(1,1,-3)
          + 3345120 \* \Sss(1,1,-2)
          - 255000 \* \Sss(1,1,1)
  \nonumber\\&& \mbox{}
          - 432000 \* \Sss(1,1,1) \* \z3
          + 871200 \* \Sss(1,1,2)
          - 2296800 \* \Sss(1,1,3)
          + 864000 \* \Sss(1,1,4)
          - 1987200 \* \Sss(1,2,-3)
  \nonumber\\&& \mbox{}
          - 705600 \* \Sss(1,2,-2)
          + 360000 \* \Sss(1,2,1)
          - 158400 \* \Sss(1,2,2)
          + 259200 \* \Sss(1,2,3)
          - 777600 \* \Sss(1,3,-2)
  \nonumber\\&& \mbox{}
          + 2570400 \* \Sss(1,3,1)
          + 86400 \* \Sss(1,3,2)
          + 1209600 \* \Sss(1,4,1)
          - 129600 \* \Sss(2,-3,1)
          + 3384000 \* \Sss(2,-2,1)
  \nonumber\\&& \mbox{}
          - 561600 \* \Sss(2,1,-3)
          - 2275200 \* \Sss(2,1,-2)
          + 938160 \* \Sss(2,1,1)
          + 21600 \* \Sss(2,1,2)
          + 648000 \* \Sss(2,1,3)
  \nonumber\\&& \mbox{}
          - 345600 \* \Sss(2,2,-2)
          - 21600 \* \Sss(2,2,1)
          - 388800 \* \Sss(2,3,1)
          + 86400 \* \Sss(3,-2,1)
          - 172800 \* \Sss(3,1,-2)
  \nonumber\\&& \mbox{}
          - 2073600 \* \Ssss(1,-2,-2,1)
          - 86400 \* \Sss(3,1,1)
          + 86400 \* \Ssss(1,-3,1,1)
          + 172800 \* \Ssss(1,-2,1,-2)
          - 86400 \* \Ssss(1,-2,1,1)
  \nonumber\\&& \mbox{}
          + 3024000 \* \Ssss(1,1,-3,1)
          - 5140800 \* \Ssss(1,1,-2,1)
          + 172800 \* \Ssss(1,1,-2,2)
          + 3542400 \* \Ssss(1,1,1,-3)
          - 576000 \* \Ssss(1,1,1,1)
  \nonumber\\&& \mbox{}
          + 115200 \* \Ssss(1,1,1,2)
          + 230400 \* \Ssss(1,1,1,-2)
          - 1036800 \* \Ssss(1,1,-2,-2)
          - 432000 \* \Ssss(1,1,1,3)
          + 518400 \* \Ssss(1,1,2,-2)
  \nonumber\\&& \mbox{}
          - 115200 \* \Ssss(1,1,2,1)
          + 259200 \* \Ssss(1,1,3,1)
          + 2764800 \* \Ssss(1,2,-2,1)
          + 518400 \* \Ssss(1,2,1,-2)
          - 86400 \* \Ssss(1,3,1,1)
  \nonumber\\&& \mbox{}
          + 259200 \* \Ssss(2,1,-2,1)
          + 518400 \* \Ssss(2,1,1,-2)
          - 172800 \* \Sssss(1,1,-2,1,1)
          - 5184000 \* \Sssss(1,1,1,-2,1)
  \nonumber\\&& \mbox{}
          - 518400 \* \Sssss(1,1,1,1,-2)
          + 4121399 \* \S(1)
          - 3672000 \* \S(1) \* \z5
          - 2016000 \* \S(1) \* \z3
          - 7154120 \* \S(2)
  \nonumber\\&& \mbox{}
          - 417600 \* \S(2) \* \z3
          + 3011136 \* \S(3)
          + 734400 \* \S(3) \* \z3
          + 2012520 \* \S(4)
          - 482400 \* \S(5))
          + {1079264 \over 675} \* \S(1)
  \nonumber\\&& \mbox{}
          - 672 \* \S(1) \* \z3
          - {53216 \over 25} \* \S(2)
          - 3968 \* \S(2) \* \z3
          + {517624 \over 225} \* \S(3)
          + {1536 \over 5} \* \S(4)
  \nonumber\\&& \mbox{}
          - {8 \over 75} \* (\Nplusthree - \Nplustwo) \* (1440 \* \Ss(1,-4)
          - 914 \* \Ss(1,-3)
          - 2727 \* \Ss(1,-2)
          + 13800 \* \Ss(1,1) \* \z3 
          - 360 \* \Ss(1,4)
  \nonumber\\&& \mbox{}
          + 5400 \* \Ss(2,-3)
          - 3074 \* \Ss(2,-2)
          + 2750 \* \Ss(2,2)
          + 5280 \* \Ss(2,3)
          + 1080 \* \Ss(3,-2)
          + 324 \* \Ss(3,1)
          - 360 \* \Ss(3,2)
  \nonumber\\&& \mbox{}
          - 10680 \* \Ss(4,1)
          - 5400 \* \Sss(1,-3,1)
          + 720 \* \Sss(1,-2,-2)
          + 3074 \* \Sss(1,-2,1)
          - 360 \* \Sss(1,-2,2)
          - 5400 \* \Sss(1,1,-3)
  \nonumber\\&& \mbox{}
          + 3074 \* \Sss(1,1,-2)
          - 2750 \* \Sss(1,1,2)
          - 5280 \* \Sss(1,1,3)
          - 360 \* \Sss(1,2,-2)
          + 2750 \* \Sss(1,2,1)
          + 5280 \* \Sss(1,3,1)
  \nonumber\\&& \mbox{}
          - 9000 \* \Sss(2,-2,1)
          - 360 \* \Sss(2,1,-2)
          + 360 \* \Sss(3,1,1)
          + 360 \* \Ssss(1,-2,1,1)
          + 9000 \* \Ssss(1,1,-2,1)
          + 360 \* \Ssss(1,1,1,-2)
  \nonumber\\&& \mbox{}
          - 13260 \* \S(1) \* \z3
          - 13800 \* \S(2) \* \z3
          - 2727 \* \S(3)
          - 914 \* \S(4)
          + 1800 \* \S(5) )
  \nonumber\\&& \mbox{}
          + {16 \over 225} \* (\Nminusthree - \Nminustwo) \* (1440 \* \Ss(1,-4)
          - 854 \* \Ss(1,-3)
          - 1897 \* \Ss(1,-2)
          + 13800 \* \Ss(1,1) \* \z3 
  \nonumber\\&& \mbox{}
          - 360 \* \Ss(1,4)
          + 720 \* \Ss(2,-3)
          + 2160 \* \Ss(2,-2)
          - 5400 \* \Sss(1,-3,1)
          + 720 \* \Sss(1,-2,-2)
          + 3014 \* \Sss(1,-2,1)
  \nonumber\\&& \mbox{}
          - 360 \* \Sss(1,-2,2)
          - 5400 \* \Sss(1,1,-3)
          + 3014 \* \Sss(1,1,-2)
          - 5280 \* \Sss(1,1,3)
          - 360 \* \Sss(1,2,-2)
          + 5280 \* \Sss(1,3,1)
  \nonumber\\&& \mbox{}
          - 720 \* \Sss(2,-2,1)
          - 720 \* \Sss(2,1,-2)
          + 360 \* \Ssss(1,-2,1,1)
          + 9000 \* \Ssss(1,1,-2,1)
          + 360 \* \Ssss(1,1,1,-2)
          + 3240 \* \S(1) \* \z3 )
  \nonumber\\&& \mbox{}
          - {8 \over 75} \* (\Nplustwo - 3) \* (5400 \* \Ss(1,-3)
          - 2354 \* \Ss(1,-2)
          - 3416 \* \Ss(1,1)
          - 16800 \* \Ss(1,1) \* \z3
          + 2390 \* \Ss(1,2)
          + 5280 \* \Ss(1,3)
  \nonumber\\&& \mbox{}
          + 10240 \* \Ss(2,-2)
          + 3056 \* \Ss(2,1)
          + 1460 \* \Ss(2,2)
          - 8400 \* \Ss(2,3)
          - 11580 \* \Ss(3,1)
          + 8400 \* \Ss(4,1)
          + 1600 \* \Sss(1,-2,1)
  \nonumber\\&& \mbox{}
          - 10960 \* \Sss(1,1,-2)
          + 360 \* \Sss(1,1,1)
          - 1100 \* \Sss(1,1,2)
          + 8400 \* \Sss(1,1,3)
          + 1100 \* \Sss(1,2,1)
          - 8400 \* \Sss(1,3,1)
  \nonumber\\&& \mbox{}
          - 360 \* \Sss(2,1,1)
          - 7961 \* \S(1)
          - 9800 \* \S(1) \* \z3
          + 11377 \* \S(2)
          + 16800 \* \S(2) \* \z3
          - 8686 \* \S(3)
          - 1080 \* \S(4) )
  \nonumber\\&& \mbox{}
          + {16 \over 225} \* (\Nminustwo - \Nminus) \* (5400 \* \Ss(1,-3)
          - 2654 \* \Ss(1,-2)
          - 6226 \* \Ss(1,1)
          - 12600 \* \Ss(1,1) \* \z3
          - 360 \* \Ss(1,2)
  \nonumber\\&& \mbox{}
          + 5280 \* \Ss(1,3)
          + 1080 \* \Ss(2,-2)
          - 6946 \* \Ss(2,1)
          - 360 \* \Ss(2,2)
          - 10680 \* \Ss(3,1)
          - 1050 \* \Sss(1,-2,1)
  \nonumber\\&& \mbox{}
          - 8310 \* \Sss(1,1,-2)
          + 360 \* \Sss(1,1,1)
          + 6300 \* \Sss(1,1,3)
          - 6300 \* \Sss(1,3,1)
          + 360 \* \Sss(2,1,1)
  \nonumber\\&& \mbox{}
          - 7431 \* \S(1)
          - 5850 \* \S(1) \* \z3
          - 4551 \* \S(2)
          + 226 \* \S(3)
          + 1800 \* \S(4) )
  \nonumber\\&& \mbox{}
          + {1 \over 1350} \* (\Nminus + 1) \* (777600 \* \Ss(1,-5)
          + 799200 \* \Ss(1,-4)
          - 379200 \* \Ss(1,-3)
          - 1139664 \* \Ss(1,-2)
  \nonumber\\&& \mbox{}
          - 1814400 \* \Ss(1,-2) \* \z3
          + 3218104 \* \Ss(1,1)
          + 2988000 \* \Ss(1,1) \* \z3
          - 329160 \* \Ss(1,2)
          + 86400 \* \Ss(1,2) \* \z3
  \nonumber\\&& \mbox{}
          - 12720 \* \Ss(1,3)
          - 540000 \* \Ss(1,4)
          - 777600 \* \Ss(1,5)
          + 259200 \* \Ss(2,-4)
          + 7200 \* \Ss(2,-3)
  \nonumber\\&& \mbox{}
          + 1146720 \* \Ss(2,-2)
          + 678072 \* \Ss(2,1)
          - 1684800 \* \Ss(2,1) \* \z3
          - 971280 \* \Ss(2,2)
          + 1612800 \* \Ss(2,3)
  \nonumber\\&& \mbox{}
          - 259200 \* \Ss(2,4)
          + 43200 \* \Ss(3,-3)
          - 158400 \* \Ss(3,-2)
          - 1926480 \* \Ss(3,1)
          + 86400 \* \Ss(3,2)
  \nonumber\\&& \mbox{}
          - 43200 \* \Ss(3,3)
          - 864000 \* \Ss(4,1)
          - 1814400 \* \Sss(1,-4,1)
          + 432000 \* \Sss(1,-3,-2)
          - 194400 \* \Sss(1,-3,1)
  \nonumber\\&& \mbox{}
          - 86400 \* \Sss(1,-3,2)
          + 950400 \* \Sss(1,-2,-3)
          - 1281600 \* \Sss(1,-2,-2)
          - 2776800 \* \Sss(1,-2,1)
          - 86400 \* \Sss(1,-2,2)
  \nonumber\\&& \mbox{}
          + 345600 \* \Sss(1,-2,3)
          - 1382400 \* \Sss(1,1,-4)
          - 1879200 \* \Sss(1,1,-3)
          + 4998240 \* \Sss(1,1,-2)
          + 20760 \* \Sss(1,1,1)
  \nonumber\\&& \mbox{}
          - 432000 \* \Sss(1,1,1) \* \z3
          + 892800 \* \Sss(1,1,2)
          - 2426400 \* \Sss(1,1,3)
          + 864000 \* \Sss(1,1,4)
          - 1987200 \* \Sss(1,2,-3)
  \nonumber\\&& \mbox{}
          - 1051200 \* \Sss(1,2,-2)
          + 338400 \* \Sss(1,2,1)
          - 158400 \* \Sss(1,2,2)
          + 259200 \* \Sss(1,2,3)
          - 777600 \* \Sss(1,3,-2)
  \nonumber\\&& \mbox{}
          + 2786400 \* \Sss(1,3,1)
          + 86400 \* \Sss(1,3,2)
          + 1209600 \* \Sss(1,4,1)
          - 129600 \* \Sss(2,-3,1)
          + 2865600 \* \Sss(2,-2,1)
  \nonumber\\&& \mbox{}
          - 561600 \* \Sss(2,1,-3)
          - 2793600 \* \Sss(2,1,-2)
          + 920880 \* \Sss(2,1,1)
          + 21600 \* \Sss(2,1,2)
          + 648000 \* \Sss(2,1,3)
  \nonumber\\&& \mbox{}
          - 345600 \* \Sss(2,2,-2)
          - 21600 \* \Sss(2,2,1)
          - 388800 \* \Sss(2,3,1)
          + 86400 \* \Sss(3,-2,1)
          - 172800 \* \Sss(3,1,-2)
  \nonumber\\&& \mbox{}
          - 86400 \* \Sss(3,1,1)
          + 86400 \* \Ssss(1,-3,1,1)
          - 2073600 \* \Ssss(1,-2,-2,1)
          + 172800 \* \Ssss(1,-2,1,-2)
          + 86400 \* \Ssss(1,-2,1,1)
  \nonumber\\&& \mbox{}
          + 3024000 \* \Ssss(1,1,-3,1)
          - 43200 \* \Ssss(1,1,-2,1)
          + 172800 \* \Ssss(1,1,-2,2)
          + 3542400 \* \Ssss(1,1,1,-3)
          - 576000 \* \Ssss(1,1,1,1)
  \nonumber\\&& \mbox{}
          - 1036800 \* \Ssss(1,1,-2,-2)
          + 576000 \* \Ssss(1,1,1,-2)
          + 115200 \* \Ssss(1,1,1,2)
          - 432000 \* \Ssss(1,1,1,3)
          + 518400 \* \Ssss(1,1,2,-2)
  \nonumber\\&& \mbox{}
          - 115200 \* \Ssss(1,1,2,1)
          + 259200 \* \Ssss(1,1,3,1)
          + 2764800 \* \Ssss(1,2,-2,1)
          + 518400 \* \Ssss(1,2,1,-2)
          - 86400 \* \Ssss(1,3,1,1)
  \nonumber\\&& \mbox{}
          + 259200 \* \Ssss(2,1,-2,1)
          + 518400 \* \Ssss(2,1,1,-2)
          - 172800 \* \Sssss(1,1,-2,1,1)
          - 5184000 \* \Sssss(1,1,1,-2,1)
  \nonumber\\&& \mbox{}
          - 518400 \* \Sssss(1,1,1,1,-2)
          + 4188519 \* \S(1)
          - 3672000 \* \S(1) \* \z5
          - 151200 \* \S(1) \* \z3
          - 7355576 \* \S(2)
  \nonumber\\&& \mbox{}
          - 158400 \* \S(2) \* \z3
          + 2709048 \* \S(3)
          + 734400 \* \S(3) \* \z3
          + 1960680 \* \S(4)
          - 482400 \* \S(5) )
          \biggr)
  \nonumber\\&& \mbox{}
       + \colour4colour{\ca^2 \* \cf}  \*  \biggl(
            256 \* \Ss(1,-4)
          + {2720 \over 3} \* \Ss(1,-3)
          - {11616 \over 5} \* \Ss(1,-2)
          - {4752 \over 5} \* \Ss(1,1)
          - {320 \over 3} \* \Ss(1,1) \* \z3
          + 352 \* \Ss(1,2)
  \nonumber\\&& \mbox{}
          + {512 \over 15} \* \Ss(1,3)
          + {23296 \over 15} \* \Ss(2,-2)
          + {23296 \over 45} \* \Ss(2,1)
          - {1504 \over 3} \* \Ss(2,3)
          - {6592 \over 15} \* \Ss(3,1)
          + {1504 \over 3} \* \Ss(4,1)
          - 1024 \* \Sss(1,-3,1)
  \nonumber\\&& \mbox{}
          + 512 \* \Sss(1,-2,-2)
          + {1312 \over 3} \* \Sss(1,-2,1)
          - 1024 \* \Sss(1,1,-3)
          - {1120 \over 3} \* \Sss(1,1,-2)
          + {1696 \over 3} \* \Sss(1,1,3)
          - {1696 \over 3} \* \Sss(1,3,1)
  \nonumber\\&& \mbox{}
          + 2048 \* \Ssss(1,1,-2,1)
          + {4 \over 2025} \* \gqq \* (64800 \* \Ss(1,-5)
          + 62100 \* \Ss(1,-4)
          - 175050 \* \Ss(1,-3)
          + 375660 \* \Ss(1,-2)
  \nonumber\\&& \mbox{}
          - 194400 \* \Ss(1,-2) \* \z3
          + 754995 \* \Ss(1,1)
          - 86400 \* \Ss(1,1) \* \z3
          - 232650 \* \Ss(1,2)
          + 32400 \* \Ss(1,2) \* \z3
          + 137610 \* \Ss(1,3)
  \nonumber\\&& \mbox{}
          - 62100 \* \Ss(1,4)
          - 64800 \* \Ss(1,5)
          + 2700 \* \Ss(2,-3)
          - 40410 \* \Ss(2,-2)
          - 215280 \* \Ss(2,1)
          - 162000 \* \Ss(2,1) \* \z3
  \nonumber\\&& \mbox{}
          + 112050 \* \Ss(2,3)
          - 35100 \* \Ss(3,-2)
          - 46260 \* \Ss(3,1)
          - 114750 \* \Ss(4,1)
          - 162000 \* \Sss(1,-4,1)
          + 64800 \* \Sss(1,-3,-2)
  \nonumber\\&& \mbox{}
          + 237600 \* \Sss(1,-3,1)
          + 129600 \* \Sss(1,-2,-3)
          - 318600 \* \Sss(1,-2,-2)
          - 416700 \* \Sss(1,-2,1)
          + 32400 \* \Sss(1,-2,3)
  \nonumber\\&& \mbox{}
          - 64800 \* \Sss(1,1,-4)
          - 86400 \* \Sss(1,1,-3)
          + 529200 \* \Sss(1,1,-2)
          + 143550 \* \Sss(1,1,1)
          - 64800 \* \Sss(1,1,1) \* \z3
  \nonumber\\&& \mbox{}
          + 29700 \* \Sss(1,1,2)
          - 207900 \* \Sss(1,1,3)
          + 64800 \* \Sss(1,1,4)
          - 129600 \* \Sss(1,2,-3)
          - 205200 \* \Sss(1,2,-2)
          - 29700 \* \Sss(1,2,1)
  \nonumber\\&& \mbox{}
          - 29700 \* \Sss(1,2,2)
          + 48600 \* \Sss(1,2,3)
          - 64800 \* \Sss(1,3,-2)
          + 324000 \* \Sss(1,3,1)
          + 81000 \* \Sss(1,4,1)
          + 356400 \* \Sss(2,-2,1)
  \nonumber\\&& \mbox{}
          - 361800 \* \Sss(2,1,-2)
          + 81000 \* \Sss(2,1,3)
          - 81000 \* \Sss(2,3,1)
          - 259200 \* \Ssss(1,-2,-2,1)
          + 259200 \* \Ssss(1,1,-3,1)
  \nonumber\\&& \mbox{}
          - 129600 \* \Ssss(1,1,-2,-2)
          - 475200 \* \Ssss(1,1,-2,1)
          + 259200 \* \Ssss(1,1,1,-3)
          + 172800 \* \Ssss(1,1,1,-2)
          + 29700 \* \Ssss(1,1,1,2)
  \nonumber\\&& \mbox{}
          - 97200 \* \Ssss(1,1,1,3)
          - 29700 \* \Ssss(1,1,2,1)
          + 97200 \* \Ssss(1,1,3,1)
          + 259200 \* \Ssss(1,2,-2,1)
          - 518400 \* \Sssss(1,1,1,-2,1)
  \nonumber\\&& \mbox{}
          + 1096495 \* \S(1)
          - 324000 \* \S(1) \* \z5
          - 655020 \* \S(1) \* \z3
          - 1330584 \* \S(2)
          + 135000 \* \S(2) \* \z3
          + 799020 \* \S(3)
  \nonumber\\&& \mbox{}
          - 7470 \* \S(4)
          - 35100 \* \S(5))
          - {203944 \over 135} \* \S(1)
          + {26176 \over 15} \* \S(1) \* \z3
          + {211376 \over 225} \* \S(2)
          + {3008 \over 3} \* \S(2) \* \z3
          - {34048 \over 45} \* \S(3)
  \nonumber\\&& \mbox{}
          + {8 \over 75} \* (\Nplusthree - \Nplustwo) \* (120 \* \Ss(1,-4)
          + 965 \* \Ss(1,-3)
          - 1869 \* \Ss(1,-2)
          + 2190 \* \Ss(1,1) \* \z3
          - 120 \* \Ss(1,4)
          + 1200 \* \Ss(2,-3)
  \nonumber\\&& \mbox{}
          - 355 \* \Ss(2,-2)
          + 1650 \* \Ss(2,2)
          + 495 \* \Ss(2,3)
          + 240 \* \Ss(3,-2)
          - 1295 \* \Ss(3,1)
          - 1695 \* \Ss(4,1)
          + 240 \* \Sss(1,-2,-2)
  \nonumber\\&& \mbox{}
          - 1200 \* \Sss(1,-3,1)
          + 355 \* \Sss(1,-2,1)
          - 1200 \* \Sss(1,1,-3)
          + 355 \* \Sss(1,1,-2)
          - 1650 \* \Sss(1,1,2)
          - 495 \* \Sss(1,1,3)
          + 1650 \* \Sss(1,2,1)
  \nonumber\\&& \mbox{}
          + 495 \* \Sss(1,3,1)
          - 2400 \* \Sss(2,-2,1)
          + 2400 \* \Ssss(1,1,-2,1)
          - 7920 \* \S(1) \* \z3
          - 2190 \* \S(2) \* \z3
          - 1869 \* \S(3)
          + 965 \* \S(4)
  \nonumber\\&& \mbox{}
          + 240 \* \S(5) )
          - {16 \over 225} \* (\Nminusthree - \Nminustwo) \* (120 \* \Ss(1,-4)
          + 965 \* \Ss(1,-3)
          - 1759 \* \Ss(1,-2)
          + 2190 \* \Ss(1,1) \* \z3 
          - 120 \* \Ss(1,4)
  \nonumber\\&& \mbox{}
          + 1320 \* \Ss(2,-2)
          - 1200 \* \Sss(1,-3,1)
          + 240 \* \Sss(1,-2,-2)
          + 355 \* \Sss(1,-2,1)
          - 1200 \* \Sss(1,1,-3)
          + 355 \* \Sss(1,1,-2)
          - 495 \* \Sss(1,1,3)
  \nonumber\\&& \mbox{}
          + 495 \* \Sss(1,3,1)
          + 2400 \* \Ssss(1,1,-2,1)
          + 1980 \* \S(1) \* \z3 )
  \nonumber\\&& \mbox{}
          + {8 \over 75} \* (\Nplustwo - 3) \* (1200 \* \Ss(1,-3)
          - 115 \* \Ss(1,-2)
          + 310 \* \Ss(1,1)
          - 4700 \* \Ss(1,1) \* \z3
          + 1650 \* \Ss(1,2)
          + 495 \* \Ss(1,3)
  \nonumber\\&& \mbox{}
          + 2810 \* \Ss(2,-2)
          - 310 \* \Ss(2,1)
          - 2350 \* \Ss(2,3)
          - 2345 \* \Ss(3,1)
          + 2350 \* \Ss(4,1)
          + 650 \* \Sss(1,-2,1)
          - 3050 \* \Sss(1,1,-2)
  \nonumber\\&& \mbox{}
          + 2350 \* \Sss(1,1,3)
          - 2350 \* \Sss(1,3,1)
          - 3304 \* \S(1)
          + 860 \* \S(1) \* \z3
          + 2994 \* \S(2)
          + 4700 \* \S(2) \* \z3
          - 2660 \* \S(3) )
  \nonumber\\&& \mbox{}
          - {8 \over 225} \* (\Nminustwo - \Nminus) \* (2400 \* \Ss(1,-3)
          - 230 \* \Ss(1,-2)
          - 2680 \* \Ss(1,1)
          - 7050 \* \Ss(1,1) \* \z3
          + 990 \* \Ss(1,3)
  \nonumber\\&& \mbox{}
          + 480 \* \Ss(2,-2)
          - 2680 \* \Ss(2,1)
          - 3390 \* \Ss(3,1)
          - 225 \* \Sss(1,-2,1)
          - 4575 \* \Sss(1,1,-2)
          + 3525 \* \Sss(1,1,3)
  \nonumber\\&& \mbox{}
          - 3525 \* \Sss(1,3,1)
          - 6388 \* \S(1)
          + 195 \* \S(1) \* \z3
          - 3748 \* \S(2)
          + 2410 \* \S(3)
          + 480 \* \S(4) )
  \nonumber\\&& \mbox{}
          - {4 \over 2025} \* (\Nminus + 1) \* (64800 \* \Ss(1,-5)
          + 126900 \* \Ss(1,-4)
          - 10350 \* \Ss(1,-3)
          - 206190 \* \Ss(1,-2)
          + 497685 \* \Ss(1,1)
  \nonumber\\&& \mbox{}
          - 194400 \* \Ss(1,-2) \* \z3
          + 140400 \* \Ss(1,1) \* \z3
          - 232650 \* \Ss(1,2)
          + 32400 \* \Ss(1,2) \* \z3
          + 119520 \* \Ss(1,3)
          - 126900 \* \Ss(1,4)
  \nonumber\\&& \mbox{}
          - 64800 \* \Ss(1,5)
          + 2700 \* \Ss(2,-3)
          + 200970 \* \Ss(2,-2)
          - 67500 \* \Ss(2,1)
          - 162000 \* \Ss(2,1) \* \z3
          + 112050 \* \Ss(2,3)
  \nonumber\\&& \mbox{}
          - 35100 \* \Ss(3,-2)
          - 30870 \* \Ss(3,1)
          - 114750 \* \Ss(4,1)
          - 162000 \* \Sss(1,-4,1)
          + 64800 \* \Sss(1,-3,-2)
          - 21600 \* \Sss(1,-3,1)
  \nonumber\\&& \mbox{}
          + 129600 \* \Sss(1,-2,-3)
          - 189000 \* \Sss(1,-2,-2)
          - 341100 \* \Sss(1,-2,1)
          + 32400 \* \Sss(1,-2,3)
          - 64800 \* \Sss(1,1,-4)
  \nonumber\\&& \mbox{}
          - 345600 \* \Sss(1,1,-3)
          + 599400 \* \Sss(1,1,-2)
          + 143550 \* \Sss(1,1,1)
          - 64800 \* \Sss(1,1,1) \* \z3
          + 29700 \* \Sss(1,1,2)
  \nonumber\\&& \mbox{}
          - 191700 \* \Sss(1,1,3)
          + 64800 \* \Sss(1,1,4)
          - 129600 \* \Sss(1,2,-3)
          - 205200 \* \Sss(1,2,-2)
          - 29700 \* \Sss(1,2,1)
  \nonumber\\&& \mbox{}
          - 29700 \* \Sss(1,2,2)
          + 48600 \* \Sss(1,2,3)
          - 64800 \* \Sss(1,3,-2)
          + 307800 \* \Sss(1,3,1)
          + 81000 \* \Sss(1,4,1)
          + 356400 \* \Sss(2,-2,1)
  \nonumber\\&& \mbox{}
          - 361800 \* \Sss(2,1,-2)
          + 81000 \* \Sss(2,1,3)
          - 81000 \* \Sss(2,3,1)
          - 259200 \* \Ssss(1,-2,-2,1)
          + 259200 \* \Ssss(1,1,-3,1)
  \nonumber\\&& \mbox{}
          - 129600 \* \Ssss(1,1,-2,-2)
          + 43200 \* \Ssss(1,1,-2,1)
          + 259200 \* \Ssss(1,1,1,-3)
          + 172800 \* \Ssss(1,1,1,-2)
          + 29700 \* \Ssss(1,1,1,2)
  \nonumber\\&& \mbox{}
          - 97200 \* \Ssss(1,1,1,3)
          - 29700 \* \Ssss(1,1,2,1)
          + 97200 \* \Ssss(1,1,3,1)
          + 259200 \* \Ssss(1,2,-2,1)
          - 518400 \* \Sssss(1,1,1,-2,1)
  \nonumber\\&& \mbox{}
          + 892516 \* \S(1)
          - 324000 \* \S(1) \* \z5
          - 259740 \* \S(1) \* \z3
          - 1254462 \* \S(2)
          + 135000 \* \S(2) \* \z3
          + 751140 \* \S(3)
  \nonumber\\&& \mbox{}
          - 7470 \* \S(4)
          - 35100 \* \S(5))
          \biggr)\biggr\}
\, .
\eea
\normalsize
The corresponding $\ar^{\, 3}$ contribution to the gluon coefficient 
function is given by 
\small
\bea
&& c^{(3)}_{L,\rm{g}}(N) \:\: = \:\: 
         \delta(N-2) \* \biggl\{
         \colour4colour{\dabcNA} \* \flg11  \*  \biggl(
            {16 \over 3}
          - {1792 \over 3} \* \z5
          + {7616 \over 15} \* \z3
          \biggr)
       + \colour4colour{\cf \* \nf^2}  \*  \biggl(
            {9031 \over 1215}
          + {256 \over 45} \* \z3
          \biggr)
  \nonumber\\&& \mbox{}
       + \colour4colour{\cf^2 \* \nf}  \*  \biggl(
            {51283 \over 1215}
          - {160 \over 3} \* \z5
          + {928 \over 45} \* \z3
          \biggr)
       + \colour4colour{\ca \* \nf^2}  \*  \biggl(
          - {5431 \over 405}
          + {4 \over 15} \* \z3
          \biggr)
       + \colour4colour{\ca \* \cf \* \nf}  \*  \biggl(
          - {71657 \over 1215}
  \nonumber\\&& \mbox{}
          + {80 \over 3} \* \z5
          - {248 \over 5} \* \z3
          \biggr)
       + \colour4colour{\ca^2 \* \nf}  \*  \biggl(
            {235283 \over 2430}
          + {64 \over 3} \* \z5
          - {148 \over 5} \* \z3
          \biggr)\biggr\}
  \nonumber\\&& \mbox{}
       + \theta(N-4) \* \biggl\{ \colour4colour{\dabcNA} \* \flg11  \*  \biggl(
            {128 \over 15} \* \z3
          + {64 \over 15} \* \gfunct1(N)
          - {64 \over 15} \* \gfunct2(N)
          + {256 \over 15} \* \Ss(-2,1)
          - {64 \over 225} \* \gqg \* (900 \* \Ss(1,-4)
  \nonumber\\&& \mbox{}
          - 1488 \* \Ss(1,-3)
          + 1383 \* \Ss(1,-2)
          - 3303 \* \Ss(1,1)
          + 12240 \* \Ss(1,1) \* \z3
          + 2475 \* \Ss(1,2)
          - 5400 \* \Ss(1,2) \* \z3
          - 1620 \* \Ss(1,3)
  \nonumber\\&& \mbox{}
          - 900 \* \Ss(1,4)
          + 900 \* \Ss(2,-3)
          - 4288 \* \Ss(2,-2)
          + 3018 \* \Ss(2,1)
          - 5400 \* \Ss(2,1) \* \z3
          + 6570 \* \Ss(2,3)
          - 1920 \* \Ss(3,-2)
  \nonumber\\&& \mbox{}
          + 5008 \* \Ss(3,1)
          - 5550 \* \Ss(4,1)
          + 900 \* \Sss(1,-2,-2)
          - 1312 \* \Sss(1,-2,1)
          - 900 \* \Sss(1,1,-3)
          + 4288 \* \Sss(1,1,-2)
          - 4950 \* \Sss(1,1,1)
  \nonumber\\&& \mbox{}
          + 10800 \* \Sss(1,1,1) \* \z3
          - 6570 \* \Sss(1,1,3)
          - 900 \* \Sss(1,2,-2)
          + 2700 \* \Sss(1,2,3)
          + 7470 \* \Sss(1,3,1)
          - 2700 \* \Sss(1,4,1)
  \nonumber\\&& \mbox{}
          - 1920 \* \Sss(2,-2,1)
          + 120 \* \Sss(2,1,-2)
          + 2700 \* \Sss(2,1,3)
          - 2700 \* \Sss(2,3,1)
          + 1800 \* \Ssss(1,1,1,-2)
          - 5400 \* \Ssss(1,1,1,3)
  \nonumber\\&& \mbox{}
          + 5400 \* \Ssss(1,1,3,1)
          - 1512 \* \S(1)
          + 13500 \* \S(1) \* \z5 
          + 1115 \* \S(1) \* \z3
          + 2985 \* \S(2)
          - 14160 \* \S(2) \* \z3 
          - 798 \* \S(3)
          - 588 \* \S(4) )
  \nonumber\\&& \mbox{}
          - {128 \over 15} \* \S(-3)
          - {128 \over 15} \* \S(-2)
          - {256 \over 75} \* (\Nplusthree - 1) \* (98 \* \Ss(1,-3)
          - 45 \* \Ss(1,-2)
          - 840 \* \Ss(1,1) \* \z3
          + 98 \* \Ss(2,-2)
          - 420 \* \Ss(2,3)
  \nonumber\\&& \mbox{}
          - 98 \* \Ss(3,1)
          + 420 \* \Ss(4,1)
          - 98 \* \Sss(1,-2,1)
          - 98 \* \Sss(1,1,-2)
          + 420 \* \Sss(1,1,3)
          - 420 \* \Sss(1,3,1)
          + 840 \* \S(2) \* \z3 - 45 \* \S(3)
          + 98 \* \S(4) )
  \nonumber\\&& \mbox{}
          - {128 \over 225} \* (\Nminusthree - \Nminustwo) \* (98 \* \Ss(1,-3)
          - 45 \* \Ss(1,-2)
          - 840 \* \Ss(1,1) \* \z3
          - 98 \* \Sss(1,-2,1)
          - 98 \* \Sss(1,1,-2)
          + 420 \* \Sss(1,1,3)
  \nonumber\\&& \mbox{}
          - 420 \* \Sss(1,3,1) )
          + {64 \over 225} \* (2 \* \Nplus + \Nminus - 3) \* (916 \* \Ss(1,-3)
          + 823 \* \Ss(1,-2)
          - 529 \* \Ss(1,1)
          - 10680 \* \Ss(1,1) \* \z3
  \nonumber\\&& \mbox{}
          + 1080 \* \Ss(1,3)
          - 1560 \* \Ss(2,-3)
          + 4616 \* \Ss(2,-2)
          - 974 \* \Ss(2,1)
          - 8040 \* \Ss(2,3)
          + 2160 \* \Ss(3,-2)
          - 3896 \* \Ss(3,1)
  \nonumber\\&& \mbox{}
          + 7680 \* \Ss(4,1)
          - 1266 \* \Sss(1,-2,1)
          - 566 \* \Sss(1,1,-2)
          + 5340 \* \Sss(1,1,3)
          - 5340 \* \Sss(1,3,1)
          + 3480 \* \Sss(2,-2,1)
  \nonumber\\&& \mbox{}
          - 360 \* \Sss(2,1,-2)
          + 81 \* \S(1)
          - 2510 \* \S(1) \* \z3
          - 1829 \* \S(2)
          + 18000 \* \S(2) \* \z3
          + 2469 \* \S(3)
          + 916 \* \S(4)
          + 240 \* \S(5) )
  \nonumber\\&& \mbox{}
          - {64 \over 225} \* (\Nminus - 1) \* (900 \* \Ss(1,-4)
          + 988 \* \Ss(1,-3)
          - 1302 \* \Ss(1,-2)
          - 4062 \* \Ss(1,1)
          - 2040 \* \Ss(1,1) \* \z3
          + 2475 \* \Ss(1,2)
  \nonumber\\&& \mbox{}
          - 5400 \* \Ss(1,2) \* \z3
          - 1020 \* \Ss(1,3)
          - 900 \* \Ss(1,4)
          - 660 \* \Ss(2,-3)
          + 328 \* \Ss(2,-2)
          + 1968 \* \Ss(2,1)
          - 5400 \* \Ss(2,1) \* \z3
  \nonumber\\&& \mbox{}
          - 1470 \* \Ss(2,3)
          + 240 \* \Ss(3,-2)
          + 1592 \* \Ss(3,1)
          + 2130 \* \Ss(4,1)
          + 900 \* \Sss(1,-2,-2)
          - 4778 \* \Sss(1,-2,1)
          - 900 \* \Sss(1,1,-3)
  \nonumber\\&& \mbox{}
          + 2802 \* \Sss(1,1,-2)
          - 4950 \* \Sss(1,1,1)
          + 10800 \* \Sss(1,1,1) \* \z3 
          + 570 \* \Sss(1,1,3)
          - 900 \* \Sss(1,2,-2)
          + 2700 \* \Sss(1,2,3)
  \nonumber\\&& \mbox{}
          + 330 \* \Sss(1,3,1)
          - 2700 \* \Sss(1,4,1)
          + 1560 \* \Sss(2,-2,1)
          - 240 \* \Sss(2,1,-2)
          + 2700 \* \Sss(2,1,3)
          - 2700 \* \Sss(2,3,1)
  \nonumber\\&& \mbox{}
          + 1800 \* \Ssss(1,1,1,-2)
          - 5400 \* \Ssss(1,1,1,3)
          + 5400 \* \Ssss(1,1,3,1)
          - 1805 \* \S(1)
          + 13500 \* \S(1) \* \z5
          - 1075 \* \S(1) \* \z3
  \nonumber\\&& \mbox{}
          + 1158 \* \S(2)
          + 3840 \* \S(2) \* \z3
          + 1797 \* \S(3)
          + 328 \* \S(4)
          + 240 \* \S(5) )
          - {128 \over 225} \* (\Nminustwo - 1) \* (98 \* \Ss(1,-2)
          + 322 \* \Ss(1,1)
  \nonumber\\&& \mbox{}
          - 900 \* \Ss(1,1) \* \z3
          - 420 \* \Ss(1,3)
          + 322 \* \Ss(2,1)
          + 420 \* \Ss(3,1)
          + 320 \* \Sss(1,-2,1)
          - 320 \* \Sss(1,1,-2)
          + 450 \* \Sss(1,1,3)
  \nonumber\\&& \mbox{}
          - 450 \* \Sss(1,3,1)
          + 53 \* \S(1)
          + 1160 \* \S(1) \* \z3
          + 53 \* \S(2)
          + 98 \* \S(3))
          \biggr)
  \nonumber\\&& \mbox{}
       + \colour4colour{\cf \* \nf^2}  \*  \biggl(
            {4 \over 2025} \* \gqg \* (6480 \* \Ss(1,-3)
          + 20736 \* \Ss(1,-2)
          + 20460 \* \Ss(1,1)
          - 25200 \* \Ss(1,2)
          + 12960 \* \Ss(2,-2)
  \nonumber\\&& \mbox{}
          - 26460 \* \Ss(2,1)
          - 2700 \* \Ss(2,2)
          - 42660 \* \Ss(3,1)
          - 2160 \* \Sss(1,-2,1)
          - 2160 \* \Sss(1,1,-2)
          + 22500 \* \Sss(1,1,1)
          - 8100 \* \Sss(1,1,2)
  \nonumber\\&& \mbox{}
          + 8100 \* \Sss(1,2,1)
          + 10800 \* \Sss(2,1,1)
          + 137689 \* \S(1)
          - 58320 \* \S(1) \* \z3
          + 9576 \* \S(2)
          + 9576 \* \S(3)
          + 71280 \* \S(4))
  \nonumber\\&& \mbox{}
          - {64 \over 225} \* (\Nplusthree - 1) \* (90 \* \Ss(1,-3)
          - 107 \* \Ss(1,-2)
          + 30 \* \Ss(2,-2)
          + 75 \* \Ss(2,2)
          - 105 \* \Ss(3,1)
          - 30 \* \Sss(1,-2,1)
          - 30 \* \Sss(1,1,-2)
  \nonumber\\&& \mbox{}
          - 75 \* \Sss(1,1,2)
          + 75 \* \Sss(1,2,1)
          - 360 \* \S(1) \* \z3
          - 107 \* \S(3)
          + 90 \* \S(4) )
          - {64 \over 675} \* (\Nminusthree - \Nminustwo) \* (45 \* \Ss(1,-3)
          - 71 \* \Ss(1,-2)
  \nonumber\\&& \mbox{}
          + 30 \* \Ss(2,-2)
          - 15 \* \Sss(1,-2,1)
          - 15 \* \Sss(1,1,-2)
          + 45 \* \S(1) \* \z3
         )
          + {4 \over 2025} \* (2 \* \Nplus + \Nminus - 3) \* ( 3870 \* \Ss(2,1)
          + 2700 \* \Ss(2,2)
  \nonumber\\&& \mbox{}
          + 18360 \* \Ss(1,-3)
          - 7548 \* \Ss(1,-2)
          - 73455 \* \Ss(1,1)
          + 20700 \* \Ss(1,2)
          + 49320 \* \Ss(2,-2)
          - 100620 \* \Ss(3,1)
          + 16200 \* \Ss(3,2)
  \nonumber\\&& \mbox{}
          + 48600 \* \Ss(4,1)
          - 6120 \* \Sss(1,-2,1)
          - 6120 \* \Sss(1,1,-2)
          - 20700 \* \Sss(1,1,1)
          - 16200 \* \Sss(1,1,2)
          + 16200 \* \Sss(1,2,1)
  \nonumber\\&& \mbox{}
          + 13500 \* \Sss(2,1,1)
          - 16200 \* \Sss(3,1,1)
          - 151547 \* \S(1)
          - 78840 \* \S(1) \* \z3
          + 128628 \* \S(2)
          - 105018 \* \S(3)
          + 216810 \* \S(4)
  \nonumber\\&& \mbox{}
          - 89100 \* \S(5) )
          - {4 \over 2025} \* (\Nminus - 1) \* (11880 \* \Ss(1,-3)
          - 64644 \* \Ss(1,-2)
          + 55635 \* \Ss(1,1)
          + 11700 \* \Ss(1,2)
  \nonumber\\&& \mbox{}
          + 36360 \* \Ss(2,-2)
          - 38790 \* \Ss(2,1)
          + 21600 \* \Ss(2,2)
          - 25560 \* \Ss(3,1)
          + 16200 \* \Ss(3,2)
          + 48600 \* \Ss(4,1)
          - 3960 \* \Sss(1,-2,1)
  \nonumber\\&& \mbox{}
          - 3960 \* \Sss(1,1,-2)
          - 9000 \* \Sss(1,1,1)
          - 8100 \* \Sss(1,1,2)
          + 8100 \* \Sss(1,2,1)
          - 13500 \* \Sss(2,1,1)
          - 16200 \* \Sss(3,1,1)
          - 2086 \* \S(1)
  \nonumber\\&& \mbox{}
          - 20520 \* \S(1) \* \z3
          - 69246 \* \S(2)
          - 39534 \* \S(3)
          + 105030 \* \S(4)
          - 89100 \* \S(5) )
          - {16 \over 2025} \* (\Nminustwo - 1) \* (
            1080 \* \Ss(1,-2)
  \nonumber\\&& \mbox{}
          - 330 \* \Ss(1,1)
          - 900 \* \Ss(1,2)
          - 180 \* \Ss(2,1)
          + 900 \* \Sss(1,1,1)
          + 1993 \* \S(1)
          - 672 \* \S(2)
          + 540 \* \S(3) )
          \biggr)
  \nonumber\\&& \mbox{}
       + \colour4colour{\cf^2 \* \nf}  \*  \biggl(
          - {8 \over 225} \* \gqg \* (6120 \* \Ss(1,-4)
          - 5550 \* \Ss(1,-3)
          - 2922 \* \Ss(1,-2)
          - 21600 \* \Ss(1,-2) \* \z3 
          + 23190 \* \Ss(1,1)
  \nonumber\\&& \mbox{}
          - 2520 \* \Ss(1,1) \* \z3
          - 675 \* \Ss(1,2)
          + 1080 \* \Ss(1,3)
          - 6120 \* \Ss(1,4)
          + 15840 \* \Ss(2,-3)
          - 17380 \* \Ss(2,-2)
          + 2760 \* \Ss(2,1)
  \nonumber\\&& \mbox{}
          + 21600 \* \Ss(2,1) \* \z3
          - 1800 \* \Ss(2,2)
          + 3420 \* \Ss(2,3)
          + 14640 \* \Ss(3,-2)
          + 10900 \* \Ss(3,1)
          - 7260 \* \Ss(4,1)
          - 10800 \* \Sss(1,-4,1)
  \nonumber\\&& \mbox{}
          - 5040 \* \Sss(1,-3,1)
          - 2160 \* \Sss(1,-2,-2)
          - 14200 \* \Sss(1,-2,1)
          + 10800 \* \Sss(1,-2,3)
          - 15840 \* \Sss(1,1,-3)
  \nonumber\\&& \mbox{}
          + 18820 \* \Sss(1,1,-2)
          + 1800 \* \Sss(1,1,1)
          + 1800 \* \Sss(1,1,2)
          - 3420 \* \Sss(1,1,3)
          - 7200 \* \Sss(1,2,-2)
          + 2700 \* \Sss(1,2,1)
  \nonumber\\&& \mbox{}
          + 7020 \* \Sss(1,3,1)
          - 7680 \* \Sss(2,-2,1)
          - 9600 \* \Sss(2,1,-2)
          + 2250 \* \Sss(2,1,1)
          - 10800 \* \Sss(2,1,3)
          + 10800 \* \Sss(2,3,1)
  \nonumber\\&& \mbox{}
          + 10080 \* \Ssss(1,1,-2,1)
          + 7200 \* \Ssss(1,1,1,-2)
          - 2250 \* \Ssss(1,1,1,1)
          + 24642 \* \S(1)
          - 54000 \* \S(1) \* \z5
          + 4370 \* \S(1) \* \z3
  \nonumber\\&& \mbox{}
          - 18432 \* \S(2)
          + 4920 \* \S(2) \* \z3
          + 3873 \* \S(3)
          + 2190 \* \S(4)
          + 1440 \* \S(5) )
  \nonumber\\&& \mbox{}
          + {32 \over 75} \* (\Nplusthree - 1) \* (120 \* \Ss(1,-4)
          - 460 \* \Ss(1,-3)
          + 533 \* \Ss(1,-2)
          - 1320 \* \Ss(1,1) \* \z3
          - 120 \* \Ss(1,4)
          + 240 \* \Ss(2,-3)
  \nonumber\\&& \mbox{}
          - 220 \* \Ss(2,-2)
          - 780 \* \Ss(2,3)
          + 240 \* \Ss(3,-2)
          + 220 \* \Ss(3,1)
          + 540 \* \Ss(4,1)
          - 240 \* \Sss(1,-3,1)
          + 240 \* \Sss(1,-2,-2)
  \nonumber\\&& \mbox{}
          + 220 \* \Sss(1,-2,1)
          - 240 \* \Sss(1,1,-3)
          + 220 \* \Sss(1,1,-2)
          + 780 \* \Sss(1,1,3)
          - 780 \* \Sss(1,3,1)
          - 480 \* \Sss(2,-2,1)
          + 480 \* \Ssss(1,1,-2,1)
  \nonumber\\&& \mbox{}
          - 360 \* \S(1) \* \z3
          + 1320 \* \S(2) \* \z3
          + 533 \* \S(3)
          - 460 \* \S(4)
          + 240 \* \S(5) )
          + {16 \over 225} \* (\Nminusthree - \Nminustwo) \* (120 \* \Ss(1,-4)
  \nonumber\\&& \mbox{}
          - 460 \* \Ss(1,-3)
          + 523 \* \Ss(1,-2)
          - 1320 \* \Ss(1,1) \* \z3
          - 120 \* \Ss(1,4)
          - 240 \* \Ss(2,-2)
          - 240 \* \Sss(1,-3,1)
          + 240 \* \Sss(1,-2,-2)
  \nonumber\\&& \mbox{}
          + 220 \* \Sss(1,-2,1)
          - 240 \* \Sss(1,1,-3)
          + 220 \* \Sss(1,1,-2)
          + 780 \* \Sss(1,1,3)
          - 780 \* \Sss(1,3,1)
          + 480 \* \Ssss(1,1,-2,1)
          - 360 \* \S(1) \* \z3 )
  \nonumber\\&& \mbox{}
          - {2 \over 225} \* (2 \* \Nplus + \Nminus - 3) \* (22560 \* \Ss(1,-4)
          - 116720 \* \Ss(1,-3)
          + 134884 \* \Ss(1,-2)
          + 5215 \* \Ss(1,1)
  \nonumber\\&& \mbox{}
          - 78960 \* \Ss(1,1) \* \z3
          - 5850 \* \Ss(1,2)
          - 1620 \* \Ss(1,3)
          - 22560 \* \Ss(1,4)
          + 28800 \* \Ss(2,-4)
          - 4080 \* \Ss(2,-3)
  \nonumber\\&& \mbox{}
          - 78480 \* \Ss(2,-2)
          + 73730 \* \Ss(2,1)
          + 136800 \* \Ss(2,1) \* \z3
          - 10800 \* \Ss(2,2)
          + 31560 \* \Ss(2,3)
          - 28800 \* \Ss(2,4)
  \nonumber\\&& \mbox{}
          + 7200 \* \Ss(3,-3)
          + 59520 \* \Ss(3,-2)
          + 96600 \* \Ss(3,1)
          - 12600 \* \Ss(3,2)
          - 7200 \* \Ss(3,3)
          - 19080 \* \Ss(4,1)
  \nonumber\\&& \mbox{}
          - 19920 \* \Sss(1,-3,1)
          + 16320 \* \Sss(1,-2,-2)
          + 60640 \* \Sss(1,-2,1)
          - 41520 \* \Sss(1,1,-3)
          + 68160 \* \Sss(1,1,-2)
  \nonumber\\&& \mbox{}
          + 4950 \* \Sss(1,1,1)
          + 3600 \* \Sss(1,1,2)
          + 35040 \* \Sss(1,1,3)
          - 14400 \* \Sss(1,2,-2)
          + 5400 \* \Sss(1,2,1)
          - 27840 \* \Sss(1,3,1)
  \nonumber\\&& \mbox{}
          - 7200 \* \Sss(2,-3,1)
          - 13440 \* \Sss(2,-2,1)
          - 50400 \* \Sss(2,1,-3)
          + 50400 \* \Sss(2,1,-2)
          + 13500 \* \Sss(2,1,1)
          + 7200 \* \Sss(2,1,2)
  \nonumber\\&& \mbox{}
          - 93600 \* \Sss(2,1,3)
          - 28800 \* \Sss(2,2,-2)
          + 10800 \* \Sss(2,2,1)
          + 108000 \* \Sss(2,3,1)
          + 14400 \* \Sss(3,-2,1)
  \nonumber\\&& \mbox{}
          - 28800 \* \Sss(3,1,-2)
          + 14400 \* \Sss(3,1,1)
          + 39840 \* \Ssss(1,1,-2,1)
          + 14400 \* \Ssss(1,1,1,-2)
          - 4500 \* \Ssss(1,1,1,1)
  \nonumber\\&& \mbox{}
          + 14400 \* \Ssss(2,1,-2,1)
          + 28800 \* \Ssss(2,1,1,-2)
          - 9000 \* \Ssss(2,1,1,1)
          - 2004 \* \S(1)
          - 82600 \* \S(1) \* \z3
  \nonumber\\&& \mbox{}
          - 199 \* \S(2)
          - 80640 \* \S(2) \* \z3
          + 83909 \* \S(3)
          + 36000 \* \S(3) \* \z3
          - 23360 \* \S(4)
          + 2220 \* \S(5) )
  \nonumber\\&& \mbox{}
          - {2 \over 225} \* (\Nminus - 1) \* (30720 \* \Ss(1,-4)
          - 72520 \* \Ss(1,-3)
          + 50948 \* \Ss(1,-2)
          - 86400 \* \Ss(1,-2) \* \z3
          + 110135 \* \Ss(1,1)
  \nonumber\\&& \mbox{}
          + 148080 \* \Ss(1,1) \* \z3
          - 8550 \* \Ss(1,2)
          - 9780 \* \Ss(1,3)
          - 30720 \* \Ss(1,4)
          - 28800 \* \Ss(2,-4)
          + 125040 \* \Ss(2,-3)
  \nonumber\\&& \mbox{}
          - 81760 \* \Ss(2,-2)
          - 27690 \* \Ss(2,1)
          - 50400 \* \Ss(2,1) \* \z3
          - 1800 \* \Ss(2,2)
          - 17880 \* \Ss(2,3)
          + 28800 \* \Ss(2,4)
  \nonumber\\&& \mbox{}
          - 7200 \* \Ss(3,-3)
          + 27840 \* \Ss(3,-2)
          - 49160 \* \Ss(3,1)
          + 12600 \* \Ss(3,2)
          + 7200 \* \Ss(3,3)
          - 9960 \* \Ss(4,1)
  \nonumber\\&& \mbox{}
          - 43200 \* \Sss(1,-4,1)
          - 7440 \* \Sss(1,-3,1)
          - 24960 \* \Sss(1,-2,-2)
          - 27360 \* \Sss(1,-2,1)
          + 43200 \* \Sss(1,-2,3)
  \nonumber\\&& \mbox{}
          - 72240 \* \Sss(1,1,-3)
          + 107120 \* \Sss(1,1,-2)
          + 12150 \* \Sss(1,1,1)
          + 10800 \* \Sss(1,1,2)
          - 113520 \* \Sss(1,1,3)
  \nonumber\\&& \mbox{}
          - 43200 \* \Sss(1,2,-2)
          + 16200 \* \Sss(1,2,1)
          + 135120 \* \Sss(1,3,1)
          + 7200 \* \Sss(2,-3,1)
          - 46080 \* \Sss(2,-2,1)
  \nonumber\\&& \mbox{}
          + 50400 \* \Sss(2,1,-3)
          - 117600 \* \Sss(2,1,-2)
          + 2700 \* \Sss(2,1,1)
          - 7200 \* \Sss(2,1,2)
          + 50400 \* \Sss(2,1,3)
  \nonumber\\&& \mbox{}
          + 28800 \* \Sss(2,2,-2)
          - 10800 \* \Sss(2,2,1)
          - 64800 \* \Sss(2,3,1)
          - 14400 \* \Sss(3,-2,1)
          + 28800 \* \Sss(3,1,-2)
  \nonumber\\&& \mbox{}
          - 14400 \* \Sss(3,1,1)
          + 14880 \* \Ssss(1,1,-2,1)
          + 43200 \* \Ssss(1,1,1,-2)
          - 13500 \* \Ssss(1,1,1,1)
          - 14400 \* \Ssss(2,1,-2,1)
  \nonumber\\&& \mbox{}
          - 28800 \* \Ssss(2,1,1,-2)
          + 9000 \* \Ssss(2,1,1,1)
          + 109092 \* \S(1)
          - 216000 \* \S(1) \* \z5
          + 11360 \* \S(1) \* \z3
  \nonumber\\&& \mbox{}
          - 70333 \* \S(2)
          + 143520 \* \S(2) \* \z3
          - 76427 \* \S(3)
          - 36000 \* \S(3) \* \z3
          + 38540 \* \S(4)
          + 3540 \* \S(5) )
  \nonumber\\&& \mbox{}
          + {16 \over 225} \* (\Nminustwo - 1) \* (240 \* \Ss(1,-3)
          + 20 \* \Ss(1,-2)
          + 760 \* \Ss(1,1)
          + 3600 \* \Ss(1,1) \* \z3
          - 780 \* \Ss(1,3)
          + 240 \* \Ss(2,-2)
  \nonumber\\&& \mbox{}
          + 760 \* \Ss(2,1)
          + 540 \* \Ss(3,1)
          - 80 \* \Sss(1,-2,1)
          - 400 \* \Sss(1,1,-2)
          - 1800 \* \Sss(1,1,3)
          + 1800 \* \Sss(1,3,1)
          + 783 \* \S(1)
  \nonumber\\&& \mbox{}
          + 1720 \* \S(1) \* \z3
          + 543 \* \S(2)
          - 220 \* \S(3)
          + 240 \* \S(4) )
          \biggr)
  \nonumber\\&& \mbox{}
       + \colour4colour{\ca \* \nf^2}  \*  \biggl(
            {16 \over 15} \* (\Nminusthree - \Nminustwo) \* \Ss(1,-2)
          + {2 \over 135} \* \gqg \* (2160 \* \Ss(1,-3)
          - 2256 \* \Ss(1,-2)
          + 15260 \* \Ss(1,1)
          - 4680 \* \Ss(1,2)
  \nonumber\\&& \mbox{}
          + 720 \* \Ss(1,3)
          + 1440 \* \Ss(2,-2)
          - 5760 \* \Ss(2,1)
          + 720 \* \Ss(2,2)
          + 720 \* \Ss(3,1)
          - 720 \* \Sss(1,-2,1)
          - 720 \* \Sss(1,1,-2)
          - 720 \* \Sss(2,1,1)
  \nonumber\\&& \mbox{}
          + 4680 \* \Sss(1,1,1)
          - 1440 \* \Sss(1,1,2)
          + 720 \* \Ssss(1,1,1,1)
          + 24961 \* \S(1)
          + 540 \* \S(1) \* \z3 
          - 12896 \* \S(2)
          + 4464 \* \S(3) )
  \nonumber\\&& \mbox{}
          + {32 \over 5} \* (\Nplusthree - 1) \* (\Ss(1,-2)
          + \S(3) )
          - {2 \over 135} \* (2 \* \Nplus + \Nminus - 3) \* (552 \* \Ss(1,-2)
          - 220 \* \Ss(1,1)
          + 240 \* \Ss(1,2)
  \nonumber\\&& \mbox{}
          - 1080 \* \Ss(2,-2)
          - 4140 \* \Ss(2,1)
          + 1440 \* \Ss(2,2)
          + 2880 \* \Ss(3,1)
          - 240 \* \Sss(1,1,1)
          - 1440 \* \Sss(2,1,1)
          - 617 \* \S(1)
          - 8832 \* \S(2)
  \nonumber\\&& \mbox{}
          + 7152 \* \S(3)
          - 4320 \* \S(4) )
          + {4 \over 135} \* (\Nminus - 1) \* (1080 \* \Ss(1,-3)
          - 552 \* \Ss(1,-2)
          + 8740 \* \Ss(1,1)
          - 2700 \* \Ss(1,2)
          + 360 \* \Ss(1,3)
  \nonumber\\&& \mbox{}
          + 180 \* \Ss(2,-2)
          - 4950 \* \Ss(2,1)
          + 1080 \* \Ss(2,2)
          + 1800 \* \Ss(3,1)
          - 360 \* \Sss(1,-2,1)
          - 360 \* \Sss(1,1,-2)
          + 2700 \* \Sss(1,1,1)
  \nonumber\\&& \mbox{}
          - 720 \* \Sss(1,1,2)
          - 1080 \* \Sss(2,1,1)
          + 360 \* \Ssss(1,1,1,1)
          + 14737 \* \S(1)
          + 270 \* \S(1) \* \z3
          - 11218 \* \S(2)
          + 5628 \* \S(3)
  \nonumber\\&& \mbox{}
          - 2160 \* \S(4) )
          - {8 \over 135} \* (\Nminustwo - 1) \* (60 \* \Ss(1,-2)
          + 250 \* \Ss(1,1)
          - 60 \* \Ss(1,2)
          + 60 \* \Sss(1,1,1)
          + 487 \* \S(1)
          - 18 \* \S(2))
          \biggr)
  \nonumber\\&& \mbox{}
       + \colour4colour{\ca \* \cf \* \nf}  \*  \biggl(
            {2 \over 675} \* \gqg \* (55080 \* \Ss(1,-4)
          - 121896 \* \Ss(1,-3)
          - 10140 \* \Ss(1,-2)
          + 556644 \* \Ss(1,1)
          - 224460 \* \Ss(1,2)
  \nonumber\\&& \mbox{}
          - 259200 \* \Ss(1,-2) \* \z3
          - 82080 \* \Ss(1,1)\* \z3
          + 119520 \* \Ss(1,3)
          - 29160 \* \Ss(1,4)
          + 151200 \* \Ss(2,-3)
          - 250896 \* \Ss(2,-2)
  \nonumber\\&& \mbox{}
          - 257904 \* \Ss(2,1)
          + 259200 \* \Ss(2,1) \* \z3
          - 5940 \* \Ss(2,2)
          + 17280 \* \Ss(2,3)
          + 171360 \* \Ss(3,-2)
          + 203556 \* \Ss(3,1)
  \nonumber\\&& \mbox{}
          - 8640 \* \Ss(3,2)
          - 121680 \* \Ss(4,1)
          - 129600 \* \Sss(1,-4,1)
          - 86400 \* \Sss(1,-3,1)
          - 49680 \* \Sss(1,-2,-2)
          - 94224 \* \Sss(1,-2,1)
  \nonumber\\&& \mbox{}
          - 8640 \* \Sss(1,-2,2)
          + 129600 \* \Sss(1,-2,3)
          - 172800 \* \Sss(1,1,-3)
          + 287616 \* \Sss(1,1,-2)
          + 272160 \* \Sss(1,1,1)
  \nonumber\\&& \mbox{}
          - 18900 \* \Sss(1,1,2)
          + 4320 \* \Sss(1,1,3)
          - 73440 \* \Sss(1,2,-2)
          - 67500 \* \Sss(1,2,1)
          + 38880 \* \Sss(1,3,1)
          - 90000 \* \Sss(2,-2,1)
  \nonumber\\&& \mbox{}
          - 91440 \* \Sss(2,1,-2)
          - 34560 \* \Sss(2,1,1)
          - 129600 \* \Sss(2,1,3)
          + 129600 \* \Sss(2,3,1)
          + 8640 \* \Sss(3,1,1)
          + 8640 \* \Ssss(1,-2,1,1)
  \nonumber\\&& \mbox{}
          + 129600 \* \Ssss(1,1,-2,1)
          + 95040 \* \Ssss(1,1,1,-2)
          + 37800 \* \Ssss(1,1,1,1)
          + 578087 \* \S(1)
          - 648000 \* \S(1) \* \z5
  \nonumber\\&& \mbox{}
          + 161340 \* \S(1) \* \z3
          - 527556 \* \S(2)
          + 56880 \* \S(2) \* \z3
          + 369768 \* \S(3)
          - 104976 \* \S(4)
          + 51840 \* \S(5) )
  \nonumber\\&& \mbox{}
          - {32 \over 225} \* (\Nplusthree - 1) \* \biggl(1620 \* \Ss(1,-4)
          - 4104 \* \Ss(1,-3)
          + 3671 \* \Ss(1,-2)
          - 5220 \* \Ss(1,1) \* \z3
          - 540 \* \Ss(1,4)
  \nonumber\\&& \mbox{}
          + 2700 \* \Ss(2,-3)
          - 2424 \* \Ss(2,-2)
          - 825 \* \Ss(2,2)
          - 2880 \* \Ss(2,3)
          + 1440 \* \Ss(3,-2)
          + 3249 \* \Ss(3,1)
  \nonumber\\&& \mbox{}
          - 360 \* \Ss(3,2)
          + 180 \* \Ss(4,1)
          - 2700 \* \Sss(1,-3,1)
          + 1080 \* \Sss(1,-2,-2)
          + 2424 \* \Sss(1,-2,1)
          - 360 \* \Sss(1,-2,2)
  \nonumber\\&& \mbox{}
          - 2700 \* \Sss(1,1,-3)
          + 2424 \* \Sss(1,1,-2)
          + 825 \* \Sss(1,1,2)
          + 2880 \* \Sss(1,1,3)
          - 360 \* \Sss(1,2,-2)
          - 825 \* \Sss(1,2,1)
  \nonumber\\&& \mbox{}
          - 2880 \* \Sss(1,3,1)
          - 3600 \* \Sss(2,-2,1)
          - 360 \* \Sss(2,1,-2)
          + 360 \* \Sss(3,1,1)
          + 360 \* \Ssss(1,-2,1,1)
          + 3600 \* \Ssss(1,1,-2,1)
  \nonumber\\&& \mbox{}
          + 360 \* \Ssss(1,1,1,-2)
          + 2430 \* \S(1) \* \z3
          + 5220 \* \S(2) \* \z3
          + 3671 \* \S(3)
          - 4104 \* \S(4)
          + 2160 \* \S(5) \biggr)
  \nonumber\\&& \mbox{}
          - {16 \over 675} \* (\Nminusthree - \Nminustwo) \* (1620 \* \Ss(1,-4)
          - 3774 \* \Ss(1,-3)
          + 3901 \* \Ss(1,-2)
          - 5220 \* \Ss(1,1) \* \z3
          - 540 \* \Ss(1,4)
  \nonumber\\&& \mbox{}
          + 720 \* \Ss(2,-3)
          - 1680 \* \Ss(2,-2)
          - 2700 \* \Sss(1,-3,1)
          + 1080 \* \Sss(1,-2,-2)
          + 2094 \* \Sss(1,-2,1)
          - 360 \* \Sss(1,-2,2)
  \nonumber\\&& \mbox{}
          - 2700 \* \Sss(1,1,-3)
          + 2094 \* \Sss(1,1,-2)
          + 2880 \* \Sss(1,1,3)
          - 360 \* \Sss(1,2,-2)
          - 2880 \* \Sss(1,3,1)
          - 720 \* \Sss(2,-2,1)
  \nonumber\\&& \mbox{}
          - 720 \* \Sss(2,1,-2)
          + 360 \* \Ssss(1,-2,1,1)
          + 3600 \* \Ssss(1,1,-2,1)
          + 360 \* \Ssss(1,1,1,-2)
          - 2520 \* \S(1) \* \z3 )
  \nonumber\\&& \mbox{}
          + {2 \over 675} \* (2 \* \Nplus + \Nminus - 3) \* (131760 \* \Ss(1,-4)
          - 454992 \* \Ss(1,-3)
          + 543512 \* \Ss(1,-2)
          + 3813 \* \Ss(1,1)
  \nonumber\\&& \mbox{}
          - 295560 \* \Ss(1,1) \* \z3
          - 22320 \* \Ss(1,2)
          + 46080 \* \Ss(1,3)
          - 58320 \* \Ss(1,4)
          + 43200 \* \Ss(2,-4)
          + 66600 \* \Ss(2,-3)
  \nonumber\\&& \mbox{}
          - 375552 \* \Ss(2,-2)
          + 176688 \* \Ss(2,1)
          + 464400 \* \Ss(2,1) \* \z3
          - 167220 \* \Ss(2,2)
          + 144360 \* \Ss(2,3)
          - 43200 \* \Ss(2,4)
  \nonumber\\&& \mbox{}
          + 10800 \* \Ss(3,-3)
          + 227520 \* \Ss(3,-2)
          + 258072 \* \Ss(3,1)
          + 18720 \* \Ss(3,2)
          - 10800 \* \Ss(3,3)
          - 57960 \* \Ss(4,1)
  \nonumber\\&& \mbox{}
          - 189000 \* \Sss(1,-3,1)
          + 73440 \* \Sss(1,-2,-2)
          + 269352 \* \Sss(1,-2,1)
          - 24480 \* \Sss(1,-2,2)
          - 221400 \* \Sss(1,1,-3)
  \nonumber\\&& \mbox{}
          + 320712 \* \Sss(1,1,-2)
          + 22320 \* \Sss(1,1,1)
          + 51300 \* \Sss(1,1,2)
          + 147240 \* \Sss(1,1,3)
          - 46080 \* \Sss(1,2,-2)
  \nonumber\\&& \mbox{}
          - 62100 \* \Sss(1,2,1)
          - 136440 \* \Sss(1,3,1)
          - 10800 \* \Sss(2,-3,1)
          - 154800 \* \Sss(2,-2,1)
          - 75600 \* \Sss(2,1,-3)
  \nonumber\\&& \mbox{}
          + 213120 \* \Sss(2,1,-2)
          + 107820 \* \Sss(2,1,1)
          - 16200 \* \Sss(2,1,2)
          - 270000 \* \Sss(2,1,3)
          - 43200 \* \Sss(2,2,-2)
  \nonumber\\&& \mbox{}
          - 5400 \* \Sss(2,2,1)
          + 291600 \* \Sss(2,3,1)
          + 21600 \* \Sss(3,-2,1)
          - 43200 \* \Sss(3,1,-2)
          - 18720 \* \Sss(3,1,1)
          + 24480 \* \Ssss(1,-2,1,1)
  \nonumber\\&& \mbox{}
          + 255600 \* \Ssss(1,1,-2,1)
          + 46080 \* \Ssss(1,1,1,-2)
          + 2700 \* \Ssss(1,1,1,1)
          + 21600 \* \Ssss(2,1,-2,1)
          + 43200 \* \Ssss(2,1,1,-2)
  \nonumber\\&& \mbox{}
          + 5400 \* \Ssss(2,1,1,1)
          + 9305 \* \S(1)
          + 4020 \* \S(1) \* \z3
          - 5880 \* \S(2)
          - 318240 \* \S(2) \* \z3
          + 427040 \* \S(3)
          + 54000 \* \S(3) \* \z3
  \nonumber\\&& \mbox{}
          - 124092 \* \S(4)
          + 29880 \* \S(5) )
          - {2 \over 675} \* (\Nminus - 1) \* (33480 \* \Ss(1,-4)
          - 12696 \* \Ss(1,-3)
          - 21172 \* \Ss(1,-2)
  \nonumber\\&& \mbox{}
          + 259200 \* \Ss(1,-2) \* \z3
          - 631881 \* \Ss(1,1)
          - 505080 \* \Ss(1,1) \* \z3
          + 274140 \* \Ss(1,2)
          - 133920 \* \Ss(1,3)
          + 14040 \* \Ss(1,4)
  \nonumber\\&& \mbox{}
          + 43200 \* \Ss(2,-4)
          - 171000 \* \Ss(2,-3)
          + 84144 \* \Ss(2,-2)
          + 341184 \* \Ss(2,1)
          + 205200 \* \Ss(2,1) \* \z3
          - 156960 \* \Ss(2,2)
  \nonumber\\&& \mbox{}
          + 127080 \* \Ss(2,3)
          - 43200 \* \Ss(2,4)
          + 10800 \* \Ss(3,-3)
          + 12960 \* \Ss(3,-2)
          + 80796 \* \Ss(3,1)
          + 27360 \* \Ss(3,2)
  \nonumber\\&& \mbox{}
          - 10800 \* \Ss(3,3)
          + 63720 \* \Ss(4,1)
          + 129600 \* \Sss(1,-4,1)
          - 91800 \* \Sss(1,-3,1)
          + 123120 \* \Sss(1,-2,-2)
          + 188376 \* \Sss(1,-2,1)
  \nonumber\\&& \mbox{}
          - 15840 \* \Sss(1,-2,2)
          - 129600 \* \Sss(1,-2,3)
          + 27000 \* \Sss(1,1,-3)
          - 272664 \* \Sss(1,1,-2)
          - 321840 \* \Sss(1,1,1)
  \nonumber\\&& \mbox{}
          + 86400 \* \Sss(1,1,2)
          + 326520 \* \Sss(1,1,3)
          + 70560 \* \Sss(1,2,-2)
          + 10800 \* \Sss(1,2,1)
          - 380520 \* \Sss(1,3,1)
          - 10800 \* \Sss(2,-3,1)
  \nonumber\\&& \mbox{}
          - 21600 \* \Sss(2,-2,1)
          - 75600 \* \Sss(2,1,-3)
          + 347760 \* \Sss(2,1,-2)
          + 138060 \* \Sss(2,1,1)
          - 16200 \* \Sss(2,1,2)
  \nonumber\\&& \mbox{}
          - 140400 \* \Sss(2,1,3)
          - 43200 \* \Sss(2,2,-2)
          - 5400 \* \Sss(2,2,1)
          + 162000 \* \Sss(2,3,1)
          + 21600 \* \Sss(3,-2,1)
          - 43200 \* \Sss(3,1,-2)
  \nonumber\\&& \mbox{}
          - 27360 \* \Sss(3,1,1)
          + 15840 \* \Ssss(1,-2,1,1)
          + 104400 \* \Ssss(1,1,-2,1)
          - 92160 \* \Ssss(1,1,1,-2)
          - 40500 \* \Ssss(1,1,1,1)
  \nonumber\\&& \mbox{}
          + 21600 \* \Ssss(2,1,-2,1)
          + 43200 \* \Ssss(2,1,1,-2)
          + 5400 \* \Ssss(2,1,1,1)
          - 687984 \* \S(1)
          + 648000 \* \S(1) \* \z5
          + 138000 \* \S(1) \* \z3
  \nonumber\\&& \mbox{}
          + 517890 \* \S(2)
          - 439920 \* \S(2) \* \z3
          + 73940 \* \S(3)
          + 54000 \* \S(3) \* \z3
          - 67716 \* \S(4)
          - 21960 \* \S(5) )
  \nonumber\\&& \mbox{}
          - {16 \over 675} \* (\Nminustwo - 1) \* (1800 \* \Ss(1,-3)
          - 924 \* \Ss(1,-2)
          + 4464 \* \Ss(1,1)
          + 10800 \* \Ss(1,1) \* \z3
          - 1710 \* \Ss(1,2)
          - 1980 \* \Ss(1,3)
  \nonumber\\&& \mbox{}
          + 1440 \* \Ss(2,-2)
          + 2994 \* \Ss(2,1)
          - 360 \* \Ss(2,2)
          + 180 \* \Ss(3,1)
          - 1950 \* \Sss(1,-2,1)
          - 210 \* \Sss(1,1,-2)
          + 1710 \* \Sss(1,1,1)
  \nonumber\\&& \mbox{}
          - 5400 \* \Sss(1,1,3)
          + 5400 \* \Sss(1,3,1)
          + 360 \* \Sss(2,1,1)
          + 6287 \* \S(1)
          + 4170 \* \S(1) \* \z3
          + 2527 \* \S(2)
          - 2334 \* \S(3)
          + 2160 \* \S(4) )
          \biggr)
  \nonumber\\&& \mbox{}
       + \colour4colour{\ca^2 \* \nf}  \*  \biggl(
            {2 \over 405} \* \gqg \* (50868 \* \Ss(1,-4)
          - 228366 \* \Ss(1,-3)
          - 19728 \* \Ss(1,-2)
          + 38880 \* \Ss(1,-2) \* \z3
          - 413139 \* \Ss(1,1)
  \nonumber\\&& \mbox{}
          + 20736 \* \Ss(1,1) \* \z3
          + 454680 \* \Ss(1,2)
          - 200070 \* \Ss(1,3)
          + 20412 \* \Ss(1,4)
          + 40176 \* \Ss(2,-3)
          - 172674 \* \Ss(2,-2)
  \nonumber\\&& \mbox{}
          + 718164 \* \Ss(2,1)
          - 38880 \* \Ss(2,1) \* \z3
          - 207360 \* \Ss(2,2)
          + 16200 \* \Ss(2,3)
          - 25704 \* \Ss(3,-2)
          - 222066 \* \Ss(3,1)
  \nonumber\\&& \mbox{}
          + 12960 \* \Ss(3,2)
          + 19224 \* \Ss(4,1)
          + 19440 \* \Sss(1,-4,1)
          - 59616 \* \Sss(1,-3,1)
          - 1944 \* \Sss(1,-2,-2)
          + 136962 \* \Sss(1,-2,1)
  \nonumber\\&& \mbox{}
          - 12960 \* \Sss(1,-2,2)
          - 19440 \* \Sss(1,-2,3)
          - 72576 \* \Sss(1,1,-3)
          + 51930 \* \Sss(1,1,-2)
          - 494280 \* \Sss(1,1,1)
          + 230040 \* \Sss(1,1,2)
  \nonumber\\&& \mbox{}
          - 48600 \* \Sss(1,1,3)
          - 19440 \* \Sss(1,2,-2)
          + 206280 \* \Sss(1,2,1)
          - 38880 \* \Sss(1,2,2)
          - 42120 \* \Sss(1,3,1)
          - 39312 \* \Sss(2,-2,1)
  \nonumber\\&& \mbox{}
          - 41040 \* \Sss(2,1,-2)
          + 227880 \* \Sss(2,1,1)
          - 38880 \* \Sss(2,1,2)
          + 19440 \* \Sss(2,1,3)
          - 38880 \* \Sss(2,2,1)
          - 19440 \* \Sss(2,3,1)
  \nonumber\\&& \mbox{}
          - 25920 \* \Sss(3,1,1)
          + 12960 \* \Ssss(1,-2,1,1)
          + 54432 \* \Ssss(1,1,-2,1)
          + 12960 \* \Ssss(1,1,1,-2)
          - 199800 \* \Ssss(1,1,1,1)
  \nonumber\\&& \mbox{}
          + 45360 \* \Ssss(1,1,1,2)
          + 45360 \* \Ssss(1,1,2,1)
          + 45360 \* \Ssss(1,2,1,1)
          + 38880 \* \Ssss(2,1,1,1)
          - 38880 \* \Sssss(1,1,1,1,1)
          - 1098464 \* \S(1)
  \nonumber\\&& \mbox{}
          + 97200 \* \S(1) \* \z5
          + 35766 \* \S(1) \* \z3
          + 55413 \* \S(2)
          + 864 \* \S(2) \* \z3
          - 863316 \* \S(3)
          + 69660 \* \S(4)
          - 5184 \* \S(5) )
  \nonumber\\&& \mbox{}
          + {8 \over 5} \* (\Nplusthree - 1) \* (16 \* \Ss(1,-4)
          + 10 \* \Ss(1,-3)
          - 19 \* \Ss(1,-2)
          - 128 \* \Ss(1,1) \* \z3
          - 16 \* \Ss(1,4)
          + 32 \* \Ss(2,-3)
          + 10 \* \Ss(2,-2)
  \nonumber\\&& \mbox{}
          - 80 \* \Ss(2,3)
          + 32 \* \Ss(3,-2)
          - 10 \* \Ss(3,1)
          + 48 \* \Ss(4,1)
          - 32 \* \Sss(1,-3,1)
          + 32 \* \Sss(1,-2,-2)
          - 10 \* \Sss(1,-2,1)
          - 32 \* \Sss(1,1,-3)
  \nonumber\\&& \mbox{}
          - 10 \* \Sss(1,1,-2)
          + 80 \* \Sss(1,1,3)
          - 80 \* \Sss(1,3,1)
          - 64 \* \Sss(2,-2,1)
          + 64 \* \Ssss(1,1,-2,1)
          + 128 \* \S(2) \* \z3
          - 19 \* \S(3)
          + 10 \* \S(4)
          + 32 \* \S(5) )
  \nonumber\\&& \mbox{}
          + {4 \over 15} \* (\Nminusthree - \Nminustwo) \* (16 \* \Ss(1,-4)
          + 10 \* \Ss(1,-3)
          - 19 \* \Ss(1,-2)
          - 128 \* \Ss(1,1) \* \z3
          - 16 \* \Ss(1,4)
          - 32 \* \Sss(1,-3,1)
  \nonumber\\&& \mbox{}
          + 32 \* \Sss(1,-2,-2)
          - 10 \* \Sss(1,-2,1)
          - 32 \* \Sss(1,1,-3)
          - 10 \* \Sss(1,1,-2)
          + 80 \* \Sss(1,1,3)
          - 80 \* \Sss(1,3,1)
          + 64 \* \Ssss(1,1,-2,1) )
  \nonumber\\&& \mbox{}
          - {2 \over 405} \* (2 \* \Nplus + \Nminus - 3) \* (7344 \* \Ss(1,-4)
          - 5346 \* \Ss(1,-3)
          + 24606 \* \Ss(1,-2)
          - 45711 \* \Ss(1,1)
          - 45792 \* \Ss(1,1) \* \z3
  \nonumber\\&& \mbox{}
          - 3960 \* \Ss(1,2)
          + 31050 \* \Ss(1,3)
          - 7344 \* \Ss(1,4)
          + 66528 \* \Ss(2,-3)
          + 137142 \* \Ss(2,-2)
          - 235854 \* \Ss(2,1)
          - 123120 \* \Ss(2,2)
  \nonumber\\&& \mbox{}
          + 77760 \* \Ss(2,1) \* \z3
          + 120420 \* \Ss(2,3)
          + 64368 \* \Ss(3,-2)
          - 152802 \* \Ss(3,1)
          + 168480 \* \Ss(3,2)
          + 204012 \* \Ss(4,1)
  \nonumber\\&& \mbox{}
          - 14688 \* \Sss(1,-3,1)
          + 14688 \* \Sss(1,-2,-2)
          + 14022 \* \Sss(1,-2,1)
          - 14688 \* \Sss(1,1,-3)
          + 22590 \* \Sss(1,1,-2)
  \nonumber\\&& \mbox{}
          + 10080 \* \Sss(1,1,1)
          - 15120 \* \Sss(1,1,2)
          + 30240 \* \Sss(1,1,3)
          - 15120 \* \Sss(1,2,1)
          - 30240 \* \Sss(1,3,1)
          - 34776 \* \Sss(2,-2,1)
  \nonumber\\&& \mbox{}
          + 144720 \* \Sss(2,1,1)
          + 57240 \* \Sss(2,1,-2)
          - 90720 \* \Sss(2,1,2)
          - 38880 \* \Sss(2,1,3)
          - 90720 \* \Sss(2,2,1)
          + 38880 \* \Sss(2,3,1)
  \nonumber\\&& \mbox{}
          - 181440 \* \Sss(3,1,1)
          + 29376 \* \Ssss(1,1,-2,1)
          + 12960 \* \Ssss(1,1,1,1)
          + 77760 \* \Ssss(2,1,1,1)
          - 31471 \* \S(1)
          - 33984 \* \S(1) \* \z3
  \nonumber\\&& \mbox{}
          + 703653 \* \S(2)
          - 92448 \* \S(2) \* \z3
          + 415476 \* \S(3)
          + 237708 \* \S(4)
          - 150552 \* \S(5) )
  \nonumber\\&& \mbox{}
          + {2 \over 405} \* (\Nminus - 1) \* (58212 \* \Ss(1,-4)
          - 251856 \* \Ss(1,-3)
          - 107352 \* \Ss(1,-2)
          + 38880 \* \Ss(1,-2) \* \z3 
          - 406092 \* \Ss(1,1)
  \nonumber\\&& \mbox{}
          - 76896 \* \Ss(1,1) \* \z3 
          + 535680 \* \Ss(1,2)
          - 257040 \* \Ss(1,3)
          + 13068 \* \Ss(1,4)
          + 106704 \* \Ss(2,-3)
          - 50004 \* \Ss(2,-2)
  \nonumber\\&& \mbox{}
          + 484398 \* \Ss(2,1)
          + 38880 \* \Ss(2,1) \* \z3 
          - 352080 \* \Ss(2,2)
          + 136620 \* \Ss(2,3)
          + 38664 \* \Ss(3,-2)
          - 356616 \* \Ss(3,1)
  \nonumber\\&& \mbox{}
          + 181440 \* \Ss(3,2)
          + 223236 \* \Ss(4,1)
          + 19440 \* \Sss(1,-4,1)
          - 74304 \* \Sss(1,-3,1)
          + 12744 \* \Sss(1,-2,-2)
  \nonumber\\&& \mbox{}
          + 139392 \* \Sss(1,-2,1)
          - 12960 \* \Sss(1,-2,2)
          - 19440 \* \Sss(1,-2,3)
          - 87264 \* \Sss(1,1,-3)
          + 18720 \* \Sss(1,1,-2)
  \nonumber\\&& \mbox{}
          - 585000 \* \Sss(1,1,1)
          + 275400 \* \Sss(1,1,2)
          + 7560 \* \Sss(1,1,3)
          - 19440 \* \Sss(1,2,-2)
          + 251640 \* \Sss(1,2,1)
  \nonumber\\&& \mbox{}
          - 38880 \* \Sss(1,2,2)
          - 98280 \* \Sss(1,3,1)
          - 74088 \* \Sss(2,-2,1)
          + 16200 \* \Sss(2,1,-2)
          + 402840 \* \Sss(2,1,1)
  \nonumber\\&& \mbox{}
          - 129600 \* \Sss(2,1,2)
          - 19440 \* \Sss(2,1,3)
          - 129600 \* \Sss(2,2,1)
          + 19440 \* \Sss(2,3,1)
          - 207360 \* \Sss(3,1,1)
  \nonumber\\&& \mbox{}
          + 12960 \* \Ssss(1,-2,1,1)
          + 83808 \* \Ssss(1,1,-2,1)
          + 12960 \* \Ssss(1,1,1,-2)
          - 238680 \* \Ssss(1,1,1,1)
          + 45360 \* \Ssss(1,1,1,2)
  \nonumber\\&& \mbox{}
          + 45360 \* \Ssss(1,1,2,1)
          + 45360 \* \Ssss(1,2,1,1)
          + 116640 \* \Ssss(2,1,1,1)
          - 38880 \* \Sssss(1,1,1,1,1)
          - 1289957 \* \S(1)
  \nonumber\\&& \mbox{}
          + 97200 \* \S(1) \* \z5
          + 70326 \* \S(1) \* \z3
          + 679896 \* \S(2)
          - 91584 \* \S(2) \* \z3
          - 487422 \* \S(3)
          + 273240 \* \S(4)
          - 155736 \* \S(5) )
  \nonumber\\&& \mbox{}
          + {4 \over 405} \* (\Nminustwo - 1) \* (9504 \* \Ss(1,-3)
          + 11394 \* \Ss(1,-2)
          + 9666 \* \Ss(1,1)
          + 6480 \* \Ss(1,1) \* \z3
          - 19260 \* \Ss(1,2)
  \nonumber\\&& \mbox{}
          + 6480 \* \Ss(1,3)
          - 1296 \* \Ss(2,-2)
          - 414 \* \Ss(2,1)
          + 4320 \* \Ss(2,2)
          + 1296 \* \Ss(3,1)
          - 6408 \* \Sss(1,-2,1)
          + 360 \* \Sss(1,1,-2)
  \nonumber\\&& \mbox{}
          + 20160 \* \Sss(1,1,1)
          - 7560 \* \Sss(1,1,2)
          - 3240 \* \Sss(1,1,3)
          - 7560 \* \Sss(1,2,1)
          + 3240 \* \Sss(1,3,1)
          - 5400 \* \Sss(2,1,1)
  \nonumber\\&& \mbox{}
          + 6480 \* \Ssss(1,1,1,1)
          + 55741 \* \S(1)
          - 144 \* \S(1) \* \z3
          - 7539 \* \S(2)
          + 1134 \* \S(3)
          + 864 \* \S(4) )
          \biggr)\biggr\}
\:\: .
\eea
\normalsize
Finally our $N$-space result for the N$^2$LO pure-singlet coefficient 
function for $F_L$ reads
\small
\bea
&& c^{(3)}_{L,\rm{ps}}(N) \:\: = \:\: 
         \delta(N-2) \* \biggl\{
         \colour4colour{\cf \* \nf^2}  \*  \biggl(
            {3364 \over 405}
          + {64 \over 15} \* \z3
          \biggr)
       + \colour4colour{\cf^2 \* \nf}  \*  \biggl(
          - {46898 \over 1215}
          + {512 \over 3} \* \z5
          - {5056 \over 45} \* \z3
          \biggr)
  \nonumber\\&& \mbox{}
       + \colour4colour{\ca \* \cf \* \nf}  \*  \biggl(
          - {48058 \over 1215}
          - {256 \over 3} \* \z5
          + {304 \over 5} \* \z3
          \biggr)\biggr\}
  \nonumber\\&& \mbox{}
       + \theta(N-4) \* \biggl\{ \colour4colour{\cf \* \nf \* \biggl(\cf-{\ca \over 2}\biggr)}
          \* {512 \over 3} \*  \biggl(
          - 2 \* \Ss(1,1) \* \z3
          - \Ss(2,3)
          + \Ss(4,1)
          + \Sss(1,1,3)
          - \Sss(1,3,1)
          + 2 \* \S(2) \* \z3
          \biggr)
  \nonumber\\&& \mbox{}
       + \colour4colour{\cf \* \nf^2}  \*  \biggl(
            {256 \over 9} \* \Ss(1,-2)
          + {128 \over 45} \* \Ss(1,-2) \* (\Nminusthree - \Nminustwo)
          - {128 \over 27} \* \Ss(1,1)
          + {128 \over 9} \* \Ss(2,1)
          + {64 \over 9} \* \Sss(1,1,1)
  \nonumber\\&& \mbox{}
          - {32 \over 405} \* \gqq \* (360 \* \Ss(1,-2)
          + 330 \* \Ss(1,1)
          - 405 \* \Ss(2,1)
          - 90 \* \Sss(1,1,1)
          + 135 \* \Sss(2,1,1)
          + 310 \* \S(1)
          - 1614 \* \S(2)
          + 2160 \* \S(3)
  \nonumber\\&& \mbox{}
          - 810 \* \S(4))
          - {10624 \over 81} \* \S(1)
          - {2624 \over 135} \* \S(2)
          + {128 \over 3} \* \S(3)
          - {64 \over 15} \* (\Nplusthree - \Nplustwo) \* (\Ss(1,-2)
          + \S(3))
  \nonumber\\&& \mbox{}
          + {32 \over 405} \* (\Nplustwo - 3) \* (180 \* \Ss(1,-2)
          + 285 \* \Ss(1,1)
          + 90 \* \Ss(2,1)
          - 90 \* \Sss(1,1,1)
          + 646 \* \S(1)
          - 771 \* \S(2)
          + 540 \* \S(3) )
  \nonumber\\&& \mbox{}
          - {32 \over 405} \* (\Nminustwo - \Nminus) \* (90 \* \Ss(1,-2)
          + 75 \* \Ss(1,1)
          - 45 \* \Sss(1,1,1)
          + 494 \* \S(1)
          - 36 \* \S(2))
          + {32 \over 405} \* (\Nminus + 1) \* (360 \* \Ss(1,-2)
  \nonumber\\&& \mbox{}
          + 645 \* \Ss(1,1)
          - 405 \* \Ss(2,1)
          - 225 \* \Sss(1,1,1)
          + 135 \* \Sss(2,1,1)
          + 1786 \* \S(1)
          - 2262 \* \S(2)
          + 2430 \* \S(3)
          - 810 \* \S(4) )
          \biggr)
  \nonumber\\&& \mbox{}
       + \colour4colour{\cf^2 \* \nf}  \*  \biggl(
            128 \* \Ss(1,-3)
          - {12544 \over 15} \* \Ss(1,-2)
          + {122944 \over 135} \* \Ss(1,1)
          - {3392 \over 9} \* \Ss(1,2)
          - {1024 \over 5} \* \Ss(1,3)
          + {1024 \over 3} \* \Ss(2,-2)
  \nonumber\\&& \mbox{}
          - {11072 \over 45} \* \Ss(2,1)
          - {256 \over 5} \* \Ss(3,1)
          - {128 \over 3} \* \Sss(1,-2,1)
          - 384 \* \Sss(1,1,-2)
          + {3488 \over 9} \* \Sss(1,1,1)
          - {128 \over 3} \* \Sss(1,1,2)
          - {128 \over 3} \* \Sss(1,2,1)
  \nonumber\\&& \mbox{}
          - {128 \over 3} \* \Sss(2,1,1)
          + {64 \over 3} \* \Ssss(1,1,1,1)
          - {16 \over 675} \* \gqq \* (3600 \* \Ss(1,-3)
          - 17090 \* \Ss(1,-2)
          + 3985 \* \Ss(1,1)
          - 10800 \* \Ss(1,1) \* \z3
  \nonumber\\&& \mbox{}
          - 7950 \* \Ss(1,2)
          - 720 \* \Ss(1,3)
          + 2700 \* \Ss(2,-3)
          - 3300 \* \Ss(2,-2)
          + 12825 \* \Ss(2,1)
          - 9450 \* \Ss(2,2)
          - 10800 \* \Ss(2,3)
  \nonumber\\&& \mbox{}
          - 900 \* \Ss(3,-2)
          - 30480 \* \Ss(3,1)
          + 5400 \* \Ss(3,2)
          + 17100 \* \Ss(4,1)
          - 3300 \* \Sss(1,-2,1)
          - 5100 \* \Sss(1,1,-2)
          + 4800 \* \Sss(1,1,1)
  \nonumber\\&& \mbox{}
          + 1800 \* \Sss(1,1,2)
          + 5400 \* \Sss(1,1,3)
          + 1800 \* \Sss(1,2,1)
          - 5400 \* \Sss(1,3,1)
          + 900 \* \Sss(2,-2,1)
          - 9900 \* \Sss(2,1,-2)
  \nonumber\\&& \mbox{}
          + 13275 \* \Sss(2,1,1)
          - 2700 \* \Sss(2,1,2)
          - 2700 \* \Sss(2,2,1)
          - 5400 \* \Sss(3,1,1)
          - 900 \* \Ssss(1,1,1,1)
          + 1350 \* \Ssss(2,1,1,1)
          + 6855 \* \S(1)
  \nonumber\\&& \mbox{}
          - 2160 \* \S(1) \* \z3
          - 19279 \* \S(2)
          + 29700 \* \S(2) \* \z3
          - 9800 \* \S(3)
          + 20175 \* \S(4)
          - 2700 \* \S(5) )
          + {31232 \over 135} \* \S(1)
  \nonumber\\&& \mbox{}
          + {2688 \over 5} \* \S(1) \* \z3
          - {12224 \over 75} \* \S(2)
          - {608 \over 3} \* \S(3)
          + 192 \* \S(4)
          + {64 \over 225} \* (\Nplusthree - \Nplustwo) \* (45 \* \Ss(1,-3)
          - 17 \* \Ss(1,-2)
  \nonumber\\&& \mbox{}
          + 360 \* \Ss(1,1) \* \z3
          - 15 \* \Ss(2,-2)
          + 180 \* \Ss(2,3)
          + 15 \* \Ss(3,1)
          - 180 \* \Ss(4,1)
          + 15 \* \Sss(1,-2,1)
          + 15 \* \Sss(1,1,-2)
          - 180 \* \Sss(1,1,3)
  \nonumber\\&& \mbox{}
          + 180 \* \Sss(1,3,1)
          + 90 \* \S(1) \* \z3
          - 360 \* \S(2) \* \z3
          - 17 \* \S(3)
          + 45 \* \S(4))
          - {128 \over 675} \* (\Nminusthree - \Nminustwo) \* (45 \* \Ss(1,-3)
          - 37 \* \Ss(1,-2)
  \nonumber\\&& \mbox{}
          + 360 \* \Ss(1,1) \* \z3
          + 60 \* \Ss(2,-2)
          + 15 \* \Sss(1,-2,1)
          + 15 \* \Sss(1,1,-2)
          - 180 \* \Sss(1,1,3)
          + 180 \* \Sss(1,3,1)
          + 90 \* \S(1) \* \z3 )
  \nonumber\\&& \mbox{}
          + {16 \over 675} \* (\Nplustwo - 3) \* (720 \* \Ss(1,-2)
          - 9005 \* \Ss(1,1)
          - 7200 \* \Ss(1,1) \* \z3
          - 300 \* \Ss(1,2)
          + 2160 \* \Ss(1,3)
          - 1800 \* \Ss(2,-2)
  \nonumber\\&& \mbox{}
          + 5580 \* \Ss(2,1)
          - 5400 \* \Ss(2,2)
          - 3600 \* \Ss(2,3)
          - 12960 \* \Ss(3,1)
          + 3600 \* \Ss(4,1)
          - 1800 \* \Sss(1,-2,1)
          + 1800 \* \Sss(1,1,-2)
  \nonumber\\&& \mbox{}
          - 2400 \* \Sss(1,1,1)
          + 1800 \* \Sss(1,1,2)
          + 3600 \* \Sss(1,1,3)
          + 1800 \* \Sss(1,2,1)
          - 3600 \* \Sss(1,3,1)
          + 4500 \* \Sss(2,1,1)
          - 900 \* \Ssss(1,1,1,1)
  \nonumber\\&& \mbox{}
          + 71 \* \S(1)
          - 9720 \* \S(1) \* \z3
          + 809 \* \S(2)
          + 7200 \* \S(2) \* \z3
          - 2700 \* \S(3)
          + 10800 \* \S(4) )
          - {16 \over 675} \* (\Nminustwo - \Nminus) \* (330 \* \Ss(1,-2)
  \nonumber\\&& \mbox{}
          - 6220 \* \Ss(1,1)
          - 3600 \* \Ss(1,1) \* \z3
          + 300 \* \Ss(1,2)
          + 1440 \* \Ss(1,3)
          - 1320 \* \Ss(2,1)
          - 1440 \* \Ss(3,1)
          - 900 \* \Sss(1,-2,1)
  \nonumber\\&& \mbox{}
          + 900 \* \Sss(1,1,-2)
          - 975 \* \Sss(1,1,1)
          + 900 \* \Sss(1,1,2)
          + 1800 \* \Sss(1,1,3)
          + 900 \* \Sss(1,2,1)
          - 1800 \* \Sss(1,3,1)
          - 450 \* \Ssss(1,1,1,1)
  \nonumber\\&& \mbox{}
          + 1904 \* \S(1)
          - 5580 \* \S(1) \* \z3
          - 416 \* \S(2)
          + 360 \* \S(3) )
          + {16 \over 675} \* (\Nminus + 1) \* (900 \* \Ss(1,-3)
          + 1270 \* \Ss(1,-2)
          - 24230 \* \Ss(1,1)
  \nonumber\\&& \mbox{}
          - 10800 \* \Ss(1,1) \* \z3
          - 300 \* \Ss(1,2)
          + 5760 \* \Ss(1,3)
          + 2700 \* \Ss(2,-3)
          - 12300 \* \Ss(2,-2)
          + 23595 \* \Ss(2,1)
          - 14850 \* \Ss(2,2)
  \nonumber\\&& \mbox{}
          - 10800 \* \Ss(2,3)
          - 900 \* \Ss(3,-2)
          - 42360 \* \Ss(3,1)
          + 5400 \* \Ss(3,2)
          + 17100 \* \Ss(4,1)
          - 4200 \* \Sss(1,-2,1)
          + 4800 \* \Sss(1,1,-2)
  \nonumber\\&& \mbox{}
          - 5775 \* \Sss(1,1,1)
          + 4500 \* \Sss(1,1,2)
          + 5400 \* \Sss(1,1,3)
          + 4500 \* \Sss(1,2,1)
          - 5400 \* \Sss(1,3,1)
          + 900 \* \Sss(2,-2,1)
          - 9900 \* \Sss(2,1,-2)
  \nonumber\\&& \mbox{}
          + 18675 \* \Sss(2,1,1)
          - 2700 \* \Sss(2,1,2)
          - 2700 \* \Sss(2,2,1)
          - 5400 \* \Sss(3,1,1)
          - 2250 \* \Ssss(1,1,1,1)
          + 1350 \* \Ssss(2,1,1,1)
  \nonumber\\&& \mbox{}
          + 2046 \* \S(1)
          - 23220 \* \S(1) \* \z3
          - 15032 \* \S(2)
          + 29700 \* \S(2) \* \z3
          - 8225 \* \S(3)
          + 26925 \* \S(4)
          - 2700 \* \S(5))
          \biggr)
  \nonumber\\&& \mbox{}
       + \colour4colour{\ca \* \cf \* \nf}  \*  \biggl(
          - {64 \over 3} \* \Ss(1,-3)
          - {11776 \over 45} \* \Ss(1,-2)
          + {86368 \over 135} \* \Ss(1,1)
          + {1984 \over 9} \* \Ss(1,2)
          + {1312 \over 5} \* \Ss(1,3)
          - {832 \over 3} \* \Ss(2,-2)
  \nonumber\\&& \mbox{}
          + {51424 \over 45} \* \Ss(2,1)
          - {896 \over 3} \* \Ss(2,2)
          - {2272 \over 5} \* \Ss(3,1)
          + {256 \over 3} \* \Sss(1,-2,1)
          + {640 \over 3} \* \Sss(1,1,-2)
          - {1888 \over 9} \* \Sss(1,1,1)
          - {320 \over 3} \* \Sss(1,1,2)
  \nonumber\\&& \mbox{}
          - {320 \over 3} \* \Sss(1,2,1)
          + {704 \over 3} \* \Sss(2,1,1)
          + {320 \over 3} \* \Ssss(1,1,1,1)
          + {16 \over 135} \* \gqq \* (2670 \* \Ss(1,-3)
          + 4839 \* \Ss(1,-2)
          - 8279 \* \Ss(1,1)
  \nonumber\\&& \mbox{}
          - 1080 \* \Ss(1,1) \* \z3
          - 3765 \* \Ss(1,2)
          + 873 \* \Ss(1,3)
          - 1710 \* \Ss(2,-3)
          + 420 \* \Ss(2,-2)
          - 4416 \* \Ss(2,1)
          + 1530 \* \Ss(2,2)
  \nonumber\\&& \mbox{}
          - 2835 \* \Ss(2,3)
          - 630 \* \Ss(3,-2)
          + 687 \* \Ss(3,1)
          - 2700 \* \Ss(3,2)
          - 3015 \* \Ss(4,1)
          - 2130 \* \Sss(1,-2,1)
          - 1050 \* \Sss(1,1,-2)
  \nonumber\\&& \mbox{}
          + 3495 \* \Sss(1,1,1)
          - 900 \* \Sss(1,1,2)
          + 540 \* \Sss(1,1,3)
          - 900 \* \Sss(1,2,1)
          - 540 \* \Sss(1,3,1)
          + 900 \* \Sss(2,-2,1)
          - 720 \* \Sss(2,1,-2)
  \nonumber\\&& \mbox{}
          - 1305 \* \Sss(2,1,1)
          + 1350 \* \Sss(2,1,2)
          + 1350 \* \Sss(2,2,1)
          + 2970 \* \Sss(3,1,1)
          + 900 \* \Ssss(1,1,1,1)
          - 1350 \* \Ssss(2,1,1,1)
          + 20761 \* \S(1)
  \nonumber\\&& \mbox{}
          - 306 \* \S(1) \* \z3
          + 3739 \* \S(2)
          + 3510 \* \S(2) \* \z3
          + 16134 \* \S(3)
          - 3315 \* \S(4)
          + 3420 \* \S(5) )
          - {6176 \over 5} \* \S(1)
          - {5312 \over 15} \* \S(1) \* \z3
  \nonumber\\&& \mbox{}
          - {282304 \over 135} \* \S(2)
          - {92192 \over 45} \* \S(3)
          + 320 \* \S(4)
          + {32 \over 15} \* (\Nplusthree - \Nplustwo) \* (4 \* \Ss(1,-3)
          - \Ss(1,-2)
          - 24 \* \Ss(1,1) \* \z3
          + 4 \* \Ss(2,-2)
  \nonumber\\&& \mbox{}
          - 12 \* \Ss(2,3)
          - 4 \* \Ss(3,1)
          + 12 \* \Ss(4,1)
          - 4 \* \Sss(1,-2,1)
          - 4 \* \Sss(1,1,-2)
          + 12 \* \Sss(1,1,3)
          - 12 \* \Sss(1,3,1)
          + 24 \* \S(2) \* \z3 
          - \S(3)
          + 4 \* \S(4) )
  \nonumber\\&& \mbox{}
          - {64 \over 45} \* (\Nminusthree - \Nminustwo) \* (4 \* \Ss(1,-3)
          - \Ss(1,-2)
          - 24 \* \Ss(1,1) \* \z3
          - 4 \* \Sss(1,-2,1)
          - 4 \* \Sss(1,1,-2)
          + 12 \* \Sss(1,1,3)
          - 12 \* \Sss(1,3,1))
  \nonumber\\&& \mbox{}
          - {16 \over 135} \* (\Nplustwo - 3) \* (1800 \* \Ss(1,-3)
          + 2928 \* \Ss(1,-2)
          - 5069 \* \Ss(1,1)
          - 720 \* \Ss(1,1) \* \z3
          - 1650 \* \Ss(1,2)
          + 1296 \* \Ss(1,3)
  \nonumber\\&& \mbox{}
          + 1980 \* \Ss(2,-2)
          - 5751 \* \Ss(2,1)
          + 1260 \* \Ss(2,2)
          - 360 \* \Ss(2,3)
          + 864 \* \Ss(3,1)
          + 360 \* \Ss(4,1)
          - 1260 \* \Sss(1,-2,1)
          - 180 \* \Sss(1,1,-2)
  \nonumber\\&& \mbox{}
          + 1515 \* \Sss(1,1,1)
          - 900 \* \Sss(1,1,2)
          + 360 \* \Sss(1,1,3)
          - 900 \* \Sss(1,2,1)
          - 360 \* \Sss(1,3,1)
          - 1260 \* \Sss(2,1,1)
          + 900 \* \Ssss(1,1,1,1)
  \nonumber\\&& \mbox{}
          + 11091 \* \S(1)
          - 1152 \* \S(1) \* \z3
          + 10103 \* \S(2)
          + 720 \* \S(2) \* \z3
          + 11526 \* \S(3) )
          + {16 \over 135} \* (\Nminustwo - \Nminus) \* (900 \* \Ss(1,-3)
  \nonumber\\&& \mbox{}
          + 777 \* \Ss(1,-2)
          - 511 \* \Ss(1,1)
          - 360 \* \Ss(1,1) \* \z3
          - 1185 \* \Ss(1,2)
          + 684 \* \Ss(1,3)
          - 180 \* \Ss(2,-2)
          - 216 \* \Ss(2,1)
          + 360 \* \Ss(2,2)
  \nonumber\\&& \mbox{}
          - 144 \* \Ss(3,1)
          - 630 \* \Sss(1,-2,1)
          - 90 \* \Sss(1,1,-2)
          + 1095 \* \Sss(1,1,1)
          - 450 \* \Sss(1,1,2)
          + 180 \* \Sss(1,1,3)
          - 450 \* \Sss(1,2,1)
  \nonumber\\&& \mbox{}
          - 180 \* \Sss(1,3,1)
          - 450 \* \Sss(2,1,1)
          + 450 \* \Ssss(1,1,1,1)
          + 4459 \* \S(1)
          - 648 \* \S(1) \* \z3
          - 716 \* \S(2)
          - 48 \* \S(3) )
  \nonumber\\&& \mbox{}
          - {16 \over 135} \* (\Nminus + 1) \* (4380 \* \Ss(1,-3)
          + 6663 \* \Ss(1,-2)
          - 10649 \* \Ss(1,1)
          - 1080 \* \Ss(1,1) \* \z3
          - 4485 \* \Ss(1,2)
          + 3276 \* \Ss(1,3)
  \nonumber\\&& \mbox{}
          - 1710 \* \Ss(2,-3)
          + 1230 \* \Ss(2,-2)
          - 5346 \* \Ss(2,1)
          + 1530 \* \Ss(2,2)
          - 2835 \* \Ss(2,3)
          - 630 \* \Ss(3,-2)
          - 366 \* \Ss(3,1)
  \nonumber\\&& \mbox{}
          - 2700 \* \Ss(3,2)
          - 3015 \* \Ss(4,1)
          - 3030 \* \Sss(1,-2,1)
          - 330 \* \Sss(1,1,-2)
          + 4125 \* \Sss(1,1,1)
          - 2250 \* \Sss(1,1,2)
          + 540 \* \Sss(1,1,3)
  \nonumber\\&& \mbox{}
          - 2250 \* \Sss(1,2,1)
          - 540 \* \Sss(1,3,1)
          + 900 \* \Sss(2,-2,1)
          - 720 \* \Sss(2,1,-2)
          - 1575 \* \Sss(2,1,1)
          + 1350 \* \Sss(2,1,2)
          + 1350 \* \Sss(2,2,1)
  \nonumber\\&& \mbox{}
          + 2970 \* \Sss(3,1,1)
          + 2250 \* \Ssss(1,1,1,1)
          - 1350 \* \Ssss(2,1,1,1)
          + 26641 \* \S(1)
          - 2952 \* \S(1) \* \z3
          + 5020 \* \S(2)
          + 3510 \* \S(2) \* \z3
  \nonumber\\&& \mbox{}
          + 19017 \* \S(3)
          - 1965 \* \S(4)
          + 3420 \* \S(5))
          \biggr)\biggr\}
\:\: .
\eea
\normalsize
Note that the (additional) bracketing of factors of $(\cf - \ca /2)$ in 
the non-singlet coefficient functions (A.5), (A.8), (A.13) and (A.18) 
and their counterparts in Appendix B has no physical significance, but 
has only been performed to shorten the formulae. 

\renewcommand{\theequation}{B.\arabic{equation}}
\setcounter{equation}{0}
\section*{Appendix B: The exact x-space results}
In this final appendix we write down the full $x$-space coefficient
functions up to the third order. These functions can be expressed in
terms of harmonic polylogarithms~\cite {Goncharov,Borwein,%
Remiddi:1999ew}, for which we adopt the notation $H_{m_1,...,m_w}(x)$,
$m_j = 0,\pm 1$ of Ref.~\cite{Remiddi:1999ew}. The $n$-loop coefficient
functions for $F_{\, 2}$ and $F_L$ involve harmonic polylogarithms up
to weight $w = 2n-1$. Below we use the short-hand notation
\beq
  H_{{\footnotesize \underbrace{0,\ldots ,0}_{\scriptstyle m} },\,
  \pm 1,\, {\footnotesize \underbrace{0,\ldots ,0}_{\scriptstyle n} },
  \, \pm 1,\, \ldots}(x) \: = \: H_{\pm (m+1),\,\pm (n+1),\, \ldots}(x)
\eeq
and suppress the argument $x$ for brevity. Furthermore we employ the 
abbreviations
\bea
  \pqq(x) &\! =\! & 2\, (1 - x)^{-1} - 1 - x \, ,\nn \\
  \pqg(x) &\! =\! & 1 - 2x + 2x^{\,2} \, , \nn \\
  \pgq(x) &\! =\! & 2x^{\,-1} -2 + x  \, , \nn \\
  \pgg(x) &\! =\! & (1-x)^{-1} + x^{\,-1} - 2 + x - x^{\,2} \:\: .
\eea
%
All divergences for $x \to 1 $ in Eq.~(B.2) and below are to be read
as the $+$-distributions ${\cal D}_k$ of Eqs.~(\ref{eq:abbr}) and 
(\ref{eq:plus}).

%
%

In this notation the one-loop coefficient functions for $F_{\, 2}$ are 
given by
\bea
c^{(1)}_{2,\rm{q}}(x) & \! = \! &  
         \colour4colour{\cf}  \*  \biggl(
            {1 \over 2} \* (9 + 5 \* x)
          - {1 \over 2} \* \pqq(x) \* (3 + 4 \* \H(0)
          + 4 \* \H(1))
          - \delta(1 - x) \* (9 + 4 \* \z2)
          \biggr)
\:\: , \\[1mm] 
c^{(1)}_{2,\rm{g}}(x) & \! = \! &  
         \colour4colour{\nf}  \*  \bigl(
            6
          - 2 \* \pqg(x) \* (4 + \H(0)
          + \H(1))
          \bigr)
\:\: .
\eea
The exact two-loop results corresponding to the approximations 
(\ref{eq:c2ns2}) -- (\ref{eq:c2gl2}) read
\small
\bea
&& c^{(2)}_{2,\rm ns}(x) \:\: = \:\: 
         \colour4colour{\cf \* \biggl(\cf-{\ca \over 2}\biggr)}  \*  \biggl(
            {8 \over 5} \* (
            9 \* \pqg( - x) \* (1 - x) 
          + \pgq( - x) \* (1 - x^{-1}) 
          + (37 + 17 \* x)) \* \Hh(-1,0) 
  \nonumber\\&& \mbox{}
          + 4 \* \pqq( - x) \* (7 \* \z3 + 6 \* \Hh(-2,0)
          + 4 \* \Hh(-1,2)
          - 8 \* \Hhh(-1,-1,0) 
            + 10 \* \Hhh(-1,0,0)
          - 3 \* \Hhh(0,0,0)
          - 8 \* \H(-1) \* \z2 - 2 \* \H(0) 
  \nonumber\\&& \mbox{}
	    + 2 \* \H(0) \* \z2 - 2 \* \H(3))
          - {72 \over 5} \* \pqg(x) \* ((1 + x) \* (\z2 - \Hh(0,0))
          - 1 - \H(0))
          + {8 \over 5} \* \pgq(x) \* (1 - \H(0))
  \nonumber\\&& \mbox{}
          + 8 \* (1 - 5 \* x) \* (\Hhh(1,0,0)
          - \H(1) \* \z2)
          - 8 \* (1 + 5 \* x) \* (2 \* \Hhh(-1,-1,0)
          - \Hhh(-1,0,0)
          + \H(-1) \* \z2)
          \biggr)
  \nonumber\\&& \mbox{}
       + \colour4colour{\cf \* \nf}  \*  \biggl(
            {1 \over 54} \* \pqq(x) \* (247 
	    - 144 \* \z2 + 180 \* \Hh(0,0)
          + 72 \* \Hh(1,0)
          + 72 \* \Hh(1,1)
          + 342 \* \H(0) 
	    + 174 \* \H(1)
          + 144 \* \H(2))
  \nonumber\\&& \mbox{}
          - {1 \over 3} \* (7 + 19 \* x) \* \H(0) 
          - {1 \over 3} \* (1 + 13 \* x) \* \H(1) 
          - {1 \over 18} \* (23 + 243 \* x)
          + \delta(1 - x) \* \biggl({457 \over 36} + {4 \over 3} \* \z3 + {38 \over 3} \* \z2\biggr)
          \biggr)
  \nonumber\\&& \mbox{}
       + \colour4colour{\cf^2}  \*  \biggl(
            16 \* \Hh(-2,0)
          + {1 \over 5} \* (33 
            + 37 \* x) \* \Hh(0,0) 
          + {1 \over 4} \* \pqq(x) \* (51 + 128 \* \z3 + 48 \* \z2 + 96 \* \Hh(-2,0)
          - 12 \* \Hh(0,0) 
  \nonumber\\&& \mbox{}
            - 72 \* \Hh(1,0)
          - 72 \* \Hh(1,1)
          - 96 \* \Hh(1,2)
          - 96 \* \Hh(2,0) 
	    - 112 \* \Hh(2,1)
          - 32 \* \Hhh(0,0,0)
          - 48 \* \Hhh(1,0,0)
          - 128 \* \Hhh(1,1,0)
          - 96 \* \Hhh(1,1,1) 
  \nonumber\\&& \mbox{}
	    + 122 \* \H(0)
          + 96 \* \H(0) \* \z2 + 54 \* \H(1) 
	    + 32 \* \H(1) \* \z2 - 48 \* \H(2)
          - 96 \* \H(3))
          - {1 \over 2} \* (43 + 63 \* x) \* \H(0) 
          + {1 \over 2} \* (59 - 109 \* x) \* \H(1)
  \nonumber\\&& \mbox{}
          - 4 \* (1 - 19 \* x) \* \z3
          + 2 \* (1 + x) \* (7 \* \Hh(1,0)
          + 2 \* \Hh(2,0)
          + 2 \* \Hh(2,1) 
	  + 5 \* \Hhh(0,0,0)
          - 4 \* \H(0) \* \z2 + 4 \* \H(3))
          + 2 \* (5 + 9 \* x) \* (\Hh(1,1) 
  \nonumber\\&& \mbox{}
	    + 2 \* \H(2))
          - {4 \over 5} \* (7 + 13 \* x) \* \z2
          - {1 \over 4} \* (93 + 209 \* x)
          + \delta(1 - x) \* \biggl({331 \over 8} - 78 \* \z3 + 69 \* \z2 + 6 \* \z2^2\biggr)
          \biggr)
  \nonumber\\&& \mbox{}
       + \colour4colour{\ca \* \cf}  \*  \biggl(
          - {1 \over 108} \* \pqq(x) \* (3155 - 216 \* \z3 - 1584 \* \z2 + 1296 \* \Hh(-2,0)
          + 1980 \* \Hh(0,0)
          + 792 \* \Hh(1,0) 
  \nonumber\\&& \mbox{}
	    + 792 \* \Hh(1,1)
          + 432 \* \Hh(1,2)
          + 648 \* \Hhh(0,0,0)
          + 864 \* \Hhh(1,0,0)
          - 432 \* \Hhh(1,1,0) 
	    + 4302 \* \H(0)
          - 432 \* \H(0) \* \z2 + 2202 \* \H(1) 
  \nonumber\\&& \mbox{}
	    - 1296 \* \H(1) \* \z2 + 1584 \* \H(2)
          + 432 \* \H(3))
          + {1 \over 6} \* (71 + 323 \* x) \* \H(0) 
          - {17 \over 6} \* (5 - 19 \* x) \* \H(1) 
          - {4 \over 5} \* (9 + 16 \* x) \* (\z2 
  \nonumber\\&& \mbox{}
	    - \Hh(0,0))
          + {1 \over 36} \* (139 + 3159 \* x)
          - 8 \* (5 \* \z3 \* x + \Hh(-2,0))
          - \delta(1 - x) \* \biggl({5465 \over 72} - {140 \over 3} \* \z3 + {251 \over 3} \* \z2 
  \nonumber\\&& \mbox{}
	    - {71 \over 5} \* \z2^2\biggr)
          \biggr)
\:\: ,
\eea
\bea
&& c^{(2)}_{2,\rm{g}}(x) \:\: = \:\: 
         \colour4colour{\cf \* \nf}  \*  \biggl(
            {4 \over 15} \* ( 
	    \pgq( - x) \* (1 - x^{-1})
          + (217 + 117 \* x)) \* \Hh(-1,0)
          + {1 \over 15} \* (639 - 1004 \* x) \* \Hh(0,0)
  \nonumber\\&& \mbox{}
          - {8 \over 5} \* \pqg( - x) \* (6 \* (1 - x)  \* \Hh(-1,0)
          - 5 \* (2 \* \Hh(-2,0) 
	    - 2 \* \Hhh(-1,-1,0)
          + \Hhh(-1,0,0)
          - \H(-1) \* \z2))
  \nonumber\\&& \mbox{}
          + {2 \over 5} \* \pqg(x) \* (6 \* (19 + 4 \* x) \* (\z2 - \Hh(0,0)) 
	    - (9 - 90 \* \z3 + 90 \* \Hh(1,0)
          + 90 \* \Hh(1,1)
          + 60 \* \Hh(1,2)
          + 40 \* \Hh(2,0)
          + 50 \* \Hh(2,1) 
  \nonumber\\&& \mbox{}
	    + 50 \* \Hhh(0,0,0)
          + 30 \* \Hhh(1,0,0)
          + 40 \* \Hhh(1,1,0)
            + 50 \* \Hhh(1,1,1)
          + 54 \* \H(0)
          - 60 \* \H(0) \* \z2 + 30 \* \H(1)
          - 40 \* \H(1) \* \z2 + 90 \* \H(2) 
  \nonumber\\&& \mbox{}
	    + 60 \* \H(3)))
          + {4 \over 15} \* \pgq(x) \* (1 - \H(0))
          + {1 \over 3} \* (16 - 61 \* x) \* \H(0)
          - 2 \* (1 - 8 \* x) \* \H(1) 
          + 4 \* (5 - 4 \* x) \* \H(2)
  \nonumber\\&& \mbox{}
          - 4 \* (1 - 18 \* x) \* \z3
          + 2 \* (1 - 2 \* x) \* (8 \* \Hh(-2,0)
          + 2 \* \Hh(2,0)
          + 2 \* \Hh(2,1)
          + 5 \* \Hhh(0,0,0) 
	    + 4 \* \Hhh(1,0,0)
          - 4 \* \H(0) \* \z2 - 4 \* \H(1) \* \z2 
  \nonumber\\&& \mbox{}
	    + 4 \* \H(3))
          - 8 \* (1 + 2 \* x) \* (2 \* \Hhh(-1,-1,0)
          - \Hhh(-1,0,0)
          + \H(-1) \* \z2)
          + 2 \* (5 + 4 \* x) \* (\Hh(1,0)
          + \Hh(1,1))
  \nonumber\\&& \mbox{}
          - {4 \over 15} \* (111 - 176 \* x) \* \z2
          - {1 \over 3} \* (117 - 121 \* x)
          \biggr)
       +  \colour4colour{\ca \* \nf}  \*  \biggl(
            {8 \over 3} \* ( 
	    \pgq( - x)
            - 3 \* (4 + 3 \* x)) \* \Hh(-1,0) 
  \nonumber\\&& \mbox{}
          + {4 \over 3} \* \Hh(0,0) \* (47 + 35 \* x)
          + 2 \* (23 - 6 \* x) \* \Hh(1,0)
          + 6 \* (7 - 2 \* x) \* \Hh(1,1)
          + 4 \* (5 + 14 \* x) \* \Hhh(0,0,0)
          + {4 \over 3} \* \pqg( - x) \* (6 \* \Hh(-2,0) 
  \nonumber\\&& \mbox{}
	    + 10 \* \Hh(-1,0)
          + 6 \* \Hh(-1,2)
          + 9 \* \Hhh(-1,0,0)
          - 6 \* \H(-1) \* \z2)
          - {1 \over 54} \* \pqg(x) \* (4493 - 648 \* \z3 - 3996 \* \z2 + 3492 \* \Hh(0,0) 
  \nonumber\\&& \mbox{}
	    + 2412 \* \Hh(1,0)
          + 2196 \* \Hh(1,1)
          + 216 \* \Hh(1,2)
          + 432 \* \Hh(2,0)
          + 432 \* \Hh(2,1) 
	    + 432 \* \Hhh(1,0,0)
          + 648 \* \Hhh(1,1,0)
          + 216 \* \Hhh(1,1,1) 
  \nonumber\\&& \mbox{}
            + 6270 \* \H(0)
          - 432 \* \H(0) \* \z2 + 4710 \* \H(1)
          - 216 \* \H(1) \* \z2 + 3996 \* \H(2)
            + 432 \* \H(3))
          + {4 \over 27} \* \pgq(x) \* (43 - 18 \* \z2 
  \nonumber\\&& \mbox{}
	    + 18 \* \Hh(1,0)
          + 18 \* \Hh(1,1)
          - 39 \* \H(1))
          + {1 \over 9} \* (1567 - 338 \* x) \* \H(0) 
          + {1 \over 3} \* (289 - 52 \* x) \* \H(1)
          + 2 \* (33 - 2 \* x) \* \H(2)
  \nonumber\\&& \mbox{}
          - 4 \* (1 - 2 \* x) \* (\Hhh(1,0,0)
          - \H(1) \* \z2)
          - 4 \* (1 + 2 \* x) \* (2 \* \Hh(-2,0)
          - 2 \* \Hhh(-1,-1,0)
          + \Hhh(-1,0,0) 
	    - \H(-1) \* \z2)
  \nonumber\\&& \mbox{}
          - 8 \* (1 + 3 \* x) \* (\z3 + 2 \* \H(0) \* \z2 - 2 \* \H(3))
          + 8 \* (1 + 4 \* x) \* (\Hh(2,0)
          + \Hh(2,1))
          + {7 \over 6} \* (105 - 46 \* x)
  \nonumber\\&& \mbox{}
          - {2 \over 3} \* (107 - 10 \* x) \* \z2
          \biggr)
\eea
\normalsize
and
\small
\bea
&& c^{(2)}_{2,\rm{ps}}(x) \:\: = \:\: 
        \colour4colour{\cf \* \nf}  \*  \biggl(
            {8 \over 3} \* (
	    - \pqg( - x)
            + \pgq( - x)) \* \Hh(-1,0)
          + {8 \over 27} \* \pqg(x) \* (28 + 27 \* \z2 - 36 \* \Hh(0,0)
          - 9 \* \Hh(1,0) 
  \nonumber\\&& \mbox{}
	    - 9 \* \Hh(1,1)
          - 24 \* \H(0)
          + 6 \* \H(1)
          - 27 \* \H(2))
          + {4 \over 27} \* \pgq(x) \* (43 - 18 \* \z2 + 18 \* \Hh(1,0)
          + 18 \* \Hh(1,1)
          - 39 \* \H(1))         
  \nonumber\\&& \mbox{}
          + {8 \over 9} \* (71 - 49 \* x) \* \H(0)
          + {4 \over 3} \* (16 - 13 \* x) \* \H(1)
          + 8 \* (1 - 2 \* x) \* \H(2)
          + 12 \* (1 - x) \* (\Hh(1,0)
          + \Hh(1,1))
  \nonumber\\&& \mbox{}
          - {2 \over 3} \* (1 + x) \* (12 \* \z3 + 12 \* \Hh(-1,0)
          - 13 \* \Hh(0,0) 
	    - 12 \* \Hh(2,0)
          - 12 \* \Hh(2,1)
          - 30 \* \Hhh(0,0,0)
          + 24 \* \H(0) \* \z2 - 24 \* \H(3))
  \nonumber\\&& \mbox{}
          + {22 \over 3} \* (3 - 5 \* x)
          - {8 \over 3} \* (5 - x) \* \z2
          \biggr)
\:\: .
\eea
\normalsize

\noindent
The full three-loop non-singlet coefficient function for $F_{\,2}$ 
underlying the parametrization (\ref{eq:c2ns3}) is 
\small
\bea
&& c^{(3)}_{2,\rm ns}(x) \:\: = \:\: 
         \colour4colour{\dabcnc} \* \fl11  \*  \biggl(
          - {64 \over 3} \* (6 - 37 \* x) \* \Hh(-2,0)
          - {256 \over 15} \* (18 - 7 \* x) \* \Hh(-2,2)
          - {128 \over 15} \* (67 + 92 \* x) \* \Hh(-1,0)
  \nonumber\\&& \mbox{}
          + {128 \over 15} \* (16 - 39 \* x) \* \Hh(-1,0) \* \z2
          - {64 \over 15} \* (317 + 542 \* x) \* \Hh(-1,2)
          - {128 \over 15} \* (101 - 130 \* x) \* \Hh(0,0)
  \nonumber\\&& \mbox{}
          + {128 \over 15} \* (149 - 129 \* x) \* \Hh(1,0) \* \z2
          - {256 \over 15} \* (9 + 4 \* x) \* \Hhh(-2,0,0)
          - {64 \over 3} \* (5 + 18 \* x) \* \Hhh(-1,-1,0)
  \nonumber\\&& \mbox{}
          - {128 \over 15} \* (73 + 113 \* x) \* \Hhh(-1,0,0)
          - {64 \over 3} \* (18 - 17 \* x) \* \Hhh(0,0,0)
          + {128 \over 15} \* (42 - 37 \* x) \* \Hhh(1,0,0)
  \nonumber\\&& \mbox{}
          + {128 \over 15} \* (27 - 142 \* x) \* \Hhh(2,0,0)
          + 32 \* \pqq( - x) \* (8 \* \z2
          - 4 \* \Hh(0,0)
          + \Hhh(0,0,0)
          - 5 \* \H(0)
          - \H(0) \* \z2
          - 8 \* \H(2))
  \nonumber\\&& \mbox{}
          + 32 \* \pqq(x) \* (3 \* \z3
          + 2 \* \Hh(-2,0)
          + 2 \* \Hh(0,0)
          - \Hhh(0,0,0)
          + \H(0) \* \z2)
          + {96 \over 5} \* \pqg( - x) \* (2 \* (3 - 13 \* x) \* \Hh(-1,0)
  \nonumber\\&& \mbox{}
          + 2 \* (17 - 10 \* x) \* \Hh(-1,2)
          + 4 \* (3 - 5 \* x) \* \Hhh(-1,0,0)
          - (29 - 30 \* x) \* \H(-1) \* \z2
          + 4 \* (1 - x) \* (4 \* \Hh(-2,2)
          + 2 \* \Hh(-1,3)
  \nonumber\\&& \mbox{}
          + 4 \* \Hh(-1,-1) \* \z2
          - 2 \* \Hh(-1,0) \* \z2
          + 2 \* \Hhh(-2,0,0)
          - 4 \* \Hhh(-1,-1,2)
          - 2 \* \Hhhh(-1,-1,0,0)
          - 4 \* \H(-2) \* \z2
          - 3 \* \H(-1) \* \z3)
  \nonumber\\&& \mbox{}
          - 10 \* (1 + 2 \* x) \* (\Hh(-2,0)
          - \Hhh(-1,-1,0)))
          + {96 \over 25} \* \pqg(x) \* (100 \* (1 + x) \* \Hhh(0,0,0)
          + 20 \* (1 - 4 \* x) \* \H(0) \* \z3
  \nonumber\\&& \mbox{}
          - 10 \* (13 + 20 \* x) \* \H(0) \* \z2
          + 25 \* (1 - 2 \* x) \* \H(1) \* \z2
          + 20 \* (4 - x) \* \H(1) \* \z3
          + 10 \* (3 + 10 \* x) \* \H(3)
  \nonumber\\&& \mbox{}
          - 20 \* (3 - 2 \* x) \* (\Hh(1,0) \* \z2
          - \Hh(1,3)
          + \Hhh(2,0,0)
          + \Hhhh(1,1,0,0))
          + 8 \* (7 - 3 \* x) \* \z2^2
          - 10 \* (23 + 13 \* x) \* (\z2 - \Hh(0,0))
  \nonumber\\&& \mbox{}
          - 5 \* (43 + 50 \* x) \* \z3
          + 10 \* (3
          - 10 \* \Hh(0,0) \* \z2
          - 4 \* \Hhh(1,0,0)
          + 13 \* \H(0)
          + 10 \* \H(1)
          + 10 \* \H(2)
          + 10 \* \H(4) ))
  \nonumber\\&& \mbox{}
          + {64 \over 15} \* \pgq( - x) \* ((3 - 13 \* x^{-1}) \* \Hh(-1,0)
          + 2 \* (1 - 5 \* x^{-1}) \* \Hh(-1,2)
          + 2 \* (3 - 5 \* x^{-1}) \* \Hhh(-1,0,0)
  \nonumber\\&& \mbox{}
          - (7 - 15 \* x^{-1}) \* \H(-1) \* \z2
          + 2 \* (1 - x^{-1}) \* (4 \* \Hh(-1,-1) \* \z2
          - 2 \* \Hh(-1,0) \* \z2
          + 2 \* \Hh(-1,3)
          - 5 \* \Hhh(-1,-1,0)
          - 4 \* \Hhh(-1,-1,2)
  \nonumber\\&& \mbox{}
          - 2 \* \Hhhh(-1,-1,0,0)
          - 3 \* \H(-1) \* \z3))
          + {64 \over 15} \* \pgq(x) \* ((1 + x^{-1}) \* (4 \* \Hh(1,0) \* \z2
          - 4 \* \Hh(1,3)
          + 4 \* \Hhhh(1,1,0,0)
          - 2 \* \H(1) \* \z3
  \nonumber\\&& \mbox{}
          - 5 \* \H(1) \* \z2 )
          + 3
          + 8 \* \z3
          + 20 \* \z2
          - 10 \* \Hh(0,0)
          - 4 \* \Hhh(1,0,0)
          - 3 \* \H(0)
          + 10 \* \H(1)
          - 10 \* \H(2))
          + {128 \over 15} \* \H(-2) \* (36 + x) \* \z2
  \nonumber\\&& \mbox{}
          + {224 \over 15} \* \H(-1) \* (87 + 142 \* x) \* \z2
          - {128 \over 15} \* \H(0) \* (9 - 134 \* x) \* \z3
          + {64 \over 15} \* \H(0) \* (109 - 269 \* x) \* \z2
  \nonumber\\&& \mbox{}
          - {32 \over 15} \* \H(0) \* (175 - 513 \* x)
          - {32 \over 3} \* \H(1) \* (5 - 18 \* x) \* \z2
          - {128 \over 15} \* \H(1) \* (117 - 67 \* x)
          - {128 \over 15} \* \H(1) \* (157 - 177 \* x) \* \z3
  \nonumber\\&& \mbox{}
          - {1664 \over 15} \* \H(2) \* (2 + 7 \* x)
          - {64 \over 15} \* \H(3) \* (19 - 184 \* x)
          - 128 \* (1 + x)
          - {512 \over 5} \* (1 + 2 \* x) \* (5 \* \Hh(-1,0) \* \z3
          + 5 \* \Hh(-1,4)
  \nonumber\\&& \mbox{}
          - 5 \* \Hh(2,0) \* \z2
          + 5 \* \Hh(2,3)
          - 5 \* \Hhh(-1,0,0) \* \z2
          - 5 \* \Hhhh(-1,2,0,0)
          - 5 \* \Hhhh(2,1,0,0)
          + 4 \* \H(-1) \* \z2^2
          + 5 \* \H(2) \* \z3)
  \nonumber\\&& \mbox{}
          - {512 \over 15} \* (2 - 3 \* x) \* (4 \* \Hh(-1,-1) \* \z2
          + 2 \* \Hh(-1,3)
          - 4 \* \Hhh(-1,-1,2)
          - 2 \* \Hhhh(-1,-1,0,0)
          - 3 \* \H(-1) \* \z3)
  \nonumber\\&& \mbox{}
          + 128 \* (3 - 10 \* x) \* (\Hh(0,0) \* \z2 - \H(4))
          + {64 \over 5} \* (63 - 65 \* x) \* \z2
          - {64 \over 25} \* (84 - 409 \* x) \* \z2^2
          + {32 \over 5} \* (122 - 637 \* x) \* \z3
  \nonumber\\&& \mbox{}
          - {128 \over 15} \* (149 - 144 \* x) \* (\Hh(1,3)
          - \Hhhh(1,1,0,0))
          + 128 \* x \* (40 \* \z5
          + 4 \* \z2 \* \z3
          + 2 \* \Hhh(-2,-1,0)
          + 2 \* \Hhh(1,-2,0)
          - \Hhhh(1,0,0,0)
  \nonumber\\&& \mbox{}
          + \Hhhh(-1,0,0,0)
          - \H(2) \* \z2)
          + \delta(1 - x) \* \biggl(64 - {1280 \over 3} \* \z5 
          + {224 \over 3} \* \z3 + 160 \* \z2 - {32 \over 5} \* \z2^2\biggr)
          \biggr)
  \nonumber\\&& \mbox{}
       +  \colour4colour{{\cf \* \biggl(\cf-{\ca \over 2}\biggr)^2}} \*  \biggl(
            {92 \over 3} \* \gfunct1(x)
          - {4 \over 3} \* \gfunct2(x)
          \biggr)
  \nonumber\\&& \mbox{}
       +  \colour4colour{\cf \* \nf \* \biggl(\cf-{\ca \over 2}\biggr)} 
          \*  \biggl(
          - {16 \over 45} \* (673 - 297 \* x) \* \Hh(-2,0)
          - {32 \over 675} \* (9661 + 8536 \* x) \* \Hh(-1,0)
  \nonumber\\&& \mbox{}
          - {64 \over 45} \* (103 + 88 \* x) \* \Hh(-1,2)
          - {32 \over 3} \* (7 + 5 \* x) \* \Hhh(-2,0,0)
          + {64 \over 45} \* (83 + 198 \* x) \* \Hhh(-1,-1,0)
  \nonumber\\&& \mbox{}
          - {128 \over 45} \* (112 + 107 \* x) \* \Hhh(-1,0,0)
          - {8 \over 405} \* \pqq( - x) \* (2700 \* \z3
          + 830 \* \z2
          + 1026 \* \z2^2
          + 4500 \* \Hh(-3,0)
  \nonumber\\&& \mbox{}
          + 3000 \* \Hh(-2,0)
          + 360 \* \Hh(-2,2)
          + 1800 \* \Hh(-1,-1) \* \z2
          + 1660 \* \Hh(-1,0)
          - 1260 \* \Hh(-1,0) \* \z2
          + 1200 \* \Hh(-1,2)
  \nonumber\\&& \mbox{}
          + 360 \* \Hh(-1,3)
          - 1910 \* \Hh(0,0)
          + 450 \* \Hh(0,0) \* \z2
          + 360 \* \Hh(3,1)
          - 3960 \* \Hhh(-2,-1,0)
          + 6120 \* \Hhh(-2,0,0)
  \nonumber\\&& \mbox{}
          - 3960 \* \Hhh(-1,-2,0)
          - 3600 \* \Hhh(-1,-1,0)
          + 4800 \* \Hhh(-1,0,0)
          - 720 \* \Hhh(-1,2,1)
          - 1245 \* \Hhh(0,0,0)
  \nonumber\\&& \mbox{}
          + 3600 \* \Hhhh(-1,-1,-1,0)
          - 6120 \* \Hhhh(-1,-1,0,0)
          + 6120 \* \Hhhh(-1,0,0,0)
          - 2070 \* \Hhhh(0,0,0,0)
          - 2340 \* \H(-2) \* \z2
  \nonumber\\&& \mbox{}
          - 1440 \* \H(-1) \* \z3
          - 3000 \* \H(-1) \* \z2
          - 1305 \* \H(0)
          + 990 \* \H(0) \* \z3
          + 345 \* \H(0) \* \z2
          - 600 \* \H(3)
          - 360 \* \H(4) )
  \nonumber\\&& \mbox{}
          - {8 \over 25} \* \pqg( - x) \* (3 \* (33 - 53 \* x) \* \Hh(-1,0)
          - 20 \* (7 + 3 \* x) \* \Hh(-1,2)
          + 180 \* (1 - x) \* \Hhh(-1,0,0)
          + 10 \* (1 + 9 \* x) \* \H(-1) \* \z2
  \nonumber\\&& \mbox{}
          + 20 \* (13 - 3 \* x) \* (\Hh(-2,0)
          - \Hhh(-1,-1,0)))
          - {8 \over 225} \* \pgq( - x) \* ((59 - 119 \* x^{-1}) \* \Hh(-1,0)
          + 30 \* (1 - x^{-1}) \* (4 \* \Hh(-2,0)
  \nonumber\\&& \mbox{}
          + 2 \* \Hh(-1,2)
          - 2 \* \Hhh(-1,-1,0)
          + 6 \* \Hhh(-1,0,0)
          - 3 \* \H(-1) \* \z2))
          + {544 \over 45} \* (17 + 22 \* x) \* \H(-1) \* \z2
  \nonumber\\&& \mbox{}
          - {32 \over 3} \* (1 + 5 \* x) \* (2 \* \Hh(-1,-1) \* \z2
          - \Hh(-1,0) \* \z2
          - 4 \* \Hhh(-2,-1,0)
          - 4 \* \Hhh(-1,-2,0)
          + 4 \* \Hhhh(-1,-1,-1,0)
          - 6 \* \Hhhh(-1,-1,0,0)
  \nonumber\\&& \mbox{}
          + 3 \* \Hhhh(-1,0,0,0)
          - 2 \* \H(-1) \* \z3)
          \biggr)
  \nonumber\\&& \mbox{}
       +  \colour4colour{\cf \* \nf^2}  \*  \biggl(
            {76 \over 27} \* (1 + 5 \* x) \* \Hh(0,0)
          - {1 \over 729} \* \pqq(x) \* (4357 
	  - 864 \* \z3 
	  - 7236 \* \z2 
          + 10980 \* \Hh(0,0)
          + 3132 \* \Hh(1,0)
  \nonumber\\&& \mbox{}
          + 3132 \* \Hh(1,1)
          + 1296 \* \Hh(1,2)
          + 2592 \* \Hh(2,0)
          + 2592 \* \Hh(2,1)
          + 4968 \* \Hhh(0,0,0)
          + 1296 \* \Hhh(1,0,0)
          + 1296 \* \Hhh(1,1,0)
  \nonumber\\&& \mbox{}
          + 1296 \* \Hhh(1,1,1)
          + 11610 \* \H(0)
          - 3888 \* \H(0) \* \z2
          + 4230 \* \H(1)
          - 1296 \* \H(1) \* \z2
          + 7236 \* \H(2)
          + 3888 \* \H(3))
  \nonumber\\&& \mbox{}
          + {2 \over 81} \* (43 + 1547 \* x) \* \H(0)
          - {2 \over 27} \* (29 - 295 \* x) \* \H(1)
          + {4 \over 9} \* (1 + 13 \* x) \* (\Hh(1,0)
          + \Hh(1,1))
          - {4 \over 9} \* (3 + 23 \* x) \* (\z2 - \H(2))
  \nonumber\\&& \mbox{}
          - {1 \over 81} \* (757 - 3599 \* x)
          - \delta(1 - x) \* \biggl({9517 \over 486}
          + {152 \over 81} \* \z3 + {860 \over 27} \* \z2 
          + {32 \over 27} \* \z2^2 \biggr)
          \biggr)
  \nonumber\\&& \mbox{}
       +  \colour4colour{\cf^2 \* \nf}  \*  \biggl(
            {1 \over 2025} \* (131057 + 525143 \* x) \* \Hh(0,0)
          - {1 \over 27} \* (1091 + 1759 \* x) \* \Hh(1,0)
          - {1 \over 9} \* (557 - 199 \* x) \* \Hh(1,1)
  \nonumber\\&& \mbox{}
          + {16 \over 3} \* (7 - 27 \* x) \* \Hh(1,2)
          - {8 \over 3} \* (33 + 25 \* x) \* \Hh(2,0)
          - {16 \over 9} \* (25 + 86 \* x) \* \Hh(2,1)
          - {4 \over 5} \* (33 + 97 \* x) \* \Hhh(0,0,0)
  \nonumber\\&& \mbox{}
          - {10 \over 9} \* (47 - 93 \* x) \* \Hhh(1,0,0)
          - {4 \over 9} \* (95 + 43 \* x) \* \Hhh(1,1,0)
          - {8 \over 3} \* (7 - 17 \* x) \* \Hhh(2,0,0)
          - {1 \over 3240} \* \pqq(x) \* (30045
  \nonumber\\&& \mbox{}
          + 580080 \* \z3
          + 92280 \* \z2
          - 89568 \* \z2^2
          + 276480 \* \Hh(-3,0)
          + 224640 \* \Hh(-2,0)
          + 138240 \* \Hh(-2,2)
  \nonumber\\&& \mbox{}
          + 64920 \* \Hh(0,0)
          + 262080 \* \Hh(0,0) \* \z2
          - 306680 \* \Hh(1,0)
          + 43200 \* \Hh(1,0) \* \z2
          - 245880 \* \Hh(1,1)
          + 57600 \* \Hh(1,1) \* \z2
  \nonumber\\&& \mbox{}
          - 368640 \* \Hh(1,2)
          - 100800 \* \Hh(1,3)
          - 333120 \* \Hh(2,0)
          - 371520 \* \Hh(2,1)
          - 218880 \* \Hh(2,2)
          - 279360 \* \Hh(3,0)
  \nonumber\\&& \mbox{}
          - 290880 \* \Hh(3,1)
          + 276480 \* \Hhh(-2,0,0)
          - 196080 \* \Hhh(0,0,0)
          + 138240 \* \Hhh(1,-2,0)
          - 301200 \* \Hhh(1,0,0)
  \nonumber\\&& \mbox{}
          - 387360 \* \Hhh(1,1,0)
          - 302400 \* \Hhh(1,1,1)
          - 172800 \* \Hhh(1,1,2)
          - 264960 \* \Hhh(1,2,0)
          - 195840 \* \Hhh(1,2,1)
  \nonumber\\&& \mbox{}
          - 204480 \* \Hhh(2,0,0)
          - 293760 \* \Hhh(2,1,0)
          - 236160 \* \Hhh(2,1,1)
          - 129600 \* \Hhhh(0,0,0,0)
          - 152640 \* \Hhhh(1,0,0,0)
  \nonumber\\&& \mbox{}
          - 262080 \* \Hhhh(1,1,0,0)
          - 236160 \* \Hhhh(1,1,1,0)
          - 172800 \* \Hhhh(1,1,1,1)
          - 138240 \* \H(-2) \* \z2
          + 287700 \* \H(0)
  \nonumber\\&& \mbox{}
          + 420480 \* \H(0) \* \z3
          + 505680 \* \H(0) \* \z2
          + 14940 \* \H(1)
          + 187200 \* \H(1) \* \z3
          + 253440 \* \H(1) \* \z2
          - 145400 \* \H(2)
  \nonumber\\&& \mbox{}
          + 92160 \* \H(2) \* \z2
          - 489360 \* \H(3)
          - 267840 \* \H(4))
          - {8 \over 25} \* \pqg(x) \* (3 \* (113 + 53 \* x) \* \Hh(0,0)
          + 180 \* (1 + x) \* \Hhh(0,0,0)
  \nonumber\\&& \mbox{}
          - 5 \* (99 + 49 \* x) \* \H(0) \* \z2
          - 5 \* (61 + 31 \* x) \* \H(1) \* \z2
          + 5 \* (87 + 37 \* x) \* \H(3)
          + 25 \* (7 + 5 \* x) \* (\Hh(1,2)
          - \Hh(2,0)
          - \Hhh(1,1,0))
  \nonumber\\&& \mbox{}
          + 25 \* (11 + 9 \* x) \* \z3
          - (94 + 159 \* x) \* \z2
          + 219 + 125 \* \Hh(1,0)
          + 154 \* \H(0)
          - 65 \* \H(1)
          - 65 \* \H(2))
  \nonumber\\&& \mbox{}
          + {8 \over 225} \* \pgq(x) \* (30 \* (1 + x^{-1}) \* \H(1) \* \z2
          - 179 
	  - 120 \* \z2
          + 180 \* \Hh(0,0)
          + 59 \* \H(0)
          - 60 \* \H(1)
          + 60 \* \H(2)) 
  \nonumber\\&& \mbox{}
          + {4 \over 3} \* (13 - 67 \* x) \* \H(0) \* \z3
          - {8 \over 45} \* (276 - 2071 \* x) \* \H(0) \* \z2
          + {1 \over 81} \* (7853 + 29421 \* x) \* \H(0)
          + {16 \over 45}\* (61 + 9 \* x) \* \H(1) \* \z2
  \nonumber\\&& \mbox{}
          - {32 \over 27} \* (162 - 271 \* x) \* \H(1)
          + {8 \over 3} \* (11 - 37 \* x) \* \H(2) \* \z2
          - {1 \over 27} \* (1873 - 627 \* x) \* \H(2) 
          + {8 \over 45} \* (168 - 2173 \* x) \* \H(3)
  \nonumber\\&& \mbox{}
          - {16 \over 3} \* (1 - 5 \* x) \* (4 \* \Hh(-2,2)
          - 7 \* \Hh(1,0) \* \z2
          - 5 \* \Hh(1,1) \* \z2
          + 5 \* \Hh(1,3)
          + 4 \* \Hhh(1,-2,0)
          + \Hhh(1,1,2)
          - \Hhh(1,2,0)
          + 6 \* \Hhhh(1,0,0,0)
  \nonumber\\&& \mbox{}
          - \Hhhh(1,1,1,0)
          - 3 \* \H(1) \* \z3)
          + {4 \over 9} \* (1 + x) \* (69 \* \Hh(0,0) \* \z2
          - 18 \* \Hh(2,2)
          - 42 \* \Hh(3,0)
          - 42 \* \Hh(3,1)
          - 18 \* \Hhh(2,1,0)
          - 18 \* \Hhh(2,1,1)
  \nonumber\\&& \mbox{}
          - 91 \* \Hhhh(0,0,0,0)
          - 69 \* \H(4))
          - {4 \over 5} \* (21 + 13 \* x) \* \z2^2
          + {4 \over 9} \* (229 - 757 \* x) \* \z3
          + {1 \over 675} \* (9601 - 260051 \* x) \* \z2
  \nonumber\\&& \mbox{}
          + {1 \over 3240} \* (199091 + 483559 \* x)
          - {32 \over 3} \* (6 \* \Hh(-3,0)
          + 7 \* \Hhh(1,1,1) \* x
          - 4 \* \H(-2) \* \z2)
          - \delta(1 - x) \* \biggl({341 \over 36} + {592 \over 9} \* \z5 
  \nonumber\\&& \mbox{}
	  + {16472 \over 135} \* \z2^2 - {1348 \over 3} \* \z3 
	  + {5491 \over 27} \* \z2 + {352 \over 9} \* \z2 \* \z3\biggr)
          \biggr)
  \nonumber\\&& \mbox{}
       +  \colour4colour{\cf^3}  \*  \biggl(
          - {128 \over 3} \* (9 - x) \* \Hh(-4,0)
          + {32 \over 3} \* (11 + 53 \* x) \* \Hh(-3,0)
          - {32 \over 3} \* (5 + 19 \* x) \* \Hh(-3,2)
  \nonumber\\&& \mbox{}
          + {32 \over 3} \* (85 + 104 \* x) \* \Hh(-2,-1) \* \z2
          - {16 \over 3} \* (101 + 151 \* x) \* \Hh(-2,0) \* \z2
          + {4 \over 75} \* (11963 + 70513 \* x) \* \Hh(-2,0)
  \nonumber\\&& \mbox{}
          + {8 \over 15} \* (1969 + 7009 \* x) \* \Hh(-2,2)
          + {16 \over 3} \* (83 + 79 \* x) \* \Hh(-2,3)
          + {8 \over 15} \* (10747 + 12357 \* x) \* \Hh(-1,-1) \* \z2
  \nonumber\\&& \mbox{}
          - {8 \over 3} \* (1953 + 2011 \* x) \* \Hh(-1,0) \* \z2
          + {4 \over 225} \* (22441 + 73581 \* x) \* \Hh(-1,0)
          + {8 \over 75} \* (28883 + 41833 \* x) \* \Hh(-1,2)
  \nonumber\\&& \mbox{}
          + {16 \over 3} \* (796 + 783 \* x) \* \Hh(-1,3)
          - {4 \over 3} \* (361 - 127 \* x) \* \Hh(0,0) \* \z3
          + {2 \over 5} \* (1039 + 5031 \* x) \* \Hh(0,0) \* \z2
  \nonumber\\&& \mbox{}
          - {1 \over 450} \* (103377 + 404653 \* x) \* \Hh(0,0)
          - 16 \* (19 - 119 \* x) \* \Hh(1,0) \* \z3
          - {16 \over 15} \* (1627 - 1102 \* x) \* \Hh(1,0) \* \z2
  \nonumber\\&& \mbox{}
          - {1 \over 6} \* (3357 - 1355 \* x) \* \Hh(1,0)
          - 48 \* (1 - 21 \* x) \* \Hh(1,1) \* \z3
          - {8 \over 15} \* (997 - 1027 \* x) \* \Hh(1,1) \* \z2
  \nonumber\\&& \mbox{}
          - {1 \over 6} \* (4691 - 3617 \* x) \* \Hh(1,1)
          + {8 \over 3} \* (88 - 211 \* x) \* \Hh(1,2)
          + {56 \over 5} \* (69 + x) \* \Hh(1,3)
          + 384 \* (1 - 4 \* x) \* \Hh(1,4)
  \nonumber\\&& \mbox{}
          - 8 \* (51 - 61 \* x) \* \Hh(2,0) \* \z2
          - {8 \over 3} \* (121 + 265 \* x) \* \Hh(2,0)
          - 16 \* (11 - 27 \* x) \* \Hh(2,1) \* \z2
          - {2 \over 3} \* (537 + 1765 \* x) \* \Hh(2,1)
  \nonumber\\&& \mbox{}
          + {4 \over 3} \* (145 + 497 \* x) \* \Hh(2,2)
          + 104 \* (3 - x) \* \Hh(2,3)
          + {8 \over 5} \* (24 + 71 \* x) \* \Hh(3,0)
          - {8 \over 15} \* (23 - 233 \* x) \* \Hh(3,1)
  \nonumber\\&& \mbox{}
          + {16 \over 3} \* (33 + 37 \* x) \* \Hh(3,2)
          + {16 \over 3} \* (41 + 29 \* x) \* \Hh(4,0)
          + {8 \over 3} \* (85 + 61 \* x) \* \Hh(4,1)
          + {32 \over 3} \* (1 - 43 \* x) \* \Hhh(-3,-1,0)
  \nonumber\\&& \mbox{}
          - {32 \over 3} \* (26 - 19 \* x) \* \Hhh(-3,0,0)
          - 64 \* (1 + 12 \* x) \* \Hhh(-2,-2,0)
          - {8 \over 15} \* (3101 - 699 \* x) \* \Hhh(-2,-1,0)
  \nonumber\\&& \mbox{}
          - {32 \over 3} \* (79 + 53 \* x) \* \Hhh(-2,-1,2)
          + {8 \over 15} \* (3169 + 4369 \* x) \* \Hhh(-2,0,0)
          - {8 \over 15} \* (1559 + 1949 \* x) \* \Hhh(-1,-2,0)
  \nonumber\\&& \mbox{}
          - {8 \over 75} \* (30243 + 22543 \* x) \* \Hhh(-1,-1,0)
          - {32 \over 3} \* (489 + 536 \* x) \* \Hhh(-1,-1,2)
          + {8 \over 75} \* (36793 + 43368 \* x) \* \Hhh(-1,0,0)
  \nonumber\\&& \mbox{}
          + {32 \over 15} \* (371 + 311 \* x) \* \Hhh(-1,2,0)
          + {32 \over 5} \* (157 + 137 \* x) \* \Hhh(-1,2,1)
          - 4 \* (57 + 5 \* x) \* \Hhh(0,0,0) \* \z2
  \nonumber\\&& \mbox{}
          - {1 \over 75} \* (63943 + 301457 \* x) \* \Hhh(0,0,0)
          - {8 \over 3} \* (439 - 497 \* x) \* \Hhh(1,-2,0)
          - 96 \* (5 - 21 \* x) \* \Hhh(1,0,0) \* \z2
  \nonumber\\&& \mbox{}
          - {1 \over 15} \* (10631 - 21901 \* x) \* \Hhh(1,0,0)
          - 64 \* (7 - 23 \* x) \* \Hhh(1,1,0) \* \z2
          + {2 \over 3} \* (305 - 823 \* x) \* \Hhh(1,1,0)
  \nonumber\\&& \mbox{}
          + 6 \* (39 - 89 \* x) \* \Hhh(1,1,1)
          + {16 \over 3} \* (3 + 61 \* x) \* \Hhh(1,1,2)
          + 96 \* (3 - 7 \* x) \* \Hhh(1,1,3)
          + {32 \over 3} \* (1 + 31 \* x) \* \Hhh(1,2,0)
  \nonumber\\&& \mbox{}
          + {4 \over 3} \* (53 + 233 \* x) \* \Hhh(1,2,1)
          + {32 \over 15} \* (22 + 83 \* x) \* \Hhh(2,0,0)
          + {4 \over 3} \* (175 + 487 \* x) \* \Hhh(2,1,0)
          + 36 \* (5 + 17 \* x) \* \Hhh(2,1,1)
  \nonumber\\&& \mbox{}
          + {16 \over 3} \* (33 + 19 \* x) \* \Hhh(3,0,0)
          + {64 \over 3} \* (11 + 10 \* x) \* \Hhh(3,1,0)
          + 64 \* (2 + 17 \* x) \* \Hhhh(-2,-1,-1,0)
  \nonumber\\&& \mbox{}
          - {16 \over 3} \* (91 + 251 \* x) \* \Hhhh(-2,-1,0,0)
          + 16 \* (1 + 51 \* x) \* \Hhhh(-2,0,0,0)
          + {16 \over 15} \* (967 + 1637 \* x) \* \Hhhh(-1,-1,-1,0)
  \nonumber\\&& \mbox{}
          - {8 \over 3} \* (1617 + 1685 \* x) \* \Hhhh(-1,-1,0,0)
          + {4 \over 3} \* (1983 + 1889 \* x) \* \Hhhh(-1,0,0,0)
          - {2 \over 5} \* (373 + 2977 \* x) \* \Hhhh(0,0,0,0)
  \nonumber\\&& \mbox{}
          + {8 \over 15} \* (1863 - 1823 \* x) \* \Hhhh(1,0,0,0)
          - {4 \over 15} \* (1163 + 647 \* x) \* \Hhhh(1,1,0,0)
          + {136 \over 3} \* (1 + 9 \* x) \* \Hhhh(1,1,1,0)
  \nonumber\\&& \mbox{}
          + 8 \* (27 - 53 \* x) \* \Hhhh(2,0,0,0)
          + 4 \* (45 - 23 \* x) \* \Hhhhh(0,0,0,0,0)
          - 32 \* (7 - 11 \* x) \* \Hhhhh(1,1,1,0,0)
  \nonumber\\&& \mbox{}
          + {2 \over 15} \* \pqq( - x) \* (4960 \* \z5
          + 150 \* \z3
          - 140 \* \z2
          - 7320 \* \z2 \* \z3
          + 816 \* \z2^2
          + 3840 \* \Hh(-4,0)
          - 280 \* \Hh(-3,0)
  \nonumber\\&& \mbox{}
          + 11040 \* \Hh(-3,2)
          + 22680 \* \Hh(-2,-1) \* \z2
          - 1700 \* \Hh(-2,0)
          - 23680 \* \Hh(-2,0) \* \z2
          + 1800 \* \Hh(-2,2)
          + 20120 \* \Hh(-2,3)
  \nonumber\\&& \mbox{}
          + 22560 \* \Hh(-1,-2) \* \z2
          + 30600 \* \Hh(-1,-1) \* \z3
          + 3960 \* \Hh(-1,-1) \* \z2
          + 240 \* \Hh(-1,0)
          - 18600 \* \Hh(-1,0) \* \z3
  \nonumber\\&& \mbox{}
          - 3180 \* \Hh(-1,0) \* \z2
          + 1240 \* \Hh(-1,2)
          + 1640 \* \Hh(-1,2) \* \z2
          + 3000 \* \Hh(-1,3)
          + 15720 \* \Hh(-1,4)
          + 550 \* \Hh(0,0)
  \nonumber\\&& \mbox{}
          + 5720 \* \Hh(0,0) \* \z3
          - 1070 \* \Hh(0,0) \* \z2
          - 240 \* \Hh(2,0)
          - 2120 \* \Hh(2,0) \* \z2
          - 240 \* \Hh(2,1)
          - 1720 \* \Hh(2,1) \* \z2
          - 60 \* \Hh(2,2)
  \nonumber\\&& \mbox{}
          + 720 \* \Hh(2,3)
          - 630 \* \Hh(3,0)
          - 960 \* \Hh(3,1)
          - 1040 \* \Hh(3,2)
          - 2440 \* \Hh(4,0)
          - 3240 \* \Hh(4,1)
          - 5920 \* \Hhh(-3,-1,0)
  \nonumber\\&& \mbox{}
          + 11200 \* \Hhh(-3,0,0)
          - 4800 \* \Hhh(-2,-2,0)
          - 840 \* \Hhh(-2,-1,0)
          - 19680 \* \Hhh(-2,-1,2)
          + 120 \* \Hhh(-2,0,0)
          + 5760 \* \Hhh(-2,2,0)
  \nonumber\\&& \mbox{}
          + 7280 \* \Hhh(-2,2,1)
          - 6560 \* \Hhh(-1,-3,0)
          - 900 \* \Hhh(-1,-2,0)
          - 19680 \* \Hhh(-1,-2,2)
          - 30240 \* \Hhh(-1,-1,-1) \* \z2
  \nonumber\\&& \mbox{}
          + 1920 \* \Hhh(-1,-1,0)
          + 35760 \* \Hhh(-1,-1,0) \* \z2
          - 2880 \* \Hhh(-1,-1,2)
          - 31120 \* \Hhh(-1,-1,3)
          - 1580 \* \Hhh(-1,0,0)
  \nonumber\\&& \mbox{}
          - 17880 \* \Hhh(-1,0,0) \* \z2
          + 1440 \* \Hhh(-1,2,0)
          + 1920 \* \Hhh(-1,2,1)
          + 2160 \* \Hhh(-1,2,2)
          + 7000 \* \Hhh(-1,3,0)
          + 9040 \* \Hhh(-1,3,1)
  \nonumber\\&& \mbox{}
          + 1195 \* \Hhh(0,0,0)
          + 4520 \* \Hhh(0,0,0) \* \z2
          - 160 \* \Hhh(2,-2,0)
          + 60 \* \Hhh(2,1,0)
          + 80 \* \Hhh(2,1,2)
          - 160 \* \Hhh(2,2,0)
          - 1280 \* \Hhh(3,0,0)
  \nonumber\\&& \mbox{}
          - 720 \* \Hhh(3,1,0)
          - 720 \* \Hhh(3,1,1)
          + 6000 \* \Hhhh(-2,-1,-1,0)
          - 18840 \* \Hhhh(-2,-1,0,0)
          + 13920 \* \Hhhh(-2,0,0,0)
  \nonumber\\&& \mbox{}
          + 5760 \* \Hhhh(-1,-2,-1,0)
          - 18800 \* \Hhhh(-1,-2,0,0)
          + 5520 \* \Hhhh(-1,-1,-2,0)
          + 2160 \* \Hhhh(-1,-1,-1,0)
  \nonumber\\&& \mbox{}
          + 27360 \* \Hhhh(-1,-1,-1,2)
          - 1800 \* \Hhhh(-1,-1,0,0)
          - 9920 \* \Hhhh(-1,-1,2,0)
          - 12480 \* \Hhhh(-1,-1,2,1)
          - 270 \* \Hhhh(-1,0,0,0)
  \nonumber\\&& \mbox{}
          + 3040 \* \Hhhh(-1,2,0,0)
          + 1360 \* \Hhhh(-1,2,1,0)
          + 1440 \* \Hhhh(-1,2,1,1)
          + 1490 \* \Hhhh(0,0,0,0)
          + 480 \* \Hhhh(2,0,0,0)
          + 480 \* \Hhhh(2,1,0,0)
  \nonumber\\&& \mbox{}
          - 80 \* \Hhhh(2,1,1,0)
          - 5760 \* \Hhhhh(-1,-1,-1,-1,0)
          + 25680 \* \Hhhhh(-1,-1,-1,0,0)
          - 21800 \* \Hhhhh(-1,-1,0,0,0)
  \nonumber\\&& \mbox{}
          + 9480 \* \Hhhhh(-1,0,0,0,0)
          - 2040 \* \Hhhhh(0,0,0,0,0)
          - 14000 \* \H(-3) \* \z2
          - 22000 \* \H(-2) \* \z3
          - 2220 \* \H(-2) \* \z2
  \nonumber\\&& \mbox{}
          - 4110 \* \H(-1) \* \z3
          - 280 \* \H(-1) \* \z2
          + 952 \* \H(-1) \* \z2^2
          + 1205 \* \H(0)
          - 10 \* \H(0) \* \z3
          + 555 \* \H(0) \* \z2
          + 772 \* \H(0) \* \z2^2
  \nonumber\\&& \mbox{}
          + 260 \* \H(2)
          - 3080 \* \H(2) \* \z3
          - 1830 \* \H(2) \* \z2
          - 1220 \* \H(3)
          - 3280 \* \H(3) \* \z2
          + 390 \* \H(4)
          - 4200 \* \H(5))
  \nonumber\\&& \mbox{}
          - {1 \over 240} \* \pqq(x) \* (15015
          + 214400 \* \z5
          - 34320 \* \z3
          + 35480 \* \z2
          + 74240 \* \z2 \* \z3
          - 3136 \* \z2^2
  \nonumber\\&& \mbox{}
          - 102400 \* \Hh(-4,0)
          - 34560 \* \Hh(-3,0)
          - 81920 \* \Hh(-3,2)
          - 311040 \* \Hh(-2,-1) \* \z2
          + 11840 \* \Hh(-2,0)
  \nonumber\\&& \mbox{}
          + 231680 \* \Hh(-2,0) \* \z2
          - 69760 \* \Hh(-2,2)
          - 186880 \* \Hh(-2,3)
          - 13480 \* \Hh(0,0)
          - 35840 \* \Hh(0,0) \* \z3
  \nonumber\\&& \mbox{}
          - 113280 \* \Hh(0,0) \* \z2
          - 55040 \* \Hh(1,-2) \* \z2
          - 47640 \* \Hh(1,0)
          + 175360 \* \Hh(1,0) \* \z3
          - 84480 \* \Hh(1,0) \* \z2
          - 33480 \* \Hh(1,1)
  \nonumber\\&& \mbox{}
          + 253440 \* \Hh(1,1) \* \z3
          - 51840 \* \Hh(1,1) \* \z2
          - 45920 \* \Hh(1,2)
          - 46080 \* \Hh(1,2) \* \z2
          + 90240 \* \Hh(1,3)
          + 11520 \* \Hh(1,4)
  \nonumber\\&& \mbox{}
          - 88800 \* \Hh(2,0)
          - 14080 \* \Hh(2,0) \* \z2
          - 60160 \* \Hh(2,1)
          - 51200 \* \Hh(2,1) \* \z2
          + 58560 \* \Hh(2,2)
          + 128000 \* \Hh(2,3)
  \nonumber\\&& \mbox{}
          + 28800 \* \Hh(3,0)
          + 61440 \* \Hh(3,1)
          + 124160 \* \Hh(3,2)
          + 84480 \* \Hh(4,0)
          + 116480 \* \Hh(4,1)
          + 276480 \* \Hhh(-3,-1,0)
  \nonumber\\&& \mbox{}
          - 230400 \* \Hhh(-3,0,0)
          + 84480 \* \Hhh(-2,-2,0)
          + 120960 \* \Hhh(-2,-1,0)
          + 258560 \* \Hhh(-2,-1,2)
          - 64640 \* \Hhh(-2,0,0)
  \nonumber\\&& \mbox{}
          - 23040 \* \Hhh(-2,2,0)
          - 23040 \* \Hhh(-2,2,1)
          - 16240 \* \Hhh(0,0,0)
          - 35840 \* \Hhh(0,0,0) \* \z2
          - 71680 \* \Hhh(1,-3,0)
  \nonumber\\&& \mbox{}
          - 17280 \* \Hhh(1,-2,0)
          + 176640 \* \Hhh(1,-2,2)
          - 44400 \* \Hhh(1,0,0)
          + 57600 \* \Hhh(1,0,0) \* \z2
          - 59520 \* \Hhh(1,1,0)
  \nonumber\\&& \mbox{}
          + 2560 \* \Hhh(1,1,0) \* \z2
          - 25920 \* \Hhh(1,1,1)
          - 46080 \* \Hhh(1,1,1) \* \z2
          + 86400 \* \Hhh(1,1,2)
          + 145920 \* \Hhh(1,1,3)
  \nonumber\\&& \mbox{}
          + 76800 \* \Hhh(1,2,0)
          + 77760 \* \Hhh(1,2,1)
          + 138240 \* \Hhh(1,2,2)
          + 131840 \* \Hhh(1,3,0)
          + 135680 \* \Hhh(1,3,1)
          + 74240 \* \Hhh(2,-2,0)
  \nonumber\\&& \mbox{}
          - 6720 \* \Hhh(2,0,0)
          + 72000 \* \Hhh(2,1,0)
          + 77760 \* \Hhh(2,1,1)
          + 147200 \* \Hhh(2,1,2)
          + 139520 \* \Hhh(2,2,0)
          + 140800 \* \Hhh(2,2,1)
  \nonumber\\&& \mbox{}
          + 44800 \* \Hhh(3,0,0)
          + 142080 \* \Hhh(3,1,0)
          + 142080 \* \Hhh(3,1,1)
          - 104960 \* \Hhhh(-2,-1,-1,0)
          + 268800 \* \Hhhh(-2,-1,0,0)
  \nonumber\\&& \mbox{}
          - 130560 \* \Hhhh(-2,0,0,0)
          + 30400 \* \Hhhh(0,0,0,0)
          + 243200 \* \Hhhh(1,-2,-1,0)
          - 48640 \* \Hhhh(1,-2,0,0)
          + 17280 \* \Hhhh(1,0,0,0)
  \nonumber\\&& \mbox{}
          + 120320 \* \Hhhh(1,1,-2,0)
          + 2880 \* \Hhhh(1,1,0,0)
          + 103680 \* \Hhhh(1,1,1,0)
          + 86400 \* \Hhhh(1,1,1,1)
          + 138240 \* \Hhhh(1,1,1,2)
  \nonumber\\&& \mbox{}
          + 161280 \* \Hhhh(1,1,2,0)
          + 126720 \* \Hhhh(1,1,2,1)
          + 67840 \* \Hhhh(1,2,0,0)
          + 158720 \* \Hhhh(1,2,1,0)
          + 126720 \* \Hhhh(1,2,1,1)
  \nonumber\\&& \mbox{}
          - 53760 \* \Hhhh(2,0,0,0)
          + 75520 \* \Hhhh(2,1,0,0)
          + 172800 \* \Hhhh(2,1,1,0)
          + 138240 \* \Hhhh(2,1,1,1)
          - 7680 \* \Hhhhh(0,0,0,0,0)
  \nonumber\\&& \mbox{}
          - 65280 \* \Hhhhh(1,0,0,0,0)
          - 94720 \* \Hhhhh(1,1,0,0,0)
          + 88320 \* \Hhhhh(1,1,1,0,0)
          + 161280 \* \Hhhhh(1,1,1,1,0)
          + 115200 \* \Hhhhh(1,1,1,1,1)
  \nonumber\\&& \mbox{}
          + 220160 \* \H(-3) \* \z2
          + 263680 \* \H(-2) \* \z3
          + 130240 \* \H(-2) \* \z2
          + 48140 \* \H(0)
          - 167040 \* \H(0) \* \z3
          + 64880 \* \H(0) \* \z2
  \nonumber\\&& \mbox{}
          + 59520 \* \H(0) \* \z2^2
          + 11220 \* \H(1)
          - 17280 \* \H(1) \* \z3
          + 15200 \* \H(1) \* \z2
          + 54528 \* \H(1) \* \z2^2
          - 31640 \* \H(2)
  \nonumber\\&& \mbox{}
          + 148480 \* \H(2) \* \z3
          - 45120 \* \H(2) \* \z2
          - 86160 \* \H(3)
          - 29440 \* \H(3) \* \z2
          + 91520 \* \H(4)
          + 46080 \* \H(5))
  \nonumber\\&& \mbox{}
          - {12 \over 25} \* \pqg( - x) \* (4 \* (188 + 197 \* x) \* \Hh(-2,0)
          + (371 + 337 \* x) \* \Hh(-1,0)
          - 4 \* (222 - 197 \* x) \* \Hh(-1,2)
  \nonumber\\&& \mbox{}
          - 4 \* (208 + 197 \* x) \* \Hhh(-1,-1,0)
          + 4 \* (13 + 237 \* x) \* \Hhh(-1,0,0)
          + 2 \* (236 - 591 \* x) \* \H(-1) \* \z2
  \nonumber\\&& \mbox{}
          - 20 \* (1 - x) \* (10 \* \Hh(-3,0)
          + 70 \* \Hh(-2,2)
          + 73 \* \Hh(-1,-1) \* \z2
          - 55 \* \Hh(-1,0) \* \z2
          + 50 \* \Hh(-1,3)
          - 6 \* \Hhh(-2,-1,0)
  \nonumber\\&& \mbox{}
          + 50 \* \Hhh(-2,0,0)
          - 6 \* \Hhh(-1,-2,0)
          - 70 \* \Hhh(-1,-1,2)
          + 6 \* \Hhh(-1,2,0)
          + 6 \* \Hhh(-1,2,1)
          + 4 \* \Hhh(1,-2,0)
          + 6 \* \Hhhh(-1,-1,-1,0)
  \nonumber\\&& \mbox{}
          - 50 \* \Hhhh(-1,-1,0,0)
          + 20 \* \Hhhh(-1,0,0,0)
          - 73 \* \H(-2) \* \z2
          - 60 \* \H(-1) \* \z3))
          - {4 \over 25} \* \pqg(x) \* (3 \* (125 - 337 \* x) \* \Hh(0,0)
  \nonumber\\&& \mbox{}
          + 10 \* (341 + 591 \* x) \* \Hh(0,0) \* \z2
          + 10 \* (11 + 261 \* x) \* \Hh(1,0) \* \z2
          - 36 \* (109 + 79 \* x) \* \Hhh(0,0,0)
          + 6 \* (153 + 788 \* x) \* \H(0) \* \z2
  \nonumber\\&& \mbox{}
          + 10 \* (373 + 123 \* x) \* \H(0) \* \z3
          + 10 \* (13 - 237 \* x) \* \H(1) \* \z3
          - 6 \* (208 - 197 \* x) \* \H(1) \* \z2
          + 6 \* (241 - 394 \* x) \* \H(3)
  \nonumber\\&& \mbox{}
          - 10 \* (281 + 531 \* x) \* \H(4)
          + 60 \* (1 + x) \* (3 \* \Hh(1,1) \* \z2
          - 6 \* \Hh(3,0)
          - 6 \* \Hh(3,1)
          - 22 \* \Hhhh(0,0,0,0)
          - 2 \* \Hhhh(1,0,0,0)
          + 3 \* \H(2) \* \z2)
  \nonumber\\&& \mbox{}
          + 10 \* (19 - 231 \* x) \* (\Hh(1,3)
          - \Hhh(2,0,0)
          - \Hhhh(1,1,0,0))
          - 3 \* (45 - 337 \* x) \* \z2
          + 30 \* (321 + 197 \* x) \* \z3
  \nonumber\\&& \mbox{}
          - 2 \* (359 + 1359 \* x) \* \z2^2
          + 3 \* (451 - 120 \* \Hh(1,0)
          - 120 \* \Hh(1,1)
          - 120 \* \Hh(2,0)
          - 120 \* \Hh(2,1)
          - 770 \* \Hhh(1,0,0)
  \nonumber\\&& \mbox{}
          + 953 \* \H(0)
          + 502 \* \H(1)
          + 382 \* \H(2)))
          - {4 \over 225} \* \pgq( - x) \* (120 \* (1 + 4 \* x^{-1}) \* \Hh(-2,0)
          + (743 + 1501 \* x^{-1}) \* \Hh(-1,0)
  \nonumber\\&& \mbox{}
          + 12 \* (173 + 177 \* x^{-1}) \* \Hh(-1,2)
          + 36 \* (49 - 59 \* x^{-1}) \* \Hhh(-1,-1,0)
          + 12 \* (33 + 217 \* x^{-1}) \* \Hhh(-1,0,0)
  \nonumber\\&& \mbox{}
          - 6 \* (199 + 531 \* x^{-1}) \* \H(-1) \* \z2
          - 60 \* (1 - x^{-1}) \* (12 \* \Hh(-2,2)
          + 73 \* \Hh(-1,-1) \* \z2
          - 55 \* \Hh(-1,0) \* \z2
          + 50 \* \Hh(-1,3)
  \nonumber\\&& \mbox{}
          - 12 \* \Hhh(-2,-1,0)
          + 12 \* \Hhh(-2,0,0)
          - 6 \* \Hhh(-1,-2,0)
          - 70 \* \Hhh(-1,-1,2)
          + 6 \* \Hhh(-1,2,0)
          + 6 \* \Hhh(-1,2,1)
          - 4 \* \Hhh(1,-2,0)
  \nonumber\\&& \mbox{}
          + 6 \* \Hhhh(-1,-1,-1,0)
          - 50 \* \Hhhh(-1,-1,0,0)
          + 20 \* \Hhhh(-1,0,0,0)
          - 18 \* \H(-2) \* \z2 - 60 \* \H(-1) \* \z3))
  \nonumber\\&& \mbox{}
          - {4 \over 225} \* \pgq(x) \* (18 \* (49 + 59 \* x^{-1}) \* \H(1) \* \z2
          + 30 \* (1 + x^{-1}) \* (87 \* \Hh(1,0) \* \z2
          + 6 \* \Hh(1,1) \* \z2
          - 77 \* \Hh(1,3)
          - 4 \* \Hhhh(1,0,0,0)
  \nonumber\\&& \mbox{}
          + 77 \* \Hhhh(1,1,0,0)
          - 79 \* \H(1) \* \z3
          + 12 \* \H(2) \* \z2)
          + 983 
	  - 1230 \* \z3
          + 222 \* \z2
          + 2004 \* \Hh(0,0)
          - 360 \* \Hh(1,0)
          - 360 \* \Hh(1,1)
  \nonumber\\&& \mbox{}
          + 360 \* \Hh(2,0)
          + 360 \* \Hh(2,1)
          + 1320 \* \Hhh(0,0,0)
          - 2310 \* \Hhh(1,0,0)
          - 743 \* \H(0)
          - 5910 \* \H(0) \* \z2
          + 1746 \* \H(1)
          - 2466 \* \H(2)
  \nonumber\\&& \mbox{}
          + 5310 \* \H(3))
          + {16 \over 3} \* (11 - 5 \* x) \* \H(-3) \* \z2
          - 8 \* (87 + 127 \* x) \* \H(-2) \* \z3
          - {4 \over 15} \* (7039 + 13319 \* x) \* \H(-2) \* \z2
  \nonumber\\&& \mbox{}
          - {8 \over 3} \* (1901 + 2140 \* x) \* \H(-1) \* \z3
          - {4 \over 75} \* (88009 + 106209 \* x) \* \H(-1) \* \z2
          - {4 \over 15} \* (151 + 15 \* x) \* \H(0) \* \z2^2
  \nonumber\\&& \mbox{}
          + {2 \over 15} \* (421 + 20889 \* x) \* \H(0) \* \z3
          + {1 \over 75} \* (59681 + 674569 \* x) \* \H(0) \* \z2
          + {1 \over 450} \* (296731 + 143711 \* x) \* \H(0)
  \nonumber\\&& \mbox{}
          + {24 \over 5} \* (25 - 61 \* x) \* \H(1) \* \z2^2
          - {4 \over 15} \* (689 - 2424 \* x) \* \H(1)
          - {4 \over 15} \* (6729 - 9769 \* x) \* \H(1) \* \z3
  \nonumber\\&& \mbox{}
          - {4 \over 75} \* (34643 - 33093 \* x) \* \H(1) \* \z2
          - {8 \over 15} \* (1913 + 1592 \* x) \* \H(2) \* \z2
          - {1 \over 30} \* (3293 - 8527 \* x) \* \H(2)
  \nonumber\\&& \mbox{}
          - {32 \over 3} \* (16 - 3 \* x) \* \H(3) \* \z2
          - {1 \over 75} \* (29713 + 504037 \* x) \* \H(3)
          - {2 \over 15} \* (2397 + 10853 \* x) \* \H(4)
          + {4 \over 3} \* (171 + 47 \* x) \* \H(5)
  \nonumber\\&& \mbox{}
          + 32 \* (1 - 5 \* x) \* (7 \* \Hh(1,-2) \* \z2
          - 3 \* \Hh(1,2) \* \z2
          - \Hhh(1,-3,0)
          - 8 \* \Hhh(1,-2,2)
          - 3 \* \Hhh(1,1,1) \* \z2
          + \Hhh(1,3,0)
          + \Hhh(1,3,1)
  \nonumber\\&& \mbox{}
          - 2 \* \Hhhh(1,-2,-1,0)
          - 3 \* \Hhhh(1,-2,0,0)
          - 4 \* \Hhhh(1,1,-2,0)
          + 5 \* \Hhhhh(1,0,0,0,0)
          + 6 \* \Hhhhh(1,1,0,0,0))
          - 48 \* (1 - x) \* (\z4
          + 2 \* \Hhhh(1,2,0,0))
  \nonumber\\&& \mbox{}
          + {16 \over 3} \* (1 + x) \* (21 \* \Hhh(2,1,2)
          + 25 \* \Hhh(2,2,0)
          + 18 \* \Hhh(2,2,1)
          + 36 \* \Hhh(3,1,1)
          + 24 \* \Hhhh(2,1,1,0)
          + 18 \* \Hhhh(2,1,1,1))
  \nonumber\\&& \mbox{}
          + {8 \over 15} \* (1 + 5 \* x) \* (1140 \* \Hh(-1,-2) \* \z2
          + 1530 \* \Hh(-1,-1) \* \z3
          - 630 \* \Hh(-1,0) \* \z3
          + 60 \* \Hh(-1,2) \* \z2
          + 660 \* \Hh(-1,4)
  \nonumber\\&& \mbox{}
          + 20 \* \Hhh(-2,2,0)
          - 300 \* \Hhh(-1,-3,0)
          - 960 \* \Hhh(-1,-2,2)
          - 1860 \* \Hhh(-1,-1,-1) \* \z2
          + 1440 \* \Hhh(-1,-1,0) \* \z2
  \nonumber\\&& \mbox{}
          - 1140 \* \Hhh(-1,-1,3)
          - 840 \* \Hhh(-1,0,0) \* \z2
          + 60 \* \Hhh(-1,3,0)
          + 60 \* \Hhh(-1,3,1)
          + 360 \* \Hhhh(-1,-2,-1,0)
  \nonumber\\&& \mbox{}
          - 900 \* \Hhhh(-1,-2,0,0)
          + 360 \* \Hhhh(-1,-1,-2,0)
          + 1680 \* \Hhhh(-1,-1,-1,2)
          - 120 \* \Hhhh(-1,-1,2,0)
          - 120 \* \Hhhh(-1,-1,2,1)
  \nonumber\\&& \mbox{}
          - 120 \* \Hhhh(-1,2,0,0)
          - 360 \* \Hhhhh(-1,-1,-1,-1,0)
          + 1500 \* \Hhhhh(-1,-1,-1,0,0)
          - 720 \* \Hhhhh(-1,-1,0,0,0)
          + 300 \* \Hhhhh(-1,0,0,0,0)
  \nonumber\\&& \mbox{}
          + 177 \* \H(-1) \* \z2^2)
          + {8 \over 3} \* (1 + 13 \* x) \* (9 \* \Hhhh(1,1,1,1)
          - 5 \* \Hhhh(2,1,0,0))
          - {128 \over 3} \* (15 + 59 \* x) \* \z5
          + {8 \over 3} \* (143 - 437 \* x) \* \z2 \* \z3
  \nonumber\\&& \mbox{}
          - {4 \over 75} \* (4164 + 1231 \* x) \* \z2^2
          + {1 \over 15} \* (28381 + 106709 \* x) \* \z3
          + {1 \over 450} \* (158091 + 142879 \* x) \* \z2
  \nonumber\\&& \mbox{}
          + {1 \over 3600} \* (1816203 + 638687 \* x)
          + 16 \* (4 \* \Hhh(-2,2,1) \* x - 8 \* \Hhh(2,-2,0)
          + 79 \* \H(2) \* \z3 \* x)
          - \delta(1 - x) \* \biggl({7255 \over 24} 
  \nonumber\\&& \mbox{}
	        - 1240 \* \z5 + {2148 \over 5} \* \z2^2
             + {950 \over 3} \* \z3 + {304 \over 3} \* \z3^2 
             + {3371 \over 6} \* \z2
             - {4184 \over 315} \* \z2^3 - 808 \* \z2 \* \z3\biggr)
          \biggr)
  \nonumber\\&& \mbox{}
       +  \colour4colour{\ca \* \cf \* \nf}  \*  \biggl(
          - {2 \over 675} \* (13561 + 138889 \* x) \* \Hh(0,0)
          - {40 \over 3} \* (1 - 7 \* x) \* \Hh(1,0) \* \z2
          + {4 \over 9} \* (47 - 213 \* x) \* \Hh(1,0)
  \nonumber\\&& \mbox{}
          + {4 \over 9} \* (116 - 333 \* x) \* \Hh(1,1)
          + {8 \over 3} \* (3 - 25 \* x) \* \Hh(1,3)
          + {2 \over 9} \* (105 - 73 \* x) \* \Hh(2,0)
          + {40 \over 3} \* (1 + x) \* \Hh(2,1)
  \nonumber\\&& \mbox{}
          - {8 \over 45} \* (162 + 253 \* x) \* \Hhh(0,0,0)
          + {16 \over 9} \* (7 - 30 \* x) \* \Hhh(1,0,0)
          + {8 \over 3} \* (5 - 11 \* x) \* \Hhh(2,0,0)
          + {1 \over 3645} \* \pqq(x) \* (402265
  \nonumber\\&& \mbox{}
          + 66690 \* \z3
          - 521010 \* \z2
          + 27216 \* \z2^2
          + 155520 \* \Hh(-3,0)
          + 126360 \* \Hh(-2,0)
          + 77760 \* \Hh(-2,2)
  \nonumber\\&& \mbox{}
          + 731250 \* \Hh(0,0)
          - 74520 \* \Hh(0,0) \* \z2
          + 193320 \* \Hh(1,0)
          - 139320 \* \Hh(1,0) \* \z2
          + 209520 \* \Hh(1,1)
          - 81000 \* \Hh(1,1) \* \z2
  \nonumber\\&& \mbox{}
          + 103680 \* \Hh(1,2)
          + 106920 \* \Hh(1,3)
          + 174150 \* \Hh(2,0)
          + 142560 \* \Hh(2,1)
          + 32400 \* \Hh(2,2)
          + 29160 \* \Hh(3,0)
  \nonumber\\&& \mbox{}
          + 35640 \* \Hh(3,1)
          + 155520 \* \Hhh(-2,0,0)
          + 318060 \* \Hhh(0,0,0)
          + 77760 \* \Hhh(1,-2,0)
          + 119880 \* \Hhh(1,0,0)
          + 38880 \* \Hhh(1,1,0)
  \nonumber\\&& \mbox{}
          + 71280 \* \Hhh(1,1,1)
          + 16200 \* \Hhh(1,1,2)
          - 29160 \* \Hhh(1,2,0)
          + 29160 \* \Hhh(1,2,1)
          + 22680 \* \Hhh(2,0,0)
          - 32400 \* \Hhh(2,1,0)
  \nonumber\\&& \mbox{}
          + 74520 \* \Hhhh(0,0,0,0)
          + 51840 \* \Hhhh(1,0,0,0)
          - 42120 \* \Hhhh(1,1,0,0)
          - 45360 \* \Hhhh(1,1,1,0)
          - 77760 \* \H(-2) \* \z2
          + 895905 \* \H(0)
  \nonumber\\&& \mbox{}
          + 119880 \* \H(0) \* \z3
          - 210870 \* \H(0) \* \z2
          + 338895 \* \H(1)
          + 12960 \* \H(1) \* \z3
          - 168480 \* \H(1) \* \z2
          + 491130 \* \H(2)
  \nonumber\\&& \mbox{}
          - 103680 \* \H(2) \* \z2
          + 220050 \* \H(3)
          + 71280 \* \H(4))
          + {4 \over 25} \* \pqg(x) \* (3 \* (63 + 53 \* x) \* \Hh(0,0)
          - 10 \* ( 32 + 27 \* x) \* \z2 \* \H(0)
  \nonumber\\&& \mbox{}
          - 20 \* (14 + 9 \* x) \* \z2 \* \H(1)
          + 10 \* (26 + 21 \* x) \* \H(3)
          + 30 \* (1 + x) \* (5 \* \Hh(1,2)
          - 5 \* \Hh(2,0)
          + 6 \* \Hhh(0,0,0)
          - 5 \* \Hhh(1,1,0))
  \nonumber\\&& \mbox{}
          + 10 \* (1 + 3 \* x) \* (4 \* \z2^2
          - 5 \* \Hh(0,0) \* \z2
          - 5 \* \Hh(1,0) \* \z2
          + 5 \* \Hh(1,3)
          - 5 \* \Hhh(2,0,0)
          - 5 \* \Hhhh(1,1,0,0)
          + 5 \* \H(0) \* \z3
          + 5 \* \H(1) \* \z3
  \nonumber\\&& \mbox{}
          + 5 \* \H(4))
          + 50 \* (1 + 6 \* x) \* \z3
          + 3 \* (27 - 53 \* x) \* \z2
          + 3 \* (73 + 50 \* \Hh(1,0)
          + 50 \* \Hhh(1,0,0)
          - 7 \* \H(0)
          - 80 \* \H(1)
          - 80 \* \H(2)))
  \nonumber\\&& \mbox{}
          - {4 \over 225} \* \pgq(x) \* (30 \* (1 + x^{-1}) \* 
            (5 \* \Hh(1,0) \* \z2
          - 5 \* \Hh(1,3)
          + 5 \* \Hhhh(1,1,0,0)
          - 5 \* \H(1) \* \z3 + \H(1) \* \z2)
          - 179
          + 150 \* \z3
  \nonumber\\&& \mbox{}
          + 30 \* \z2
          + 180 \* \Hh(0,0)
          - 150 \* \Hhh(1,0,0)
          + 59 \* \H(0)
          - 150 \* \H(0) \* \z2
          + 90 \* \H(1)
          - 90 \* \H(2)
          + 150 \* \H(3))
          - 8 \* (1 - 8 \* x) \* \H(0) \* \z3
  \nonumber\\&& \mbox{}
          + {2 \over 45} \* (837 - 1117 \* x) \* \H(0) \* \z2
          + {1 \over 81} \* (2615 - 56333 \* x) \* \H(0)
          - {40 \over 3} \* (1 - 3 \* x) \* \H(1) \* \z3
          - {4 \over 45} \* (177 - 602 \* x) \* \H(1) \* \z2
  \nonumber\\&& \mbox{}
          + {1 \over 27} \* (5243 - 13441 \* x) \* \H(1)
          - {8 \over 3} \* (4 - 19 \* x) \* \H(2) \* \z2
          + {10 \over 9} \* (69 - 229 \* x) \* \H(2)
          - {2 \over 45} \* (621 - 1321 \* x) \* \H(3)
  \nonumber\\&& \mbox{}
          + {8 \over 3} \* (1 - 5 \* x) \* (4 \* \Hh(-2,2)
          - 5 \* \Hh(1,1) \* \z2
          + 4 \* \Hhh(1,-2,0)
          + \Hhh(1,1,2)
          - \Hhh(1,2,0)
          + 6 \* \Hhhh(1,0,0,0)
          - \Hhhh(1,1,1,0))
  \nonumber\\&& \mbox{}
          - {8 \over 15} \* (1 - 2 \* x) \* \z2^2
          + {8 \over 3} \* (3 - 2 \* x) \* (\Hh(0,0) \* \z2
          - \H(4))
          - {2 \over 3} \* (13 - 237 \* x) \* \z3
          - {4 \over 9} \* (31 - 38 \* x) \* (\Hh(1,2)
          - \Hhh(1,1,0))
  \nonumber\\&& \mbox{}
          - {2 \over 675} \* (16569 - 146969 \* x) \* \z2
          + {2 \over 405} \* (34157 - 136987 \* x)
          + {16 \over 3} \* (6 \* \Hh(-3,0)
          + \Hhhh(1,1,0,0)
          - 4 \* \H(-2) \* \z2)
  \nonumber\\&& \mbox{}
          + {8 \over 3} \* x \* (\Hh(2,2) 
	  - \Hh(3,0)
          - \Hhh(2,1,0))
          + \delta(1 - x) \* \biggl({142883 \over 486} + {8 \over 3} \* \z5 
               - {2488 \over 135} \* \z2^2
               - {18314 \over 81} \* \z3 + {40862 \over 81} \* \z2 
  \nonumber\\&& \mbox{}
	       - {56 \over 3} \* \z2 \* \z3\biggr)
          \biggr)
  \nonumber\\&& \mbox{}
       +  \colour4colour{\ca \* \cf^2}  \*  \biggl(
            {16 \over 3} \* (41 - 5 \* x) \* \Hh(-4,0)
          + {8 \over 15} \* (427 - 1033 \* x) \* \Hh(-3,0)
          + {32 \over 3} \* (8 + 13 \* x) \* \Hh(-3,2)
  \nonumber\\&& \mbox{}
          - 304 \* (3 + 2 \* x) \* \Hh(-2,-1) \* \z2
          + {8 \over 3} \* (209 + 163 \* x) \* \Hh(-2,0) \* \z2
          + {2 \over 225} \* (130141 - 392209 \* x) \* \Hh(-2,0)
  \nonumber\\&& \mbox{}
          - {4 \over 15} \* (1489 + 14829 \* x) \* \Hh(-2,2)
          - {688 \over 3} \* (2 + x) \* \Hh(-2,3)
          - {4 \over 15} \* (21367 + 21047 \* x) \* \Hh(-1,-1) \* \z2
  \nonumber\\&& \mbox{}
          + {4 \over 15} \* (17871 + 17711 \* x) \* \Hh(-1,0) \* \z2
          + {2 \over 135} \* (151139 + 62435 \* x) \* \Hh(-1,0)
  \nonumber\\&& \mbox{}
          - {4 \over 225} \* (116459 + 219859 \* x) \* \Hh(-1,2)
          - 8 \* (488 + 473 \* x) \* \Hh(-1,3)
          - {14 \over 15} \* (673 + 3017 \* x) \* \Hh(0,0) \* \z2
  \nonumber\\&& \mbox{}
          - {1 \over 4050} \* (1771723 + 5003917 \* x) \* \Hh(0,0)
          + 8 \* (47 - 283 \* x) \* \Hh(1,0) \* \z3
          + {4 \over 5} \* (1701 - 1181 \* x) \* \Hh(1,0) \* \z2
  \nonumber\\&& \mbox{}
          + {1 \over 54} \* (30283 + 4811 \* x) \* \Hh(1,0)
          + 8 \* (11 - 151 \* x) \* \Hh(1,1) \* \z3
          + {4 \over 15} \* (1017 - 107 \* x) \* \Hh(1,1) \* \z2
  \nonumber\\&& \mbox{}
          + {7 \over 18} \* (2437 - 1847 \* x) \* \Hh(1,1)
          - {4 \over 3} \* (340 - 853 \* x) \* \Hh(1,2)
          - {8 \over 15} \* (936 - 11 \* x) \* \Hh(1,3)
          - 64 \* (5 - 19 \* x) \* \Hh(1,4)
  \nonumber\\&& \mbox{}
          + {8 \over 3} \* (89 + 17 \* x) \* \Hh(2,0) \* \z2
          + {4 \over 3} \* \Hh(2,0) \* (501 + 682 \* x)
          + {16 \over 3} \* (1 - 56 \* x) \* \Hh(2,1) \* \z2
          + {4 \over 9} \* (1201 + 3992 \* x) \* \Hh(2,1)
  \nonumber\\&& \mbox{}
          + {2 \over 3} \* (73 + 21 \* x) \* \Hh(2,2)
          - {16 \over 3} \* (26 + 47 \* x) \* \Hh(2,3)
          + {4 \over 15} \* (1048 + 2057 \* x) \* \Hh(3,0)
          + {4 \over 15} \* (1423 + 2367 \* x) \* \Hh(3,1)
  \nonumber\\&& \mbox{}
          + 16 \* (3 + 5 \* x) \* \Hh(4,0)
          + 16 \* (5 + 7 \* x) \* \Hh(4,1)
          - {224 \over 3} \* (1 - 4 \* x) \* \Hhh(-3,-1,0)
          + {32 \over 3} \* (20 - 13 \* x) \* \Hhh(-3,0,0)
  \nonumber\\&& \mbox{}
          - {16 \over 3} \* (1 - 79 \* x) \* \Hhh(-2,-2,0)
          + {4 \over 15} \* (3901 - 6679 \* x) \* \Hhh(-2,-1,0)
          + {32 \over 3} \* (85 + 29 \* x) \* \Hhh(-2,-1,2)
  \nonumber\\&& \mbox{}
          - {4 \over 15} \* (2689 + 5649 \* x) \* \Hhh(-2,0,0)
          - {32 \over 3} \* (1 + 2 \* x) \* \Hhh(-2,2,0)
          + {4 \over 15} \* (1399 - 1311 \* x) \* \Hhh(-1,-2,0)
  \nonumber\\&& \mbox{}
          + {4 \over 75} \* (39113 + 8863 \* x) \* \Hhh(-1,-1,0)
          + {16 \over 3} \* (1009 + 1024 \* x) \* \Hhh(-1,-1,2)
  \nonumber\\&& \mbox{}
          - {4 \over 225} \* (83949 + 128224 \* x) \* \Hhh(-1,0,0)
          - {16 \over 15} \* (571 + 511 \* x) \* \Hhh(-1,2,0)
          - {16 \over 5} \* (257 + 237 \* x) \* \Hhh(-1,2,1)
  \nonumber\\&& \mbox{}
          - 24 \* (4 + 9 \* x) \* \Hhh(0,0,0) \* \z2
          + {2 \over 225} \* (64641 + 404134 \* x) \* \Hhh(0,0,0)
          + {4 \over 3} \* (1099 - 1661 \* x) \* \Hhh(1,-2,0)
  \nonumber\\&& \mbox{}
          + 48 \* (9 - 37 \* x) \* \Hhh(1,0,0) \* \z2
          + {1 \over 45} \* (44737 - 81507 \* x) \* \Hhh(1,0,0)
          + {2 \over 9} \* (397 + 2141 \* x) \* \Hhh(1,1,0)
  \nonumber\\&& \mbox{}
          - {8 \over 3} \* (72 - 289 \* x) \* \Hhh(1,1,1)
          + {8 \over 3} \* (17 - 46 \* x) \* \Hhh(1,1,2)
          - 48 \* (3 + x) \* \Hhh(1,1,3)
          + {4 \over 3} \* (1 + 93 \* x) \* \Hhh(1,2,0)
  \nonumber\\&& \mbox{}
          - {4 \over 3} \* (21 - 31 \* x) \* \Hhh(1,2,1)
          + {32 \over 3} \* (19 + x) \* \Hhh(2,-2,0)
          + {2 \over 3} \* (47 + 91 \* x) \* \Hhh(2,1,0)
          + {8 \over 3} \* (3 + 17 \* x) \* \Hhh(3,0,0)
  \nonumber\\&& \mbox{}
          - {32 \over 3} \* (1 + 56 \* x) \* \Hhhh(-2,-1,-1,0)
          + {16 \over 3} \* (79 + 140 \* x) \* \Hhhh(-2,-1,0,0)
          - {8 \over 3} \* (13 + 167 \* x) \* \Hhhh(-2,0,0,0)
  \nonumber\\&& \mbox{}
          - {8 \over 15} \* (1187 + 567 \* x) \* \Hhhh(-1,-1,-1,0)
          + {4 \over 3} \* (2645 + 1667 \* x) \* \Hhhh(-1,-1,0,0)
          - {2 \over 15} \* (14087 + 9737 \* x) \* \Hhhh(-1,0,0,0)
  \nonumber\\&& \mbox{}
          + {2 \over 9} \* (1649 + 6389 \* x) \* \Hhhh(0,0,0,0)
          - {2 \over 15} \* (6263 - 3353 \* x) \* \Hhhh(1,0,0,0)
          + {4 \over 15} \* (937 + 153 \* x) \* \Hhhh(1,1,0,0)
  \nonumber\\&& \mbox{}
          - {4 \over 3} \* (13 - 61 \* x) \* \Hhhh(1,1,1,0)
          + 48 \* (1 + 3 \* x) \* \Hhhh(1,2,0,0)
          - {32 \over 3} \* (13 - 23 \* x) \* \Hhhh(2,0,0,0)
          + {16 \over 3} \* (23 + 98 \* x) \* \Hhhh(2,1,0,0)
  \nonumber\\&& \mbox{}
          + 16 \* (7 + 13 \* x) \* \Hhhhh(1,1,1,0,0)
          - {1 \over 405} \* \pqq( - x) \* (197640 \* \z5
          - 98550 \* \z3
          - 92260 \* \z2
          - 338040 \* \z2 \* \z3
  \nonumber\\&& \mbox{}
          + 33102 \* \z2^2
          + 125280 \* \Hh(-4,0)
          - 183960 \* \Hh(-3,0)
          + 583200 \* \Hh(-3,2)
          + 1143720 \* \Hh(-2,-1) \* \z2
  \nonumber\\&& \mbox{}
          - 208860 \* \Hh(-2,0)
          - 1185840 \* \Hh(-2,0) \* \z2
          + 32760 \* \Hh(-2,2)
          + 1011960 \* \Hh(-2,3)
          + 1134000 \* \Hh(-1,-2) \* \z2
  \nonumber\\&& \mbox{}
          + 1608120 \* \Hh(-1,-1) \* \z3
          + 27720 \* \Hh(-1,-1) \* \z2
          - 78680 \* \Hh(-1,0)
          - 860760 \* \Hh(-1,0) \* \z3
          - 33660 \* \Hh(-1,0) \* \z2
  \nonumber\\&& \mbox{}
          + 25320 \* \Hh(-1,2)
          + 113400 \* \Hh(-1,2) \* \z2
          + 65160 \* \Hh(-1,3)
          + 735480 \* \Hh(-1,4)
          + 137620 \* \Hh(0,0)
  \nonumber\\&& \mbox{}
          + 232200 \* \Hh(0,0) \* \z3
          - 133470 \* \Hh(0,0) \* \z2
          - 6480 \* \Hh(2,0)
          - 126360 \* \Hh(2,0) \* \z2
          - 6480 \* \Hh(2,1)
          - 98280 \* \Hh(2,1) \* \z2
  \nonumber\\&& \mbox{}
          - 1620 \* \Hh(2,2)
          + 49680 \* \Hh(2,3)
          - 17010 \* \Hh(3,0)
          - 41760 \* \Hh(3,1)
          - 36720 \* \Hh(3,2)
          - 91800 \* \Hh(4,0)
          - 130680 \* \Hh(4,1)
  \nonumber\\&& \mbox{}
          - 142560 \* \Hhh(-3,-1,0)
          + 453600 \* \Hhh(-3,0,0)
          - 120960 \* \Hhh(-2,-2,0)
          + 151560 \* \Hhh(-2,-1,0)
  \nonumber\\&& \mbox{}
          - 1062720 \* \Hhh(-2,-1,2)
          - 272520 \* \Hhh(-2,0,0)
          + 280800 \* \Hhh(-2,2,0)
          + 369360 \* \Hhh(-2,2,1)
          - 194400 \* \Hhh(-1,-3,0)
  \nonumber\\&& \mbox{}
          + 149940 \* \Hhh(-1,-2,0)
          - 1062720 \* \Hhh(-1,-2,2)
          - 1594080 \* \Hhh(-1,-1,-1) \* \z2
          + 244800 \* \Hhh(-1,-1,0)
  \nonumber\\&& \mbox{}
          + 1864080 \* \Hhh(-1,-1,0) \* \z2
          - 77760 \* \Hhh(-1,-1,2)
          - 1643760 \* \Hhh(-1,-1,3)
          - 276180 \* \Hhh(-1,0,0)
  \nonumber\\&& \mbox{}
          - 854280 \* \Hhh(-1,0,0) \* \z2
          + 38880 \* \Hhh(-1,2,0)
          + 83520 \* \Hhh(-1,2,1)
          + 79920 \* \Hhh(-1,2,2)
          + 294840 \* \Hhh(-1,3,0)
  \nonumber\\&& \mbox{}
          + 416880 \* \Hhh(-1,3,1)
          + 143385 \* \Hhh(0,0,0)
          + 191160 \* \Hhh(0,0,0) \* \z2
          - 12960 \* \Hhh(2,-2,0)
          + 6480 \* \Hhh(2,0,0)
  \nonumber\\&& \mbox{}
          + 1620 \* \Hhh(2,1,0)
          + 6480 \* \Hhh(2,1,2)
          - 12960 \* \Hhh(2,2,0)
          - 56160 \* \Hhh(3,0,0)
          - 10800 \* \Hhh(3,1,0)
          - 19440 \* \Hhh(3,1,1)
  \nonumber\\&& \mbox{}
          + 162000 \* \Hhhh(-2,-1,-1,0)
          - 795960 \* \Hhhh(-2,-1,0,0)
          + 568080 \* \Hhhh(-2,0,0,0)
          + 142560 \* \Hhhh(-1,-2,-1,0)
  \nonumber\\&& \mbox{}
          - 792720 \* \Hhhh(-1,-2,0,0)
          + 123120 \* \Hhhh(-1,-1,-2,0)
          - 100080 \* \Hhhh(-1,-1,-1,0)
          + 1516320 \* \Hhhh(-1,-1,-1,2)
  \nonumber\\&& \mbox{}
          + 220680 \* \Hhhh(-1,-1,0,0)
          - 509760 \* \Hhhh(-1,-1,2,0)
          - 682560 \* \Hhhh(-1,-1,2,1)
          - 273330 \* \Hhhh(-1,0,0,0)
  \nonumber\\&& \mbox{}
          + 142560 \* \Hhhh(-1,2,0,0)
          + 15120 \* \Hhhh(-1,2,1,0)
          + 38880 \* \Hhhh(-1,2,1,1)
          + 173970 \* \Hhhh(0,0,0,0)
          + 30240 \* \Hhhh(2,0,0,0)
  \nonumber\\&& \mbox{}
          + 30240 \* \Hhhh(2,1,0,0)
          - 6480 \* \Hhhh(2,1,1,0)
          - 155520 \* \Hhhhh(-1,-1,-1,-1,0)
          + 1134000 \* \Hhhhh(-1,-1,-1,0,0)
  \nonumber\\&& \mbox{}
          - 916920 \* \Hhhhh(-1,-1,0,0,0)
          + 381240 \* \Hhhhh(-1,0,0,0,0)
          - 81000 \* \Hhhhh(0,0,0,0,0)
          - 654480 \* \H(-3) \* \z2
  \nonumber\\&& \mbox{}
          - 1088640 \* \H(-2) \* \z3
          + 43020 \* \H(-2) \* \z2
          - 47610 \* \H(-1) \* \z3
          + 97080 \* \H(-1) \* \z2
          + 35856 \* \H(-1) \* \z2^2
  \nonumber\\&& \mbox{}
          + 138825 \* \H(0)
          - 99990 \* \H(0) \* \z3
          - 17655 \* \H(0) \* \z2
          + 38124 \* \H(0) \* \z2^2
          + 52920 \* \H(2)
          - 182520 \* \H(2) \* \z3
  \nonumber\\&& \mbox{}
          - 107730 \* \H(2) \* \z2
          - 28860 \* \H(3)
          - 183600 \* \H(3) \* \z2
          + 84690 \* \H(4)
          - 178200 \* \H(5))
          + {1 \over 6480} \* \pqq(x) \* (2292435
  \nonumber\\&& \mbox{}
          + 4812480 \* \z5
          + 2680080 \* \z3
          + 1050360 \* \z2
          - 1347840 \* \z2 \* \z3
          - 1746432 \* \z2^2
          - 1589760 \* \Hh(-4,0)
  \nonumber\\&& \mbox{}
          + 1382400 \* \Hh(-3,0)
          + 829440 \* \Hh(-3,2)
          - 7724160 \* \Hh(-2,-1) \* \z2
          + 1853280 \* \Hh(-2,0)
          + 4371840 \* \Hh(-2,0) \* \z2
  \nonumber\\&& \mbox{}
          - 1218240 \* \Hh(-2,2)
          - 3144960 \* \Hh(-2,3)
          + 1045320 \* \Hh(0,0)
          + 2056320 \* \Hh(0,0) \* \z3
          - 188640 \* \Hh(0,0) \* \z2
  \nonumber\\&& \mbox{}
          - 4821120 \* \Hh(1,-2) \* \z2
          - 3855640 \* \Hh(1,0)
          + 9849600 \* \Hh(1,0) \* \z3
          + 501120 \* \Hh(1,0) \* \z2
          - 3033000 \* \Hh(1,1)
  \nonumber\\&& \mbox{}
          + 10108800 \* \Hh(1,1) \* \z3
          + 1229760 \* \Hh(1,1) \* \z2
          - 4279680 \* \Hh(1,2)
          + 1520640 \* \Hh(1,2) \* \z2
          - 1005120 \* \Hh(1,3)
  \nonumber\\&& \mbox{}
          - 2108160 \* \Hh(1,4)
          - 4427520 \* \Hh(2,0)
          + 1866240 \* \Hh(2,0) \* \z2
          - 4648320 \* \Hh(2,1)
          + 1503360 \* \Hh(2,1) \* \z2
  \nonumber\\&& \mbox{}
          - 2472480 \* \Hh(2,2)
          + 915840 \* \Hh(2,3)
          - 2904480 \* \Hh(3,0)
          - 2888640 \* \Hh(3,1)
          - 622080 \* \Hh(3,2)
          - 535680 \* \Hh(4,0)
  \nonumber\\&& \mbox{}
          - 535680 \* \Hh(4,1)
          + 7050240 \* \Hhh(-3,-1,0)
          - 3525120 \* \Hhh(-3,0,0)
          + 2246400 \* \Hhh(-2,-2,0)
          + 3499200 \* \Hhh(-2,-1,0)
  \nonumber\\&& \mbox{}
          + 6255360 \* \Hhh(-2,-1,2)
          + 25920 \* \Hhh(-2,0,0)
          + 241920 \* \Hhh(-2,2,0)
          + 380160 \* \Hhh(-2,2,1)
          - 851520 \* \Hhh(0,0,0)
  \nonumber\\&& \mbox{}
          + 1192320 \* \Hhh(0,0,0) \* \z2
          + 691200 \* \Hhh(1,-3,0)
          + 1391040 \* \Hhh(1,-2,0)
          + 7914240 \* \Hhh(1,-2,2)
          - 3673200 \* \Hhh(1,0,0)
  \nonumber\\&& \mbox{}
          + 4008960 \* \Hhh(1,0,0) \* \z2
          - 5107680 \* \Hhh(1,1,0)
          + 1831680 \* \Hhh(1,1,0) \* \z2
          - 3741120 \* \Hhh(1,1,1)
          + 1313280 \* \Hhh(1,1,1) \* \z2
  \nonumber\\&& \mbox{}
          - 2030400 \* \Hhh(1,1,2)
          + 1693440 \* \Hhh(1,1,3)
          - 2914560 \* \Hhh(1,2,0)
          - 2257920 \* \Hhh(1,2,1)
          - 380160 \* \Hhh(1,2,2)
  \nonumber\\&& \mbox{}
          + 241920 \* \Hhh(1,3,0)
          + 414720 \* \Hhh(1,3,1)
          + 3628800 \* \Hhh(2,-2,0)
          - 2897280 \* \Hhh(2,0,0)
          - 3114720 \* \Hhh(2,1,0)
  \nonumber\\&& \mbox{}
          - 2597760 \* \Hhh(2,1,1)
          - 207360 \* \Hhh(2,1,2)
          - 120960 \* \Hhh(2,2,0)
          - 328320 \* \Hhh(2,2,1)
          - 1866240 \* \Hhh(3,0,0)
  \nonumber\\&& \mbox{}
          - 138240 \* \Hhh(3,1,0)
          - 311040 \* \Hhh(3,1,1)
          - 2937600 \* \Hhhh(-2,-1,-1,0)
          + 6462720 \* \Hhhh(-2,-1,0,0)
  \nonumber\\&& \mbox{}
          - 2453760 \* \Hhhh(-2,0,0,0)
          - 99360 \* \Hhhh(0,0,0,0)
          + 6186240 \* \Hhhh(1,-2,-1,0)
          + 1900800 \* \Hhhh(1,-2,0,0)
  \nonumber\\&& \mbox{}
          - 2158560 \* \Hhhh(1,0,0,0)
          + 5080320 \* \Hhhh(1,1,-2,0)
          - 3841920 \* \Hhhh(1,1,0,0)
          - 2364480 \* \Hhhh(1,1,1,0)
  \nonumber\\&& \mbox{}
          - 1900800 \* \Hhhh(1,1,1,1)
          - 69120 \* \Hhhh(1,1,1,2)
          + 103680 \* \Hhhh(1,1,2,0)
          - 241920 \* \Hhhh(1,1,2,1)
          - 2661120 \* \Hhhh(1,2,0,0)
  \nonumber\\&& \mbox{}
          + 103680 \* \Hhhh(1,2,1,0)
          - 311040 \* \Hhhh(1,2,1,1)
          - 4320000 \* \Hhhh(2,0,0,0)
          - 2488320 \* \Hhhh(2,1,0,0)
          + 535680 \* \Hhhh(2,1,1,0)
  \nonumber\\&& \mbox{}
          - 1296000 \* \Hhhhh(0,0,0,0,0)
          - 4337280 \* \Hhhhh(1,0,0,0,0)
          - 6272640 \* \Hhhhh(1,1,0,0,0)
          - 2937600 \* \Hhhhh(1,1,1,0,0)
  \nonumber\\&& \mbox{}
          + 622080 \* \Hhhhh(1,1,1,1,0)
          + 2695680 \* \H(-3) \* \z2
          + 6048000 \* \H(-2) \* \z3
          + 2967840 \* \H(-2) \* \z2
          + 5132100 \* \H(0)
  \nonumber\\&& \mbox{}
          - 1625760 \* \H(0) \* \z3
          + 4330560 \* \H(0) \* \z2
          + 271296 \* \H(0) \* \z2^2
          + 1001340 \* \H(1)
          + 1838880 \* \H(1) \* \z3
  \nonumber\\&& \mbox{}
          + 2321280 \* \H(1) \* \z2
          + 1565568 \* \H(1) \* \z2^2
          - 1679800 \* \H(2)
          + 7568640 \* \H(2) \* \z3
          + 1260000 \* \H(2) \* \z2
  \nonumber\\&& \mbox{}
          - 5074800 \* \H(3)
          + 1762560 \* \H(3) \* \z2
          - 591840 \* \H(4)
          - 984960 \* \H(5))
  \nonumber\\&& \mbox{}
          + {2 \over 25} \* \pqg( - x) \* (2 \* (3583 + 1537 \* x) \* \Hh(-2,0)
          + (5081 - 2727 \* x) \* \Hh(-1,0)
          - 2 \* (2337 - 1537 \* x) \* \Hh(-1,2)
  \nonumber\\&& \mbox{}
          - 2 \* (3943 + 1537 \* x) \* \Hhh(-1,-1,0)
          + 2 \* (2243 + 457 \* x) \* \Hhh(-1,0,0)
          + (731 - 4611 \* x) \* \H(-1) \* \z2
  \nonumber\\&& \mbox{}
          - 180 \* (1 - x) \* (6 \* \Hh(-3,0)
          + 50 \* \Hh(-2,2)
          + 51 \* \Hh(-1,-1) \* \z2
          - 33 \* \Hh(-1,0) \* \z2
          + 30 \* \Hh(-1,3)
          - 2 \* \Hhh(-2,-1,0)
  \nonumber\\&& \mbox{}
          + 30 \* \Hhh(-2,0,0)
          - 2 \* \Hhh(-1,-2,0)
          - 50 \* \Hhh(-1,-1,2)
          + 2 \* \Hhh(-1,2,0)
          + 2 \* \Hhh(-1,2,1)
          + 4 \* \Hhh(1,-2,0)
          + 2 \* \Hhhh(-1,-1,-1,0)
  \nonumber\\&& \mbox{}
          - 30 \* \Hhhh(-1,-1,0,0)
          + 8 \* \Hhhh(-1,0,0,0)
          - 51 \* \H(-2) \* \z2 - 40 \* \H(-1) \* \z3))
          + {2 \over 25} \* \pqg(x) \* (1680 \* (2 + 7 \* x) \* \Hh(0,0) \* \z2
  \nonumber\\&& \mbox{}
          + 101 \* (113 + 27 \* x) \* \Hh(0,0)
          - 10 \* (421 + 275 \* x) \* \Hh(2,0)
          - 2 \* (997 + 457 \* x) \* \Hhh(0,0,0)
          + 60 \* (163 + 23 \* x) \* \H(0) \* \z3
  \nonumber\\&& \mbox{}
          - 2 \* (4091 - 1699 \* x) \* \H(0) \* \z2
          - (7793 + 1213 \* x) \* \H(1) \* \z2
          + 12 \* (938 - 27 \* x) \* \H(3)
          - 120 \* (19 + 89 \* x) \* \H(4)
  \nonumber\\&& \mbox{}
          + 180 \* (1 + x) \* (\Hh(1,1) \* \z2
          - 2 \* \Hh(3,0)
          - 2 \* \Hh(3,1)
          - 10 \* \Hhhh(0,0,0,0)
          - 2 \* \Hhhh(1,0,0,0)
          + \H(2) \* \z2)
          + 550 \* (7 + 5 \* x) \* (\Hh(1,2)
  \nonumber\\&& \mbox{}
          - \Hhh(1,1,0))
          + 240 \* (13 - 22 \* x) \* (\Hh(1,3)
          - \Hhh(2,0,0)
          - \Hhhh(1,1,0,0))
          - 60 \* (43 - 97 \* x) \* (\Hh(1,0) \* \z2
          - \H(1) \* \z3)
  \nonumber\\&& \mbox{}
          + 6 \* (263 - 857 \* x) \* \z2^2
          + 5 \* (5391 + 3187 \* x) \* \z3
          - ( 5783 + 2727 \* x) \* \z2
          + 7961 
	  + 2390 \* \Hh(1,0)
  \nonumber\\&& \mbox{}
          - 360 \* \Hh(1,1)
          - 360 \* \Hh(2,1)
          - 5280 \* \Hhh(1,0,0)
          + 11377 \* \H(0)
          + 3416 \* \H(1)
          + 3056 \* \H(2))
  \nonumber\\&& \mbox{}
          + {2 \over 225} \* \pgq( - x) \* (1080 \* 
            (3 - 2 \* x^{-1}) \* \Hh(-2,0)
          + (3831 - 1357 \* x^{-1}) \* \Hh(-1,0)
          + 2 \* (3083 + 1417 \* x^{-1}) \* \Hh(-1,2)
  \nonumber\\&& \mbox{}
          + 2 \* (1237 - 1417 \* x^{-1}) \* \Hhh(-1,-1,0)
          + 2 \* (2363 + 337 \* x^{-1}) \* \Hhh(-1,0,0)
          - 3 \* (1643 + 1417 \* x^{-1}) \* \H(-1) \* \z2 
  \nonumber\\&& \mbox{}
	  - 180 \* (1 - x^{-1}) \* (4 \* \Hh(-2,2)
          + 51 \* \Hh(-1,-1) \* \z2
          - 33 \* \Hh(-1,0) \* \z2
          + 30 \* \Hh(-1,3)
          - 4 \* \Hhh(-2,-1,0)
          + 4 \* \Hhh(-2,0,0)
  \nonumber\\&& \mbox{}
          - 2 \* \Hhh(-1,-2,0)
          - 50 \* \Hhh(-1,-1,2)
          + 2 \* \Hhh(-1,2,0)
          + 2 \* \Hhh(-1,2,1)
          - 4 \* \Hhh(1,-2,0)
          + 2 \* \Hhhh(-1,-1,-1,0)
          - 30 \* \Hhhh(-1,-1,0,0)
  \nonumber\\&& \mbox{}
          + 8 \* \Hhhh(-1,0,0,0)
          - 6 \* \H(-2) \* \z2
          - 40 \* \H(-1) \* \z3))
          + {2 \over 225} \* \pgq(x) \* ((1237 + 1417 \* x^{-1}) \* \H(1) \* \z2
  \nonumber\\&& \mbox{}
          + 60 \* (1 + x^{-1}) \* (97 \* \Hh(1,0) \* \z2
          + 3 \* \Hh(1,1) \* \z2
          - 88 \* \Hh(1,3)
          - 6 \* \Hhhh(1,0,0,0)
          + 88 \* \Hhhh(1,1,0,0)
          - 97 \* \H(1) \* \z3
          + 6 \* \H(2) \* \z2)
  \nonumber\\&& \mbox{}
          + 6711
          - 1380 \* \z3
          + 4652 \* \z2
          - 406 \* \Hh(0,0)
          - 360 \* \Hh(1,0)
          - 360 \* \Hh(1,1)
          + 360 \* \Hh(2,0)
          + 360 \* \Hh(2,1)
  \nonumber\\&& \mbox{}
          + 1800 \* \Hhh(0,0,0)
          - 5280 \* \Hhh(1,0,0)
          - 3831 \* \H(0)
          - 11760 \* \H(0) \* \z2
          + 6406 \* \H(1)
          - 7126 \* \H(2)
          + 10680 \* \H(3))
  \nonumber\\&& \mbox{}
          - {16 \over 3} \* (23 - 2 \* x) \* \H(-3) \* \z2
          + {4 \over 3} \* (511 + 419 \* x) \* \H(-2) \* \z3
          + {2 \over 15} \* (6879 + 22979 \* x) \* \H(-2) \* \z2
  \nonumber\\&& \mbox{}
          + {4 \over 3} \* (3649 + 3534 \* x) \* \H(-1) \* \z3
          + {2 \over 225} \* (350257 + 466307 \* x) \* \H(-1) \* \z2
          + {32 \over 15} \* (45 + 52 \* x) \* \H(0) \* \z2^2
  \nonumber\\&& \mbox{}
          - {2 \over 15} \* (4793 + 15387 \* x) \* \H(0) \* \z3
          - {2 \over 225} \* (22362 + 1067413 \* x) \* \H(0) \* \z2
  \nonumber\\&& \mbox{}
          - {2 \over 2025} \* (1105108 + 1150983 \* x) \* \H(0)
          - {4 \over 5} \* (35 + 209 \* x) \* \H(1) \* \z2^2
          + {1 \over 54} \* (5409 - 80143 \* x) \* \H(1)
  \nonumber\\&& \mbox{}
          + {4 \over 15} \* (5507 - 8662 \* x) \* \H(1) \* \z3
          + {2 \over 75} \* (56113 - 51513 \* x) \* \H(1) \* \z2
          - {4 \over 3} \* (77 + 839 \* x) \* \H(2) \* \z3
  \nonumber\\&& \mbox{}
          - {1 \over 270} \* (911 + 353151 \* x) \* \H(2)
          + {4 \over 15} \* (1768 + 3287 \* x) \* \H(2) \* \z2
          - {8 \over 3} \* (25 + 59 \* x) \* \H(3) \* \z2
  \nonumber\\&& \mbox{}
          - {8 \over 75} \* (622 - 71447 \* x) \* \H(3)
          + {2 \over 15} \* (4063 + 16987 \* x) \* \H(4)
          + {8 \over 3} \* (36 + 71 \* x) \* \H(5)
          - 16 \* (1 - 5 \* x) \* (23 \* \Hh(1,-2) \* \z2
  \nonumber\\&& \mbox{}
          - 3 \* \Hh(1,2) \* \z2
          - 5 \* \Hhh(1,-3,0)
          - 24 \* \Hhh(1,-2,2)
          - 3 \* \Hhh(1,1,1) \* \z2
          + \Hhh(1,3,0)
          + \Hhh(1,3,1)
          - 2 \* \Hhhh(1,-2,-1,0)
          - 11 \* \Hhhh(1,-2,0,0)
  \nonumber\\&& \mbox{}
          - 12 \* \Hhhh(1,1,-2,0)
          + 9 \* \Hhhhh(1,0,0,0,0)
          + 10 \* \Hhhhh(1,1,0,0,0))
          + {4 \over 5} \* (1 - x) \* (90 \* \z4
          + 437 \* \Hhh(2,0,0))
          + {4 \over 3} \* (1 + x) \* (33 \* \Hhh(2,1,1)
  \nonumber\\&& \mbox{}
          - 8 \* \Hhh(2,2,0)
          + 12 \* \Hhh(2,2,1)
          - 12 \* \Hhhh(2,1,1,0))
          - {4 \over 5} \* (1 + 5 \* x) \* (700 \* \Hh(-1,-2) \* \z2
          + 990 \* \Hh(-1,-1) \* \z3
          - 450 \* \Hh(-1,0) \* \z3
  \nonumber\\&& \mbox{}
          + 20 \* \Hh(-1,2) \* \z2
          + 420 \* \Hh(-1,4)
          - 180 \* \Hhh(-1,-3,0)
          - 640 \* \Hhh(-1,-2,2)
          - 1260 \* \Hhh(-1,-1,-1) \* \z2
          + 880 \* \Hhh(-1,-1,0) \* \z2
  \nonumber\\&& \mbox{}
          - 700 \* \Hhh(-1,-1,3)
          - 560 \* \Hhh(-1,0,0) \* \z2
          + 20 \* \Hhh(-1,3,0)
          + 20 \* \Hhh(-1,3,1)
          + 120 \* \Hhhh(-1,-2,-1,0)
          - 460 \* \Hhhh(-1,-2,0,0)
  \nonumber\\&& \mbox{}
          + 120 \* \Hhhh(-1,-1,-2,0)
          + 1200 \* \Hhhh(-1,-1,-1,2)
          - 40 \* \Hhhh(-1,-1,2,0)
          - 40 \* \Hhhh(-1,-1,2,1)
          - 80 \* \Hhhh(-1,2,0,0)
  \nonumber\\&& \mbox{}
          - 120 \* \Hhhhh(-1,-1,-1,-1,0)
          + 820 \* \Hhhhh(-1,-1,-1,0,0)
          - 320 \* \Hhhhh(-1,-1,0,0,0)
          + 180 \* \Hhhhh(-1,0,0,0,0)
          + 51 \* \H(-1) \* \z2^2)
  \nonumber\\&& \mbox{}
          + {32 \over 3} \* (3 - 7 \* x) \* (4 \* \Hh(0,0) \* \z3
          + 9 \* \Hhh(1,1,0) \* \z2)
          - {64 \over 3} \* (5 - 68 \* x) \* \z2 \* \z3
          + {8 \over 3} \* (11 + 3 \* x) \* (\Hh(3,2)
          - \Hhh(3,1,0))
  \nonumber\\&& \mbox{}
          + 2 \* (201 + 985 \* x) \* \z5
          + {2 \over 75} \* (4496 + 21049 \* x) \* \z2^2
          + {1 \over 270} \* (47939 + 701327 \* x) \* \z2
  \nonumber\\&& \mbox{}
          - {2 \over 45} \* (50257 + 100858 \* x) \* \z3
          - {1 \over 32400} \* (20611421 + 47144809 \* x)
          - 8  \* x \* (4 \* \Hhh(-2,2,1)
          - 21 \* \Hhhhh(0,0,0,0,0))
  \nonumber\\&& \mbox{}
          + \delta(1 - x) \* \biggl({9161 \over 12} - {4952 \over 9} \* \z5 
                + {87632 \over 135} \* \z2^2 - 2141 \* \z3 
                + {1016 \over 3} \* \z3^2 + {104045 \over 54} \* \z2
                - {33556 \over 315} \* \z2^3 
  \nonumber\\&& \mbox{}
		- {6644 \over 9} \* \z2 \* \z3\biggl)
          \biggr)
  \nonumber\\&& \mbox{}
       +  \colour4colour{\ca^2 \* \cf}  \*  \biggl(
          - {8 \over 3} \* (5 - x) \* \Hh(-4,0)
          - {4 \over 15} \* (537 - 503 \* x) \* \Hh(-3,0)
          - {8 \over 3} \* (11 + 7 \* x) \* \Hh(-3,2)
  \nonumber\\&& \mbox{}
          + {16 \over 3} \* (43 + 5 \* x) \* \Hh(-2,-1) \* \z2
          - 16 \* (9 + x) \* \Hh(-2,0) \* \z2
          - {2 \over 45} \* (16603 - 18067 \* x) \* \Hh(-2,0)
  \nonumber\\&& \mbox{}
          - {8 \over 3} \* (24 - 391 \* x) \* \Hh(-2,2)
          + {4 \over 3} \* (89 + 7 \* x) \* \Hh(-2,3)
          + {4 \over 3} \* (1062 + 869 \* x) \* \Hh(-1,-1) \* \z2
  \nonumber\\&& \mbox{}
          - {4 \over 5} \* (1351 + 1276 \* x) \* \Hh(-1,0) \* \z2
          - {2 \over 675} \* (411509 + 266459 \* x) \* \Hh(-1,0)
          + {4 \over 45} \* (2981 + 9436 \* x) \* \Hh(-1,2)
  \nonumber\\&& \mbox{}
          + {16 \over 3} \* (167 + 159 \* x) \* \Hh(-1,3)
          + {4 \over 15} \* (163 + 2437 \* x) \* \Hh(0,0) \* \z2
          + {2 \over 675} \* (91133 + 621237 \* x) \* \Hh(0,0)
  \nonumber\\&& \mbox{}
          - 16 \* (7 - 41 \* x) \* \Hh(1,0) \* \z3
          - {4 \over 15} \* (1081 - 336 \* x) \* \Hh(1,0) \* \z2
          - {1 \over 9} \* (1947 - 3857 \* x) \* \Hh(1,0)
  \nonumber\\&& \mbox{}
          - 32 \* (1 - 11 \* x) \* \Hh(1,1) \* \z3
          - {4 \over 3} \* (17 + 161 \* x) \* \Hh(1,1) \* \z2
          - {5 \over 9} \* (795 - 1429 \* x) \* \Hh(1,1)
          + {4 \over 9} \* (190 - 173 \* x) \* \Hh(1,2)
  \nonumber\\&& \mbox{}
          + {32 \over 15} \* (46 + 39 \* x) \* \Hh(1,3)
          + 32 \* (2 - 7 \* x) \* \Hh(1,4)
          - {8 \over 3} \* (17 + 65 \* x) \* \Hh(2,0) \* \z2
          - {2 \over 9} \* (921 - 145 \* x) \* \Hh(2,0)
  \nonumber\\&& \mbox{}
          - {4 \over 3} \* (140 + 141 \* x) \* \Hh(2,1)
          + {10 \over 3} \* (1 + 3 \* x) \* \Hh(2,2)
          + 20 \* (1 + 9 \* x) \* \Hh(2,3)
          - {2 \over 3} \* (75 + 89 \* x) \* \Hh(3,0)
  \nonumber\\&& \mbox{}
          - 4 \* (21 + 20 \* x) \* \Hh(3,1)
          - {56 \over 3} \* (2 - x) \* \Hhh(-3,0,0)
          - {8 \over 3} \* (40 - 299 \* x) \* \Hhh(-2,-1,0)
          - {8 \over 3} \* (91 + 5 \* x) \* \Hhh(-2,-1,2)
  \nonumber\\&& \mbox{}
          - {64 \over 3} \* (3 - 8 \* x) \* \Hhh(-2,0,0)
          + {8 \over 3} \* (8 + 163 \* x) \* \Hhh(-1,-2,0)
          - {4 \over 15} \* (887 - 1368 \* x) \* \Hhh(-1,-1,0)
  \nonumber\\&& \mbox{}
          - {64 \over 3} \* (65 + 61 \* x) \* \Hhh(-1,-1,2)
          - {4 \over 45} \* (2643 + 188 \* x) \* \Hhh(-1,0,0)
          + 8 \* (3 + 4 \* x) \* \Hhh(0,0,0) \* \z2
  \nonumber\\&& \mbox{}
          + {2 \over 15} \* (547 - 2497 \* x) \* \Hhh(0,0,0)
          - 8 \* (55 - 97 \* x) \* \Hhh(1,-2,0)
          - 96 \* (1 - 4 \* x) \* \z2 \* \Hhh(1,0,0)
  \nonumber\\&& \mbox{}
          - {2 \over 45} \* (7399 - 14709 \* x) \* \Hhh(1,0,0)
          - {8 \over 9} \* (95 - 88 \* x) \* \Hhh(1,1,0)
          - {4 \over 3} \* (5 - 54 \* x) \* \Hhh(1,1,2)
          + {16 \over 3} \* (1 - 13 \* x) \* \Hhh(1,2,0)
  \nonumber\\&& \mbox{}
          - {4 \over 3} \* (2 + 3 \* x) \* \Hhh(1,2,1)
          - {16 \over 3} \* (13 + x) \* \Hhh(2,-2,0)
          - {2 \over 15} \* (1143 - 2333 \* x) \* \Hhh(2,0,0)
          + {2 \over 3} \* (1 - 15 \* x) \* \Hhh(2,1,0)
  \nonumber\\&& \mbox{}
          - {4 \over 3} \* (67 + 29 \* x) \* \Hhhh(-2,-1,0,0)
          + {8 \over 3} \* (5 + 7 \* x) \* \Hhhh(-2,0,0,0)
          + {8 \over 3} \* (22 - 107 \* x) \* \Hhhh(-1,-1,-1,0)
  \nonumber\\&& \mbox{}
          - {4 \over 3} \* (514 - 9 \* x) \* \Hhhh(-1,-1,0,0)
          + {4 \over 15} \* (1043 + 73 \* x) \* \Hhhh(-1,0,0,0)
          - {4 \over 15} \* (72 + 733 \* x) \* \Hhhh(0,0,0,0)
  \nonumber\\&& \mbox{}
          + {4 \over 15} \* (683 + 287 \* x) \* \Hhhh(1,0,0,0)
          - {4 \over 15} \* (198 - 403 \* x) \* \Hhhh(1,1,0,0)
          + {4 \over 3} \* (7 - 51 \* x) \* \Hhhh(1,1,1,0)
          + {16 \over 3} \* (7 + x) \* \Hhhh(2,0,0,0)
  \nonumber\\&& \mbox{}
          - 4 \* (7 + 47 \* x) \* \Hhhh(2,1,0,0)
          + {1 \over 405} \* \pqq( - x) \* (31860 \* \z5
          - 51300 \* \z3
          - 44240 \* \z2
          - 70200 \* \z2 \* \z3
          + 5535 \* \z2^2
  \nonumber\\&& \mbox{}
          + 10800 \* \Hh(-4,0)
          - 88200 \* \Hh(-3,0)
          + 142560 \* \Hh(-3,2)
          + 265680 \* \Hh(-2,-1) \* \z2
          - 81480 \* \Hh(-2,0)
  \nonumber\\&& \mbox{}
          - 273240 \* \Hh(-2,0) \* \z2
          - 7920 \* \Hh(-2,2)
          + 234360 \* \Hh(-2,3)
          + 262440 \* \Hh(-1,-2) \* \z2
          + 390960 \* \Hh(-1,-1) \* \z3
  \nonumber\\&& \mbox{}
          - 39600 \* \Hh(-1,-1) \* \z2
          - 42580 \* \Hh(-1,0)
          - 179280 \* \Hh(-1,0) \* \z3
          + 26100 \* \Hh(-1,0) \* \z2
          - 4080 \* \Hh(-1,2)
  \nonumber\\&& \mbox{}
          + 34560 \* \Hh(-1,2) \* \z2
          - 7920 \* \Hh(-1,3)
          + 155520 \* \Hh(-1,4)
          + 61385 \* \Hh(0,0)
          + 38880 \* \Hh(0,0) \* \z3
          - 52290 \* \Hh(0,0) \* \z2
  \nonumber\\&& \mbox{}
          - 34560 \* \Hh(2,0) \* \z2
          - 25920 \* \Hh(2,1) \* \z2
          + 15120 \* \Hh(2,3)
          - 7920 \* \Hh(3,1)
          - 4320 \* \Hh(3,2)
          - 12960 \* \Hh(4,0)
  \nonumber\\&& \mbox{}
          - 21600 \* \Hh(4,1)
          + 8640 \* \Hhh(-3,-1,0)
          + 75600 \* \Hhh(-3,0,0)
          + 4320 \* \Hhh(-2,-2,0)
          + 87120 \* \Hhh(-2,-1,0)
  \nonumber\\&& \mbox{}
          - 265680 \* \Hhh(-2,-1,2)
          - 137880 \* \Hhh(-2,0,0)
          + 62640 \* \Hhh(-2,2,0)
          + 86400 \* \Hhh(-2,2,1)
          - 8640 \* \Hhh(-1,-3,0)
  \nonumber\\&& \mbox{}
          + 87120 \* \Hhh(-1,-2,0)
          - 265680 \* \Hhh(-1,-2,2)
          - 388800 \* \Hhh(-1,-1,-1) \* \z2
          + 96480 \* \Hhh(-1,-1,0)
  \nonumber\\&& \mbox{}
          + 449280 \* \Hhh(-1,-1,0) \* \z2
          - 401760 \* \Hhh(-1,-1,3)
          - 116760 \* \Hhh(-1,0,0)
          - 185760 \* \Hhh(-1,0,0) \* \z2
          + 15840 \* \Hhh(-1,2,1)
  \nonumber\\&& \mbox{}
          + 10800 \* \Hhh(-1,2,2)
          + 52920 \* \Hhh(-1,3,0)
          + 86400 \* \Hhh(-1,3,1)
          + 55560 \* \Hhh(0,0,0)
          + 34560 \* \Hhh(0,0,0) \* \z2
  \nonumber\\&& \mbox{}
          - 4320 \* \Hhh(2,-2,0)
          + 3240 \* \Hhh(2,0,0)
          + 2160 \* \Hhh(2,1,2)
          - 4320 \* \Hhh(2,2,0)
          - 10800 \* \Hhh(3,0,0)
          + 4320 \* \Hhh(3,1,0)
  \nonumber\\&& \mbox{}
          - 143640 \* \Hhhh(-2,-1,0,0)
          + 96120 \* \Hhhh(-2,0,0,0)
          - 6480 \* \Hhhh(-1,-2,-1,0)
          - 142560 \* \Hhhh(-1,-2,0,0)
  \nonumber\\&& \mbox{}
          - 12960 \* \Hhhh(-1,-1,-2,0)
          - 79200 \* \Hhhh(-1,-1,-1,0)
          + 388800 \* \Hhhh(-1,-1,-1,2)
          + 134640 \* \Hhhh(-1,-1,0,0)
  \nonumber\\&& \mbox{}
          - 120960 \* \Hhhh(-1,-1,2,0)
          - 172800 \* \Hhhh(-1,-1,2,1)
          - 133020 \* \Hhhh(-1,0,0,0)
          + 30240 \* \Hhhh(-1,2,0,0)
  \nonumber\\&& \mbox{}
          - 10800 \* \Hhhh(-1,2,1,0)
          + 66870 \* \Hhhh(0,0,0,0)
          + 8640 \* \Hhhh(2,0,0,0)
          + 8640 \* \Hhhh(2,1,0,0)
          - 2160 \* \Hhhh(2,1,1,0)
  \nonumber\\&& \mbox{}
          + 220320 \* \Hhhhh(-1,-1,-1,0,0)
          - 164160 \* \Hhhhh(-1,-1,0,0,0)
          + 62640 \* \Hhhhh(-1,0,0,0,0)
          - 12960 \* \Hhhhh(0,0,0,0,0)
  \nonumber\\&& \mbox{}
          - 138240 \* \H(-3) \* \z2
          - 247320 \* \H(-2) \* \z3
          + 51480 \* \H(-2) \* \z2
          + 31680 \* \H(-1) \* \z3
          + 52320 \* \H(-1) \* \z2
  \nonumber\\&& \mbox{}
          + 5076 \* \H(-1) \* \z2^2
          + 53145 \* \H(0)
          - 49860 \* \H(0) \* \z3
          - 16320 \* \H(0) \* \z2
          + 8640 \* \H(0) \* \z2^2
          + 22950 \* \H(2)
  \nonumber\\&& \mbox{}
          - 49680 \* \H(2) \* \z3
          - 29160 \* \H(2) \* \z2
          + 2040 \* \H(3)
          - 47520 \* \H(3) \* \z2
          + 37080 \* \H(4)
          - 32400 \* \H(5))
  \nonumber\\&& \mbox{}
          - {1 \over 7290} \* \pqq(x) \* (2996875
          + 651240 \* \z5
          - 797580 \* \z3
          - 3438810 \* \z2
          - 252720 \* \z2 \* \z3
          - 20412 \* \z2^2
  \nonumber\\&& \mbox{}
          - 116640 \* \Hh(-4,0)
          + 1040040 \* \Hh(-3,0)
          + 1088640 \* \Hh(-3,2)
          - 1982880 \* \Hh(-2,-1) \* \z2
          + 952560 \* \Hh(-2,0)
  \nonumber\\&& \mbox{}
          + 699840 \* \Hh(-2,0) \* \z2
          - 155520 \* \Hh(-2,2)
          - 349920 \* \Hh(-2,3)
          + 4599450 \* \Hh(0,0)
          + 699840 \* \Hh(0,0) \* \z3
  \nonumber\\&& \mbox{}
          - 1607040 \* \Hh(0,0) \* \z2
          - 2293920 \* \Hh(1,-2) \* \z2
          + 1062450 \* \Hh(1,0)
          + 3285360 \* \Hh(1,0) \* \z3
          - 1474200 \* \Hh(1,0) \* \z2
  \nonumber\\&& \mbox{}
          + 1255230 \* \Hh(1,1)
          + 2799360 \* \Hh(1,1) \* \z3 
          - 891000 \* \Hh(1,1) \* \z2 
          + 826200 \* \Hh(1,2)
          - 97200 \* \Hh(1,2) \* \z2 
  \nonumber\\&& \mbox{}
          + 1146960 \* \Hh(1,3)
          + 291600 \* \Hh(1,4)
          + 1070820 \* \Hh(2,0)
          - 213840 \* \Hh(2,0) \* \z2 
          + 784080 \* \Hh(2,1)
          + 38880 \* \Hh(2,1) \* \z2 
  \nonumber\\&& \mbox{}
          + 341820 \* \Hh(2,2)
          + 913680 \* \Hh(2,3)
          + 306180 \* \Hh(3,0)
          + 362880 \* \Hh(3,1)
          + 116640 \* \Hh(4,0)
          + 194400 \* \Hh(4,1)
  \nonumber\\&& \mbox{}
          + 1866240 \* \Hhh(-3,-1,0)
          - 233280 \* \Hhh(-3,0,0)
          + 622080 \* \Hhh(-2,-2,0)
          + 1049760 \* \Hhh(-2,-1,0)
  \nonumber\\&& \mbox{}
          + 1555200 \* \Hhh(-2,-1,2)
          + 505440 \* \Hhh(-2,0,0)
          + 311040 \* \Hhh(-2,2,0)
          + 388800 \* \Hhh(-2,2,1)
          + 2502900 \* \Hhh(0,0,0)
  \nonumber\\&& \mbox{}
          - 77760 \* \Hhh(0,0,0) \* \z2 
          + 933120 \* \Hhh(1,-3,0)
          + 913680 \* \Hhh(1,-2,0)
          + 3110400 \* \Hhh(1,-2,2)
          + 887760 \* \Hhh(1,0,0)
  \nonumber\\&& \mbox{}
          + 252720 \* \Hhh(1,0,0) \* \z2
          - 42120 \* \Hhh(1,1,0)
          - 349920 \* \Hhh(1,1,0) \* \z2 
          + 392040 \* \Hhh(1,1,1)
          + 77760 \* \Hhh(1,1,1) \* \z2 
  \nonumber\\&& \mbox{}
          + 178200 \* \Hhh(1,1,2)
          + 1205280 \* \Hhh(1,1,3)
          - 291600 \* \Hhh(1,2,0)
          + 320760 \* \Hhh(1,2,1)
          + 38880 \* \Hhh(1,2,2)
  \nonumber\\&& \mbox{}
          + 505440 \* \Hhh(1,3,0)
          + 777600 \* \Hhh(1,3,1)
          + 1477440 \* \Hhh(2,-2,0)
          + 220320 \* \Hhh(2,0,0)
          - 312660 \* \Hhh(2,1,0)
  \nonumber\\&& \mbox{}
          - 38880 \* \Hhh(2,1,2)
          - 58320 \* \Hhh(2,2,0)
          + 19440 \* \Hhh(2,2,1)
          - 427680 \* \Hhh(3,0,0)
          - 855360 \* \Hhhh(-2,-1,-1,0)
  \nonumber\\&& \mbox{}
          + 1594080 \* \Hhhh(-2,-1,0,0)
          - 388800 \* \Hhhh(-2,0,0,0)
          + 1203660 \* \Hhhh(0,0,0,0)
          + 1632960 \* \Hhhh(1,-2,-1,0)
  \nonumber\\&& \mbox{}
          + 1438560 \* \Hhhh(1,-2,0,0)
          + 541080 \* \Hhhh(1,0,0,0)
          + 1944000 \* \Hhhh(1,1,-2,0)
          - 463320 \* \Hhhh(1,1,0,0)
  \nonumber\\&& \mbox{}
          - 498960 \* \Hhhh(1,1,1,0)
          - 77760 \* \Hhhh(1,1,1,2)
          - 116640 \* \Hhhh(1,1,2,0)
          + 77760 \* \Hhhh(1,1,2,1)
          - 1049760 \* \Hhhh(1,2,0,0)
  \nonumber\\&& \mbox{}
          - 38880 \* \Hhhh(1,2,1,0)
          - 1263600 \* \Hhhh(2,0,0,0)
          - 1088640 \* \Hhhh(2,1,0,0)
          + 19440 \* \Hhhh(2,1,1,0)
          - 233280 \* \Hhhhh(0,0,0,0,0)
  \nonumber\\&& \mbox{}
          - 1127520 \* \Hhhhh(1,0,0,0,0)
          - 2021760 \* \Hhhhh(1,1,0,0,0)
          - 1555200 \* \Hhhhh(1,1,1,0,0)
          - 155520 \* \H(-3) \* \z2
  \nonumber\\&& \mbox{}
          + 1399680 \* \H(-2) \* \z3 
          + 680400 \* \H(-2) \* \z2 
          + 5966955 \* \H(0)
          - 1296000 \* \H(0) \* \z3 
          - 1808460 \* \H(0) \* \z2 
  \nonumber\\&& \mbox{}
          + 252720 \* \H(0) \* \z2^2 
          + 2281005 \* \H(1)
          - 324000 \* \H(1) \* \z3 
          - 1694520 \* \H(1) \* \z2 
          + 1183896 \* \H(1) \* \z2^2 
  \nonumber\\&& \mbox{}
          + 3055590 \* \H(2)
          + 2099520 \* \H(2) \* \z3 
          - 1125900 \* \H(2) \* \z2 
          + 1551420 \* \H(3)
          - 77760 \* \H(3) \* \z2 
          + 1333260 \* \H(4)
  \nonumber\\&& \mbox{}
          + 116640 \* \H(5))
          - {1 \over 25} \* \pqg( - x) \* (10 \* (491 + 71 \* x) \* \Hh(-2,0)
          + 2 \* (1984 - 1869 \* x) \* \Hh(-1,0)
  \nonumber\\&& \mbox{}
          - 10 \* (201 - 71 \* x) \* \Hh(-1,2)
          - 10 \* (539 + 71 \* x) \* \Hhh(-1,-1,0)
          + 10 \* (433 - 193 \* x) \* \Hhh(-1,0,0)
  \nonumber\\&& \mbox{}
          - 5 \* (137 + 213 \* x) \* \H(-1) \* \z2 
          - 240 \* (1 - x) \* (2 \* \Hh(-3,0)
          + 20 \* \Hh(-2,2)
          + 20 \* \Hh(-1,-1) \* \z2 
          - 11 \* \Hh(-1,0) \* \z2 
  \nonumber\\&& \mbox{}
          + 10 \* \Hh(-1,3)
          + 10 \* \Hhh(-2,0,0)
          - 20 \* \Hhh(-1,-1,2)
          + 2 \* \Hhh(1,-2,0)
          - 10 \* \Hhhh(-1,-1,0,0)
          + \Hhhh(-1,0,0,0)
          - 20 \* \H(-2) \* \z2 
  \nonumber\\&& \mbox{}
          - 15 \* \H(-1) \* \z3))
          + {1 \over 25} \* \pqg(x) \* (10 \* (83 - 387 \* x) \* \Hh(0,0) \* \z2
          - 14 \* (647 + 267 \* x) \* \Hh(0,0)
  \nonumber\\&& \mbox{}
          + 10 \* (347 - 123 \* x) \* \Hh(1,0) \* \z2 
          + 10 \* (657 + 188 \* x) \* \H(0) \* \z2 
          - 10 \* (683 + 213 \* x) \* \H(0) \* \z3 
  \nonumber\\&& \mbox{}
          - 10 \* (323 - 147 \* x) \* \H(1) \* \z3 
          + 5 \* (1199 + 589 \* x) \* \H(1) \* \z2 
          - 70 \* (104 + 37 \* x) \* \H(3)
          - 10 \* (131 - 339 \* x) \* \H(4)
  \nonumber\\&& \mbox{}
          - 10 \* (1 + x) \* (330 \* \Hh(1,2)
          - 330 \* \Hh(2,0)
          + 193 \* \Hhh(0,0,0)
          - 330 \* \Hhh(1,1,0)
          - 48 \* \Hhhh(0,0,0,0)
          - 24 \* \Hhhh(1,0,0,0))
  \nonumber\\&& \mbox{}
          - 40 \* (73 - 21 \* x) \* \z2^2 
          - 10 \* (371 - 99 \* x) \* (\Hh(1,3)
          - \Hhh(2,0,0)
          - \Hhhh(1,1,0,0))
          + 2 \* (1559 + 1869 \* x) \* \z2 
  \nonumber\\&& \mbox{}
          - 5 \* (2739 + 2335\* x) \* \z3 
          - 2 \* (3304 
          + 1650 \* \Hh(1,0)
          - 495 \* \Hhh(1,0,0)
          + 2994 \* \H(0)
          - 310 \* \H(1)
          - 310 \* \H(2)  ))
  \nonumber\\&& \mbox{}
          - {1 \over 225} \* \pgq( - x) \* (240 \* 
            (13 - 11 \* x^{-1}) \* \Hh(-2,0)
          + 2 \* (1544 - 1429 \* x^{-1}) \* \Hh(-1,0)
          + 10 \* (409 + 71 \* x^{-1}) \* \Hh(-1,2)
  \nonumber\\&& \mbox{}
          + 10 \* (433 - 193 \* x^{-1}) \* \Hhh(-1,0,0)
          - 15 \* (249 + 71 \* x^{-1}) \* \H(-1) \* \z2 
          - 10 \* (1 - x^{-1}) \* (480 \* \Hh(-1,-1) \* \z2
  \nonumber\\&& \mbox{}
          - 264 \* \Hh(-1,0) \* \z2 
          + 240 \* \Hh(-1,3)
          - 71 \* \Hhh(-1,-1,0)
          - 480 \* \Hhh(-1,-1,2)
          - 48 \* \Hhh(1,-2,0)
          - 240 \* \Hhhh(-1,-1,0,0)
  \nonumber\\&& \mbox{}
          + 24 \* \Hhhh(-1,0,0,0)
          - 360 \* \H(-1) \* \z3))
          - {1 \over 225} \* \pgq(x) \* (5 \* (1 + x^{-1}) \* 
            (246 \* \Hh(1,0) \* \z2 
          - 198 \* \Hh(1,3)
          - 48 \* \Hhhh(1,0,0,0)
  \nonumber\\&& \mbox{}
          + 198 \* \Hhhh(1,1,0,0)
          - 294 \* \H(1) \* \z3 
          + 71 \* \H(1) \* \z2)
          + 2 \* (2864 
          - 1065 \* \z3 
          + 1225 \* \z2 
          - 1205 \* \Hh(0,0)
          + 240 \* \Hhh(0,0,0)
  \nonumber\\&& \mbox{}
          - 495 \* \Hhh(1,0,0)
          - 1544 \* \H(0)
          - 1935 \* \H(0) \* \z2 
          + 1340 \* \H(1)
          - 1340 \* \H(2)
          + 1695 \* \H(3) ))
          + {4 \over 3} \* (35 + x) \* \H(-3) \* \z2
  \nonumber\\&& \mbox{}
          + {4 \over 3} \* (8 - 483 \* x) \* \H(-2) \* \z2
          - {4 \over 3} \* (125 + 19 \* x) \* \H(-2) \* \z3
          - {4 \over 3} \* (874 + 697 \* x) \* \H(-1) \* \z3
  \nonumber\\&& \mbox{}
          - {2 \over 45} \* (8623 + 14768 \* x) \* \H(-1) \* \z2
          - {8 \over 15} \* (20 + 51 \* x) \* \H(0) \* \z2^2
          - {4 \over 3} \* (77 - 1339 \* x) \* \H(0) \* \z2
  \nonumber\\&& \mbox{}
          + {2 \over 15} \* (1879 + 611 \* x) \* \H(0) \* \z3
          - {1 \over 810} \* (16723 - 1625125 \* x) \* \H(0)
          - {16 \over 5} \* (5 - 49 \* x) \* \H(1) \* \z2^2
  \nonumber\\&& \mbox{}
          - {4 \over 15} \* (1081 - 1486 \* x) \* \H(1) \* \z3
          - {2 \over 45} \* (4561 + 2374 \* x) \* \H(1) \* \z2
          - {1 \over 270} \* (121321 - 456491 \* x) \* \H(1)
  \nonumber\\&& \mbox{}
          + {64 \over 3} \* (1 + 10 \* x) \* \H(2) \* \z3
          - {2 \over 3} \* (85 + 613 \* x) \* \H(2) \* \z2
          - {1 \over 45} \* (9553 - 68567 \* x) \* \H(2)
          + {4 \over 3} \* (15 + 11 \* x) \* \H(3) \* \z2
  \nonumber\\&& \mbox{}
          + {14 \over 9} \* (87 - 913 \* x) \* \H(3)
          - {4 \over 15} \* (91 + 1934 \* x) \* \H(4)
          - {8 \over 3} \* (9 + 11 \* x) \* \H(5)
          + 32 \* (1 - 5 \* x) \* (4 \* \Hh(1,-2) \* \z2 
          - \Hhh(1,-3,0)
  \nonumber\\&& \mbox{}
          - 4 \* \Hhh(1,-2,2)
          - 2 \* \Hhhh(1,-2,0,0)
          - 2 \* \Hhhh(1,1,-2,0)
          + \Hhhhh(1,0,0,0,0)
          + \Hhhhh(1,1,0,0,0))
          - {8 \over 3} \* (1 - x) \* (9 \* \z4 
          + \Hh(3,2)
  \nonumber\\&& \mbox{}
          - 13 \* \Hhh(-3,-1,0)
          - 7 \* \Hhh(-2,-2,0)
          - \Hhh(-2,2,0)
          - \Hhh(3,1,0)
          + 10 \* \Hhhh(-2,-1,-1,0))
          + {4 \over 3} \* (1 + x) \* (10 \* \Hh(2,1) \* \z2 
          - 9 \* \Hh(4,0)
  \nonumber\\&& \mbox{}
          - 15 \* \Hh(4,1)
          + 80 \* \Hhh(-1,2,0)
          + 120 \* \Hhh(-1,2,1)
          - 24 \* \Hhh(1,1,0) \* \z2 
          + \Hhh(2,2,0)
          - \Hhh(3,0,0))
          + {16 \over 5} \* (1 + 5 \* x) \* (40 \* \Hh(-1,-2) \* \z2 
  \nonumber\\&& \mbox{}
          + 60 \* \Hh(-1,-1) \* \z3 
          - 30 \* \Hh(-1,0) \* \z3 
          + 25 \* \Hh(-1,4)
          - 10 \* \Hhh(-1,-3,0)
          - 40 \* \Hhh(-1,-2,2)
          - 80 \* \Hhh(-1,-1,-1) \* \z2 
  \nonumber\\&& \mbox{}
          + 50 \* \Hhh(-1,-1,0) \* \z2 
          - 40 \* \Hhh(-1,-1,3)
          - 35 \* \Hhh(-1,0,0) \* \z2 
          - 20 \* \Hhhh(-1,-2,0,0)
          + 80 \* \Hhhh(-1,-1,-1,2)
          - 5 \* \Hhhh(-1,2,0,0)
  \nonumber\\&& \mbox{}
          + 40 \* \Hhhhh(-1,-1,-1,0,0)
          - 10 \* \Hhhhh(-1,-1,0,0,0)
          + 10 \* \Hhhhh(-1,0,0,0,0)
          - \H(-1) \* \z2^2)
          - {4 \over 3} \* (5 + 299 \* x) \* \z2 \* \z3
  \nonumber\\&& \mbox{}
          + {101 \over 15} \* (9 - 25 \* x) \* \z2^2
          - {8 \over 3} \* (15 + 133 \* x) \* \z5
          + {1 \over 45} \* (21999 + 14831 \* x) \* \z3
  \nonumber\\&& \mbox{}
          + {1 \over 675} \* (43749 - 1565269 \* x) \* \z2
          - {2 \over 405} \* (116833 - 467843 \* x)
          + {8 \over 3} \* x \* (19 \* \Hh(0,0) \* \z3
          + 72 \* \Hhh(1,1,3)
  \nonumber\\&& \mbox{}
          - 36 \* \Hhhh(1,2,0,0)
          - 6 \* \Hhhhh(0,0,0,0,0)
          - 72 \* \Hhhhh(1,1,1,0,0) )
          - \delta(1 - x) \* \biggl({1909753 \over 1944} + {416 \over 3} \* \z5
                - {25184 \over 135} \* \z2^2 
  \nonumber\\&& \mbox{}
		- {105739 \over 81} \* \z3 
	        + {248 \over 3} \* \z3^2 + {143228 \over 81} \* \z2
	        + {3512 \over 63} \* \z2^3 - 540 \* \z2 \* \z3\biggr)
          \biggr)
\:\: ,
\eea
\normalsize
where the functions $g_i(x)$ are specified in Eqs.~(\ref{eq:xn1fac1}) -- 
(\ref{eq:xn3fac1}). The gluon coefficient function reads
\small
\bea
&& c^{(3)}_{2,\rm{g}}(x) \:\: = \:\: 
        \colour4colour{\dabcNA} \* \flg11  \*  \biggl(
            {1472 \over 45} \* \gfunct1(x)
          - {64 \over 15} \* \gfunct2(x)
          + {64 \over 45} \* \gfunct3(x)
          - {128 \over 15} \* (83 + 147 \* x) \* \Hh(-3,0)
  \nonumber\\&& \mbox{}
          + {32 \over 225} \* (4801 + 16646 \* x) \* \Hh(-2,0)
          - {16 \over 225} \* (22893 + 21383 \* x) \* \Hh(-1,0)
          - {32 \over 225} \* (14698 + 3663 \* x) \* \Hh(-1,2)
  \nonumber\\&& \mbox{}
          - {64 \over 15} \* (49 - 342 \* x) \* \Hh(0,0) \* \z2
          - {32 \over 45} \* (727 - 4357 \* x) \* \Hh(0,0)
          + {64 \over 15} \* (64 - 249 \* x) \* \Hh(1,0) \* \z2
  \nonumber\\&& \mbox{}
          - {64 \over 15} \* (134 - 269 \* x) \* \Hh(1,3)
          - {128 \over 15} \* (10 - 223 \* x) \* \Hhh(-2,0,0)
          + {32 \over 225} \* (1568 + 11003 \* x) \* \Hhh(-1,-1,0)
  \nonumber\\&& \mbox{}
          - {32 \over 225} \* (8133 + 7333 \* x) \* \Hhh(-1,0,0)
          - {128 \over 75} \* (147 - 457 \* x) \* \Hhh(0,0,0)
          - 128 \* (3 + 10 \* x) \* \Hhh(1,-2,0)
  \nonumber\\&& \mbox{}
          - {64 \over 15} \* (273 - 253 \* x) \* \Hhh(1,0,0)
          - {64 \over 15} \* (153 - 298 \* x) \* \Hhh(2,0,0)
          - {128 \over 15} \* (14 + 19 \* x) \* \Hhhh(0,0,0,0)
  \nonumber\\&& \mbox{}
          + {128 \over 3} \* (7 - 2 \* x) \* \Hhhh(1,0,0,0)
          + {64 \over 15} \* (184 - 249 \* x) \* \Hhhh(1,1,0,0)
          - {16 \over 225} \* \pqg( - x) \* (6 \* (1043 + 270 \* x) \* \Hh(-1,0)
  \nonumber\\&& \mbox{}
          - 2 \* (8329 - 1764 \* x) \* \Hh(-1,2)
          + 36 \* (27 + 98 \* x) \* \Hhh(-1,0,0)
          + 7 \* (1051 - 756 \* x) \* \H(-1) \* \z2 
  \nonumber\\&& \mbox{}
          + 2 \* (9301 + 1764 \* x) \* (\Hh(-2,0)
          - \Hhh(-1,-1,0))
          - 120 \* (83 \* \Hh(-3,0)
          - 110 \* \Hh(-2,2)
          - 65 \* \Hh(-1,-1) \* \z2 
  \nonumber\\&& \mbox{}
          + 55 \* \Hh(-1,0) \* \z2 
          + 34 \* \Hhh(-2,-1,0)
          - 72 \* \Hhh(-2,0,0)
          + 65 \* \Hhh(-1,-2,0)
          + 45 \* \Hhh(1,-2,0)
          - 130 \* \Hhhh(-1,-1,-1,0)
  \nonumber\\&& \mbox{}
          + 65 \* \Hhhh(-1,-1,0,0)
          - 55 \* \Hhhh(-1,0,0,0)
          + 127 \* \H(-2) \* \z2 
          + 65 \* \H(-1) \* \z3))
          + {16 \over 225} \* \pqg(x) \* (4 \* \Hh(0,0) \* (2027 + 405 \* x)
  \nonumber\\&& \mbox{}
          + 180 \* (131 - 84 \* x) \* \Hh(1,0) \* \z2 
          - 60 \* (283 - 252 \* x) \* \Hh(1,3)
          + 3528 \* (1 + x) \* \Hhh(0,0,0)
  \nonumber\\&& \mbox{}
          - 1680 \* (2 - 9 \* x) \* \H(0) \* \z3 
          + 2 \* (10597 - 3528 \* x) \* \H(0) \* \z2 
          + 180 \* (59 + 84 \* x) \* \H(1) \* \z3 
          + (9301 - 1764 \* x) \* \H(1) \* \z2 
  \nonumber\\&& \mbox{}
          - 2 \* (12361 - 1764 \* x) \* \H(3)
          + 420 \* (11 - 36 \* x) \* (\Hh(0,0) \* \z2 
          - \H(4))
          + 540 \* (17 - 28 \* x) \* (\Hhh(2,0,0)
          + \Hhhh(1,1,0,0))
  \nonumber\\&& \mbox{}
          + 6 \* (367 + 2016 \* x) \* \z2^2
          - 35 \* (535 + 252 \* x) \* \z3 
          - 4 \* (4372 + 405 \* x) \* \z2 
          + 2 \* ( 2436 
          + 6825 \* \Hh(1,0)
  \nonumber\\&& \mbox{}
          - 8100 \* \Hh(1,0) \* \z3 
          + 13650 \* \Hh(1,1)
          - 16200 \* \Hh(1,1) \* \z3 
          + 3900 \* \Hh(1,1) \* \z2 
          - 8100 \* \Hh(1,4)
          + 8100 \* \Hh(2,0) \* \z2
  \nonumber\\&& \mbox{}
          - 8100 \* \Hh(2,3)
          + 5310 \* \Hhh(1,0,0)
          + 8100 \* \Hhh(1,0,0) \* \z2 
          + 16200 \* \Hhh(1,1,0) \* \z2 
          - 16200 \* \Hhh(1,1,3)
          - 3300 \* \Hhhh(1,0,0,0)
  \nonumber\\&& \mbox{}
          + 8100 \* \Hhhh(1,2,0,0)
          + 8100 \* \Hhhh(2,1,0,0)
          + 16200 \* \Hhhhh(1,1,1,0,0)
          + 6740 \* \H(0)
          + 7734 \* \H(1)
          - 6480 \* \H(1) \* \z2^2 
          + 7934 \* \H(2)
  \nonumber\\&& \mbox{}
          - 8100 \* \H(2) \* \z3 
          + 1020 \* \H(2) \* \z2))
          + {16 \over 225} \* \pgq( - x) \* ((53 + 45 \* x^{-1}) \* \Hh(-1,0)
          - 49 \* (1 - x^{-1}) \* (2 \* \Hh(-1,2)
  \nonumber\\&& \mbox{}
          - 2 \* \Hhh(-1,-1,0)
          + 2 \* \Hhh(-1,0,0)
          - 3 \* \H(-1) \* \z2)
          - 840 \* (\Hh(-3,0)
          + 2 \* \Hh(-2,2)
          + \Hhh(-2,0,0)
          - 2 \* \H(-2) \* \z2))
  \nonumber\\&& \mbox{}
          + {16 \over 225} \* \pgq(x) \* (7 \* (1 + x^{-1}) \* (60 \* \Hh(1,0) \* \z2 
          - 60 \* \Hh(1,3)
          + 60 \* \Hhhh(1,1,0,0)
          - 60 \* \H(1) \* \z3 
          + 7 \* \H(1) \* \z2)
          + 53 
          + 420 \* \z3 
  \nonumber\\&& \mbox{}
          + 64 \* \z2 
          + 1050 \* \z2^2 
          + 178 \* \Hh(0,0)
          + 1680 \* \Hh(0,0) \* \z2 
          - 420 \* \Hhh(1,0,0)
          - 840 \* \Hhhh(0,0,0,0)
          - 133 \* \H(0)
          + 2100 \* \H(0) \* \z3 
  \nonumber\\&& \mbox{}
          - 420 \* \H(0) \* \z2 
          + 322 \* \H(1)
          - 162 \* \H(2)
          + 420 \* \H(3)
          - 840 \* \H(4))
          - {64 \over 225} \* \pgg( - x) \* (80 \* \z2 
          - 441 \* \z2^2 
          - 420 \* \Hh(-3,0)
  \nonumber\\&& \mbox{}
          - 40 \* \Hh(0,0)
          - 210 \* \Hh(0,0) \* \z2 
          + 210 \* \Hhhh(0,0,0,0)
          + 40 \* \H(0)
          - 420 \* \H(0) \* \z3 
          - 80 \* \H(2))
          - {896 \over 75} \* \pgg(x) \* (2 \* \z2^2 
  \nonumber\\&& \mbox{}
          + 20 \* \Hh(-2,2)
          + 15 \* \Hh(0,0) \* \z2 
          + 10 \* \Hhh(-2,0,0)
          - 5 \* \Hhhh(0,0,0,0)
          - 20 \* \H(-2) \* \z2 
          + 15 \* \H(0) \* \z3 
          - 10 \* \H(4))
  \nonumber\\&& \mbox{}
          + {256 \over 15} \* (5 - 223 \* x) \* \H(-2) \* \z2
          + {16 \over 225} \* (30964 + 18329 \* x) \* \H(-1) \* \z2
          + {64 \over 15} \* (112 + 111 \* x) \* \H(0) \* \z3
  \nonumber\\&& \mbox{}
          - {208 \over 225} \* (1078 - 1657 \* x) \* \H(0)
          - {32 \over 225} \* (11227 - 7952 \* x) \* \H(0) \* \z2
          - {64 \over 15} \* (394 - 309 \* x) \* \H(1) \* \z3
  \nonumber\\&& \mbox{}
          - {32 \over 45} \* (978 + 689 \* x) \* \H(1)
          + {16 \over 225} \* (1568 - 11003 \* x) \* \H(1) \* \z2
          - {32 \over 45} \* (1618 - 603 \* x) \* \H(2)
  \nonumber\\&& \mbox{}
          + {32 \over 225} \* (12991 - 13436 \* x) \* \H(3)
          + {64 \over 15} \* (77 - 304 \* x) \* \H(4)
          + {128 \over 3} \* (5 - 2 \* x) \* (\Hh(-1,-1) \* \z2 
          - \Hhh(-1,-2,0)
  \nonumber\\&& \mbox{}
          + 2 \* \Hhhh(-1,-1,-1,0)
          - \Hhhh(-1,-1,0,0)
          - \H(-1) \* \z3)
          + {128 \over 15} \* (5 + 2 \* x) \* (15 \* \Hh(1,0) \* \z3 
          + 30 \* \Hh(1,1) \* \z3 
          - 5 \* \Hh(1,1) \* \z2 
  \nonumber\\&& \mbox{}
          + 15 \* \Hh(1,4)
          - 15 \* \Hh(2,0) \* \z2 
          + 15 \* \Hh(2,3)
          - 15 \* \Hhh(1,0,0) \* \z2 
          - 30 \* \Hhh(1,1,0) \* \z2 
          + 30 \* \Hhh(1,1,3)
          - 15 \* \Hhhh(1,2,0,0)
  \nonumber\\&& \mbox{}
          - 15 \* \Hhhh(2,1,0,0)
          - 30 \* \Hhhhh(1,1,1,0,0)
          + 12 \* \H(1) \* \z2^2 
          + 15 \* \H(2) \* \z3)
          - {128 \over 3} \* (7 + 2 \* x) \* (\Hh(-1,0) \* \z2 
          - \Hhhh(-1,0,0,0))
  \nonumber\\&& \mbox{}
          - {32 \over 3} \* (53 + 26 \* x) \* (\Hh(1,0)
          + 2 \* \Hh(1,1))
          + {32 \over 75} \* (221 + 2221 \* x) \* \z2^2
          + {112 \over 45} \* (571 - 1382 \* x) \* \z3
  \nonumber\\&& \mbox{}
          + {64 \over 225} \* (4401 - 10781 \* x) \* \z2
          - {16 \over 225} \* (5034 - 1219 \* x)
          + {256 \over 15} \* (223 \* \Hh(-2,2) \* x + 10 \* \Hhh(-2,-1,0)
          + 5 \* \H(2) \* \z2)
  \nonumber\\&& \mbox{}
          + \delta(1 - x) \* \biggl({256 \over 45} \* \z3 
            + {256 \over 45} \* \z2)
          \biggr)
  \nonumber\\&& \mbox{}
       +  \colour4colour{\nf^2 \* \biggl(\cf-{\ca \over 2}\biggr)}  \*  \biggl(
          - {32 \over 3} \* (1 - 6 \* x) \* \Hhh(1,-2,0)
          \biggr)
  \nonumber\\&& \mbox{}
       +  \colour4colour{\cf \* \nf^2}  \*  \biggl(
          - {8 \over 45} \* (1533 - 1432 \* x) \* \Hh(-2,0)
          - {32 \over 3} \* (3 - 10 \* x) \* \Hh(-2,2)
          - {4 \over 675} \* (101209 + 76209 \* x) \* \Hh(-1,0)
  \nonumber\\&& \mbox{}
          - {8 \over 45} \* (682 + 447 \* x) \* \Hh(-1,2)
          - 4 \* (43 - 140 \* x) \* \Hh(0,0) \* \z2
          + {2 \over 675} \* (325421 - 247471 \* x) \* \Hh(0,0)
  \nonumber\\&& \mbox{}
          + {2 \over 27} \* (3655 - 5424 \* x) \* \Hh(1,0)
          + {10 \over 27} \* (787 - 1098 \* x) \* \Hh(1,1)
          + {8 \over 3} \* (11 - 35 \* x) \* \Hh(1,2)
          + {4 \over 9} \* (449 + 56 \* x) \* \Hh(2,0)
  \nonumber\\&& \mbox{}
          + {4 \over 9} \* (451 + 8 \* x) \* \Hh(2,1)
          - {16 \over 3} \* (19 - 58 \* x) \* \Hhh(-2,0,0)
          + {8 \over 45} \* (872 + 827 \* x) \* \Hhh(-1,-1,0)
  \nonumber\\&& \mbox{}
          - {8 \over 45} \* (2141 + 1531 \* x) \* \Hhh(-1,0,0)
          + {2 \over 135} \* (42919 - 66224 \* x) \* \Hhh(0,0,0)
          + {8 \over 9} \* (9 - 43 \* x) \* \Hhh(1,0,0)
  \nonumber\\&& \mbox{}
          + {8 \over 3} \* (13 - 19 \* x) \* \Hhh(1,1,0)
          + {8 \over 9} \* (40 - 81 \* x) \* \Hhh(1,1,1)
          + 24 \* \Hhh(2,0,0)
          + {20 \over 3} \* (35 - 136 \* x) \* \Hhhh(0,0,0,0)
  \nonumber\\&& \mbox{}
          - {4 \over 675} \* \pqg( - x) \* (30 \* (19 + 36 \* x) \* \Hh(-2,0)
          - 2 \* (8461 - 1926 \* x) \* \Hh(-1,0)
          - 30 \* (191 - 36 \* x) \* \Hh(-1,2)
  \nonumber\\&& \mbox{}
          + 30 \* (1 - 36 \* x) \* \Hhh(-1,-1,0)
          - 60 \* (239 - 54 \* x) \* \Hhh(-1,0,0)
          + 15 \* (383 - 108 \* x) \* \H(-1) \* \z2 
          + 450 \* (24 \* \Hh(-3,0)
  \nonumber\\&& \mbox{}
          + 8 \* \Hh(-2,2)
          + 8 \* \Hh(-1,-1) \* \z2 
          - 4 \* \Hh(-1,0) \* \z2 
          + 2 \* \Hh(-1,3)
          - 12 \* \Hhh(-2,-1,0)
          + 26 \* \Hhh(-2,0,0)
          - 8 \* \Hhh(-1,-2,0)
  \nonumber\\&& \mbox{}
          - 4 \* \Hhh(-1,-1,2)
          + 4 \* \Hhh(1,-2,0)
          + 8 \* \Hhhh(-1,-1,-1,0)
          - 14 \* \Hhhh(-1,-1,0,0)
          + 6 \* \Hhhh(-1,0,0,0)
          - 14 \* \H(-2) \* \z2 
          - 7 \* \H(-1) \* \z3))
  \nonumber\\&& \mbox{}
          - {1 \over 36450} \* \pqg(x) \* (72 \* (5999 - 11556 \* x) \* \Hh(0,0)
          - 64800 \* (59 + 9 \* x) \* \Hh(1,2)
          - 64800 \* (40 - 9 \* x) \* \Hh(2,0)
  \nonumber\\&& \mbox{}
          - 25920 \* (32 + 27 \* x) \* \Hhh(0,0,0)
          - 194400 \* (14 - 3 \* x) \* \Hhh(1,1,0)
          + 19440 \* (149 + 54 \* x) \* \H(0) \* \z2 
  \nonumber\\&& \mbox{}
          + 3240 \* (1181 + 216 \* x) \* \H(1) \* \z2 
          - 19440 \* (137 + 42 \* x) \* \H(3)
          + 48600 \* (115 - 24 \* x) \* \z3 
  \nonumber\\&& \mbox{}
          + 72 \* (63421 + 11556 \* x) \* \z2 
          - 16071037 
	  + 511920 \* \z2^2
          - 972000 \* \Hh(0,0) \* \z2 
          - 6013800 \* \Hh(1,0)
  \nonumber\\&& \mbox{}
          + 972000 \* \Hh(1,0) \* \z2 
          - 5261400 \* \Hh(1,1)
          + 194400 \* \Hh(1,1) \* \z2 
          - 1166400 \* \Hh(1,3)
          - 2667600 \* \Hh(2,1)
  \nonumber\\&& \mbox{}
          - 194400 \* \Hh(2,2)
          + 194400 \* \Hh(3,0)
          + 194400 \* \Hh(3,1)
          - 3607200 \* \Hhh(1,0,0)
          - 2797200 \* \Hhh(1,1,1)
          - 583200 \* \Hhh(1,1,2)
  \nonumber\\&& \mbox{}
          - 972000 \* \Hhh(1,2,0)
          - 972000 \* \Hhh(1,2,1)
          - 194400 \* \Hhh(2,1,0)
          - 162000 \* \Hhh(2,1,1)
          + 1879200 \* \Hhhh(0,0,0,0)
  \nonumber\\&& \mbox{}
          - 907200 \* \Hhhh(1,0,0,0)
          - 388800 \* \Hhhh(1,1,0,0)
          - 583200 \* \Hhhh(1,1,1,0)
          - 550800\* \Hhhh(1,1,1,1)
          - 9442692 \* \H(0)
  \nonumber\\&& \mbox{}
          + 388800 \* \H(0) \* \z3 
          - 7615380 \* \H(1)
          + 550800 \* \H(1) \* \z3 
          - 3734280 \* \H(2)
          - 388800 \* \H(2) \* \z2 
          + 972000 \* \H(4))
  \nonumber\\&& \mbox{} 
          - {4 \over 675} \* \pgq( - x) \* (60  \* (1 - x^{-1})\* \Hh(-2,0)
          - (1503 + 127 \* x^{-1}) \* \Hh(-1,0)
          + 15 \* (21 - x^{-1}) \* (2 \* \Hh(-1,2)
  \nonumber\\&& \mbox{}
          - 2 \* \Hhh(-1,-1,0)
          + 6 \* \Hhh(-1,0,0)
          - 3 \* \H(-1) \* \z2))
          + {4 \over 18225} \* \pgq(x) \* (405 \* (61 + x^{-1}) \* \H(1) \* \z2 
          + 137536 
  \nonumber\\&& \mbox{}
          + 10800 \* \z3 
          - 44820 \* \z2 
          + 2430 \* \Hh(0,0)
          + 43200 \* \Hh(1,0)
          + 43200 \* \Hh(1,1)
          - 16200 \* \Hh(1,2)
          - 16200 \* \Hhh(1,0,0)
  \nonumber\\&& \mbox{}
          - 16200 \* \Hhh(1,1,0)
          - 16200 \* \Hhh(1,1,1)
          + 2619 \* \H(0)
          - 78210 \* \H(1)
          + 810 \* \H(2))
          + {32 \over 3} \* (5 - 14 \* x) \* \H(-2) \* \z2
  \nonumber\\&& \mbox{}
          + {4 \over 45} \* (2236 + 1721 \* x) \* \H(-1) \* \z2
          - {4 \over 9} \* (293 - 796 \* x) \* \H(0) \* \z3
          - {4 \over 45} \* (4616 - 5441 \* x) \* \H(0) \* \z2
  \nonumber\\&& \mbox{}
          + {2 \over 243} \* (180821 - 109969 \* x) \* \H(0)
          + {4 \over 45} \* (542 + 223 \* x) \* \H(1) \* \z2
          + {2 \over 81} \* (35110 - 43389 \* x) \* \H(1)
  \nonumber\\&& \mbox{}
          - {8 \over 3} \* (3 - 20 \* x) \* \H(2) \* \z2
          + {4 \over 27} \* (4021 - 2037 \* x) \* \H(2)
          + {4 \over 45} \* (4688 - 2793 \* x) \* \H(3)
          + 4 \* \H(4) \* (43 - 124 \* x)
  \nonumber\\&& \mbox{}
          - {8 \over 15} \* (1 - 2 \* x) \* (45 \* \z5 
          - 60 \* \z2 \* \z3 
          + 180 \* \Hh(-3,0)
          + 95 \* \Hh(0,0) \* \z3 
          - 30 \* \Hh(1,0) \* \z2 
          - 30 \* \Hh(1,1) \* \z2 
          + 20 \* \Hh(1,3)
  \nonumber\\&& \mbox{}
          - 30 \* \Hh(3,2)
          - 90 \* \Hh(4,0)
          - 90 \* \Hh(4,1)
          - 80 \* \Hhh(-2,-1,0)
          + 165 \* \Hhh(0,0,0) \* \z2 
          + 10 \* \Hhh(1,1,2)
          - 10 \* \Hhh(1,2,0)
          - 30 \* \Hhh(3,0,0)
  \nonumber\\&& \mbox{}
          - 30 \* \Hhh(3,1,0)
          - 30 \* \Hhh(3,1,1)
          + 30 \* \Hhhh(1,0,0,0)
          + 10 \* \Hhhh(1,1,0,0)
          - 10 \* \Hhhh(1,1,1,0)
          - 225 \* \Hhhhh(0,0,0,0,0)
          - 21 \* \H(0) \* \z2^2
  \nonumber\\&& \mbox{}
          - 5 \* \H(1) \* \z3 
          + 30 \* \H(3) \* \z2 
          - 165 \* \H(5))
          - {8 \over 3} \* (1 + 2 \* x) \* (8 \* \Hh(-1,-1) \* \z2 
          - 4 \* \Hh(-1,0) \* \z2 
          + 2 \* \Hh(-1,3)
          - 8 \* \Hhh(-1,-2,0)
  \nonumber\\&& \mbox{}
          - 4 \* \Hhh(-1,-1,2)
          + 8 \* \Hhhh(-1,-1,-1,0)
          - 14 \* \Hhhh(-1,-1,0,0)
          + 6 \* \Hhhh(-1,0,0,0)
          - 7 \* \H(-1) \* \z3)
          + {8 \over 3} \* (11 - 4 \* x) \* (\Hh(2,2)
  \nonumber\\&& \mbox{}
          + \Hhh(2,1,0)
          + \Hhh(2,1,1))
          - {4 \over 15} \* (23 + 668 \* x) \* \z2^2
          + {8 \over 3} \* (35 - 64 \* x) \* (\Hh(3,0)
          + \Hh(3,1))
          - {4 \over 9} \* (514 - 763 \* x) \* \z3
  \nonumber\\&& \mbox{}
          - {4 \over 675} \* (99933 - 983 \* x) \* \z2
          + {1 \over 540} \* (1016383 - 1276718 \* x)
          \biggr)
  \nonumber\\&& \mbox{}
       +  \colour4colour{\cf^2 \* \nf}  \*  \biggl(
          - {64 \over 3} \* (7 + 64 \* x) \* \Hh(-4,0)
          + {8 \over 15} \* (1861 - 1214 \* x) \* \Hh(-3,0)
          - 32 \* (1 + 6 \* x) \* \Hh(-3,2)
  \nonumber\\&& \mbox{}
          + {64 \over 3} \* (12 + 19 \* x) \* \Hh(-2,0) \* \z2
          + {4 \over 45} \* (31779 + 13114 \* x) \* \Hh(-2,0)
          + {32 \over 5} \* (81 - 199 \* x) \* \Hh(-2,2)
  \nonumber\\&& \mbox{}
          - 64 \* (3 + 2 \* x) \* \Hh(-2,3)
          - {4 \over 15} \* (67 - 798 \* x) \* \Hh(-1,-1) \* \z2
          - 32 \* (1 - 22 \* x) \* \Hh(-1,0) \* \z3
  \nonumber\\&& \mbox{}
          + {4 \over 15} \* (529 - 626 \* x) \* \Hh(-1,0) \* \z2
          + {2 \over 225} \* (477779 + 556129 \* x) \* \Hh(-1,0)
          + {4 \over 45} \* (11363 + 21528 \* x) \* \Hh(-1,2)
  \nonumber\\&& \mbox{}
          - {32 \over 15} \* (97 - 3 \* x) \* \Hh(-1,3)
          - 256 \* (1 - x) \* \Hh(-1,4)
          - {4 \over 3} \* (251 - 358 \* x) \* \Hh(0,0) \* \z3
          - {4 \over 3} \* (1043 - 1751 \* x) \* \Hh(0,0) \* \z2
  \nonumber\\&& \mbox{}
          + {8 \over 225} \* (16063 - 157358 \* x) \* \Hh(0,0)
          - 128 \* (1 - 12 \* x) \* \Hh(1,-2) \* \z2
          + {1 \over 3} \* (1541 + 90 \* x) \* \Hh(1,0)
  \nonumber\\&& \mbox{}
          - {8 \over 5} \* (1759 - 884 \* x) \* \Hh(1,0) \* \z2
          - {4 \over 3} \* (471 - 256 \* x) \* \Hh(1,1) \* \z2
          + {1 \over 3} \* (2119 - 150 \* x) \* \Hh(1,1)
          + {8 \over 3} \* (13 + 174 \* x) \* \Hh(1,2)
  \nonumber\\&& \mbox{}
          + {4 \over 15} \* (10307 - 4702 \* x) \* \Hh(1,3)
          + {8 \over 3} \* (9 - 1282 \* x) \* \Hh(2,0) \* \z2
          + {26 \over 3} \* (45 + 58 \* x) \* \Hh(2,0)
          - {40 \over 3} \* (11 + 18 \* x) \* \Hh(2,1) \* \z2
  \nonumber\\&& \mbox{}
          + {2 \over 3} \* (595 + 818 \* x) \* \Hh(2,1)
          + {8 \over 3} \* (123 - 74 \* x) \* \Hh(2,2)
          + {8 \over 3} \* (15 + 1178 \* x) \* \Hh(2,3)
          + {4 \over 3} \* (347 - 490 \* x) \* \Hh(3,0)
  \nonumber\\&& \mbox{}
          + {8 \over 3} \* (189 - 247 \* x) \* \Hh(3,1)
          + {16 \over 3} \* (41 - 94 \* x) \* \Hh(4,0)
          + {8 \over 3} \* (95 - 214 \* x) \* \Hh(4,1)
          - {32 \over 3} \* (3 - 122 \* x) \* \Hhh(-3,-1,0)
  \nonumber\\&& \mbox{}
          - {64 \over 3} \* (7 + 79 \* x) \* \Hhh(-3,0,0)
          - {128 \over 3} \* (1 - 20 \* x) \* \Hhh(-2,-2,0)
          - 8 \* (139 - 118 \* x) \* \Hhh(-2,-1,0)
  \nonumber\\&& \mbox{}
          + 32 \* (7 - 2 \* x) \* \Hhh(-2,-1,2)
          + {4 \over 15} \* (5627 - 3738 \* x) \* \Hhh(-2,0,0)
          - {8 \over 3} \* (311 + 318 \* x) \* \Hhh(-1,-2,0)
  \nonumber\\&& \mbox{}
          - {4 \over 45} \* (30611 + 27276 \* x) \* \Hhh(-1,-1,0)
          + {16 \over 15} \* (463 + 288 \* x) \* \Hhh(-1,-1,2)
          + 64 \* (5 - 2 \* x) \* \Hhh(-1,0,0) \* \z2
  \nonumber\\&& \mbox{}
          + {4 \over 45} \* (36119 + 41814 \* x) \* \Hhh(-1,0,0)
          + {2 \over 45} \* (8139 - 59354 \* x) \* \Hhh(0,0,0)
          + {64 \over 3} \* (3 - 82 \* x) \* \Hhh(1,-3,0)
  \nonumber\\&& \mbox{}
          + {32 \over 3} \* (51 - 125 \* x) \* \Hhh(1,-2,0)
          + 128 \* (1 - 10 \* x) \* \Hhh(1,-2,2)
          + {2 \over 15} \* (1523 + 702 \* x) \* \Hhh(1,0,0)
  \nonumber\\&& \mbox{}
          + {4 \over 3} \* (47 + 324 \* x) \* \Hhh(1,1,0)
          + {4 \over 3} \* (63 + 328 \* x) \* \Hhh(1,1,1)
          + {8 \over 3} \* (57 + 67 \* x) \* \Hhh(1,1,2)
          + {16 \over 3} \* (19 + 44 \* x) \* \Hhh(1,2,0)
  \nonumber\\&& \mbox{}
          + {4 \over 3} \* (115 + 176 \* x) \* \Hhh(1,2,1)
          + {128 \over 3} \* (4 - 13 \* x) \* \Hhh(2,-2,0)
          - {16 \over 15} \* (309 - 2519 \* x) \* \Hhh(2,0,0)
  \nonumber\\&& \mbox{}
          + {8 \over 3} \* (117 - 88 \* x) \* \Hhh(2,1,0)
          + {4 \over 3} \* (239 - 152 \* x) \* \Hhh(2,1,1)
          + {8 \over 3} \* (41 - 74 \* x) \* \Hhh(2,1,2)
          + {64 \over 3} \* (5 - 11 \* x) \* \Hhh(2,2,0)
  \nonumber\\&& \mbox{}
          + {16 \over 3} \* (35 - 6 \* x) \* \Hhh(3,0,0)
          + {32 \over 3} \* (7 - 82 \* x) \* \Hhhh(-2,-1,-1,0)
          + {32 \over 3} \* (9 + 122 \* x) \* \Hhhh(-2,-1,0,0)
  \nonumber\\&& \mbox{}
          - {112 \over 3} \* (5 + 26 \* x) \* \Hhhh(-2,0,0,0)
          + 8 \* (119 + 130 \* x) \* \Hhhh(-1,-1,-1,0)
          - {4 \over 15} \* (3569 + 4074 \* x) \* \Hhhh(-1,-1,0,0)
  \nonumber\\&& \mbox{}
          + {96 \over 5} \* (26 + 41 \* x) \* \Hhhh(-1,0,0,0)
          + {4 \over 15} \* (2067 - 4007 \* x) \* \Hhhh(0,0,0,0)
          + 512 \* x \* \Hhhh(1,-2,-1,0)
  \nonumber\\&& \mbox{}
          + {64 \over 3} \* (3 - 86 \* x) \* \Hhhh(1,-2,0,0)
          + {4 \over 15} \* (1887 - 1252 \* x) \* \Hhhh(1,0,0,0)
          + {128 \over 3} \* (3 - 26 \* x) \* \Hhhh(1,1,-2,0)
  \nonumber\\&& \mbox{}
          - {16 \over 5} \* (621 - 346 \* x) \* \Hhhh(1,1,0,0)
          + {8 \over 3} \* (48 + 83 \* x) \* \Hhhh(1,1,1,0)
          + {4 \over 3} \* (115 + 148 \* x) \* \Hhhh(1,1,1,1)
  \nonumber\\&& \mbox{}
          + {8 \over 3} \* (15 + 194 \* x) \* \Hhhh(2,0,0,0)
          + {8 \over 3} \* (89 - 1170 \* x) \* \Hhhh(2,1,0,0)
          + {8 \over 3} \* (35 - 78 \* x) \* \Hhhh(2,1,1,0)
  \nonumber\\&& \mbox{}
          + {4 \over 3} \* (135 - 406 \* x) \* \Hhhhh(0,0,0,0,0)
          + {2 \over 75} \* \pqg( - x) \* (20 \* (319 - 144 \* x) \* \Hh(-3,0)
          - 30 \* (271 + 88 \* x) \* \Hh(-2,0)
  \nonumber\\&& \mbox{}
          + 360 \* (11 - 16 \* x) \* \Hh(-2,2)
          + 90 \* (9 - 64 \* x) \* \Hh(-1,-1) \* \z2 
          - 90 \* (23 - 48 \* x) \* \Hh(-1,0) \* \z2 
  \nonumber\\&& \mbox{}
          + 2 \* (13843 - 3198 \* x) \* \Hh(-1,0)
          + 30 \* (1361 - 88 \* x) \* \Hh(-1,2)
          + 240 \* (17 - 12 \* x) \* \Hh(-1,3)
  \nonumber\\&& \mbox{}
          + 30 \* (367 + 88 \* x) \* \Hhh(-1,-1,0)
          + 120 \* (7 + 48 \* x) \* \Hhh(-1,-1,2)
          + 10 \* (2339 - 552 \* x) \* \Hhh(-1,0,0)
  \nonumber\\&& \mbox{}
          - 30 \* (107 - 192 \* x) \* \H(-2) \* \z2 
          - 30 \* (79 - 144 \* x) \* \H(-1) \* \z3 
          - 45 \* (785 - 88 \* x) \* \H(-1) \* \z2 
  \nonumber\\&& \mbox{}
          + 10 \* (113 - 288 \* x) \* (\Hhh(-2,0,0)
          - \Hhhh(-1,-1,0,0))
          + 40 \* (131 - 36 \* x) \* (2 \* \Hhh(1,-2,0)
          + \Hhhh(-1,0,0,0))
  \nonumber\\&& \mbox{}
          + 20 \* (1280 \* \Hh(-4,0)
          + 480 \* \Hh(-3,2)
          - 40 \* \Hh(-2,-1) \* \z2 
          - 180 \* \Hh(-2,0) \* \z2 
          + 180 \* \Hh(-2,3)
          + 220 \* \Hh(-1,-2) \* \z2 
  \nonumber\\&& \mbox{}
          + 110 \* \Hh(-1,-1) \* \z3 
          + 1250 \* \Hh(-1,0) \* \z3
          + 120 \* \Hh(-1,2) \* \z2 
          + 1200 \* \Hh(-1,4)
          - 600 \* \Hh(1,-2) \* \z2 
          - 680 \* \Hhh(-3,-1,0)
  \nonumber\\&& \mbox{}
          + 1460 \* \Hhh(-3,0,0)
          - 400 \* \Hhh(-2,-2,0)
          + 75 \* \Hhh(-2,-1,0)
          + 240 \* \Hhh(-2,-1,2)
          + 120 \* \Hhh(-2,2,0)
          + 120 \* \Hhh(-2,2,1)
  \nonumber\\&& \mbox{}
          - 520 \* \Hhh(-1,-3,0)
          - 35 \* \Hhh(-1,-2,0)
          + 40 \* \Hhh(-1,-1,-1) \* \z2 
          + 180 \* \Hhh(-1,-1,0) \* \z2 
          - 180 \* \Hhh(-1,-1,3)
  \nonumber\\&& \mbox{}
          - 1080 \* \Hhh(-1,0,0) \* \z2
          + 90 \* \Hhh(-1,2,0)
          + 90 \* \Hhh(-1,2,1)
          + 60 \* \Hhh(-1,3,0)
          + 60 \* \Hhh(-1,3,1)
          + 760 \* \Hhh(1,-3,0)
  \nonumber\\&& \mbox{}
          + 480 \* \Hhh(1,-2,2)
          + 400 \* \Hhh(2,-2,0)
          + 400 \* \Hhhh(-2,-1,-1,0)
          - 600 \* \Hhhh(-2,-1,0,0)
          + 720 \* \Hhhh(-2,0,0,0)
  \nonumber\\&& \mbox{}
          + 440 \* \Hhhh(-1,-2,-1,0)
          - 660 \* \Hhhh(-1,-2,0,0)
          + 400 \* \Hhhh(-1,-1,-2,0)
          + 165 \* \Hhhh(-1,-1,-1,0)
          - 240 \* \Hhhh(-1,-1,-1,2)
  \nonumber\\&& \mbox{}
          - 120 \* \Hhhh(-1,-1,2,0)
          - 120 \* \Hhhh(-1,-1,2,1)
          - 1220 \* \Hhhh(-1,2,0,0)
          - 240 \* \Hhhh(1,-2,-1,0)
          + 800 \* \Hhhh(1,-2,0,0)
  \nonumber\\&& \mbox{}
          + 400 \* \Hhhh(1,1,-2,0)
          - 400 \* \Hhhhh(-1,-1,-1,-1,0)
          + 600 \* \Hhhhh(-1,-1,-1,0,0)
          - 720 \* \Hhhhh(-1,-1,0,0,0)
          + 50 \* \Hhhhh(-1,0,0,0,0)
  \nonumber\\&& \mbox{}
          - 820 \* \H(-3) \* \z2 
          - 110 \* \H(-2) \* \z3 
          + 788 \* \H(-1) \* \z2^2))
          + {1 \over 75} \* \pqg(x) \* 
            (200 \* (557 + 36 \* x) \* \Hh(0,0) \* \z2 
  \nonumber\\&& \mbox{}
          - 8 \* (6226 - 1599 \* x) \* \Hh(0,0)
          + 20 \* (5747 + 792 \* x) \* \Hh(1,0) \* \z2 
          - 40 \* (2773 + 468 \* x) \* \Hh(1,3)
  \nonumber\\&& \mbox{}           
          - 60 \* (521 - 184 \* x) \* \Hhh(0,0,0)
          - 80 \* (503 - 72 \* x) \* \Hhhh(0,0,0,0)
          - 960 \* (37 - 3 \* x) \* \Hhhh(1,0,0,0)
  \nonumber\\&& \mbox{}
          - 20 \* (121 + 1656 \* x) \* \H(0) \* \z3 
          + 30 \* (3027 - 352 \* x) \* \H(0) \* \z2 
          + 40 \* (133 - 612 \* x) \* \H(1) \* \z3 
  \nonumber\\&& \mbox{}
          + 240 \* (204 - 11 \* x) \* \H(1) \* \z2 
          - 30 \* (3203 - 176 \* x) \* \H(3)
          - 40 \* (2929 + 324 \* x) \* \H(4)
          - 10 \* (331 + 1320 \* x) \* \z3 
  \nonumber\\&& \mbox{}
          + 20 \* (1411 + 936 \* x) \* (\Hhh(2,0,0)
          + \Hhhh(1,1,0,0))
          - 8 \* (8681 + 2376 \* x) \* \z2^2 
          + 2 \* (11659 - 6396 \* x) \* \z2 
  \nonumber\\&& \mbox{}
          - 57793 
          + 13800 \* \z5 
          - 31200 \* \z2 \* \z3 
          + 46400 \* \Hh(0,0) \* \z3 
          - 44500 \* \Hh(1,0)
          + 53200 \* \Hh(1,0) \* \z3 
          - 54350 \* \Hh(1,1)
  \nonumber\\&& \mbox{}
          + 44000 \* \Hh(1,1) \* \z3 
          + 44000 \* \Hh(1,1) \* \z2 
          - 37950 \* \Hh(1,2)
          + 23600 \* \Hh(1,2) \* \z2 
          - 27600 \* \Hh(1,4)
          - 40850 \* \Hh(2,0)
  \nonumber\\&& \mbox{}
          - 30000 \* \Hh(2,0) \* \z2 
          - 45450 \* \Hh(2,1)
          + 20400 \* \Hh(2,1) \* \z2 
          - 47300 \* \Hh(2,2)
          + 30000 \* \Hh(2,3)
          - 48000 \* \Hh(3,0)
  \nonumber\\&& \mbox{}
          - 56400 \* \Hh(3,1)
          - 32400 \* \Hh(3,2)
          - 31600 \* \Hh(4,0)
          - 39200 \* \Hh(4,1)
          + 32800 \* \Hhh(0,0,0) \* \z2 
          - 19520 \* \Hhh(1,0,0)
  \nonumber\\&& \mbox{}
          + 32400 \* \Hhh(1,0,0) \* \z2 
          - 40850 \* \Hhh(1,1,0)
          + 12800 \* \Hhh(1,1,0) \* \z2 
          - 45450 \* \Hhh(1,1,1)
          + 20400 \* \Hhh(1,1,1) \* \z2 
  \nonumber\\&& \mbox{}
          - 47300 \* \Hhh(1,1,2)
          - 12800 \* \Hhh(1,1,3)
          - 44400 \* \Hhh(1,2,0)
          - 52800 \* \Hhh(1,2,1)
          - 32400 \* \Hhh(1,2,2)
          - 29200 \* \Hhh(1,3,0)
  \nonumber\\&& \mbox{}
          - 37200 \* \Hhh(1,3,1)
          - 47500 \* \Hhh(2,1,0)
          - 51200 \* \Hhh(2,1,1)
          - 28400 \* \Hhh(2,1,2)
          - 25600 \* \Hhh(2,2,0)
          - 32800 \* \Hhh(2,2,1)
  \nonumber\\&& \mbox{}
          - 19600 \* \Hhh(3,0,0)
          - 30000 \* \Hhh(3,1,0)
          - 35200 \* \Hhh(3,1,1)
          - 47500 \* \Hhhh(1,1,1,0)
          - 51200 \* \Hhhh(1,1,1,1)
          - 28400 \* \Hhhh(1,1,1,2)
  \nonumber\\&& \mbox{}
          - 25600 \* \Hhhh(1,1,2,0)
          - 33200 \* \Hhhh(1,1,2,1)
          - 25600 \* \Hhhh(1,2,0,0)
          - 29600 \* \Hhhh(1,2,1,0)
          - 35200 \* \Hhhh(1,2,1,1)
  \nonumber\\&& \mbox{}
          - 11600 \* \Hhhh(2,0,0,0)
          - 77600 \* \Hhhh(2,1,0,0)
          - 27200 \* \Hhhh(2,1,1,0)
          - 30000 \* \Hhhh(2,1,1,1)
          - 18000 \* \Hhhhh(0,0,0,0,0)
  \nonumber\\&& \mbox{}
          - 16000 \* \Hhhhh(1,0,0,0,0)
          - 11600 \* \Hhhhh(1,1,0,0,0)
          - 34800 \* \Hhhhh(1,1,1,0,0)
          - 26800 \* \Hhhhh(1,1,1,1,0)
          - 30000 \* \Hhhhh(1,1,1,1,1)
  \nonumber\\&& \mbox{}
          - 92853 \* \H(0)
          - 3360 \* \H(0) \* \z2^2 
          - 111885 \* \H(1)
          - 5840 \* \H(1) \* \z2^2 
          - 36110 \* \H(2)
          + 87200 \* \H(2) \* \z3 
          + 48800 \* \H(2) \* \z2 
  \nonumber\\&& \mbox{}
          + 18800 \* \H(3) \* \z2 
          - 32800 \* \H(5))  
          + {2 \over 225} \* \pgq( - x) \* (240 \* x^{-1} \* \Hh(-2,0) 
          - (483 - 463 \* x^{-1}) \* \Hh(-1,0)
  \nonumber\\&& \mbox{}
          + 20 \* (13 + 11 \* x^{-1}) \* \Hh(-1,2)
          - 20 \* (11 - 23 \* x^{-1}) \* \Hhh(-1,0,0)
          - 30 \* (5 + 11 \* x^{-1}) \* \H(-1) \* \z2
  \nonumber\\&& \mbox{}
          - 20 \* (1 - x^{-1}) \* (24 \* \Hh(-1,-1) \* \z2 
          - 18 \* \Hh(-1,0) \* \z2
          + 12 \* \Hh(-1,3)
          - 11 \* \Hhh(-1,-1,0)
          - 24 \* \Hhh(-1,-1,2)
          - 12 \* \Hhh(1,-2,0)
  \nonumber\\&& \mbox{}
          - 12 \* \Hhhh(-1,-1,0,0)
          + 6 \* \Hhhh(-1,0,0,0)
          - 18 \* \H(-1) \* \z3 ))
          - {2 \over 225} \* \pgq(x) \* (10 \* (1 + x^{-1}) \* 
            (66 \* \Hh(1,0) \* \z2 
          - 78 \* \Hh(1,3)
  \nonumber\\&& \mbox{}
          + 12 \* \Hhhh(1,0,0,0)
          + 78 \* \Hhhh(1,1,0,0)
          - 102 \* \H(1) \* \z3
          - 11 \* \H(1) \* \z2)
          + 723 
	  + 1380 \* \z3 
          + 740 \* \z2 
          - 220 \* \Hh(0,0)
  \nonumber\\&& \mbox{}
          - 240 \* \Hhh(0,0,0)
          - 780 \* \Hhh(1,0,0)
          - 483 \* \H(0)
          - 300 \* \H(0) \* \z2 
          + 760 \* \H(1)
          - 760 \* \H(2)
          + 540 \* \H(3))
  \nonumber\\&& \mbox{}
          + {16 \over 3} \* (3 + 158 \* x) \* \H(-3) \* \z2
          + {16 \over 3} \* (29 + 82 \* x) \* \H(-2) \* \z3
          - {4 \over 5} \* (1343 - 2182 \* x) \* \H(-2) \* \z2
  \nonumber\\&& \mbox{}
          - {32 \over 5} \* (17 - 62 \* x) \* \H(-1) \* \z2^2
          - {4 \over 5} \* (147 + 422 \* x) \* \H(-1) \* \z3
          - {2 \over 15} \* (17779 + 23444 \* x) \* \H(-1) \* \z2
  \nonumber\\&& \mbox{}
          + {4 \over 15} \* (197 + 1390 \* x) \* \H(0) \* \z2^2
          + {4 \over 15} \* (1246 - 13941 \* x) \* \H(0) \* \z3
          - {1 \over 45} \* (49713 - 188198 \* x) \* \H(0) \* \z2
  \nonumber\\&& \mbox{}
          + {1 \over 45} \* (58817 - 89022 \* x) \* \H(0)
          + {8 \over 15} \* (2954 - 1069 \* x) \* \H(1) \* \z3
          - {2 \over 45} \* (31391 - 16836 \* x) \* \H(1) \* \z2
  \nonumber\\&& \mbox{}
          + {1 \over 45} \* (99429 - 35804 \* x) \* \H(1)
          - 16 \* (7 - 142 \* x) \* \H(2) \* \z3
          - {4 \over 3} \* (663 + 206 \* x) \* \H(2) \* \z2
  \nonumber\\&& \mbox{}
          + {1 \over 45} \* (26872 + 70787 \* x) \* \H(2)
          - {16 \over 3} \* (37 + 54 \* x) \* \H(3) \* \z2
          + {1 \over 45} \* (52689 - 161494 \* x) \* \H(3)
  \nonumber\\&& \mbox{}
          + {4 \over 15} \* (5503 - 10483 \* x) \* \H(4)
          + {8 \over 15} \* (1 - 2 \* x) \* (480 \* \Hh(1,0) \* \z3 
          + 600 \* \Hh(1,1) \* \z3 
          + 60 \* \Hh(1,4)
          + 340 \* \Hh(3,2)
  \nonumber\\&& \mbox{}
          + 60 \* \Hhh(1,0,0) \* \z2 
          - 240 \* \Hhh(1,1,0) \* \z2
          + 360 \* \Hhh(1,1,3)
          + 215 \* \Hhh(2,2,1)
          + 310 \* \Hhh(3,1,0)
          + 335 \* \Hhh(3,1,1)
          + 360 \* \Hhhh(-1,2,0,0)
  \nonumber\\&& \mbox{}
          - 180 \* \Hhhh(1,2,0,0)
          + 185 \* \Hhhh(2,1,1,1)
          - 120 \* \Hhhhh(1,0,0,0,0)
          - 120 \* \Hhhhh(1,1,0,0,0)
          - 360 \* \Hhhhh(1,1,1,0,0)
          + 228 \* \H(1) \* \z2^2)
  \nonumber\\&& \mbox{}
          - {8 \over 3} \* (1 + 2 \* x) \* (70 \* \Hh(-2,-1) \* \z2 
          + 48 \* \Hh(-1,-2) \* \z2 
          + 72 \* \Hh(-1,-1) \* \z3 
          + 12 \* \Hhh(-2,2,0)
          + 12 \* \Hhh(-2,2,1)
          - 24 \* \Hhh(-1,-3,0)
  \nonumber\\&& \mbox{}
          - 48 \* \Hhh(-1,-2,2)
          - 96 \* \Hhh(-1,-1,-1) \* \z2 
          + 72 \* \Hhh(-1,-1,0) \* \z2 
          - 48 \* \Hhh(-1,-1,3)
          - 24 \* \Hhhh(-1,-2,0,0)
          + 96 \* \Hhhh(-1,-1,-1,2)
  \nonumber\\&& \mbox{}
          + 48 \* \Hhhhh(-1,-1,-1,0,0)
          - 24 \* \Hhhhh(-1,-1,0,0,0)
          + 24 \* \Hhhhh(-1,0,0,0,0)
          + 9 \* \H(0) \* \z4)
          + 16 \* (1 + 4 \* x) \* (\Hhh(-1,2,0)
          + \Hhh(-1,2,1))
  \nonumber\\&& \mbox{}
          + {64 \over 3} \* (12 - 35 \* x) \* \z2 \* \z3
          - {4 \over 3} \* (197 - 490 \* x) \* (\Hhh(0,0,0) \* \z2 - \H(5))
          - {4 \over 3} \* (231 + 5482 \* x) \* \z5
  \nonumber\\&& \mbox{}
          + {4 \over 75} \* (18722 - 26587 \* x) \* \z2^2
          - {1 \over 75} \* (31968 - 338243 \* x) \* \z2
          + {1 \over 45} \* (34203 + 272182 \* x) \* \z3
  \nonumber\\&& \mbox{}
          + {1 \over 120} \* (120731 - 24406 \* x)
          \biggr)
  \nonumber\\&& \mbox{}
       +  \colour4colour{\ca \* \nf^2}  \*  \biggl(
            {16 \over 3} \* (3 + 2 \* x) \* \Hh(-3,0)
          + {8 \over 9} \* (59 + 51 \* x) \* \Hh(-2,0)
          + {2 \over 135} \* (6272 + 5877 \* x) \* \Hh(-1,0)
  \nonumber\\&& \mbox{}
          + {8 \over 3} \* (7 + 6 \* x) \* \Hh(-1,2)
          - {2 \over 405} \* (29127 + 111148 \* x) \* \Hh(0,0)
          - {4 \over 9} \* (305 - 46 \* x) \* \Hh(1,0)
          - {8 \over 27} \* (461 - 75 \* x) \* \Hh(1,1)
  \nonumber\\&& \mbox{}
          - {4 \over 9} \* (43 - 18 \* x) \* \Hh(1,2)
          - {4 \over 9} \* (99 + 179 \* x) \* \Hh(2,0)
          - {4 \over 9} \* (99 + 184 \* x) \* \Hh(2,1)
          - {64 \over 9} \* (5 + 7 \* x) \* \Hhh(-1,-1,0)
  \nonumber\\&& \mbox{}
          + {4 \over 9} \* (149 + 146 \* x) \* \Hhh(-1,0,0)
          - {16 \over 27} \* (92 + 525 \* x) \* \Hhh(0,0,0)
          - {16 \over 9} \* (13 + 5 \* x) \* \Hhh(1,0,0)
          - {4 \over 9} \* (95 - 18 \* x) \* \Hhh(1,1,0)
  \nonumber\\&& \mbox{}
          - {4 \over 9} \* (71 - 18 \* x) \* \Hhh(1,1,1)
          - {8 \over 3} \* (1 + 10 \* x) \* \Hhh(2,0,0)
          - {8 \over 9} \* (1 + 102 \* x) \* \Hhhh(0,0,0,0)
  \nonumber\\&& \mbox{}
          - {2 \over 405} \* \pqg( - x) \* ((5323 + 972 \* x) \* \Hh(-1,0)
          + 30 \* ( 36 \* \Hh(-3,0)
          + 102 \* \Hh(-2,0)
          + 36 \* \Hh(-2,2)
          - 18 \* \Hh(-1,-1) \* \z2 
  \nonumber\\&& \mbox{}
          - 90 \* \Hh(-1,0) \* \z2 
          + 102 \* \Hh(-1,2)
          + 90 \* \Hh(-1,3)
          + 90 \* \Hhh(-2,0,0)
          + 36 \* \Hhh(-1,-2,0)
          + 12 \* \Hhh(-1,-1,0)
          + 221 \* \Hhh(-1,0,0)
  \nonumber\\&& \mbox{}
          + 36 \* \Hhh(-1,2,0)
          + 36 \* \Hhh(-1,2,1)
          - 36 \* \Hhh(1,-2,0)
          - 36 \* \Hhhh(-1,-1,-1,0)
          + 18 \* \Hhhh(-1,-1,0,0)
          + 138 \* \Hhhh(-1,0,0,0)
  \nonumber\\&& \mbox{}
          - 36 \* \H(-2) \* \z2
          - 9 \* \H(-1) \* \z3 
          - 96 \* \H(-1) \* \z2))
          + {1 \over 7290} \* \pqg(x) \* (36 \* (26407 + 972 \* x) \* \Hh(0,0)
  \nonumber\\&& \mbox{}
          - 17496 \* (77 + 2 \* x) \* \z2 
          + 4708987 
          - 403920 \* \z3 
          + 25272 \* \z2^2 
          - 38880 \* \Hh(0,0) \* \z2 
          + 1031400 \* \Hh(1,0)
  \nonumber\\&& \mbox{}
          + 32400 \* \Hh(1,0) \* \z2 
          + 1052640 \* \Hh(1,1)
          - 71280 \* \Hh(1,1) \* \z2 
          + 115560 \* \Hh(1,2)
          - 32400 \* \Hh(1,3)
          + 340200 \* \Hh(2,0)
  \nonumber\\&& \mbox{}
          + 340200 \* \Hh(2,1)
          + 38880 \* \Hh(3,0)
          + 38880 \* \Hh(3,1)
          + 253800 \* \Hhh(0,0,0)
          + 328320 \* \Hhh(1,0,0)
          + 448200 \* \Hhh(1,1,0)
  \nonumber\\&& \mbox{}
          + 266760 \* \Hhh(1,1,1)
          + 51840 \* \Hhh(1,1,2)
          + 51840 \* \Hhh(1,2,0)
          - 6480 \* \Hhh(1,2,1)
          + 58320 \* \Hhh(2,0,0)
          + 77760 \* \Hhh(2,1,0)
  \nonumber\\&& \mbox{}
          + 38880 \* \Hhh(2,1,1)
          + 77760 \* \Hhhh(1,0,0,0)
          + 129600 \* \Hhhh(1,1,0,0)
          + 64800 \* \Hhhh(1,1,1,0)
          + 45360 \* \Hhhh(1,1,1,1)
  \nonumber\\&& \mbox{}
          + 2569392 \* \H(0)
          - 58320 \* \H(0) \* \z3 
          - 385560 \* \H(0) \* \z2 
          + 2855640 \* \H(1)
          + 68040 \* \H(1) \* \z3 
          - 122040 \* \H(1) \* \z2 
  \nonumber\\&& \mbox{}
          + 1312200 \* \H(2)
          + 385560 \* \H(3)
          + 38880 \* \H(4))
          - {2 \over 135} \* \pgq( - x) \* ((209 - 9 \* x^{-1}) \* \Hh(-1,0)
          + 60 \* (2 \* \Hh(-1,2)
  \nonumber\\&& \mbox{}
          - 2 \* \Hhh(-1,-1,0)
          + 6 \* \Hhh(-1,0,0)
          - 3 \* \H(-1) \* \z2 ))
          - {2 \over 3645} \* \pgq(x) \* (958 
          - 2160 \* \z3 
          - 5400 \* \z2 
          + 5400 \* \Hh(1,0)
  \nonumber\\&& \mbox{}
          + 5400 \* \Hh(1,1)
          + 3240 \* \Hh(1,2)
          + 3240 \* \Hhh(1,0,0)
          + 3240 \* \Hhh(1,1,0)
          + 3240 \* \Hhh(1,1,1)
          - 243 \* \H(0)
          - 2700 \* \H(1)
  \nonumber\\&& \mbox{}
          - 4860 \* \H(1) \* \z2 )
          - {8 \over 9} \* (41 + 46 \* x) \* \H(-1) \* \z2
          + {8 \over 9} \* (1 + 70 \* x) \* \H(0) \* \z3
          + {4 \over 9} \* (126 + 461 \* x) \* \H(0) \* \z2
  \nonumber\\&& \mbox{}
          - {2 \over 243} \* (48632 + 50747 \* x) \* \H(0)
          + {4 \over 9} \* (3 + 38 \* x) \* \H(1) \* \z2
          - {2 \over 81} \* (15601 - 1476 \* x) \* \H(1)
          + 32 \* x \* \H(2) \* \z2
  \nonumber\\&& \mbox{}
          - {76 \over 27} \* (60 + 73 \* x) \* \H(2)
          - {28 \over 9} \* (18 + 61 \* x) \* \H(3)
          - {4 \over 3} \* (1 - 2 \* x) \* (6 \* \Hh(1,0) \* \z2 
          + 6 \* \Hh(1,1) \* \z2 
          - 4 \* \Hh(1,3)
          - 2 \* \Hhh(1,1,2)
  \nonumber\\&& \mbox{}
          + 2 \* \Hhh(1,2,0)
          - 6 \* \Hhhh(1,0,0,0)
          - 2 \* \Hhhh(1,1,0,0)
          + 2 \* \Hhhh(1,1,1,0)
          + \H(1) \* \z3)
          + {4 \over 3} \* (1 + 2 \* x) \* (4 \* \Hh(-2,2)
          + 8 \* \Hh(-1,-1) \* \z2 
  \nonumber\\&& \mbox{}
          - 4 \* \Hh(-1,0) \* \z2 
          + 2 \* \Hh(-1,3)
          - 8 \* \Hhh(-2,-1,0)
          + 14 \* \Hhh(-2,0,0)
          - 8 \* \Hhh(-1,-2,0)
          - 4 \* \Hhh(-1,-1,2)
          + 8 \* \Hhhh(-1,-1,-1,0)
  \nonumber\\&& \mbox{}
          - 14 \* \Hhhh(-1,-1,0,0)
          + 6 \* \Hhhh(-1,0,0,0)
          - 8 \* \H(-2) \* \z2 
          - 7 \* \H(-1) \* \z3)
          - {16 \over 3} \* (1 + 4 \* x) \* (\Hh(2,2)
          + \Hhh(2,1,0)
          + \Hhh(2,1,1))
  \nonumber\\&& \mbox{}
          - {16 \over 3} \* (1 + 8 \* x) \* (\Hh(3,0)
          + \Hh(3,1))
          + {8 \over 3} \* (1 + 26 \* x) \* (\Hh(0,0) \* \z2 - \H(4))
          + {4 \over 45} \* (3 - 94 \* x) \* \z2^2
          + {4 \over 9} \* (63 + 440 \* x) \* \z3
  \nonumber\\&& \mbox{}
          + {4 \over 405} \* (18186 + 23339 \* x) \* \z2
          - {1 \over 162} \* (115221 - 17222 \* x)
          \biggr)
  \nonumber\\&& \mbox{}
       +  \colour4colour{\ca \* \cf \* \nf}  \*  \biggl(
          - {640 \over 3} \* (2 - 3 \* x) \* \Hh(-4,0)
          - {8 \over 15} \* (3237 + 2372 \* x) \* \Hh(-3,0)
          - {16 \over 3} \* (29 + 250 \* x) \* \Hh(-3,2)
  \nonumber\\&& \mbox{}
          + 16 \* (31 - 14 \* x) \* \Hh(-2,-1) \* \z2
          - {8 \over 3} \* (179 - 222 \* x) \* \Hh(-2,0) \* \z2
          - {2 \over 225} \* (293407 + 126682 \* x) \* \Hh(-2,0)
  \nonumber\\&& \mbox{}
          - {8 \over 15} \* (71 - 3734 \* x) \* \Hh(-2,2)
          + {8 \over 3} \* (173 - 382 \* x) \* \Hh(-2,3)
          + {4 \over 15} \* (7259 + 7864 \* x) \* \Hh(-1,-1) \* \z2
  \nonumber\\&& \mbox{}
          - 32 \* (7 + 38 \* x) \* \Hh(-1,0) \* \z3
          - {16 \over 15} \* (1924 + 1719 \* x) \* \Hh(-1,0) \* \z2
          - {2 \over 675} \* (1213559 + 1602219 \* x) \* \Hh(-1,0)
  \nonumber\\&& \mbox{}
          - {2 \over 225} \* (7549 + 134124 \* x) \* \Hh(-1,2)
          + {28 \over 3} \* (211 + 182 \* x) \* \Hh(-1,3)
          + 64 \* (7 + 2 \* x) \* \Hh(-1,4)
  \nonumber\\&& \mbox{}
          - 16 \* (9 - 47 \* x) \* \Hh(0,0) \* \z3
          - {4 \over 45} \* (7507 - 18512 \* x) \* \Hh(0,0) \* \z2
          - {1 \over 2025} \* (30724 - 6952739 \* x) \* \Hh(0,0)
  \nonumber\\&& \mbox{}
          + 32 \* (5 - 32 \* x) \* \Hh(1,-2) \* \z2
          + {8 \over 15} \* (5234 - 2969 \* x) \* \Hh(1,0) \* \z2
          - {2 \over 27} \* (7657 - 1173 \* x) \* \Hh(1,0)
  \nonumber\\&& \mbox{}
          + {8 \over 15} \* (363 - 268 \* x) \* \Hh(1,1) \* \z2
          - {4 \over 27} \* (5857 - 90 \* x) \* \Hh(1,1)
          + {1 \over 9} \* (3257 - 1862 \* x) \* \Hh(1,2)
  \nonumber\\&& \mbox{}
          - {4 \over 5} \* (3593 - 2148 \* x) \* \Hh(1,3)
          + 24 \* (1 + 166 \* x) \* \Hh(2,0) \* \z2
          + {5 \over 27} \* (3697 - 9566 \* x) \* \Hh(2,0)
  \nonumber\\&& \mbox{}
          - {8 \over 3} \* (13 - 118 \* x) \* \Hh(2,1) \* \z2
          + {4 \over 27} \* (4178 - 11347 \* x) \* \Hh(2,1)
          + {4 \over 3} \* (593 - 554 \* x) \* \Hh(2,2)
          - {8 \over 3} \* (15 + 1334 \* x) \* \Hh(2,3)
  \nonumber\\&& \mbox{}
          + {4 \over 45} \* (11527 - 15272 \* x) \* \Hh(3,0)
          + {4 \over 45} \* (12227 - 18352 \* x) \* \Hh(3,1)
          + {8 \over 3} \* (85 + 226 \* x) \* \Hh(3,2)
  \nonumber\\&& \mbox{}
          + {8 \over 3} \* (89 + 226 \* x) \* \Hh(4,0)
          + {8 \over 3} \* (97 + 258 \* x) \* \Hh(4,1)
          + {16 \over 3} \* (61 - 166 \* x) \* \Hhh(-3,-1,0)
          - {128 \over 3} \* (11 - 5 \* x) \* \Hhh(-3,0,0)
  \nonumber\\&& \mbox{}
          + {32 \over 3} \* (19 - 126 \* x) \* \Hhh(-2,-2,0)
          + {64 \over 15} \* (268 - 7 \* x) \* \Hhh(-2,-1,0)
          - {16 \over 3} \* (109 - 182 \* x) \* \Hhh(-2,-1,2)
  \nonumber\\&& \mbox{}
          - {4 \over 15} \* (5837 - 318 \* x) \* \Hhh(-2,0,0)
          + {16 \over 3} \* (13 - 54 \* x) \* \Hhh(-2,2,0)
          + {64 \over 3} \* (3 - 16 \* x) \* \Hhh(-2,2,1)
  \nonumber\\&& \mbox{}
          + {8 \over 15} \* (1831 + 896 \* x) \* \Hhh(-1,-2,0)
          + {2 \over 225} \* (206879 + 171404 \* x) \* \Hhh(-1,-1,0)
          - {16 \over 3} \* (472 + 451 \* x) \* \Hhh(-1,-1,2)
  \nonumber\\&& \mbox{}
          - 192 \* (3 + 2 \* x) \* \Hhh(-1,0,0) \* \z2
          - {4 \over 225} \* (107627 + 155552 \* x) \* \Hhh(-1,0,0)
          + {8 \over 15} \* (599 + 494 \* x) \* \Hhh(-1,2,0)
  \nonumber\\&& \mbox{}
          + {8 \over 15} \* (659 + 584 \* x) \* \Hhh(-1,2,1)
          - {32 \over 3} \* (23 + 56 \* x) \* \Hhh(0,0,0) \* \z2
          + {1 \over 675} \* (943193 - 480718 \* x) \* \Hhh(0,0,0)
  \nonumber\\&& \mbox{}
          - {64 \over 3} \* (3 - 49 \* x) \* \Hhh(1,-3,0)
          - {16 \over 3} \* (171 - 131 \* x) \* \Hhh(1,-2,0)
          - {64 \over 3} \* (9 - 62 \* x) \* \Hhh(1,-2,2)
  \nonumber\\&& \mbox{}
          - {4 \over 45} \* (1699 - 1544 \* x) \* \Hhh(1,0,0)
          + {1 \over 9} \* (2339 - 2438 \* x) \* \Hhh(1,1,0)
          + {2 \over 9} \* (1097 - 1290 \* x) \* \Hhh(1,1,1)
  \nonumber\\&& \mbox{}
          + {4 \over 3} \* (291 - 124 \* x) \* \Hhh(1,1,2)
          + {8 \over 3} \* (158 - 45 \* x) \* \Hhh(1,2,0)
          + {4 \over 3} \* (295 - 148 \* x) \* \Hhh(1,2,1)
          - {64 \over 3} \* (11 - 27 \* x) \* \Hhh(2,-2,0)
  \nonumber\\&& \mbox{}
          + {8 \over 15} \* (2156 - 7981 \* x) \* \Hhh(2,0,0)
          + 16 \* (41 - 39 \* x) \* \Hhh(2,1,0)
          + {8 \over 9} \* (761 - 757 \* x) \* \Hhh(2,1,1)
          + 24 \* (5 + 18 \* x) \* \Hhh(2,1,2)
  \nonumber\\&& \mbox{}
          + {8 \over 3} \* (47 + 194 \* x) \* \Hhh(2,2,1)
          + 96 \* (1 - x) \* \Hhh(3,0,0)
          + {8 \over 3} \* (77 + 214 \* x) \* \Hhh(3,1,1)
          - {128 \over 3} \* (4 - 35 \* x) \* \Hhhh(-2,-1,-1,0)
  \nonumber\\&& \mbox{}
          - {8 \over 3} \* (13 + 498 \* x) \* \Hhhh(-2,-1,0,0)
          + {8 \over 3} \* (19 + 262 \* x) \* \Hhhh(-2,0,0,0)
          - {8 \over 15} \* (2181 + 1156 \* x) \* \Hhhh(-1,-1,-1,0)
  \nonumber\\&& \mbox{}
          - {4 \over 3} \* (101 + 510 \* x) \* \Hhhh(-1,-1,0,0)
          + {16 \over 15} \* (184 + 219 \* x) \* \Hhhh(-1,0,0,0)
          + {4 \over 45} \* (10177 - 24182 \* x) \* \Hhhh(0,0,0,0)
  \nonumber\\&& \mbox{}
          - {64 \over 3} \* (3 - 28 \* x) \* \Hhhh(1,-2,-1,0)
          - 64 \* (1 - 15 \* x) \* \Hhhh(1,-2,0,0)
          - {8 \over 15} \* (794 - 569 \* x) \* \Hhhh(1,0,0,0)
  \nonumber\\&& \mbox{}
          - {64 \over 3} \* (9 - 70 \* x) \* \Hhhh(1,1,-2,0)
          + {8 \over 15} \* (5347 - 2847 \* x) \* \Hhhh(1,1,0,0)
          + {4 \over 3} \* (295 - 108 \* x) \* \Hhhh(1,1,1,0)
  \nonumber\\&& \mbox{}
          + {8 \over 3} \* (129 - 71 \* x) \* \Hhhh(1,1,1,1)
          - {16 \over 3} \* (13 + 212 \* x) \* \Hhhh(2,0,0,0)
          + {88 \over 3} \* (5 + 118 \* x) \* \Hhhh(2,1,0,0)
  \nonumber\\&& \mbox{}
          + 8 \* (15 + 62 \* x) \* \Hhhh(2,1,1,1)
          + {8 \over 3} \* (45 + 146 \* x) \* \Hhhhh(0,0,0,0,0)
          + {1 \over 225} \* \pqg( - x) \* (120 \* 
            (1657 + 288 \* x) \* \Hh(-3,0)
  \nonumber\\&& \mbox{}
          + 2 \* (34447 + 29088 \* x) \* \Hh(-2,0)
          + 600 \* (125 + 144 \* x) \* \Hh(-2,2)
          + 3360 \* (58 + 27 \* x) \* \Hh(-1,-1) \* \z2 
  \nonumber\\&& \mbox{}
          - 120 \* (1021 + 684 \* x) \* \Hh(-1,0) \* \z2 
          - 12 \* (16429 - 7342 \* x) \* \Hh(-1,0)
          - 6 \* (38221 - 9696 \* x) \* \Hh(-1,2)
  \nonumber\\&& \mbox{}
          + 300 \* (185 + 216 \* x) \* \Hh(-1,3)
          - 240 \* (1009 + 36 \* x) \* \Hhh(-2,-1,0)
          + 2700 \* (103 + 24 \* x) \* \Hhh(-2,0,0)
  \nonumber\\&& \mbox{}
          - 360 \* (451 + 24 \* x) \* \Hhh(-1,-2,0)
          - 2 \* (50107 + 29088 \* x) \* \Hhh(-1,-1,0)
          - 600 \* (173 + 144 \* x) \* \Hhh(-1,-1,2)
  \nonumber\\&& \mbox{}
          - 4 \* (35399 - 24624 \* x) \* \Hhh(-1,0,0)
          + 240 \* (79 + 36 \* x) \* \Hhh(-1,2,0)
          + 480 \* (47 + 18 \* x) \* \Hhh(-1,2,1)
  \nonumber\\&& \mbox{}
          - 480 \* (49 - 54 \* x) \* \Hhh(1,-2,0)
          + 720 \* (253 + 12 \* x) \* \Hhhh(-1,-1,-1,0)
          - 300 \* (875 + 216 \* x) \* \Hhhh(-1,-1,0,0)
  \nonumber\\&& \mbox{}
          + 180 \* (649 + 216 \* x) \* \Hhhh(-1,0,0,0)
          - 240 \* (817 + 378 \* x) \* \H(-2) \* \z2 
          - 150 \* (1163 + 504 \* x) \* \H(-1) \* \z3 
  \nonumber\\&& \mbox{}
          + (179219 - 87264 \* x) \* \H(-1) \* \z2 
          - 120 \* (560 \* \Hh(-4,0)
          - 760 \* \Hh(-3,2)
          - 2430 \* \Hh(-2,-1) \* \z2 
          + 2085 \* \Hh(-2,0) \* \z2 
  \nonumber\\&& \mbox{}
          - 1645 \* \Hh(-2,3)
          - 1605 \* \Hh(-1,-2) \* \z2 
          - 2230 \* \Hh(-1,-1) \* \z3 
          + 2665 \* \Hh(-1,0) \* \z3 
          + 60 \* \Hh(-1,2) \* \z2 
          + 260 \* \Hh(-1,4)
  \nonumber\\&& \mbox{}
          - 330 \* \Hh(1,-2) \* \z2 
          + 500 \* \Hhh(-3,-1,0)
          - 290 \* \Hhh(-3,0,0)
          + 380 \* \Hhh(-2,-2,0)
          + 2130 \* \Hhh(-2,-1,2)
          - 350 \* \Hhh(-2,2,0)
  \nonumber\\&& \mbox{}
          - 420 \* \Hhh(-2,2,1)
          + 200 \* \Hhh(-1,-3,0)
          + 1380 \* \Hhh(-1,-2,2)
          + 2575 \* \Hhh(-1,-1,-1) \* \z2 
          - 2240 \* \Hhh(-1,-1,0) \* \z2 
  \nonumber\\&& \mbox{}
          + 1735 \* \Hhh(-1,-1,3)
          + 55 \* \Hhh(-1,0,0) \* \z2 
          - 170 \* \Hhh(-1,2,2)
          - 330 \* \Hhh(-1,3,0)
          - 390 \* \Hhh(-1,3,1)
          + 430 \* \Hhh(1,-3,0)
  \nonumber\\&& \mbox{}
          + 440 \* \Hhh(1,-2,2)
          + 500 \* \Hhh(2,-2,0)
          - 600 \* \Hhhh(-2,-1,-1,0)
          + 1615 \* \Hhhh(-2,-1,0,0)
          - 825 \* \Hhhh(-2,0,0,0)
  \nonumber\\&& \mbox{}
          - 450 \* \Hhhh(-1,-2,-1,0)
          + 1145 \* \Hhhh(-1,-2,0,0)
          - 490 \* \Hhhh(-1,-1,-2,0)
          - 2310 \* \Hhhh(-1,-1,-1,2)
          + 350 \* \Hhhh(-1,-1,2,0)
  \nonumber\\&& \mbox{}
          + 420 \* \Hhhh(-1,-1,2,1)
          - 1430 \* \Hhhh(-1,2,0,0)
          - 150 \* \Hhhh(-1,2,1,0)
          - 160 \* \Hhhh(-1,2,1,1)
          + 220 \* \Hhhh(1,-2,-1,0)
  \nonumber\\&& \mbox{}
          + 390 \* \Hhhh(1,-2,0,0)
          + 520 \* \Hhhh(1,1,-2,0)
          + 530 \* \Hhhhh(-1,-1,-1,-1,0)
          - 1960 \* \Hhhhh(-1,-1,-1,0,0)
          + 1035 \* \Hhhhh(-1,-1,0,0,0)
  \nonumber\\&& \mbox{}
          - 675 \* \Hhhhh(-1,0,0,0,0)
          + 1010 \* \H(-3) \* \z2 
          + 2130 \* \H(-2) \* \z3 
          + 662 \* \H(-1) \* \z2^2))
  \nonumber\\&& \mbox{}
          + {1 \over 2025} \* \pqg(x) \* (720 \* 
            (2143 + 378 \* x) \* \Hh(0,0) \* \z2 
          - 27 \* (12205 + 29368 \* x) \* \Hh(0,0)
  \nonumber\\&& \mbox{}
          - 180 \* (9257 + 2592 \* x) \* \Hh(1,0) \* \z2 
          + 180 \* (1421 + 216 \* x) \* \Hh(1,1) \* \z2 
          - 75 \* (12181 + 2376 \* x) \* \Hh(1,2)
  \nonumber\\&& \mbox{}
          + 360 \* (2953 + 1728 \* x) \* \Hh(1,3)
          - 165 \* (8729 - 1080 \* x) \* \Hh(2,0)
          - 360 \* (6221 + 216 \* x) \* \Hh(3,0)
  \nonumber\\&& \mbox{}
          - 360 \* (6781 + 216 \* x) \* \Hh(3,1)
          - 36 \* (71779 + 24624 \* x) \* \Hhh(0,0,0)
          - 75 \* (8287 - 2376 \* x) \* \Hhh(1,1,0)
  \nonumber\\&& \mbox{}
          - 69120 \* (44 + 9 \* x) \* \Hhh(2,0,0)
          - 3240 \* (679 + 144 \* x) \* \Hhhh(0,0,0,0)
          + 180 \* (277 - 648 \* x) \* \Hhhh(1,0,0,0)
  \nonumber\\&& \mbox{}
          - 360 \* (7283 + 1728 \* x) \* \Hhhh(1,1,0,0)
          + 360 \* (11113 + 4428 \* x) \* \H(0) \* \z3 
          + 9 \* (203197 + 136152 \* x) \* \H(0) \* \z2 
  \nonumber\\&& \mbox{}
          + 90 \* (1517 + 10152 \* x) \* \H(1) \* \z3 
          + 12 \* (38551 + 36666 \* x) \* \H(1) \* \z2 
          + 38880 \* (21 + x) \* \H(2) \* \z2 
  \nonumber\\&& \mbox{}
          - 9 \* (145021 + 77976 \* x) \* \H(3)
          - 720 \* (1711 - 54 \* x) \* \H(4)
          + 720 \* (1429 + 675 \* x) \* \z2^2 
  \nonumber\\&& \mbox{}
          + 90 \* (26377 + 8604 \* x) \* \z3 
          + (269975 + 792936 \* x) \* \z2 
          + 4855533 
          + 240300 \* \z5 
          - 251100 \* \z4 
  \nonumber\\&& \mbox{}
          - 540000 \* \z2 \* \z3 
          + 75600 \* \Hh(0,0) \* \z3 
          + 453215 \* \Hh(1,0)
          - 1587600 \* \Hh(1,0) \* \z3 
          + 1423215 \* \Hh(1,1)
  \nonumber\\&& \mbox{}
          - 1803600 \* \Hh(1,1) \* \z3 
          - 10800 \* \Hh(1,2) \* \z2 
          + 351000 \* \Hh(1,4)
          + 1225800 \* \Hh(2,0) \* \z2 
          - 1205310 \* \Hh(2,1)
  \nonumber\\&& \mbox{}
          - 10800 \* \Hh(2,1) \* \z2 
          - 1906200 \* \Hh(2,2)
          - 1701000 \* \Hh(2,3)
          - 270000 \* \Hh(3,2)
          - 162000 \* \Hh(4,0)
          - 183600 \* \Hh(4,1)
  \nonumber\\&& \mbox{}
          + 54000 \* \Hhh(0,0,0) \* \z2 
          - 770520 \* \Hhh(1,0,0)
          - 691200 \* \Hhh(1,0,0) \* \z2
          - 91800 \* \Hhh(1,1,0) \* \z2 
          - 400050 \* \Hhh(1,1,1)
  \nonumber\\&& \mbox{}
          - 75600 \* \Hhh(1,1,1) \* \z2 
          - 1075500 \* \Hhh(1,1,2)
          - 453600 \* \Hhh(1,1,3)
          - 1307700 \* \Hhh(1,2,0)
          - 1047600 \* \Hhh(1,2,1)
  \nonumber\\&& \mbox{}
          - 232200 \* \Hhh(1,2,2)
          - 118800 \* \Hhh(1,3,0)
          + 135000 \* \Hhh(1,3,1)
          - 1671300 \* \Hhh(2,1,0)
          - 1665000 \* \Hhh(2,1,1)
  \nonumber\\&& \mbox{}
          - 313200 \* \Hhh(2,1,2)
          - 329400 \* \Hhh(2,2,0)
          - 178200 \* \Hhh(2,2,1)
          - 399600 \* \Hhh(3,0,0)
          - 378000 \* \Hhh(3,1,0)
  \nonumber\\&& \mbox{}
          - 248400 \* \Hhh(3,1,1)
          - 1083600 \* \Hhhh(1,1,1,0)
          - 766800 \* \Hhhh(1,1,1,1)
          - 210600 \* \Hhhh(1,1,1,2)
          - 372600 \* \Hhhh(1,1,2,0)
  \nonumber\\&& \mbox{}
          - 64800 \* \Hhhh(1,1,2,1)
          - 172800 \* \Hhhh(1,2,0,0)
          - 259200 \* \Hhhh(1,2,1,0)
          - 91800 \* \Hhhh(2,0,0,0)
          + 1339200 \* \Hhhh(2,1,0,0)
  \nonumber\\&& \mbox{}
          - 329400 \* \Hhhh(2,1,1,0)
          - 162000 \* \Hhhh(2,1,1,1)
          + 243000 \* \Hhhhh(1,0,0,0,0)
          + 124200 \* \Hhhhh(1,1,0,0,0)
          + 108000 \* \Hhhhh(1,1,1,0,0)
  \nonumber\\&& \mbox{}
          - 275400 \* \Hhhhh(1,1,1,1,0)
          + 2771054 \* \H(0)
          - 211680 \* \H(0) \* \z2^2 
          + 4195196 \* \H(1)
          - 133920 \* \H(1) \* \z2^2 
          + 522961 \* \H(2)
  \nonumber\\&& \mbox{}
          - 2705400 \* \H(2) \* \z3 
          - 54000 \* \H(5))
          + {2 \over 675} \* \pgq( - x) \* (240 \* 
            (1 - 7 \* x^{-1}) \* \Hh(-2,0)
          + 180 \* (61 + 19 \* x^{-1}) \* \Hh(-1,0) \* \z2 
  \nonumber\\&& \mbox{}
          + (4987 - 3481 \* x^{-1}) \* \Hh(-1,0)
          + 6 \* (619 - 319 \* x^{-1}) \* \Hh(-1,2)
          - 900 \* (5 + 3 \* x^{-1}) \* \Hh(-1,3)
  \nonumber\\&& \mbox{}
          - 6 \* (3759 - 319 \* x^{-1}) \* \Hhh(-1,-1,0)
          + 3600 \* (3 + x^{-1}) \* \Hhh(-1,-1,2)
          + 6 \* (2349 - 599 \* x^{-1}) \* \Hhh(-1,0,0)
  \nonumber\\&& \mbox{}
          + 900 \* (29 + 3 \* x^{-1}) \* \Hhhh(-1,-1,0,0)
          - 180 \* (71 + 9 \* x^{-1}) \* \Hhhh(-1,0,0,0)
          - 3 \* (4997 - 957 \* x^{-1}) \* \H(-1) \* \z2 
  \nonumber\\&& \mbox{}
          + 360 \* (1 - x^{-1}) \* (2 \* \Hh(-2,2)
          - 2 \* \Hhh(-2,-1,0)
          + 2 \* \Hhh(-2,0,0)
          + \Hhh(-1,2,0)
          + \Hhh(-1,2,1)
          - 3 \* \Hhh(1,-2,0)
          - 3 \* \H(-2) \* \z2 )
  \nonumber\\&& \mbox{}
          - 90 \* (33 + 7 \* x^{-1}) \* (6 \* \Hh(-1,-1) \* \z2
          - 5 \* \H(-1) \* \z3)
          + 360 \* (39 + x^{-1}) \* (\Hhh(-1,-2,0)
          - \Hhhh(-1,-1,-1,0)))
  \nonumber\\&& \mbox{}
          + {1 \over 6075} \* \pgq(x) \* (4320 \* 
            (19 + 9 \* x^{-1}) \* \Hh(1,0) \* \z2 
          - 1080 \* (103 + 3 \* x^{-1}) \* \Hh(1,1) \* \z2 
          + 17280 \* (2 - 3 \* x^{-1}) \* \Hh(1,3)
  \nonumber\\&& \mbox{}
          - 1080 \* (121 - 9 \* x^{-1}) \* \Hhhh(1,0,0,0)
          + 8640 \* (31 + 6 \* x^{-1}) \* \Hhhh(1,1,0,0)
          + 540 \* (139 - 141 \* x^{-1}) \* \H(1) \* \z3 
  \nonumber\\&& \mbox{}
          - 18 \* (3677 + 957 \* x^{-1}) \* \H(1) \* \z2 
          - 6480 \* (1 + x^{-1}) \* \H(2) \* \z2 
          + 181071 
          - 137160 \* \z3 
          + 23544 \* \z2 
  \nonumber\\&& \mbox{}
          + 116640 \* \z2^2 
          - 38772 \* \Hh(0,0)
          + 45780 \* \Hh(1,0)
          + 17430 \* \Hh(1,1)
          - 136800 \* \Hh(1,2)
          - 6480 \* \Hh(2,0)
          - 6480 \* \Hh(2,1)
  \nonumber\\&& \mbox{}
          - 38880 \* \Hhh(0,0,0)
          + 57960 \* \Hhh(1,0,0)
          - 39600 \* \Hhh(1,1,0)
          - 88200 \* \Hhh(1,1,1)
          + 237600 \* \Hhh(1,1,2)
          + 237600 \* \Hhh(1,2,0)
  \nonumber\\&& \mbox{}
          + 259200 \* \Hhh(1,2,1)
          + 237600 \* \Hhhh(1,1,1,0)
          + 248400 \* \Hhhh(1,1,1,1)
          - 41166 \* \H(0)
          + 22680 \* \H(0) \* \z2 
          + 33487 \* \H(1)
  \nonumber\\&& \mbox{}
          - 50652 \* \H(2)
          + 3240 \* \H(3))
          + {8 \over 3} \* (119 + 334 \* x) \* \H(-3) \* \z2
          - {92 \over 3} \* (13 - 6 \* x) \* \H(-2) \* \z3
  \nonumber\\&& \mbox{}
          + {24 \over 5} \* (127 - 418 \* x) \* \H(-2) \* \z2
          + {8 \over 5} \* (101 - 182 \* x) \* \H(-1) \* \z2^2
          - {8 \over 3} \* (579 + 631 \* x) \* \H(-1) \* \z3
  \nonumber\\&& \mbox{}
          + {1 \over 225} \* (221977 + 439652 \* x) \* \H(-1) \* \z2
          + {32 \over 5} \* (5 - 8 \* x) \* \H(0) \* \z2^2
          - {4 \over 45} \* (20581 - 114206 \* x) \* \H(0) \* \z3
  \nonumber\\&& \mbox{}
          - {7 \over 225} \* (29741 + 7834 \* x) \* \H(0) \* \z2
          - {1 \over 6075} \* (10709929 - 809411 \* x) \* \H(0)
          - {8 \over 15} \* (4222 - 2387 \* x) \* \H(1) \* \z3
  \nonumber\\&& \mbox{}
          + {2 \over 75} \* (20909 - 20809 \* x) \* \H(1) \* \z2
          - {1 \over 405} \* (1199371 - 413046 \* x) \* \H(1)
          + {4 \over 3} \* (5 - 1482 \* x) \* \H(2) \* \z3
  \nonumber\\&& \mbox{}
          - {4 \over 15} \* (821 - 2826 \* x) \* \H(2) \* \z2
          - {1 \over 81} \* (56969 + 190382 \* x) \* \H(2)
          - 32 \* (2 + 5 \* x) \* \H(3) \* \z2
  \nonumber\\&& \mbox{}
          + {1 \over 225} \* (151931 + 54494 \* x) \* \H(3)
          + {4 \over 45} \* (5779 - 9404 \* x) \* \H(4)
          + {32 \over 3} \* (23 + 60 \* x) \* \H(5)
  \nonumber\\&& \mbox{}
          - {8 \over 5} \* (1 - 2 \* x) \* (340 \* \Hh(1,0) \* \z3 
          + 390 \* \Hh(1,1) \* \z3 
          + 60 \* \Hh(1,2) \* \z2 
          - 80 \* \Hh(1,4)
          + 160 \* \Hhh(1,0,0) \* \z2 
          + 20 \* \Hhh(1,1,0) \* \z2
  \nonumber\\&& \mbox{}
          + 60 \* \Hhh(1,1,1) \* \z2 
          + 100 \* \Hhh(1,1,3)
          - 20 \* \Hhh(1,3,0)
          - 20 \* \Hhh(1,3,1)
          + 120 \* \Hhhh(-1,2,0,0)
          - 80 \* \Hhhh(1,2,0,0)
          - 120 \* \Hhhhh(1,0,0,0,0)
  \nonumber\\&& \mbox{}
          - 140 \* \Hhhhh(1,1,0,0,0)
          - 140 \* \Hhhhh(1,1,1,0,0)
          + 59 \* \H(1) \* \z2^2)
          + {8 \over 3} \* (1 + 2 \* x) \* (204 \* \Hh(-1,-2) \* \z2 
          + 270 \* \Hh(-1,-1) \* \z3
  \nonumber\\&& \mbox{}
          + 12 \* \Hh(-1,2) \* \z2 
          - 72 \* \Hhh(-1,-3,0)
          - 168 \* \Hhh(-1,-2,2)
          - 324 \* \Hhh(-1,-1,-1) \* \z2 
          + 276 \* \Hhh(-1,-1,0) \* \z2 
          - 204 \* \Hhh(-1,-1,3)
  \nonumber\\&& \mbox{}
          + 12 \* \Hhh(-1,3,0)
          + 12 \* \Hhh(-1,3,1)
          + 79 \* \Hhh(3,1,0)
          + 72 \* \Hhhh(-1,-2,-1,0)
          - 168 \* \Hhhh(-1,-2,0,0)
          + 72 \* \Hhhh(-1,-1,-2,0)
  \nonumber\\&& \mbox{}
          + 288 \* \Hhhh(-1,-1,-1,2)
          - 24 \* \Hhhh(-1,-1,2,0)
          - 24 \* \Hhhh(-1,-1,2,1)
          - 72 \* \Hhhhh(-1,-1,-1,-1,0)
          + 276 \* \Hhhhh(-1,-1,-1,0,0)
  \nonumber\\&& \mbox{}
          - 156 \* \Hhhhh(-1,-1,0,0,0)
          + 72 \* \Hhhhh(-1,0,0,0,0))
          + {16 \over 3} \* (17 - 173 \* x) \* \z2 \* \z3
          + 8 \* (17 + 58 \* x) \* (\Hhh(2,2,0)
          + \Hhhh(2,1,1,0))
  \nonumber\\&& \mbox{}
          + {2 \over 3} \* (285 + 12658 \* x) \* \z5
          - {22 \over 135} \* (6893 + 15152 \* x) \* \z3
          - {4 \over 45} \* (6359 - 19705 \* x) \* \z2^2
  \nonumber\\&& \mbox{}
          + {1 \over 2025} \* (613217 - 6837442 \* x) \* \z2
          - {1 \over 48600} \* (167724077 - 57125642 \* x)
          \biggr)
  \nonumber\\&& \mbox{}
       +  \colour4colour{\ca^2 \* \nf}  \*  \biggl(
          - 8 \* (19 - 6 \* x) \* \Hh(-4,0)
          - {4 \over 45} \* (1397 + 11002 \* x) \* \Hh(-3,0)
          - {8 \over 3} \* (47 - 98 \* x) \* \Hh(-3,2)
  \nonumber\\&& \mbox{}
          - {32 \over 3} \* (23 + 52 \* x) \* \Hh(-2,-1) \* \z2
          + {16 \over 3} \* (67 - 20 \* x) \* \Hh(-2,0) \* \z2
          - {1 \over 27} \* (7163 - 12062 \* x) \* \Hh(-2,0)
  \nonumber\\&& \mbox{}
          - {8 \over 15} \* (1034 + 2939 \* x) \* \Hh(-2,2)
          - {4 \over 3} \* (259 - 214 \* x) \* \Hh(-2,3)
          - {2 \over 15} \* (7647 + 7262 \* x) \* \Hh(-1,-1) \* \z2
  \nonumber\\&& \mbox{}
          + 24 \* (5 + 18 \* x) \* \Hh(-1,0) \* \z3
          + {2 \over 15} \* (11309 + 7494 \* x) \* \Hh(-1,0) \* \z2
          + {1 \over 270} \* (74807 + 127717 \* x) \* \Hh(-1,0)
  \nonumber\\&& \mbox{}
          - {1 \over 45} \* (7846 - 11009 \* x) \* \Hh(-1,2)
          - {4 \over 15} \* (5288 + 3343 \* x) \* \Hh(-1,3)
          - 32 \* (5 + 4 \* x) \* \Hh(-1,4)
  \nonumber\\&& \mbox{}
          - {16 \over 3} \* (71 + 277 \* x) \* \Hh(0,0) \* \z3
          - {4 \over 45} \* (9766 + 29899 \* x) \* \Hh(0,0) \* \z2
          + {1 \over 405} \* (1465864 + 1126431 \* x) \* \Hh(0,0)
  \nonumber\\&& \mbox{}
          - 16 \* (3 - 20 \* x) \* \Hh(1,-2) \* \z2
          - {2 \over 5} \* (3979 - 1734 \* x) \* \Hh(1,0) \* \z2
          + {4 \over 27} \* (16187 + 1381 \* x) \* \Hh(1,0)
  \nonumber\\&& \mbox{}
          - {4 \over 3} \* (579 - 152 \* x) \* \Hh(1,1) \* \z2
          + {1 \over 27} \* (64571 + 11738 \* x) \* \Hh(1,1)
          + {1 \over 9} \* (16163 - 1426 \* x) \* \Hh(1,2)
  \nonumber\\&& \mbox{}
          + {4 \over 3} \* (1267 - 601 \* x) \* \Hh(1,3)
          - {16 \over 3} \* (39 + 346 \* x) \* \Hh(2,0) \* \z2
          + {1 \over 9} \* (18353 - 446 \* x) \* \Hh(2,0)
  \nonumber\\&& \mbox{}
          - {8 \over 3} \* (49 + 302 \* x) \* \Hh(2,1) \* \z2
          + {4 \over 27} \* (13375 + 157 \* x) \* \Hh(2,1)
          + {2 \over 3} \* (1031 + 1738 \* x) \* \Hh(2,2)
  \nonumber\\&& \mbox{}
          + {20 \over 3} \* (33 + 250 \* x) \* \Hh(2,3)
          + 4 \* (161 + 505 \* x) \* \Hh(3,0)
          + {10 \over 9} \* (647 + 1988 \* x) \* \Hh(3,1)
          + {8 \over 3} \* (51 + 338 \* x) \* \Hh(3,2)
  \nonumber\\&& \mbox{}
          + 4 \* (29 + 274 \* x) \* \Hh(4,0)
          + 52 \* (3 + 22 \* x) \* \Hh(4,1)
          + {8 \over 3} \* (1 + 270 \* x) \* \Hhh(-3,-1,0)
          - {16 \over 3} \* (31 + 15 \* x) \* \Hhh(-3,0,0)
  \nonumber\\&& \mbox{}
          - {8 \over 3} \* (21 - 218 \* x) \* \Hhh(-2,-2,0)
          + {4 \over 3} \* (31 + 830 \* x) \* \Hhh(-2,-1,0)
          + {8 \over 3} \* (101 + 70 \* x) \* \Hhh(-2,-1,2)
  \nonumber\\&& \mbox{}
          - {194 \over 45} \* (107 + 412 \* x) \* \Hhh(-2,0,0)
          - {8 \over 3} \* (35 - 62 \* x) \* \Hhh(-2,2,0)
          - {8 \over 3} \* (64 + 13 \* x) \* \Hhh(-1,-2,0)
  \nonumber\\&& \mbox{}
          - {1 \over 9} \* (5706 + 3349 \* x) \* \Hhh(-1,-1,0)
          + {8 \over 15} \* (2218 + 1913 \* x) \* \Hhh(-1,-1,2)
          + 16 \* (13 + 14 \* x) \* \Hhh(-1,0,0) \* \z2
  \nonumber\\&& \mbox{}
          - {1 \over 15} \* (466 - 5479 \* x) \* \Hhh(-1,0,0)
          - {8 \over 3} \* (131 + 53 \* x) \* \Hhh(-1,2,0)
          - {16 \over 3} \* (69 + 26 \* x) \* \Hhh(-1,2,1)
  \nonumber\\&& \mbox{}
          - 40 \* (1 + 28 \* x) \* \Hhh(0,0,0) \* \z2
          + {2 \over 135} \* (110399 - 11334 \* x) \* \Hhh(0,0,0)
          + {16 \over 3} \* (3 - 82 \* x) \* \Hhh(1,-3,0)
  \nonumber\\&& \mbox{}
          + {4 \over 9} \* (587 - 770 \* x) \* \Hhh(1,-2,0)
          + {64 \over 3} \* (3 - 13 \* x) \* \Hhh(1,-2,2)
          + {17 \over 9} \* (1087 - 190 \* x) \* \Hhh(1,0,0)
  \nonumber\\&& \mbox{}
          + {1 \over 9} \* (17489 - 2634 \* x) \* \Hhh(1,1,0)
          + {2 \over 9} \* (7871 - 844 \* x) \* \Hhh(1,1,1)
          + {2 \over 3} \* (913 - 226 \* x) \* \Hhh(1,1,2)
  \nonumber\\&& \mbox{}
          + {44 \over 3} \* (46 - 15 \* x) \* \Hhh(1,2,0)
          + {2 \over 3} \* (881 - 218 \* x) \* \Hhh(1,2,1)
          + {16 \over 3} \* (15 - 34 \* x) \* \Hhh(2,-2,0)
  \nonumber\\&& \mbox{}
          + {2 \over 9} \* (2387 + 10172 \* x) \* \Hhh(2,0,0)
          + {4 \over 9} \* (1951 + 2182 \* x) \* \Hhh(2,1,0)
          + {4 \over 3} \* (493 + 894 \* x) \* \Hhh(2,1,1)
  \nonumber\\&& \mbox{}
          + {32 \over 3} \* (10 + 41 \* x) \* \Hhh(2,1,2)
          + {4 \over 3} \* (97 + 386 \* x) \* \Hhh(2,2,0)
          + {4 \over 3} \* (127 + 862 \* x) \* \Hhh(3,0,0)
          + {8 \over 3} \* (69 + 382 \* x) \* \Hhh(3,1,0)
  \nonumber\\&& \mbox{}
          + {64 \over 3} \* (6 + 41 \* x) \* \Hhh(3,1,1)
          + 16 \* (3 - 46 \* x) \* \Hhhh(-2,-1,-1,0)
          + {4 \over 3} \* (89 + 558 \* x) \* \Hhhh(-2,-1,0,0)
  \nonumber\\&& \mbox{}
          - {8 \over 3} \* (87 - 98 \* x) \* \Hhhh(-2,0,0,0)
          + {4 \over 3} \* (245 + 78 \* x) \* \Hhhh(-1,-1,-1,0)
          + {2 \over 15} \* (5381 + 4956 \* x) \* \Hhhh(-1,-1,0,0)
  \nonumber\\&& \mbox{}
          - {16 \over 15} \* (1011 + 526 \* x) \* \Hhhh(-1,0,0,0)
          + {112 \over 45} \* (121 + 1374 \* x) \* \Hhhh(0,0,0,0)
          + {32 \over 3} \* (3 + 8 \* x) \* \Hhhh(1,-2,-1,0)
  \nonumber\\&& \mbox{}
          + 16 \* (1 - 22 \* x) \* \Hhhh(1,-2,0,0)
          + {4 \over 15} \* (3221 - 1196 \* x) \* \Hhhh(1,0,0,0)
          + 64 \* (1 - 7 \* x) \* \Hhhh(1,1,-2,0)
  \nonumber\\&& \mbox{}
          + {2 \over 3} \* (53 + 368 \* x) \* \Hhhh(1,1,0,0)
          + 4 \* (151 - 42 \* x) \* \Hhhh(1,1,1,0)
          + {4 \over 3} \* (347 - 78 \* x) \* \Hhhh(1,1,1,1)
  \nonumber\\&& \mbox{}
          + {64 \over 3} \* (10 + 51 \* x) \* \Hhhh(2,0,0,0)
          + {4 \over 3} \* (69 - 134 \* x) \* \Hhhh(2,1,0,0)
          + {32 \over 3} \* (10 + 39 \* x) \* \Hhhh(2,1,1,0)
  \nonumber\\&& \mbox{}
          - 48 \* (5 - 22 \* x) \* \Hhhhh(0,0,0,0,0)
          + {1 \over 810} \* \pqg( - x) \* (72 \* (5717 - 432 \* x) \* \Hh(-3,0)
          + 6 \* (7939 + 1620 \* x) \* \Hh(-2,0)
  \nonumber\\&& \mbox{}
          + 20736 \* (23 - 3 \* x) \* \Hh(-2,2)
          + 36 \* (4073 - 1728 \* x) \* \Hh(-1,-1) \* \z2 
          + 38 \* (12347 + 486 \* x) \* \Hh(-1,0)
  \nonumber\\&& \mbox{}
          - 36 \* (16501 - 1296 \* x) \* \Hh(-1,0) \* \z2 
          + 6 \* (58127 + 1620 \* x) \* \Hh(-1,2)
          + 216 \* (2879 - 144 \* x) \* \Hh(-1,3)
  \nonumber\\&& \mbox{}
          + 36 \* (18479 - 864 \* x) \* \Hhh(-2,0,0)
          + 30 \* (3133 - 324 \* x) \* \Hhh(-1,-1,0)
          - 2304 \* (97 - 27 \* x) \* \Hhh(-1,-1,2)
  \nonumber\\&& \mbox{}
          + 18 \* (21937 + 540 \* x) \* \Hhh(-1,0,0)
          + 72 \* (697 - 432 \* x) \* \Hhh(1,-2,0)
          - 108 \* (2053 - 288 \* x) \* \Hhhh(-1,-1,0,0)
  \nonumber\\&& \mbox{}
          + 72 \* (8831 - 216 \* x) \* \Hhhh(-1,0,0,0)
          - 108 \* (5171 - 576 \* x) \* \H(-2) \* \z2 
          - 324 \* (809 - 144 \* x) \* \H(-1) \* \z3 
  \nonumber\\&& \mbox{}
          - 3 \* (100589 + 4860 \* x) \* \H(-1) \* \z2 
          + 72 \* (480 \* \Hh(-4,0)
          + 480 \* \Hh(-3,2)
          - 480 \* \Hh(-2,-1) \* \z2 
          - 930 \* \Hh(-2,0) \* \z2 
  \nonumber\\&& \mbox{}
          + 1470 \* \Hh(-2,3)
          - 1320 \* \Hh(-1,-2) \* \z2 
          - 1950 \* \Hh(-1,-1) \* \z3 
          + 1140 \* \Hh(-1,0) \* \z3 
          - 90 \* \Hh(-1,2) \* \z2 
          + 3450 \* \Hh(-1,4)
  \nonumber\\&& \mbox{}
          - 630 \* \Hh(1,-2) \* \z2 
          - 1200 \* \Hhh(-3,-1,0)
          + 840 \* \Hhh(-3,0,0)
          - 60 \* \Hhh(-2,-2,0)
          - 2265 \* \Hhh(-2,-1,0)
          + 300 \* \Hhh(-2,-1,2)
  \nonumber\\&& \mbox{}
          + 780 \* \Hhh(-2,2,0)
          + 960 \* \Hhh(-2,2,1)
          - 360 \* \Hhh(-1,-3,0)
          + 940 \* \Hhh(-1,-2,0)
          + 900 \* \Hhh(-1,-2,2)
          + 3330 \* \Hhh(-1,-1,-1) \* \z2 
  \nonumber\\&& \mbox{}
          - 180 \* \Hhh(-1,-1,0) \* \z2 
          - 690 \* \Hhh(-1,-1,3)
          - 3060 \* \Hhh(-1,0,0) \* \z2 
          + 3270 \* \Hhh(-1,2,0)
          + 3590 \* \Hhh(-1,2,1)
          + 300 \* \Hhh(-1,2,2)
  \nonumber\\&& \mbox{}
          + 1440 \* \Hhh(-1,3,0)
          + 1680 \* \Hhh(-1,3,1)
          + 1140 \* \Hhh(1,-3,0)
          + 420 \* \Hhh(1,-2,2)
          + 1140 \* \Hhh(2,-2,0)
          - 360 \* \Hhhh(-2,-1,-1,0)
  \nonumber\\&& \mbox{}
          - 690 \* \Hhhh(-2,-1,0,0)
          + 960 \* \Hhhh(-2,0,0,0)
          - 840 \* \Hhhh(-1,-2,-1,0)
          + 690 \* \Hhhh(-1,-2,0,0)
          - 840 \* \Hhhh(-1,-1,-2,0)
  \nonumber\\&& \mbox{}
          - 2135 \* \Hhhh(-1,-1,-1,0)
          - 2700 \* \Hhhh(-1,-1,-1,2)
          - 1260 \* \Hhhh(-1,-1,2,0)
          - 1320 \* \Hhhh(-1,-1,2,1)
          - 1440 \* \Hhhh(-1,2,0,0)
  \nonumber\\&& \mbox{}
          + 420 \* \Hhhh(-1,2,1,0)
          + 120 \* \Hhhh(-1,2,1,1)
          - 420 \* \Hhhh(1,-2,-1,0)
          + 900 \* \Hhhh(1,-2,0,0)
          + 900 \* \Hhhh(1,1,-2,0)
  \nonumber\\&& \mbox{}
          + 1260 \* \Hhhhh(-1,-1,-1,-1,0)
          - 2220 \* \Hhhhh(-1,-1,-1,0,0)
          - 1260 \* \Hhhhh(-1,-1,0,0,0)
          + 1020 \* \Hhhhh(-1,0,0,0,0)
  \nonumber\\&& \mbox{}
          - 1080 \* \H(-3) \* \z2 
          - 210 \* \H(-2) \* \z3 
          + 1113 \* \H(-1) \* \z2^2 ))
          + {1 \over 72900} \* \pqg(x) \* 
            (12960 \* (4703 + 108 \* x) \* \Hh(0,0) \* \z2 
  \nonumber\\&& \mbox{}
          - 540 \* (347977 + 3078 \* x) \* \Hh(0,0)
          + 3240 \* (21953 + 1728 \* x) \* \Hh(1,0) \* \z2 
          - 194400 \* (353 + 36 \* x) \* \Hh(1,3)
  \nonumber\\&& \mbox{}
          - 10800 \* (19180 + 81 \* x) \* \Hhh(0,0,0)
          - 16200 \* (1871 - 432 \* x) \* \Hhh(2,0,0)
          - 174960 \* (59 - 16 \* x) \* \Hhhh(0,0,0,0)
  \nonumber\\&& \mbox{}
          - 6480 \* (8069 - 216 \* x) \* \Hhhh(1,0,0,0)
          - 16200 \* (2665 - 432 \* x) \* \Hhhh(1,1,0,0)
          + 32400 \* (2005 - 432 \* x) \* \H(0) \* \z3 
  \nonumber\\&& \mbox{}
          + 2160 \* (110989 + 810 \* x) \* \H(0) \* \z2 
          + 3240 \* (17401 - 3024 \* x) \* \H(1) \* \z3 
          + 4050 \* (34153 + 108 \* x) \* \H(1) \* \z2 
  \nonumber\\&& \mbox{}
          - 2160 \* (110584 + 405 \* x) \* \H(3)
          - 12960 \* (4919 + 324 \* x) \* \H(4)
          - 15552 \* (2261 + 486 \* x) \* \z2^2 
  \nonumber\\&& \mbox{}
          + 2430 \* (83699 + 900 \* x) \* \z3 
          + 180 \* (1249121 + 9234 \* x) \* \z2 
          - 5 \* (79819747 
          + 4237920 \* \z5 
  \nonumber\\&& \mbox{}
          - 1807920 \* \z4 
          + 1671840 \* \z2 \* \z3 
          + 39153960 \* \Hh(1,0)
          - 6026400 \* \Hh(1,0) \* \z3 
          + 43750620 \* \Hh(1,1)
  \nonumber\\&& \mbox{}
          - 6181920 \* \Hh(1,1) \* \z3 
          - 10568880 \* \Hh(1,1) \* \z2 
          + 26818020 \* \Hh(1,2)
          - 1905120 \* \Hh(1,2) \* \z2 
          + 3188160 \* \Hh(1,4)
  \nonumber\\&& \mbox{}
          + 35764740 \* \Hh(2,0)
          + 427680 \* \Hh(2,0) \* \z2 
          + 37847520 \* \Hh(2,1)
          - 1788480 \* \Hh(2,1) \* \z2 
          + 9438120 \* \Hh(2,2)
  \nonumber\\&& \mbox{}
          - 1127520 \* \Hh(2,3)
          + 9447840 \* \Hh(3,0)
          + 10967400 \* \Hh(3,1)
          + 1555200 \* \Hh(3,2)
          + 1166400 \* \Hh(4,0)
  \nonumber\\&& \mbox{}
          + 2099520 \* \Hh(4,1)
          - 1399680 \* \Hhh(0,0,0) \* \z2 
          + 27245700 \* \Hhh(1,0,0)
          - 3693600 \* \Hhh(1,0,0) \* \z2
          + 29059020 \* \Hhh(1,1,0)
  \nonumber\\&& \mbox{}
          - 2838240 \* \Hhh(1,1,0) \* \z2 
          + 26653320 \* \Hhh(1,1,1)
          - 2177280 \* \Hhh(1,1,1) \* \z2 
          + 9185400 \* \Hhh(1,1,2)
          + 1710720 \* \Hhh(1,1,3)
  \nonumber\\&& \mbox{}
          + 9259920 \* \Hhh(1,2,0)
          + 8472600 \* \Hhh(1,2,1)
          + 1360800 \* \Hhh(1,2,2)
          + 1944000 \* \Hhh(1,3,0)
          + 2604960 \* \Hhh(1,3,1)
  \nonumber\\&& \mbox{}
          + 12214800 \* \Hhh(2,1,0)
          + 8754480 \* \Hhh(2,1,1)
          + 1555200 \* \Hhh(2,1,2)
          + 1283040 \* \Hhh(2,2,0)
          + 1671840 \* \Hhh(2,2,1)
  \nonumber\\&& \mbox{}
          + 233280 \* \Hhh(3,0,0)
          + 1244160 \* \Hhh(3,1,0)
          + 1399680 \* \Hhh(3,1,1)
          + 9434880 \* \Hhhh(1,1,1,0)
          + 7464960 \* \Hhhh(1,1,1,1)
  \nonumber\\&& \mbox{}
          + 1360800 \* \Hhhh(1,1,1,2)
          + 1671840 \* \Hhhh(1,1,2,0)
          + 1477440 \* \Hhhh(1,1,2,1)
          + 1788480 \* \Hhhh(1,2,0,0)
          + 1710720 \* \Hhhh(1,2,1,0)
  \nonumber\\&& \mbox{}
          + 1555200 \* \Hhhh(1,2,1,1)
          + 855360 \* \Hhhh(2,0,0,0)
          + 4510080 \* \Hhhh(2,1,0,0)
          + 1671840 \* \Hhhh(2,1,1,0)
          + 1399680 \* \Hhhh(2,1,1,1)
  \nonumber\\&& \mbox{}
          + 2177280 \* \Hhhhh(1,0,0,0,0)
          + 2954880 \* \Hhhhh(1,1,0,0,0)
          + 2877120 \* \Hhhhh(1,1,1,0,0)
          + 1205280 \* \Hhhhh(1,1,1,1,0)
  \nonumber\\&& \mbox{}
          + 1166400 \* \Hhhhh(1,1,1,1,1)
          + 79559484 \* \H(0)
          + 894240 \* \H(0) \* \z2^2 
          + 77840592 \* \H(1)
          + 58320 \* \H(1) \* \z2^2 
  \nonumber\\&& \mbox{}
          + 44635932 \* \H(2)
          - 7231680 \* \H(2) \* \z3 
          - 7970400 \* \H(2) \* \z2 
          - 777600 \* \H(3) \* \z2 
          + 1399680 \* \H(5) ))
  \nonumber\\&& \mbox{}
          + {1 \over 270} \* \pgq( - x) \* (192 \* (67 
	  + 3 \* x^{-1}) \* \Hh(-1,-1) \* \z2
          + 48 \* (79 - 9 \* x^{-1}) \* \Hh(-1,0) \* \z2 
          - 3 \* (15589 + 57 \* x^{-1}) \* \Hh(-1,0)
  \nonumber\\&& \mbox{}
          + 2 \* (17053 - 45 \* x^{-1}) \* \Hh(-1,2)
          - 288 \* (21 - x^{-1}) \* \Hh(-1,3)
          - 10 \* (1801 - 9 \* x^{-1}) \* \Hhh(-1,-1,0)
  \nonumber\\&& \mbox{}
          - 192 \* (47 + 3 \* x^{-1}) \* \Hhh(-1,-1,2)
          + 2 \* (33589 - 45 \* x^{-1}) \* \Hhh(-1,0,0)
          - 96 \* (17 + 3 \* x^{-1}) \* \Hhh(1,-2,0)
  \nonumber\\&& \mbox{}
          - 288 \* (39 + x^{-1}) \* \Hhhh(-1,-1,0,0)
          - 144 \* (71- x^{-1}) \* \Hhhh(-1,0,0,0)
          - 432 \* (14 + x^{-1}) \* \H(-1) \* \z3 
  \nonumber\\&& \mbox{}
          - (43111 - 135 \* x^{-1}) \* \H(-1) \* \z2 
          - 192 \* (31 \* \Hh(-2,0)
          - 30 \* \Hh(-2,2)
          + 30 \* \Hhh(-2,-1,0)
          - 60 \* \Hhh(-2,0,0)
  \nonumber\\&& \mbox{}
          + 35 \* \Hhh(-1,-2,0)
          + 30 \* \Hhh(-1,2,0)
          + 40 \* \Hhh(-1,2,1)
          - 40 \* \Hhhh(-1,-1,-1,0)
          + 45 \* \H(-2) \* \z2))
  \nonumber\\&& \mbox{}
          - {1 \over 7290} \* \pgq(x) \* 
            (7776 \* (57 + 2 \* x^{-1}) \* \Hh(1,0) \* \z2 
          - 6480 \* (59 + 3 \* x^{-1}) \* \Hh(1,3)
          - 1296 \* (337 - 3 \* x^{-1}) \* \Hhhh(1,0,0,0)
  \nonumber\\&& \mbox{}
          - 6480 \* (53 - 3 \* x^{-1}) \* \Hhhh(1,1,0,0)
          + 1296 \* (274 - 21 \* x^{-1}) \* \H(1) \* \z3 
          + 135 \* (3713 + 9 \* x^{-1}) \* \H(1) \* \z2 
          + 5017249 
  \nonumber\\&& \mbox{}
          - 257040 \* \z3 
          - 725292 \* \z2 
          - 321408 \* \z2^2 
          + 10206 \* \Hh(0,0)
          + 1128780 \* \Hh(1,0)
          + 1210140 \* \Hh(1,1)
  \nonumber\\&& \mbox{}
          + 362880 \* \Hh(1,1) \* \z2 
          - 258120 \* \Hh(1,2)
          + 336960 \* \Hh(2,0)
          + 442800 \* \Hh(2,1)
          - 194400 \* \Hh(2,2)
          - 7776 \* \Hhh(0,0,0)
  \nonumber\\&& \mbox{}
          - 165240 \* \Hhh(1,0,0)
          - 130680 \* \Hhh(1,1,0)
          - 292680 \* \Hhh(1,1,1)
          - 259200 \* \Hhh(1,1,2)
          - 311040 \* \Hhh(1,2,0)
  \nonumber\\&& \mbox{}
          - 233280 \* \Hhh(1,2,1)
          - 155520 \* \Hhh(2,0,0)
          - 194400 \* \Hhh(2,1,0)
          - 207360 \* \Hhh(2,1,1)
          - 259200 \* \Hhhh(1,1,1,0)
  \nonumber\\&& \mbox{}
          - 168480 \* \Hhhh(1,1,1,1)
          + 586731 \* \H(0)
          + 51840 \* \H(0) \* \z3 
          - 172368 \* \H(0) \* \z2 
          - 2756286 \* \H(1)
          - 542034 \* \H(2)
  \nonumber\\&& \mbox{}
          + 272160 \* \H(2) \* \z2 
          + 11664 \* \H(3))
          + {4 \over 3} \* (95 + 74 \* x) \* \H(-3) \* \z2
          + 4 \* (61 + 94 \* x) \* \H(-2) \* \z3
  \nonumber\\&& \mbox{}
          + {2 \over 15} \* (4291 + 15906 \* x) \* \H(-2) \* \z2
          - {4 \over 5} \* (67 - 58 \* x) \* \H(-1) \* \z2^2
          + {2 \over 15} \* (7379 + 6184 \* x) \* \H(-1) \* \z3
  \nonumber\\&& \mbox{}
          - {1 \over 90} \* (12838 + 38763 \* x) \* \H(-1) \* \z2
          - {8 \over 15} \* (37 - 1016 \* x) \* \H(0) \* \z2^2
          - {4 \over 9} \* (2141 + 8828 \* x) \* \H(0) \* \z3
  \nonumber\\&& \mbox{}
          - {28 \over 135} \* (12001 - 5486 \* x) \* \H(0) \* \z2
          + {1 \over 810} \* (2142802 + 105755 \* x) \* \H(0)
          - {2 \over 15} \* (449 + 2406 \* x) \* \H(1) \* \z3
  \nonumber\\&& \mbox{}
          - {1 \over 18} \* (38032 - 6201 \* x) \* \H(1) \* \z2
          + {1 \over 405} \* (1948816 - 55581 \* x) \* \H(1)
          - {8 \over 3} \* (31 + 58 \* x) \* \H(2) \* \z3
  \nonumber\\&& \mbox{}
          - {16 \over 3} \* (125 + 321 \* x) \* \H(2) \* \z2
          + {1 \over 135} \* (506631 + 419981 \* x) \* \H(2)
          - {4 \over 3} \* (101 + 946 \* x) \* \H(3) \* \z2
  \nonumber\\&& \mbox{}
          + {4 \over 27} \* (17018 - 3709 \* x) \* \H(3)
          + {4 \over 45} \* (10198 + 29467 \* x) \* \H(4)
          + {40 \over 3} \* (3 + 94 \* x) \* \H(5)
  \nonumber\\&& \mbox{}
          + {4 \over 5} \* (1 - 2 \* x) \* (260 \* \Hh(1,0) \* \z3 
          + 290 \* \Hh(1,1) \* \z3 
          + 60 \* \Hh(1,2) \* \z2 
          - 90 \* \Hh(1,4)
          - 120 \* \Hhh(-2,2,1)
          + 150 \* \Hhh(1,0,0) \* \z2
  \nonumber\\&& \mbox{}
          + 60 \* \Hhh(1,1,0) \* \z2 
          + 60 \* \Hhh(1,1,1) \* \z2 
          + 40 \* \Hhh(1,1,3)
          - 20 \* \Hhh(1,3,0)
          - 20 \* \Hhh(1,3,1)
          + 60 \* \Hhhh(-1,2,0,0)
          - 50 \* \Hhhh(1,2,0,0)
  \nonumber\\&& \mbox{}
          - 100 \* \Hhhhh(1,0,0,0,0)
          - 120 \* \Hhhhh(1,1,0,0,0)
          - 80 \* \Hhhhh(1,1,1,0,0)
          + 21 \* \H(1) \* \z2^2 )
          - 8 \* (1 + 2 \* x) \* (30 \* \Hh(-1,-2) \* \z2 
  \nonumber\\&& \mbox{}
          + 39 \* \Hh(-1,-1) \* \z3 
          + 2 \* \Hh(-1,2) \* \z2 
          - 10 \* \Hhh(-1,-3,0)
          - 24 \* \Hhh(-1,-2,2)
          - 46 \* \Hhh(-1,-1,-1) \* \z2 
          + 40 \* \Hhh(-1,-1,0) \* \z2 
  \nonumber\\&& \mbox{}
          - 30 \* \Hhh(-1,-1,3)
          + 2 \* \Hhh(-1,3,0)
          + 2 \* \Hhh(-1,3,1)
          + 12 \* \Hhhh(-1,-2,-1,0)
          - 26 \* \Hhhh(-1,-2,0,0)
          + 12 \* \Hhhh(-1,-1,-2,0)
  \nonumber\\&& \mbox{}
          + 40 \* \Hhhh(-1,-1,-1,2)
          - 4 \* \Hhhh(-1,-1,2,0)
          - 4 \* \Hhhh(-1,-1,2,1)
          - 12 \* \Hhhhh(-1,-1,-1,-1,0)
          + 42 \* \Hhhhh(-1,-1,-1,0,0)
  \nonumber\\&& \mbox{}
          - 24 \* \Hhhhh(-1,-1,0,0,0)
          + 10 \* \Hhhhh(-1,0,0,0,0))
          + {16 \over 3} \* (1 + 4 \* x) \* (18 \* \Hhh(2,2,1)
          + 13 \* \Hhhh(2,1,1,1))
          + {4 \over 3} \* (247 + 914 \* x) \* \z2 \* \z3
  \nonumber\\&& \mbox{}
          - {4 \over 3} \* (435 + 1654 \* x) \* \z5
          + {2 \over 225} \* (62009 - 2214 \* x) \* \z2^2
          - {1 \over 270} \* (829687 - 580202 \* x) \* \z3
  \nonumber\\&& \mbox{}
          - {1 \over 405} \* (1388313 + 580970 \* x) \* \z2
          + {1 \over 4860} \* (22628627 + 6387628 \* x)
          \biggr)
\:\: .
\eea
\normalsize
The exact three-loop pure-singlet coefficient function corresponding
to Eq.~(\ref{eq:c2ps3}) is given by
\small
\bea
&& c^{(3)}_{2,\rm{ps}}(x) \:\: = \:\: 
        \colour4colour{\cf \* \nf^2}  \*  \biggl(
            {64 \over 9} \* (5 - 2 \* x) \* \Hh(-2,0)
          + {16 \over 45} \* (153 + 163 \* x) \* \Hh(-1,0)
          - {8 \over 405} \* (9353 + 8192 \* x) \* \Hh(0,0)
  \nonumber\\&& \mbox{}
          - {8 \over 9} \* (22 - 19 \* x) \* \Hh(1,0)
          - {16 \over 9} \* (7 + 13 \* x) \* \Hh(2,0)
          + {40 \over 27} \* (1 - 17 \* x) \* \Hh(2,1)
          - {8 \over 27} \* (271 + 451 \* x) \* \Hhh(0,0,0)
  \nonumber\\&& \mbox{}
          + {16 \over 45} \* \pqg( - x) \* ((16 + 9 \* x) \* \Hh(-1,0)
          + 20 \* (\Hh(-2,0)
          + \Hhh(-1,0,0)))
          + {8 \over 3645} \* \pqg(x) \* (27 \* (491 - 54 \* x) \* \Hh(0,0)
  \nonumber\\&& \mbox{}
          - 81 \* (37- 18 \* x) \* \z2 
          + 9587 
          + 1350 \* \z3 
          + 2565 \* \Hh(1,0)
          + 2655 \* \Hh(1,1)
          - 810 \* \Hh(1,2)
          - 810 \* \Hh(2,0)
          - 3240 \* \Hh(2,1)
  \nonumber\\&& \mbox{}
          + 6480 \* \Hhh(0,0,0)
          - 810 \* \Hhh(1,1,0)
          - 1080 \* \Hhh(1,1,1)
          + 26262 \* \H(0)
          + 810 \* \H(0) \* \z2
          + 5760 \* \H(1)
          + 810 \* \H(1) \* \z2 
  \nonumber\\&& \mbox{}
          + 4455 \* \H(2)
          - 810 \* \H(3))
          + {16 \over 45} \* \pgq( - x) \* ((4 + x^{-1}) \* \Hh(-1,0)
          - 20 \* \Hhh(-1,0,0))
          + {8 \over 3645} \* \pgq(x) \* (6748 
  \nonumber\\&& \mbox{}
          + 1080 \* \z3 
          - 810 \* \z2 
          - 1350 \* \Hh(1,0)
          - 2655 \* \Hh(1,1)
          + 810 \* \Hh(1,2)
          + 810 \* \Hhh(1,1,0)
          + 1080 \* \Hhh(1,1,1)
          + 162 \* \H(0)
  \nonumber\\&& \mbox{}
          + 315 \* \H(1)
          - 810 \* \H(1) \* \z2)
          - {8 \over 243} \* (7225 - 2207 \* x) \* \H(0)
          - {8 \over 81} \* (649 - 514 \* x) \* \H(1)
          - {8 \over 81} \* (328 + 229 \* x) \* \H(2)
  \nonumber\\&& \mbox{}
          + {8 \over 27} \* (1 - x) \* (16 \* \Hh(1,1)
          + 27 \* \Hh(1,2)
          + 27 \* \Hhh(1,1,0)
          + 36 \* \Hhh(1,1,1)
          - 27 \* \H(1) \* \z2)
          + {8 \over 45} \* (1 + x) \* (21 \* \z2^2 
  \nonumber\\&& \mbox{}
          + 20 \* \Hh(0,0) \* \z2 
          + 30 \* \Hh(2,2)
          + 70 \* \Hh(3,1)
          + 120 \* \Hhh(-1,0,0)
          + 30 \* \Hhh(2,1,0)
          + 40 \* \Hhh(2,1,1)
          - 230 \* \Hhhh(0,0,0,0)
          - 10 \* \H(0) \* \z3 
  \nonumber\\&& \mbox{}
          - 30 \* \H(2) \* \z2 
          - 20 \* \H(4) )
          + {32 \over 27} \* (25 + 34 \* x) \* (\H(0) \* \z2 - \H(3))
          + {8 \over 27} \* (47 + 29 \* x) \* \z3
          + {8 \over 405} \* (1298 + 4997 \* x) \* \z2
  \nonumber\\&& \mbox{}
          - {8 \over 81} \* (2513 - 2150 \* x)
          \biggr)
  \nonumber\\&& \mbox{}
       +  \colour4colour{\cf^2 \* \nf}  \*  \biggl(
          - {32 \over 3} \* (38 + 15 \* x) \* \Hh(-3,0)
          - 128 \* (2 + x) \* \Hh(-3,2)
          - {16 \over 45} \* (2423 + 1808 \* x) \* \Hh(-2,0)
  \nonumber\\&& \mbox{}
          - {32 \over 3} \* (40 - 17 \* x) \* \Hh(-2,2)
          - {8 \over 675} \* (88958 + 87033 \* x) \* \Hh(-1,0)
          - {32 \over 45} \* (1294 + 1329 \* x) \* \Hh(-1,2)
  \nonumber\\&& \mbox{}
          - {128 \over 3} \* (5 + 2 \* x) \* \Hh(0,0) \* \z3
          - {16 \over 45} \* (1282 + 463 \* x) \* \Hh(0,0) \* \z2
          - {2 \over 2025} \* (39593 - 70273 \* x) \* \Hh(0,0)
  \nonumber\\&& \mbox{}
          + {16 \over 15} \* (347 - 327 \* x) \* \Hh(1,0) \* \z2
          + {8 \over 9} \* (789 - 905 \* x) \* \Hh(1,0)
          + {2 \over 27} \* (13598 - 14405 \* x) \* \Hh(1,1)
  \nonumber\\&& \mbox{}
          + {4 \over 9} \* (1121 - 1139 \* x) \* \Hh(1,2)
          - {64 \over 15} \* (38 - 33 \* x) \* \Hh(1,3)
          + {16 \over 27} \* (1289 - 820 \* x) \* \Hh(2,0)
  \nonumber\\&& \mbox{}
          + {4 \over 27} \* (6251 - 2728 \* x) \* \Hh(2,1)
          + {8 \over 3} \* (123 - 83 \* x) \* \Hh(2,2)
          + {8 \over 9} \* (427 - 173 \* x) \* \Hh(3,0)
          + {32 \over 9} \* (137 - 19 \* x) \* \Hh(3,1)
  \nonumber\\&& \mbox{}
          + 128 \* (3 - 2 \* x) \* \Hhh(-3,-1,0)
          - 64 \* (9 - x) \* \Hhh(-3,0,0)
          + {32 \over 3} \* (44 + 15 \* x) \* \Hhh(-2,-1,0)
          - {32 \over 3} \* (74 + 13 \* x) \* \Hhh(-2,0,0)
  \nonumber\\&& \mbox{}
          + {64 \over 45} \* (662 + 657 \* x) \* \Hhh(-1,-1,0)
          - {16 \over 15} \* (1572 + 1607 \* x) \* \Hhh(-1,0,0)
          + {4 \over 45} \* (7803 + 9497 \* x) \* \Hhh(0,0,0)
  \nonumber\\&& \mbox{}
          + {8 \over 45} \* (1981 - 2356 \* x) \* \Hhh(1,0,0)
          + {4 \over 9} \* (953 - 935 \* x) \* \Hhh(1,1,0)
          + {16 \over 15} \* (206 - 381 \* x) \* \Hhh(2,0,0)
  \nonumber\\&& \mbox{}
          + {8 \over 3} \* (103 - 79 \* x) \* \Hhh(2,1,0)
          + {32 \over 9} \* (83 - 43 \* x) \* \Hhh(2,1,1)
          + {584 \over 9} \* (1 - 2 \* x) \* \Hhhh(0,0,0,0)
          + {32 \over 15} \* (181 - 171 \* x) \* \Hhhh(1,1,0,0)
  \nonumber\\&& \mbox{}
          + {8 \over 225} \* \pqg( - x) \* (6 \* (93 + 17 \* x) \* \Hh(-1,0)
          - 10 \* (106 + 9 \* x) \* \Hh(-1,2)
          - 270 \* (6 - x) \* \Hhh(-1,0,0)
  \nonumber\\&& \mbox{}
          + 5 \* (358 + 27 \* x) \* \H(-1) \* \z2 
          - 10 \* (146 + 9 \* x) \* (\Hh(-2,0)
          - \Hhh(-1,-1,0))
          + 150 \* (10 \* \Hh(-3,0)
          + 2 \* \Hh(-2,2)
  \nonumber\\&& \mbox{}
          + 9 \* \Hh(-1,-1) \* \z2
          - \Hh(-1,0) \* \z2 
          - 2 \* \Hh(-1,3)
          - 10 \* \Hhh(-2,-1,0)
          + 10 \* \Hhh(-2,0,0)
          - 6 \* \Hhh(-1,-2,0)
          - 6 \* \Hhh(-1,-1,2)
  \nonumber\\&& \mbox{}
          - 2 \* \Hhh(-1,2,0)
          - 2 \* \Hhh(-1,2,1)
          + 6 \* \Hhhh(-1,-1,-1,0)
          - 10 \* \Hhhh(-1,-1,0,0)
          + 2 \* \Hhhh(-1,0,0,0)
          - 7 \* \H(-2) \* \z2 
          - 6 \* \H(-1) \* \z3 ))
  \nonumber\\&& \mbox{}
          + {2 \over 2025} \* \pqg(x) \* (2 \* (889 - 1836 \* x) \* \Hh(0,0)
          + 360 \* (97 - 108 \* x) \* \Hh(1,0) \* \z2 
          - 720 \* (71 - 54 \* x) \* \Hh(1,3)
  \nonumber\\&& \mbox{}
          - 360 \* (332 + 27 \* x) \* \Hhh(0,0,0)
          - 6480 \* (11 + 6 \* x) \* \Hhh(2,0,0)
          - 720 \* (29 + 54 \* x) \* \Hhhh(1,1,0,0)
  \nonumber\\&& \mbox{}
          + 720 \* (139 + 54 \* x) \* \H(0) \* \z3 
          + 360 \* (395 - 18 \* x) \* \H(0) \* \z2 
          - 1440 \* (23 - 27 \* x) \* \H(1) \* \z3 
          - 60 \* (62 + 27 \* x) \* \H(1) \* \z2 
  \nonumber\\&& \mbox{}
          - 360 \* (404 - 9 \* x) \* \H(3)
          + 720 \* (401 - 54 \* x) \* (\Hh(0,0) \* \z2 - \H(4))
          - 72 \* (1043 - 432 \* x) \* \z2^2 
  \nonumber\\&& \mbox{}
          - 60 \* (1808 + 135 \* x) \* \z3 
          + 2 \* (38281 + 1836 \* x) \* \z2 
          + 207453 
          - 16200 \* \z4 
          + 68750 \* \Hh(1,0)
          + 26050 \* \Hh(1,1)
  \nonumber\\&& \mbox{}
          + 28800 \* \Hh(1,1) \* \z2 
          + 30000 \* \Hh(1,2)
          - 69600 \* \Hh(2,0)
          - 80400 \* \Hh(2,1)
          - 135000 \* \Hh(2,2)
          - 187200 \* \Hh(3,0)
  \nonumber\\&& \mbox{}
          - 196200 \* \Hh(3,1)
          + 52080 \* \Hhh(1,0,0)
          - 2400 \* \Hhh(1,1,0)
          - 3600 \* \Hhh(1,1,1)
          - 45000 \* \Hhh(1,1,2)
          - 50400 \* \Hhh(1,2,0)
  \nonumber\\&& \mbox{}
          - 48600 \* \Hhh(1,2,1)
          - 113400 \* \Hhh(2,1,0)
          - 104400 \* \Hhh(2,1,1)
          - 214200 \* \Hhhh(0,0,0,0)
          + 12600 \* \Hhhh(1,0,0,0)
  \nonumber\\&& \mbox{}
          - 45000 \* \Hhhh(1,1,1,0)
          - 39600 \* \Hhhh(1,1,1,1)
          + 75838 \* \H(0)
          - 15440 \* \H(1)
          - 72890 \* \H(2)
          + 108000 \* \H(2) \* \z2)
  \nonumber\\&& \mbox{}
          - {8 \over 675} \* \pgq( - x) \* (120 \* (1 - x^{-1}) \* \Hh(-2,0)
          - (601 + 44 \* x^{-1}) \* \Hh(-1,0)
          - 30 \* (26 - x^{-1}) \* \Hh(-1,2)
  \nonumber\\&& \mbox{}
          + 30 \* (156 - x^{-1}) \* \Hhh(-1,-1,0)
          - 90 \* (29 + x^{-1}) \* \Hhh(-1,0,0)
          + 15 \* (208 - 3 \* x^{-1}) \* \H(-1) \* \z2 
          + 450 \* (9 \* \Hh(-1,-1) \* \z2 
  \nonumber\\&& \mbox{}
          - \Hh(-1,0) \* \z2 
          - 2 \* \Hh(-1,3)
          - 6 \* \Hhh(-1,-2,0)
          - 6 \* \Hhh(-1,-1,2)
          - 2 \* \Hhh(-1,2,0)
          - 2 \* \Hhh(-1,2,1)
          + 6 \* \Hhhh(-1,-1,-1,0)
  \nonumber\\&& \mbox{}
          - 10 \* \Hhhh(-1,-1,0,0)
          + 2 \* \Hhhh(-1,0,0,0)
          - 6 \* \H(-1) \* \z3))
          - {1 \over 2025} \* \pgq(x) \* 
            (720 \* (37 + 12 \* x^{-1}) \* \Hh(1,0) \* \z2 
  \nonumber\\&& \mbox{}
          - 1440 \* (41 + 6 \* x^{-1}) \* \Hh(1,3)
          - 1440 \* (59 - 6 \* x^{-1}) \* \Hhhh(1,1,0,0)
          - 2880 \* (8 + 3 \* x^{-1}) \* \H(1) \* \z3 
  \nonumber\\&& \mbox{}
          + 120 \* (103 + 3 \* x^{-1}) \* \H(1) \* \z2 
          - 36219 
          + 98640 \* \z3 
          + 23400 \* \z2 
          - 38880 \* \z2^2 
          - 2160 \* \Hh(0,0)
          - 71300 \* \Hh(1,0)
  \nonumber\\&& \mbox{}
          - 68950 \* \Hh(1,1)
          + 57600 \* \Hh(1,1) \* \z2 
          + 43800 \* \Hh(1,2)
          - 30840 \* \Hhh(1,0,0)
          + 11400 \* \Hhh(1,1,0)
          - 7200 \* \Hhh(1,1,1)
  \nonumber\\&& \mbox{}
          - 90000 \* \Hhh(1,1,2)
          - 100800 \* \Hhh(1,2,0)
          - 97200 \* \Hhh(1,2,1)
          + 25200 \* \Hhhh(1,0,0,0)
          - 90000 \* \Hhhh(1,1,1,0)
          - 79200 \* \Hhhh(1,1,1,1)
  \nonumber\\&& \mbox{}
          - 1776 \* \H(0)
          - 8640 \* \H(0) \* \z2 
          + 129095 \* \H(1)
          - 7920 \* \H(2)
          + 8640 \* \H(3)) 
          + {16 \over 3} \* (124 - 19 \* x) \* \H(-2) \* \z2
  \nonumber\\&& \mbox{}
          + {64 \over 15} \* (326 + 331 \* x) \* \H(-1) \* \z2
          + {8 \over 3} \* (29 + 41 \* x) \* \H(0) \* \z2^2
          - {16 \over 45} \* (353 - 1243 \* x) \* \H(0) \* \z3  
  \nonumber\\&& \mbox{}
          - {4 \over 45} \* (13491 + 15499 \* x) \* \H(0) \* \z2
          + {2 \over 405} \* (36651 - 296056 \* x) \* \H(0)
          - {4 \over 45} \* (309 - 439 \* x) \* \H(1) \* \z2
  \nonumber\\&& \mbox{}
          + {8 \over 15} \* (521 - 561 \* x) \* \H(1) \* \z3
          + {1 \over 405} \* (548474 - 516479 \* x) \* \H(1)
          - {8 \over 3} \* (35 - 53 \* x) \* \H(2) \* \z2
  \nonumber\\&& \mbox{}
          + {868 \over 405} \* (464 - 391 \* x) \* \H(2)
          - 32 \* (3 + 5 \* x) \* \H(3) \* \z2
          + {4 \over 45} \* (13527 + 7043 \* x) \* \H(3)
          + {16 \over 45} \* (1282 + 313 \* x) \* \H(4)
  \nonumber\\&& \mbox{}
          + {8 \over 9} \* (1 - x) \* (81 \* \z4 
          - 288 \* \Hh(-2,-1) \* \z2 
          + 216 \* \Hh(-2,0) \* \z2 
          - 72 \* \Hh(-2,3)
          + 9 \* \Hh(1,1) \* \z2 
          + 288 \* \Hhh(-2,-2,0)
  \nonumber\\&& \mbox{}
          + 144 \* \Hhh(-2,-1,2)
          + 500 \* \Hhh(1,1,1)
          + 225 \* \Hhh(1,1,2)
          + 252 \* \Hhh(1,2,0)
          + 243 \* \Hhh(1,2,1)
          - 288 \* \Hhhh(-2,-1,-1,0)
  \nonumber\\&& \mbox{}
          + 504 \* \Hhhh(-2,-1,0,0)
          - 288 \* \Hhhh(-2,0,0,0)
          - 279 \* \Hhhh(1,0,0,0)
          + 225 \* \Hhhh(1,1,1,0)
          + 198 \* \Hhhh(1,1,1,1)
          + 252 \* \H(-2) \* \z3)
  \nonumber\\&& \mbox{}
          + {16 \over 3} \* (1 + x) \* (33 \* \z5 
          - 69 \* \Hh(-1,-1) \* \z2 
          + 69 \* \Hh(-1,0) \* \z2 
          - 30 \* \Hh(-1,3)
          + 40 \* \Hh(2,0) \* \z2 
          - \Hh(2,1) \* \z2 
          - 16 \* \Hh(2,3)
  \nonumber\\&& \mbox{}
          + 54 \* \Hh(3,2)
          + 67 \* \Hh(4,0)
          + 75 \* \Hh(4,1)
          + 78 \* \Hhh(-1,-2,0)
          + 30 \* \Hhh(-1,-1,2)
          - 6 \* \Hhh(-1,2,0)
          - 6 \* \Hhh(-1,2,1)
          - 83 \* \Hhh(0,0,0) \* \z2 
  \nonumber\\&& \mbox{}
          + 25 \* \Hhh(2,1,2)
          + 28 \* \Hhh(2,2,0)
          + 27 \* \Hhh(2,2,1)
          + 34 \* \Hhh(3,0,0)
          + 46 \* \Hhh(3,1,0)
          + 44 \* \Hhh(3,1,1)
          - 78 \* \Hhhh(-1,-1,-1,0)
  \nonumber\\&& \mbox{}
          + 138 \* \Hhhh(-1,-1,0,0)
          - 90 \* \Hhhh(-1,0,0,0)
          - 31 \* \Hhhh(2,0,0,0)
          + 44 \* \Hhhh(2,1,0,0)
          + 25 \* \Hhhh(2,1,1,0)
          + 22 \* \Hhhh(2,1,1,1)
  \nonumber\\&& \mbox{}
          + 45 \* \Hhhhh(0,0,0,0,0)
          + 66 \* \H(-1) \* \z3 
          + 9 \* \H(0) \* \z4 
          + 31 \* \H(2) \* \z3 
          + 83 \* \H(5))
          + 16 \* (9 + x) \* \z2 \* \z3
  \nonumber\\&& \mbox{}
          + {8 \over 225} \* (4606 + 2389 \* x) \* \z2^2
          - {4 \over 135} \* (12986 + 88109 \* x) \* \z3
          - {1 \over 405} \* (31094 + 59131 \* x)
  \nonumber\\&& \mbox{}
          - {4 \over 2025} \* (513016 + 73939 \* x) \* \z2
          - 64 \* (6 \* \Hh(-4,0)
          - 7 \* \H(-3) \* \z2)
          \biggr)
  \nonumber\\&& \mbox{}
       +  \colour4colour{\ca \* \cf \* \nf}  \*  \biggl(
          - {16 \over 3} \* (31 - 7 \* x) \* \Hh(-4,0)
          + {40 \over 9} \* (47 - 41 \* x) \* \Hh(-3,0)
          - {16 \over 3} \* (11 - 17 \* x) \* \Hh(-3,2)
  \nonumber\\&& \mbox{}
          - {8 \over 135} \* (8582 - 5423 \* x) \* \Hh(-2,0)
          + {8 \over 3} \* (59 - 123 \* x) \* \Hh(-2,2)
          - {4 \over 135} \* (4456 + 11281 \* x) \* \Hh(-1,0)
  \nonumber\\&& \mbox{}
          + {8 \over 45} \* (1182 + 1597 \* x) \* \Hh(-1,2)
          - {32 \over 3} \* (32 + 33 \* x) \* \Hh(0,0) \* \z3
          - {8 \over 45} \* (643 + 1567 \* x) \* \Hh(0,0) \* \z2
  \nonumber\\&& \mbox{}
          + {4 \over 405} \* (197909 + 122612 \* x) \* \Hh(0,0)
          - {16 \over 15} \* (391 - 381 \* x) \* \Hh(1,0) \* \z2
          - {2 \over 27} \* (2342 - 3515 \* x) \* \Hh(1,0)
  \nonumber\\&& \mbox{}
          - {2 \over 27} \* (5584 - 7003 \* x) \* \Hh(1,1)
          + {4 \over 9} \* (151 - 25 \* x) \* \Hh(1,2)
          + {4 \over 15} \* (1309 - 1269 \* x) \* \Hh(1,3)
          - {112 \over 3} \* (5 + 11 \* x) \* \Hh(2,0)
  \nonumber\\&& \mbox{}
          - {76 \over 9} \* (43 + 49 \* x) \* \Hh(2,1)
          + {8 \over 3} \* (29 - 33 \* x) \* \Hh(2,2)
          + {4 \over 3} \* (53 + 55 \* x) \* \Hh(3,0)
          + {4 \over 9} \* (19 + 61 \* x) \* \Hh(3,1)
  \nonumber\\&& \mbox{}
          + {16 \over 3} \* (13 + 33 \* x) \* \Hh(3,2)
          + 8 \* (1 + 29 \* x) \* \Hh(4,0)
          + {8 \over 3} \* (11 + 95 \* x) \* \Hh(4,1)
          - {16 \over 3} \* (11 - 23 \* x) \* \Hhh(-3,-1,0)
  \nonumber\\&& \mbox{}
          - {16 \over 3} \* (20 - 7 \* x) \* \Hhh(-3,0,0)
          - 8 \* (27 - 11 \* x) \* \Hhh(-2,-1,0)
          + {16 \over 9} \* (205 - 181 \* x) \* \Hhh(-2,0,0)
  \nonumber\\&& \mbox{}
          - {8 \over 45} \* (1942 + 1907 \* x) \* \Hhh(-1,-1,0)
          + {32 \over 45} \* (403 + 563 \* x) \* \Hhh(-1,0,0)
          + {16 \over 3} \* (20 - 43 \* x) \* \Hhh(0,0,0) \* \z2
  \nonumber\\&& \mbox{}
          - {4 \over 135} \* (13349 + 12071 \* x) \* \Hhh(0,0,0)
          + {56 \over 45} \* (166 - 91 \* x) \* \Hhh(1,0,0)
          + 4 \* (31 - 21 \* x) \* \Hhh(1,1,0)
  \nonumber\\&& \mbox{}
          + {4 \over 9} \* (175 - 67 \* x) \* \Hhh(1,1,1)
          + {4 \over 45} \* (1499 + 731 \* x) \* \Hhh(2,0,0)
          + {8 \over 9} \* (137 - 133 \* x) \* \Hhh(2,1,0)
          + {40 \over 9} \* (23 - 22 \* x) \* \Hhh(2,1,1)
  \nonumber\\&& \mbox{}
          + {8 \over 3} \* (39 + 83 \* x) \* \Hhh(3,0,0)
          + {16 \over 3} \* (23 + 39 \* x) \* \Hhh(3,1,0)
          + {16 \over 3} \* (19 + 37 \* x) \* \Hhh(3,1,1)
          + {8 \over 9} \* (427 + 670 \* x) \* \Hhhh(0,0,0,0)
  \nonumber\\&& \mbox{}
          + {4 \over 15} \* (221 - 261 \* x) \* \Hhhh(1,1,0,0)
          - {4 \over 135} \* \pqg( - x) \* (2 \* (1783 - 108 \* x) \* \Hh(-2,0)
          + (1183 - 54 \* x) \* \Hh(-1,0)
  \nonumber\\&& \mbox{}
          + 2 \* (743 - 108 \* x) \* \Hh(-1,2)
          - 2 \* (1123 - 108 \* x) \* \Hhh(-1,-1,0)
          + 2 \* (2093 - 108 \* x) \* \Hhh(-1,0,0)
  \nonumber\\&& \mbox{}
          - (2609 - 324 \* x) \* \H(-1) \* \z2 
          - 30 \* (104 \* \Hh(-3,0)
          + 60 \* \Hh(-2,2)
          - 38 \* \Hh(-1,-1) \* \z2 
          - 56 \* \Hh(-1,0) \* \z2 
          + 60 \* \Hh(-1,3)
  \nonumber\\&& \mbox{}
          - 12 \* \Hhh(-2,-1,0)
          + 112 \* \Hhh(-2,0,0)
          + 16 \* \Hhh(-1,-2,0)
          + 28 \* \Hhh(-1,-1,2)
          + 36 \* \Hhh(-1,2,0)
          + 44 \* \Hhh(-1,2,1)
          - 8 \* \Hhh(1,-2,0)
  \nonumber\\&& \mbox{}
          - 20 \* \Hhhh(-1,-1,-1,0)
          + 12 \* \Hhhh(-1,-1,0,0)
          + 78 \* \Hhhh(-1,0,0,0)
          - 66 \* \H(-2) \* \z2 
          + 3 \* \H(-1) \* \z3 ))
  \nonumber\\&& \mbox{}
          + {1 \over 18225} \* \pqg(x) \* (180 \* (2639 - 162 \* x) \* \Hh(0,0)
          + 38880 \* (14 + 9 \* x) \* \Hh(1,0) \* \z2 
          - 12960 \* (37 + 27 \* x) \* \Hh(1,3)
  \nonumber\\&& \mbox{}
          - 4860 \* (1769 + 24 \* x) \* \Hhh(0,0,0)
          - 6480 \* (191 - 54 \* x) \* \Hhh(2,0,0)
          - 12960 \* (73 - 27 \* x) \* \Hhhh(1,1,0,0)
  \nonumber\\&& \mbox{}
          + 6480 \* (491 - 54 \* x) \* \H(0) \* \z3 
          + 540\* (11899 + 432 \* x) \* \H(0)  \* \z2 
          + 9720 \* (89 - 36 \* x) \* \H(1) \* \z3 
  \nonumber\\&& \mbox{}
          + 1080 \* (1969 + 54 \* x) \* \H(1) \* \z2 
          - 540 \* (11683 + 216 \* x) \* \H(3)
          + 12960 \* (17 + 27 \* x) \* (\Hh(0,0)\* \z2 - \H(4))
  \nonumber\\&& \mbox{}
          + 1944 \* (241 - 144 \* x) \* \z2^2 
          - 360 \* (7652 - 81\* x) \* \z2 
          + 540 \* (7733 + 540 \* x) \* \z3 
          - 5 \* (255239 
	  - 58320 \* \z4
  \nonumber\\&& \mbox{}
          - 304920 \* \Hh(1,0)
          - 277200 \* \Hh(1,1)
          - 142560 \* \Hh(1,1) \* \z2 
          + 304020 \* \Hh(1,2)
          + 785700 \* \Hh(2,0)
          + 712260 \* \Hh(2,1)
  \nonumber\\&& \mbox{}
          + 136080 \* \Hh(2,2)
          + 155520 \* \Hh(3,0)
          + 168480 \* \Hh(3,1)
          + 411804 \* \Hhh(1,0,0)
          + 181980 \* \Hhh(1,1,0)
          + 227340 \* \Hhh(1,1,1)
  \nonumber\\&& \mbox{}
          + 110160 \* \Hhh(1,1,2)
          + 116640 \* \Hhh(1,2,0)
          + 97200 \* \Hhh(1,2,1)
          + 200880 \* \Hhh(2,1,0)
          + 162000 \* \Hhh(2,1,1)
  \nonumber\\&& \mbox{}
          + 181440 \* \Hhhh(1,0,0,0)
          + 110160 \* \Hhhh(1,1,1,0)
          + 90720 \* \Hhhh(1,1,1,1)
          + 1798638 \* \H(0)
          + 592494 \* \H(1)
          - 556776 \* \H(2)
  \nonumber\\&& \mbox{}
          - 116640 \* \H(2) \* \z2))
          - {8 \over 135} \* \pgq( - x) \* ((2791 - 3 \* x^{-1}) \* \Hh(-1,0)
          - 4 \* (527 + 3 \* x^{-1}) \* \Hh(-1,2)
  \nonumber\\&& \mbox{}
          + 2 \* (569 + 6 \* x^{-1}) \* \Hhh(-1,-1,0)
          - (4133 + 12 \* x^{-1}) \* \Hhh(-1,0,0)
          + (2677 + 18 \* x^{-1}) \* \H(-1) \* \z2 
          + 15 \* (26 \* \Hh(-2,0)
  \nonumber\\&& \mbox{}
          - 24 \* \Hh(-2,2)
          - 38 \* \Hh(-1,-1) \* \z2 
          - 56 \* \Hh(-1,0) \* \z2 
          + 60 \* \Hh(-1,3)
          + 24 \* \Hhh(-2,-1,0)
          - 48 \* \Hhh(-2,0,0)
          + 16 \* \Hhh(-1,-2,0)
  \nonumber\\&& \mbox{}
          + 28 \* \Hhh(-1,-1,2)
          + 36 \* \Hhh(-1,2,0)
          + 44 \* \Hhh(-1,2,1)
          + 8 \* \Hhh(1,-2,0)
          - 20 \* \Hhhh(-1,-1,-1,0)
          + 12 \* \Hhhh(-1,-1,0,0)
  \nonumber\\&& \mbox{}
          + 78 \* \Hhhh(-1,0,0,0)
          + 36 \* \H(-2) \* \z2 + 3 \* \H(-1) \* \z3 ))
          - {2 \over 3645} \* \pgq(x) \* (3888 \* \Hh(1,0) \* 
            (19 - x^{-1}) \* \z2 
  \nonumber\\&& \mbox{}
          - 1296 \* \Hh(1,3) \* (52 - 3 \* x^{-1})
          - 1296 \* \Hhhh(1,1,0,0) \* (58 + 3 \* x^{-1})
          + 972 \* \H(1) \* (69 + 4 \* x^{-1}) \* \z3 
  \nonumber\\&& \mbox{}
          + 54 \* \H(1) \* (2063 - 12 \* x^{-1}) \* \z2 
          + 1213133 
          - 80028 \* \z3 
          - 170496 \* \z2 
          - 80352 \* \z2^2
          - 1296 \* \Hh(0,0)
  \nonumber\\&& \mbox{}
          + 310815 \* \Hh(1,0)
          + 330165 \* \Hh(1,1)
          + 71280 \* \Hh(1,1) \* \z2 
          - 49950 \* \Hh(1,2)
          + 84240 \* \Hh(2,0)
          + 110700 \* \Hh(2,1)
  \nonumber\\&& \mbox{}
          - 48600 \* \Hh(2,2)
          - 35802 \* \Hhh(1,0,0)
          - 18090 \* \Hhh(1,1,0)
          - 26190 \* \Hhh(1,1,1)
          - 55080 \* \Hhh(1,1,2)
          - 58320 \* \Hhh(1,2,0)
  \nonumber\\&& \mbox{}
          - 48600 \* \Hhh(1,2,1)
          - 38880 \* \Hhh(2,0,0)
          - 48600 \* \Hhh(2,1,0)
          - 51840 \* \Hhh(2,1,1)
          - 90720 \* \Hhhh(1,0,0,0)
          - 55080 \* \Hhhh(1,1,1,0)
  \nonumber\\&& \mbox{}
          - 45360 \* \Hhhh(1,1,1,1)
          + 149052 \* \H(0)
          + 12960 \* \H(0) \* \z3 
          - 38232 \* \H(0) \* \z2 
          - 753222 \* \H(1)
          - 130608 \* \H(2)
  \nonumber\\&& \mbox{}
          + 68040 \* \H(2) \* \z2 
          - 3888 \* \H(3))
          - {4 \over 3} \* (199 - 279 \* x) \* \H(-2) \* \z2
          - {4 \over 45} \* (4306 + 5101 \* x) \* \H(-1) \* \z2
  \nonumber\\&& \mbox{}
          - {8 \over 15} \* (105 - 229 \* x) \* \H(0) \* \z2^2
          - {8 \over 45} \* (1262 + 2213 \* x) \* \H(0) \* \z3
          + {4 \over 27} \* (2996 + 6149 \* x) \* \H(0) \* \z2
  \nonumber\\&& \mbox{}
          - {2 \over 1215} \* (614933 + 126185 \* x) \* \H(0)
          - {4 \over 15} \* (971 - 1011 \* x) \* \H(1) \* \z3
          - {4 \over 45} \* (2697 - 2032 \* x) \* \H(1) \* \z2
  \nonumber\\&& \mbox{}
          - {2 \over 405} \* (23126 + 27649 \* x) \* \H(1)
          - {4 \over 3} \* (139 - 33 \* x) \* \H(2) \* \z2
          + {4 \over 405} \* (81331 + 59866 \* x) \* \H(2)
  \nonumber\\&& \mbox{}
          - {8 \over 3} \* (37 + 89 \* x) \* \H(3) \* \z2
          - {4 \over 135} \* (13636 + 26599 \* x) \* \H(3)
          + {8 \over 45} \* (643 + 1582 \* x) \* \H(4)
          - {160 \over 3} \* (2 - 5 \* x) \* \H(5)
  \nonumber\\&& \mbox{}
          - {4 \over 9} \* (1 - x) \* (162 \* \z4 
          - 216 \* \Hh(-2,-1) \* \z2 
          - 156 \* \Hh(-2,0) \* \z2 
          + 258 \* \Hh(-2,3)
          + 504 \* \Hh(1,1) \* \z2 
          + 228 \* \Hhh(-2,-2,0)
  \nonumber\\&& \mbox{}
          + 84 \* \Hhh(-2,-1,2)
          + 132 \* \Hhh(-2,2,0)
          + 144 \* \Hhh(-2,2,1)
          + 28 \* \Hhh(1,-2,0)
          - 306 \* \Hhh(1,1,2)
          - 333 \* \Hhh(1,2,0)
          - 270 \* \Hhh(1,2,1)
  \nonumber\\&& \mbox{}
          - 264 \* \Hhhh(-2,-1,-1,0)
          + 258 \* \Hhhh(-2,-1,0,0)
          + 240 \* \Hhhh(-2,0,0,0)
          - 792 \* \Hhhh(1,0,0,0)
          - 306 \* \Hhhh(1,1,1,0)
          - 252 \* \Hhhh(1,1,1,1)
  \nonumber\\&& \mbox{}
          + 540 \* \Hhhhh(0,0,0,0,0)
          - 66 \* \H(-3) \* \z2 
          + 96 \* \H(-2) \* \z3)
          + {4 \over 3} \* (1 + x) \* (12 \* \Hh(-1,-1) \* \z2 
          + 126 \* \Hh(-1,0) \* \z2 
          - 177 \* \Hh(-1,3)
  \nonumber\\&& \mbox{}
          - 196 \* \Hh(2,0) \* \z2 
          - 112 \* \Hh(2,1) \* \z2 
          + 162 \* \Hh(2,3)
          - 114 \* \Hhh(-1,-2,0)
          + 54 \* \Hhh(-1,-1,2)
          - 66 \* \Hhh(-1,2,0)
          - 72 \* \Hhh(-1,2,1)
  \nonumber\\&& \mbox{}
          + 8 \* \Hhh(2,-2,0)
          + 68 \* \Hhh(2,1,2)
          + 74 \* \Hhh(2,2,0)
          + 60 \* \Hhh(2,2,1)
          + 132 \* \Hhhh(-1,-1,-1,0)
          - 81 \* \Hhhh(-1,-1,0,0)
          - 120 \* \Hhhh(-1,0,0,0)
  \nonumber\\&& \mbox{}
          + 176 \* \Hhhh(2,0,0,0)
          + 42 \* \Hhhh(2,1,0,0)
          + 68 \* \Hhhh(2,1,1,0)
          + 56 \* \Hhhh(2,1,1,1)
          + 24 \* \H(-1) \* \z3 
          - 36 \* \H(0) \* \z4 
          - 126 \* \H(2) \* \z3)
  \nonumber\\&& \mbox{}
          + {8 \over 3} \* (111 + 113 \* x) \* \z2 \* \z3
          - 4 \* (155 + 33 \* x) \* \z5
          + {4 \over 225} \* (8254 - 2489 \* x) \* \z2^2
          - {4 \over 135} \* (23077 - 35477 \* x) \* \z3
  \nonumber\\&& \mbox{}
          - {4 \over 405} \* (48037 + 100465 \* x) \* \z2
          + {1 \over 405} \* (303133 - 5188 \* x)
          \biggr)
\:\: .
\eea
\normalsize

\noindent
Finally we turn to the $x$-space analogues of the results (A.11) --
(A.18) for the longitudinal structure function. The one- and two-loop
results are
\bea
c^{(1)}_{L,\rm{q}}(x) & \! = \! &  
        \colour4colour{\cf}  \*  (
            4 \* x
          )
\:\: , \\[1mm] 
c^{(1)}_{L,\rm{g}}(x) & \! = \! &  
       \colour4colour{\nf}  \*  (
            8 \* x \* (1 - x)
          )
\:\: , 
\eea
and
\small
\bea
&& c^{(2)}_{L,\rm ns}(x) \:\: = \:\: 
          \colour4colour{\cf \* \biggl(\cf-{\ca \over 2}\biggr)}  \*  \biggl(
	    {16 \over 5} \* ( 
            3 \* \pqg( - x) \* (1 - x)
          + 2 \* \pgq( - x) \* (1 - x^{-1})
          + (19 + 9 \* x)) \* \Hh(-1,0) 
  \nonumber\\&& \mbox{}
          - {48 \over 5} \* \pqg(x) \* ((1 + x) \* (\z2 - \Hh(0,0))
          - (1 + \H(0)))
          + {32 \over 5} \* \pgq(x) \* (1 - \H(0))
          \biggr)
       +  \colour4colour{\cf \* \nf}  \*  \biggl(
            {4 \over 9} \* (6 - 25 \* x)
  \nonumber\\&& \mbox{}
          - {8 \over 3} \* x \* (2 \* \H(0)
          + \H(1))
          \biggr)
       +  \colour4colour{\cf^2}  \*  \biggl(
          - 4 \* (2 + 7 \* x) \* \H(1)
          - {8 \over 5} \* (3 + 2 \* x) \* (2 \* \Hh(0,0)
          + 5 \* \H(0))
          + 2 \* (6 - 31 \* x)
  \nonumber\\&& \mbox{}
          + {8 \over 5} \* (6 - x) \* \z2
          + 8  \* x \* (8 \* \z3 + 2 \* \Hh(1,0)
          + 2 \* \Hh(1,1)
          - 8 \* \Hhh(-1,-1,0) 
            + 4 \* \Hhh(-1,0,0)
          - 4 \* \Hhh(1,0,0)
          - 4 \* \H(-1) \* \z2 + 4 \* \H(1) \* \z2 
  \nonumber\\&& \mbox{}
	    + 3 \* \H(2))
          \biggr)
       + \colour4colour{\ca \* \cf}  \*  \biggl(
            {8 \over 3} \* (3 + 14 \* x) \* \H(0) 
          - {8 \over 5} \* (3 + 7 \* x) \* (\z2 - \Hh(0,0))
          - {2 \over 9} \* (66 - 317 \* x)
          - {4 \over 3} \* x \* (24 \* \z3 
  \nonumber\\&& \mbox{}
	    - 24 \* \Hhh(-1,-1,0)
          + 12 \* \Hhh(-1,0,0) 
	    - 12 \* \Hhh(1,0,0)
          - 12 \* \H(-1) \* \z2 - 23 \* \H(1) 
	    + 12 \* \H(1) \* \z2)
          \biggr)
\:\: , 
\eea
\bea
&& c^{(2)}_{L,\rm{g}}(x) \:\: = \:\:  
          \colour4colour{\cf \* \nf}  \*  \biggl(
	    {16 \over 15} \* (
          - 6 \* \pqg( - x) \* (1 - x)
          + \pgq( - x) \* (1 - x^{-1})
          + (7 - 3 \* x)) \* \Hh(-1,0)
  \nonumber\\&& \mbox{}
          + {32 \over 15} \* (3 - 13 \* x) \* \Hh(0,0) 
          + {8 \over 5} \* \pqg(x) \* (4 \* (1 + x) \* (\z2 - \Hh(0,0)) 
	    + (21 + 6 \* \H(0)
          + 10 \* \H(1)))
  \nonumber\\&& \mbox{}
          + {16 \over 15} \* \pgq(x) \* (1 - \H(0))
          - {8 \over 3} \* (7 + 8 \* x) \* \H(0)
          - 8 \* (3 - x) \* \H(1) 
          - 16 \* x \* \H(2)
          - {16 \over 15} \* (6 - 11 \* x) \* \z2
          - {8 \over 3} \* (15 - 2 \* x)
          \biggr)
  \nonumber\\&& \mbox{}
       +  \colour4colour{\ca \* \nf}  \*  \biggl(
            16 \* \pqg( - x) \* \Hh(-1,0)
          - {8 \over 9} \* \pqg(x) \* (53 - 18 \* \z2 + 18 \* \Hh(1,0)
          + 18 \* \Hh(1,1) 
	      + 117 \* \H(0)
          + 87 \* \H(1)
          + 18 \* \H(2))
  \nonumber\\&& \mbox{}    
          - {8 \over 9} \* \pgq(x) \* (1 + 3 \* \H(1))
          + 40 \* (3 - 2 \* x) \* \H(0)
          + 8 \* (11 - x) \* \H(1) 
          + 16 \* (1 + 4 \* x) \* \H(2)
          - 16 \* (1 + 2 \* x) \* \z2
  \nonumber\\&& \mbox{}
          + {8 \over 3} \* (19 - x)
          - 16 \* (\Hh(-1,0)
          - 6 \* \Hh(0,0) \* x - \Hh(1,0)
          - \Hh(1,1))
          \biggr)
\:\: , 
\eea
\bea
&& c^{(2)}_{L,\rm{ps}}(x) \:\: = \:\: 
         \colour4colour{\cf \* \nf}  \*  \biggl(
            {16 \over 9} \* \pqg(x) \* (5 - 9 \* \H(0)
          - 3 \* \H(1))
          - {8 \over 9} \* \pgq(x) \* (1 + 3 \* \H(1))
          + 16 \* (2 - 3 \* x) \* \H(0)
  \nonumber\\&& \mbox{}
          + 8 \* (2 - x) \* \H(1)
          - {8 \over 3} \* (2 + x)
          - 16 \* x \* (\z2 - 2 \* \Hh(0,0)
          - \H(2))
          \biggr)
\:\: .
\eea
\normalsize
 
\noindent
The third-order (N$^2$LO) non-singlet coefficient function is given by
\small
\bea
&& c^{(3)}_{L,\rm{q}}(x) \:\: = \:\: 
          \colour4colour{\dabcnc} \* \fl11  \*  \biggl(
            {128 \over 3} \* (3 + 14 \* x) \* \Hh(-2,0)
          - {256 \over 15} \* (12 - 13 \* x) \* \Hh(-2,2)
          + {128 \over 15} \* (4 - 51 \* x) \* \Hh(-1,0) \* \z2
  \nonumber\\&& \mbox{}
          - {128 \over 15} \* (13 + 38 \* x) \* \Hh(-1,0)
          - {64 \over 15} \* (8 + 233 \* x) \* \Hh(-1,2)
          - {128 \over 15} \* (89 - 70 \* x) \* \Hh(0,0)
          + {512 \over 15} \* (14 - 9 \* x) \* \Hh(1,0) \* \z2
  \nonumber\\&& \mbox{}
          - {256 \over 15} \* (6 + x) \* \Hhh(-2,0,0)
          - {64 \over 3} \* (8 + 21 \* x) \* \Hhh(-1,-1,0)
          + {1024 \over 15} \* (1 - 4 \* x) \* \Hhh(-1,0,0)
          - {256 \over 3} \* (3 - 2 \* x) \* \Hhh(0,0,0)
  \nonumber\\&& \mbox{}
          + {128 \over 15} \* (18 - 13 \* x) \* \Hhh(1,0,0)
          + {256 \over 15} \* (9 - 29 \* x) \* \Hhh(2,0,0)
          + {64 \over 5} \* \pqg( - x) \* (2 \* (3 - 13 \* x) \* \Hh(-1,0)
  \nonumber\\&& \mbox{}
          + 2 \* (17 - 10 \* x) \* \Hh(-1,2)
          + 4 \* (3 - 5 \* x) \* \Hhh(-1,0,0)
          - (29 - 30 \* x) \*  \H(-1) \* \z2 
          + 4 \* (1 - x) \* (4 \* \Hh(-2,2)
          + 4 \* \Hh(-1,-1) \* \z2
  \nonumber\\&& \mbox{}
          - 2 \* \Hh(-1,0) \* \z2 
          + 2 \* \Hh(-1,3)
          + 2 \* \Hhh(-2,0,0)
          - 4 \* \Hhh(-1,-1,2)
          - 2 \* \Hhhh(-1,-1,0,0)
          - 4 \* \H(-2) \* \z2 
          - 3 \* \H(-1) \* \z3 )
  \nonumber\\&& \mbox{}
          - 10 \* (1 + 2 \* x) \* (\Hh(-2,0)
          - \Hhh(-1,-1,0) ))
          + {64 \over 25} \* \pqg(x) \* (100 \* (1 + x) \* \Hhh(0,0,0)
          + 20 \* (1 - 4 \* x) \* \H(0) \* \z3 
  \nonumber\\&& \mbox{}
          - 10 \* (13 + 20 \* x) \* \H(0) \* \z2 
          + 25 \* (1 - 2 \* x) \* \H(1) \* \z2 
          + 20 \* (4 - x) \* \H(1) \* \z3 
          + 10 \* (3 + 10 \* x) \* \H(3)
  \nonumber\\&& \mbox{}
          - 20 \* (3 - 2 \* x) \* (\Hh(1,0) \* \z2 
          - \Hh(1,3)
          + \Hhh(2,0,0)
          + \Hhhh(1,1,0,0))
          + 8 \* (7 - 3 \* x) \* \z2^2 
          - 10 \* (23 + 13 \* x) \* (\z2 - \Hh(0,0))
  \nonumber\\&& \mbox{}
          - 5 \* (43 + 50 \* x) \* \z3 
          + 10 \* (3 - 10 \* \Hh(0,0) \* \z2 
          - 4 \* \Hhh(1,0,0)
          + 13 \* \H(0)
          + 10 \* \H(1)
          + 10 \* \H(2)
          + 10 \* \H(4) ))
  \nonumber\\&& \mbox{}
          + {32 \over 15} \* \pgq( - x) \* (8 \* (3 - 13 \* x^{-1}) \* \Hh(-1,0)
          + 2 \* (53 - 40 \* x^{-1}) \* \Hh(-1,2)
          + 10 \* (1 + 8 \* x^{-1}) \* \Hhh(-1,-1,0)
  \nonumber\\&& \mbox{}
          + 16 \* (3 - 5 \* x^{-1}) \* \Hhh(-1,0,0)
          - (101 - 120 \* x^{-1}) \* \H(-1) \* \z2 
          + 16 \* (1 - x^{-1}) \* (4 \* \Hh(-1,-1) \* \z2 
          - 2 \* \Hh(-1,0) \* \z2 
  \nonumber\\&& \mbox{}
          + 2 \* \Hh(-1,3)
          - 4 \* \Hhh(-1,-1,2)
          - 2 \* \Hhhh(-1,-1,0,0)
          - 3 \* \H(-1) \* \z3))
          + {32 \over 15} \* \pgq(x) \* (5 \* (1 - 8 \* x^{-1}) \* \H(1) \* \z2 
  \nonumber\\&& \mbox{}
          + 4 \* (11 - 4 \* x^{-1}) \* \H(1) \* \z3 
          - 4 \* (7 - 8 \* x^{-1}) \* (\Hh(1,0) \* \z2
          - \Hh(1,3)
          + \Hhhh(1,1,0,0))
          + 8 \* (3 
          + 8 \* \z3 
          + 20 \* \z2 
          - 10 \* \Hh(0,0)
  \nonumber\\&& \mbox{}
          - 4 \* \Hhh(1,0,0)
          - 3 \* \H(0)
          + 10 \* \H(1)
          - 10 \* \H(2) ))
          + {128 \over 15} \* (24 - 11 \* x) \* \H(-2) \* \z2
          - {32 \over 15} \* (24 - 361 \* x) \* \H(-1) \* \z2
  \nonumber\\&& \mbox{}
          - {256 \over 15} \* (3 - 28 \* x) \* \H(0) \* \z3
          + {128 \over 15} \* (53 - 43 \* x) \* \H(0) \* \z2
          - {128 \over 15} \* (55 - 63 \* x) \* \H(0)
          - {32 \over 3} \* (8 - 21 \* x) \* \H(1) \* \z2
  \nonumber\\&& \mbox{}
          - {256 \over 15} \* (29 - 39 \* x) \* \H(1) \* \z3
          - {128 \over 15} \* (63 - 13 \* x) \* \H(1)
          + {128 \over 15} \* (1 - 64 \* x) \* \H(2)
          - {2944 \over 15} \* (1 - x) \* \H(3)
  \nonumber\\&& \mbox{}
          - {256 \over 15} \* (1 - 9 \* x) \* (4 \* \Hh(-1,-1) \* \z2 
          + 2 \* \Hh(-1,3)
          - 4 \* \Hhh(-1,-1,2)
          - 2 \* \Hhhh(-1,-1,0,0)
          - 3 \* \H(-1) \* \z3)
  \nonumber\\&& \mbox{}
          + 256 \* (1 - 2 \* x) \* (\Hh(0,0) \* \z2 - \H(4))
          - 128 \* (1 + x)
          + {832 \over 5} \* (3 - 13 \* x) \* \z3
          + {256 \over 5} \* (13 - 10 \* x) \* \z2
  \nonumber\\&& \mbox{}
          - {64 \over 25} \* (56 - 181 \* x) \* \z2^2
          - {128 \over 15} \* (56 - 51 \* x) \* (\Hh(1,3)
          - \Hhhh(1,1,0,0) )
          + {128 \over 5} \* x \* (100 \* \z5
          + 10 \* \z2 \* \z3 
          - 20 \* \Hh(-1,0) \* \z3 
  \nonumber\\&& \mbox{}
          - 20 \* \Hh(-1,4) 
          + 20 \* \Hh(2,0) \* \z2 
          - 20 \* \Hh(2,3) 
          + 10 \* \Hhh(-2,-1,0) 
          + 20 \* \Hhh(-1,0,0) \* \z2 
          + 10 \* \Hhh(1,-2,0) 
          + 5 \* \Hhhh(-1,0,0,0)
  \nonumber\\&& \mbox{}
          + 20 \* \Hhhh(-1,2,0,0) 
          - 5 \* \Hhhh(1,0,0,0) 
          + 20 \* \Hhhh(2,1,0,0) 
          - 16 \* \H(-1) \* \z2^2 
          - 20 \* \H(2) \* \z3 
          - 5 \* \H(2) \* \z2) 
	  \biggr)
  \nonumber\\&& \mbox{}
       +  \colour4colour{\cf \* \biggl(\cf-{\ca \over 2}\biggr)^2}  \*  (
            32 \* \gfunct1(x)
          )
  \nonumber\\&& \mbox{}
       +  \colour4colour{\cf \* \nf \* \biggl(\cf-{\ca \over 2}\biggr)} 
            \*  \biggl(
          - {64 \over 15} \* (31 - 14 \* x) \* \Hh(-2,0)
          - {16 \over 75} \* (1647 + 1397 \* x) \* \Hh(-1,0) 
          - {64 \over 15} \* (19 + 14 \* x) \* \Hh(-1,2) 
  \nonumber\\&& \mbox{}
          + {64 \over 45} \* (57 + 172 \* x) \* \Hhh(-1,-1,0)
          - {64 \over 45} \* (171 + 161 \* x) \* \Hhh(-1,0,0)
          - {16 \over 75} \* \pqg( - x) \* (3 \* (33 - 53 \* x) \* \Hh(-1,0)
  \nonumber\\&& \mbox{}
          - 20 \* (7 + 3 \* x) \* \Hh(-1,2)
          + 180 \* (1 - x) \* \Hhh(-1,0,0)
          + 10 \* (1 + 9 \* x) \* \H(-1) \* \z2 
          + 20 \* (13 - 3 \* x) \* (\Hh(-2,0)
  \nonumber\\&& \mbox{}
          - \Hhh(-1,-1,0)))
          - {32 \over 225} \* \pgq( - x) \* ((89 - 149 \* x^{-1}) \* \Hh(-1,0)
          - 30 \* (3 + 2 \* x^{-1}) \* \Hh(-1,2)
  \nonumber\\&& \mbox{}
          - 30 \* (7 - 2 \* x^{-1}) \* \Hhh(-1,-1,0)
          - 15 \* (1 - 6 \* x^{-1}) \* \H(-1) \* \z2 
          + 60 \* (1 - x^{-1}) \* (2 \* \Hh(-2,0)
          + 3 \* \Hhh(-1,0,0)))
  \nonumber\\&& \mbox{}
          + {32 \over 45} \* (171 + 256 \* x) \* \H(-1) \* \z2
          \biggr)
  \nonumber\\&& \mbox{}
       +  \colour4colour{\cf \* \nf^2}  \*  \biggl(
          - {32 \over 27} \* (3 - 25 \* x) \* \H(0)
          - {16 \over 27} \* (6 - 25 \* x) \* \H(1)
          - {8 \over 81} \* (114 - 317 \* x)
          - {32 \over 9} \* x \* (2 \* \z2
          - 3 \* \Hh(0,0)
          - \Hh(1,0)
  \nonumber\\&& \mbox{}
          - \Hh(1,1)
          - 2 \* \H(2) )
          \biggr)
  \nonumber\\&& \mbox{}
       +  \colour4colour{\cf^2 \* \biggl(\cf-{\ca \over 2}\biggr)}  \*  \biggl(
            {64 \over 5} \* (19 + 9 \* x) \* (\Hhh(-1,2,0)
          + \Hhh(-1,2,1))
          \biggr)
  \nonumber\\&& \mbox{}
       +  \colour4colour{\cf^2 \* \nf}  \*  \biggl(
            {8 \over 225} \* (3429 + 5296 \* x) \* \Hh(0,0)
          + {8 \over 9} \* (60 - 83 \* x) \* \Hh(1,0)
          + {8 \over 9} \* (30 - 7 \* x) \* \Hh(1,1)
          + {16 \over 3} \* (7 - 15 \* x) \* \Hh(1,2)
  \nonumber\\&& \mbox{}
          - {16 \over 3} \* (7 + 4 \* x) \* \Hh(2,0)
          + {8 \over 15} \* (72 + 13 \* x) \* \Hhh(0,0,0)
          - {16 \over 3} \* (7 + 3 \* x) \* \Hhh(1,1,0)
          - {16 \over 75} \* \pqg(x) \* (3 \* (113 + 53 \* x) \* \Hh(0,0)
  \nonumber\\&& \mbox{}
          + 180 \* (1 + x) \* \Hhh(0,0,0)
          - 5 \* (99 + 49 \* x) \* \H(0) \* \z2 
          - 5 \* (61 + 31 \* x) \* \H(1) \* \z2 
          + 5 \* (87 + 37 \* x) \* \H(3)
  \nonumber\\&& \mbox{}
          + 25 \* (7 + 5 \* x) \* (\Hh(1,2)
          - \Hh(2,0)
          - \Hhh(1,1,0) )
          + 25 \* (11 + 9 \* x) \* \z3 
          - (94 + 159 \* x) \* \z2 
          + 219 + 125 \* \Hh(1,0)
  \nonumber\\&& \mbox{}
          + 154 \* \H(0)
          - 65 \* \H(1)
          - 65 \* \H(2))
          + {32 \over 225} \* \pgq(x) \* (15 \* (7 + 2 \* x^{-1}) \* \H(1) * \z2 
          - 209
          - 120 \* \z2 
          + 180 \* \Hh(0,0)
  \nonumber\\&& \mbox{}
          + 89 \* \H(0)
          - 60 \* \H(1)
          + 60 \* \H(2)) 
          - {16 \over 15} \* (99 - 194 \* x) \* \H(0) \* \z2
          + {8 \over 27} \* (252 + 965 \* x) \* \H(0)
          + {16 \over 45} \* (9 - 119 \* x) \* \H(1) \* \z2
  \nonumber\\&& \mbox{}
          - {16 \over 27} \* (96 - 383 \* x) \* \H(1)
          + 8 \* (6 - x) \* \H(2)
          + {16 \over 15} \* (87 - 202 \* x) \* \H(3)
          + {16 \over 9} \* (33 - 119 \* x) \* \z3
  \nonumber\\&& \mbox{}
          - {8 \over 225} \* (2784 + 6841 \* x) \* \z2
          - {1 \over 135} \* (5144 - 26129 \* x)
          - {16 \over 45} \* x \* (12 \* \z2^2 
          - 240 \* \Hh(-2,2) 
          + 240 \* \Hh(-1,-1) \* \z2
  \nonumber\\&& \mbox{}
          - 120 \* \Hh(-1,0) \* \z2 
          + 420 \* \Hh(1,0) \* \z2 
          + 300 \* \Hh(1,1) \* \z2 
          - 300 \* \Hh(1,3) 
          + 195 \* \Hh(2,1) 
          - 480 \* \Hhh(-2,-1,0) 
          + 120 \* \Hhh(-2,0,0) 
  \nonumber\\&& \mbox{}
          - 480 \* \Hhh(-1,-2,0) 
          - 240 \* \Hhh(1,-2,0) 
          - 245 \* \Hhh(1,0,0) 
          + 120 \* \Hhh(1,1,1) 
          - 60 \* \Hhh(1,1,2) 
          + 60 \* \Hhh(1,2,0)
          - 120 \* \Hhh(2,0,0) 
  \nonumber\\&& \mbox{}
          + 480 \* \Hhhh(-1,-1,-1,0) 
          - 720 \* \Hhhh(-1,-1,0,0) 
          + 360 \* \Hhhh(-1,0,0,0) 
          - 360 \* \Hhhh(1,0,0,0) 
          + 60 \* \Hhhh(1,1,1,0) 
          - 240 \* \H(-1) \* \z3 
  \nonumber\\&& \mbox{}
          + 240 \* \H(0) \* \z3 
          + 180 \* \H(1) \* \z3 
          + 240 \* \H(2) \* \z2)
          \biggr)
  \nonumber\\&& \mbox{}
       +  \colour4colour{\cf^3}  \*  \biggl(
            {64 \over 3} \* (9 + 7 \* x) \* \Hh(-3,0)
          + {32 \over 75} \* (1024 + 4849 \* x) \* \Hh(-2,0) 
          + {64 \over 15} \* (87 + 422 \* x) \* \Hh(-2,2) 
  \nonumber\\&& \mbox{}
          + {32 \over 15} \* (1667 + 1977 \* x) \* \Hh(-1,-1) \* \z2
          - {32 \over 3} \* (259 + 270 \* x) \* \Hh(-1,0) \* \z2
          - {8 \over 225} \* (11203 - 14427 \* x) \* \Hh(-1,0) 
  \nonumber\\&& \mbox{}
          + {16 \over 75} \* (8286 + 14761 \* x) \* \Hh(-1,2)
          + {32 \over 3} \* (212 + 201 \* x) \* \Hh(-1,3)
          + {16 \over 15} \* (341 + 424 \* x) \* \Hh(0,0) \* \z2
  \nonumber\\&& \mbox{}
          - {4 \over 225} \* (22446 + 68669 \* x) \* \Hh(0,0)
          + 4 \* (48 - 83 \* x) \* \Hh(1,0)
          - {16 \over 5} \* (271 - 146 \* x) \* \Hh(1,0) \* \z2
  \nonumber\\&& \mbox{}
          - {32 \over 5} \* (39 - 49 \* x) \* \Hh(1,1) \* \z2
          + 4 \* (40 - 73 \* x) \* \Hh(1,1)
          - 96 \* (1 + 4 \* x) \* \Hh(1,2)
          + {112 \over 15} \* (49 + 36 \* x) \* \Hh(1,3)
  \nonumber\\&& \mbox{}
          - 16 \* (9 + 16 \* x) \* \Hh(2,0)
          - 80 \* (2 + 5 \* x) \* \Hh(2,1)
          - {16 \over 5} \* (12 - 47 \* x) \* \Hh(3,0)
          - {192 \over 5} \* (1 - 6 \* x) \* \Hh(3,1)
  \nonumber\\&& \mbox{}
          - {64 \over 5} \* (61 - 39 \* x) \* \Hhh(-2,-1,0)
          + {32 \over 15} \* (354 + 359 \* x) \* \Hhh(-2,0,0)
          - {64 \over 5} \* (39 + 44 \* x) \* \Hhh(-1,-2,0)
  \nonumber\\&& \mbox{}
          - {16 \over 25} \* (2802 + 1477 \* x) \* \Hhh(-1,-1,0)
          - {64 \over 3} \* (155 + 174 \* x) \* \Hhh(-1,-1,2)
          + {32 \over 75} \* (4703 + 6328 \* x) \* \Hhh(-1,0,0)
  \nonumber\\&& \mbox{}
          - {8 \over 75} \* (5339 + 14511 \* x) \* \Hhh(0,0,0)
          - {64 \over 3} \* (26 - 41 \* x) \* \Hhh(1,-2,0)
          + {8 \over 15} \* (52 + 1323 \* x) \* \Hhh(1,0,0)
  \nonumber\\&& \mbox{}
          - 16 \* (8 + 19 \* x) \* \Hhh(1,1,0)
          - 48 \* (2 + 7 \* x) \* \Hhh(1,1,1)
          - {16 \over 15} \* (19 + 26 \* x) \* \Hhh(2,0,0)
          + {64 \over 5} \* (39 + 79 \* x) \* \Hhhh(-1,-1,-1,0)
  \nonumber\\&& \mbox{}
          - {32 \over 3} \* (248 + 243 \* x) \* \Hhhh(-1,-1,0,0)
          + {32 \over 3} \* (116 + 99 \* x) \* \Hhhh(-1,0,0,0)
          - {8 \over 15} \* (264 + 391 \* x) \* \Hhhh(0,0,0,0)
  \nonumber\\&& \mbox{}
          + {64 \over 15} \* (119 - 129 \* x) \* \Hhhh(1,0,0,0)
          - {16 \over 15} \* (223 + 282 \* x) \* \Hhhh(1,1,0,0)
          - {8 \over 25} \* \pqg( - x) \* (4 \* (188 + 197 \* x) \* \Hh(-2,0)
  \nonumber\\&& \mbox{}
          + (371 + 337 \* x) \* \Hh(-1,0)
          - 4 \* (222 - 197 \* x) \* \Hh(-1,2)
          - 4 \* (208 + 197 \* x) \* \Hhh(-1,-1,0)
          + 4 \* (13 + 237 \* x) \* \Hhh(-1,0,0)
  \nonumber\\&& \mbox{}
          + 2 \* (236 - 591 \* x) \* \H(-1) \* \z2 
          - 20 \* (1 - x) \* (10 \* \Hh(-3,0)
          + 70 \* \Hh(-2,2)
          + 73 \* \Hh(-1,-1) \* \z2 
          - 55 \* \Hh(-1,0) \* \z2 
  \nonumber\\&& \mbox{}
          + 50 \* \Hh(-1,3)
          - 6 \* \Hhh(-2,-1,0)
          + 50 \* \Hhh(-2,0,0)
          - 6 \* \Hhh(-1,-2,0)
          - 70 \* \Hhh(-1,-1,2)
          + 6 \* \Hhh(-1,2,0)
          + 6 \* \Hhh(-1,2,1)
  \nonumber\\&& \mbox{}
          + 4 \* \Hhh(1,-2,0)
          + 6 \* \Hhhh(-1,-1,-1,0)
          - 50 \* \Hhhh(-1,-1,0,0)
          + 20 \* \Hhhh(-1,0,0,0)
          - 73 \* \H(-2) \* \z2 
          - 60 \* \H(-1) \* \z3))
  \nonumber\\&& \mbox{}
          - {8 \over 75} \* \pqg(x) \* (3 \* (125 - 337 \* x) \* \Hh(0,0)
          + 10 \* (341 + 591 \* x) \* \Hh(0,0) \* \z2 
          + 10 \* (11 + 261 \* x) \* \Hh(1,0) \* \z2 
  \nonumber\\&& \mbox{}
          - 36 \* (109 + 79 \* x) \* \Hhh(0,0,0)
          + 6 \* (153 + 788 \* x) \* \H(0) \* \z2 
          + 10 \* (373 + 123 \* x) \* \H(0) \* \z3 
          + 10 \* (13 - 237 \* x) \* \H(1) \* \z3 
  \nonumber\\&& \mbox{}
          - 6 \* (208 - 197 \* x) \* \H(1) \* \z2 
          + 6 \* (241 - 394 \* x) \* \H(3)
          - 10 \* (281 + 531 \* x) \* \H(4)
          + 60 \* (1 + x) \* (3 \* \Hh(1,1) \* \z2 
          - 6 \* \Hh(3,0)
  \nonumber\\&& \mbox{}
          - 6 \* \Hh(3,1)
          - 22 \* \Hhhh(0,0,0,0)
          - 2 \* \Hhhh(1,0,0,0)
          + 3 \* \H(2) \* \z2)
          + 10 \* (19 - 231 \* x) \* (\Hh(1,3)
          - \Hhh(2,0,0)
          - \Hhhh(1,1,0,0))
  \nonumber\\&& \mbox{}
          - 3 \* (45 - 337 \* x) \* \z2 
          + 30 \* (321 + 197 \* x) \* \z3 
          - 2 \* (359 + 1359 \* x) \* \z2^2 
          + 3 \* (451 - 120 \* \Hh(1,0)
          - 120 \* \Hh(1,1)
  \nonumber\\&& \mbox{}
          - 120 \* \Hh(2,0)
          - 120 \* \Hh(2,1)
          - 770 \* \Hhh(1,0,0)
          + 953 \* \H(0)
          + 502 \* \H(1)
          + 382 \* \H(2)))
  \nonumber\\&& \mbox{}
          - {8 \over 225} \* \pgq( - x) \* (240 \* (1 + 4 \* x^{-1}) \* \Hh(-2,0)
          + 2 \* (803 + 1621 \* x^{-1}) \* \Hh(-1,0)
          - 6 \* (493 - 768 \* x^{-1}) \* \Hh(-1,2)
  \nonumber\\&& \mbox{}
          - 18 \* (159 + 256 \* x^{-1}) \* \Hhh(-1,-1,0)
          + 48 \* (9 + 116 \* x^{-1}) \* \Hhh(-1,0,0)
          + 3 \* (509 - 2304 \* x^{-1}) \* \H(-1) \* \z2
  \nonumber\\&& \mbox{}
          - 120 \* (1 - x^{-1}) \* (12 \* \Hh(-2,2)
          + 73 \* \Hh(-1,-1) \* \z2 
          - 55 \* \Hh(-1,0) \* \z2 
          + 50 \* \Hh(-1,3)
          - 12 \* \Hhh(-2,-1,0)
          + 12 \* \Hhh(-2,0,0)
  \nonumber\\&& \mbox{}
          - 6 \* \Hhh(-1,-2,0)
          - 70 \* \Hhh(-1,-1,2)
          + 6 \* \Hhh(-1,2,0)
          + 6 \* \Hhh(-1,2,1)
          - 4 \* \Hhh(1,-2,0)
          + 6 \* \Hhhh(-1,-1,-1,0)
          - 50 \* \Hhhh(-1,-1,0,0)
  \nonumber\\&& \mbox{}
          + 20 \* \Hhhh(-1,0,0,0)
          - 18 \* \H(-2) \* \z2 
          - 60 \* \H(-1) \* \z3))
          - {8 \over 225} \* \pgq(x) \* 
            (30 \* (49 + 174 \* x^{-1}) \* \Hh(1,0) \* \z2 
  \nonumber\\&& \mbox{}
          - 30 \* (33 + 158 \* x^{-1}) \* \H(1) \* \z3 
          - 9 \* (159 - 256 \* x^{-1}) \* \H(1) \* \z2 
          + 120 \* (1 + x^{-1}) \* (3 \* \Hh(1,1) \* \z2 
          - 2 \* \Hhhh(1,0,0,0)
  \nonumber\\&& \mbox{}
          + 6 \* \H(2) \* \z2 )
          - 30 \* (29 + 154 \* x^{-1}) \* (\Hh(1,3)
          - \Hhhh(1,1,0,0) )
          + 2 \* (1043 
          - 1230 \* \z3 
          - 138 \* \z2 
          + 2184 \* \Hh(0,0)
  \nonumber\\&& \mbox{}
          - 360 \* \Hh(1,0)
          - 360 \* \Hh(1,1)
          + 360 \* \Hh(2,0)
          + 360 \* \Hh(2,1)
          + 1320 \* \Hhh(0,0,0)
          - 2310 \* \Hhh(1,0,0)
          - 803 \* \H(0)
          - 5910 \* \H(0) \* \z2 
  \nonumber\\&& \mbox{}
          + 1566 \* \H(1)
          - 2286 \* \H(2)
          + 5310 \* \H(3) ))
          - {32 \over 15} \* (357 + 727 \* x) \* \H(-2) \* \z2
          - 16 \* (182 + 211 \* x) \* \H(-1) \* \z3
  \nonumber\\&& \mbox{}
          - {8 \over 75} \* (24978 + 33953 \* x) \* \H(-1) \* \z2
          + {16 \over 15} \* (373 + 732 \* x) \* \H(0) \* \z3
          + {16 \over 25} \* (1623 + 6677 \* x) \* \H(0) \* \z2
  \nonumber\\&& \mbox{}
          + {8 \over 225} \* (16294 + 41789 \* x) \* \H(0)
          - {16 \over 15} \* (469 - 1389 \* x) \* \H(1) \* \z3
          - {8 \over 15} \* (2263 - 3268 \* x) \* \H(1)
  \nonumber\\&& \mbox{}
          - {8 \over 25} \* (2502 - 2677 \* x) \* \H(1) \* \z2
          - {32 \over 5} \* (61 + 84 \* x) \* \H(2) \* \z2
          - {4 \over 15} \* (304 + 5239 \* x) \* \H(2)
  \nonumber\\&& \mbox{}
          - {16 \over 75} \* (3287 + 15113 \* x) \* \H(3)
          - {16 \over 15} \* (281 + 284 \* x) \* \H(4)
          - {16 \over 75} \* (359 - 1069 \* x) \* \z2^2
          + {43 \over 450} \* (1008 + 4657 \* x)
  \nonumber\\&& \mbox{}
          + {16 \over 15} \* (1249 + 3046 \* x) \* \z3
          + {4 \over 225} \* (30018 + 75467 \* x) \* \z2
          - {16 \over 5} \* x \* (615 \* \z5 
          + 300 \* \z2 \* \z3 
          + 40 \* \Hh(-3,2) 
  \nonumber\\&& \mbox{}
          - 240 \* \Hh(-2,-1) \* \z2 
          + 140 \* \Hh(-2,0) \* \z2 
          - 60 \* \Hh(-2,3) 
          - 760 \* \Hh(-1,-2) \* \z2 
          - 1020 \* \Hh(-1,-1) \* \z3
          + 420 \* \Hh(-1,0) \* \z3
  \nonumber\\&& \mbox{}
          - 40 \* \Hh(-1,2) \* \z2
          - 440 \* \Hh(-1,4)
          - 40 \* \Hh(0,0) \* \z3
          + 280 \* \Hh(1,-2) \* \z2
          - 500 \* \Hh(1,0) \* \z3
          - 300 \* \Hh(1,1) \* \z3
          - 120 \* \Hh(1,2) \* \z2
  \nonumber\\&& \mbox{}
          + 360 \* \Hh(1,4)
          - 100 \* \Hh(2,0) \* \z2
          - 120 \* \Hh(2,1) \* \z2
          - 90 \* \Hh(2,2)
          + 20 \* \Hh(2,3)
          + 80 \* \Hhh(-3,-1,0)
          - 20 \* \Hhh(-3,0,0)
  \nonumber\\&& \mbox{}
          + 160 \* \Hhh(-2,-2,0)
          + 120 \* \Hhh(-2,-1,2)
          + 200 \* \Hhh(-1,-3,0)
          + 640 \* \Hhh(-1,-2,2)
          + 1240 \* \Hhh(-1,-1,-1) \* \z2
  \nonumber\\&& \mbox{}
          - 960 \* \Hhh(-1,-1,0) \* \z2
          + 760 \* \Hhh(-1,-1,3)
          + 560 \* \Hhh(-1,0,0) \* \z2
          - 40 \* \Hhh(-1,3,0)
          - 40 \* \Hhh(-1,3,1)
          - 40 \* \Hhh(1,-3,0)
  \nonumber\\&& \mbox{}
          - 320 \* \Hhh(1,-2,2)
          - 480 \* \Hhh(1,0,0) \* \z2
          - 320 \* \Hhh(1,1,0) \* \z2
          - 120 \* \Hhh(1,1,1) \* \z2
          - 60 \* \Hhh(1,1,2)
          + 120 \* \Hhh(1,1,3)
          - 60 \* \Hhh(1,2,0)
  \nonumber\\&& \mbox{}
          - 70 \* \Hhh(1,2,1)
          + 40 \* \Hhh(1,3,0)
          + 40 \* \Hhh(1,3,1)
          - 100 \* \Hhh(2,1,0)
          - 90 \* \Hhh(2,1,1)
          + 20 \* \Hhh(3,0,0)
          - 240 \* \Hhhh(-2,-1,-1,0)
  \nonumber\\&& \mbox{}
          + 260 \* \Hhhh(-2,-1,0,0)
          - 120 \* \Hhhh(-2,0,0,0)
          - 240 \* \Hhhh(-1,-2,-1,0)
          + 600 \* \Hhhh(-1,-2,0,0)
          - 240 \* \Hhhh(-1,-1,-2,0)
  \nonumber\\&& \mbox{}
          - 1120 \* \Hhhh(-1,-1,-1,2)
          + 80 \* \Hhhh(-1,-1,2,0)
          + 80 \* \Hhhh(-1,-1,2,1)
          + 80 \* \Hhhh(-1,2,0,0)
          - 80 \* \Hhhh(1,-2,-1,0)
  \nonumber\\&& \mbox{}
          - 120 \* \Hhhh(1,-2,0,0)
          - 160 \* \Hhhh(1,1,-2,0)
          - 80 \* \Hhhh(1,1,1,0)
          - 60 \* \Hhhh(1,1,1,1)
          + 120 \* \Hhhh(2,0,0,0)
          + 100 \* \Hhhh(2,1,0,0)
  \nonumber\\&& \mbox{}
          + 240 \* \Hhhhh(-1,-1,-1,-1,0)
          - 1000 \* \Hhhhh(-1,-1,-1,0,0)
          + 480 \* \Hhhhh(-1,-1,0,0,0)
          - 200 \* \Hhhhh(-1,0,0,0,0)
          + 200 \* \Hhhhh(1,0,0,0,0)
  \nonumber\\&& \mbox{}
          + 240 \* \Hhhhh(1,1,0,0,0)
          - 40 \* \Hhhhh(1,1,1,0,0)
          + 210 \* \H(-2) \* \z3
          - 118 \* \H(-1) \* \z2^2
          + 12 \* \H(0) \* \z2^2
          + 54 \* \H(1) \* \z2^2
          - 310 \* \H(2) \* \z3
  \nonumber\\&& \mbox{}
          - 40 \* \H(3) \* \z2 )
          \biggr)
  \nonumber\\&& \mbox{}
       +  \colour4colour{\ca \* \cf \* \nf}  \*  \biggl(
          - {8 \over 225} \* (1227 + 7223 \* x) \* \Hh(0,0)
          + {16 \over 3} \* (1 + 14 \* x) \* \Hh(1,0) \* \z2
          - {16 \over 9} \* (9 + 22 \* x) \* \Hh(1,0)
  \nonumber\\&& \mbox{}
          - {16 \over 3} \* (1 + 10 \* x) \* \Hh(1,3)
          - {32 \over 15} \* (9 + 16 \* x) \* \Hhh(0,0,0)
          + {16 \over 9} \* (3 - 26 \* x) \* \Hhh(1,0,0)
          + {16 \over 3} \* (1 - 5 \* x) \* \Hhh(2,0,0)
  \nonumber\\&& \mbox{}
          + {8 \over 75} \* \pqg(x) \* (3 \* (63 + 53 \* x) \* \Hh(0,0)
          - 10 \* (32 + 27 \* x) \* \H(0) \* \z2 
          - 20 \* (14 + 9 \* x) \* \H(1) \* \z2 
          + 10 \* (26 + 21 \* x) \* \H(3)
  \nonumber\\&& \mbox{}
          + 30 \* (1 + x) \* (5 \* \Hh(1,2)
          - 5 \* \Hh(2,0)
          + 6 \* \Hhh(0,0,0)
          - 5 \* \Hhh(1,1,0) )
          + 10 \* (1 + 3 \* x) \* (4 \* \z2^2 
          - 5 \* \Hh(0,0) \* \z2 
          - 5 \* \Hh(1,0) \* \z2 
  \nonumber\\&& \mbox{}
          + 5 \* \Hh(1,3)
          - 5 \* \Hhh(2,0,0)
          - 5 \* \Hhhh(1,1,0,0)
          + 5 \* \H(0) \* \z3 
          + 5 \* \H(1) \* \z3 
          + 5 \* \H(4))
          + 50 \* (1 + 6 \* x) \* \z3 
          + 3 \* (27 - 53 \* x) \* \z2
  \nonumber\\&& \mbox{} 
          + 3 \* (73 
          + 50 \* \Hh(1,0)
          + 50 \* \Hhh(1,0,0)
          - 7 \* \H(0)
          - 80 \* \H(1)
          - 80 \* \H(2) ))
          - {16 \over 225} \* \pgq(x) \* 
            (15 \* (7 + 2 \* x^{-1}) \* \H(1) \* \z2 
  \nonumber\\&& \mbox{}
          + 75 \* (1 + 2 \* x^{-1}) \* (\Hh(1,0) \* \z2 
          - \Hh(1,3)
          + \Hhhh(1,1,0,0)
          - \H(1) \* \z3)
          - 209 
          + 150 \* \z3 
          + 30 \* \z2 
          + 180 \* \Hh(0,0)
  \nonumber\\&& \mbox{}
          - 150 \* \Hhh(1,0,0)
          + 89 \* \H(0)
          - 150 \* \H(0) \* \z2 
          + 90 \* \H(1)
          - 90 \* \H(2)
          + 150 \* \H(3))
          - {16 \over 3} \* (1 - 9 \* x) \* \H(0) \* \z3
  \nonumber\\&& \mbox{}
          + {16 \over 15} \* (52 - 27 \* x) \* \H(0) \* \z2
          + {8 \over 27} \* (129 - 1670 \* x) \* \H(0)
          - {16 \over 3} \* (1 - 6 \* x) \* \H(1) \* \z3
          - {16 \over 45} \* (12 - 157 \* x) \* \H(1) \* \z2
  \nonumber\\&& \mbox{}
          + {8 \over 27} \* (348 - 1045 \* x) \* \H(1)
          + {32 \over 9} \* (9 - 40 \* x) \* \H(2)
          - {16 \over 15} \* (46 - 31 \* x) \* \H(3)
          + {16 \over 3} \* (1 - x) \* (\Hh(0,0) \* \z2 
          + 3 \* \Hh(2,0)
  \nonumber\\&& \mbox{}
          - \H(4) )
          - {32 \over 15} \* (2 - 3 \* x) \* \z2^2
          - {16 \over 3} \* (3 - x) \* (\Hh(1,2)
          - \Hhh(1,1,0))
          - {16 \over 9} \* (15 - 49 \* x) \* \z3
          - {8 \over 75} \* (61 - 2511 \* x) \* \z2
  \nonumber\\&& \mbox{}
          + {8 \over 405} \* (8142 - 23507 \* x)
          - {16 \over 9} \* x \* (24 \* \Hh(-2,2) 
          - 24 \* \Hh(-1,-1) \* \z2 
          + 12 \* \Hh(-1,0) \* \z2 
          + 40 \* \Hh(1,1) 
          - 30 \* \Hh(1,1) \* \z2 
  \nonumber\\&& \mbox{}
          + 48 \* \Hhh(-2,-1,0) 
          - 12 \* \Hhh(-2,0,0) 
          + 48 \* \Hhh(-1,-2,0) 
          + 24 \* \Hhh(1,-2,0) 
          + 6 \* \Hhh(1,1,2) 
          - 6 \* \Hhh(1,2,0) 
          - 48 \* \Hhhh(-1,-1,-1,0) 
  \nonumber\\&& \mbox{}
          + 72 \* \Hhhh(-1,-1,0,0) 
          - 36 \* \Hhhh(-1,0,0,0) 
          + 36 \* \Hhhh(1,0,0,0) 
          - 6 \* \Hhhh(1,1,1,0) 
          + 24 \* \H(-1) \* \z3 
          - 24 \* \H(2) \* \z2 )
          + {16 \over 3} \* \Hhhh(1,1,0,0)
          \biggr)
  \nonumber\\&& \mbox{}
       +  \colour4colour{\ca \* \cf^2}  \*  \biggl(
          - {352 \over 15} \* (3 + 8 \* x) \* \Hh(-3,0)
          + {8 \over 225} \* (16911 - 54139 \* x) \* \Hh(-2,0)
          + {32 \over 15} \* (33 - 962 \* x) \* \Hh(-2,2)
  \nonumber\\&& \mbox{}
          - {16 \over 15} \* (3507 + 3337 \* x) \* \Hh(-1,-1) \* \z2
          + {16 \over 15} \* (2451 + 2396 \* x) \* \Hh(-1,0) \* \z2
          + {4 \over 45} \* (27311 + 12515 \* x) \* \Hh(-1,0)
  \nonumber\\&& \mbox{}
          - {4136 \over 225} \* (59 + 159 \* x) \* \Hh(-1,2)
          - 48 \* (44 + 41 \* x) \* \Hh(-1,3)
          - {32 \over 15} \* (84 + 361 \* x) \* \Hh(0,0) \* \z2
  \nonumber\\&& \mbox{}
          - {4 \over 225} \* (22556 + 12599 \* x) \* \Hh(0,0)
          + {16 \over 15} \* (533 - 148 \* x) \* \Hh(1,0) \* \z2
          - {4 \over 9} \* (804 - 1087 \* x) \* \Hh(1,0)
  \nonumber\\&& \mbox{}
          + {16 \over 15} \* (117 + 163 \* x) \* \Hh(1,1) \* \z2
          - {4 \over 9} \* (474 - 281 \* x) \* \Hh(1,1)
          - {8 \over 3} \* (83 - 243 \* x) \* \Hh(1,2)
          - {496 \over 15} \* (2 + 13 \* x) \* \Hh(1,3)
  \nonumber\\&& \mbox{}
          + {8 \over 3} \* (89 + 98 \* x) \* \Hh(2,0)
          + {8 \over 3} \* (12 + 227 \* x) \* \Hh(2,1)
          + {32 \over 15} \* (183 - 787 \* x) \* \Hhh(-2,-1,0)
          - {16 \over 15} \* (234 + 229 \* x) \* \Hhh(-2,0,0)
  \nonumber\\&& \mbox{}
          + {32 \over 15} \* (117 - 248 \* x) \* \Hhh(-1,-2,0)
          + {8 \over 225} \* (32663 - 13837 \* x) \* \Hhh(-1,-1,0)
          + 32 \* (113 + 114 \* x) \* \Hhh(-1,-1,2)
  \nonumber\\&& \mbox{}
          - {8 \over 225} \* (9683 + 31483 \* x) \* \Hhh(-1,0,0)
          + {4 \over 225} \* (11022 + 64553 \* x) \* \Hhh(0,0,0)
          + {32 \over 3} \* (78 - 155 \* x) \* \Hhh(1,-2,0)
  \nonumber\\&& \mbox{}
          - {8 \over 45} \* (234 + 5521 \* x) \* \Hhh(1,0,0)
          + {8 \over 3} \* (83 + 99 \* x) \* \Hhh(1,1,0)
          + {64 \over 15} \* (39 - 74 \* x) \* \Hhh(2,0,0)
  \nonumber\\&& \mbox{}
          - {32 \over 15} \* (117 - 83 \* x) \* \Hhhh(-1,-1,-1,0)
          + 48 \* (48 + 19 \* x) \* \Hhhh(-1,-1,0,0)
          - {16 \over 5} \* (272 + 87 \* x) \* \Hhhh(-1,0,0,0)
  \nonumber\\&& \mbox{}
          + {16 \over 3} \* (18 + 49 \* x) \* \Hhhh(0,0,0,0)
          - {16 \over 5} \* (158 - 33 \* x) \* \Hhhh(1,0,0,0)
          + {16 \over 15} \* (2 + 213 \* x) \* \Hhhh(1,1,0,0)
  \nonumber\\&& \mbox{}
          + {4 \over 75} \* \pqg( - x) \* (2 \* (3583 + 1537 \* x) \* \Hh(-2,0)
          + (5081 - 2727 \* x) \* \Hh(-1,0)
          - 2 \* (2337 - 1537 \* x) \* \Hh(-1,2)
  \nonumber\\&& \mbox{}
          - 2 \* (3943 + 1537 \* x) \* \Hhh(-1,-1,0)
          + 2 \* (2243 + 457 \* x) \* \Hhh(-1,0,0)
          + (731 - 4611 \* x) \* \H(-1) \* \z2 
  \nonumber\\&& \mbox{}
          - 180 \* (1 - x) \* (6 \* \Hh(-3,0)
          + 50 \* \Hh(-2,2)
          + 51 \* \Hh(-1,-1) \* \z2 
          - 33 \* \Hh(-1,0) \* \z2 
          + 30 \* \Hh(-1,3)
          - 2 \* \Hhh(-2,-1,0)
  \nonumber\\&& \mbox{}
          + 30 \* \Hhh(-2,0,0)
          - 2 \* \Hhh(-1,-2,0)
          - 50 \* \Hhh(-1,-1,2)
          + 2 \* \Hhh(-1,2,0)
          + 2 \* \Hhh(-1,2,1)
          + 4 \* \Hhh(1,-2,0)
          + 2 \* \Hhhh(-1,-1,-1,0)
  \nonumber\\&& \mbox{}
          - 30 \* \Hhhh(-1,-1,0,0)
          + 8 \* \Hhhh(-1,0,0,0)
          - 51 \* \H(-2) \* \z2 
          - 40 \* \H(-1) \* \z3 ))
          + {4 \over 75} \* \pqg(x) \* (1680 \* (2 + 7 \* x) \* \Hh(0,0) \* \z2
  \nonumber\\&& \mbox{}
          + 101 \* (113 + 27 \* x) \* \Hh(0,0)
          - 10 \* (421 + 275 \* x) \* \Hh(2,0)
          - 2 \* (997 + 457 \* x) \* \Hhh(0,0,0)
          + 60 \* (163 + 23 \* x) \* \H(0) \* \z3 
  \nonumber\\&& \mbox{}
          - 2 \* (4091 - 1699 \* x) \* \H(0) \* \z2 
          - (7793 + 1213 \* x) \* \H(1) \* \z2 
          + 12 \* (938 - 27 \* x) \* \H(3)
          - 120 \* (19 + 89 \* x) \* \H(4)
  \nonumber\\&& \mbox{}
          + 180 \* (1 + x) \* (\Hh(1,1) \* \z2 
          - 2 \* \Hh(3,0)
          - 2 \* \Hh(3,1)
          - 10 \* \Hhhh(0,0,0,0)
          - 2 \* \Hhhh(1,0,0,0)
          + \H(2) \* \z2)
          + 550 \* (7 + 5 \* x) \* (\Hh(1,2)
  \nonumber\\&& \mbox{}
          - \Hhh(1,1,0))
          + 240 \* (13 - 22 \* x) \* (\Hh(1,3)
          - \Hhh(2,0,0)
          - \Hhhh(1,1,0,0))
          - 60 \* (43 - 97 \* x) \* (\Hh(1,0) \* \z2 - \H(1) \* \z3)
  \nonumber\\&& \mbox{}
          + 6 \* (263 - 857 \* x) \* \z2^2 
          + 5 \* (5391 + 3187 \* x) \* \z3 
          - (5783 + 2727 \* x) \* \z2 
          + (7961 + 2390 \* \Hh(1,0)
  \nonumber\\&& \mbox{}
          - 360 \* \Hh(1,1)
          - 360 \* \Hh(2,1)
          - 5280 \* \Hhh(1,0,0)
          + 11377 \* \H(0)
          + 3416 \* \H(1)
          + 3056 \* \H(2)))
  \nonumber\\&& \mbox{}
          + {8 \over 225} \* \pgq( - x) \* 
            (1080 \* (3 - 2 \* x^{-1}) \* \Hh(-2,0)
          + (4551 - 1897 \* x^{-1}) \* \Hh(-1,0)
          - 2 \* (982 - 1507 \* x^{-1}) \* \Hh(-1,2)
  \nonumber\\&& \mbox{}
          - 2 \* (2648 + 1507 \* x^{-1}) \* \Hhh(-1,-1,0)
          + 2 \* (2273 + 427 \* x^{-1}) \* \Hhh(-1,0,0)
          - 3 \* (228 + 1507 \* x^{-1}) \* \H(-1) \* \z2 
  \nonumber\\&& \mbox{}
          - 180 \* (1 - x^{-1}) \* (4 \* \Hh(-2,2)
          + 51 \* \Hh(-1,-1) \* \z2 
          - 33 \* \Hh(-1,0) \* \z2 
          + 30 \* \Hh(-1,3)
          - 4 \* \Hhh(-2,-1,0)
          + 4 \* \Hhh(-2,0,0)
  \nonumber\\&& \mbox{}
          - 2 \* \Hhh(-1,-2,0)
          - 50 \* \Hhh(-1,-1,2)
          + 2 \* \Hhh(-1,2,0)
          + 2 \* \Hhh(-1,2,1)
          - 4 \* \Hhh(1,-2,0)
          + 2 \* \Hhhh(-1,-1,-1,0)
          - 30 \* \Hhhh(-1,-1,0,0)
  \nonumber\\&& \mbox{}
          + 8 \* \Hhhh(-1,0,0,0)
          - 6 \* \H(-2) \* \z2 
          - 40 \* \H(-1) \* \z3 ))
          - {8 \over 225} \* \pgq(x) \* 
            ((2648 - 1507 \* x^{-1}) \* \H(1) \* \z2 
  \nonumber\\&& \mbox{}
          - 180 \* (1 + x^{-1}) \* (\Hh(1,1) \* \z2 
          - 2 \* \Hhhh(1,0,0,0)
          + 2 \* \H(2) \* \z2 )
          + 60 \* (8 - 97 \* x^{-1}) \* (\Hh(1,0) \* \z2 - \H(1) \* \z3)
  \nonumber\\&& \mbox{}
          - 60 \* (17 - 88 \* x^{-1}) \* (\Hh(1,3)
          - \Hhhh(1,1,0,0))
          - (7431
          - 1380 \* \z3 
          + 4292 \* \z2 
          - 226 \* \Hh(0,0)
          - 360 \* \Hh(1,0)
  \nonumber\\&& \mbox{}
          - 360 \* \Hh(1,1)
          + 360 \* \Hh(2,0)
          + 360 \* \Hh(2,1)
          + 1800 \* \Hhh(0,0,0)
          - 5280 \* \Hhh(1,0,0)
          - 4551 \* \H(0)
          - 11760 \* \H(0) \* \z2 
  \nonumber\\&& \mbox{}
          + 6226 \* \H(1)
          - 6946 \* \H(2)
          + 10680 \* \H(3) ))
          + {16 \over 5} \* (39 + 379 \* x) \* \H(-2) \* \z2
          + {8 \over 3} \* (1098 + 1009 \* x) \* \H(-1) \* \z3
  \nonumber\\&& \mbox{}
          + {4 \over 225} \* (93669 + 150569 \* x) \* \H(-1) \* \z2
          - {16 \over 15} \* (489 + 56 \* x) \* \H(0) \* \z3
          - {8 \over 225} \* (9327 + 125948 \* x) \* \H(0) \* \z2
  \nonumber\\&& \mbox{}
          - {4 \over 675} \* (182187 + 457387 \* x) \* \H(0)
          + {8 \over 15} \* (1184 - 2649 \* x) \* \H(1) \* \z3
          + {4 \over 27} \* (8250 - 19157 \* x) \* \H(1)
  \nonumber\\&& \mbox{}
          + {4 \over 225} \* (45113 - 22613 \* x) \* \H(1) \* \z2
          - {4 \over 45} \* (76 - 8449 \* x) \* \H(2)
          + {16 \over 15} \* (183 + 772 \* x) \* \H(2) \* \z2
  \nonumber\\&& \mbox{}
          + {8 \over 225} \* (2556 + 100369 \* x) \* \H(3)
          + {32 \over 5} \* (19 + 91 \* x) \* \H(4)
          + {32 \over 5} \* (3 + 7 \* x) \* (\Hh(3,0)
          + \Hh(3,1))
  \nonumber\\&& \mbox{}
          - {4 \over 15} \* (137 - 2044 \* x) \* \z2
          - {8 \over 25} \* (263 - 83 \* x) \* \z2^2
          - {4 \over 5} \* (2021 + 2269 \* x) \* \z3
  \nonumber\\&& \mbox{}
          + {1 \over 1350} \* (73064 - 2185319 \* x)
          + {8 \over 15} \* x \* (2685 \* \z5
          + 1980 \* \z2 \* \z3
          + 120 \* \Hh(-3,2)
          - 720 \* \Hh(-2,-1) \* \z2
  \nonumber\\&& \mbox{}
          + 420 \* \Hh(-2,0) \* \z2
          - 180 \* \Hh(-2,3)
          - 4200 \* \Hh(-1,-2) \* \z2
          - 5940 \* \Hh(-1,-1) \* \z3
          + 2700 \* \Hh(-1,0) \* \z3
          - 120 \* \Hh(-1,2) \* \z2
  \nonumber\\&& \mbox{}
          - 2520 \* \Hh(-1,4)
          - 120 \* \Hh(0,0) \* \z3
          + 2760 \* \Hh(1,-2) \* \z2
          - 3540 \* \Hh(1,0) \* \z3
          - 2100 \* \Hh(1,1) \* \z3
          - 360 \* \Hh(1,2) \* \z2
  \nonumber\\&& \mbox{}
          + 1680 \* \Hh(1,4)
          + 300 \* \Hh(2,0) \* \z2
          - 360 \* \Hh(2,1) \* \z2
          + 30 \* \Hh(2,2)
          - 540 \* \Hh(2,3)
          + 240 \* \Hhh(-3,-1,0)
          - 60 \* \Hhh(-3,0,0)
  \nonumber\\&& \mbox{}
          + 480 \* \Hhh(-2,-2,0)
          + 360 \* \Hhh(-2,-1,2)
          + 1080 \* \Hhh(-1,-3,0)
          + 3840 \* \Hhh(-1,-2,2)
          + 7560 \* \Hhh(-1,-1,-1) \* \z2
  \nonumber\\&& \mbox{}
          - 5280 \* \Hhh(-1,-1,0) \* \z2
          + 4200 \* \Hhh(-1,-1,3)
          + 3360 \* \Hhh(-1,0,0) \* \z2
          - 120 \* \Hhh(-1,3,0)
          - 120 \* \Hhh(-1,3,1)
          - 600 \* \Hhh(1,-3,0)
  \nonumber\\&& \mbox{}
          - 2880 \* \Hhh(1,-2,2)
          - 2520 \* \Hhh(1,0,0) \* \z2
          - 720 \* \Hhh(1,1,0) \* \z2
          + 800 \* \Hhh(1,1,1)
          - 360 \* \Hhh(1,1,1) \* \z2
          - 160 \* \Hhh(1,1,2)
  \nonumber\\&& \mbox{}
          - 360 \* \Hhh(1,1,3)
          + 220 \* \Hhh(1,2,0)
          + 120 \* \Hhh(1,3,0)
          + 120 \* \Hhh(1,3,1)
          - 30 \* \Hhh(2,1,0)
          + 60 \* \Hhh(3,0,0)
          - 720 \* \Hhhh(-2,-1,-1,0)
  \nonumber\\&& \mbox{}
          + 780 \* \Hhhh(-2,-1,0,0)
          - 360 \* \Hhhh(-2,0,0,0)
          - 720 \* \Hhhh(-1,-2,-1,0)
          + 2760 \* \Hhhh(-1,-2,0,0)
          - 720 \* \Hhhh(-1,-1,-2,0)
  \nonumber\\&& \mbox{}
          - 7200 \* \Hhhh(-1,-1,-1,2)
          + 240 \* \Hhhh(-1,-1,2,0)
          + 240 \* \Hhhh(-1,-1,2,1)
          + 480 \* \Hhhh(-1,2,0,0)
          - 240 \* \Hhhh(1,-2,-1,0)
  \nonumber\\&& \mbox{}
          - 1320 \* \Hhhh(1,-2,0,0)
          - 1440 \* \Hhhh(1,1,-2,0)
          + 160 \* \Hhhh(1,1,1,0)
          + 360 \* \Hhhh(1,2,0,0)
          + 360 \* \Hhhh(2,0,0,0)
          + 900 \* \Hhhh(2,1,0,0)
  \nonumber\\&& \mbox{}
          + 720 \* \Hhhhh(-1,-1,-1,-1,0)
          - 4920 \* \Hhhhh(-1,-1,-1,0,0)
          + 1920 \* \Hhhhh(-1,-1,0,0,0)
          - 1080 \* \Hhhhh(-1,0,0,0,0)
  \nonumber\\&& \mbox{}
          + 1080 \* \Hhhhh(1,0,0,0,0)
          + 1200 \* \Hhhhh(1,1,0,0,0)
          + 600 \* \Hhhhh(1,1,1,0,0)
          + 630 \* \H(-2) \* \z3
          - 306 \* \H(-1) \* \z2^2
          + 36 \* \H(0) \* \z2^2
  \nonumber\\&& \mbox{}
          - 366 \* \H(1) \* \z2^2
          - 1530 \* \H(2) \* \z3
          - 120 \* \H(3) \* \z2 )
          \biggr)
  \nonumber\\&& \mbox{}
       +  \colour4colour{\ca^2 \* \cf}  \*  \biggl(
          - {16 \over 15} \* (12 - 53 \* x) \* \Hh(-3,0)
          - {4 \over 45} \* (4611 - 5009 \* x) \* \Hh(-2,0)
          - 64 \* \Hh(-2,2) \* (2 - 9 \* x)
  \nonumber\\&& \mbox{}
          + {128 \over 3} \* (23 + 17 \* x) \* \Hh(-1,-1) \* \z2
          - {16 \over 15} \* (578 + 523 \* x) \* \Hh(-1,0) \* \z2
          - {4 \over 225} \* (62676 + 38501 \* x) \* \Hh(-1,0)
  \nonumber\\&& \mbox{}
          + {4 \over 45} \* (1129 + 7584 \* x) \* \Hh(-1,2)
          - {4 \over 15} \* (83 - 903 \* x) \* \Hh(0,0) \* \z2
          + {4 \over 225} \* (13087 + 48843 \* x) \* \Hh(0,0)
  \nonumber\\&& \mbox{}
          + {88 \over 9} \* (9 + 11 \* x) \* \Hh(1,0)
          - {4 \over 15} \* (319 + 376 \* x) \* \Hh(1,0) \* \z2
          - {4 \over 15} \* (153 - 788 \* x) \* \Hh(1,3)
  \nonumber\\&& \mbox{}
          - 88 \* (1 - x) \* \Hh(2,0)
          - {16 \over 3} \* (12 + 13 \* x) \* \Hhh(-2,0,0)
          - {4 \over 45} \* (1489 - 5426 \* x) \* \Hhh(-1,-1,0)
  \nonumber\\&& \mbox{}
          - {4 \over 45} \* (3707 + 1297 \* x) \* \Hhh(-1,0,0)
          + {4 \over 45} \* (387 - 317 \* x) \* \Hhh(0,0,0)
          - {32 \over 3} \* (26 - 57 \* x) \* \Hhh(1,-2,0)
  \nonumber\\&& \mbox{}
          - {4 \over 45} \* (303 - 4048 \* x) \* \Hhh(1,0,0)
          - {4 \over 15} \* (371 - 951 \* x) \* \Hhh(2,0,0)
          - {64 \over 3} \* (23 - 9 \* x) \* \Hhhh(-1,-1,0,0)
  \nonumber\\&& \mbox{}
          + {16 \over 15} \* (118 - 117 \* x) \* \Hhhh(-1,0,0,0)
          - {16 \over 15} \* (12 + 53 \* x) \* \Hhhh(0,0,0,0)
          + {16 \over 15} \* (118 + 117 \* x) \* \Hhhh(1,0,0,0)
  \nonumber\\&& \mbox{}
          + {12 \over 5} \* (17 + 8 \* x) \* \Hhhh(1,1,0,0)
          - {2 \over 75} \* \pqg( - x) \* (10 \* (491 + 71 \* x) \* \Hh(-2,0)
          + 2 \* (1984 - 1869 \* x) \* \Hh(-1,0)
  \nonumber\\&& \mbox{}
          - 10 \* (201 - 71 \* x) \* \Hh(-1,2)
          - 10 \* (539 + 71 \* x) \* \Hhh(-1,-1,0)
          + 10 \* (433 - 193 \* x) \* \Hhh(-1,0,0)
  \nonumber\\&& \mbox{}
          - 5 \* (137 + 213 \* x) \* \H(-1) \* \z2 
          - 240 \* (1 - x) \* (2 \* \Hh(-3,0)
          + 20 \* \Hh(-2,2)
          + 20 \* \Hh(-1,-1) \* \z2 
          - 11 \* \Hh(-1,0) \* \z2 
  \nonumber\\&& \mbox{}
          + 10 \* \Hh(-1,3)
          + 10 \* \Hhh(-2,0,0)
          - 20 \* \Hhh(-1,-1,2)
          + 2 \* \Hhh(1,-2,0)
          - 10 \* \Hhhh(-1,-1,0,0)
          + \Hhhh(-1,0,0,0)
          - 20 \* \H(-2) \* \z2 
  \nonumber\\&& \mbox{}
          - 15 \* \H(-1) \* \z3 ))
          + {2 \over 75} \* \pqg(x) \* (10 \* (83 - 387 \* x) \* \Hh(0,0) \* \z2
          - 14 \* (647 + 267 \* x) \* \Hh(0,0)
  \nonumber\\&& \mbox{}
          + 10 \* (347 - 123 \* x) \* \Hh(1,0) \* \z2 
          + 10 \* (657 + 188 \* x) \* \H(0) \* \z2 
          - 10 \* (683 + 213 \* x) \* \H(0) \* \z3 
  \nonumber\\&& \mbox{}
          - 10 \* (323 - 147 \* x) \* \H(1) \* \z3 
          + 5 \* (1199 + 589 \* x) \* \H(1) \* \z2 
          - 70 \* (104 + 37 \* x) \* \H(3)
          - 10 \* (131 - 339 \* x) \* \H(4)
  \nonumber\\&& \mbox{}
          - 10 \* (1 + x) \* (330 \* \Hh(1,2)
          - 330 \* \Hh(2,0)
          + 193 \* \Hhh(0,0,0)
          - 330 \* \Hhh(1,1,0)
          - 48 \* \Hhhh(0,0,0,0)
          - 24 \* \Hhhh(1,0,0,0))
  \nonumber\\&& \mbox{}
          - 40 \* (73 - 21 \* x) \* \z2^2 
          - 10 \* (371 - 99 \* x) \* (\Hh(1,3)
          - \Hhh(2,0,0)
          - \Hhhh(1,1,0,0))
          + 2 \* (1559 + 1869 \* x) \* \z2 
  \nonumber\\&& \mbox{}
          - 5 \* (2739 + 2335 \* x) \* \z3 
          - 2 \* (3304 + 1650 \* \Hh(1,0)
          - 495 \* \Hhh(1,0,0)
          + 2994 \* \H(0)
          - 310 \* \H(1)
          - 310 \* \H(2) ))
  \nonumber\\&& \mbox{}
          - {2 \over 225} \* \pgq( - x) \* 
            (480 \* (13 - 11 \* x^{-1}) \* \Hh(-2,0)
          + 4 \* (1874 - 1759 \* x^{-1}) \* \Hh(-1,0)
          - 10 \* (97 - 142 \* x^{-1}) \* \Hh(-1,2)
  \nonumber\\&& \mbox{}
          - 10 \* (773 + 142 \* x^{-1}) \* \Hhh(-1,-1,0)
          + 20 \* (433 - 193 \* x^{-1}) \* \Hhh(-1,0,0)
          - 15 \* (193 + 142 \* x^{-1}) \* \H(-1) \* \z2 
  \nonumber\\&& \mbox{}
          - 480 \* (1 - x^{-1}) \* (20 \* \Hh(-1,-1) \* \z2 
          - 11 \* \Hh(-1,0) \* \z2 
          + 10 \* \Hh(-1,3)
          - 20 \* \Hhh(-1,-1,2)
          - 2 \* \Hhh(1,-2,0)
          - 10 \* \Hhhh(-1,-1,0,0)
  \nonumber\\&& \mbox{}
          + \Hhhh(-1,0,0,0)
          - 15 \* \H(-1) \* \z3))
          + {2 \over 225} \* \pgq(x) \* (30 \* 
            (153 - 82 \* x^{-1}) \* \Hh(1,0) \* \z2 
          + 480 \* (1 + x^{-1}) \* \Hhhh(1,0,0,0)
  \nonumber\\&& \mbox{}
          - 30 \* (137 - 98 \* x^{-1}) \* \H(1) \* \z3 
          + 5 \* (773 - 142 \* x^{-1}) \* \H(1) \* \z2 
          - 30 \* (169 - 66 \* x^{-1}) \* (\Hh(1,3)
          - \Hhhh(1,1,0,0))
  \nonumber\\&& \mbox{}
          - 4 \* (3194 
          - 1065 \* \z3
          + 1225 \* \z2 
          - 1205 \* \Hh(0,0)
          + 240 \* \Hhh(0,0,0)
          - 495 \* \Hhh(1,0,0)
          - 1874 \* \H(0)
          - 1935 \* \H(0) \* \z2 
  \nonumber\\&& \mbox{}
          + 1340 \* \H(1)
          - 1340 \* \H(2)
          + 1695 \* \H(3) ))
          + {16 \over 3} \* (24 - 41 \* x) \* \H(-2) \* \z2
          - {32 \over 3} \* (69 + 47 \* x) \* \H(-1) \* \z3
  \nonumber\\&& \mbox{}
          - {2 \over 45} \* (3747 + 9742 \* x) \* \H(-1) \* \z2
          - {4 \over 9} \* (153 - 1784 \* x) \* \H(0) \* \z2
          + {4 \over 15} \* (683 - 893 \* x) \* \H(0) \* \z3
  \nonumber\\&& \mbox{}
          + {16 \over 135} \* (1203 + 16225 \* x) \* \H(0)
          - {436 \over 135} \* (192 - 497 \* x) \* \H(1)
          - {4 \over 15} \* (649 - 984 \* x) \* \H(1) \* \z3
  \nonumber\\&& \mbox{}
          - {2 \over 45} \* (3469 + 4766 \* x) \* \H(1) \* \z2
          - {8 \over 45} \* (1199 - 2496 \* x) \* \H(2)
          + {4 \over 9} \* (234 - 1351 \* x) \* \H(3)
  \nonumber\\&& \mbox{}
          + {4 \over 15} \* (131 - 691 \* x) \* \H(4)
          + {88 \over 3} \* (3 - x) \* (\Hh(1,2)
          - \Hhh(1,1,0))
          + {64 \over 3} \* (23 + 21 \* x) \* (\Hh(-1,3)
          - 2 \* \Hhh(-1,-1,2))
  \nonumber\\&& \mbox{}
          + {4 \over 15} \* (292 - 233 \* x) \* \z2^2
          + {4 \over 75} \* (2281 - 20211 \* x) \* \z2
          + {2 \over 15} \* (3571 + 759 \* x) \* \z3
  \nonumber\\&& \mbox{}
          - {4 \over 405} \* (45951 - 159431 \* x)
          - {8 \over 45} \* x \* (1260 \* \z5
          + 1620 \* \z2 \* \z3
          - 2880 \* \Hh(-1,-2) \* \z2
          - 4320 \* \Hh(-1,-1) \* \z3
  \nonumber\\&& \mbox{}
          + 2160 \* \Hh(-1,0) \* \z3
          - 1800 \* \Hh(-1,4)
          + 2880 \* \Hh(1,-2) \* \z2
          - 3060 \* \Hh(1,0) \* \z3
          - 1595 \* \Hh(1,1)
          - 1800 \* \Hh(1,1) \* \z3
  \nonumber\\&& \mbox{}
          + 1290 \* \Hh(1,1) \* \z2
          + 900 \* \Hh(1,4)
          + 900 \* \Hh(2,0) \* \z2
          - 900 \* \Hh(2,3)
          - 4020 \* \Hhh(-2,-1,0)
          + 720 \* \Hhh(-1,-3,0)
  \nonumber\\&& \mbox{}
          - 2280 \* \Hhh(-1,-2,0)
          + 2880 \* \Hhh(-1,-2,2)
          + 5760 \* \Hhh(-1,-1,-1) \* \z2
          - 3600 \* \Hhh(-1,-1,0) \* \z2
          + 2880 \* \Hhh(-1,-1,3)
  \nonumber\\&& \mbox{}
          + 2520 \* \Hhh(-1,0,0) \* \z2
          - 720 \* \Hhh(1,-3,0)
          - 2880 \* \Hhh(1,-2,2)
          - 1620 \* \Hhh(1,0,0) \* \z2
          + 360 \* \Hhh(1,1,0) \* \z2
          - 330 \* \Hhh(1,1,2)
  \nonumber\\&& \mbox{}
          - 1080 \* \Hhh(1,1,3)
          + 330 \* \Hhh(1,2,0)
          + 1440 \* \Hhhh(-1,-2,0,0)
          + 1920 \* \Hhhh(-1,-1,-1,0)
          - 5760 \* \Hhhh(-1,-1,-1,2)
  \nonumber\\&& \mbox{}
          + 360 \* \Hhhh(-1,2,0,0)
          - 1440 \* \Hhhh(1,-2,0,0)
          - 1440 \* \Hhhh(1,1,-2,0)
          + 330 \* \Hhhh(1,1,1,0)
          + 540 \* \Hhhh(1,2,0,0)
          + 900 \* \Hhhh(2,1,0,0)
  \nonumber\\&& \mbox{}
          - 2880 \* \Hhhhh(-1,-1,-1,0,0)
          + 720 \* \Hhhhh(-1,-1,0,0,0)
          - 720 \* \Hhhhh(-1,0,0,0,0)
          + 720 \* \Hhhhh(1,0,0,0,0)
          + 720 \* \Hhhhh(1,1,0,0,0)
  \nonumber\\&& \mbox{}
          + 1080 \* \Hhhhh(1,1,1,0,0)
          + 72 \* \H(-1) \* \z2^2
          - 792 \* \H(1) \* \z2^2
          - 900 \* \H(2) \* \z3
          + 2010 \* \H(2) \* \z2)
          \biggr)
\:\: .
\eea
\normalsize
The corresponding correction to the gluon coefficient function reads
\small
\bea
&& c^{(3)}_{L,\rm{g}}(x) \:\: = \:\: 
        \colour4colour{\dabcNA} \* \flg11  \*  \biggl(
            {64 \over 15} \* (\gfunct1(x)
          - \gfunct2(x))
          - {2048 \over 15} \* (4 + 9 \* x) \* \Hh(-3,0)
          + {256 \over 225} \* (778 + 1573 \* x) \* \Hh(-2,0)
  \nonumber\\&& \mbox{}
          + {1024 \over 15} \* (8 + 29 \* x) \* \Hh(-2,2)
          - {64 \over 225} \* (2318 + 2943 \* x) \* \Hh(-1,0)
          - {128 \over 225} \* (1753 + 93 \* x) \* \Hh(-1,2)
  \nonumber\\&& \mbox{}
          - {128 \over 15} \* (17 - 76 \* x) \* \Hh(0,0) \* \z2
          - {64 \over 45} \* (155 - 1247 \* x) \* \Hh(0,0)
          + {128 \over 15} \* (23 - 93 \* x) \* \Hh(1,0) \* \z2
  \nonumber\\&& \mbox{}
          - {128 \over 15} \* (53 - 93 \* x) \* \Hh(1,3)
          + {512 \over 15} \* (1 + 6 \* x) \* \Hhh(-2,-1,0)
          + {256 \over 15} \* (15 + 52 \* x) \* \Hhh(-2,0,0)
  \nonumber\\&& \mbox{}
          - {128 \over 225} \* (407 - 1113 \* x) \* \Hhh(-1,-1,0)
          - {128 \over 225} \* (673 + 603 \* x) \* \Hhh(-1,0,0)
          - {256 \over 225} \* (147 - 277 \* x) \* \Hhh(0,0,0)
  \nonumber\\&& \mbox{}
          - 256 \* (1 + 4 \* x) \* \Hhh(1,-2,0) 
          - {256 \over 15} \* (63 - 43 \* x) \* \Hhh(1,0,0)
          - {128 \over 15} \* (51 - 116 \* x) \* \Hhh(2,0,0)
  \nonumber\\&& \mbox{}
          + {128 \over 15} \* (83 - 93 \* x) \* \Hhhh(1,1,0,0)
          - {64 \over 225} \* \pqg( - x) \* (3 \* (281 + 90 \* x) \* \Hh(-1,0)
          - 4 \* (622 - 147 \* x) \* \Hh(-1,2) 
  \nonumber\\&& \mbox{}
          + 12 \* (26 + 49 \* x) \* \Hhh(-1,0,0)
          + 2 \* (466 - 441 \* x) \* \H(-1) \* \z2 
          + 4 \* (778 + 147 \* x) \* (\Hh(-2,0)
          - \Hhh(-1,-1,0))
  \nonumber\\&& \mbox{}
          - 60 \* (32 \* \Hh(-3,0)
          - 32 \* \Hh(-2,2)
          - 15 \* \Hh(-1,-1) \* \z2 
          + 15 \* \Hh(-1,0) \* \z2 
          - 2 \* \Hhh(-2,-1,0)
          - 15 \* \Hhh(-2,0,0)
          + 15 \* \Hhh(-1,-2,0)
  \nonumber\\&& \mbox{}
          + 15 \* \Hhh(1,-2,0)
          - 30 \* \Hhhh(-1,-1,-1,0)
          + 15 \* \Hhhh(-1,-1,0,0)
          - 15 \* \Hhhh(-1,0,0,0)
          + 31 \* \H(-2) \* \z2 
          + 15 \* \H(-1) \* \z3 ))
  \nonumber\\&& \mbox{}
          + {64 \over 225} \* \pqg(x) \* (3 \* (451 + 90 \* x) \* \Hh(0,0) 
          + 90 \* (37 - 28 \* x) \* \Hh(1,0) \* \z2 
          - 90 \* (27 - 28 \* x) \* \Hh(1,3)
          + 588 \* (1 + x) \* \Hhh(0,0,0)
  \nonumber\\&& \mbox{}
          - 30 \* (23 - 84 \* x) \* \H(0) \* \z3 
          + 4 \* (811 - 294 \* x) \* \H(0) \* \z2 
          + 90 \* (13 + 28 \* x) \* \H(1) \* \z3 
          + 2 \* (778 - 147 \* x) \* \H(1) \* \z2 
  \nonumber\\&& \mbox{}
          - 4 \* (958 - 147 \* x) \* \H(3)
          + 84 \* (7 + 24 \* x) \* \z2^2
          + 30 \* (17 - 84 \* x) \* (\Hh(0,0) \* \z2 - \H(4))
          + 90 \* (17 - 28 \* x) \* (\Hhh(2,0,0)
  \nonumber\\&& \mbox{}
          + \Hhhh(1,1,0,0))
          - 6 \* (553 + 45 \* x) \* \z2
          - 5 \* (743 + 294 \* x) \* \z3 
          + 3 \* (504 + 825 \* \Hh(1,0)
          - 900 \* \Hh(1,0) \* \z3 
          + 1650 \* \Hh(1,1)
  \nonumber\\&& \mbox{}
          - 1800 \* \Hh(1,1) \* \z3 
          + 300 \* \Hh(1,1) \* \z2 
          - 900 \* \Hh(1,4)
          + 900 \* \Hh(2,0) \* \z2 
          - 900 \* \Hh(2,3)
          + 540 \* \Hhh(1,0,0)
          + 900 \* \Hhh(1,0,0) \* \z2 
  \nonumber\\&& \mbox{}
          + 1800 \* \Hhh(1,1,0) \* \z2 
          - 1800 \* \Hhh(1,1,3)
          - 300 \* \Hhhh(1,0,0,0)
          + 900 \* \Hhhh(1,2,0,0)
          + 900 \* \Hhhh(2,1,0,0)
          + 1800 \* \Hhhhh(1,1,1,0,0)
  \nonumber\\&& \mbox{}
          + 990 \* \H(0)
          + 1101 \* \H(1)
          - 720 \* \H(1) \* \z2^2 
          + 1016 \* \H(2)
          - 900 \* \H(2) \* \z3 
          - 20 \* \H(2) \* \z2 ))
  \nonumber\\&& \mbox{}
          + {64 \over 225} \* \pgq( - x) \* ((53 + 45 \* x^{-1}) \* \Hh(-1,0)
          + 2 \* (111 + 49 \* x^{-1}) \* \Hh(-1,2)
          + 2 \* (209 - 49 \* x^{-1}) \* \Hhh(-1,-1,0)
  \nonumber\\&& \mbox{}
          - 98 \* (1 - x^{-1}) \* \Hhh(-1,0,0)
          - (13 + 147 \* x^{-1}) \* \H(-1) \* \z2)
          + {64 \over 225} \* \pgq(x) \* ((209 + 49 \* x^{-1}) \* \H(1) \* \z2 
  \nonumber\\&& \mbox{}
          + 30 \* (29 + 14 \* x^{-1}) \* (\Hh(1,0) \* \z2 
          - \Hh(1,3)
          + \Hhhh(1,1,0,0)
          - \H(1) \* \z3 )
          + 53 
          + 420 \* \z3 
          + 194 \* \z2 
          + 113 \* \Hh(0,0)
  \nonumber\\&& \mbox{}
          - 420 \* \Hhh(1,0,0)
          - 68 \* \H(0)
          - 420 \* \H(0) \* \z2 
          + 322 \* \H(1)
          - 292 \* \H(2)
          + 420 \* \H(3))
          - {128 \over 15} \* \pgg( - x) \* (2 \* \z2 
          - \Hh(0,0)
          + \H(0)
  \nonumber\\&& \mbox{}
          - 2 \* \H(2) )
          - {256 \over 15} \* (31 + 110 \* x) \* \H(-2) \* \z2
          + {64 \over 75} \* (1033 + 433 \* x) \* \H(-1) \* \z2
          + {128 \over 15} \* (23 - 40 \* x) \* \H(0) \* \z3
  \nonumber\\&& \mbox{}
          - {256 \over 225} \* (1351 - 1041 \* x) \* \H(0) \* \z2
          - {128 \over 225} \* (1622 - 1833 \* x) \* \H(0)
          - {128 \over 15} \* (173 - 93 \* x) \* \H(1) \* \z3
  \nonumber\\&& \mbox{}
          - {64 \over 225} \* (407 + 1113 \* x) \* \H(1) \* \z2
          - {64 \over 45} \* (449 + 276 \* x) \* \H(1)
          + {256 \over 15} \* (1 - 6 \* x) \* \H(2) \* \z2
          - {256 \over 45} \* (207 - 118 \* x) \* \H(2)
  \nonumber\\&& \mbox{}
          + {512 \over 225} \* (749 - 659 \* x) \* \H(3)
          + {128 \over 15} \* (17 - 92 \* x) \* \H(4)
          + {3008 \over 9} \* (5 - 6 \* x) \* \z3
          - {128 \over 75} \* (98 - 149 \* x) \* \z2^2
  \nonumber\\&& \mbox{}
          - {64 \over 225} \* (1674 - 109 \* x)
          + {128 \over 225} \* (2107 - 3742 \* x) \* \z2
          + {64 \over 15} \* (60 \* \Hh(-1,-1) \* \z2 
          - 60 \* \Hh(-1,0) \* \z2 
          - 165 \* \Hh(1,0)
  \nonumber\\&& \mbox{}
          + 180 \* \Hh(1,0) \* \z3 
          - 330 \* \Hh(1,1)
          + 360 \* \Hh(1,1) \* \z3 
          - 60 \* \Hh(1,1) \* \z2 
          + 180 \* \Hh(1,4)
          - 180 \* \Hh(2,0) \* \z2 
          + 180 \* \Hh(2,3)
  \nonumber\\&& \mbox{}
          - 60 \* \Hhh(-1,-2,0)
          - 180 \* \Hhh(1,0,0) \* \z2 
          - 360 \* \Hhh(1,1,0) \* \z2 
          + 360 \* \Hhh(1,1,3)
          + 120 \* \Hhhh(-1,-1,-1,0)
          - 60 \* \Hhhh(-1,-1,0,0)
  \nonumber\\&& \mbox{}
          + 60 \* \Hhhh(-1,0,0,0)
          + 32 \* \Hhhh(0,0,0,0) \* x 
          + 60 \* \Hhhh(1,0,0,0)
          - 180 \* \Hhhh(1,2,0,0)
          - 180 \* \Hhhh(2,1,0,0)
          - 360 \* \Hhhhh(1,1,1,0,0)
          - 60 \* \H(-1) \* \z3 
  \nonumber\\&& \mbox{}
          + 144 \* \H(1) \* \z2^2 
          + 180 \* \H(2) \* \z3 )
          + \delta(1 - x) \* {64 \over 15} \* (\z3 + \z2)
          \biggr)
  \nonumber\\&& \mbox{}
       +  \colour4colour{\cf \* \nf^2}  \*  \biggl(
            {704 \over 45} \* (1 + 11 \* x) \* \Hh(-2,0)
          - {32 \over 675} \* (2732 + 957 \* x) \* \Hh(-1,0)
          + {16 \over 675} \* (4533 - 14383 \* x) \* \Hh(0,0)
  \nonumber\\&& \mbox{}
          + {32 \over 3} \* (3 - 8 \* x) \* \Hh(1,0)
          + {16 \over 3} \* (7 - 16 \* x) \* \Hh(1,1)
          + {16 \over 3} \* (11 - 8 \* x) \* \Hh(2,0)
          + {160 \over 3} \* (1 - x) \* \Hh(2,1)
  \nonumber\\&& \mbox{}
          + {16 \over 15} \* (183 - 743 \* x) \* \Hhh(0,0,0)
          - {32 \over 225} \* \pqg( - x) \* (30 \* (4 + x) \* \Hh(-2,0)
          - (502 - 107 \* x) \* \Hh(-1,0) 
  \nonumber\\&& \mbox{}
          - 15 \* (1 - x) \* (2 \* \Hh(-1,2)
          - 2 \* \Hhh(-1,-1,0)
          + 6 \* \Hhh(-1,0,0)
          - 3 \* \H(-1) \* \z2 ))
          + {4 \over 2025} \* \pqg(x) \* (72 \* (347 + 107 \* x) \* \Hh(0,0)
  \nonumber\\&& \mbox{}
          - 2700 \* (5 + 2 \* x) \* \Hh(2,0)
          - 6480 \* (9 - x) \* \Hhh(0,0,0)
          + 540 \* (47 - 18 \* x) \* \H(0) \* \z2 
          - 540 \* (7 + 12 \* x) \* \H(1) \* \z2 
  \nonumber\\&& \mbox{}
          - 540 \* (51 - 14 \* x) \* \H(3)
          + 2700 \* (1 + 2 \* x) \* (\Hh(1,2)
          - \Hhh(1,1,0) )
          - 2700 \* (3 - 4 \* x) \* \z3 
          - 36 \* (949 + 214 \* x) \* \z2 
  \nonumber\\&& \mbox{}
          + 137689 
          + 25200 \* \Hh(1,0)
          + 22500 \* \Hh(1,1)
          - 10800 \* \Hh(2,1)
          - 9576 \* \H(0)
          + 20460 \* \H(1)
          + 26460 \* \H(2))
  \nonumber\\&& \mbox{}
          + {16 \over 675} \* \pgq( - x) \* (2 \* 
            (19 + 71 \* x^{-1}) \* \Hh(-1,0)
          - 15 \* (1 - x^{-1}) \* (4 \* \Hh(-2,0)
          + 2 \* \Hh(-1,2)
          - 2 \* \Hhh(-1,-1,0)
          + 6 \* \Hhh(-1,0,0)
  \nonumber\\&& \mbox{}
          - 3 \* \H(-1) \* \z2))
          + {8 \over 2025} \* \pgq(x) \* (90 \* (1 + x^{-1}) \* \H(1) \* \z2 
          + 1993
          - 1260 \* \z2 
          + 540 \* \Hh(0,0)
          + 900 \* \Hh(1,0)
  \nonumber\\&& \mbox{}
          + 900 \* \Hh(1,1)
          + 672 \* \H(0)
          - 330 \* \H(1)
          + 180 \* \H(2))
          - {16 \over 45} \* (321 - 1276 \* x) \* \H(0) \* \z2
          + {8 \over 27} \* (1355 - 1724 \* x) \* \H(0) 
  \nonumber\\&& \mbox{}
          + {176 \over 45} \* (2 + 3 \* x) \* \H(1) \* \z2
          + {8 \over 9} \* (281 - 325 \* x) \* \H(1)
          + {16 \over 9} \* (49 - 9 \* x) \* \H(2)
          + {16 \over 45} \* (333 - 908 \* x) \* \H(3)
  \nonumber\\&& \mbox{}
          - {16 \over 3} \* (1 + 2 \* x) \* (\Hh(1,2)
          - \Hhh(1,1,0))
          - {16 \over 45} \* (7 - 3 \* x) \* (2 \* \Hh(-1,2)
          - 2 \* \Hhh(-1,-1,0)
          + 6 \* \Hhh(-1,0,0)
          - 3 \* \H(-1) \* \z2 )
  \nonumber\\&& \mbox{}
          - {16 \over 9} \* (9 - 202 \* x) \* \z3
          - {16 \over 675} \* (3393 - 3643 \* x) \* \z2
          + {4 \over 135} \* (11027 - 20472 \* x)
          + {32 \over 5}  \* x \* (\z2^2
          + 30 \* \Hh(0,0) \* \z2 
  \nonumber\\&& \mbox{}
          - 10 \* \Hh(3,0) 
          - 10 \* \Hh(3,1) 
          - 55 \* \Hhhh(0,0,0,0) 
          + 20 \* \H(0) \* \z3 
          - 30 \* \H(4) )
          \biggr)
  \nonumber\\&& \mbox{}
       +  \colour4colour{\cf^2 \* \nf}  \*  \biggl(
            {128 \over 15} \* (79 + 94 \* x) \* \Hh(-3,0)
          + {32 \over 45} \* (1907 + 1632 \* x) \* \Hh(-2,0)
          + {256 \over 15} \* (19 - 16 \* x) \* \Hh(-2,2)
  \nonumber\\&& \mbox{}
          + {128 \over 15} \* (61 + 36 \* x) \* \Hh(-1,-1) \* \z2
          - {64 \over 5} \* (23 + 18 \* x) \* \Hh(-1,0) \* \z2
          + {8 \over 225} \* (39459 + 50909 \* x) \* \Hh(-1,0)
  \nonumber\\&& \mbox{}
          + {32 \over 45} \* (289 + 1179 \* x) \* \Hh(-1,2)
          - {32 \over 3} \* (41 - 73 \* x) \* \Hh(0,0) \* \z2
          - {4 \over 225} \* (7291 + 52369 \* x) \* \Hh(0,0)
  \nonumber\\&& \mbox{}
          + 8 \* (16 - 13 \* x) \* \Hh(1,0)
          - {32 \over 15} \* (572 - 207 \* x) \* \Hh(1,0) \* \z2
          + 8 \* (19 - 11 \* x) \* \Hh(1,1)
          + {32 \over 15} \* (496 - 141 \* x) \* \Hh(1,3)
  \nonumber\\&& \mbox{}
          + 128 \* (3 - 17 \* x) \* \Hh(2,0) \* \z2
          - 16 \* (7 - 12 \* x) \* \Hh(2,0)
          - 48 \* (3 - 5 \* x) \* \Hh(2,1) 
          - 384 \* (1 - 5 \* x) \* \Hh(2,3)
  \nonumber\\&& \mbox{}
          - {896 \over 3} \* (2 - 3 \* x) \* \Hhh(-2,-1,0)
          + {64 \over 15} \* (228 - 77 \* x) \* \Hhh(-2,0,0)
          - {32 \over 45} \* (2128 + 1383 \* x) \* \Hhh(-1,-1,0)
  \nonumber\\&& \mbox{}
          + {16 \over 45} \* (4229 + 4434 \* x) \* \Hhh(-1,0,0)
          - {8 \over 45} \* (1671 - 376 \* x) \* \Hhh(0,0,0)
          + {128 \over 3} \* (2 + 13 \* x) \* \Hhh(1,-2,0)
  \nonumber\\&& \mbox{}
          + {16 \over 15} \* (9 + x) \* \Hhh(1,0,0)
          - {32 \over 15} \* (213 - 653 \* x) \* \Hhh(2,0,0)
          - {64 \over 15} \* (211 + 111 \* x) \* \Hhhh(-1,-1,0,0)
  \nonumber\\&& \mbox{}
          + {448 \over 15} \* (14 + 9 \* x) \* \Hhhh(-1,0,0,0)
          - {16 \over 15} \* (48 - 203 \* x) \* \Hhhh(0,0,0,0)
          + {448 \over 15} \* (14 - 9 \* x) \* \Hhhh(1,0,0,0)
  \nonumber\\&& \mbox{}
          - {32 \over 15} \* (376 - 81 \* x) \* \Hhhh(1,1,0,0) 
          + 128 \* (3 - 13 \* x) \* \Hhhh(2,1,0,0)
          - {16 \over 225} \* \pqg( - x) \* (120 \* (49 + 6 \* x) \* \Hh(-3,0)
  \nonumber\\&& \mbox{}
          + 110 \* (67 + 6 \* x) \* \Hh(-2,0)
          + 480 \* (2 + 3 \* x) \* \Hh(-2,2)
          + 360 \* (11 + 4 \* x) \* \Hh(-1,-1) \* \z2 
          - 3 \* (1553 - 533 \* x) \* \Hh(-1,0)
  \nonumber\\&& \mbox{}
          - 20 \* (421 - 33 \* x) \* \Hh(-1,2)
          - 10 \* (809 + 66 \* x) \* \Hhh(-1,-1,0)
          + 15 \* (1 + 92 \* x) \* \Hhh(-1,0,0)
  \nonumber\\&& \mbox{}
          + 180 \* (13 + 2 \* x ) \* \Hhhh(-1,0,0,0)
          - 480 \* (7 + 3 \* x) \* \H(-2) \* \z2 
          - 180 \* (19 + 6 \* x) \* \H(-1) \* \z3
          + 5 \* (875 - 198 \* x) \* \H(-1) \* \z2 
  \nonumber\\&& \mbox{}
          - 180 \* (3 + 2 \* x) \* (3 \* \Hh(-1,0) \* \z2 
          - 2 \* \Hh(-1,3)
          + 4 \* \Hhh(-1,-1,2)
          - 2 \* \Hhh(1,-2,0))
          + 720 \* (9 + x) \* (\Hhh(-2,0,0)
  \nonumber\\&& \mbox{}
          - \Hhhh(-1,-1,0,0))
          - 120 \* (45 \* \Hh(-1,0) \* \z3 
          + 45 \* \Hh(-1,4)
          + 40 \* \Hhh(-2,-1,0)
          + 30 \* \Hhh(-1,-2,0)
          - 45 \* \Hhh(-1,0,0) \* \z2 
  \nonumber\\&& \mbox{}
          - 30 \* \Hhhh(-1,-1,-1,0)
          - 45 \* \Hhhh(-1,2,0,0)
          + 36 \* \H(-1) \* \z2^2))
          + {8 \over 225} \* \pqg(x) \* (300 \* (41 + 6 \* x) \* \Hh(0,0) \* \z2
  \nonumber\\&& \mbox{}
          + 3 \* (841 + 1066 \* x) \* \Hh(0,0)
          + 180 \* (97 + 22 \* x) \* \Hh(1,0) \* \z2 
          - 2340 \* (7 + 2 \* x) \* \Hh(1,3)
          + 30 \* (257 + 92 \* x) \* \Hhh(0,0,0)
  \nonumber\\&& \mbox{}
          + 1440 \* (1 + x) \* \Hhhh(0,0,0,0)
          - 360 \* (13 - 2 \* x) \* \Hhhh(1,0,0,0)
          - 60 \* (343 + 138 \* x) \* \H(0) \* \z3 
          + 20 \* (347 - 132 \* x) \* \H(0) \* \z2
  \nonumber\\&& \mbox{}
          + 110 \* (49 - 6 \* x) \* \H(1) \* \z2 
          - 180 \* (59 + 34 \* x) \* \H(1) \* \z3 
          - 20 \* (413 - 66 \* x) \* \H(3)
          - 60 \* (229 + 54 \* x) \* \H(4)
  \nonumber\\&& \mbox{}
          + 180 \* (71 + 26 \* x) \* (\Hhh(2,0,0)
          + \Hhhh(1,1,0,0))
          - 110 \* (209 + 30 \* x) \* \z3 
          - 6 \* (993 + 533 \* x) \* \z2 
  \nonumber\\&& \mbox{}
          - 12 \* (1421 + 396 \* x) \* \z2^2 
          - 3 \* (8214 + 225 \* \Hh(1,0)
          + 600 \* \Hh(1,1)
          - 1200 \* \Hh(1,1) \* \z2 
          - 900 \* \Hh(1,2)
          - 600 \* \Hh(2,0)
  \nonumber\\&& \mbox{}
          + 3600 \* \Hh(2,0) \* \z2 
          - 750 \* \Hh(2,1)
          - 3600 \* \Hh(2,3)
          + 360 \* \Hhh(1,0,0)
          - 600 \* \Hhh(1,1,0)
          - 750 \* \Hhh(1,1,1)
  \nonumber\\&& \mbox{}
          + 3600 \* \Hhhh(2,1,0,0)
          + 6144 \* \H(0)
          + 7730 \* \H(1)
          - 920 \* \H(2)
          - 3600 \* \H(2) \* \z3 
          - 1600 \* \H(2) \* \z2 ))
  \nonumber\\&& \mbox{}
          + {8 \over 225} \* \pgq( - x) \* (240 \* x^{-1} \* \Hh(-2,0)
          - (543 - 523 \* x^{-1}) \* \Hh(-1,0)
          - 20 \* (7 - 11 \* x^{-1}) \* \Hh(-1,2)
  \nonumber\\&& \mbox{}
          - 20 \* (9 + 11 \* x^{-1}) \* \Hhh(-1,-1,0)
          - 20 \* (11 - 23 \* x^{-1}) \* \Hhh(-1,0,0)
          + 10 \* (5 - 33 \* x^{-1}) \* \H(-1) \* \z2 
  \nonumber\\&& \mbox{}
          - 120 \* (1 - x^{-1}) \* (4 \* \Hh(-1,-1) \* \z2 
          - 3 \* \Hh(-1,0) \* \z2 
          + 2 \* \Hh(-1,3)
          - 4 \* \Hhh(-1,-1,2)
          - 2 \* \Hhh(1,-2,0)
          - 2 \* \Hhhh(-1,-1,0,0)
  \nonumber\\&& \mbox{}
          + \Hhhh(-1,0,0,0)
          - 3 \* \H(-1) \* \z3 ))
          + {8 \over 225} \* \pgq(x) \* (60 \* 
            (19 - 11 \* x^{-1}) \* \Hh(1,0) \* \z2 
          - 120 \* (1 + x^{-1}) \* \Hhhh(1,0,0,0)
  \nonumber\\&& \mbox{}
          - 10 \* (9 - 11 \* x^{-1}) \* \H(1) \* \z2 
          - 60 \* (13 - 17 \* x^{-1}) \* \H(1) \* \z3 
          - 60 \* (17 - 13 \* x^{-1}) \* (\Hh(1,3)
          - \Hhhh(1,1,0,0))
          - 783 
  \nonumber\\&& \mbox{}
          - 1380 \* \z3 
          - 740 \* \z2
          + 220 \* \Hh(0,0)
          + 240 \* \Hhh(0,0,0)
          + 780 \* \Hhh(1,0,0)
          + 543 \* \H(0)
          + 300 \* \H(0) \* \z2 
          - 760 \* \H(1)
          + 760 \* \H(2)
  \nonumber\\&& \mbox{}
          - 540 \* \H(3))
          - {64 \over 15} \* (146 - 169 \* x) \* \H(-2) \* \z2
          - {32 \over 5} \* (71 + 41 \* x) \* \H(-1) \* \z3
          - {16 \over 15} \* (902 + 1247 \* x) \* \H(-1) \* \z2
  \nonumber\\&& \mbox{}
          - {32 \over 45} \* (311 - 2096 \* x) \* \H(0) \* \z2
          + {32 \over 15} \* (343 - 1343 \* x) \* \H(0) \* \z3
          + {4 \over 45} \* (7922 - 257 \* x) \* \H(0)
  \nonumber\\&& \mbox{}
          + {32 \over 15} \* (179 + 96 \* x) \* \H(1) \* \z3
          - {16 \over 45} \* (1588 - 1113 \* x) \* \H(1) \* \z2
          + {4 \over 45} \* (10319 - 739 \* x) \* \H(1)
          - 192 \* (2 - 5 \* x) \* \H(2) \* \z3
  \nonumber\\&& \mbox{}
          - {128 \over 3} \* (7 + 6 \* x) \* \H(2) \* \z2
          - {8 \over 45} \* (1694 - 7221 \* x) \* \H(2)
          + {32 \over 45} \* (353 - 2058 \* x) \* \H(3)
          + {32 \over 15} \* (229 - 429 \* x) \* \H(4)
  \nonumber\\&& \mbox{}
          - 16 \* (2 - x) \* (8 \* \Hh(1,1) \* \z2 + 6 \* \Hh(1,2)
          + 4 \* \Hhh(1,1,0)
          + 5 \* \Hhh(1,1,1))
          - 256 \* (2 + x) \* (\Hhh(-1,-2,0)
          - \Hhhh(-1,-1,-1,0))
  \nonumber\\&& \mbox{}
          + {64 \over 15} \* (31 + 21 \* x) \* (\Hh(-1,3)
          - 2 \* \Hhh(-1,-1,2))
          + {224 \over 75} \* (203 - 223 \* x) \* \z2^2
          + {8 \over 15} \* (1576 + 119 \* x)
  \nonumber\\&& \mbox{}
          + {16 \over 45} \* (2857 + 3833 \* x) \* \z3
          + {8 \over 225} \* (11708 + 27567 \* x) \* \z2
          - {32 \over 5} \* x \* (625 \* \z5
          + 20 \* \z2 \* \z3
          - 40 \* \Hh(-3,2)
  \nonumber\\&& \mbox{}
          + 80 \* \Hh(-2,-1) \* \z2
          - 60 \* \Hh(-2,0) \* \z2
          + 20 \* \Hh(-2,3)
          + 40 \* \Hh(0,0) \* \z3
          + 40 \* \Hh(2,1) \* \z2
          + 30 \* \Hh(2,2)
          + 35 \* \Hh(3,0)
          + 40 \* \Hh(3,1)
  \nonumber\\&& \mbox{}
          - 80 \* \Hhh(-3,-1,0)
          + 20 \* \Hhh(-3,0,0)
          - 80 \* \Hhh(-2,-2,0)
          - 40 \* \Hhh(-2,-1,2)
          + 20 \* \Hhh(2,1,0)
          + 25 \* \Hhh(2,1,1)
          - 20 \* \Hhh(3,0,0)
  \nonumber\\&& \mbox{}
          + 80 \* \Hhhh(-2,-1,-1,0)
          - 140 \* \Hhhh(-2,-1,0,0)
          + 80 \* \Hhhh(-2,0,0,0)
          - 80 \* \Hhhh(2,0,0,0) 
          - 70 \* \H(-2) \* \z3 
          - 12 \* \H(0) \* \z2^2 
          + 40 \* \H(3) \* \z2 )
  \nonumber\\&& \mbox{}
          - {384 \over 5} \* (
	    5 \* \Hh(-1,0) \* \z3
          + 5 \* \Hh(-1,4)
          - 5 \* \Hhh(-1,0,0) \* \z2 
          - 5 \* \Hhhh(-1,2,0,0)
          + 4 \* \H(-1) \* \z2^2 )
          \biggr)
  \nonumber\\&& \mbox{}
       +  \colour4colour{\ca \* \nf^2}  \*  \biggl(
            {32 \over 3} \* (2 + 3 \* x) \* \Hh(-2,0)
          + {8 \over 45} \* \pgq( - x) \* (7 + 3 \* x^{-1}) \* \Hh(-1,0)
          + {8 \over 45} \* (139 + 9 \* x) \* \Hh(-1,0) 
  \nonumber\\&& \mbox{}
          - {16 \over 45} \* (189 + 551 \* x) \* \Hh(0,0)
          - {16 \over 45} \* \pqg( - x) \* ((76 + 9 \* x) \* \Hh(-1,0)
          + 15 \* (4 \* \Hh(-2,0)
          + 2 \* \Hh(-1,2)
          - 2 \* \Hhh(-1,-1,0)
  \nonumber\\&& \mbox{}
          + 6 \* \Hhh(-1,0,0)
          - 3 \* \H(-1) \* \z2))
          + {2 \over 135} \* \pqg(x) \* (72 \* (68 + 3 \* x) \* \Hh(0,0)
          - 72 \* (83 + 3 \* x) \* \z2 
          + 24961 - 1620 \* \z3 
  \nonumber\\&& \mbox{}
          + 4680 \* \Hh(1,0)
          + 4680 \* \Hh(1,1)
          + 720 \* \Hh(2,0)
          + 720 \* \Hh(2,1)
          + 720 \* \Hhh(1,0,0)
          + 1440 \* \Hhh(1,1,0)
          + 720 \* \Hhh(1,1,1)
  \nonumber\\&& \mbox{}
          + 12896 \* \H(0)
          - 720 \* \H(0) \* \z2 
          + 15260 \* \H(1)
          + 360 \* \H(1) \* \z2 
          + 5760 \* \H(2)
          + 720 \* \H(3))
          + {4 \over 135} \* \pgq(x) \* (487 
          - 60 \* \z2 
  \nonumber\\&& \mbox{}
          + 60 \* \Hh(1,0)
          + 60 \* \Hh(1,1)
          + 18 \* \H(0)
          + 250 \* \H(1))
          + {32 \over 3} \* (1 + 7 \* x) \* \H(0) \* \z2
          - {8 \over 27} \* (673 + 885 \* x) \* \H(0)
  \nonumber\\&& \mbox{}
          - {8 \over 27} \* (785 + 3 \* x) \* \H(1)
          - {16 \over 3} \* (16 + 23 \* x) \* \H(2)
          - {32 \over 3} \* (1 + 8 \* x) \* \H(3)
          - {32 \over 3} \* (1 + 4 \* x) \* (\Hh(2,0)
          + \Hh(2,1) )
  \nonumber\\&& \mbox{}
          + {8 \over 3} \* (9 + 10 \* x) \* \z3
          - {16 \over 9} \* (43 - 3 \* x) \* (\Hh(1,0)
          + \Hh(1,1) )
          + {16 \over 45} \* (239 + 181 \* x) \* \z2
          - {26 \over 27} \* (403 - 4 \* x)
  \nonumber\\&& \mbox{}
          + {16 \over 3} \* (2 \* \Hh(-1,2)
          - 2 \* \Hhh(-1,-1,0)
          + 6 \* \Hhh(-1,0,0)
          - 24 \* \Hhh(0,0,0) \* x 
          - 2 \* \Hhh(1,0,0)
          - 4 \* \Hhh(1,1,0)
          - 2 \* \Hhh(1,1,1)
          - 3 \* \H(-1) \* \z2 
  \nonumber\\&& \mbox{}
          - \H(1) \* \z2)
          \biggr)
  \nonumber\\&& \mbox{}
       +  \colour4colour{\ca \* \cf \* \nf}  \*  \biggl(
          - {64 \over 15} \* (101 + 196 \* x) \* \Hh(-3,0)
          - {32 \over 225} \* (7493 + 9568 \* x) \* \Hh(-2,0)
          + {32 \over 15} \* (59 + 174 \* x) \* \Hh(-2,2)
  \nonumber\\&& \mbox{}
          - {32 \over 5} \* (11 + 41 \* x) \* \Hh(-1,-1) \* \z2
          - {16 \over 15} \* (131 - 204 \* x) \* \Hh(-1,0) \* \z2
          - {8 \over 675} \* (96914 + 158649 \* x) \* \Hh(-1,0)
  \nonumber\\&& \mbox{}
          + {16 \over 225} \* (523 - 10302 \* x) \* \Hh(-1,2)
          + 128 \* (1 - x) \* \Hh(-1,3)
          + {224 \over 15} \* (19 - 29 \* x) \* \Hh(0,0) \* \z2
  \nonumber\\&& \mbox{}
          - {8 \over 675} \* (53559 - 105724 \* x) \* \Hh(0,0)
          + {16 \over 5} \* (281 - 86 \* x) \* \Hh(1,0) \* \z2
          - {8 \over 3} \* (299 - 42 \* x) \* \Hh(1,0)
  \nonumber\\&& \mbox{}
          + {32 \over 15} \* (83 - 33 \* x) \* \Hh(1,1) \* \z2
          - {16 \over 3} \* (176 - 21 \* x) \* \Hh(1,1)
          - {8 \over 3} \* (37 - 28 \* x) \* \Hh(1,2)  
          - {32 \over 15} \* (427 - 87 \* x) \* \Hh(1,3)
  \nonumber\\&& \mbox{}
          - {104 \over 3} \* (3 + 20 \* x) \* \Hh(2,0)
          - 64 \* (6 - 29 \* x) \* \Hh(2,0) \* \z2
          - {16 \over 3} \* (20 + 119 \* x) \* \Hh(2,1)
          + 192 \* (2 - 9 \* x) \* \Hh(2,3)
  \nonumber\\&& \mbox{}
          + {32 \over 15} \* (159 - 656 \* x) \* \Hhh(-2,-1,0)
          - {32 \over 15} \* (146 - 269 \* x) \* \Hhh(-2,0,0)
          + {64 \over 15} \* (68 + 33 \* x) \* \Hhh(-1,-2,0)
  \nonumber\\&& \mbox{}
          + {16 \over 225} \* (19667 + 14292 \* x) \* \Hhh(-1,-1,0)
          - {64 \over 3} \* (5 - 9 \* x) \* \Hhh(-1,-1,2)
          - {16 \over 225} \* (10792 + 17067 \* x) \* \Hhh(-1,0,0)
  \nonumber\\&& \mbox{}
          + {16 \over 25} \* (491 - 1091 \* x) \* \Hhh(0,0,0)
          - 160 \* (1 + 6 \* x) \* \Hhh(1,-2,0)
          - {32 \over 15} \* (294 - 139 \* x) \* \Hhh(1,0,0)
          - {8 \over 3} \* (83 + 4 \* x) \* \Hhh(1,1,0)
  \nonumber\\&& \mbox{}
          - 16 \* (8 - x) \* \Hhh(1,1,1)
          + {32 \over 15} \* (216 - 881 \* x) \* \Hhh(2,0,0)
          - {64 \over 15} \* (83 + 33 \* x) \* \Hhhh(-1,-1,-1,0)
  \nonumber\\&& \mbox{}
          + 320 \* (1 + x) \* \Hhhh(-1,-1,0,0)
          - {16 \over 5} \* (13 + 58 \* x) \* \Hhhh(-1,0,0,0)
          + {32 \over 15} \* (72 - 277 \* x) \* \Hhhh(0,0,0,0)
  \nonumber\\&& \mbox{}
          - {16 \over 5} \* (41 - 46 \* x) \* \Hhhh(1,0,0,0)
          + {32 \over 15} \* (337 - 57 \* x) \* \Hhhh(1,1,0,0)
          - 64 \* (6 - 25 \* x) \* \Hhhh(2,1,0,0)
  \nonumber\\&& \mbox{}
          + {8 \over 225} \* \pqg( - x) \* (120 \* (71 + 24 \* x) \* \Hh(-3,0)
          + 4 \* (2803 + 1212 \* x) \* \Hh(-2,0)
          - 300 \* (23 - 24 \* x) \* \Hh(-2,2)
  \nonumber\\&& \mbox{}
          - 360 \* (1 - 21 \* x) \* \Hh(-1,-1) \* \z2 
          + 90 \* (61 - 76 \* x) \* \Hh(-1,0) \* \z2
          - (15529 - 7342 \* x) \* \Hh(-1,0)
  \nonumber\\&& \mbox{}
          - 4 \* (4387 - 1212 \* x) \* \Hh(-1,2)
          - 1800 \* (2 - 3 \* x) \* \Hh(-1,3)
          - 60 \* (103 + 12 \* x) \* \Hhh(-2,-1,0)
  \nonumber\\&& \mbox{}
          + 1800 \* (1 + 3 \* x) \* \Hhh(-2,0,0)
          - 360 \* (13 + 2 \* x) \* \Hhh(-1,-2,0)
          - 16 \* (892 + 303 \* x) \* \Hhh(-1,-1,0)
  \nonumber\\&& \mbox{}
          + 3600 \* (1 - 2 \* x) \* \Hhh(-1,-1,2)
          - 6 \* (1043 - 1368 \* x) \* \Hhh(-1,0,0)
          + 180 \* (23 + 12 \* x) \* \Hhh(1,-2,0)
  \nonumber\\&& \mbox{}
          + 720 \* (9 + x) \* \Hhhh(-1,-1,-1,0)
          - 1800 \* (2 + 3 \* x) \* \Hhhh(-1,-1,0,0)
          - 270 \* (7 - 12 \* x) \* \Hhhh(-1,0,0,0)
  \nonumber\\&& \mbox{}
          + 30 \* (127 - 252 \* x) \* \H(-2) \* \z2 
          + 4 \* (2603 - 1818 \* x) \* \H(-1) \* \z2 
          - 720 \* (1 - x) \* (\Hhh(-1,2,0)
          + \Hhh(-1,2,1))
  \nonumber\\&& \mbox{}
          - 180 \* (60 \* \Hh(-1,0) \* \z3 
          + 60 \* \Hh(-1,4)
          - 60 \* \Hhh(-1,0,0) \* \z2 
          - 60 \* \Hhhh(-1,2,0,0)
          + 35 \* \H(-1) \* \z3 \* x 
          + 48 \* \H(-1) \* \z2^2 ))
  \nonumber\\&& \mbox{}
          - {2 \over 675} \* \pqg(x) \* (5040 \* 
            (19 - 6 \* x) \* \Hh(0,0) \* \z2 
          - 24 \* (8065 - 3671 \* x) \* \Hh(0,0)
          + 1080 \* (143 + 48 \* x) \* \Hh(1,0) \* \z2 
  \nonumber\\&& \mbox{}
          + 4320 \* (9 - x) \* \Hh(1,1) \* \z2 
          - 900 \* (31 - 22 \* x) \* \Hh(1,2)
          - 4320 \* (41 + 16 \* x) \* \Hh(1,3)
          - 1980 \* (17 + 10 \* x) \* \Hh(2,0)
  \nonumber\\&& \mbox{}
          + 1296 \* (71 + 76 \* x) \* \Hhh(0,0,0)
          - 900 \* (65 + 22 \* x) \* \Hhh(1,1,0)
          + 17280 \* (9 + 4 \* x) \* \Hhh(2,0,0)
          - 3240 \* (1 - 4 \* x) \* \Hhhh(1,0,0,0)
  \nonumber\\&& \mbox{}
          + 4320 \* (31 + 16 \* x) \* \Hhhh(1,1,0,0)
          - 29520 \* (11 + 6 \* x) \* \H(0) \* \z3 
          - 12 \* (881 + 11346 \* x) \* \H(0) \* \z2 
  \nonumber\\&& \mbox{}
          - 1080 \* (179 + 94 \* x) \* \H(1) \* \z3 
          + 12 \* (9461 - 4074 \* x) \* \H(1) \* \z2 
          + 360 \* (103 - 12 \* x) \* \H(2) \* \z2 
  \nonumber\\&& \mbox{}
          - 12 \* (3967 - 6498 \* x) \* \H(3)
          - 720 \* (181 + 6 \* x) \* \H(4)
          + 8640 \* (1 + x) \* (\Hh(3,0)
          + \Hh(3,1)
          + 6 \* \Hhhh(0,0,0,0) )
  \nonumber\\&& \mbox{}
          - 360 \* (593 + 150 \* x) \* \z2^2 
          - 120 \* (2881 + 717 \* x) \* \z3 
          + 24 \* (7075 - 3671 \* x) \* \z2 
          - 578087 
  \nonumber\\&& \mbox{}
          - 224460 \* \Hh(1,0)
          - 272160 \* \Hh(1,1)
          - 129600 \* \Hh(2,0) \* \z2 
          - 34560 \* \Hh(2,1)
          + 129600 \* \Hh(2,3)
          - 119520 \* \Hhh(1,0,0)
  \nonumber\\&& \mbox{}
          - 37800 \* \Hhh(1,1,1)
          - 129600 \* \Hhhh(2,1,0,0)
          - 527556 \* \H(0)
          - 556644 \* \H(1)
          - 257904 \* \H(2)
          + 129600 \* \H(2) \* \z3 ) 
  \nonumber\\&& \mbox{}
          + {8 \over 675} \* \pgq( - x) \* 
            (240 \* (1 - 7 \* x^{-1}) \* \Hh(-2,0)
          + (2977 - 3901 \* x^{-1}) \* \Hh(-1,0)
          + 6 \* (24 - 349 \* x^{-1}) \* \Hh(-1,2)
  \nonumber\\&& \mbox{}
          - 6 \* (314 - 349 \* x^{-1}) \* \Hhh(-1,-1,0)
          + 6 \* (329 - 629 \* x^{-1}) \* \Hhh(-1,0,0)
          - 3 \* (362 - 1047 \* x^{-1}) \* \H(-1) \* \z2
  \nonumber\\&& \mbox{}
          + 90 \* (1 - x^{-1}) \* (8 \* \Hh(-2,2)
          + 42 \* \Hh(-1,-1) \* \z2 
          - 38 \* \Hh(-1,0) \* \z2 
          + 30 \* \Hh(-1,3)
          - 8 \* \Hhh(-2,-1,0)
          + 8 \* \Hhh(-2,0,0)
  \nonumber\\&& \mbox{}
          - 4 \* \Hhh(-1,-2,0)
          - 40 \* \Hhh(-1,-1,2)
          + 4 \* \Hhh(-1,2,0)
          + 4 \* \Hhh(-1,2,1)
          - 12 \* \Hhh(1,-2,0)
          + 4 \* \Hhhh(-1,-1,-1,0)
          - 30 \* \Hhhh(-1,-1,0,0)
  \nonumber\\&& \mbox{}
          + 18 \* \Hhhh(-1,0,0,0)
          - 12 \* \H(-2) \* \z2 
	     - 35 \* \H(-1) \* \z3 ))
          - {8 \over 675} \* \pgq(x) \* 
            (1080 \* (3 - 2 \* x^{-1}) \* \Hh(1,0) \* \z2 
  \nonumber\\&& \mbox{}
          - 90 \* (13 - 47 \* x^{-1}) \* \H(1) \* \z3 
          + 3 \* (314 + 349 \* x^{-1}) \* \H(1) \* \z2 
          + 180 \* (1 + x^{-1}) \* (\Hh(1,1) \* \z2 
          - 3 \* \Hhhh(1,0,0,0)
          + 2 \* \H(2) \* \z2 )
  \nonumber\\&& \mbox{}
          - 360 \* (7 - 8 \* x^{-1}) \* (\Hh(1,3)
          - \Hhhh(1,1,0,0))
          - 6287 
          - 5580 \* \z3 
          - 3918 \* \z2 
          + 2334 \* \Hh(0,0)
          - 1710 \* \Hh(1,0)
  \nonumber\\&& \mbox{}
          - 1710 \* \Hh(1,1)
          + 360 \* \Hh(2,0)
          + 360 \* \Hh(2,1)
          + 2160 \* \Hhh(0,0,0)
          + 1980 \* \Hhh(1,0,0)
          + 2527 \* \H(0)
          - 1260 \* \H(0) \* \z2 
  \nonumber\\&& \mbox{}
          - 4464 \* \H(1)
          + 2994 \* \H(2)
          - 180 \* \H(3))
          + {16 \over 15} \* (41 - 1004 \* x) \* \H(-2) \* \z2
          + {224 \over 3} \* (1 + 3 \* x) \* \H(-1) \* \z3
  \nonumber\\&& \mbox{}
          + {8 \over 75} \* (6207 + 11632 \* x) \* \H(-1) \* \z2
          - {32 \over 15} \* (451 - 1691 \* x) \* \H(0) \* \z3
          - {8 \over 225} \* (2351 + 17174 \* x) \* \H(0) \* \z2
  \nonumber\\&& \mbox{}
          - {4 \over 675} \* (282101 - 10934 \* x) \* \H(0)
          - {16 \over 15} \* (689 + 96 \* x) \* \H(1) \* \z3
          + {16 \over 225} \* (11221 - 8196 \* x) \* \H(1) \* \z2
  \nonumber\\&& \mbox{}
          - {4 \over 45} \* (18809 + 341 \* x) \* \H(1)
          + 96 \* (4 - 13 \* x) \* \H(2) \* \z3
          + {16 \over 15} \* (159 + 686 \* x) \* \H(2) \* \z2
  \nonumber\\&& \mbox{}
          - {16 \over 9} \* (354 + 569 \* x) \* \H(2)
          - {8 \over 225} \* (1537 - 10462 \* x) \* \H(3)
          - {32 \over 15} \* (181 - 191 \* x) \* \H(4)
          + {128 \over 15} \* (3 - 28 \* x) \* (\Hh(3,0)
  \nonumber\\&& \mbox{}
          + \Hh(3,1))
          + {64 \over 15} \* (7 - 3 \* x) \* (\Hhh(-1,2,0)
          + \Hhh(-1,2,1))
          - {16 \over 15} \* (593 - 804 \* x) \* \z2^2
          - {16 \over 45} \* (2743 + 2742 \* x) \* \z3
  \nonumber\\&& \mbox{}
          + {16 \over 675} \* (16461 - 63161 \* x) \* \z2
          - {2 \over 675} \* (596697 + 6538 \* x)
          + {16 \over 5} \* x \* (1225 \* \z5
          + 80 \* \z2 \* \z3 
          - 40 \* \Hh(-3,2) 
  \nonumber\\&& \mbox{}
          + 80 \* \Hh(-2,-1) \* \z2 
          - 60 \* \Hh(-2,0) \* \z2 
          + 20 \* \Hh(-2,3) 
          + 40 \* \Hh(0,0) \* \z3 
          + 40 \* \Hh(2,1) \* \z2 
          - 10 \* \Hh(2,2) 
          - 80 \* \Hhh(-3,-1,0) 
  \nonumber\\&& \mbox{}
          + 20 \* \Hhh(-3,0,0) 
          - 80 \* \Hhh(-2,-2,0) 
          - 40 \* \Hhh(-2,-1,2) 
          - 30 \* \Hhh(2,1,0)
          - 10 \* \Hhh(2,1,1) 
          - 20 \* \Hhh(3,0,0) 
          + 80 \* \Hhhh(-2,-1,-1,0)
  \nonumber\\&& \mbox{}
          - 140 \* \Hhhh(-2,-1,0,0) 
          + 80 \* \Hhhh(-2,0,0,0) 
          - 80 \* \Hhhh(2,0,0,0) 
          - 70 \* \H(-2) \* \z3 
          - 12 \* \H(0) \* \z2^2 
          + 40 \* \H(3) \* \z2 )
  \nonumber\\&& \mbox{}
          + {384 \over 5} \* (
            5 \* \Hh(-1,0) \* \z3
          + 5 \* \Hh(-1,4)
          - 5 \* \Hhh(-1,0,0) \* \z2 
          - 5 \* \Hhhh(-1,2,0,0)
          + 4 \* \H(-1) \* \z2^2 )
          \biggr)
  \nonumber\\&& \mbox{}
       +  \colour4colour{\ca^2 \* \nf}  \*  \biggr(
            {16 \over 15} \* (71 + 476 \* x) \* \Hh(-3,0)
          - {4 \over 9} \* (2101 + 2972 \* x) \* \Hh(-2,0)
          - {32 \over 15} \* (139 - 41 \* x) \* \Hh(-2,2)
  \nonumber\\&& \mbox{}
          - {32 \over 15} \* (193 - 12 \* x) \* \Hh(-1,-1) \* \z2
          + {8 \over 5} \* (263 - 12 \* x) \* \Hh(-1,0) \* \z2
          + {2 \over 45} \* (10339 + 5969 \* x) \* \Hh(-1,0)
  \nonumber\\&& \mbox{}
          - {4 \over 15} \* (1782 - 793 \* x) \* \Hh(-1,2)
          - {64 \over 15} \* (82 - 3 \* x) \* \Hh(-1,3)
          - {16 \over 15} \* (137 + 1423 \* x) \* \Hh(0,0) \* \z2
  \nonumber\\&& \mbox{}
          + {4 \over 45} \* (50251 - 46956 \* x) \* \Hh(0,0)
          - {56 \over 15} \* (91 - 6 \* x) \* \Hh(1,0) \* \z2
          + {8 \over 9} \* (2570 + 63 \* x) \* \Hh(1,0)
          + 16 \* (73 - 7 \* x) \* \Hh(1,2)
  \nonumber\\&& \mbox{}
          + {16 \over 3} \* (77 - 3 \* x) \* \Hh(1,3)
          + {64 \over 3} \* (49 + 58 \* x) \* \Hh(2,0)
          + 16 \* (73 + 91 \* x) \* \Hh(2,1)
          + 64 \* (1 + 26 \* x) \* \Hh(3,0)
  \nonumber\\&& \mbox{}
          + 128 \* (1 + 14 \* x) \* \Hh(3,1)
          + {32 \over 3} \* (19 + 53 \* x) \* \Hhh(-2,-1,0)
          - {32 \over 15} \* (117 - 248 \* x) \* \Hhh(-2,0,0)
  \nonumber\\&& \mbox{}
          - {4 \over 3} \* (22 + 197 \* x) \* \Hhh(-1,-1,0)
          + {64 \over 15} \* (89 - 6 \* x) \* \Hhh(-1,-1,2)
          - {4 \over 15} \* (3796 - 529 \* x) \* \Hhh(-1,0,0)
  \nonumber\\&& \mbox{}
          + {8 \over 15} \* (391 + 4319 \* x) \* \Hhh(0,0,0)
          - {16 \over 3} \* (1 - 16 \* x) \* \Hhh(1,-2,0)
          + {4 \over 3} \* (971 - 206 \* x) \* \Hhh(1,0,0)
  \nonumber\\&& \mbox{}
          + {16 \over 3} \* (241 - 21 \* x) \* \Hhh(1,1,0)
          + {16 \over 3} \* (209 - 18 \* x) \* \Hhh(1,1,1)
          - {16 \over 3} \* (9 - 283 \* x) \* \Hhh(2,0,0)
          + 192 \* (1 + 4 \* x) \* \Hhh(2,1,1)
  \nonumber\\&& \mbox{}
          + {64 \over 15} \* (97 - 3 \* x) \* \Hhhh(-1,-1,0,0)
          - {8 \over 15} \* (523 - 12 \* x) \* \Hhhh(-1,0,0,0)
          - {32 \over 15} \* (12 - 757 \* x) \* \Hhhh(0,0,0,0)
  \nonumber\\&& \mbox{}
          + {8 \over 15} \* (137 - 12 \* x) \* \Hhhh(1,0,0,0)
          + {16 \over 3} \* (7 + 3 \* x) \* \Hhhh(1,1,0,0)
          - {2 \over 45} \* \pqg( - x) \* (24 \* (71 + 24 \* x) \* \Hh(-3,0)
  \nonumber\\&& \mbox{}
          - 2 \* (9413 + 90 \* x) \* \Hh(-2,0)
          - 48 \* (139 - 24 \* x) \* \Hh(-2,2)
          - 144 \* (63 - 8 \* x) \* \Hh(-1,-1) \* \z2 
  \nonumber\\&& \mbox{}
          + 36 \* (259 - 24 \* x) \* \Hh(-1,0) \* \z2 
          + 2 \* (1438 - 171 \* x) \* \Hh(-1,0)
          - 2 \* (7429 + 90 \* x) \* \Hh(-1,2)
          - 288 \* (27 - 2 \* x) \* \Hh(-1,3)
  \nonumber\\&& \mbox{}
          - 144 \* (39 - 4 \* x) \* \Hhh(-2,0,0)
          + 10 \* (541 + 18 \* x) \* \Hhh(-1,-1,0)
          + 288 \* (29 - 4 \* x) \* \Hhh(-1,-1,2)
  \nonumber\\&& \mbox{}
          - 6 \* (4169 + 30 \* x) \* \Hhh(-1,0,0)
          - 72 \* (3 - 8 \* x) \* \Hhh(1,-2,0)
          + 576 \* (16 - x) \* \Hhhh(-1,-1,0,0)
  \nonumber\\&& \mbox{}
          - 36 \* (173 - 8 \* x) \* \Hhhh(-1,0,0,0)
          + 24 \* (373 - 48 \* x) \* \H(-2) \* \z2 
          + 288 \* (28 - 3 \* x) \* \H(-1) \* \z3 
  \nonumber\\&& \mbox{}
          + (17563 + 270 \* x) \* \H(-1) \* \z2 
          - 48 \* (45 \* \Hh(-1,0) \* \z3 
          + 45 \* \Hh(-1,4)
          - 95 \* \Hhh(-2,-1,0)
          - 45 \* \Hhh(-1,-2,0)
  \nonumber\\&& \mbox{}
          - 45 \* \Hhh(-1,0,0) \* \z2 
          + 30 \* \Hhh(-1,2,0)
          + 30 \* \Hhh(-1,2,1)
          + 30 \* \Hhhh(-1,-1,-1,0)
          - 45 \* \Hhhh(-1,2,0,0)
          + 36 \* \H(-1) \* \z2^2))
  \nonumber\\&& \mbox{}
          + {2 \over 2025} \* \pqg(x) \* (1080 \* 
            (137 + 12 \* x) \* \Hh(0,0) \* \z2 
          - 90 \* (48304 + 171 \* x) \* \Hh(0,0)
          + 1620 \* (167 + 32 \* x) \* \Hh(1,0) \* \z2
  \nonumber\\&& \mbox{}
          - 16200 \* (21 + 4 \* x) \* \Hh(1,3)
          - 8100 \* (45 + x) \* \Hhh(0,0,0)
          + 16200 \* (3 + 4 \* x) \* \Hhh(2,0,0)
          + 25920 \* (1 + x) \* \Hhhh(0,0,0,0)
  \nonumber\\&& \mbox{}
          - 1620 \* (47 - 8 \* x) \* \Hhhh(1,0,0,0)
          - 16200 \* (7 - 4 \* x) \* \Hhhh(1,1,0,0)
          - 2700 \* (95 + 48 \* x) \* \H(0) \* \z3 
  \nonumber\\&& \mbox{}
          + 90 \* (12607 + 180 \* x) \* \H(0) \* \z2 
          + 11340 \* (7 - 8 \* x) \* \H(1) \* \z3 
          + 225 \* (4043 + 18 \* x) \* \H(1) \* \z2 
  \nonumber\\&& \mbox{}
          - 90 \* (12517 + 90 \* x) \* \H(3) 
          - 1080 \* (161 + 36 \* x) \* \H(4)
          - 54 \* (6091 + 1296 \* x) \* \z2^2 
          + 45 \* (37027 + 450 \* x) \* \z3 
  \nonumber\\&& \mbox{}
          + 90 \* (40069 + 171 \* x) \* \z2 
          - 5 \* (1098464 
          + 454680 \* \Hh(1,0)
          + 494280 \* \Hh(1,1)
          - 38880 \* \Hh(1,1) \* \z2 
  \nonumber\\&& \mbox{}
          + 206280 \* \Hh(1,2)
          + 207360 \* \Hh(2,0)
          + 19440 \* \Hh(2,0) \* \z2 
          + 227880 \* \Hh(2,1)
          + 38880 \* \Hh(2,2)
          - 19440 \* \Hh(2,3)
  \nonumber\\&& \mbox{}
          + 12960 \* \Hh(3,0)
          + 25920 \* \Hh(3,1)
          + 200070 \* \Hhh(1,0,0)
          + 230040 \* \Hhh(1,1,0)
          + 199800 \* \Hhh(1,1,1)
          + 45360 \* \Hhh(1,1,2)
  \nonumber\\&& \mbox{}
          + 38880 \* \Hhh(1,2,0)
          + 45360 \* \Hhh(1,2,1)
          + 38880 \* \Hhh(2,1,0)
          + 38880 \* \Hhh(2,1,1)
          + 45360 \* \Hhhh(1,1,1,0)
          + 38880 \* \Hhhh(1,1,1,1)
  \nonumber\\&& \mbox{}
          + 19440 \* \Hhhh(2,1,0,0)
          + 55413 \* \H(0)
          + 413139 \* \H(1)
          + 718164 \* \H(2)
          - 19440 \* \H(2) \* \z3 
          - 18360 \* \H(2) \* \z2))
  \nonumber\\&& \mbox{}
          - {2 \over 45} \* \pgq( - x) \* (144 \* \Hh(-2,0)
          + 3 \* (403 + 19 \* x^{-1}) \* \Hh(-1,0)
          - 2 \* (371 - 15 \* x^{-1}) \* \Hh(-1,2)
  \nonumber\\&& \mbox{}
          - 10 \* (1 + 3 \* x^{-1}) \* \Hhh(-1,-1,0)
          - 6 \* (181 - 5 \* x^{-1}) \* \Hhh(-1,0,0)
          + (737 - 45 \* x^{-1}) \* \H(-1) \* \z2 
  \nonumber\\&& \mbox{}
          + 48 \* (1 - x^{-1}) \* (4 \* \Hh(-1,-1) \* \z2 
          - 3 \* \Hh(-1,0) \* \z2 
          + 2 \* \Hh(-1,3)
          - 4 \* \Hhh(-1,-1,2)
          - 2 \* \Hhh(1,-2,0)
          - 2 \* \Hhhh(-1,-1,0,0)
  \nonumber\\&& \mbox{}
          + \Hhhh(-1,0,0,0)
          - 3 \* \H(-1) \* \z3 ))
          + {2 \over 405} \* \pgq(x) \* (216 \* 
            (7 - 8 \* x^{-1}) \* \Hh(1,0) \* \z2 
          - 432 \* (1 + x^{-1}) \* \Hhhh(1,0,0,0)
  \nonumber\\&& \mbox{}
          - 216 \* (1 - 14 \* x^{-1}) \* \H(1) \* \z3 
          + 45 \* (169 - 3 \* x^{-1}) \* \H(1) \* \z2 
          - 1080 \* (1 - 2 \* x^{-1}) \* (\Hh(1,3)
          - \Hhhh(1,1,0,0))
  \nonumber\\&& \mbox{}
          - 55741 
          + 6480 \* \z3 
          + 11808 \* \z2 
          - 1134 \* \Hh(0,0)
          - 19260 \* \Hh(1,0)
          - 20160 \* \Hh(1,1)
          - 7560 \* \Hh(1,2)
  \nonumber\\&& \mbox{}
          - 4320 \* \Hh(2,0)
          - 5400 \* \Hh(2,1)
          + 864 \* \Hhh(0,0,0)
          - 6480 \* \Hhh(1,0,0)
          - 7560 \* \Hhh(1,1,0)
          - 6480 \* \Hhh(1,1,1)
          - 7539 \* \H(0)
  \nonumber\\&& \mbox{}
          + 2592 \* \H(0) \* \z2 
          - 9666 \* \H(1)
          - 414 \* \H(2)
          - 1296 \* \H(3))
          + {16 \over 15} \* (373 + 183 \* x) \* \H(-2) \* \z2
          + {96 \over 5} \* (19 - x) \* \H(-1) \* \z3
  \nonumber\\&& \mbox{}
          + {2 \over 15} \* (3454 - 2571 \* x) \* \H(-1) \* \z2
          + {8 \over 3} \* (95 - 878 \* x) \* \H(0) \* \z3
          - {4 \over 45} \* (11161 + 12454 \* x) \* \H(0) \* \z2
  \nonumber\\&& \mbox{}
          - {2 \over 135} \* (17971 - 471615 \* x) \* \H(0)
          + {8 \over 15} \* (1 - 6 \* x) \* \H(1) \* \z3
          - {2 \over 3} \* (1774 - 365 \* x) \* \H(1) \* \z2
  \nonumber\\&& \mbox{}
          + {2 \over 135} \* (107239 + 33696 \* x) \* \H(1)
          - {272 \over 3} \* (1 + 13 \* x) \* \H(2) \* \z2
          + {4 \over 45} \* (39922 - 26183 \* x) \* \H(2)
  \nonumber\\&& \mbox{}
          + {4 \over 9} \* (2243 + 3320 \* x) \* \H(3)
          + {16 \over 15} \* (161 + 1709 \* x) \* \H(4)
          + 96 \* (1 - 4 \* x) \* (\Hh(2,0) \* \z2 
          - \Hh(2,3)
          + \Hhhh(2,1,0,0)
          - \H(2) \* \z3 )
  \nonumber\\&& \mbox{}
          + 64 \* (3 + 14 \* x) \* (\Hh(2,2)
          + \Hhh(2,1,0) )
          + {4 \over 75} \* (6091 - 11606 \* x) \* \z2^2
          - {4 \over 45} \* (38827 - 26605 \* x) \* \z2
  \nonumber\\&& \mbox{}
          - {2 \over 45} \* (42979 - 34184 \* x) \* \z3
          + {2 \over 45} \* (115058 + 13187 \* x)
          - {16 \over 45} \* (2700 \* \z5 \* x 
          + 270 \* \z2 \* \z3 \* x 
          + 270 \* \Hh(-1,0) \* \z3 
  \nonumber\\&& \mbox{}
          + 270 \* \Hh(-1,4)
          - 7145 \* \Hh(1,1)
          + 540 \* \Hh(1,1) \* \z2 
          - 270 \* \Hhh(-1,-2,0)
          - 270 \* \Hhh(-1,0,0) \* \z2 
          + 180 \* \Hhh(-1,2,0)
  \nonumber\\&& \mbox{}
          + 180 \* \Hhh(-1,2,1)
          - 630 \* \Hhh(1,1,2)
          - 540 \* \Hhh(1,2,0)
          - 630 \* \Hhh(1,2,1)
          + 180 \* \Hhhh(-1,-1,-1,0)
          - 270 \* \Hhhh(-1,2,0,0)
  \nonumber\\&& \mbox{}
         - 630 \* \Hhhh(1,1,1,0) 
         - 540 \* \Hhhh(1,1,1,1)
          + 216 \* \H(-1) \* \z2^2 )
          \biggr)
\:\: , 
\eea
\normalsize
and the $\ar^{\,3}$ contribution to the pure-singlet coefficient function 
for $F_L$ is
\small
\bea
&& c^{(3)}_{L,\rm{ps}}(x) \:\: = \:\:  
        \colour4colour{\cf \* \nf \* \biggl(\cf-{\ca \over 2}\biggr)} 
            \*  \biggl(
            {64 \over 15} \* (2 + 3 \* x) \* (\Hh(1,0) \* \z2 - \Hh(1,3)
          + \Hhhh(1,1,0,0)
          - \H(1) \* \z3)
          \biggr)
  \nonumber\\&& \mbox{}
       +  \colour4colour{\cf \* \nf^2}  \*  \biggl(
            {32 \over 45} \* ( 
	    \pqg( - x) \* (7 + 3 \* x)
          + \pgq( - x) \* (3 + 2 \* x^{-1})
          + (1 + 6 \* x)) \* \Hh(-1,0)
          - {32 \over 15} \* (19 + 41 \* x) \* \Hh(0,0) 
  \nonumber\\&& \mbox{}
          + {16 \over 3} \* (2 - x) \* \Hh(1,1) 
          + {16 \over 405} \* \pqg(x) \* (54 \* \Hh(0,0) \* (9 - x)
          + 18 \* (8 + 3 \* x) \* \z2 
          + 646 - 90 \* \Hh(1,1)
          + 771 \* \H(0)
  \nonumber\\&& \mbox{}
          + 285 \* \H(1)
          - 90 \* \H(2))
          + {16 \over 405} \* \pgq(x) \* (494 
          - 90 \* \z2 
          - 45 \* \Hh(1,1)
          + 36 \* \H(0)
          + 75 \* \H(1))
          - {16 \over 27} \* (133 + 12 \* x) \* \H(0)
  \nonumber\\&& \mbox{}
          - {16 \over 9} \* (17 - 9 \* x) \* \H(1)
          + {32 \over 9} \* (1 - 13 \* x) \* \H(2)
          - {32 \over 15} \* (6 - 31 \* x) \* \z2
          - {32 \over 27} \* (87 - 49 \* x)
          - {32 \over 3} \* x \* (\z3 
          - \Hh(2,1) 
  \nonumber\\&& \mbox{}
          + 6 \* \Hhh(0,0,0) )
          \biggr)
  \nonumber\\&& \mbox{}
       +  \colour4colour{\cf^2 \* \nf}  \*  \biggl(
          - {32 \over 45} \* (341 + 21 \* x) \* \Hh(-2,0)
          - {16 \over 675} \* (18436 + 18411 \* x) \* \Hh(-1,0)
          + {32 \over 45} \* (29 - 6 \* x) \* \Hh(-1,2)
  \nonumber\\&& \mbox{}
          - {128 \over 15} \* (2 + 33 \* x) \* \Hh(0,0) \* \z2
          + {16 \over 675} \* (237 + 4118 \* x) \* \Hh(0,0) 
          + {512 \over 3} \* (1 - x) \* \Hh(1,0)
          + 8 \* (32 - 27 \* x) \* \Hh(1,1) 
  \nonumber\\&& \mbox{}
          + 32 \* (6 - x) \* \Hh(2,0)
          + {16 \over 3} \* (34 + 19 \* x) \* \Hh(2,1) 
          + {32 \over 45} \* (331 + 276 \* x) \* \Hhh(-1,-1,0)
          - {32 \over 15} \* (29 + 39 \* x) \* \Hhh(-1,0,0)
  \nonumber\\&& \mbox{}
          + {16 \over 45} \* (828 - 113 \* x) \* \Hhh(0,0,0)
          - {128 \over 15} \* (17 - 12 \* x) \* \Hhh(1,0,0)
          - {128 \over 15} \* (2 + 13 \* x) \* \Hhh(2,0,0)
  \nonumber\\&& \mbox{}
          + {16 \over 225} \* \pqg( - x) \* (2 \* (43 + 17 \* x) \* \Hh(-1,0)
          - 30 \* (9 + x) \* \Hh(-1,2)
          - 90 \* (1 - x) \* \Hhh(-1,0,0)
          + 15 \* (7 + 3 \* x) \* \H(-1) \* \z2 
  \nonumber\\&& \mbox{}
          + 30 \* (11 - x) \* (\Hh(-2,0)
          - \Hhh(-1,-1,0) ))
          - {8 \over 675} \* \pqg(x) \* (12 \* (242 + 17 \* x) \* \Hh(0,0)
          + 540 \* (21 + x) \* \Hhh(0,0,0)
  \nonumber\\&& \mbox{}
          - 360 \* (35 - x) \* \H(0) \* \z2 
          - 90 \* (9 - x)  \* \H(1) \* \z2 
          + 180 \* (71 - x) \* \H(3)
          + 144 \* (2 - 3 \* x) \* (4 \* \z2^2 
          - 5 \* \Hh(0,0) \* \z2 
  \nonumber\\&& \mbox{}
          - 5 \* \Hh(1,0) \* \z2 
          + 5 \* \Hh(1,3)
          - 5 \* \Hhh(2,0,0)
          - 5 \* \Hhhh(1,1,0,0)
          + 5 \* \H(0) \* \z3 
          + 5 \* \H(1) \* \z3 
          + 5 \* \H(4))
          + 90 \* (9 + 5 \* x) \* \z3 
  \nonumber\\&& \mbox{}
          - 12 \* (482 + 17 \* x) \* \z2 
          - 71 - 300 \* \Hh(1,0)
          + 2400 \* \Hh(1,1)
          + 1800 \* \Hh(1,2)
          + 5400 \* \Hh(2,0)
          + 4500 \* \Hh(2,1)
  \nonumber\\&& \mbox{}
          - 2160 \* \Hhh(1,0,0)
          + 1800 \* \Hhh(1,1,0)
          + 900 \* \Hhh(1,1,1)
          + 809 \* \H(0)
          + 9005 \* \H(1)
          + 5580 \* \H(2))
  \nonumber\\&& \mbox{}
          + {16 \over 675} \* \pgq( - x) \* ((17 + 148 \* x^{-1}) \* \Hh(-1,0)
          - 30 \* (13 + 2 \* x^{-1}) \* \Hh(-1,2)
          - 30 \* (17 - 2 \* x^{-1}) \* \Hhh(-1,-1,0)
  \nonumber\\&& \mbox{}
          + 45 \* (3 + 2 \* x^{-1})  \* \H(-1) \* \z2 
          - 60 \* (1 - x^{-1})\* (4 \* \Hh(-2,0)
          + 3 \* \Hhh(-1,0,0) ))
          + {8 \over 675} \* \pgq(x) \* 
            (30 \* (13 - 2 \* x^{-1})  \* \H(1) \* \z2 
  \nonumber\\&& \mbox{}
          + 360 \* (1 - 4 \* x^{-1}) \* (\Hh(1,0) \* \z2 
          - \Hh(1,3)
          + \Hhhh(1,1,0,0)
          - \H(1) \* \z3)
          + 1904 
          - 3240 \* \z3 
          - 1650 \* \z2 
          + 360 \* \Hh(0,0)
  \nonumber\\&& \mbox{}
          - 300 \* \Hh(1,0)
          - 975 \* \Hh(1,1)
          - 900 \* \Hh(1,2)
          + 1440 \* \Hhh(1,0,0)
          - 900 \* \Hhh(1,1,0)
          - 450 \* \Hhh(1,1,1)
          + 416 \* \H(0)
          + 1440 \* \H(0) \* \z2 
  \nonumber\\&& \mbox{}
          - 6220 \* \H(1)
          + 1320 \* \H(2)
          - 1440 \* \H(3))
          + {16 \over 15} \* (91 + 96 \* x) \* \H(-1) \* \z2
          + {32 \over 15} \* (8 + 87 \* x) \* \H(0) \* \z3
  \nonumber\\&& \mbox{}
          - {64 \over 15} \* (93 + 22 \* x) \* \H(0) \* \z2
          + {8 \over 135} \* (2027 - 8442 \* x) \* \H(0) 
          + {16 \over 45} \* (151 - 186 \* x) \* \H(1) \* \z2
  \nonumber\\&& \mbox{}
          + {8 \over 45} \* (3533 - 2518 \* x) \* \H(1)
          + {16 \over 45} \* (992 + 67 \* x) \* \H(2)
          + {32 \over 45} \* (561 + 179 \* x) \* \H(3)
          + {64 \over 15} \* (4 + 61 \* x) \* \H(4)
  \nonumber\\&& \mbox{}
          + 16 \* (2 - x) \* (2 \* \Hh(1,2)
          + 2 \* \Hhh(1,1,0)
          + \Hhh(1,1,1))
          - {208 \over 45} \* (9 + 106 \* x) \* \z3
          + {16 \over 75} \* (64 + 221 \* x) \* \z2^2
  \nonumber\\&& \mbox{}
          + {8 \over 135} \* (2671 - 3066 \* x)
          - {64 \over 675} \* (3828 + 4637 \* x) \* \z2
          - {32 \over 3}  \* x \* (2 \* \Hh(-3,0) 
          + 2 \* \Hh(-2,2) 
          - 6 \* \Hh(2,2) 
          - 12 \* \Hh(3,0) 
  \nonumber\\&& \mbox{}
          - 12 \* \Hh(3,1) 
          + 22 \* \Hhh(-2,-1,0) 
          - 6 \* \Hhh(-2,0,0)
          - 6 \* \Hhh(2,1,0) 
          - 3 \* \Hhh(2,1,1) 
          - 6 \* \Hhhh(0,0,0,0) 
          + 9 \* \H(-2) \* \z2 
          - 5 \* \H(2) \* \z2 )
          \biggr)
  \nonumber\\&& \mbox{}
       +  \colour4colour{\ca \* \cf \* \nf}  \*  \biggl(
          - {32 \over 45} \* (324 + 599 \* x) \* \Hh(-2,0)
          + {8 \over 45} \* (1676 + 421 \* x) \* \Hh(-1,0)
          - {16 \over 45} \* (296 + 21 \* x) \* \Hh(-1,2)
  \nonumber\\&& \mbox{}
          + {16 \over 15} \* (8 - 333 \* x) \* \Hh(0,0) \* \z2
          + {16 \over 45} \* (2901 - 2317 \* x) \* \Hh(0,0)
          + {8 \over 3} \* (16 + 47 \* x) \* \Hh(1,0)
          + {8 \over 3} \* (13 + 45 \* x) \* \Hh(1,1)
  \nonumber\\&& \mbox{}
          + 32  \* (1 - 3 \* x)\* \Hh(2,0)
          + {32 \over 3} \* (5 - 11 \* x) \* \Hh(2,1)
          - {16 \over 45} \* (244 + 249 \* x) \* \Hhh(-1,-1,0)
          - {32 \over 45} \* (283 + 78 \* x) \* \Hhh(-1,0,0)
  \nonumber\\&& \mbox{}
          - {16 \over 45} \* (438 - 1093 \* x) \* \Hhh(0,0,0)
          + {16 \over 15} \* (263 - 153 \* x) \* \Hhh(1,0,0)
          + {16 \over 15} \* (8 + 247 \* x) \* \Hhh(2,0,0)
  \nonumber\\&& \mbox{}
          + {16 \over 45} \* \pqg( - x) \* (6 \* (53 + 2 \* x) \* \Hh(-2,0)
          - (491 - 3 \* x) \* \Hh(-1,0)
          + 6 \* (33 + 2 \* x) \* \Hh(-1,2)
          - 6 \* (3 + 2 \* x) \* \Hhh(-1,-1,0)
  \nonumber\\&& \mbox{}
          + 12 \* (24 + x) \* \Hhh(-1,0,0)
          - 9 \* (23 + 2 \* x) \* \H(-1) \* \z2)
          - {8 \over 675} \* \pqg(x) \* (30 \* (1924 + 3 \* x) \* \Hh(0,0) 
          + 360 \* (1 + x) \* \Hhh(0,0,0)
  \nonumber\\&& \mbox{}
          - 720 \* (7 + x) \* \H(0) \* \z2
          - 90 \* (47 + 2 \* x) \* \H(1) \* \z2
          + 360 \* (13 + x) \* \H(3)
          - 72 \* (2 - 3 \* x) \* (4 \* \z2^2
          - 5 \* \Hh(0,0) \* \z2 
  \nonumber\\&& \mbox{}
          - 5 \* \Hh(1,0) \* \z2 
          + 5 \* \Hh(1,3)
          - 5 \* \Hhh(2,0,0)
          - 5 \* \Hhhh(1,1,0,0)
          + 5 \* \H(0) \* \z3
          + 5 \* \H(1) \* \z3
          + 5 \* \H(4))
          - 90 \* (217 + 10 \* x) \* \z3
  \nonumber\\&& \mbox{}
          - 45 \* (641 + 2 \* x) \* \z2
          + 5 \* (11091
          + 1650 \* \Hh(1,0)
          + 1515 \* \Hh(1,1)
          + 900 \* \Hh(1,2)
          + 1260 \* \Hh(2,0)
          + 1260 \* \Hh(2,1)
  \nonumber\\&& \mbox{}
          + 1296 \* \Hhh(1,0,0)
          + 900 \* \Hhh(1,1,0)
          + 900 \* \Hhh(1,1,1)
          - 10103 \* \H(0)
          - 5069 \* \H(1)
          + 5751 \* \H(2)))
  \nonumber\\&& \mbox{}
          - {8 \over 45} \* \pgq( - x) \* (60 \* \Hh(-2,0)
          + (263 - 4 \* x^{-1}) \* \Hh(-1,0)
          - 2 \* (97 + 8 \* x^{-1}) \* \Hh(-1,2)
          + 2 \* (7 + 8 \* x^{-1}) \* \Hhh(-1,-1,0)
  \nonumber\\&& \mbox{}
          - 4 \* (71 + 4 \* x^{-1}) \* \Hhh(-1,0,0)
          + 3 \* (67 + 8 \* x^{-1}) \* \H(-1) \* \z2)
          + {8 \over 135} \* \pgq(x) \* 
            (3 \* (143 + 8 \* x^{-1}) \* \H(1) \* \z2
  \nonumber\\&& \mbox{}
          - 36 \* (1 - 4 \* x^{-1}) \* (\Hh(1,0) \* \z2
          - \Hh(1,3)
          + \Hhhh(1,1,0,0)
          - \H(1) \* \z3)
          - 4459 
          + 1044 \* \z3 
          + 993 \* \z2 
          + 48 \* \Hh(0,0)
  \nonumber\\&& \mbox{}
          - 1185 \* \Hh(1,0)
          - 1095 \* \Hh(1,1)
          - 450 \* \Hh(1,2)
          - 360 \* \Hh(2,0)
          - 450 \* \Hh(2,1)
          - 684 \* \Hhh(1,0,0)
          - 450 \* \Hhh(1,1,0)
  \nonumber\\&& \mbox{}
          - 450 \* \Hhh(1,1,1)
          - 716 \* \H(0)
          + 36 \* \H(0) \* \z2 
          + 511 \* \H(1)
          - 216 \* \H(2)
          + 144 \* \H(3))
          + {8 \over 15} \* (116 - 69 \* x) \* \H(-1) \* \z2
  \nonumber\\&& \mbox{}
          - {32 \over 15} \* (4 + 231 \* x) \* \H(0) \* \z3
          + {16 \over 9} \* (39 - 31 \* x)  \* \H(0) \* \z2
          - {8 \over 135} \* (14097 - 33650 \* x) \* \H(0)
  \nonumber\\&& \mbox{}
          - {8 \over 45} \* (1144 - 699 \* x) \* \z2 \* \H(1)
          - {8 \over 45} \* (2929 - 1069 \* x) \* \H(1)
          + {8 \over 45} \* (2393 - 4652 \* x) \* \H(2) 
  \nonumber\\&& \mbox{}
          - {16 \over 45} \* (147 + 383 \* x) \* \H(3) 
          - {16 \over 15} \* (8 - 403 \* x) \* \H(4)
          + 80 \* (2 - x) \* (\Hh(1,2)
          + \Hhh(1,1,0)
          + \Hhh(1,1,1))
  \nonumber\\&& \mbox{}
          - {8 \over 75} \* (64 + 641 \* x) \* \z2^2
          - {16 \over 45} \* (1329 - 824 \* x) \* \z3
          - {8 \over 45} \* (1881 - 3125 \* x) \* \z2
          + {8 \over 135} \* (13933 + 1617 \* x)
  \nonumber\\&& \mbox{}
          + {32 \over 3}  \* x \* (7 \* \Hh(-3,0)
          + 10 \* \Hh(-2,2) 
          + 15 \* \Hh(2,2) 
          + 30 \* \Hh(3,0) 
          + 33 \* \Hh(3,1) 
          + 8 \* \Hhh(-2,-1,0) 
          + 19 \* \Hhh(-2,0,0)
          + 15 \* \Hhh(2,1,0) 
  \nonumber\\&& \mbox{}
          + 15 \* \Hhh(2,1,1) 
          + 38 \* \Hhhh(0,0,0,0) 
          - 6 \* \H(-2) \* \z2 
          - 19 \* \H(2) \* \z2 )
          \biggr)
\:\: .
\eea
\normalsize
Compact parametrizaztions of Eqs.~(B.11) -- (B.18) have been provided
in Ref.~\cite{Moch:2004xu}.
%

\end{document}